\documentclass[a4paper,fleqn]{cas-sc}

\usepackage[authoryear]{natbib}
\usepackage[authoryear,]{natbib}

\usepackage[most]{tcolorbox}
    \tcbuselibrary{theorems}
    \newtcolorbox{answerbox}{sharp corners=all, colframe=black, colback=black!5!white, boxrule=1.5pt, width = 1\textwidth, valign=center}

\definecolor{md_orange}{rgb}{1,0.55,0.1}
\definecolor{md_blue}{rgb}{0.1,0.55,1}

\newtcolorbox{orangebox}[2][]{
    colback=md_orange!5!white,
    colframe=md_orange!75!black,
    colbacktitle=md_orange!10!white,
    coltitle=md_orange!70!black,
    title={#2},fonttitle=\bfseries,#1}

\newtcolorbox{bluebox}[2][]{
    colback=md_blue!5!white,
    colframe=md_blue!75!black,
    colbacktitle=md_blue!10!white,
    coltitle=md_blue!70!black,
    title={#2},fonttitle=\bfseries,#1}

\def\tsc#1{\csdef{#1}{\textsc{\lowercase{#1}}\xspace}}
\tsc{WGM}
\tsc{QE}
\tsc{EP}
\tsc{PMS}
\tsc{BEC}
\tsc{DE}

\newcommand\hubble{\textit{HST}\xspace}
\newcommand\jwst{\textit{JWST}\xspace}
\newcommand\dracoii{Draco~{\sc II}}
\newcommand\msun{$M_\odot$}

\newcommand {\A}[1]{And~{\sc #1}}
\newcommand\hi{H\,{\sc i}}
\newcommand\hii{H\,{\sc ii}}

\makeatletter
\let\jnl@style=\rmfamily
\def\ref@jnl#1{{\jnl@style#1}}%
\newcommand\aj{\ref@jnl{AJ}}%
\newcommand\psj{\ref@jnl{PSJ}}%
\newcommand\araa{\ref@jnl{ARA\&A}}%
\newcommand\apj{\ref@jnl{ApJ}}%
\newcommand\apjl{\ref@jnl{ApJL}}
\newcommand\apjs{\ref@jnl{ApJS}}%
\newcommand\ao{\ref@jnl{ApOpt}}%
\newcommand\apss{\ref@jnl{Ap\&SS}}%
\newcommand\aap{\ref@jnl{A\&A}}%
\newcommand\aapr{\ref@jnl{A\&A~Rv}}%
\newcommand\aaps{\ref@jnl{A\&AS}}%
\newcommand\azh{\ref@jnl{AZh}}%
\newcommand\baas{\ref@jnl{BAAS}}%
\newcommand\icarus{\ref@jnl{Icarus}}%
\newcommand\jaavso{\ref@jnl{JAAVSO}}
\newcommand\jrasc{\ref@jnl{JRASC}}%
\newcommand\memras{\ref@jnl{MmRAS}}%
\newcommand\mnras{\ref@jnl{MNRAS}}%
\newcommand\pra{\ref@jnl{PhRvA}}%
\newcommand\prb{\ref@jnl{PhRvB}}%
\newcommand\prc{\ref@jnl{PhRvC}}%
\newcommand\prd{\ref@jnl{PhRvD}}%
\newcommand\pre{\ref@jnl{PhRvE}}%
\newcommand\prl{\ref@jnl{PhRvL}}%
\newcommand\pasp{\ref@jnl{PASP}}%
\newcommand\pasj{\ref@jnl{PASJ}}%
\newcommand\qjras{\ref@jnl{QJRAS}}%
\newcommand\skytel{\ref@jnl{S\&T}}%
\newcommand\solphys{\ref@jnl{SoPh}}%
\newcommand\sovast{\ref@jnl{Soviet~Ast.}}%
\newcommand\ssr{\ref@jnl{SSRv}}%
\newcommand\zap{\ref@jnl{ZA}}%
\newcommand\nat{\ref@jnl{Nature}}%
\newcommand\iaucirc{\ref@jnl{IAUC}}%
\newcommand\aplett{\ref@jnl{Astrophys.~Lett.}}%
\newcommand\apspr{\ref@jnl{Astrophys.~Space~Phys.~Res.}}%
\newcommand\bain{\ref@jnl{BAN}}%
\newcommand\fcp{\ref@jnl{FCPh}}%
\newcommand\gca{\ref@jnl{GeoCoA}}%
\newcommand\grl{\ref@jnl{Geophys.~Res.~Lett.}}%
\newcommand\jcp{\ref@jnl{JChPh}}%
\newcommand\jgr{\ref@jnl{J.~Geophys.~Res.}}%
\newcommand\jqsrt{\ref@jnl{JQSRT}}%
\newcommand\memsai{\ref@jnl{MmSAI}}%
\newcommand\nphysa{\ref@jnl{NuPhA}}%
\newcommand\physrep{\ref@jnl{PhR}}%
\newcommand\physscr{\ref@jnl{PhyS}}%
\newcommand\planss{\ref@jnl{Planet.~Space~Sci.}}%
\newcommand\procspie{\ref@jnl{Proc.~SPIE}}%

\newcommand\actaa{\ref@jnl{AcA}}%
\newcommand\caa{\ref@jnl{ChA\&A}}%
\newcommand\cjaa{\ref@jnl{ChJA\&A}}%
\newcommand\jcap{\ref@jnl{JCAP}}%
\newcommand\na{\ref@jnl{NewA}}%
\newcommand\nar{\ref@jnl{NewAR}}%
\newcommand\pasa{\ref@jnl{PASA}}%
\newcommand\rmxaa{\ref@jnl{RMxAA}}%

\newcommand\maps{\ref@jnl{M\&PS}}%
\newcommand\aas{\ref@jnl{AAS Meeting Abstracts}}%
\newcommand\dps{\ref@jnl{AAS/DPS Meeting Abstracts}}%

\makeatother

\usepackage{amssymb}
\usepackage{amsmath}
\usepackage{xspace}

\usepackage{rotating}

\usepackage{threeparttable}
\usepackage{booktabs}
\usepackage{longtable}
\usepackage{array}
\usepackage{caption}
\usepackage{pdflscape}

\ExplSyntaxOn
\cs_set:Npn \__first_footerline:
{
  \group_begin:
  \small
  \sffamily
  \ifnum\theblind>0\relax
  \else
    \__short_authors: :~
  \fi
  { \rmfamily \itshape Accepted~ for~ publication~ in~ Physics~ Reports }
  \group_end:
}
\ExplSyntaxOff

\begin{document}
\let\WriteBookmarks\relax

\shorttitle{Resolved Stellar Populations}

\shortauthors{Weisz et~al.}

\title [mode = title]{The Stellar Populations of Resolved Galaxies}

\author[1]{Daniel R.\ Weisz}[orcid=0000-0002-6442-6030]

\cormark[1]

\ead{dan.weisz@berkeley.edu}

\ead[url]{https://dweisz.github.io}

\credit{Daniel Weisz: Conceptualization, Writing -- original draft, Writing -- review \& editing, Visualization; Alessandro Savino: Conceptualization, Writing -- original draft, Writing -- review \& editing, Visualization; Evan Skillman: Conceptualization, Writing -- original draft, Writing -- review \& editing}

\affiliation[1]{organization={Department of Astronomy, University of California},
    addressline={501 Campbell Hall \#3411},
    city={Berkeley},
    postcode={94720},
     state={CA},
    country={USA}}

\author[1]{Alessandro Savino}[orcid=0000-0002-1445-4877]

\author[2]{Evan D. Skillman}[orcid=0000-0003-0605-8732
   ]

\affiliation[2]{organization={University of Minnesota, Minnesota Institute for Astrophysics, School of Physics and Astronomy},
     addressline={116 Church Street, S.E.},
    city={Minneapolis},
    postcode={55455},
    state={MN},
    country={USA}}

\cortext[1]{Corresponding author}

\begin{abstract}
The study of galaxies outside the Milky Way using observations of resolved stellar populations has experienced extraordinary growth over the past two decades.  There now exist panchromatic photometry and precise astrometry for hundreds of millions of individually resolved stars in hundreds of galaxies that span $\sim10^9$ in luminosity, $\sim10^3$ in metallicity, and sample a wide range of environments.  Here, we review advances in extragalactic astronomy enabled by resolved stellar population imaging. We summarize major observational programs and focus on four interconnected science areas: the star formation histories of galaxies, `bursty' star formation, the stellar initial mass function, and the orbital histories of Local Group galaxies.  We review the literature in each area, synthesize key findings, identify emerging trends, and discuss how new and future facilities (e.g., \textit{JWST}, LSST, \textit{Euclid}, \textit{Roman}, \textit{CSST}, ELTs, \textit{HWO}) will enable further progress.  Finally, we highlight shortcomings in current datasets and discuss promising directions for new areas of investigation.
\end{abstract}


\maketitle
\tableofcontents

\section{Introduction}
\label{sec:intro}

\subsection{The Growth of Resolved Stellar Population Studies of Galaxies}
\label{sec:history}

While most extragalactic studies are based on observations of integrated light, galaxies in the local Universe can be resolved into individual stars, enabling exquisitely detailed studies of their stellar populations. Spectroscopic and photometric measurements of resolved stars facilitate the study of entire galaxies on a star-by-star basis.  The luminosities, colors, velocities, and chemical compositions of resolved stars in nearby galaxies have produced unique, diverse, and foundational science in areas including galaxy evolution and chemical enrichment, star formation and stellar physics, dark matter, dark energy, and cosmology \citep[as illustrated by papers such as][]{Leavitt:1912, Hubble:1925, Hubble:1926, Baade:1944, Hodge:1971, Sandage:1974, deVaucouleurs:1978, Aaronson:1983, Hodge:1989, Madore:1991, Olszewski:1996, Mateo:1998, Mould:2000, Freedman:2001, Grebel:2003, Massey:2003, Willman:2005a, Belokurov:2006a, Simon:2007, Tolstoy:2009, Bastian:2010, Kirby:2011, Dalcanton:2012b, Kennicutt:2012, McConnachie:2012, Kroupa:2013, Weisz:2014a, Frebel:2015, Gallart:2015, Bullock:2017, Freedman:2019, Krumholz:2019, Simon:2019, Freedman:2020, Sales:2022, Riess:2022, Hennebelle:2024, Verde:2024, Savino:2025, Kroupa:2026}.

The earliest observations of extragalactic resolved stars on record focused on very luminous stars such as supergiants and Cepheids in Local Group (LG) galaxies \citep[e.g.,][]{Dunlop:1828, Herschel:1847, Herschel:1864, Leavitt:1912, Hubble:1925}.  Increasingly sensitive and powerful telescopes revealed fainter stars of varying spectral types and provided access to individual stars in more distant galaxies.  Though bright variable stars in LG galaxies remained common targets throughout the early to mid 20th century \citep[e.g.,][]{Hubble:1926, Baade:1939, Baade:1963, Payne-Gaposchkin:1966, Baade:1965, Swope:1968, Sandage:1971}, by the 1960s, several studies of Milky Way (MW) satellite galaxies and more distant dwarf irregular (dIrr) galaxies were able to construct color-magnitude diagrams (CMDs) that included main sequence (MS), red giant branch (RGB), asymptotic giant branch (AGB), and horizontal branch (HB) stars, enabling the study of the ages and chemical compositions of their stellar populations \citep[e.g.,][]{Baade:1961, Hodge:1965, Swope:1967, Kayser:1967, vanAgt:1967}.   Subsequent decades of observations resulted in deep CMDs in MW satellites and an increased number of stars with spectroscopically measured properties (see \citealt{Hodge:1989, Mateo:1998} and references therein).  These types of datasets were central to a number of ground-breaking discoveries, including showing that MW dwarf galaxies had different stellar populations from MW globular clusters (GCs) \citep[e.g.,][]{Aaronson:1980, Aaronson:1983, Mould:1983b}. A handful of systems also had their bulk proper motions measured from photographic plates observed many decades apart \citep[e.g.,][]{Meatheringham:1988, Jones:1994, Schweitzer:1995, Kroupa:1997, Irwin:1999}.

Starting in the 1970s, several technological developments revolutionized the study of resolved stars in external galaxies.  The first was the widespread adoption of charge-coupled devices (CCDs; e.g., \citealt{Boyle:1970, Young:1978, Ford:1979, Westphal:1980}).  These digitized detectors enabled more precise measurements of stellar positions and fluxes compared to analog devices (e.g., photographic plates; e.g., \citealt{King:1983, Walker:1984, McClure:1985, Smith:1986}).  Precise measurements from CCDs made it possible to study fainter stars in increasingly ``crowded'' regions (e.g., dense star clusters, more distant galaxies; \citealt{Tody:1980, Mould:1983a, Mould:1983b, Irwin:1985, Mateo:1985, Tody:1986, Mould:1986, Stetson:1987, Schechter:1993}).%

A second major advance was the launch of the Hubble Space Telescope (\hubble).  The excellent stability, sensitivity, and high-angular resolution of \hubble\ made it possible to routinely detect stars at fainter flux levels, in more crowded environments, with higher astrometric precision, and to greater distances than ever before.  The power of \hubble\ for resolved star studies was demonstrated very early in its lifetime through the construction of CMDs of unprecedented depth and quality for galaxies in and around the LG \citep[e.g.,][]{Elson:1991, Holtzman:1995, Hunter:1995, Tolstoy:1998, Dohm-Palmer:1998, Gallart:1999}. The installation of the Advanced Camera for Surveys (ACS; \citealt{Ford:1998}) during Servicing Mission 3B in 2002 was particularly revolutionary as it provided a significant upgrade to \hubble{}'s imaging capabilities.  More than 35 years after \hubble's launch, resolved stellar populations continue to be at the forefront of \hubble-based science.

A third major pillar in the growth of resolved star studies outside the MW was the ground-based astronomy revolution.  In the late 1990s and early 2000s, increasingly wide-area digital sky surveys covered large areas of the sky to unprecedented depths, transforming our knowledge of the MW and LG through observations of resolved stars \citep[e.g.,][]{Ibata:1994, Ibata:2001, Ferguson:2002, Willman:2005a,  Willman:2005b, Belokurov:2006a}.  At the same time, the spectroscopic capabilities of 10m class facilities (e.g., Keck, the Very Large Telescope [VLT]; \citealt{McCarthy:1995, Cote:1999b, Cote:1999a, Venn:1999, Guhathakurta:2000, Reitzel:2002, Reitzel:2004}) enabled the measurement of velocities and chemical abundances throughout the LG and, for the brightest stars, out to several Mpc \citep[e.g.,][]{Bresolin:2001}.

Of course, some of the greatest gains in resolved star science have been driven by hardware advances.  Larger diameter telescopes, adaptive optics, and the stability of telescopes in space are examples of hardware improvements that have opened new avenues for resolved star studies in external galaxies.  The projected sensitivity, stability, and angular resolution of new and upcoming ground-based facilities, such as the Rubin Observatory and the Extremely Large Telescopes (ELTs), and current and future space missions (e.g., \textit{Euclid}, \textit{Roman}, the Chinese Space Station Survey Telescope [\textit{CSST}], and the Habitable Worlds Observatory [\textit{HWO}]), promise to be equally transformational for resolved stellar populations studies \citep[e.g.,][]{Davies:2018, Ivezic:2019, Anand:2025b, Hunt:2025, Smercina:2025, CSST:2026, Sanderson:2026}.  Similarly, the increase in computational power over the past several decades has made it possible to perform complex calculations and develop sophisticated models in ways that were previously unimaginable.  The growth of computational power has touched virtually every aspect of resolved star science from the availability of large, searchable, curated datasets to a more sophisticated understanding of noise properties to the generation of complex models of stellar evolution that underpin most scientific interpretation in resolved galaxy studies.  The continued growth of hardware and computational power is central to future resolved star studies.

\subsection{Observational Challenges}
\label{sec:challenges}

\begin{figure}[pos=tbp]
    \centering
    \includegraphics[width=1.\linewidth]{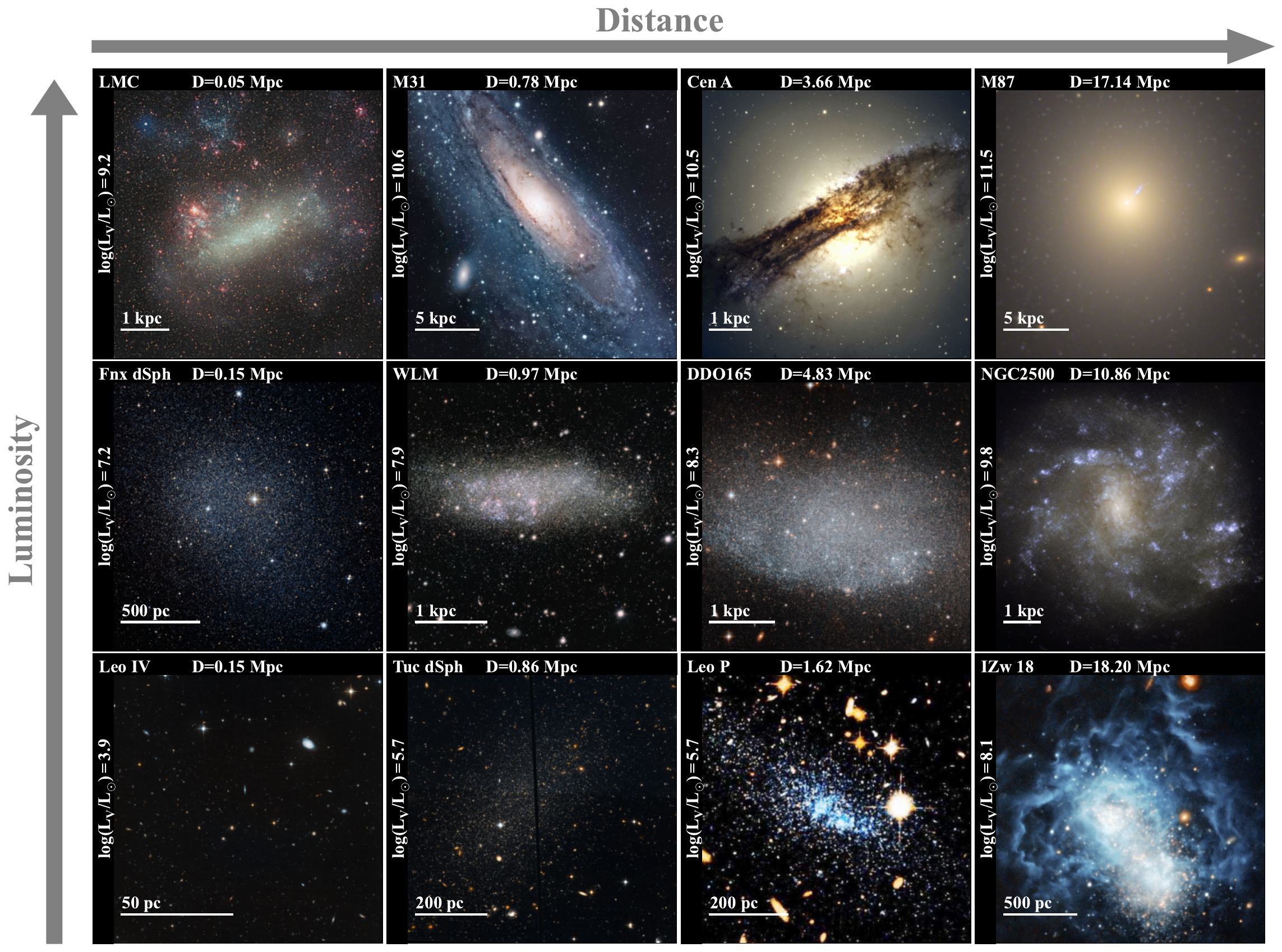}
    \caption{An illustrative gallery of galaxies that have been resolved into individual stars. The galaxies are ordered approximately by luminosity (y-axis) and distance (x-axis), and span a wide-range of morphologies. Image credits: NASA, ESA, ESO, DSS2, A. Aloisi, A. Block, T. Brown, R. Gendler, E. Slawik. \label{fig:Mosaic_Images}}

\end{figure}

Figure~\ref{fig:Mosaic_Images} shows a gallery of images for galaxies that have been resolved into individual stars.  Organized by luminosity and distance, these images illustrate the various types of galaxies that have been (and can be) resolved by modern facilities.  They range from ultra-faint dwarf galaxies (UFDs), that are so faint they have historically only been detected and studied near the MW, to massive disk and elliptical galaxies at $D\gtrsim20$~Mpc, for which bright, evolved stars have only been recently resolved with the James Webb Space Telescope (\jwst).

\begin{figure}[pos=tbp]
    \centering
    \includegraphics[width=1.\linewidth]{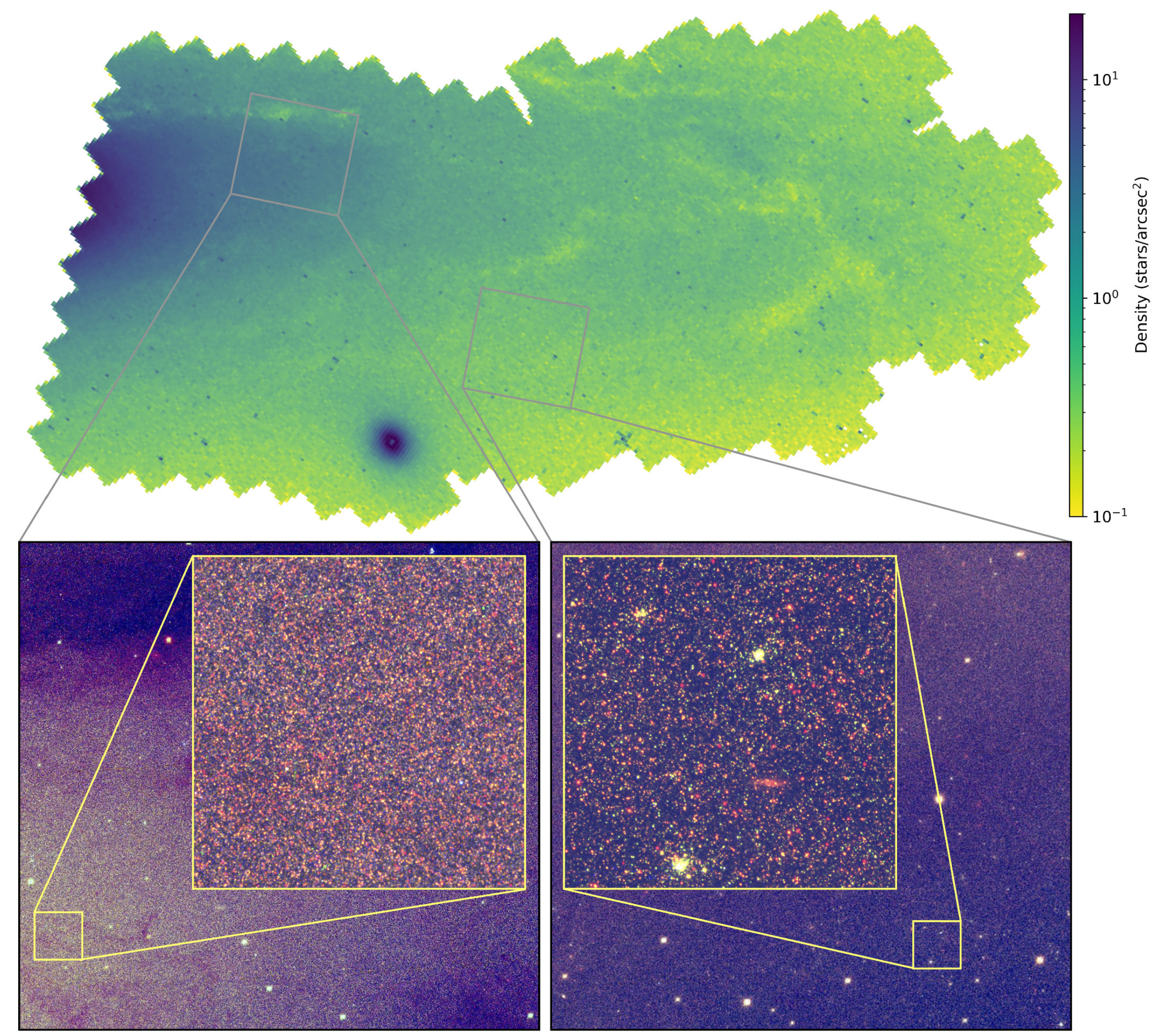}
    \caption{An illustration of resolved stars in M31 from the \hubble\ PHAST survey, adapted from \citet{Chen:2025}. \textbf{Top:} A density map of stars detected over the magnitude range $21.5<m_{\rm F814W}<23$ color-coded by stellar density in $10^{\prime\prime}\times10^{\prime\prime}$ bins. \textbf{Bottom:} A three color-composite (F336W, blue; F475W, green; F814W, red) showing a closer look at M31's disk and the rich detail revealed by observations of resolved stars.  The two $0.5^{\prime}~\times~0.5^{\prime}$ insets show regions of high stellar density (left; $\sim2$ stars arcsec$^{-2}$) and moderate stellar density (right; $\sim0.3$ stars arcsec$^{-2}$).  The ability to accurately measure stellar fluxes and positions in crowded regions is foundational to extragalactic resolved stellar populations science.}
 \label{fig:m31_disk_density}
\end{figure}

\begin{figure}[pos=tbp]
    \centering
    \includegraphics[width=1.\linewidth]{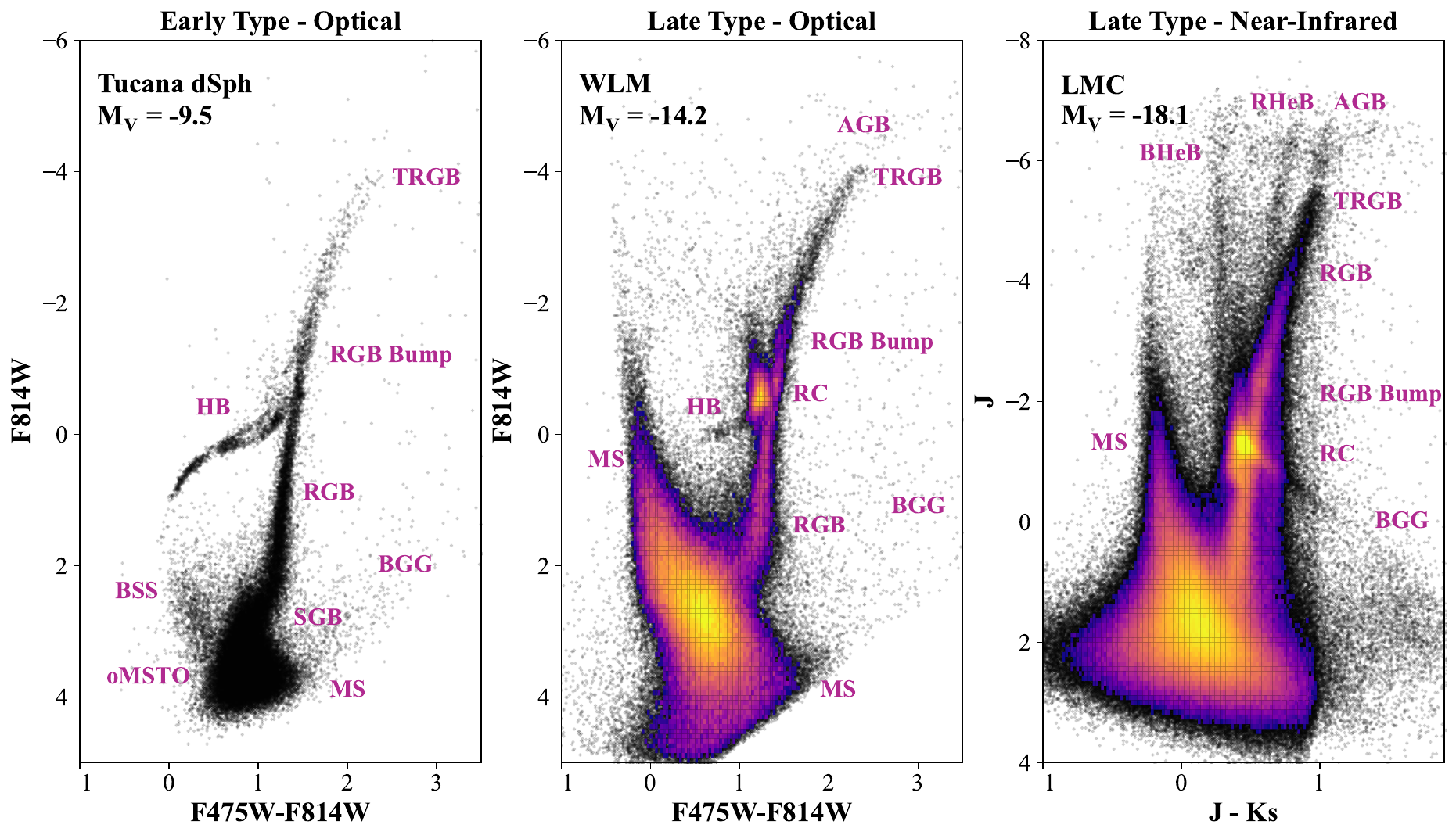}
    \caption{Example optical and near-IR color-magnitude diagrams (CMDs) of early and late-type galaxies that highlight various major stellar evolutionary sequences, including: main sequence (MS), red giant branch (RGB), asymptotic giant branch (AGB), horizontal branch (HB), red clump (RC), blue and red massive helium burners (BHeB and RHeB), tip-of-the-RGB (TRGB), and blue straggler stars (BSSs).  See \S \ref{sec:stellar_evolution} for a brief description of each evolutionary phase. Background galaxies (BGG) that are consistent with stellar photometric culling criteria are also labeled. Data originally from \citet{Monelli:2010_b}, \citet{Albers:2019}, and VISTA Magellanic Cloud Survey \citep{Cioni:2011}.}
    \label{fig:cmd_pedagogy}
\end{figure}

There have been two central, persistent challenges for resolved star studies: crowding and contamination (e.g., from MW foreground stars and background galaxies).  Crowding refers to the challenge of accurately and self-consistently measuring the positions and fluxes of individual stars in regions of high surface brightness or high stellar density, i.e., where the point spread functions (PSFs) of neighboring stars overlap.\footnote{While there are analogous technical challenges, and solutions, for resolved star spectroscopy, we do not summarize them here as the focus of this review is on imaging, as described in \S \ref{sec:scope}.} To illustrate this challenge, Figure~\ref{fig:m31_disk_density} provides examples of varying densities of resolved stars in the disk of M31 as seen by \hubble\ \citep{Chen:2025}.  In regions of low stellar density, it is straightforward to accurately determine the precise location and total flux of each star.  However, in regions of high stellar density, in which the PSFs of neighboring stars substantially overlap (i.e., the stars are crowded), measuring the position and flux of a single star requires modeling the positions and fluxes of all its neighboring stars at the same time, many of which are separated by $\lesssim1$~pixel on the detector.  This technically and computationally challenging problem can be further compounded by observational and instrumental effects such as variable seeing, under-sampled PSFs, imperfect knowledge of the PSF at all points on the focal plane, geometric distortions, detector effects (e.g., persistence), and the astrometric (mis-)alignment of images.  The net effect of crowding is to (i) limit the depth to which resolved stars can be detected; (ii) introduce non-Poissonian noise into the flux measurement of a star (e.g., the flux of a star can be correlated with spatially neighboring stars); and (iii) introduce complicated selection effects that can depend on position, stellar type, and properties of an instrument. For example, near the center of M31, \hubble\ can detect only the brightest stars (e.g., upper MS, RGB, AGB; see Figure~\ref{fig:cmd_pedagogy} for examples of these stellar sequences and \S \ref{sec:stellar_evolution} for a description) while crowding prevents fainter stars from being detected \citep[e.g.,][]{Dalcanton:2012b, Rosenfield:2012, Dong:2018, Chen:2025}. In the low density outskirts of M31, \hubble\ can reach the oldest main-sequence turnoff (oMSTO) before a combination of crowding and faintness of stars significantly affects the data \citep[e.g.,][]{Brown:2003}. Because crowding depends on both surface brightness and angular resolution, not just on galaxy distance, no single facility resolves stars across the full extent of a typical galaxy. This has motivated the multi-facility observing strategies we describe in \S\ref{sec:future}.  The effects of crowding, including covariant fluxes, are typically quantified by injecting synthetic stars into real images and analyzing the differences between their intrinsic and recovered properties (i.e., artificial star tests, ASTs; \citealt{Stetson:1988}).

Driven by the foundational importance of resolving stars in crowded-fields, there has been considerable technical development of crowded-field stellar photometry and astrometry software over the past $\sim40$ years.  It is now a mature field and several software packages exist that can routinely provide photometry and/or astrometry of stars in crowded (and sparse) fields for a wide variety of telescopes, astronomical scenes, and science goals.  For example, the centroids of stars can be measured to precisions of $\ll1$~pixel on an \hubble\ camera \citep[e.g.,][]{Anderson:2000}. Other software routines are capable of high-precision multi-wavelength photometry that can use positions and fluxes from deeper, higher-angular resolution images (e.g., in the optical) to extract improved (i.e., deeper, more precise) photometry at other wavelengths (e.g., in the near-infrared [near-IR]) from images obtained with different cameras \citep[e.g.,][]{Williams:2014}.  High-precision astrometry and photometry generally require different technical approaches to PSF modeling as astrometry focuses on precisely modeling the ``core'' in the central region of the PSF \citep[e.g.,][]{Anderson:2000}, at the expense of the total flux, while photometry routines model the entire PSF to ensure accurate fluxes, at the expense of an extremely precise stellar centroid \citep[e.g.,][]{Stetson:1987, Schechter:1993, Lauer:1999, Dolphin:2000}.

\begin{figure}[pos=tbp]
    \centering
    \includegraphics[width=0.95\linewidth]{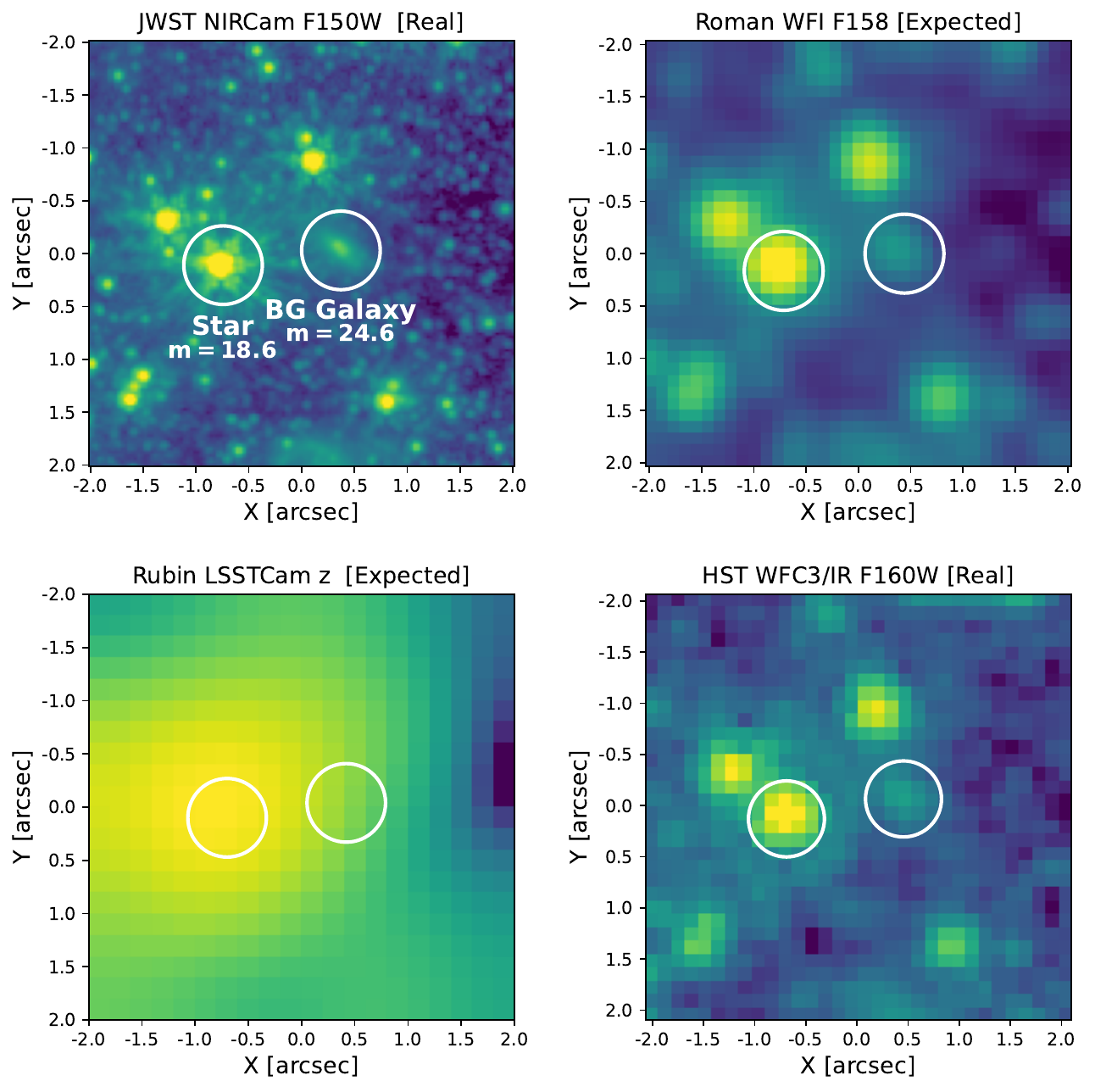}
    \caption{A pedagogical example of the challenge in distinguishing stars from galaxies.  \textbf{Top Left:} A zoomed-in region of WLM from a \jwst/NIRCam F150W image with a bright star and a faint background (BG) galaxy indicated in the white circles. \textbf{Top Right:}  A simulated \textit{Roman} image in the F158 band of the same scene in WLM with the star and BG galaxy indicated. \textbf{Bottom Left:}  A simulated Rubin $z$-band image of the same scene in WLM with the star and BG galaxy indicated. \textbf{Bottom Right:}  The same zoomed-in region of WLM as imaged by \hubble/Wide Field Camera 3 (WFC3) in the F160W band.  This \hubble\ image has a comparable integration time to the \jwst\ image in the top left panel and a comparable angular resolution to \textit{Roman} shown in the top right. Although the star and BG galaxy classification is visually obvious in the \jwst\ image, it is not as straightforward to make this differentiation even in the \hubble\ image.  While \textit{Roman} will clearly provide improvement over Rubin, for example, it may still face challenges at fainter magnitudes.}
    \label{fig:sg_separation}
\end{figure}

The second challenge is differentiating between \textit{bona fide} member stars of a nearby galaxy and contaminants that have a similar appearance such as MW foreground stars, compact background galaxies, and photometric artifacts (e.g., \hii{} regions in extended background galaxies, spurious sources detected on diffraction spikes). Figure~\ref{fig:sg_separation} illustrates the challenge for a star and a faint background galaxy in WLM for various facilities.  One long-standing aspect of contamination is the photometric classification of stars and galaxies, i.e., star-galaxy separation, which dates back decades to when observational studies of the faint, distant galaxies were still in their infancy and faint MW stars were the main contaminant of galaxy samples \citep[e.g.,][]{MacGillivray:1976, Kron:1978, Sebok:1979, Kron:1980, Jarvis:1981}.  Over the years, a variety of approaches have been developed or refined that use some combination of apparent angular sizes (e.g., relative to a PSF), colors, curves of growth, and/or morphologies \citep[e.g.,][]{Maddox:1990, Yee:1991, Odewahn:1992, Weir:1995, Bertin:1996}.  The importance of star-galaxy separation was underscored by the first generations of wide-area sky surveys (e.g., the Sloan Digital Sky Survey [SDSS]) which found that the increasingly faint flux limits and larger sky coverage led to more confusion between stars and galaxies \citep[e.g.,][]{Lupton:2001, Strauss:2002, Abazajian:2004}.  It, along with newer facilities and surveys, served as the basis for identifying large, statistical samples of known stars, galaxies and quasars, which have been used to train increasingly sophisticated techniques for classification \citep[e.g.,][]{Soumagnac:2015, Kim:2017, Tachibana:2018, Bailer-Jones:2019, Slater:2020, Zeraatgari:2024}.

The ability to identify and remove, or even statistically model, these types of contaminants has long been recognized as an issue in resolved star studies, particularly in galaxies or regions of galaxies with low stellar densities and/or at very faint fluxes (e.g., stellar halos, ultra-faint galaxies; \citealt{Willman:2005a, Koposov:2008, Walsh:2009, Radburn-Smith:2011, Fadely:2012, Bechtol:2015, Drlica-Wagner:2015, Smercina:2017, Gennaro:2018b, Simon:2019, Smercina:2020, Bell_e:2022, Doliva-Dolinsky:2022, Savino:2023, Cohen:2026}).  These contaminants pose a range of challenges from erroneous identification of new ultra-faint galaxies to inaccurate characterization of nearby galaxy stellar populations to a reduction in the photometric depth to which a given survey is sensitive.  While there has been marked progress, particularly in the application of machine learning, in the ability to separate stars from galaxies based on imaging \citep[e.g.,][]{Gatto:2026}, present-day solutions are not yet as mature as those for crowded-field photometry.  In practice, MW foreground stars are the dominant contaminant for nearby low-surface-brightness systems such as UFDs, while unresolved background galaxies dominate at the faint magnitudes required for the low-mass initial mass function (IMF) and for halos throughout the Local Volume (LV; i.e., $D\lesssim10$~Mpc). As we discuss in \S \ref{sec:future}, an improved ability to differentiate between stars and background galaxies will be increasingly important for resolved galaxy studies, particularly for faint and/or low-surface brightness systems beyond the LG.

\subsection{Resolved Stars in the Nearby Universe}
\label{sec:nearby}

Well-measured photometry and astrometry of individual stars in nearby galaxies enable a wide variety of science.  For example, the motions of stars in the plane of the sky, relative to static sources (e.g., quasars or background galaxies) enable the measurement of a galaxy's proper motion, which is necessary for determining its orbital history.  Similarly, the measurement of stellar fluxes in two or more filters in a galaxy allows for the construction of CMDs, which depict stars in various phases of stellar evolution (i.e., they are an observer's Hertzsprung-Russell Diagram; \citealt{Russell:1914}).  CMDs are the foundation upon which most resolved galaxy science is built (e.g., measurements of galaxy star formation histories, the IMF, stellar physics, the distance scale).

\begin{figure}[pos=tbp]
    \centering
    \includegraphics[width=1.\linewidth]{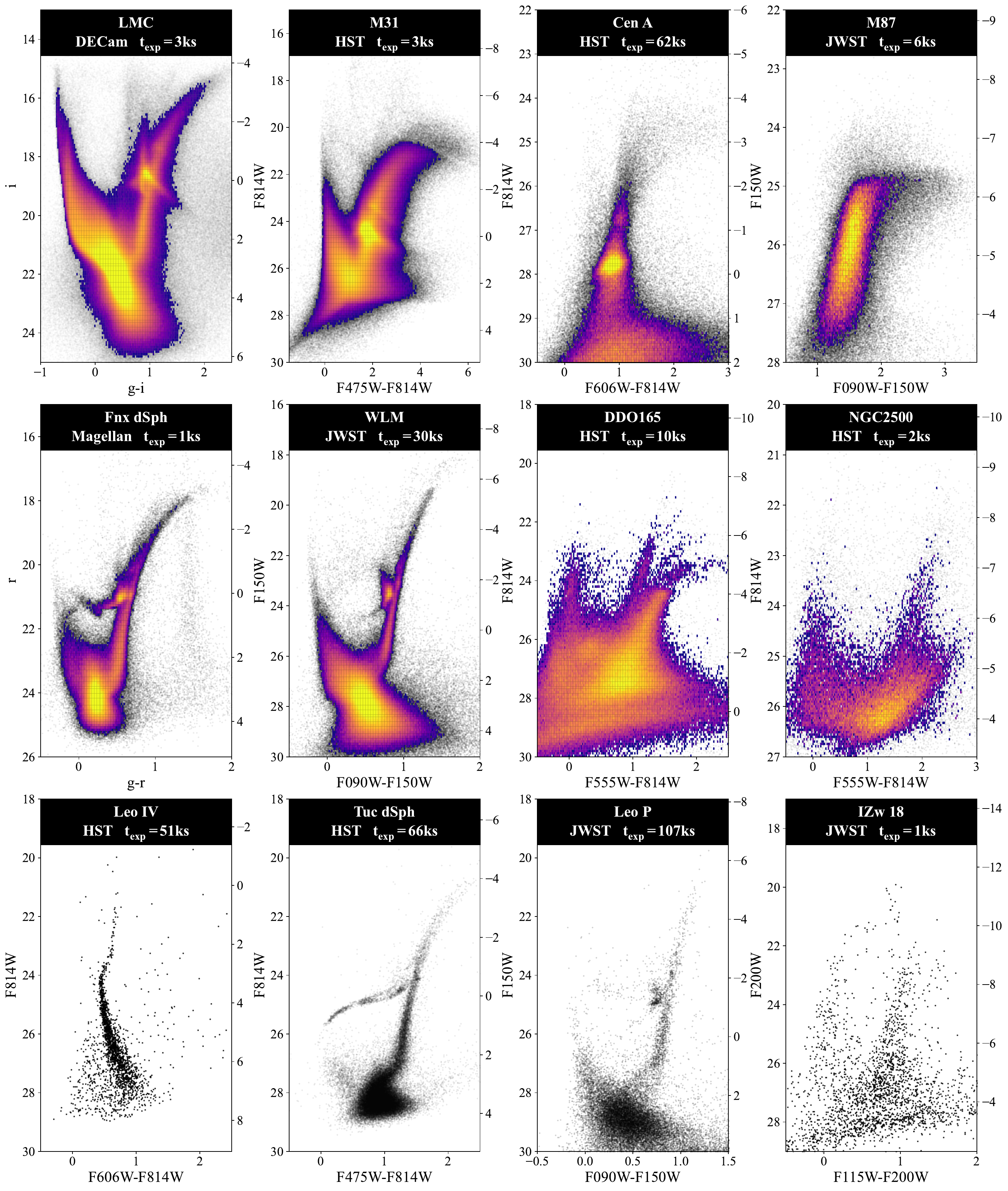}
    \caption{Some of the deepest available CMDs for each of the galaxies shown in Figure~\ref{fig:Mosaic_Images}.  CMDs with large numbers of stars are shown as density plots, i.e., Hess diagrams.  The label in each panel indicates the facility that acquired the data along with the total exposure time.  The y-axes reflect both apparent and absolute magnitude for the reddest filter. Data from \citet{Nidever:2017}, \citet{Williams:2014}, \citet{Rejkuba:2005}, \citet{Anand:2025}, \citet{Munoz:2018}, \citet{Weisz:2024}, \citet{Weisz:2008}, \citet{Calzetti:2015}, \citet{Durbin:2025}, \citet{Monelli:2010_b}, \citet{McQuinn:2024b}, and \citet{Hirschauer:2024}. }
    \label{fig:cmd_mosaic}
\end{figure}

Figure~\ref{fig:cmd_pedagogy} shows three example CMDs from LG galaxies with select phases of stellar evolution indicated. Comparisons of a galaxy's CMD with predictions from theoretical stellar isochrones provide insight into a galaxy's formation and enrichment history as well as constraints on stellar physics and star formation processes (see \citealt{Cignoni:2010}, for example). The positions of member stars relative to background or foreground objects provide a measure of motion in the plane of the sky. Spectroscopic observations of these resolved stars provide their radial velocities, which can be used to measure their host galaxy's mass and mass profile, and chemical abundance of stars, which trace a galaxy's enrichment and formation history \citep[e.g.,][] {Venn:1999, Guhathakurta:2000, Shetrone:2003, Venn:2004, Gilbert:2006, Kalirai:2006, Simon:2007, Walker:2007, Carrera:2008, Tolstoy:2009, Geha:2009, Frebel:2010, Letarte:2010, Starkenburg:2010, Kirby:2011, Tollerud:2012, Collins:2013, Ibata:2013, Ji:2016, Li_t:2017, Hill:2019, Escala:2020, Dey:2023, Walker:2023, Shetrone:2025, Geha:2026}.

Figure~\ref{fig:cmd_mosaic} shows CMDs for the galaxy images displayed in Figure~\ref{fig:Mosaic_Images}.  These CMDs illustrate the type of data (e.g., depth, quality, wavelengths) available from contemporary facilities and highlight the characteristic types of stellar populations for galaxies of different types and at a variety of distances.   A comparison of this gallery with Figure~\ref{fig:cmd_pedagogy} and the definitions provided in \S \ref{sec:stellar_evolution} gives a qualitative sense of the types of stellar populations present in each galaxy.  For example, stellar populations older than $\sim1$~Gyr are observed in every CMD through features like the RGB or main sequence turnoff position, whereas only a subset of the galaxies have very recent star formation ($\sim10$-$100$~Myr) which is evidenced by the presence of luminous MS and core helium burning stars.  As discussed in \S \ref{sec:stellar_evolution}, and illustrated in Figure~\ref{fig:cmd_pedagogy}, one of the most important features on a CMD is the oMSTO, which traces the age and metallicity of the first stellar population in a galaxy.  Data that include this feature are considered the `gold standard' for CMD analysis.

\begin{figure}[pos=tbp]
    \centering
    \includegraphics[width=1.0\linewidth]{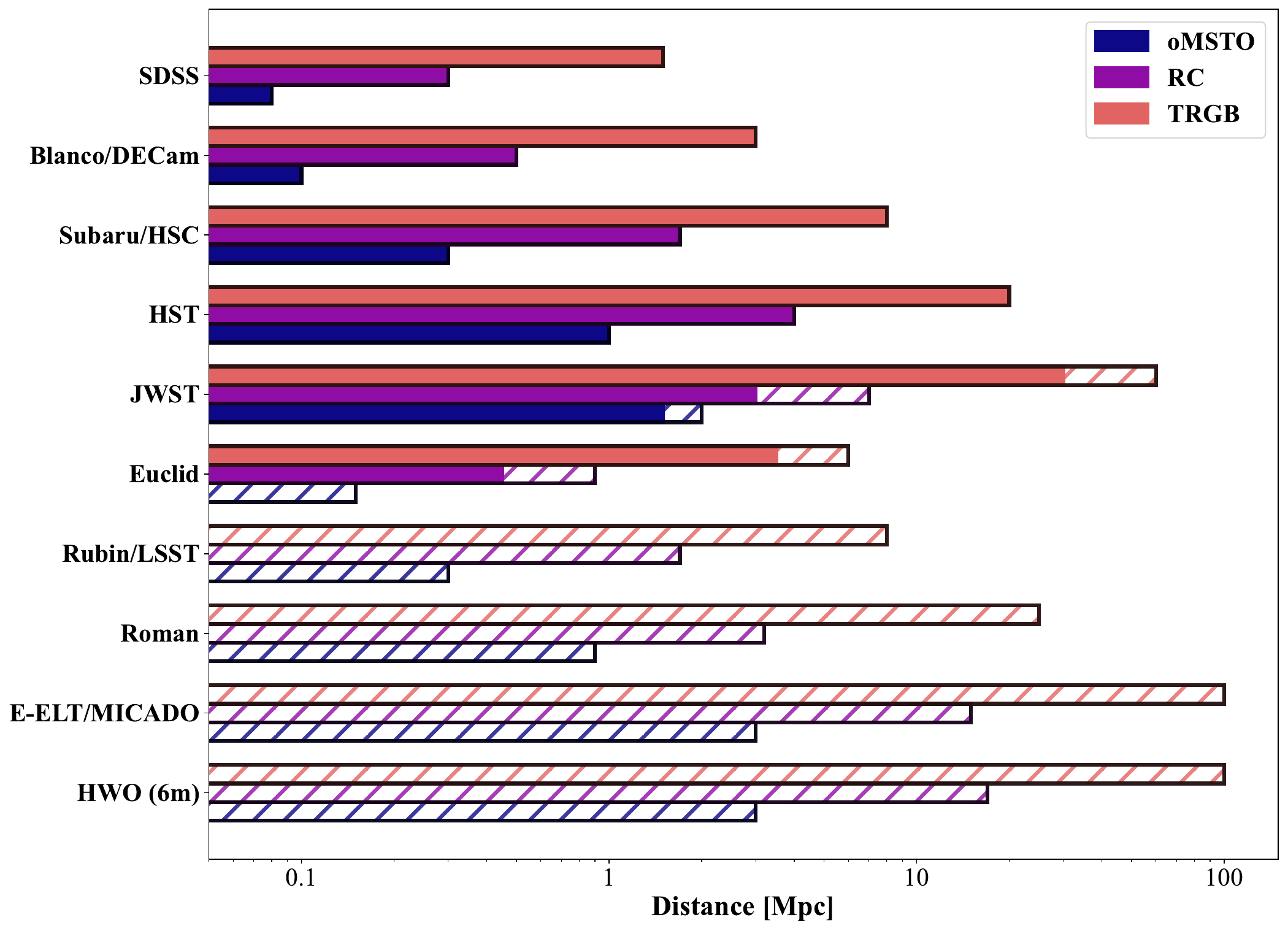}
    \caption{Our current best estimates of the maximum distances to which select telescopes or surveys can detect different, well-populated CMD features, assuming reasonable integration times. The shaded portion of the bar indicates maximum distances reached by existing datasets.  The hatched regions reflect our best estimate at approximate limiting distance given reported literature depths and rough estimations of crowding effects. Our estimates do not account for some details such as dust, filter choice, field of view, etc.  We did not include AGB or JAGB features in the near-IR as their use for stellar populations work requires CMDs that are similar in depth to the TRGB. The \textit{CSST} will have capabilities similar to \hubble\ only over a much wider field.}
    \label{fig:Telescopes}
\end{figure}

To date, there are hundreds of galaxies outside the MW that have at least some portion of their stellar populations resolved into individual stars.  As illustrated in Figure~\ref{fig:Telescopes}, the technological capabilities of even the most powerful observatories to date have limited the majority of resolved galaxy studies to systems within $\sim10$~Mpc, with a small, but growing, number with $D\gtrapprox20$~Mpc \citep[e.g.,][]{Aloisi:2007, Williams:2007, Shen:2021, Lee:2025, Li:2025}.
The hatched areas in Figure~\ref{fig:Telescopes} highlight the potential of current and future facilities for resolving stars in galaxies as a function of their distance.  While it will be impossible to duplicate the data quality available in the LG (e.g., CMDs that reach the oMSTO) at these distances, new approaches to resolved star analyses, long time-baselines for proper motions, and observational access to $\sim10^4$ galaxies in these distance ranges \citep[e.g.,][]{Mutlu:2021, Ohlson:2024, Shread:2026} promise to expand resolved stellar population science in new directions that have been limited by current observational capabilities (e.g., resolved star studies in giant elliptical galaxies and massive galaxy clusters, extensive mapping of stellar halos for a statistically robust galaxy sample).  These same facilities will also enable even more detailed observations of closer systems (e.g., LG galaxies), making routine what has previously been near-impossible or at least come at great observational cost (e.g., resolving the hydrogen burning limit outside the MW, precision proper motions for all LG galaxies, uncovering the stellar populations in M31's central bulge).

The study of resolved stars in nearby galaxies is undergoing an exciting phase of rapid growth.  \textit{Gaia} is providing accurate photometry and high-fidelity proper motions and orbital histories of LG galaxies, which continue to increase in precision with each data release \citep[e.g.,][]{Perryman:2026}.  Over the past decade, sensitive, wide-area ground-based telescopes and instruments (e.g., the Visible and Infrared Survey Telescope for Astronomy [VISTA], the Dark Energy Camera [DECam]) have significantly increased our ability to detect faint and extremely low-surface brightness galaxies, while also providing exquisite new detail for known galaxies.   \jwst\ is rapidly building on the legacy of \hubble\ by extending distance measurements, age characterizations, and insights into star formation to systems that were even too distant or crowded for \hubble.  The recently launched \textit{Euclid} and \textit{Roman} missions, the upcoming launch of \textit{CSST}, the start of the Rubin Observatory Legacy Survey of Space and Time (LSST) in 2026, and new ground-based instruments (e.g., VLT/MAVIS, ELT/MICADO) will provide high-quality, deep, multi-wavelength imaging from space and the ground over a large fraction of the extragalactic sky, including the potential to study millions (and perhaps billions) of newly resolved stars in thousands of nearby galaxies, many of which will be discovered by these new facilities.  It is the dawn of a new era for the study of galaxies with resolved stellar populations.

\subsection{An Illustrative Example of Galaxy Archaeology}
\label{sec:archaeology}

Before turning to the details of each science area, we find it useful to provide a concrete picture of what ``galaxy archaeology'' with resolved stars means in practice.  Every star encodes a record of the conditions under which it formed, i.e., its age, its chemical composition, and, in a statistical sense, the mass function from which it was drawn \citep[e.g.,][]{Weisz:2013}.  The stellar populations of a galaxy comprise a fossil record of its star formation history (SFH), its chemical enrichment, and, through the present-day positions and motions of its stars, its assembly history.  In principle, all of this information is accessible if the stars can be individually resolved.  In practice, how much of, or perhaps how well, a galaxy's formation and assembly history can be reconstructed depends on distance, crowding, and the depth and field of view of the specific facility.

\begin{figure}[pos=tbp]
    \centering
    \includegraphics[width=1.\linewidth]{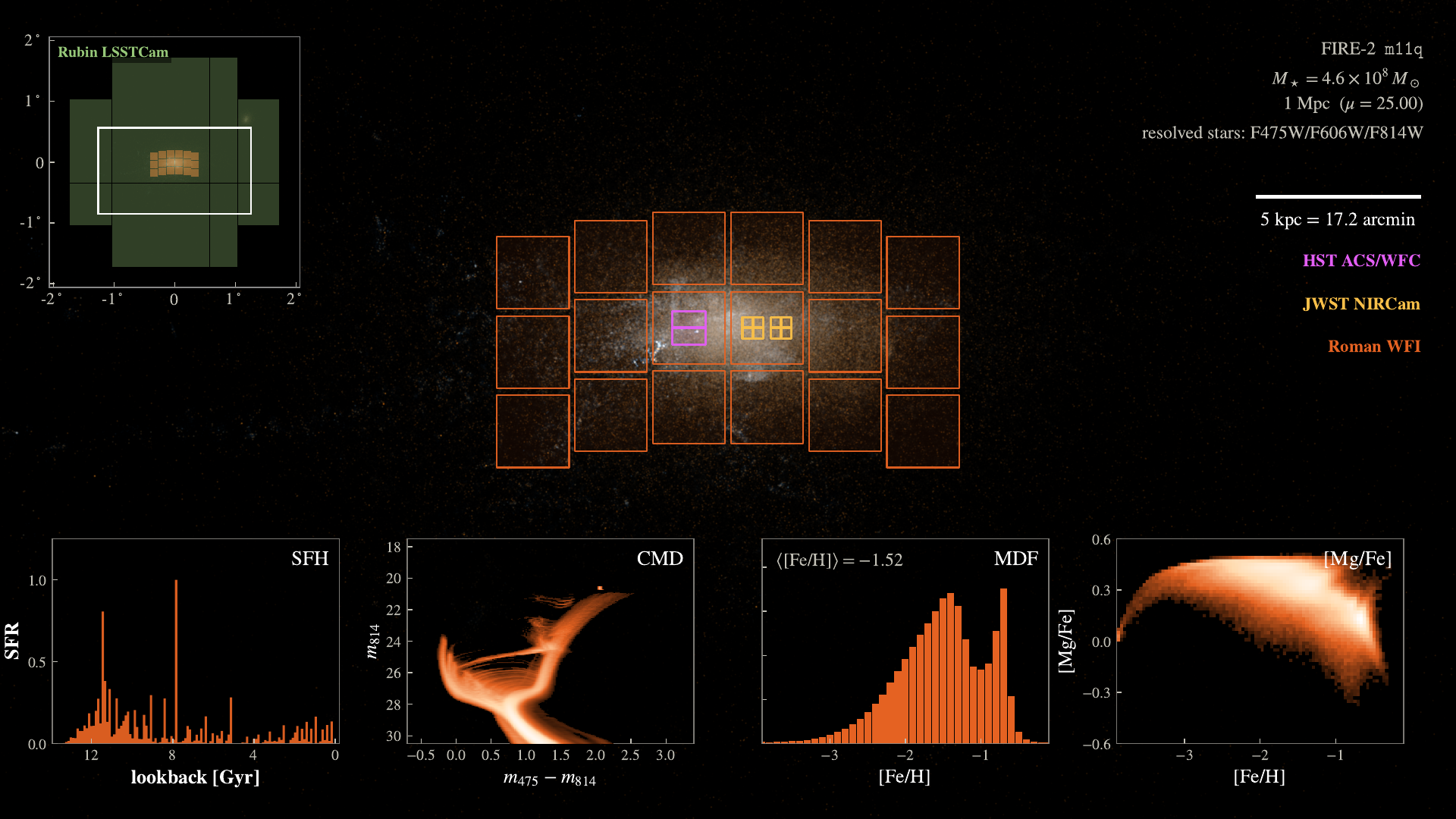}
    \caption{An illustrative example of galaxy archaeology in practice.  The movie version of this figure (\url{https://doi.org/10.5281/zenodo.23627360}) shows the evolution of the main image and each panel over cosmic time.  \textbf{Main Panel:} A rendering of the resolved stars (in \hubble/ACS F475W/F606W/F814W filters) of the FIRE-2 dwarf galaxy \texttt{m11q} \citep[$M_{\star}\sim4.6\times10^{8}$~\msun;][]{Hopkins:2018b, Wetzel:2023}, placed at a distance of 1~Mpc ($\mu=25$).  Individual stars are sampled from each star particle using a Kroupa IMF and MIST isochrones. Overlaid are the footprints of a single pointing of \hubble/ACS/WFC (magenta), \jwst/NIRCam (yellow), and \textit{Roman}/WFI (orange) in illustrative positions.  \textbf{Upper Inset:} The inset shows the Rubin/LSSTCam field of view (green) and \textit{Roman}/WFI (orange) on the same angular scale; the white box marks the extent of the main panel.  A small companion galaxy $\sim25$~kpc from \texttt{m11q} is visible in the upper right of the Rubin footprint but is not within extent of the main panel, highlighting the importance of very wide-field imaging.  \textbf{Lower Insets:} The intrinsic SFH, MDF, and [Mg/Fe]--[Fe/H] distribution of the simulated galaxy, computed directly from its star particles, alongside the CMD that an observer would measure under ideal conditions (e.g., no crowding).  The CMD only reflects photometric scatter and differential extinction for young stars \citep[following][]{Dolphin:2003} and does not include unresolved binaries or incompleteness.  In the accompanying movie, each frame shows the stars older than the given lookback time at the ages they had at that epoch, but at their present-day positions, illustrating how a galaxy's archaeological record is built.}
    \label{fig:fire_m11q}
\end{figure}

Figure~\ref{fig:fire_m11q} illustrates the concept of galaxy archaeology based on resolved stars using a simulated dwarf galaxy, \texttt{m11q}, from the FIRE-2 suite \citep{Hopkins:2018b, Wetzel:2023}. It has a stellar mass of $\sim4.6\times10^{8}$~\msun, making it a few times more massive than LG dIrrs such as NGC~6822 and IC~1613.  Briefly, we describe how this illustrative example is constructed.  First, from the $z=0$ snapshot data, we acquired the properties of star particles (e.g., present-day locations, ages, metallicities/abundances) for \texttt{m11q} from  \href{https://flathub.flatironinstitute.org/fire}{flathub}.  Given the relatively coarse mass resolution in the simulations compared to individual stars (1 star particle is 880 $M_{\odot}$), we next populate each star particle with individual stars drawn from a Kroupa IMF \citep{Kroupa:2001}, assign them magnitudes from MIST isochrones \citep{Dotter:2016, Choi:2016, Dotter:2026}, and render an image of the galaxy as it would appear to \hubble/ACS at a distance of 1~Mpc, i.e., comparable to the isolated dwarfs of the LG.

The bottom panels show the quantities that a resolved star study aims to recover.  The SFH, metallicity distribution function (MDF), and [Mg/Fe]--[Fe/H] distribution are taken directly from the simulation and represent the ``truth,'' whereas the CMD is what an observer would measure.  An accompanying movie (\url{https://doi.org/10.5281/zenodo.23627360}) shows how the same galaxy forms, and the inset panels change, as a function of lookback time.  Spectroscopy is usually required to measure the abundance information shown in the insets.

The figure highlights two recurring points of this review.  First, the CMD encodes a remarkable amount of information. Ancient, metal-poor stars and younger, more metal-rich stars occupy distinct regions of the CMD. The morphology and relative densities of these features are what enable SFH and photometric MDF reconstruction (\S \ref{sec:formation}).  The IMF enters as a fundamental input into every CMD.  In turn, CMDs like this are the primary means by which the IMF can be measured outside the MW, though low-mass IMF measurements require photometry that extends several magnitudes below the oMSTO, i.e., deeper than the CMD shown here (\S \ref{sec:imf}).  The CMD features also track the enhancements and lulls in star formation over time, which motivates the discussion of `bursty' star formation.

Second, the footprints overlaid on the image show how differently current and future facilities sample a nearby galaxy.  A single \textit{Roman} Wide Field Instrument (WFI) pointing covers nearly the entire galaxy, whereas a single \hubble/ACS or \jwst\ Near Infrared Camera (NIRCam) pointing covers only a small fraction of it.  Even wider field imaging (e.g., Rubin) can identify sub-structures (e.g., halos, streams, satellites), though often at reduced depth compared to space-based imaging due to seeing, crowding, and/or sensitivity limitations.  Targeted follow-up is often needed to fully characterize what is discovered in wide-area ground-based imaging.  Overall, this example highlights the interplay between depth, angular resolution, and field of view that we return to in \S \ref{sec:future}.

\subsection{Scope and Organization}
\label{sec:scope}

Since the general concept of a stellar population was first introduced by \citet{Baade:1944}, there have been numerous excellent review articles covering the science and technical aspects of resolved stellar populations in galaxies outside the MW \citep[e.g.,][]{Hodge:1971, VandenBergh:1975, Hodge:1989, Olszewski:1996, Mateo:1998, Grebel:2003, Massey:2003, Gallart:2005, Cignoni:2010, Tolstoy:2009, Bastian:2010, McConnachie:2012, Willman:2012, Frebel:2015, Simon:2019, Annibali:2022, Belokurov:2022, Pace:2024}.

Much of the progress from the past $\sim20$~years was due to the foundational work enabled by the launch of \hubble\ in 1990, and particularly the period from $\sim1995$--$2005$ driven by the Wide Field and Planetary Camera 2 (WFPC2). For the first time, photometry of galaxies beyond the immediate halo of the MW reached a depth and fidelity at which quantitative comparison with stellar evolution models was both possible and informative \citep[e.g.,][]{Holtzman:1995, Tolstoy:1998, Dohm-Palmer:1998, Gallart:1999}.  At the same time, the analysis techniques used to make those comparisons were in rapid flux, often heterogeneous between groups, and not always straightforward to reproduce \citep[e.g.,][]{Skillman:2002}.  The maturity of those methods, the accumulation of much deeper and wider archival data sets, and the broad homogenization of CMD modeling that followed over the subsequent two decades define the body of work we summarize here.

This review focuses on areas of extragalactic resolved star science that have experienced remarkable growth over the past $\sim20$~years but have not been the primary focus of other reviews during this period.  These criteria alone yield a tremendous number of topics.  To further refine the scope, we have made several judicious choices.  First, we limit our scope to science based on imaging rather than spectroscopy.  Spectroscopy plays a central role in areas such as kinematics and dark matter constraints, as well as stellar abundances and chemical evolution.  Spectroscopy is a wide-ranging topic that is clearly worthy of its own contemporary review article.  For the purposes of this review, we refer to results from spectroscopic studies, but do not attempt to review the state of the field.  Second, we focus on stellar populations, not individual stars or binary systems.  Third, we do not specifically summarize results from variable stars (e.g., RR Lyrae), which are primarily, though not exclusively, used for distances; we also exclude studies that focus on extragalactic distance measurements.  Fourth, to the extent possible, we emphasize results from quantitative modeling (e.g., fitting a CMD) rather than qualitative comparisons (e.g., visual comparisons of data to isochrones).  Finally, we do not specifically focus on the discovery of new galaxies (e.g., ultra-faint systems around the MW).  While these generally make use of information from resolved stars and CMDs, such new galaxy discovery has already been discussed in sufficient depth in the recent literature \citep[e.g.,][]{Simon:2019}.

Within these parameters, we have identified four science themes as our focus: (1) the SFHs of galaxies; (2) evidence for bursty star formation; (3) the stellar IMF; (4) galaxy orbital histories.   We summarize progress in these areas over the last $\sim20$ years and then discuss exciting future directions for each, particularly in light of new, powerful facilities that have recently come online or are expected to see first light in the near future.

These four themes are connected and, in our view, the SFH of a galaxy is a common theme that ties them together.  Measuring a SFH from a CMD requires adopting an IMF, while measuring an IMF from a CMD also requires some knowledge of a SFH.  Describing a SFH requires a common vocabulary for episodes of enhanced star formation relative to a galaxy's long-term average, which motivates our discussion of bursty star formation.  Finally, interpreting a SFH requires connecting it to the external processes that shape a galaxy over its lifetime, of which the orbital history is particularly important.  Although each section of this review is meant to be reasonably self-contained, we encourage the reader to view them as facets of a single question: what do the resolved stars of nearby galaxies tell us about how those galaxies formed and evolved?

This review is designed to fulfill three goals: (i) to summarize, in one place, the major observational programs that have either surveyed nearby galaxies or targeted particularly important individual systems over the past $\sim20$~years; (ii) to provide an overview of the current state of our four selected science areas based on the literature from the past $\sim20$ years; (iii) to preview several future science directions enabled by observations of resolved stellar populations. We generally focus on literature published since the last major review(s) that comprehensively summarized observational studies on each topic \citep[e.g.,][]{Tolstoy:2009, Bastian:2010}.

This review is organized as follows.  We begin in \S \ref{sec:formation} with the measurement of SFHs from CMDs, including a summary of methodology and a synthesis of results for MW dwarf galaxies, the M31 system, isolated LG galaxies, and galaxies beyond the LG.  In \S \ref{sec:sfh_trends}, we examine the broad trends that emerge from the ensemble of published SFHs, including the impact of reionization, the diversity of low-mass galaxy formation, environmental trends in quenching, and evidence for bursty star formation (\S \ref{sec:bursts}).  In \S \ref{sec:imf}, we review measurements of the low- and high-mass IMF in nearby galaxies.  In \S \ref{sec:orbital_histories}, we review proper motions and the orbital histories of LG galaxies and their connection to SFHs.  Finally, in \S \ref{sec:future}, we consider future directions in each of these areas.  To keep the main text focused on synthesis, we have placed two extensive compilations in appendices.  Appendix~\ref{sec:important_obs} summarizes the major observational programs of the past $\sim20$~years, including efforts from ground- and space-based facilities and surveys (e.g., \textit{Gaia}) that impacted the field without resolved stellar populations being a primary science aim.  We expand the acronyms of these programs and surveys there, rather than in the main text.  Appendix~\ref{sec:individual_sfhs} provides galaxy-by-galaxy summaries of published SFHs, which complement the tabulated compilation in Appendix~\ref{sec:sfh_table}.  We reference these appendices throughout the main text.
Throughout this paper, we adopt a \citet{Planck:2020} cosmology when converting between age and redshift, and do not account for uncertainties in the age-redshift relationship \citep[e.g.,][]{Boylan-Kolchin:2021}.

\subsection{Definitions and Introduction to Key Concepts}
\label{sec:concepts}

A number of terms and concepts appear throughout this review, sometimes well before the sections that treat them in detail.  Here, we provide brief working definitions for several of these:  the key phases of stellar evolution seen on CMDs, and the terms we use to describe SFHs, bursty star formation, and orbital histories.  Fuller definitions are given in the relevant sections.  We also state the conventions we adopt to classify nearby galaxies and discuss the ambiguities inherent in them.

\subsubsection{Key Phases of Stellar Evolution} \label{sec:stellar_evolution}

Throughout this review, we refer to several phases of stellar evolution.  Although detailed information on stellar evolution is widely available in numerous textbooks \citep[e.g.,][]{Kippenhahn:1990, Cassisi:2013, Lamers:2017, Pinsonneault:2023}, for reference we provide a brief qualitative summary of the evolutionary phases that appear in this review.  We organize the discussion by the dominant evolutionary phase and group important diagnostic features as sub-items within each phase.  The approximate stellar age range over which each phase is most useful for resolved population studies is indicated in square brackets next to the phase header.

\begin{itemize}
    \item \textbf{Main sequence (MS) [all ages]:} A star spends the majority of its lifetime in this stable phase, generating energy by fusing hydrogen into helium in its core.  For unevolved MS stars there is a nearly one-to-one mapping between stellar mass and luminosity (and effective temperature), which in principle allows resolved star CMDs to test both stellar evolution theory and the stellar IMF.  The main sequence itself is not a strong indicator of age and is only moderately sensitive to metallicity, but two diagnostic features along it are widely used:
    \begin{itemize}
        \item \textit{Main sequence turnoff (MSTO):} Observationally, the MSTO is defined as the bluest extent of the main sequence in old, metal-poor populations and as the brightest concentration of MS stars in younger populations.  It marks the approximate point at which a population's most massive surviving MS stars are about to leave the main sequence; the actual exhaustion of core hydrogen occurs slightly later, at the terminal-age main sequence for stars with convective cores and at a less obvious inflection point for stars with radiative cores.  The MSTO serves as a critical observational tool for determining the age and metallicity of a stellar population.
        \item \textit{Main sequence kink} (also referred to as the MS ``knee'')\textit{:} A distinct, observable bend in the lower MS, typically occurring near $0.5$\msun.  It is caused by changes in atmospheric opacity and the structural transition to a fully convective stellar interior, and is sensitive to metallicity.
    \end{itemize}

    \item \textbf{Subgiant branch (SGB) [$\sim1$~Gyr to $\sim13$~Gyr]:} Once core hydrogen is depleted, a star's core contracts while its outer layers expand and cool.  Its luminosity is powered by a new hydrogen-burning shell surrounding the inert helium core.  The SGB is sensitive to age and metallicity almost as much as the MSTO.

    \item \textbf{Red Giant Branch (RGB) [$\sim1$~Gyr to $\sim13$~Gyr, low and intermediate mass]:} As the envelope greatly expands and cools, the star becomes highly luminous and red, with luminosity entirely generated by a hydrogen-burning shell surrounding a growing helium core.  The morphology of the RGB depends strongly on stellar mass.  In low-mass stars ($M\lesssim2$\,\msun), the helium core becomes electron-degenerate, which dramatically slows core evolution and produces the extended, vertical RGB seen in old and intermediate-age populations.  Intermediate-mass stars ($M\sim2$--$8$\,\msun) never develop a degenerate helium core; they traverse a much more compact, ``stubby'' RGB on a much shorter timescale.  RGB star properties depend weakly on age but more strongly on metallicity.  Two features within the RGB are particularly useful as diagnostics:
    \begin{itemize}
        \item \textit{RGB bump:} A brief evolutionary stutter during the red giant phase in which an expanding hydrogen-burning shell crosses a chemical boundary left by earlier convection.  This produces a temporary dip and recovery in the star's increasing luminosity.
        \item \textit{Tip of the Red Giant Branch (TRGB):} In low-mass stars with a degenerate helium core, the RGB ends abruptly when the core reaches a critical temperature and helium ignites in a rapid ``helium flash.''  Because this occurs at nearly constant helium-core mass, the TRGB has a well-defined peak luminosity, making it an excellent standard candle \citep[e.g.,][]{Lee:1993, Beaton:2016, McQuinn:2019, Freedman:2019}.  No helium flash occurs above the degeneracy boundary ($M\gtrsim2$\,\msun), and so the TRGB is purely a feature of the low-mass red giant population.
    \end{itemize}

    \item \textbf{Core helium burning (CHeB) phases [varies with mass and age]:} After helium ignition---either through a flash in low-mass stars or quiescently in more massive stars---stars enter a phase of stable core helium fusion.  The CMD morphology of this phase depends sensitively on mass, metallicity, and prior mass loss, and goes by several different names depending on the regime:
    \begin{itemize}
        \item \textit{Horizontal branch (HB) [$\gtrsim10$~Gyr, metal-poor, low-mass]:} Following the helium flash, low-mass, metal-poor stars enter a phase of stable core helium fusion and hydrogen-shell fusion.  Stars in this phase span a wide range of surface temperatures, forming a horizontal feature on the Hertzsprung-Russell diagram.  The HB is a good proxy for both age and metallicity, but also depends on other parameters such as RGB mass loss.  RR Lyrae variables occupy the instability strip portion of the HB.
        \item \textit{Red clump (RC) [$\sim1$~Gyr to $\sim13$~Gyr metal-rich, low-mass]:} Following the helium flash, more metal-rich stars settle into a long-lived, narrow concentration in the CMD characterized by steady core helium fusion and a very specific, predictable luminosity.  The RC typically contains stars that span a wide range of ages and metallicities and have similar luminosities and temperatures, making it a coarse indicator of age and metallicity.
        \item \textit{Red helium burner (RHeB) [$\sim100$~Myr to $\sim1$~Gyr, intermediate and high mass]:} For high-mass stars ($M\gtrsim8$\,\msun), this corresponds to the red supergiant (RSG) phase, in which the star steadily fuses core helium beneath a distended, cool envelope.  For intermediate-mass stars ($M\sim2$--$8$\,\msun), the picture is more complex: after helium ignition the star moves backwards down a stubby, non-degenerate RGB track, with the core gradually becoming dominant as the hydrogen-burning shell loses influence; once envelope opacity drops sufficiently, the star rapidly deflates onto the blue side (see BHeB).  As a result, observationally distinguishing a $\sim3$--$5$\,\msun\ star ascending its non-degenerate RGB from one that has already ignited helium and is backtracking is difficult or impossible from photometry alone, and the brightest red helium burners below $\sim8$\,\msun\ correspond to short-lived double-shell-burning episodes that account for only a small fraction of the post-helium-ignition lifetime.  In principle, RHeBs trace age and metallicity for younger stellar populations.  The color/temperature extension and morphology of the model tracks remain poorly constrained owing to complex physics (mass loss, rotation, convection, overshooting), but the underlying luminosity--age relation is comparatively robust and is what makes RHeBs useful as age tracers in practice.
        \item \textit{Blue helium burner (BHeB) [$\sim10$~Myr to $\sim1$~Gyr, intermediate and high mass]:} Closely related to the RHeB phase, the BHeB corresponds to the ``blue loop'' excursion during which intermediate- and high-mass core-helium-burning stars contract their envelopes and become significantly hotter and bluer.  For intermediate-mass stars this excursion is not a brief temporary contraction; it can last as long as, or longer than, the corresponding red phase, depending on mass and metallicity.  BHeBs are typically more numerous than RHeBs and are slightly better age and metallicity proxies.  As with RHeBs, the morphology and color extent of the blue-loop is sensitive to poorly modeled physics, while the luminosity--age relation is comparatively robust and is what enables their use as burst chronometers in practice (see \S\ref{sec:bursts}).  Classical Cepheids appear where the BHeB (and intermediate-mass RHeB) tracks cross the instability strip.
    \end{itemize}

    \item \textbf{Asymptotic Giant Branch (AGB) [$\sim100$~Myr to $\sim13$~Gyr, low and intermediate mass]:} After exhausting its core helium, a star expands once more, becoming a highly luminous giant powered by hydrogen- and helium-burning shells surrounding an inert carbon-oxygen core.  In principle, AGB stars (particularly TP-AGB) are good tracers of age and metallicity, but they are physically complex systems (e.g., mass loss, thermal pulses, dredge-up), and most TP-AGB stars are long-period pulsators, which complicates determination of their mean magnitudes and colors.  Recent empirical calibrations provide a path toward improved age and metallicity measurements.  The AGB is conventionally divided into two sub-phases, plus an important near-IR diagnostic feature:
    \begin{itemize}
        \item \textit{Early AGB (eAGB):} The phase immediately following the end of core helium burning, during which the hydrogen-burning shell is the primary energy source and the helium-burning shell remains dormant.  Stars with initial masses as low as $\sim0.5$~\msun can evolve to be eAGB stars and their ages can be as old as $\sim13$ Gyr \citep[e.g.,][]{Marigo:2017}.
        \item \textit{Thermally pulsing AGB (TP-AGB):} Repeated thermal instabilities in which the helium-burning shell rapidly ignites, drives powerful convective events, and extinguishes before the cycle repeats.  TP-AGB stars are luminous, exhibit significant mass loss, and include the Mira-type long-period variables (LPVs).  The youngest TP-AGB stars can be only $\sim100$~Myr in age.
        \item \textit{J-band AGB (JAGB):} A relatively narrow color and magnitude locus in near-IR CMDs dominated by carbon-rich TP-AGB stars, which has emerged as a useful standard candle and age tracer \citep[e.g.,][]{Madore:2020}.
    \end{itemize}
\end{itemize}

\subsubsection{Star Formation Histories}
\label{sec:concepts_sfh}

Throughout this review, we use the term SFH to mean the star formation rate (SFR) of a galaxy as a function of lookback time and metallicity, i.e., SFH $=$ SFR$(t, Z)$.  In the language of \citet{Hodge:1989}, this is a galaxy's ``population box.''  A SFH is most commonly presented in one of two forms: the \textit{differential SFH}, which is the SFR as a function of time, marginalized over metallicity; and the \textit{cumulative SFH}, which is the fraction of a galaxy's total stellar mass that had formed prior to a given lookback time.  From the cumulative SFH, it is convenient to define characteristic timescales such as $\tau_{50}$ and $\tau_{80}$, i.e., the lookback times by which a galaxy had formed 50\% and 80\% of its stellar mass.  We use $\tau_{50}$ as a proxy for the median stellar age of a galaxy and, in galaxies that are no longer forming stars, $\tau_{80}$ as a proxy for its quenching time, i.e., the epoch at which a galaxy effectively stopped forming stars.  The third component of a SFH is the age-metallicity relation (AMR), which depicts the change in metallicity over time.  The measurement of SFHs from CMDs, including the choice of these metrics and their limitations, is described in \S \ref{sec:formation}.

\subsubsection{Bursty Star Formation}
\label{sec:concepts_bursts}

The vocabulary used to describe enhanced star formation is only loosely defined in the literature. We adopt three working definitions in this review.  A \textit{burst} of star formation is a temporally localized episode of star formation that significantly exceeds the long-term average of the host galaxy.  A \textit{starburst galaxy} is one whose current SFR is elevated by a factor of $\gtrsim10$ relative to its long-term mean, such that it would exhaust its gas reservoir on a timescale much shorter than the age of the Universe.  \textit{Bursty star formation} describes a galaxy whose SFH consists of repeated bursts separated by periods of relative quiescence, as opposed to a smoothly varying history.  None of these labels has a uniquely well-defined boundary. Moreover, the degree to which a SFH appears bursty depends on the time resolution of the measurement.  We discuss these definitions, and how bursts are measured from resolved stars, in \S \ref{sec:bursts}.

\subsubsection{Orbital Histories}
\label{sec:concepts_orbits}

The orbital history of a galaxy refers to the reconstruction of its past trajectory through a gravitational potential.  Most cases in practice so far are of a low-mass galaxy moving through the potential of a host system (e.g., the MW or M31).  An orbital measurement requires a galaxy's full six-dimensional phase-space information: a position on the sky, a distance, a line-of-sight velocity from spectroscopy, and a proper motion, i.e., its angular motion in the plane of the sky, from multi-epoch astrometry.  Given an assumed potential, this information can be integrated backward in time to estimate quantities such as pericentric and apocentric distances, the time of infall into the host halo, and the number of pericentric passages.  These quantities are central to interpreting SFHs, as they determine when, and how strongly, a galaxy was subject to environmental processes such as tidal and ram pressure stripping.  As we discuss in \S \ref{sec:orbital_histories}, uncertainties in orbital histories are increasingly dominated by the adopted potential (e.g., the masses of the MW and the Large Magellanic Cloud [LMC]) rather than by the proper motion measurements themselves.

\subsubsection{Galaxy Classifications}
\label{sec:concepts_classification}

Throughout this review, and following literature convention, we group nearby galaxies into categories such as satellite versus isolated, classical versus ultra-faint, and dwarf spheroidal (dSph) versus dIrr.  These categories can provide useful shorthand for broad discussion or organizing a large body of literature.  However, their boundaries are porous. The classification of an individual galaxy can depend on a number of factors including field-specific convention, the available data, and historical precedent.  Here, we state the conventions we adopt.  We note explicitly where a particular choice affects our interpretation.

\textit{Satellites and isolated galaxies.}  There is no single definition of a satellite galaxy in the literature.  Observational studies typically adopt a fixed radius, often 300~kpc \citep[e.g.,][]{Gallart:2015, McQuinn:2023} or, beyond the LG, 300~kpc in projection, sometimes with a velocity cut \citep[e.g.,][]{Mao:2021, Carlsten:2022}.  Simulations typically define satellites as subhalos within the host's virial radius \citep[e.g.,][]{Garrison-Kimmel:2014}, although the splashback radius has been suggested as a more physical halo boundary \citep{More:2015}.  Dynamical studies consider whether a galaxy is gravitationally bound to its host \citep[e.g.,][]{Boylan-Kolchin:2013}.  For the sake of consistency and simplicity, we consider a galaxy to be a satellite if it currently lies within 300~kpc of the MW or M31, approximately the virial radius of a $\sim1.5\times10^{12}$~\msun\ halo \citep{Bryan:1998}.  Otherwise, a galaxy is considered isolated.  This division is necessarily approximate.  Host masses are uncertain, adopted virial radii vary between studies (e.g., 266~kpc versus 300~kpc for M31; \citealt{Savino:2022, McQuinn:2023}), and revised distances can move a galaxy across the boundary.  Moreover, present-day position may not reflect a galaxy's orbital history. A galaxy beyond the virial radius today may have passed within it in the past, i.e., it may be a ``backsplash'' galaxy \citep[e.g.,][]{Gill:2005, Kravtsov:2004, Sales:2007b, Teyssier:2012}.

We organize our discussion of the LG by host: the MW's dwarf galaxies (\S \ref{sec:mw_dwarfs}), the M31 ecosystem (\S \ref{sec:m31_ecosystem}), and isolated galaxies (\S \ref{sec:isolated}).  This organization largely follows our satellite definition, with three caveats.  First, a satellite by our definition need not be a long-term satellite. The LMC and the Small Magellanic Cloud (SMC) are likely on their first infall \citep[e.g.,][]{Besla:2007, Kallivayalil:2013}, several UFDs likely arrived with the LMC (\S \ref{sec:LMC_Satellites}), and some M31 satellites, such as Pisces~I, may once have been companions of M33 \citep[e.g.,][]{Patel:2026}.  Second, we discuss a few isolated galaxies that lie just beyond 300~kpc alongside their nearest host (e.g., Eri~II, Peg~W).  Third, we follow historical convention in a few cases: \A{XXVIII} and \A{XXXIII} are located 340--370~kpc from M31 \citep{Savino:2025} but are typically referred to as M31 satellites, and we do the same.

Among isolated galaxies, we reserve the term ``truly isolated'' for those whose proper motion-based orbits indicate that they are on first infall into the LG \citep[e.g.,][]{Bennet:2024, Bennet:2025}, or that are located far enough beyond the LG, and not near another group, that a past passage through a massive host within a Hubble time is implausible (e.g., Tucana~B; \citealt{Sand:2022}).  In practice, orbits cannot be reliably reconstructed beyond $\sim6$~Gyr ago \citep{Bennet:2025}, and interactions with other low-mass galaxies are virtually impossible to rule out (e.g., Leo~A; \citealt{Cole:2007}), always leaving a sense of ambiguity.  Some isolated galaxies may themselves be members of small associations, such as that of NGC~3109, whose membership is also uncertain \citep{Tully:2006}.

\textit{Classical dwarfs and UFDs.}  Following \citet{Simon:2019}, we refer to galaxies with $M_V>-7.7$ as UFDs and to those with $M_V\le-7.7$ as classical dwarfs.  However, luminosity uncertainties and revisions (e.g., from updated distances) can move galaxies across the threshold in either direction.  For example, \A{XXX} lies at exactly $M_V=-7.7$ \citep{Savino:2022}, the luminosity of \A{XXIV} was revised from $M_V=-8.4$ to $-7.6$ \citep{Martin:2016a, Savino:2022}, and Leo~T has $M_V=-8.0\pm0.5$ \citep{McConnachie:2012}.  As we discuss in \S \ref{sec:sfh_trends}, the SFHs of LG galaxies show no notable change at this threshold, and the distinction between the faintest galaxies and star clusters is also increasingly blurred \citep[e.g.,][]{Willman:2012, Cerny:2023c}.

\textit{Morphology.}  Following convention \citep[e.g.,][]{Mateo:1998}, we use dSph, dwarf elliptical (dE), dIrr, and transition dwarf (dTrans) to describe present-day morphology, which largely reflects gas content and current star formation.  These labels do not map uniquely onto SFHs \citep[e.g.,][]{Weisz:2011, Gallart:2015}.

\section{Star Formation History Measurements}

\label{sec:formation}

A galaxy’s resolved stellar populations encode its entire SFH. By modeling various age and metallicity sensitive features of stellar populations on CMDs (see \S \ref{sec:stellar_evolution}), it is possible to reconstruct the star formation and chemical enrichment history of a galaxy in great detail. SFHs measured from deep CMDs are widely considered the `gold standard’ approach to measuring a galaxy’s SFH as it involves the most direct approach (effectively counting stars) and far fewer assumptions (e.g., no functional form for the SFH needs to be adopted) than techniques involving integrated light \citep[e.g.,][]{Kennicutt:2012, Conroy:2013}.

Though qualitative translation of a CMD into a SFH has been understood for decades \citep[e.g.,][]{Sandage:1952, Hoyle:1955, Schwarzschild:1958, Hodge:1971, Hodge:1989}, it was only in the 1990s that advances in stellar models, instrumentation and observations, and computational power enabled the quantitative measurement of SFHs from resolved stars \citep[e.g.,][]{Tosi:1989, Tosi:1991}.  The literature now consists of SFHs (with varying look back time fidelity) measured from stellar populations for $>100$ galaxies that span $\sim9$ decades in stellar mass.%

In this section, we have two goals. First, we summarize some of the important methodological details related to SFH measurements from CMDs including measurement techniques, uncertainty calculations, visualization, and other select topics. Second, we synthesize the published SFHs from the past $\sim20$~years that have been measured by quantitatively modeling a galaxy's CMD, organized by environment (MW dwarf galaxies, the M31 system, isolated LG galaxies, and galaxies beyond the LG). In some cases, older studies have been included to provide context.  For the most prominent systems (e.g., the Magellanic Clouds, M31, M33) and for populations that are best understood as an ensemble (e.g., UFDs), we discuss the literature directly in the main text.  For systems with a rich astronomical history as individual objects (e.g., classical MW satellites, isolated dwarfs), we provide a population-level synthesis in the main text and defer the galaxy-by-galaxy summaries, including the relevant details of each observing setup (e.g., filters, depth, area), to Appendix~\ref{sec:individual_sfhs}.  A tabulated compilation of all published lifetime SFHs is given in Appendix~\ref{sec:sfh_table}.

This overview of recent progress in CMD-based SFHs builds on a number of excellent review articles from the past several decades that cover the resolved stellar populations, SFHs of nearby galaxies, and the intricacies of CMD modeling \citep[e.g.,][]{Hodge:1989, Mateo:1998, Gallart:2005, Tolstoy:2009, Cignoni:2010, Simon:2019}.  We refer the reader to these papers for further reading.

\subsection{Definitions and Introduction to Concepts}

As first defined by \citet{Hodge:1989}, a primary goal of modeling a CMD is to measure a galaxy's ``population box''—the relationship between SFR and metallicity evolution over cosmic time. \citet{Hodge:1989} introduced the population box as a successor to the Pop I/Pop II classification scheme of \citet{Baade:1944}, which was deemed too simple to describe the complex stellar populations being uncovered in dwarf galaxies. This two-dimensional plane of age versus metallicity defines a galaxy's SFH, where SFH $=$ SFR(t,Z).  For visualization purposes, the SFR can be projected into the third dimension, producing a ``box''.  The process of measuring a population box (i.e., SFH) from a CMD is the topic of many technical papers \citep[e.g.,][]{Tosi:1991, Tolstoy:1996, Harris:2001, Dolphin:2002b, Aparicio:2009, Cole:2014, Garling:2025, Zheng:2025} and has been covered in several review articles cited above, and we only briefly summarize the process here.

The data being modeled are in the form of a CMD, i.e., a color and magnitude for each star. Figures~\ref{fig:cmd_pedagogy} and \ref{fig:cmd_mosaic} show examples of CMDs that are typically used to measure a galaxy's SFH.  Uncertainties in the data are generally characterized using ASTs, which involve injecting synthetic stars into the raw images and re-running the photometric reduction. The distribution of recovered vs. input magnitudes is used to quantify the completeness and noise properties of the CMD. In general, uncertainties produced by ASTs are no smaller than those reported from photometry (i.e., Poisson noise), as ASTs capture virtually all noise effects in the image (e.g., crowding), not just counting statistics.   ASTs also encode information beyond the completeness function.  In crowded or low-signal-to-noise ratio (SNR) regimes, recovered stars are systematically up-scattered in brightness relative to their input magnitudes and can also exhibit systematic color shifts.  Both effects must be propagated into CMD modeling. While single-color CMD fits remain the standard, recent work has expanded to simultaneous multi-band CMD fits (e.g., ultraviolet [UV] and optical; e.g., \citealt{Williams:2017, Cignoni:2018, Choi:2025}).

Constructing a model CMD involves populating a stellar IMF, accounting for unresolved binary stars, and adopting a stellar library to map initial stellar masses to observable properties.  Model basis functions typically consist of single or simple stellar populations that are generated over a grid of age and metallicity.  The basis models are linearly combined to form a model composite CMD.  The relative weight for each basis function in this linear combination ultimately is the SFH measurement.

Next, the synthetic CMD is shifted into the observational frame by applying the galaxy's distance modulus and an extinction model (foreground and/or internal to the galaxy) to account for the effects of dust. These are often done via the computation of extinction coefficients that can be applied directly to stellar models, but these are less accurate in regions of high-extinction.  In such cases, applying extinction directly to stellar atmospheres and projecting the resulting extinguished spectra through filter curves yields a better result \citep[e.g.,][]{Gordon:2016}.

Finally, the composite model CMD is convolved with the uncertainties and completeness function from the ASTs.  ASTs define the observational selection function that maps an intrinsic synthetic CMD into the observed CMD. They are particularly important when comparing CMDs across fields with different crowding, surface brightness, filters, instruments, and/or spatial coverage.   Typically, the 50\% completeness limit is used to define the faint end of the CMD being modeled, but this choice is an arbitrary empirical convention rather than a principled approach.  However, depending on crowding, photometric bias, non-Gaussian errors, blending, and/or the science goal a limit brighter or fainter than the 50\% completeness limit may be more appropriate.  Moreover, in cases of extreme crowding, e.g., resolved extragalactic star clusters, more sophisticated approaches such as forward modeling entire stellar systems at the image level and re-running the entire photometric reduction process may provide for more realistic selection functions \citep[e.g.,][]{Larsen:2011}.  While extremely informative about the selection function, such an approach is also computationally expensive and may not always be necessary.

The result of the above steps is a synthetic CMD that can be quantitatively compared to an observed CMD using a likelihood function and a set of implicit or explicit priors, depending on the fitting technique.  The iterative process of generating model CMDs and comparing them to observed CMDs is repeated until some stopping criterion is satisfied (e.g., minimizing $\chi^2$, maximizing a likelihood function).  Most SFHs in the literature typically represent a `best fit'. What gets reported as SFH uncertainties varies widely, as we discuss in \S \ref{sec:sfh_errors}.

Prior to their comparison, most CMD modeling approaches use binning to turn the observed and synthetic CMDs into two-dimensional histograms in color and magnitude, called ``Hess diagrams'' \citep{Hess:1924}.  Generally, the aim of this approach is to increase computational efficiency and/or easily exclude regions of the CMD that are less reliable (e.g., less reliably modeled phases of evolution such as the HB).  Although binning of data is generally considered a lossy process, there appears to be little effect on the accuracy and precision of recovered SFHs when binning a CMD in color and magnitude space.  One advantage to binning is that it suppresses shot noise. Each Hess bin integrates the underlying color-magnitude probability density over a finite interval, which smooths over the Poisson fluctuations that arise from the finite sampling of the IMF used in synthetic CMD construction.

\begin{figure}[pos=tbp]
    \centering
    \includegraphics[width=1\linewidth]{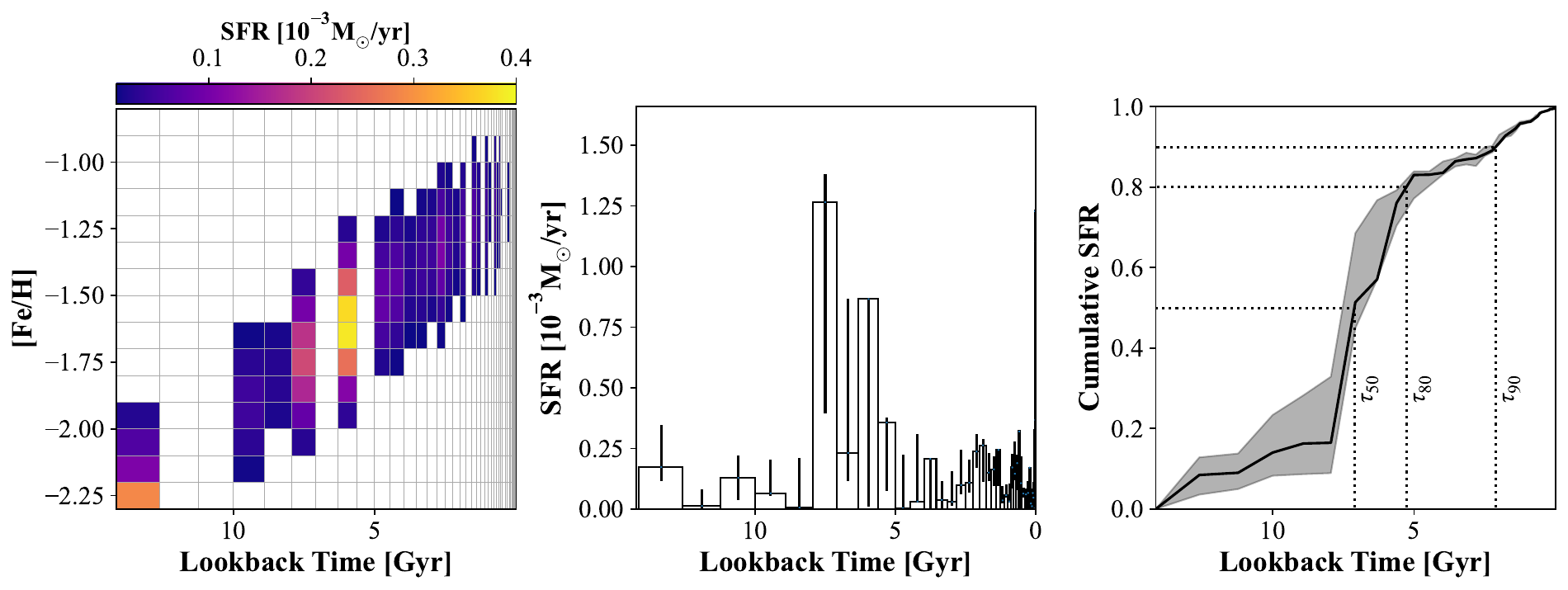}
    \caption{An example SFH measured from fitting the deep \hubble\ CMD of a nearby galaxy, Aquarius (DDO~210) \citep[Photometry from][]{Cole:2014}. \textbf{Left:} The population box, i.e., a plot of the age and metallicity of the SFH bins, color-coded by the star formation rate (SFR) in that bin. This diagram simultaneously depicts stellar mass growth and chemical enrichment. \textbf{Center:} Differential SFH, i.e., the galaxy's star formation rate as a function of lookback time. This plot is obtained by marginalizing the population box over metallicity. Nearby time bins have covariant SFRs, which result in large uncertainties in the differential SFH. \textbf{Right:} Cumulative SFH, i.e., the fraction of the total stellar mass formed as a function of time, normalized by the total mass formed. We indicate common SFH timescale metrics, $\tau_{50}$, $\tau_{80}$, and $\tau_{90}$. Note smaller uncertainties (grey shaded region) compared to the differential SFH.  In the cumulative SFH, the SFR covariance terms largely cancel out. }
    \label{fig:BoxPlot}
\end{figure}

Figure~\ref{fig:BoxPlot} shows an example SFH for the LG galaxy Aquarius (DDO~210).  This SFH is based on the deep, SGB-depth data presented in \citet{Cole:2014}.  We discuss the visual aspects of plotting SFHs later in this section and detail the SFH of Aquarius in Appendix~\ref{sec:Aquarius_sfhs}.%

\subsubsection{CMD Depth, Filter Choice, and Age Diagnostics} \label{sec:cmd_depth}

At present, there is only a loose consensus for what CMD characteristics (filters, depth, SNR, etc.) are ``best'' for SFH measurements.  The point of broadest agreement is that CMDs that reach $\sim1$ mag below the oMSTO provide the best possible SFH and AMR measurements.  This is because (i) MSTOs and SGBs are highly sensitive to age and metallicity with minimal overlap on the CMD; (ii) CMDs of this depth include the MSTO and SGB populations for stars of all ages and metallicities; and (iii) the MSTO and SGB phases of evolution are reasonably well-reproduced by current generations of stellar models.  We discuss age and metallicity constraints from brighter CMD features in \S \ref{sec:shallow_cmds}.

Such deep CMDs are still subject to the well-known age-metallicity degeneracy \citep[e.g.,][]{Renzini:1988, Worthey:1994, Gallart:2005}.  The MSTOs and SGBs of populations with different ages and metallicities have some degree of overlap on CMDs, meaning that there is some degeneracy between age and metallicity.  This degeneracy effectively limits the precision to which a galaxy's SFH and AMR can be measured. We note that this is a practical, not a formal mathematical, degeneracy.  In principle, perfect photometry of a population with a strictly one-to-one mapping between age and metallicity would resolve both quantities exactly.  In practice, the intrinsic dispersion of metallicities at fixed age, differential reddening, and finite photometric precision result in a residual degeneracy.  However, because the oMSTO and the RGB respond differently to age and metallicity (i.e., the luminosity of the oMSTO is primarily sensitive to age, while the color of the RGB is primarily sensitive to metallicity), CMDs that include both features substantially reduce this degeneracy, though they do not entirely eliminate it.

Beyond the age-metallicity degeneracy, recovered SFHs and IMFs depend on several modeling ingredients that are partially degenerate with age and metallicity and are often assumed rather than jointly fit.  For example, many CMD modeling approaches assume a galaxy's distance is constrained externally by the TRGB or by variable stars (e.g., RR Lyrae, Cepheids). Otherwise, it must be left as a free parameter in the CMD fit \citep[e.g.,][]{Brown:2014, Durbin:2025}.  Similarly, foreground and internal extinction can vary with stellar age and spatial position.  Any unmodeled differential reddening broadens CMD features in a way that mimics intrinsic age and metallicity dispersion \citep[e.g.,][]{Dolphin:2003, Lewis:2015}.  Unresolved binary stars are usually incorporated through a simple companion-mass distribution, but a fuller treatment of binary evolution (e.g., mass transfer, common-envelope evolution, mergers, blue straggler formation) is much more involved and can systematically bias the recovered SFH and IMF \citep[e.g.,][]{DalTio:2021}.

Beyond community consensus about CMD depth, there is less agreement over optimal SNR and filter combinations for SFH measurements. Even for oMSTO-depth CMDs, some approaches prefer broad optical color baselines (e.g., g-i) with lower SNR \citep[$\sim10$;][]{Cole:2007, Monelli:2010_a, Skillman:2017}, while others use a shorter color baseline (e.g., r-i) with higher SNR \citep[$\sim80$;][]{Brown:2014, Sacchi:2021}.  Several papers have demonstrated the fidelity of SFH recovered in the near-IR \citep[e.g.,][]{Rubele:2012}.  Although near-IR CMD features are more compressed in color due to a decreased temperature baseline, SFHs from near-IR CMDs that reach the oMSTO appear to be in good agreement with those derived from optical CMDs \citep[e.g.,][]{McQuinn:2024a}.  Several papers have also explored the addition of UV filters for SFH recovery, with the general finding that they may marginally improve the time resolution of very recent SFHs and possibly improve constraints on dust \citep[e.g.,][]{Cignoni:2018, Cignoni:2019, Chen:2025}, though with caveats.  For example, some stellar populations (e.g., BHeBs) do not appear to be well captured by current stellar models in the UV \citep{Choi:2025}, which can adversely affect inferred SFHs.  Similarly, the adopted extinction curve matters more for dust attenuation in the UV.

Underlying these debates is a single physical criterion: what matters for SFH recovery is the separation of diagnostic CMD features in units of the photometric uncertainty at the relevant magnitude, not the absolute color or magnitude difference between them.  The optimal trade-off between color baseline and SNR therefore depends on the instrument and the observational scene (i.e., the target galaxy and astrophysical backgrounds).  What can at times appear as a methodological disagreement between groups may simply reflect different effective feature separations on the underlying datasets.

Although optimal trade-offs between SNR and color baseline remain debated, there is some consensus around other observational requirements. For example, the literature has clearly established that at least $\sim100$ MSTO stars are needed for a robust SFH measurement and SNRs $\gtrsim100$ at the oMSTO do not appear to improve the precision of a galaxy's SFH \citep[e.g.,][]{Brown:2014, Durbin:2025}.  \citet{Durbin:2025} quantified this requirement empirically using uniform SFHs of MW UFDs.  They found that $\sim100$ stars within $\pm1$~mag of the oMSTO corresponds to an effective luminosity, i.e., the luminosity within the observed field of view, of $M_V\sim-2.5$.  Below this limit, the precision of the SFH is set by the number of stars rather than by the photometric quality of the CMD.  This result places a practical lower bound on the luminosity of a galaxy for which a precise CMD-based SFH can be measured.  It may be possible that the inclusion of the MS kink in CMD fitting relaxes requirements on the number of MSTO stars, but that has yet to be explored in the literature.

\subsubsection{Star Formation Histories from Shallow CMDs} \label{sec:shallow_cmds}

oMSTO stars are faint ($M_V \sim +4$) and can be subject to crowding.  Unfortunately, these effects have restricted even \hubble\ to only resolving the oMSTO for galaxies within the LG.  \jwst's improved sensitivity and angular resolution have already resolved the oMSTO outside the LG in Leo~P \citep[$D\sim1.6$~Mpc;][]{McQuinn:2024b}.  However, even \jwst{} is unlikely to reach the oMSTO stars in galaxies beyond 2--3~Mpc owing to faintness and crowding (e.g., see Figure~\ref{fig:Telescopes}).

SFHs from CMDs that do not reach the oMSTO still use information from younger MSTO stars (e.g., 5 Gyr old MSTO stars).  However, they also rely more heavily on evolved stellar phases.  Relative to the oMSTO, the SGB is the next brightest CMD feature that is sensitive to age and metallicity.  \citet{Cole:2014} demonstrated that SGB-depth observations in Aquarius produced a SFH that is nearly as accurate and precise as one measured from oMSTO depth imaging.  However, the SGB spans a range in magnitude, and its faint end for the oldest populations ($M_V\sim+3.5$) is only marginally brighter than the oMSTO, such that it provides limited observational advantages in practice.

After the SGB, the next brightest age-sensitive CMD features are the HB ($M_V \sim +0.5$) and RC ($-0.5 \leq M_V \leq +0.5$).  The morphology of the HB is known to be highly sensitive to age and metallicity, and the mere presence of blue HB stars is an unambiguous indicator of an ancient ($\gtrsim10$~Gyr) population \citep[e.g.,][]{Cassisi:2013}.  However, sizable uncertainties in mass-loss on the RGB, helium abundance, a paucity of HB stars, and potential CMD overlap with the young MS in star-forming galaxies limit its use in practice \citep[e.g.,][]{Salaris:2002, Gallart:2005, Cassisi:2013}.  Including mass-loss as a free-parameter when measuring a SFH from the HB appears necessary for robust results \citep[e.g.,][]{Savino:2018}.  Many standard stellar models do not produce realistic HB morphologies, which can lead to inaccurate SFHs compared to those measured from the oMSTO \citep[e.g.,][]{Weisz:2019, Savino:2025}.

Even with \hubble, the faintness of the HB, compounded by crowding, makes it observationally prohibitive to access in galaxies beyond a few Mpc.  For example, reaching just below the HB in two Sculptor Group dwarfs at $\sim2$~Mpc required $\sim11$~hr of ACS imaging per galaxy \citep{Yang:2014}.  At $\sim4$~Mpc, the HB is $\sim1.5$~mag fainter, which would require roughly an order of magnitude more exposure time.  Crowding may further limit the achievable depth in practice (\S \ref{sec:crowding_future}).  The situation does not improve in the near-IR, as capturing the blue HB stars, which are hotter than MSTO stars, can require nearly as much integration time as reaching the oMSTO.

CMDs that reach some portion of the RC are common for resolved studies of galaxies within $\sim5$~Mpc \citep[e.g.,][]{Dalcanton:2009, Weisz:2011}.  CMDs of this depth enable the study of recent star formation ($\lesssim 1$–$2$ Gyr) via the corresponding MS and MSTO populations, along with RHeBs and BHeBs ($M_\star \gtrsim 1.5~M_{\odot}$).  The RC itself is not a particularly good age indicator \citep[e.g.,][]{Girardi:2016}. It is a complex mixture of stars spanning a wide range of ages and metallicities that share similar colors and magnitudes. The age and mass distribution functions are skewed toward younger ($\sim1$--$4$ Gyr), metal-rich populations \citep[{[Fe/H] $\gtrsim -1$};][]{Bovy:2014}.

In RC-depth CMDs, age information beyond a few Gyr generally relies on the ratio of RC to RGB stars.  If present, the HB also provides age information, subject to the caveats provided above. The RC-to-RGB star ratio is highly model-dependent, meaning that SFHs from CMDs of this depth can differ substantially depending on the choice of underlying models \citep[e.g.,][]{Gallart:2005, Weisz:2011}.  Several papers have shown that it is essential that the entire RC (and preferably slightly deeper) be included on the CMD for SFH modeling as an incomplete census of RC stars can lead to substantial variations in derived ancient and intermediate-age SFHs \citep[e.g.,][]{McQuinn:2010a, Weisz:2011, Dolphin:2012, Weisz:2014a}.  SFHs from CMDs that just reach the RC, or only partially capture it, often have poor age resolution at older epochs \citep[e.g., a single old-age time bin that extends from $\sim8$–$14$ Gyr ago;][]{McQuinn:2010a, Crnojevic:2011, Weisz:2011, Crnojevic:2012, Weisz:2014a, Williams:2017}.

The RGB ($M_I\sim0$ to $-4$) is bright, ubiquitous, well-populated, and contains stars over a wide range of ages and metallicities. However, it is known to be a very poor age indicator as RGB color is generally more sensitive to metallicity than age \citep[e.g.,][]{Salaris:2002, Gallart:2005}.  SFHs based on CMDs that only include the RGB, and do not reach the RC, are subject to very large uncertainties for ages older than $\sim0.5$–$1$ Gyr \citep[e.g.,][]{McQuinn:2010a, Weisz:2011, Weisz:2014a} owing to both significant age-metallicity degeneracies and sizable variations in the predicted RGB color and shape from different stellar models and underlying physics \citep[e.g., due to dredge-up, mass-loss;][]{Salaris:2002}.

AGB stars are luminous ($M_V \lesssim -2$, $M_J \lesssim -5$) and are present in populations of all ages ($\sim0.1$--$13$~Gyr) and metallicities \citep[e.g.,][]{Marigo:2017}. In principle, the locations of AGB stars on CMDs are  sensitive to both age and metallicity.  \citet{Gallart:1994} and \citet{Gallart:1996} were the first papers to demonstrate the potential of AGB stars for measuring a galaxy's quantitative SFH up to $\sim10$~Gyr ago.  Despite their well-understood potential, AGB stars have historically been less favored for SFH measurements.  Observationally, in optical CMDs, AGB stars substantially overlap with age-insensitive RGB stars, limiting their practical utility for SFHs, and due to self-extinction, some of the most extreme AGB stars may not be detected in optical filters \citep[e.g.,][]{Jackson:2007a, Boyer:2011}.  Theoretically, AGB stars are known to have a variety of complex physics (e.g., dredge up, mass-loss, thermal pulses, hot bottom burning) that have proven challenging to accurately model \citep[e.g.,][]{Herwig:2005, Hofner:2018}.  AGB stars are also a short-lived phase of evolution.  Their low frequency per unit stellar mass formed \citep[e.g.,][]{Marigo:2017} presents another challenge to their use as SFH tracers.  Many AGB stars are long-period pulsators, which complicates determination of their mean magnitudes and colors and, by extension, the comparison of observed CMDs to static stellar isochrones.

Despite these challenges, there have been several efforts over the past decades to model SFHs from AGB stars on optical and near-IR CMDs.  These range from estimating quenching epochs \citep[e.g.,][]{Rejkuba:2022, Harmsen:2023, Velguth:2024} to more formal modeling of luminosity functions, CMDs, and even LPVs in the near-IR \citep[e.g.,][]{Javadi:2011, Rezaeikh:2014, HamedaniGolshan:2017, Javadi:2017, Ren:2024}.

In the past several years, there have been two significant advances that have catalyzed high-fidelity AGB star-based SFHs.  The first is a substantial improvement in stellar models.  Notably, concerted efforts by the stellar theory group in Padova have produced increasingly realistic stellar models for AGB stars that include detailed treatment of many of the complex physics involved \citep[e.g.,][]{Marigo:2013, Rosenfield:2014, Marigo:2017}.

The second is the increased prevalence of near-IR imaging of resolved stellar populations \citep[e.g.,][]{Cioni:2011, Dalcanton:2012a, Ren:2021a, Ren:2021b}. It has long been known that near-IR CMDs enable the identification of carbon stars and cleaner separation of oxygen-rich and carbon-rich AGB stars compared to the optical \citep[e.g.,][]{Cohen:1981}. The larger number of near-IR CMDs is useful for calibrating AGB star models as well as for SFH measurements.

Recently, \citet{Pastorelli:2019} and \citet{Pastorelli:2020} used SFHs of the LMC and SMC from CMDs that reach the oMSTO to calibrate the COLIBRI AGB star models. For SFH measurement purposes, these studies effectively placed the AGB star ages on the same scale as the PARSEC-based ages for MSTO stars, providing an important new foundation for AGB star-based SFHs.  Several recent studies have shown that using the COLIBRI models to fit AGB stars on near-IR CMDs can produce SFHs comparable to deeper optical CMDs, though with much less exposure time \citep[e.g.,][]{Ren:2024, Lee:2024, Lee:2026}. However, \citet{Ventura:2026} provide caution about the large variation in the predicted number of AGB stars between different stellar models. Further refinements to AGB star models over a wider range of metallicities will continue to improve their accuracy for SFH measurements.

For recent star formation, luminous MS and MSTO stars ($M_V \lesssim 0$), along with BHeBs and RHeBs ($-10 \lesssim M_V \lesssim 0$), constrain SFH measurements over the past $\lesssim1$--$2$~Gyr \citep[e.g.,][]{Dohm-Palmer:1997, Dohm-Palmer:1998, Dolphin:2003, Weisz:2008, McQuinn:2009}.  In addition to their good age and metallicity sensitivity, BHeBs have also been used to measure spatially resolved SFHs in several nearby galaxies.  There are a number of challenges associated with measuring SFHs from BHeBs and RHeBs including a paucity of stars and stellar models that are not always in good agreement with the observed locations and number densities of BHeBs and RHeBs in optical CMDs \citep[e.g.,][]{McQuinn:2011, Gull:2022, Choi:2025}.  Further improvements to the BHeB and RHeB models are of great importance, but they are challenging owing to poor constraints on their complex, metallicity-dependent underlying physics \citep[e.g., mass loss, rotation, convection, overshooting, binary interactions;][]{Langer:1995, Maeder:2001, Dohm-Palmer:2002, Eggenberger:2002, Eldridge:2008, Saio:2013, Rosenfield:2017, Costa:2019, Dorn-Wallenstein:2020}.

\subsubsection{SFH Uncertainties}
\label{sec:sfh_errors}

Random uncertainties on CMD-based SFHs stem from the finite number of stars on the CMD, CMD noise properties, selected filter combination, age-metallicity degeneracies, etc. Systematic uncertainties can arise from uncertainties in the photometric zero points, distance, the adopted extinction law, choice of parametric enrichment model (if used), and limitations of the underlying stellar evolution models (e.g., single star, 1-D, non-rotating, stellar abundance patterns).

Uncertainties on CMD-based SFHs in the literature are computed using a variety of techniques. For random uncertainties, the historical standard is bootstrapping, in which CMDs are resampled and SFHs are measured again \citep[e.g.,][]{Cole:2007, Weisz:2008, Monelli:2010_a, deBoer:2012}. These are sometimes combined with shifts in the age-metallicity grid and binning scheme, as first systematized in the IAC-pop framework of \citet{Aparicio:2009} and adopted by other CMD fitting approaches \citep[e.g.,][]{deBoer:2012}. A set of resulting SFHs (e.g., those within $1\sigma$ of the best fit solution) is then used to assign random uncertainties. \citet{Dolphin:2013} introduced a Markov chain Monte Carlo (MCMC) approach to estimate random uncertainties in CMD fitting. They showed that an MCMC approach to uncertainty estimation provides more realistic uncertainties than bootstrapping, which can be overly precise.  Several codes employ MCMC techniques for random uncertainty estimation \citep[e.g.,][]{Brown:2014, Garling:2025, Zheng:2025}.

Systematic issues in modeling CMDs due to the limitations of stellar models were discussed extensively in \citet{Gallart:2005}.  To account for these issues, in some cases, a CMD is modeled with different stellar models resulting in multiple SFHs \citep[e.g.,][]{Monelli:2010_a, Skillman:2017}.  Differences in the resulting SFHs give a broad sense of systematics. Generally, CMDs that reach below the oMSTO only show modest variations due to the choice of underlying stellar model. There are at least two important limitations to the multiple stellar library approach: (i) There are a very limited number of stellar evolution models, and even fewer that cover a large range of evolutionary phases or relative chemical abundance patterns; (ii) there are poorly documented correlations between stellar libraries (e.g., many use comparable input physics).

The situation is more complicated for shallower CMDs, as stellar models can predict broader ranges of SFHs when CMD fitting relies more heavily on certain evolved phases of evolution (e.g., RC, HB, RGB). By generating and recovering SFHs from CMDs with different stellar models, \citet{Weisz:2011} and \citet{Dolphin:2012} introduced a method to compute systematic uncertainties on SFHs as a function of CMD depth. This approach was designed to enable comparisons of SFHs derived from CMDs of different depth (e.g., shallower CMDs from ANGST vs.~deeper CMDs available only in the LG). Due to the large number of stars in many shallow CMDs, the resulting SFHs can be highly precise but show modest to sizable variations when different stellar models are adopted. The inclusion of the systematic estimates from \citet{Dolphin:2012} provided uncertainties that were designed to capture these differences as a function of CMD depth. Not all studies adopt this approach to systematics and many do not report systematics at all.

\subsubsection{Temporal Resolution} \label{sec:time_resolution}

In addition to formal uncertainties on the SFH, there is a question of what the correct temporal resolution is for a SFH given an observed CMD.  Perhaps the most common definition is $\frac{\Delta t}{T}$, which specifies the fractional time resolution at a given epoch.  This definition is similar in spirit to the resolving power of a spectrograph, $R = \frac{\lambda}{\Delta \lambda}$, in which $\Delta \lambda$ is, for a Gaussian line profile, the full width at half maximum (FWHM).  For the time resolution of a SFH, $\Delta t$ is then the FWHM of an instantaneous burst of star formation that occurred at a lookback time of $T$. Factors that affect $\Delta t$ include photometric precision/depth, crowding, number of stars, filter choice, CMD depth, the adopted time binning, and complexity of the SFH itself.

Historically, there has been an anecdotal rule-of-thumb that the achievable time resolution from an oMSTO-depth CMD is $\frac{\Delta t}{T} \approx 10$\%, or $\sigma \approx 5$\% for a Gaussian.  Practically, this means that under the best conditions, the details of star formation 10~Gyr ago could not be discerned more finely than $\sim1$~Gyr.  The origins of this guidance are unclear to the authors, though our best guess is that it is due to time resolution provided by most stellar evolution models, which is $\log(t)\sim0.05$~dex.  This value corresponds to a fractional resolution of $\sim10$\%.  This value reflects the fact that most stars do not evolve substantially on shorter timescales than $\log(t)\sim0.05$~dex and that producing stellar models is computationally expensive.

Given the increase in computational power, several studies measure SFHs using finer age and metallicity resolution achieved through interpolation \citep[e.g.,][]{Savino:2025, Durbin:2025}, which could in principle lead to better than 10\% time resolution.  Similarly,  there are scenarios in which the 10\% rule is likely incorrect, e.g., a high SNR oMSTO depth CMD with millions of oMSTO stars should allow for the recovery of a SFH with higher resolution than $\frac{\Delta t}{T} \approx 10$\%.   We therefore urge caution in adopting a general rule-of-thumb for SFH time resolution as it should greatly depend on the quality of the data.

In practice, several studies using oMSTO-depth CMDs have estimated the time resolution of their recovered SFHs by injecting a delta function input SFH, adopting the noise model of their dataset, and then re-running their CMD analysis code to measure the spread in the resulting SFH \citep[e.g.,][]{Aparicio:2009, deBoer:2012}. For example, using CMDs that reached the oMSTO, \citet{Hidalgo:2011} injected single Gaussian bursts of star formation into their \hubble\ LCID LGS-3 dataset and found a temporal resolution of $\sigma \sim 1$~Gyr at the oldest ages, where $\sigma$ is approximately half of the FWHM. \citet{deBoer:2012} and \citet{deBoer:2012b} performed a similar exercise for their ground-based data of Sculptor and Fornax and found $\sigma\approx1.5$~Gyr and $\sigma\approx2.5$~Gyr, respectively, for bursts injected $\sim12.5$~Gyr ago.  \citet{Savino:2018} showed that including the HB in CMD fitting and marginalizing over mass-loss could achieve a resolution of $\sigma \sim 0.5$~Gyr at the oldest ages. \citet{Bettinelli:2018} and \citet{Bettinelli:2019} reported age resolution of $\sim0.5$ and $\sim0.6$~Gyr at $12$–$13$ Gyr ago from their recovery of Gaussian bursts of star formation for their ground-based Sextans and Sculptor CMDs, respectively.

That many of these results produced time resolutions in excess of 10\% is likely the result of lower than optimal SNR at the oMSTO. The SNR of the LCID data was typically $\sim10$ near the oMSTO, and slightly lower for the \citet{deBoer:2012} and \citet{deBoer:2012b} datasets. Recent ground-based datasets of MW satellites often feature higher SNR data \citep[e.g., $\sim50$;][]{Bettinelli:2018, Bettinelli:2019, Sato:2025}, which should help improve age resolution of a SFH.

For shallower CMDs, the time resolution at older ages is notably worse. Several papers that derived lifetime SFH from CMDs that reach the RC, for example, adopted broad time bins for older ages \citep[e.g., 6--14 Gyr as a single ancient time bin;][]{Weisz:2008, McQuinn:2010a, Weisz:2011, Williams:2017}. As discussed in \citet{Weisz:2011}, CMDs that do not reach the oMSTO but still contain a large number of stars can produce SFHs with random uncertainties that are comparable to those from oMSTO-depth CMDs.  However, they are known to have poor accuracy as changing the underlying stellar model results in a vastly different SFH that would also depend on CMD depth. This motivated the introduction of the stellar evolution-based systematic uncertainties into some SFH measurements \citep[e.g.,][]{Weisz:2011, Dolphin:2012}.  Adopting time resolution based on some degree of agreement is another way to think about how well a SFH can be resolved, but it remains largely unexplored in the literature.

For recent star formation ($\lesssim$~1 Gyr ago), a time resolution of $10$–$20\%$ of lookback time is commonly used in the literature \citep[e.g.,][]{Weisz:2008, McQuinn:2010a, Lewis:2015, Cignoni:2019, Lazzarini:2022, Choi:2025, Wainer:2026}, though the exact values vary from study-to-study. The time resolution at these ages is primarily driven by the number of stars in the upper MS, BHeB, and RHeB sequences as well as the SNR of the data, assuming a fixed IMF, distance, extinction, etc. BHeBs and RHeBs are known to be poorly-modeled phases of evolution \citep[e.g.,][]{McQuinn:2011}, but these effects are not particularly well-quantified in terms of their impact on patterns of recent star formation and time resolution. Further work in this area is needed.

\subsubsection{Visualizing Star Formation Histories}

SFHs measured from CMDs are visually displayed several different ways in the literature. Building on the two-population classification of \citet{Baade:1944}, \citet{Hodge:1989} introduced the concept of population planes (lookback time vs. enrichment and lookback time vs. SFR) and the population box, a 3D representation of lookback time, SFR, and metallicity. These types of visualizations remain widely used today, with modest variations accounting for stylistic preferences. The most significant recent addition to visualizing SFHs is the cumulative SFH plot, which shows the fraction of stellar mass formed prior to a given lookback time and is discussed below.

Figure~\ref{fig:BoxPlot} shows an example SFH visualization for the LG galaxy Aquarius, including the population box (left), a differential SFR (middle), and a cumulative SFH (right). Each type of visualization has strengths and weaknesses, and it is the combination of them that permits a holistic picture of a galaxy's SFH. The population box provides a comprehensive view of how a galaxy's star formation and enrichment progressed over cosmic time. Aquarius's steady buildup of metals toward the present is evident, along with periods of strongly enhanced star formation (orange and yellow boxes) and lulls between star formation episodes.  It also provides a clear sense of the dispersion in metallicity at a fixed age (columns) and the range of ages at a fixed metallicity (rows).  The main downsides to the population box are challenges in displaying uncertainties, a requisite level of existing familiarity for interpretation, and challenges in directly comparing population boxes for larger galaxy samples.

The differential SFH plot (middle panel) is constructed by integrating the population box over metallicity. This differential plot is historically the most common way to visualize SFHs measured from CMDs. It provides a high degree of readability and is generally intuitive, even to non-experts. It enables a clear identification of features in the SFH (e.g., enhancements, lulls), and gives a visual sense of how star formation changed over cosmic time. Though shown here as a linear plot, the differential SFH can also be plotted versus logarithmic lookback time, which has the advantage of approximately uniformly spaced time bins. The main drawback to the differential SFH plot is the uncertainties.  In this example, which is measured from a well-populated, SGB-depth CMD, the uncertainties appear quite large and often significant stretches of time (e.g., from 8 to 12 Gyr ago) have SFRs that are almost all consistent with zero.  This gives the visual impression that little is known about star formation at these epochs.  The appearance of such large uncertainties is the result of SFRs in nearby time bins that can be highly correlated.  The resulting uncertainties on SFRs appear to be extremely large owing to these large covariances.  At times, this can make interpretation of the differential plot non-intuitive. For example, in galaxies with low SFRs, the SFR uncertainties at all times may be consistent with zero even though the galaxy clearly had a non-zero amount of star formation.   A second challenge with differential SFHs is in comparing multiple galaxies if the dynamic range of the SFRs is large (e.g., the LMC and a UFD might have SFRs that differ by $\sim10^3$ making it challenging to put them on the same plot).

One solution to both challenges with the differential SFH was the introduction of cumulative SFH plots\footnote{To the best of the authors' knowledge, the use of cumulative SFH plots was first suggested by Jon Holtzman at the ``Resolved Stellar Populations'' conference in Cancun, Mexico in 2005.}. An example cumulative SFH is illustrated in the right-hand panel of Figure~\ref{fig:BoxPlot}. Such plots show the fraction of stellar mass formed prior to a specified lookback time, and, by definition, are set to zero at the oldest lookback time and to one at the present day. Cumulative SFHs enable comparisons across populations and are less affected by bin-to-bin covariances than traditional SFH plots as evidenced by the grey envelope in the right-hand panel for Figure~\ref{fig:BoxPlot}.  This uncertainty envelope was constructed using the same SFR uncertainties as the central panel.  Finally, cumulative plots allow for the direct calculation of quantities such as $\tau_{50}$ and $\tau_{90}$.

One downside to cumulative SFH plots is that they are generally less intuitive to non-experts than differential SFH figures and require explanation and pedagogy to interpret clearly. Moreover, while cumulative SFHs allow for normalized comparison among larger samples, they remove the absolute scale of star formation, which can be a quantity of interest.

While the population box (left panel of Figure~\ref{fig:BoxPlot}) displays the AMR, we have not plotted the 1D AMR, which is typically a line plot of (or fit to) [Fe/H] vs. lookback time that is constructed by integrating the population box over SFR.

\subsubsection{Incorporation of Independent Metallicity Information}

The age-metallicity degeneracy can significantly affect the fidelity of CMD fitting. For certain features such as the MSTO, SGB, and RGB, younger ages or lower metallicities each make a stellar population bluer and brighter, while older ages or higher metallicities make them redder and fainter.

CMDs of galaxies typically have stellar populations that span multiple ages and metallicities, the more complex combination of which exacerbates age-metallicity degeneracies.  CMDs that reach below the oMSTO suffer the least from these effects \citep[e.g.,][]{Tolstoy:1996, Tolstoy:1998, Gallart:2005}, as the morphology of the MSTOs and SGBs, along with the color of the RGB, provides the best available, though not perfect, discrimination between age and metallicity effects.  Filter choice also matters as some combinations (e.g., g-i) are more sensitive to metallicity than others (e.g., r-i). Finally, the effects of the age-metallicity degeneracy are severe for shallower CMDs (e.g., those that only reach the HB or RC), as there is (a) limited discriminating power and (b) larger reliance on less certain phases of stellar evolution.

In principle, independent age or metallicity information should reduce the age-metallicity degeneracy.  The most common approach is to pair CMD fitting with spectroscopic metallicities of RGB stars, either by fitting the CMD and MDF simultaneously or by using the MDF as a prior on the metallicities \citep[e.g.,][]{Cole:2009, deBoer:2012, deBoer:2012b, Brown:2014}, though its efficacy is debated.  Using synthetic data, \citet{Dolphin:2016} finds only modest improvement in the recovered SFH and AMR relative to CMD fitting alone.  That study also highlights several practical challenges: spectroscopic samples are small compared to the number of stars on a CMD, selection effects are hard to model (e.g., spectra exist only for the brightest stars and often cover a different area than the imaging), and spectroscopic and CMD-based metallicities can be systematically offset.  These challenges can be overcome with appropriate care.

A complementary approach is to couple CMD-based SFHs with chemical evolution models.  \citet{Vincenzo:2016} generate synthetic CMDs from a chemical evolution model of Sculptor and compare them to the observed CMD.  \citet{Garling:2025b} take the inverse approach, using the CMD-based SFH as input to a model that links the metallicity of the star-forming gas to the cumulative stellar mass formed. Applying this framework to deep \hubble\ and \jwst\ imaging of WLM, Aquarius, Leo~A, and Leo~P, they validate the resulting enrichment histories against spectroscopic metallicities of RGB stars in the first three galaxies and quantify the systematic uncertainties due to the choice of stellar library.

\citet{Weisz:2008} introduced a different approach to mitigating the age-metallicity degeneracy by (a) parameterizing the functional form of the AMR (e.g., linear enrichment vs. linear time) and (b) requiring a monotonically increasing AMR with some allowed dispersion at each time step. This approach has been shown to produce consistent SFHs and AMRs compared to non-parametric solutions for CMDs that reach below the oMSTO \citep[e.g.,][]{Cole:2014, Weisz:2014a, Skillman:2017, Savino:2023, McQuinn:2024a}. It has been widely used to improve the fitting of shallower CMDs \citep[e.g.,][]{Gogarten:2010, McQuinn:2010a, Weisz:2011, Williams:2011a, Lewis:2015, McQuinn:2015a, Lazzarini:2022, Savino:2025, Cohen:2025, Choi:2025, Lee:2026, Smercina:2026}, and is even adopted for CMDs that reach below the oMSTO to reduce spurious fits to contamination (e.g., poorly modeled background galaxies such as those in \citealt{Durbin:2025}) that can cause ancient metal-rich populations to be fit to known metal-poor systems.

\subsubsection{Determining a Quenching Time} \label{sec:quenching}

A key measurement in CMD-based SFHs is the time at which a galaxy stops forming stars (i.e., the quenching time). Ideally, the quenching time of a galaxy is defined as the time at which 100\% of a galaxy's present day stellar mass has formed ($\tau_{100}$), at which point there are no obvious prospects of forming more stars (e.g., it is devoid of gas).

In practice, measuring $\tau_{100}$ to a high degree of confidence is challenging owing to small number statistics (e.g., low SFR, effects of a steep IMF) and/or contaminants (e.g., unresolved binary stars, compact background galaxies). A particularly troublesome contaminant is blue straggler stars (BSSs), which are highlighted in Figure~\ref{fig:cmd_pedagogy}. BSSs are the products of binary star formation or evolution (e.g., mass transfer, collisions), though the formation pathways appear to vary with environment \citep[e.g.,][]{Mathieu:2025}. Their colors and magnitudes are virtually identical to young MS stars that formed $\sim1$--$5$~Gyr ago. Both occupy the ``blue plume'' region of the CMD. When modeled as young MS stars, these blue plume populations are often assigned very low metallicities (e.g., [M/H] $\lesssim -2$) that are inconsistent with the expected AMR \citep[e.g.,][]{Monelli:2012}. Though there appear to be some differences between bulk properties of young MS stars and BSSs (e.g., young MS stars are usually more concentrated in a galaxy's center; \citealt{Mapelli:2006, Momany:2007, Santana:2013, Wyse:2020}), there is no systematic, quantitative way to discern how many stars on the blue plume are young MS stars versus BSSs. Accordingly, at present, it is not possible to reliably interpret $\tau_{100}$ as a quenching time.

Instead, it has become common to adopt a proxy for quenching time by measuring the time at which 90\% ($\tau_{90}$) or even 80\% ($\tau_{80}$) of a galaxy's total stellar mass formed \citep[e.g.,][]{Brown:2014, Weisz:2015a, Skillman:2017, Sacchi:2021, McQuinn:2023, Savino:2023, McQuinn:2024c}. Either of these metrics is highly correlated with a galaxy's true quenching time, but they are more robust to small number statistics and contaminants when modeling a CMD.  For consistency, throughout this review, we adopt $\tau_{80}$ as our standard metric for a galaxy's quenching time. 
We caution that $\tau_{80}$ is only a proxy for quenching time in galaxies that have actually stopped forming stars.  For a galaxy that is still forming stars, $\tau_{100}$ is by definition the present day, and $\tau_{80}$ instead indicates how recently the bulk of its stellar mass formed.  For example, the starburst IC~10 has $\tau_{80}\sim2$~Gyr \citep{Weisz:2014a}, despite being far from quenched.  For gas-poor systems with no ongoing star formation, e.g., most MW and M31 satellites, $\tau_{80}$ closely tracks the end of star formation and can reasonably be described as a quenching time.

\subsubsection{Spatial Considerations} \label{sec:spatial}

It is important to note that modeling the CMD of a galaxy, particularly if the CMD does not spatially cover an entire galaxy, may not reflect the true SFH of the system \citep[e.g.,][]{Rix:2013, Graus:2019}. We illustrate this through an example. Consider the common case in which a \hubble-based CMD is used to infer the SFH, but the field does not subtend the entire system, such as shown in Figure \ref{fig:fire_m11q}. Due to the dynamical evolution of stars in the galaxy, the stars located in that field at the present day may have been (or likely were) in a different region of the galaxy when they formed. In practice, what is actually measured from a CMD in this case is the distribution of ages and metallicities that fall within the \hubble\ field at the present day, as(see e.g., Figure~\ref{fig:fire_m11q}). Spatial coverage biases vary as a function of stellar population. For example, incomplete coverage of a galaxy may miss young stars, which tend to cluster, often centrally. Observing a single central field could also lead to biases for older stellar populations, which have had more time to migrate from their birth locations. We note that this spatial effect is not unique to resolved galaxy observations and it is equally important for any unresolved galaxy observations in which archaeological analysis is being used.

This spatial coverage effect can be mitigated by modeling CMDs that include the entire spatial extent of the galaxy, or a large fraction of it. In the absence of this, strategic sparser coverage (e.g., sampling of a central and outer field, radial strips) can provide a sense of the population variations.  The importance of this effect is underscored by \citet{Cohen:2026b}, who measured radial age gradients from CMD-based SFHs for 42 LV dwarf galaxies. They found that the strength of the outside-in gradient correlates with the global SFH of the galaxy.  This finding confirmed predictions from the FIRE \citep{Graus:2019} and Marvel/DC Justice League \citep{Riggs:2024} simulations.  Importantly, because the two suites disagree on when feedback-driven stellar migration operates, age gradients also offer a test of feedback models, and the bias from a single central field is largest in galaxies with extended SFHs.

The  challenge of spatial coverage notwithstanding, following literature convention, we continue to use the term SFH, even if a particular study did not have observations that cover most or all of a galaxy.

For the nearest systems (e.g., LMC and SMC) there is also a significant line-of-sight depth component \citep[e.g.,][]{Hatzidimitriou:1989, Cioni:2000, VanderMarel:2001, Nidever:2013b, Ripepi:2017, Choi:2018a, Murray:2024a}.  In these cases, a CMD is effectively a sampling through the three-dimensional structure of the system, and may include multiple kinematic, chemical, or distance components superposed in projection.  Unlike for more distant systems, these components may also be resolvable in 3D due to the close proximity of these galaxies.  For SFH determinations,  it is possible to statistically include distance and extinction along the line of sight as free parameters \citep[e.g.,][]{Cohen:2024b, Cohen:2024a} to account for the depth effects.  However, disentangling contributions from specific sub-populations requires combining CMD-based SFHs and AMRs with proper motions, radial velocities, abundances, and external distance constraints (e.g., from variable stars).  Future \textit{Gaia} data releases and the multi-tracer programs discussed in \S\ref{sec:orbital_histories} and \S\ref{sec:future} provide a possible path forward, particularly for the Magellanic Clouds and the classical MW satellites.

\subsection{Milky Way Dwarf Galaxies}
\label{sec:mw_dwarfs}

Here, we summarize the SFHs of the MW's dwarf satellite galaxies. We discuss the Magellanic Clouds and the population of UFDs directly, and provide a synthesis of the classical satellites, most of which have been the subject of extensive studies in the literature. The corresponding galaxy-by-galaxy summaries are given in Appendix~\ref{sec:app_mw_classical}.

\subsubsection{Large Magellanic Cloud} \label{sec:lmc}

First known observations of the LMC date back at least a millennium to people native to the southern hemisphere \citep[e.g.,][]{Stanbridge:1857, Norris:2016}.  Located at a distance of $\sim55$~kpc \citep[e.g.,][]{Pietrzynski:2013}, the LMC is a star-forming irregular, gas-rich galaxy \citep[e.g.,][]{McConnachie:2012}.  It is among the best-studied extragalactic systems in the Universe and is host to some of the most vigorous star formation in the local Universe.  Its proximity and modest dark matter mass are thought to have a measurable influence on the dynamics of the outer MW and many MW satellites.  Along with the SMC, the LMC is thought to be on first infall onto the MW halo \citep[e.g.,][]{Besla:2007}.

The seminal work of \citet{Harris:2009} measured the first wide-area SFH of the LMC.  They used $B-V$ and $V-I$ CMDs of over $\sim20$~million stars that cover the inner $\sim64$~deg$^2$.  This area was spatially subdivided into 1380 individual regions and CMDs.  The limiting depth of the CMDs ranges from $m_v\sim20-21$, which is moderately to slightly brighter than the expected depth of the oMSTO.  They characterized the global SFH of the LMC as an initial burst of star formation, followed by a quiescent
epoch from  $\sim12- 5$~Gyr ago.  They reported that 50\% of the LMC's stellar mass formed by 5~Gyr ago, at which point the SFH increased to a nearly constant rate to the present day.  They identified modest peaks in the SFH (at the factor of $\sim2$ level) approximately 2 Gyr, 500 Myr, 100 Myr, and 12 Myr ago.  They noted that peaks at 2 and 0.5~Gyr ago are also seen in their SFH of the SMC, which they suggested may indicate a degree of synchronization.
They measured a global AMR that featured rapid enrichment  reaching [Fe/H] $ = -1.2$ by 12 Gyr ago followed by little enrichment during the quiescent period.  Beginning $\sim5$~Gyr ago, the metallicity steadily increased until it reached a value of [Fe/H] $= -0.5$ at the present day.
\citet{Harris:2009} also explored SFHs of selected regions in the LMC such as the Bar, 30 Dor, and putative spiral arms.  In a global sense, they found only modest differences in the ancient SFHs of most regions. They provided detailed commentary on the patterns of recent SFH in each of these regions.

\citet{Rubele:2012} measured the SFHs of select regions in the LMC using deep near-IR CMDs from the VMC survey.  Specifically, they constructed CMDs in $Y, J, K_s$ filters for three $\sim1.4$~deg$^2$ fields located $\sim3.5^{\circ}$ from the center of the LMC, which they divided into $\sim40$ sub-regions.  They added a small number of sub-regions near 30 Dor.  Though the imaging formally reached the oMSTO in less crowded regions, their adopted completeness limits set the faint limit of their CMDs to be $K_s\sim20.5$, which is brighter than the oMSTO.  Their CMD modeling included distance and extinction as free parameters as well as SFH.
On aggregate, their SFHs revealed that 15-30\% of the LMC's stellar mass formed $\gtrsim10$~Gyr ago.  They identified peaks in the LMC's SFH $\sim$  6.5, 5.5, and 2~Gyr ago.  They found that more recent star formation was generally higher toward the center of the LMC, but that it showed a high degree of spatial variation.  They noted that recent star formation is generally more intense in regions near 30 Dor.
They measured an AMR for each region, but they did not assemble an aggregate AMR.  In general, their AMR had a mean metallicity of [Fe/H] $\sim -1.3$ more than 10~Gyr ago, which rapidly enriches until $\sim8$--$9$~Gyr ago when it reaches a value of [Fe/H] $\sim -0.4$.  There is modest, and sometimes no, enrichment from that point to the present day.

\citet{Weisz:2013b} measured the SFHs of 8 \hubble/WFPC2 fields (F555W, F814W) spread through the LMC from the bar to the outskirts.  Three of their fields are located in the crowded bar of the LMC.  All 8 CMDs reached $\sim2$~mag below the oMSTO.  Averaging over the fields, they found a mean SFH of the LMC such that $\sim10$\% of its stellar mass formed $>12$~Gyr ago followed by a period of nearly constant star formation from $\sim12$--$4$~Gyr ago.  There is a minor peak in the SFH $\sim5$~Gyr ago.  For ages younger than $\sim4$~Gyr, they found a significant increase in star formation such that $\sim50$\% of the stellar mass formed within the most recent $\sim4$~Gyr.  They did not analyze recent star formation given the sparsity of their fields.  Similarly, they did not publish an AMR or analyze spatial trends.

\citet{Meschin:2014} used deep V and I band imaging of 3 fields to study spatial trends in the SFH of the LMC.  The fields ranged from 3.5 to 6~kpc from the center of the LMC.  The CMD of each field reached at least 1~mag below the oMSTO.  They identified two main epochs of star formation.  The first was an ancient event that started 13~Gyr ago with a peak from $8$--$10$~Gyr ago.  This was found in all fields.  Following a lull in star formation, they reported a second event from $\sim4$--$7$~Gyr ago that lasted until the most recent $\sim1$~Gyr.  Within the last Gyr, they identified a gradient such that the inner field had the most star formation and the outer field had the least.  They reported an overall age gradient such that $\tau_{50}$ varies from $\sim 4$~Gyr in the inner field to $\sim 9$~Gyr in the outer field.  They measured AMRs for each field, but did not feature them in their analysis.

\citet{Monteagudo:2018} measured the SFH in select regions in the LMC with a focus on assessing formation scenarios for the LMC bar.  Specifically, they used 11 $B-R$ CMDs that reach $\sim0.5$~mag below the oMSTO based on imaging acquired with the Visible Multi-Object Spectrograph on the VLT.  Two fields covered the bar, while the others were spatially distributed out to $3.2^{\circ}$.  They also analyzed archival \hubble/WFPC2 data that overlapped these regions to assess the robustness of their solutions.   They focused on SFH variations within the bar and between the disk and the bar.  They found that the central and outer bar fields have virtually identical SFHs.  Compared to the disk fields, they found at most, modest differences with the bar for most epochs.  The main differences appeared to be at very young ages.  They concluded that there is no clear star formation event that can distinguish the bar from their disk fields.

\citet{Ruiz-Lara:2020b} used deep $g-i$ CMDs from the SMASH survey to measure the SFHs of the northwest and southeast regions of the LMC located $\sim4$~kpc from the galaxy's center in order to study the spiral arm structures.   Both CMDs reach $\sim2$~mag below the oMSTO.  They found that the SFHs of the two regions were virtually identical until $\sim2$~Gyr ago, indicating a well-mixed older population.  Within the last $\sim2$~Gyr, the northern (arm) region exhibited a high rate of star formation compared to the southern (non-arm) region.  They identified low-level peaks in the arm region at $\sim$2, 1, 0.5, and 0.1 Gyr ago, with hints of similar features in the non-arm region. They reported that their general trends are in good agreement with past SFH studies \citep[e.g.,][]{Harris:2009, Weisz:2013b, Meschin:2014, Monteagudo:2018}, but with improved precision due to the superior data quality of SMASH. They did not report an AMR.

\citet{Mazzi:2021} derived the SFH of the LMC over a 96~deg$^2$ area using near-IR imaging from the VMC survey.  They divided this area into 756 sub-regions and measured the SFH from $J-K_s$ and $Y-K_s$ CMDs of each region.   They fit for SFH, distance, and extinction in each region.  CMDs outside the most crowded regions extended to the oMSTO.  In their fitting, they adopted a fiducial AMR from \citet{Carrera:2008b} and allowed their code to compute shifts at each age relative to this reference AMR. In general, they found a global SFH that is in reasonable agreement with the previous wide-area SFH of \citet{Harris:2009}, with a few differences.  First, they found a smoother variation in a SFH at most epochs and a lower level of star formation over the most recent $\sim1$~Gyr.  Second, they reported that the most intense period of star formation occurred $\sim4$ to $\sim0.5$~Gyr ago.  Third, they noted that star formation within the last $\sim100$~Myr is patchy in a spatial sense, while the structure of arms and the bar are not visible in their SFH maps for ages $\gtrsim1.6$~Gyr ago.  They did not find evidence for a significant metallicity gradient over their entire survey area.

As part of the Scylla program, \citet{Cohen:2024a} measured the SFH of the LMC from 111 new and archival \hubble\ fields that were spatially distributed across the LMC from the central bar to the inner stellar halo.  The archival nature of the data means that several bluer filters were used (F475W, F555W, F606W) paired with a common red filter (F814W).  In all cases, the CMDs reached at least 1 mag below the oMSTO.  They fit for extinction and distance in each field along with the SFH.  They also used multiple stellar models as a check on the robustness of their solutions.  For the global SFH of the LMC, they reported good agreement with most past studies, including an early enhancement of star formation $>12$~Gyr ago followed by a quieter phase of nearly constant star formation from $\sim$3.5-12~Gyr ago, with minor peaks in the SFH at $\sim8$ and $\sim5$~Gyr ago.  They found a dramatic increase in the SFH beginning around $\sim3.5$~Gyr ago, that continues until very recent ages, depending on spatial location. They commented that the timing of particular bursts depended on the underlying stellar model used.  They also noted that the largest level of disagreement, which is modest, was with the SFHs determined from the VMC data \citep{Mazzi:2021}.  They commented that their global AMR is consistent with results from past spectroscopic studies \citep[e.g.,][]{Carrera:2008b}, despite not having imposed any assumptions about the AMR in their CMD fitting.

A major finding from \citet{Cohen:2024a} is the inversion of an age gradient.  Specifically, they reported a
present-day inside-out age gradient in the inner disk and an outside-in age gradient for radii larger than $\sim2$ disk scale lengths.  They noted that although this gradient is less pronounced toward older lookback times, it is detectable to ages of at least $\sim10$~Gyr ago.  In addition to this work, \cite{Burhenne:2026} measured SFHs of the LMC and SMC using Scylla data in search of bursts.  In the LMC, they identified one global burst $\sim5$~Gyr ago.

\subsubsection{Small Magellanic Cloud} \label{sec:smc}

Along with the LMC, observations of the SMC date back at least a millennium by people native to the southern hemisphere \citep[e.g.,][]{Stanbridge:1857, Norris:2016}.  Located at a distance of $\sim60$~kpc \citep[e.g.,][]{Graczyk:2020}, the SMC is a gas-rich, star-forming, irregular galaxy \citep[e.g.,][]{McConnachie:2012} that appears to extend $\sim10$~kpc along the line-of-sight \citep[e.g.,][]{Hatzidimitriou:1989}.  Recently, \citet{Murray:2024a} suggested that the SMC may be two star-forming systems that are superimposed along our line-of-sight.

The stellar populations of the SMC have been the target of numerous studies.  There are several quantitative SFH measurements that either pre-date \citep[e.g.,][]{Dolphin:2001, McCumber:2005, Chiosi:2006, Chiosi:2007} our review period or focus on clusters \citep[e.g.,][]{Sabbi:2007, Cignoni:2009, Rubele:2010}.  Other studies have undertaken a more qualitative approach to estimating the stellar age distribution in the SMC \citep[e.g.,][]{Noel:2007, Sabbi:2009}.  Here, we focus on summarizing studies that have used quantitative CMD modeling to measure SFHs of the SMC's field population.

Although it pre-dates the period of this review, we believe it important to summarize the first wide-area SFH of the SMC, which was presented in \citet{Harris:2004}.  This work used UBVI CMDs that covered the central $\sim18$~deg$^2$.  The deepest CMDs they analyzed were $\sim2$~mag brighter than the expected luminosity of the oMSTO.
Their global SFH was characterized by 50\% of the stellar mass forming prior to 8.4~Gyr ago, a lull from $\sim8$--$3$~Gyr ago, and an increasing SFH beginning $\sim3$~Gyr ago.  They identified bursts at 2.4, 0.4, and 0.06 Gyr ago.  They reported a global AMR with a mean metallicity that changes from [Fe/H] $\sim -1.0$ at $\sim12$~Gyr ago to [Fe/H] $\sim -0.5$ at $<1$~Gyr ago.  They noted slow enrichment until $\sim3$~Gyr ago, followed by a rapid increase.  They also highlighted a decrease in metallicity from $\sim7$--$4$~Gyr ago. They noted the existence of a spatial ``ring'' of stars with ages of $\sim2.5$~Gyr in their maps.  Their spatially resolved SFH maps have proven invaluable to the community over the past two decades.

\citet{Noel:2009} used ground-based imaging of 12 fields for a combined area of $\sim0.25$~deg$^2$ to measure the SFH of the SMC.  The fields were spatially distributed from $\sim1.1$ to $\sim4.5$~kpc from the center of the SMC and do not include any regions in the center due to crowding. The $B-R$ CMDs for each field reached the oMSTO. Each field was analyzed separately to preserve spatial information. They identified four main enhancements in the SFH of the SMC: (i) A peak at $\sim10$~Gyr in all fields, except the western field which shows peaks at $\sim12$ and $\sim8$~Gyr ago; (ii) an intermediate-age peak in all fields at $\sim4$--$5$~Gyr ago located in all fields; (iii) a peak from 200--500~Myr ago located primarily in the inner fields; (iv) a very young peak located in the fields closest to the SMC.  They note general agreement with previous SFHs that reach the oMSTO in similar locations in the SMC \citep[e.g.,][]{Dolphin:2001, Chiosi:2006, Chiosi:2007}, and disagreement with the \citet{Harris:2004} SFH at older and intermediate ages.  They suggest this disagreement could be due to the shallower CMDs used by \citet{Harris:2004}.   They report AMRs for the eastern, western, and central portions of the galaxy that broadly increase from [Fe/H]$\sim-1.5$ at the oldest ages to [Fe/H]$\sim-0.6$ at the youngest ages.  They note good agreement with spectroscopic stellar metallicities in the literature.

\citet{Cignoni:2012} measured the SFHs of two \hubble/ACS fields located in the central bar of the SMC, based on data presented in \citet{Sabbi:2009}.  Their F555W-F814W CMDs reach several magnitudes below the oMSTO.  For both fields they found a long quiescent phase characterized
by low SFRs until 5--6 Gyr
ago, when both fields experience a rapid increase in star formation.  Following this increase, one bar field exhibits a rising SFH to the present day, while the other shows a declining SFH.  In both fields, they report an AMR that increases from [Fe/H] $\sim-1.7$ at $\sim12$~Gyr ago to $\sim-0.6$ at the present day.  They note similarities in their SFH to those of \citet{Chiosi:2006} and \citet{Chiosi:2007}.

\citet{Cignoni:2013} built on their previous study of the SFHs in the SMC bar by measuring the SFHs of 4 additional \hubble/ACS fields that sample different locations in the SMC out to $\sim2^{\circ}$ from its center.  Each F555-F814W CMD reached well-below the oMSTO. They found that all fields formed $\lesssim20$\% of their stellar mass prior to 10 Gyr ago, and that there is a shift in median age such that it shifts from $\sim4$ to $\sim$6~Gyr ago from the inner to the outer fields.  They found little difference in the lifetime SFHs of the bar vs slightly off-bar central fields.  They reported an average AMR  over all fields that increases from [Fe/H] $\sim-1.7$ at $\sim12$~Gyr ago to $\sim-0.7$ at the present day along with a nearly flat metallicity gradient.

As part of a comparative study between the SFHs of the SMC and LMC, \citet{Weisz:2013b} measured the SFH of the SMC using 7 \hubble/WFPC2 fields (F555W, F814W).  Three of these fields were in the main body (i.e., within 1~kpc) of the SMC and four were in the outer disk.  The CMDs all reached $>2$~mag below the oMSTO.  From the combined 7 fields, \citet{Weisz:2013b} found that the SFH of the SMC was delayed at early times such that 50\% of its stellar mass formed only $\sim4$~Gyr ago.  Prior to this, the star formation levels were low and nearly constant, with small enhancements at 4.5 and 9 Gyr ago.  They reported a rapid rise in the SFH beginning $\sim4.5$~Gyr ago that continues to the present day. Their overall SFH of the SMC was broadly similar to that of \citet{Noel:2009}, even though the spatial locations of their fields did not generally overlap. They did not publish an AMR or consider spatial information.

\citet{Rubele:2015} measured the first spatially comprehensive SFH of the SMC using CMDs that reached below the oMSTO using data from the VMC survey \citep{Cioni:2011}.  Their data cover 14~deg$^2$ of the main body and wing of the SMC, which they divided into 120 regions for the purposes of this study in order to characterize spatial trends.  In each region they model the near-IR CMDs ($K_s$ vs.\  $Y - K_s$ and $K_s$ vs.\ $J - K_s$) to measure the SFH, distance, and line-of-sight extinction.  They model all CMDs with $K_s < 20.5$ to ensure that the completeness is always $>50$\%.  Although this depth generally includes the $\sim10$~Gyr MSTO in most fields, the expected oMSTO is not part of their modeling.

For the global SFH, they find a modest level of ancient star formation, similar to conclusions from previous work \citep[e.g.,][]{Noel:2009, Cignoni:2012, Cignoni:2013, Weisz:2013b}.  They report that 50\% of the stellar mass formed by $\sim5$~Gyr ago.  They reported a strong enhancement within star formation $\sim5$~Gyr ago, similar to other SFHs. They detected enhancements of star formation at $\sim5$ and $\sim1.5$~Gyr ago.  They detected recent bursts of star formation $\sim0.2$~Gyr ago in the wing and $0.04$~Gyr ago in the bar.  All of these features are similar to those reported in \citet{Harris:2004}, but with slightly different ages.  They did not find evidence for the $\sim2.5$~Gyr old ring reported by \citet{Harris:2004}, and suggest that it may be an artifact of shallower photometry.  They also note that their SFRs at intermediate ages agree with SFRs derived from LPVs \citep{Rezaeikh:2014}.

\citet{Rubele:2015} reported a global AMR that increases from a mean [Fe/H] $\sim -1.6$ at $\sim12$~Gyr ago to [Fe/H] $\sim -0.3$ at $<1$~Gyr ago.  They noted a decrease in the metallicity $\sim0.8$~Gyr ago, which they suggested could be due to the accretion of lower-metallicity gas $>1$~Gyr ago.  They note good agreement with spectroscopic metallicities.

\citet{Rubele:2018} built on their previous work by increasing the areal coverage to cover a contiguous area of 23.57~deg$^2$ that extends to an angular distance of $\sim3.5^{\circ}$ from the center of the SMC.  Though this work supersedes that of \citet{Rubele:2015}, they noted that there are only modest differences in the SFH and AMR, which are generally more precise in the updated analysis.   They noted that, in this new SFH, 50\% of the stellar mass formed by $\sim6.3$~Gyr ago and that there remains a marked upturn in star formation beginning $\sim5$~Gyr ago.  The new AMR is consistent with the previous AMR within the reported uncertainties.  They noted increased tensions with the \citet{Harris:2004} SFH both in terms of timing, duration, and total mass formed, which they attribute to a combination of different choices in IMF and stellar models, along with shallower CMDs.

\citet{Massana:2022} measured the SFH of the SMC using data from the SMASH survey \citep{Nidever:2017}.  Outside the very center, their CMDs reach $>1$~mag below the oMSTO and cover 31~deg$^2$ and extend to an angular distance of $\sim4^{\circ}$ from the SMC's center.  These are the uniformly deepest ground-based optical CMDs of the SMC to date.  They measured the SFH of the SMC by dividing their data into 71 regions and measured the SFH of each region using a $g-i$ CMD.

Overall, they found a global SFH that is characterized by low-level star formation $>10$~Gyr ago, enhanced star formation from 6--10 Gyr ago followed by a lull, and a notable increase in star formation beginning $\sim3.5$~Gyr ago.  They also noted several distinct episodes of star formation including $3.5$, $\sim3$, $\sim2$, $\sim1.1$, and $\sim0.45$ Gyr ago, as well as one occurring at the present.  They contrasted their highly episodic star formation with that of \citet{Rubele:2018}, which shows smoother variations, and attributed differences primarily to their superior depth of data.   They did not undertake a detailed discussion of spatial trends or present an AMR.

\citet{Sakowska:2024} used SMASH $g-i$ CMDs that reach the oMSTO to measure the SFH of the northeastern shell in the SMC.  This region is known to feature a large population of young stars. Their SFH is consistent with no ancient ($>12$~Gyr ago) star formation and 40\% of the stellar mass having formed within the most recent $\sim1$~Gyr. They reported most stars formed within the most recent $\sim500$~Myr, with significant star formation enhancements at $\sim450$ and $\sim250$ Myr, which they  attributed to the interaction history of the SMC with the LMC and MW.  They also report a significant increase in metallicity $\sim450$~Myr ago.

\citet{Cohen:2024b} measured the spatially resolved lifetime SFH of the SMC using 83 \hubble\ imaging fields taken as part of the Scylla survey \citep{Murray:2024b} and using deep archival imaging.  The optical CMDs in all cases reached $>1$~mag below the oMSTO providing excellent constraints on the SFH of each field.  The spatial distribution of the 83 fields includes the crowded bar, the wing, and fields in the outer disk at a distance of $\sim4$--$5$~kpc from the SMC's center.

The general SFH of the SMC from \citet{Cohen:2024b} is in good agreement with other studies, particularly the SMASH results of \citet{Massana:2022}. The spatial sampling and excellent temporal resolution from \citet{Cohen:2024b} reveal an outside-in age gradient in the SMC with gradients of $\delta \tau_{50} = 0.82^{+0.12}_{-0.16}$ Gyr kpc$^{-1}$ and $\delta \tau_{90} = 0.61^{+0.08}_{-0.07}$ Gyr kpc$^{-1}$, which they argued are driven primarily by internal processes, with some individual star forming events that appear to have common timing with the LMC.  \citet{Cohen:2024b} also measured a global AMR that ranges from [Fe/H] $\sim -1.5$ at $\sim12$~Gyr ago to [Fe/H] $\sim -0.3$ at $<1$~Gyr ago, in good agreement with other CMD-based AMRs.  \citet{Cohen:2024b} also considered the effects of different stellar models on their SFH results and tested for the impact of line-of-sight distance effects.

\citet{Burhenne:2026} used deep CMDs from the Scylla survey to search for local and global bursts in the SMC and LMC.  They adopted the SFH methodology described in \citet{Cohen:2024a}.  In the SMC, they identified two global bursts at $\sim1.5$ and $\sim5$~Gyr ago, and one burst that was spatially localized $\sim3$~Gyr ago.

\citet{Ficara:2026} measured the first SFH of the Magellanic Bridge over 14~deg$^2$ using $g$ and $i$ band imaging from the STEP program.  All their CMDs reached below the oMSTO.  They reported a clear peak in the recent SFH at $\sim100$~Myr in the Bridge, that increases in intensity toward the SMC.  They identify this young population as anomalously low in metallicity and suggest it may have formed from gas stripped from the SMC.  They also identified two older peaks in star formation at $\sim2$ and $\sim10$~Gyr ago, that are near the LMC.  They estimated a total stellar mass of the Bridge to be $\sim5\times10^5$\msun.

In general, published CMD-based AMRs and metallicities are in good agreement with spectroscopic metallicities reported for the SMC \citep[e.g.,][]{Carrera:2008, Dobbie:2014, Nidever:2020, Hasselquist:2021, Navabi:2025}.

\subsubsection{Classical Satellites} \label{sec:mw_classical}

Historically, there have been 9 ``classical'' satellite galaxies ($D\lesssim 300$~kpc, $M_V\le-7.7$, excluding the LMC and SMC) that orbit the MW.  Within the past $\sim20$ years, three more MW satellites have been discovered that formally qualify as ``classical'' satellites: Canes Venatici~I (\citealt{Zucker:2006b}; $M_V\sim-8.8$, \citealt{munoz:2018b}), Crater~II ($M_V \sim -8.0$; \citealt{Torrealba:2016}), Antlia~II ($M_V \sim -9.0$; \citealt{Torrealba:2019}), bringing the count to 12 classical MW satellites.  Of these, Crater~II and Antlia~II are very low-surface brightness and do not have published SFHs. We do not discuss them further.

MW satellites are among the best studied dwarf galaxies in the Universe, owing to their close proximity and long time since discovery.  Because these galaxies are so well-studied, we summarize our knowledge of each galaxy's SFH from the literature on a galaxy-by-galaxy basis in Appendix~\ref{sec:app_mw_classical}.  Here, we synthesize the main results.

Virtually every classical MW satellite has been the subject of multiple SFH measurements from CMDs that reach below the oMSTO. In many cases, these include both pencil-beam \hubble\ fields near the galaxy center \citep[e.g.,][]{Weisz:2014a, Ruiz-Lara:2021, Rusakov:2021} and wide-area ground-based imaging that extends to, or beyond, the tidal radius \citep[e.g.,][]{Aparicio:2001, Carrera:2002, deBoer:2012, deBoer:2012b, deBoer:2014, Bettinelli:2018, Bettinelli:2019, Sato:2025}.

Taken together, these studies establish that the classical satellites are not a homogeneous population.  Instead, they span nearly the full range of SFHs seen among low-mass galaxies.  At one extreme, Ursa~Minor, Draco, and Sextans are almost entirely ancient, having formed $\gtrsim80$--$90$\% of their stellar mass by $\sim10$--$11.5$~Gyr ago.  CVn~I is also predominantly ancient, having formed $\sim90$\% of its stellar mass by $\sim10$~Gyr ago, but its star formation peaked later, $\sim11.8$~Gyr ago, i.e., after the end of reionization (\citealt{Perez-Garcia:2026}; Appendix~\ref{sec:CVnI_sfhs}).  At the other extreme, Fornax and Leo~I formed the bulk of their stars at intermediate ages, with $\tau_{50}\sim5$--$9$~Gyr and had star formation persist until $\lesssim1$~Gyr ago.  Sculptor, Leo~II, Carina, and Sagittarius fall between these extremes.  Notably, several of these galaxies (e.g., Carina, Leo~I, Leo~II, Sagittarius) show clearly episodic SFHs, with distinct periods of enhanced star formation separated by lulls of several Gyr.

There are several trends across this sample.  First, where wide-area data exist, nearly every galaxy shows a radial age gradient in the sense that younger and/or more metal-rich populations are more centrally concentrated \citep[e.g.,][]{deBoer:2012, deBoer:2012b, deBoer:2014, Santana:2016, Ruiz-Lara:2021, Rusakov:2021}.  This underscores the importance of spatial coverage as SFHs from a single central \hubble\ field can differ from wide-area measurements of the same galaxy, in some cases by several Gyr in $\tau_{80}$ (e.g., Ursa~Minor; \citealt{Weisz:2014a, Sato:2025}).  Second, the AMRs measured from CMDs are in good agreement (i.e., within $\sim0.3$~dex) with spectroscopic metallicities where the two have been compared over matched areas \citep[e.g.,][]{deBoer:2012, deBoer:2014, Santana:2016, Sato:2025} and when $\alpha$-enhancements are considered.  Third, for the predominantly ancient systems, there remains a persistent ambiguity in interpreting the ``blue plume'' as either BSSs or a sparse population of $\sim1$--$3$~Gyr old stars.  Deep CMDs generally favor BSSs \citep[e.g.,][]{Mapelli:2007, Bettinelli:2018, Sato:2025, Perez-Garcia:2026}, but claims of young CHeB stars from \textit{Gaia} \citep[e.g.,][]{Yang:2024} provide an intriguing counterpoint.  This distinction affects our knowledge of when, and perhaps how, these galaxies quenched as discussed in \S \ref{sec:quenching}.  Finally, only a handful of MW satellites have deep CMDs across their full spatial extent.  There is still a need for wide-field imaging that reaches high SNR (e.g., $\gtrsim80$) at the oMSTO over the full extent of each galaxy.  As we discuss in \S \ref{sec:future_formation}, spatially complete SFHs of the highest quality for all classical satellites are within reach of current wide-field facilities.

\subsubsection{Ultra-Faint Dwarf Galaxies} \label{sec:mw_ufds}

There are $\sim60$ known UFDs ($M_V>-7.7$) that reside in the halo of the MW at $D\lesssim 300$~kpc.  Most appear to be MW satellites, though some may be associated with the LMC or on first-infall \citep[e.g.,][]{Patel:2020}. Of the $\sim60$ known UFDs, $\sim60\%$ currently have published CMD-based SFHs.

The discovery of UFDs around the MW starting in 2005 was a watershed moment in our understanding of the dark matter and galaxy formation at extremely low-masses \citep[e.g.,][]{Willman:2005a, Willman:2005b, Belokurov:2006a, Zucker:2006a, Zucker:2006b, Belokurov:2007}.  Theoretical studies of structure formation suggested there should be a wealth of sub-structure in the form of dark matter halos \citep[e.g.,][]{Klypin:1999, Moore:1999}, and possibly faint galaxies, around the MW.  But as of the early 2000s only the 11 ``classical'' MW satellites (including the LMC and SMC) were known.  This ``missing satellites'' problem was widely viewed as a credible challenge to the $\Lambda$ cold dark matter ($\Lambda$CDM) cosmology on small scales and motivated numerous theoretical studies on the possibility that reionization could suppress galaxy formation in such low-mass halos \citep[e.g.,][]{Efstathiou:1992, Bullock:2000, Benson:2002, Somerville:2002}.  The discovery history of UFDs is covered in detail in \citet{Simon:2019}.  Here, we focus on the subsequent measurements of their SFHs.

The first quantitative CMD-based SFHs of UFDs around the MW were presented by \citet{deJong:2008}.  This study applied the CMD fitting code \texttt{MATCH} to the SDSS discovery data of 12 MW UFDs.  The code was used to constrain distance, age, and metallicity of each system.  The SDSS data is quite shallow, only reaching the oMSTO for the nearest objects.  Moreover, the CMDs contain large amounts of contamination from MW foreground stars and compact background galaxies.  When fixing the distance, \citet{deJong:2008} found  that Bootes~I, CVn~II, Coma~Berenices, and Hercules have stellar populations with ages $>10$~Gyr and metallicities near [Fe/H] = $-$2.  For a fixed distance, they were able to make similar conclusions for Segue~I, Ursa~Major~I, Ursa~Major~II, and Willman~I, though in some cases they could not rule out the possibility of slightly more metal-rich ([Fe/H] = $-$1.6) populations.

The systematic measurement of LG dwarf galaxy SFHs using archival \hubble/WFPC2 data by \citet{Weisz:2014a} includes three MW UFDs (CVn~II, Hercules, Leo~IV).  All data reach the oMSTO, though generally with modest SNR.  They found that Hercules and Leo~IV are almost entirely ancient, modulo contributions from BSSs, but that CVn~II shows evidence for younger populations.  Their SFH of CVn~II extends to ages of $\sim8$~Gyr ago.

The first systematic SFH measurements for MW UFDs using deep \hubble/ACS data were conducted by \citet{Brown:2014}. They obtained deep imaging of 6 MW UFDs in F606W and F814W such that the typical SNR at the oMSTO was $\sim100$.  The resulting CMDs are of far higher quality and depth than the SDSS and WFPC2 data analyzed in other studies.  \citet{Brown:2014} found that the SFHs for 5 of their UFDs are remarkably similar with at least 75\% of their stellar mass forming by 13.3~Gyr ago, and all 6 galaxies consistent with 80\% of their stellar mass forming by 12.8 Gyr ago.  All galaxies quenched by 11.6~Gyr ago.  The SFH of Ursa~Major~I shows a modestly younger mass-weighted age ($12.7\pm0.3$~Gyr ago) than the other five galaxies (all $\ge13.1$~Gyr ago).  Unlike \citet{Weisz:2014a}, they find no evidence for a significantly extended SFH in CVn~II.  Overall, \citet{Brown:2014} concluded that the comparably ancient ages of these galaxies are consistent with a global and external quenching event, such as cosmic reionization.

Using deep F606W and F814W ACS data from the \hubble\ MW Halo Treasury survey, \citet{Sacchi:2021} undertook the second systematic SFH measurements of MW UFDs.  They analyzed 7 UFDs (Hor~I, Hydra~II, Phoenix~II, Reticulum~II, Sagittarius~II, Triangulum~II, and Tucana~II)\footnote{There are some suggestions in the literature, based on size and chemistry, that Sag~II and Tri~II may be GCs and not genuine UFDs. Observationally distinguishing between the two types of objects can be quite challenging \citep[e.g.,][]{Simon:2019}.}.  The CMDs of these systems all reach well-below the oMSTO.  The main aims of this study were to (a) compare the timing of star formation with the epoch of reionization and (b) to search for putative differences in the SFHs of UFDs kinematically associated with the MW and the LMC.

\citet{Sacchi:2021} found that all seven systems have SFHs that are consistent with at least 80\% of the stellar mass having formed by 12.6~Gyr ago and all quenched by 11.5~Gyr ago.  Like \citet{Brown:2014}, this paper concluded these SFHs are consistent with external quenching due to reionization.  However, we note that whereas \citet{Brown:2014} adopted an oldest allowable age of 14.1~Gyr, \citet{Sacchi:2021} adopted a value of 13.7~Gyr.  This subtle difference complicates a direct comparison between the studies.

\begin{figure}[pos=tbp]
    \centering
\includegraphics[width=1.0\linewidth]{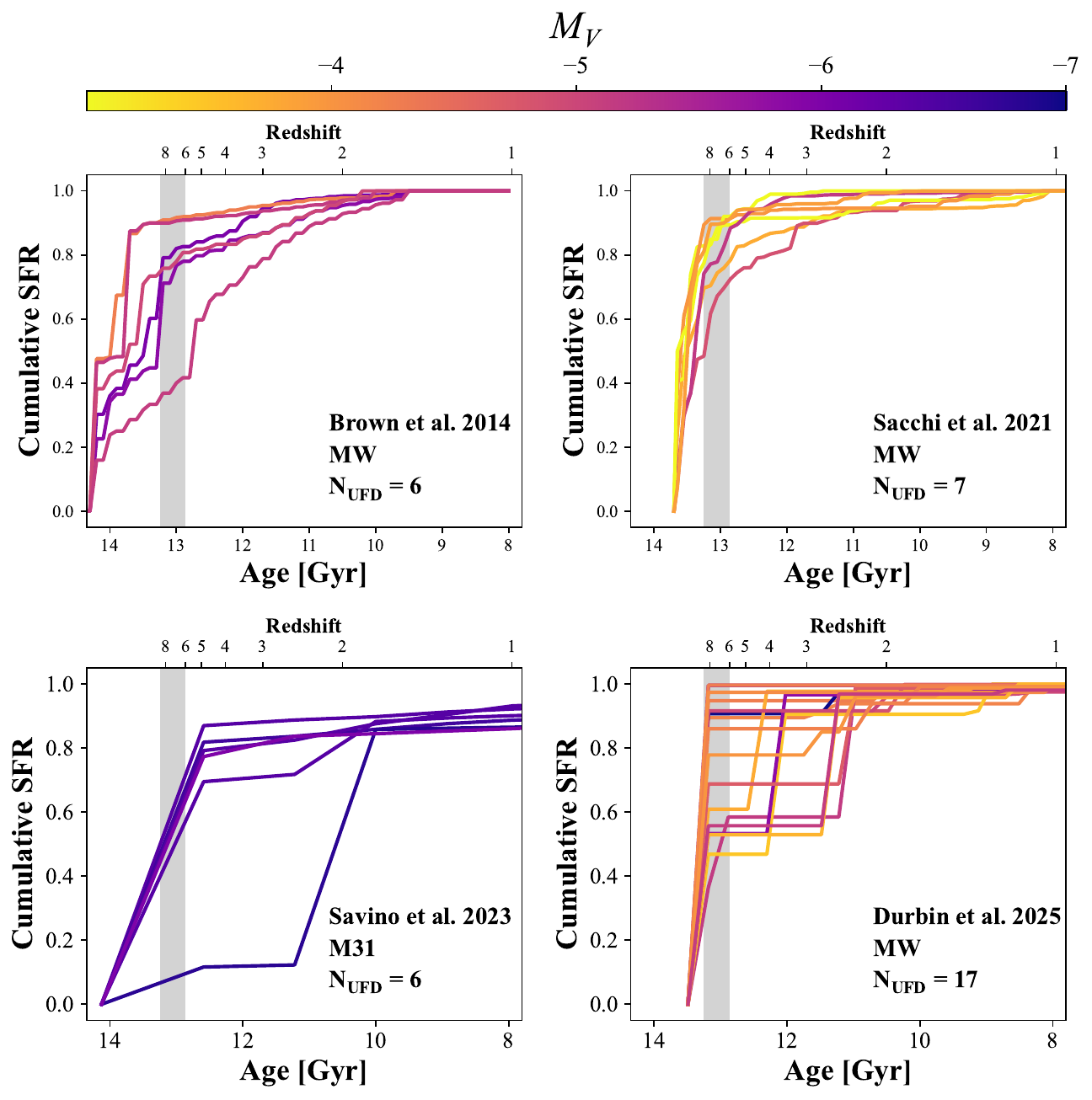}
    \caption{The SFHs of MW and M31 UFDs  from the literature, color-coded by present-day luminosity. The epoch of reionization is indicated by the grey band.  We note that Eri~II, located just beyond 300~kpc from the MW, is included.  The SFHs are grouped by the original publication and ordered chronologically from upper left to lower right.  The \citet{Brown:2014} and \citet{Sacchi:2021} solutions are the median from a set of SFH realizations, while the \citet{Savino:2023} and \citet{Durbin:2025} solutions are best-fit SFHs.  The \citet{Durbin:2025} panel shows the 17 UFDs with effective $M_V\le-2.5$, the subset with the most robust SFHs.  Uncertainties have been omitted for clarity. }%
    \label{fig:UFDs}
\end{figure}
\citet{Sacchi:2021} found that the mean quenching time of UFDs associated with the LMC is $\sim0.6$~Gyr younger than UFDs associated with the MW, though at low statistical significance.    They tentatively suggested it may be possible that the LMC-associated UFDs experienced a weaker ionization field than those associated with the MW.  There is some ambiguity as to which UFDs were associated with the LMC, MW, or neither at the time of reionization, as the uncertainties on orbital histories can be large at such old lookback times (see \S \ref{sec:orbital_histories}).

As part of the \hubble\ Legacy Archival Uniform Reduction of Local Group Imaging (LAURELIN), \citet{Durbin:2025} measured uniform SFHs of 36 UFDs located within $\sim400$~kpc of the MW.  This study performed uniform stellar photometry in the F606W and F814W bands on hundreds of existing \hubble\ orbits of data, measured distances using a self-consistent isochrone fitting method, and measured homogeneous SFHs for all 36 systems. As illustrated in Figure~\ref{fig:UFDs},   they found good agreement between their SFHs and those from previous studies \citep[e.g.,][]{Brown:2014, Sacchi:2021} for the $\sim12$ overlapping galaxies.  Across their entire sample, \citet{Durbin:2025} found a mean quenching time of $12.48  \pm  0.18$~Gyr ago ($z=4.6^{+0.6}_{-0.5}$) which is compatible with scenarios in which reionization plays an important role in quenching.  They measured a systematic difference in the quenching times of UFDs thought to be satellites of the LMC or those on their first infall versus long-term MW satellites.  They discussed how these measurements can be used to constrain various scenarios for non-uniform reionization.

Beyond establishing broad trends, a handful of studies focus on other astrophysical applications of UFD SFHs. One example is that of Eri~{\sc II}, which is a distant and luminous UFD that hosts a central GC.  \citet{Simon:2021} and \citet{Gallart:2021} each measured the SFH of Eri~{\sc II} using different \hubble\ data that reach below the oMSTO.  They both found that Eri~{\sc II} is dominated by stars older than $\sim12$--$13$ Gyr, with very low levels of star formation that extended to $\sim9$--$10$~Gyr, but are consistent with no star formation.  \citet{Weisz:2023b} re-analyzed the same \hubble\ data and reached the same conclusion, but assigned a slightly higher significance ($\sim1 \sigma$) to the 9~Gyr population.  They also reported that the GC and field population had the same age and metallicity. \citet{Durbin:2025} reported a SFH similar to others in the literature, with low-significance star formation from $\sim9$--$11$~Gyr ago.

Another example is Ret~{\sc II}, whose stellar population is largely dominated by stars highly enriched in $r$-process elements.  \citet{Simon:2023} measured a detailed SFH from multiple deep \hubble\ fields in order to place constraints on plausible physical mechanisms responsible for $r$-process enhancements.  From the SFH and spectroscopic data, they constrained the delay time between the formation of the first stars and the $r$-process production to be $\lesssim500$~Myr, which is incompatible with r-process sources with delay times of several Gyr (e.g., GW170817).

A persistent challenge in measuring the SFHs, and other characteristics, of UFDs is their paucity of stars.  UFDs often have a few hundred observationally accessible stars, and, in some cases only a few dozen.  As a result, even basic parameters such as distance and dependent characteristics (e.g., luminosity, SFH), can have modest to large uncertainties.  In terms of SFHs, both \citet{Brown:2014} and \citet{Durbin:2025} find that at least $\sim100$ MSTO stars are needed to measure a robust SFH, which \citet{Durbin:2025} translate to a minimum luminosity over the field-of-view to be $M_V \le -2.5$.  One possibility for improving SFH precision is to obtain much deeper imaging that reaches the MS kink.  The MS kink is known to be metallicity sensitive and will have more stars than the MSTO. In theory, such deep imaging could help improve the SFHs of very sparse systems, though it remains untested in practice.

\subsection{The M31 Ecosystem}
\label{sec:m31_ecosystem}

\begin{figure}[pos=tbp]
    \centering
    \includegraphics[width=1.0\linewidth]{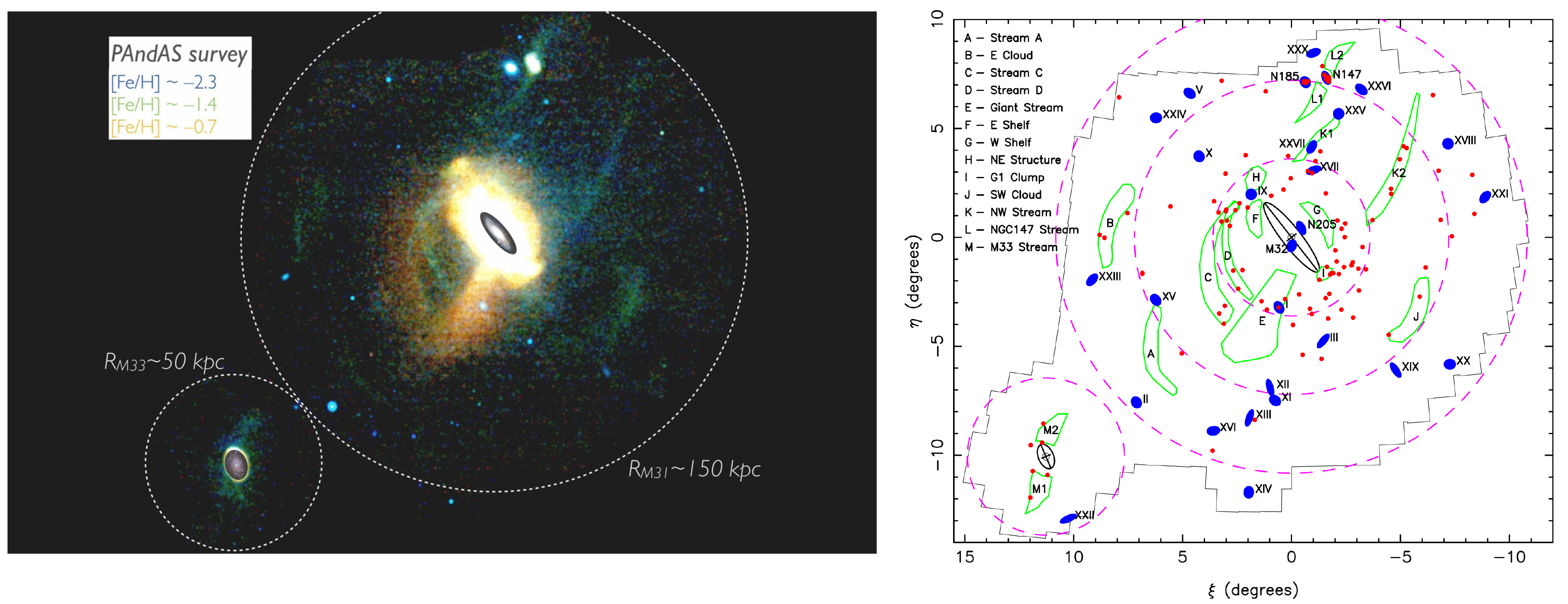}
    \caption{Maps of the M31 ecosystem produced by the PAndAS survey.  \textbf{Left:} A stellar density map of RGB stars color-coded by metallicity \citep{Martin:2013}.  Metallicities were estimated from broadband colors of RGB stars.  Note the rich structures in the M31 system from blue point sources (metal-poor satellites galaxies or globular clusters) to complex linear features to shells and metal-rich tidal debris. \textbf{Right:}  A map of known structures in the M31 system as of 2018 \citep{McConnachie:2018}. As we discuss in \S \ref{sec:m31_ufds}, since the publication of these figures a number of new, faint systems in the M31 halos have been discovered and the 3D distances of the satellites have been updated.}
    \label{fig:m31_map}
\end{figure}

M31 is the nearest massive galaxy to the MW.  It is close enough that it can be resolved into individual stars using a combination of ground- and space-based facilities.  Much of the inner halos of M31 and M33 has been mapped into resolved RGB stars by the PAndAS survey, with pencil beam \hubble pointings resolving stars down to the oMSTO in a handful of regions.  The resolved stellar populations in the disks of M31 and M33 have been mapped with \hubble, as have most of the known satellite galaxies.  We describe the results of these programs in detail below.  As a point of reference for the following discussion, Figure~\ref{fig:m31_map} displays two maps of the M31 ecosystem produced by the PAndAS survey.

\subsubsection{M31} \label{sec:m31}

M31, commonly referred to as the Andromeda galaxy, is the closest large spiral galaxy to the MW at $\sim 770$ kpc \citep[e.g.,][]{deGrijs:2014,Li:2021b,Savino:2022} and the most massive member of the LG, with a total stellar mass of $\sim 10^{11} M_{\odot}$ \citep{Tamm:2012}, and a neutral hydrogen (\hi{}) mass  of $\sim 7 \times 10^9 M_{\odot}$ \citep{Braun:2009}.  The oldest surviving record of an observation of M31 is the description by Abd al-Rahman al-Sufi in 964~AD, who reported it as a ``small cloud'' visible to the naked eye \citep[e.g.,][]{Jones:1991}.

Despite the broad morphological similarities to the MW, there is a large body of evidence demonstrating that M31 has followed a somewhat different evolutionary path, with meaningful differences in, e.g., the number, timing, and strength of its environmental interactions; its metal enrichment history; and the rate of stellar mass build-up across cosmic time \citep[e.g.,][]{Ferguson:2002,Brown:2003, Brown:2006, Kalirai:2006, McConnachie:2009, Bernard:2012, Gilbert:2014,Williams:2017,DSouza:2018, Hammer:2018, McConnachie:2018}.

The first quantitative SFH for any stellar population in the field of M31 comes from the work of \cite{Williams:2002} who leveraged the power of \hubble\ to analyze 27 archival WFPC2 fields (a mix of F439W, F555W and F555W, F814W images) within $\sim 1.5^\circ$ from the M31 nucleus.

These archival datasets were originally taken for a variety of science purposes and not specifically for SFH measurements. Due to the heterogeneous nature of these data, the depth of this photometric dataset varies across the different fields, but it generally ranges from the upper RGB to below the RC. From the analysis of these fields, \citet{Williams:2002} concludes that the average SFR of M31 was significantly higher in the past ($t > 10$ Gyr) than it is today. This result, albeit with some spatial variation, holds across the entire disk. The global SFR of M31 also appears to have decreased substantially in the last 1 Gyr, with the exception of the actively star forming regions like the spiral arms. The recovered metallicity evolution is roughly consistent across the disk of M31, showing a rapid early increase up to slightly sub-Solar metallicity.

The analysis of \citet{Olsen:2006} looked at the stellar populations in 13 additional fields of M31, this time concentrated in the bulge and inner disk. The data consist of a mix of JHK images obtained with the Near InfraRed Imager (NIRI) on Gemini North and \hubble\ Near Infrared Camera and Multi-Object Spectrometer (NICMOS) images in the F110W (J) and F222M (K) filters. The resulting photometry reaches approximately 2 magnitudes below the TRGB. Quantitative modeling of the K band luminosity function in these fields allowed the authors to recover a coarse SFH showing that the inner region of M31 is dominated by old ($t > 6$ Gyr) stars. The recovered metallicity is approximately Solar in the bulge of M31, while it is $\sim 0.5$ dex lower in the inner disk.

The first SFH obtained from observations of the oMSTO comes from the work of \citet{Brown:2003, Brown:2006}, who obtained deep \hubble/ACS (F606W, F814W) imaging of three fields in M31: an inner spheroid field (at 11 kpc along the minor axis of M31), an outer disk field (25 kpc along the major axis), and a third field targeting the Giant Stellar Stream (GSS). The superb depth of the imaging resulted in photometry reaching $\sim 2$ mag below the oMSTO, which in turn allowed a detailed reconstruction of the SFH down to the oldest epochs. The SFHs show that the average [Fe/H] of the stellar population in the spheroid field is $-$0.6 and $\sim 40\%$ of the stars are younger than 10 Gyr ($\sim 26\%$ younger than 8 Gyr), which is in stark contrast to the halo of the MW. The stellar population in the GSS field is approximately 1 Gyr younger (mean age of 8.8 Gyr, versus 9.7 Gyr in the spheroid field) but has similar metallicity. The stellar population in the outer disk field is younger and more metal rich than both the spheroid and the GSS fields (mean age of 7.5 Gyr, mean [Fe/H] of $-$0.2), showing a range of ages and metallicities comparable to thick disk stars in the solar neighborhood. The spheroid and GSS fields from this work have been recently re-analyzed by \citet{Savino:2025}, who found comparable results.

In \citet{Brown:2007}, a fourth ACS field was analyzed, this time probing the stellar population of M31 at 21 kpc along the minor axis. The data consist of the same F606W and F814W filters as the other fields and, albeit having a lower photometric depth, the photometry also reaches below the oMSTO. Compared to the spheroid field at 11 kpc, the population in this new 21 kpc field was found to have lower metallicity (by $\sim$ 0.2 dex) and older mean age (by $\sim 1.3$ Gyr). Both fields show a substantial range in ages and metallicities, but the outer field has a much lower fraction ($\lesssim 5\%$) of stars younger than 8 Gyr. These new data help paint the picture of a remarkable complexity in the stellar population of the M31 stellar spheroid, which \citet{Brown:2007} interpret as supporting evidence of hierarchical halo mass growth.  This interpretation is considerably strengthened when the positions of these ACS fields are placed in the context of the wide-field maps of \citet{Ferguson:2002} and PAndAS \citep{McConnachie:2009, McConnachie:2018}, which show unambiguous evidence of hierarchical growth in the form of streams and shells.  This combination also illustrates the value of coordinating pointed \hubble\ observations with wide-field imaging (see \S \ref{sec:future}). These findings are further supported by two additional ACS fields, analyzed in \citet{Brown:2008}. The two adjacent fields target the population at 35 kpc along the minor axis, use the same F606W and F814W filters as the previous fields, and have similar photometric depth as the 21 kpc field. Compared to the 21 kpc field, the stellar population at 35 kpc is marginally more metal-poor (by $\sim 0.1$ dex) and has slightly younger mean age (by $\sim 0.5$ Gyr). Like the other halo fields, the stellar population has a large range of ages and metallicities, with approximately a third of the stars being younger than 10 Gyr.

\citet{Faria:2007} analyzed an \hubble/ACS (F606W, F814W) field centered on the G1 clump, a stellar over-density in the outer disk of M31. The field covers an approximate physical area of 0.8 kpc $\times$ 0.8 kpc and the resulting photometry approximately reaches the level of the SGB. Although the stellar population is not modeled through a statistical CMD-fitting, the authors explore fiducial stellar population models built by using synthetic CMDs as a guide. The conclusion of this analysis is that the stellar population in the G1 clump is consistent with high metallicity ([Fe/H] $\sim -0.4$) and has a large range of ages (from 250 Myr to 10 Gyr), with most of the stars being older than 6 Gyr. The favored interpretation of these results is that the G1 clump is consistent with the stellar population of the M31 disk, which has plausibly been perturbed by a past accretion event.

The work of \cite{Bernard:2012} presented deep \hubble/ACS (F606W, F814W) photometry of two fields: an outer disk field, placed at 26.5 kpc along the major axis, and a second field at 25.5 kpc centered on the observed warp of M31's disk. In both fields, the photometry reaches well below the oMSTO. While the outer disk field  was found to have complex differential reddening that hindered a quantitative modeling, CMD-fitting of the warp field revealed a SFH in which $\sim 75\%$ of the mass was formed by 5 Gyr ago, at a roughly constant SFR, followed by a lull and then by a strong star formation event between 4 and 2 Gyr ago, in which most of the remaining stellar mass was formed. This star formation event is in contrast to the more recent SFR in this field, which was found to be low relative to the lifetime average. The AMR shows metallicity increasing, from [Fe/H] $\sim -1.2$ to supersolar values, until roughly 2 Gyr ago followed by a decline to [Fe/H] $\sim -0.3$, suggesting inflow of metal-poor gas. This field was later re-analyzed in \citet{Bernard:2015b}, along with two additional \hubble/ACS (F606W, F814W) fields, of similar depth, at 20 and 23 kpc along the southern M31 disk. The two new disk fields show a similar decline in the recent SFR as seen in the warp field, but lack the 2 Gyr burst reported in \citet{Bernard:2012} or in other concurrent studies \citep{Williams:2015,Bernard:2015a}. The AMRs in these fields show a slow increase, from [Fe/H] $\sim -0.6$ 14 Gyr ago to approximately solar 3 Gyr ago, followed by a sharp decline down to [Fe/H] $\sim -0.4$ at younger ages.

In a subsequent study, \citet{Bernard:2015a} derived the SFHs of 14 archival \hubble/ACS (F606W, F814W) fields, originally described in \citet{Richardson:2008}. The fields are placed on known stellar substructures in the outer disk and inner halo of M31, and the photometric depth approximately reaches the oMSTO. Regardless of position, \citet{Bernard:2015a} classified the fields as ``disk-like'' and ``stream-like'' depending on the SFH features, although select fields were reported to resemble a superposition of the two classes. Generally speaking, this study found that the disk-like fields have little star formation before 10 Gyr ago, and have median ages of $\sim7$~Gyr. In all disk-like fields, a star formation enhancement is observed 2 Gyr ago, similar to \citet{Bernard:2012}, although the strength of the event varies from field to field. The stream-like fields have predominantly old and intermediate stellar populations, relatively low star formation more recent than 4--5 Gyr ago, and their median age is $\sim 1.5$ Gyr older than the disk-like fields. The stream-like fields show a rapid metallicity evolution, enriching by $\sim1$~dex from 15 to 5 Gyr ago, when the stellar metallicity is approximately solar. In contrast, the disk-like fields show milder evolution, with [Fe/H] at the oldest ages of $\sim -0.5$, reaching approximately solar values 2 Gyr ago, and then declining to $\sim -0.3$. The overall conclusion of this study is the presence of stellar populations with disk-like SFH throughout the stellar substructures of M31's inner halo, which suggests the material from the thin disk was ejected into the halo by a recent dynamical perturbation.

\begin{figure}[pos=tbp]
    \centering
    \includegraphics[width=0.85\linewidth]{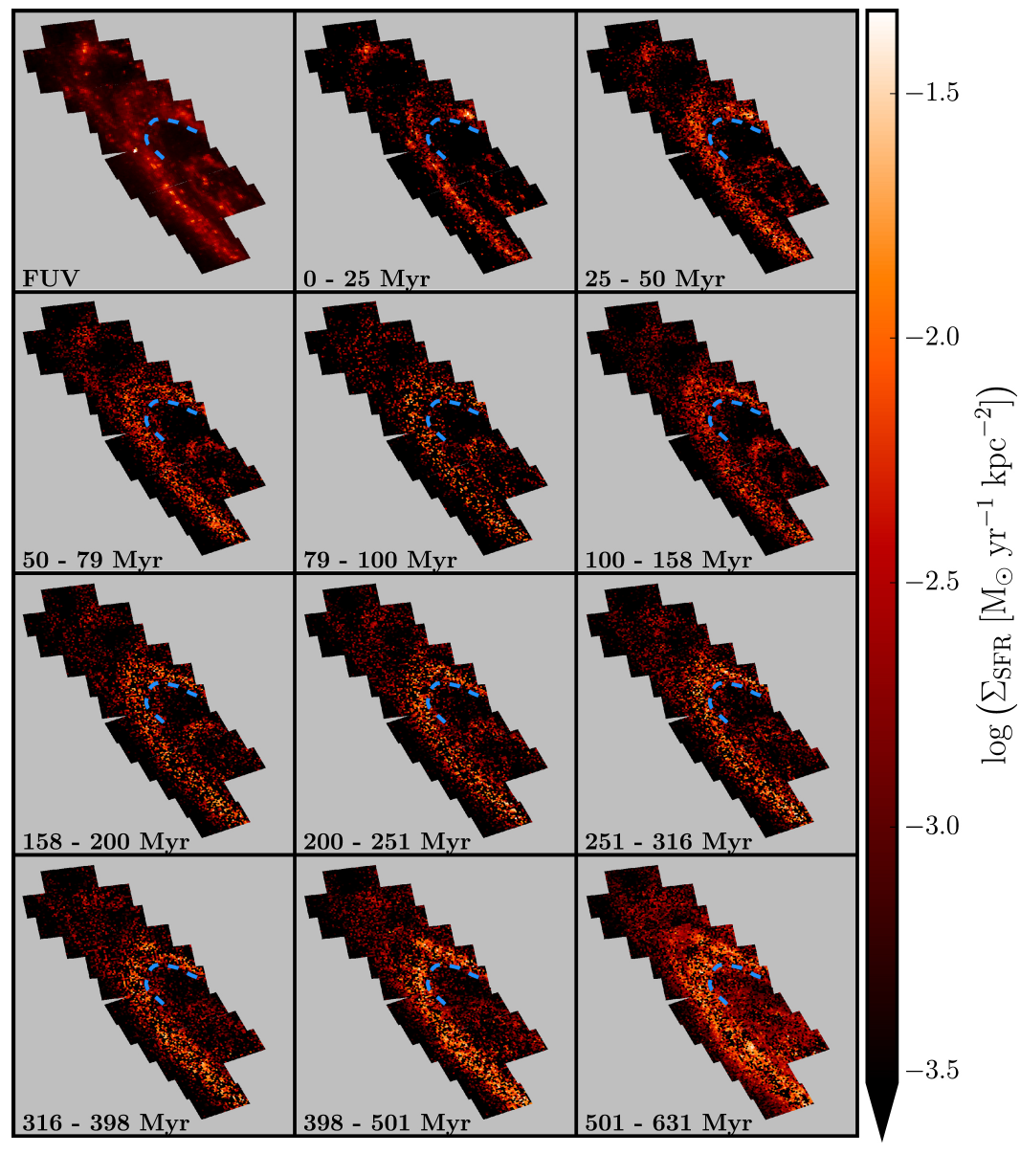}
    \caption{The spatially resolved recent SFH of the PHAT region of M31 as presented in \citet{Lewis:2015}.  The upper left panel shows the observed \textit{Galaxy Evolution Explorer} (\textit{GALEX}) map of M31. The remaining panels show the SFR over the indicated time range as measured from the \hubble\ F475W-F814W CMDs of each of the $\sim 9000$ regions ($100~{\rm pc} \times 100~{\rm pc}$ in area) modeled by \citet{Lewis:2015}. The spatially resolved SFH maps show both remarkable correspondence between to the \textit{GALEX} integrated light images, as well as the persistence of M31's famous star-forming rings over time scales of several hundred Myr. The construction of comparable maps for a larger sample of LV disk galaxies would shed new insight into the star formation and dynamics of massive galaxies.}
    \label{fig:m31_sfh_map}
\end{figure}

As illustrated in Figure~\ref{fig:m31_sfh_map},  \citet{Lewis:2015} reconstructed the recent SFH of M31 over a large portion of the stellar disk. The work made use of \hubble/ACS (F475W, F814W) photometry from the full PHAT footprint, which covers $\sim$1/3 of the stellar disk. This area was segmented into 9000 regions of $\sim100\times100$~pc. The SFH was measured from the blue portion of the CMD, which mostly contained young MS stars and allowed them to reach lookback times between 400 and 600 Myr, depending on location. The result of this analysis is a spatially resolved star SFH map which highlights how the morphology of M31's disk changes as function of stellar age (Figure~\ref{fig:m31_sfh_map}). The recent star formation of M31, over the survey area, is primarily confined within three ring structures, placed at  $\sim$5, 10, and 15 kpc from the nucleus. The inner stellar ring shows most prominently for stars with ages between 25 Myr and 200 Myr, while it is less defined for older ages. By contrast, the 10 kpc ring is well visible over the whole range of ages covered by the SFH. Finally, the outer 15 kpc ring is predominantly visible in the most recent SFH, showing the strongest signal at ages younger than 80 Myr. Over the whole PHAT region, the global SFR is found to be roughly constant, with a slow decline in the last 400 Myr.

Using the same PHAT photometry, \citet{Williams:2015} determined the intermediate-age SFH of 8 low-extinction regions of M31's disk. The selected regions span deprojected distances from 3 to 18 kpc from the nucleus of M31 and encompass a total area of approximately $\rm 1\,kpc^2$. They modeled the full PHAT filter set, from the UV to the near-IR. The photometric depth varies based on filter and location, but approximately ranges from the level of the RC to the base of the RGB. The recovered SFH spans ages between 5 Gyr and 500 Myr ago. The primary conclusion of this study is the presence of a strong event of star formation occurring between 2 and 3.5 Gyr ago, followed by a significant decline in star formation during the last 2 Gyr, which corroborates the results of previous investigations. Crucially, the fact that such an episode is observed in all the analyzed regions is strongly suggestive that the event was global across the whole disk of M31.

This study was followed up by \citet{Williams:2017}, who derived the lifetime SFH over the entire PHAT footprint. They divided the area into $83^{\prime\prime}\times 83^{\prime\prime}$ (0.3 kpc $\times$ 1.4 kpc, deprojected) regions. The SFH was derived using the ACS (F475W, F814W) and Wide Field Camera 3 (WFC3; F110W, F160W) CMDs. The solution ranges back to the earliest epochs, but the limited photometric depth of the dataset only allows the authors to place a single age bin between 8 and 14 Gyr ago. On a global scale, most of the stellar mass appears to have formed prior to 8 Gyr ago. The SFR then declined between 8 and 4 Gyr ago, after which a new intense star formation episode took place between 4 and 2 Gyr ago. From the SFH maps, there does not appear to be a significant change in the age of the stellar population as a function of radius, except for stellar populations younger than 1 Gyr, for which the previously discussed 10~kpc stellar ring manifests as a prominent feature. The AMR obtained from these data shows that [Fe/H] values between $-$0.5 and 0 are predominant across the disk at all ages and all radii, although a slight radial gradient is present, with the more external regions being on average more metal poor than the inner disk.

\citet{Lee:2024} used wide-field J and K imaging from the Wide Field Camera (WFCAM) on the United Kingdom Infrared Telescope (UKIRT) to study the bright stars in M31's halo, spanning radii up to 30 kpc in projected size. They used updated AGB star models \citep{Marigo:2017, Pastorelli:2019,Pastorelli:2020} to measure the SFH of 6 regions in the inner halo of M31, back to 8 Gyr ago, using only AGB stars on their near-IR CMD. As previously observed in the disk, most of M31's inner halo has relatively low star formation between 1 and 3 Gyr ago, and significantly higher star formation between 3 and 5 Gyr ago. The region in the vicinity of the dwarf elliptical NGC~205 shows additional star formation activity at both 4.5 Gyr and 1.4 Gyr ago, in accordance with the SFH previously derived for NGC 205 itself \citep{Savino:2025}.

Finally, building on the work from \citet{Lewis:2015}, \citet{Wainer:2026} combined nearly 1000 \hubble\ orbits of imaging to measure the spatially resolved recent SFH of M31 over the PHAST area (southern disk of M31).  They used \hubble/ACS F475W and F814W CMDs reaching a faint limit that varies across the disk but enables a robust measurement of the SFH for the most recent $\sim500$~Myr in $\sim6500$ regions.  They combined the PHAT and PHAST measurements to construct a spatially comprehensive recent SFH of M31.  They reported a total mean SFR of M31 over the last $\sim100$~Myr and $\sim20$~Myr to be $\sim0.67$ \msun~yr$^{-1}$ and $\sim0.43$ \msun~yr$^{-1}$, respectively.  Combined with other studies, they concluded that this is part of a several Gyr decline in star formation in M31.  They noted that the region near M32 shows a distinct SFH relative to the rest of M31 over the past $\sim60$~Myr, such that this region appears to exhibit localized quenching in the last $\sim15$~Myr.  Finally, they noted that the widely-used integrated light tracer of recent star formation, far-UV (FUV) plus 24~$\mu$m, under-predicts the CMD-based SFRs by a factor of $\sim2$, likely due to short-term variations inferred in the CMD-based SFH.  We discuss this later point more in \S \ref{sec:bursts}.

\subsubsection{M33} \label{sec:m33}
M33, also known as the Triangulum galaxy, is a small star-forming spiral galaxy and the third most massive member of the LG. It has a stellar mass of $\sim 5\times 10^9 M_{\odot}$ and a M$_{HI}$/M$_*$ of $\sim 0.35$ \citep{Corbelli:2014,Kam:2017}. At a distance of $\sim860$~kpc from the MW and $\sim230$~kpc from M31 \citep{Savino:2022}, it is currently orbiting within the virial radius of M31. M33 was first discovered in 1654 by Giovanni Battista Hodierna and independently re-discovered by Charles Messier in 1764 \citep[e.g.,][]{Fodera-Serio:1985, Jones:1991}.

Despite its brightness and importance in the LG context, the large angular size of M33 ($R_{25} \sim 20^\prime \times 35^\prime$) and the high apparent surface density of its stars remain a challenge for studying the star formation record of this galaxy on a global scale. To date, most published results on the old ($t > 1$ Gyr) SFH of M33 are based on deep pencil-beam fields obtained with \hubble. Despite this challenge, the range of locations probed by these \hubble\ fields, and evidence coming from larger-area surveys of bright age indicators, already provide a breadth of insights into the evolutionary history of this galaxy.

While resolved studies of bright field stars in M33, primarily focused on metallicity estimation, date back more than 40 years \citep[e.g.,][]{Mould:1986,Cuillandre:1999,Davidge:2003,Tiede:2004, McQuinn:2007}, the first quantitative SFH determination for a stellar population in M33 comes from the work of \citet{Barker:2007a,Barker:2007b}. This study acquired and analyzed three \hubble/ACS (F606W, F814W) imaging fields along the minor axis of M33. The fields were placed between $20^\prime$ and $30^\prime$ (corresponding to 9 and 13 kpc, projected) from the center of M33. The depth of the imaging and the stellar density at these locations resulted in photometry reaching down to the MSTO of an $\sim 8$ Gyr population. The resulting SFH reveals a radial age trend, such that the mean age of the stellar population increases from $\sim 6$ Gyr to $\sim 8$ Gyr from the inner to the outer field, while the mean metallicity decreases from [Fe/H] $\sim -0.7$ to [Fe/H] $\sim -0.9$. All three fields are dominated by old and intermediate stars, with a fraction of stars older than 4.5 Gyr ranging from 65\% to 80\%.

\citet{Williams:2009a} focused on 4 ACS/WFC (F606W/F814W) fields along the major axis of M33, spanning distances between 1 and 6 kpc from the center, projected. Due to crowding, the depth of the photometry varied with radius, reaching below the oMSTO in the outer field but only reaching the base of the RGB in the innermost field. The SFHs of these fields reveal a negative age gradient with radius, which the authors interpret as a clear indication of inside-out growth in M33. The radial age trend in these fields is opposite to what was observed in the outer disk by \citet{Barker:2007b}. \citet{Williams:2009a} confirmed this result by re-analyzing the \citet{Barker:2007b} fields with the same methodology of the 4 major-axis fields. This suggests that the age gradient reversal is a real feature that happens roughly between 6 and 9 kpc in the disc, which is in accordance with the observed surface brightness break in the disk of M33, at 8 kpc from the center \citep{Ferguson:2007}. The analysis of \cite{Williams:2009a} also shows that most of the stellar mass outside the inner kpc has formed at $z<1$, in contrast to what was observed in the outer disk of more massive spirals \citep[e.g.,][]{Brown:2006,Williams:2009}. Finally, by tracing the mass assembly as a function of time and radius, the authors report that the disk scale-length of M33 has approximately doubled in the last 10~Gyr.%

To further sample M33's radial age profile, \cite{Barker:2011} analyzed two additional \hubble/ACS/WFC (F606W, F814W) fields on the major axis of M33, located at 9 and 11.5 kpc from the center, projected. The SFH of the 9 kpc field revealed a relatively young stellar population, with a mean age of $\sim 3$ Gyr and almost no star formation older than 4.5 Gyr. The mean [Fe/H] of this field is reported at $-$0.5. The properties of this field continue the inside-out age gradient reported by \cite{Williams:2009a}. By contrast, the 11.5 kpc field hosts an older, more metal-poor population, with a mean age of $\sim 7$ Gyr and a mean [Fe/H] of $-$0.8. This study further supports the existence of a positive age gradient in the outer disk of M33, which the authors attribute to either radial migration, a change in the efficiency of star formation at low gas densities, or the transition from the thin disk of M33 to another stellar component (i.e., thick disk or halo).

The same 9 kpc field of \citet{Barker:2011} was later reanalyzed by \citet{Bernard:2012}. The higher-resolution SFH shows the same qualitative features as the solution of \cite{Barker:2011} but also reveals a strong episode of star formation approximately 2 Gyr ago, which appears to be roughly coeval with a star formation burst observed in the outer disk of M31 (analyzed in the same study). The authors tentatively associate the two events. This association was further supported at the time by dynamical models of the M31-M33 system, which placed the last pericenter between the two galaxies at $\sim 2.5$ Gyr ago \citep{McConnachie:2009,Putman:2009}. We note that more recent measurements of the proper motion of the M31/M33 pair have since disfavored such a recent interaction between the two galaxies \citep[e.g.,][]{Sohn:2012,Patel:2017a,VanderMarel:2019,Corbelli:2024}.

Using a different approach, \citet{Javadi:2011} used UKIRT/UIST JHK photometry to study the LPV population in the central $\rm kpc^2$ ($4^\prime\times4^\prime$) of M33. By modeling the K-band luminosity function of the LPVs, they reconstructed the lifetime SFH in this region. Their solution shows a main epoch of star formation between 10 and 4 Gyr ago, with a mean age of $\sim 6$ Gyr. They also identify a second burst of star formation approximately 2 Gyr ago. In a follow-up paper \citep{Javadi:2017}, the analysis was extended to the central $\rm deg^2$ (corresponding to the entire disk of M33 out to 7 kpc and partial coverage out to 15 kpc), this time using JHK photometry from UKIRT/WFCAM. They revise the previous SFH, measuring a main star formation event in this region between 6 and 3 Gyr ago, with a mean age of 5 Gyr, and responsible for 71\% of the total stellar mass. This star formation episode is slightly older than the event measured at larger radii by \cite{Barker:2011}, further supporting the inside-out growth scenario. The LPV-based SFH also shows a second star formation episode occurring between 200 and 300 Myr ago, when star formation reached the highest peak in this region of M33. The spatially resolved SFH solution shows a negative age gradient inside 8 kpc, reversing to a positive age gradient outside of this radius, in accordance with the \hubble-based studies mentioned above. The analysis of the sample spatial distribution shows that only LPVs younger than 100 Myr are spatially concentrated around the spiral arms and known high-density gas regions, while older stars present a smoother distribution.

\begin{figure}[pos=tbp]
    \centering
    \includegraphics[width=1.0\linewidth]{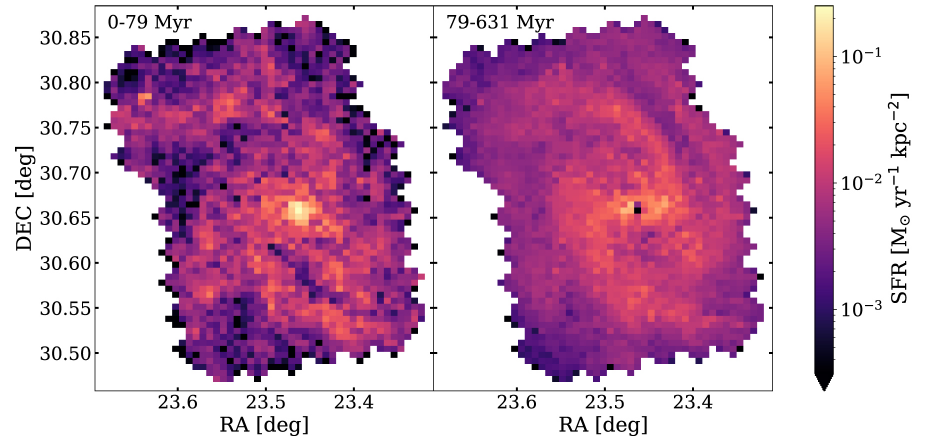}
    \caption{The spatially resolved recent SFH of M33 from PHATTER \citep{Lazzarini:2022} showing the change in morphological structure of M33 starting 80 Myr ago.}
    \label{fig:m33_sfh_map}
\end{figure}

Another study of the star formation over a large area of M33 was carried out by \cite{Lazzarini:2022}, using \hubble/ACS/WFC (F475W, F814W) photometry from the PHATTER program. The data used cover a de-projected area of 38 $\rm kpc^2$ in M33's central region, extending to $\sim3.5$ and $\sim2$~kpc from the center along the major and minor axes, respectively. The depth of the photometry varies depending on the local stellar densities but reaches down to the 630 Myr MSTO over the whole region. CMD modeling of these data resulted in a spatially resolved recent SFH over 2005 regions, each $100\times100$~pc, that reach lookback times of 630~Myr. The authors reported that the global SFR of M33 has been roughly constant over the most recent 500 Myr, with a slight increase at very recent times ($t<100$ Myr). As illustrated in Figure~\ref{fig:m33_sfh_map}, stars younger than 80 Myr appear to trace the flocculent structure of M33, while stars older than 80 Myr follow a smoother grand-design spiral pattern, and also outline the presence of a bar. This result suggests a time evolution of the spiral structure.

Finally, as part of a larger \hubble\ study of galaxies in the M31 system, \cite{Savino:2025} re-analyzed the SFH of M33 using four fields from \cite{Williams:2009a}, two fields from \cite{Barker:2011}, and two additional ACS/WFC (F606W, F814W) fields of comparable depth to the others located at $\sim 2$ kpc from M33's center. The resulting SFHs are qualitatively comparable to the results of \citet{Williams:2009a} and \citet{Barker:2011}, with the inside-out growth in the inner disk, and the age reversal in the outer disk, clearly appreciable as a function of galactocentric radius.

\subsubsection{Classical Satellites and Dwarf Ellipticals} \label{sec:m31_classical}
The M31 system contains a rich population of 20 known ``classical'' (i.e., $M_V\le-7.7$) dSph galaxies, based on the luminosities of \citet{Savino:2022}. These objects have $M_{\star} \sim$ $2\times 10^5$~$M_{\odot}$ to $\sim 3\times10^7 M_{\odot}$ \citep{Martin:2016a,McConnachie:2006b,Slater:2011,Bell:2011,Martin:2013c,Martin:2013b}, have a distance between $\sim 560$ kpc and $\sim 830$ kpc \citep{Savino:2022}, and do not possess detectable amounts of \hi{} \citep{Putman:2021}.

The first SFH measured from CMD fitting among M31's dSph satellites was performed by \citet{Weisz:2014a}, who analyzed \hubble/WFPC2 imaging (F450W, F555W or F606W, and F814W) of \A{I}, \A{II}, \A{III}, \A{V}, \A{VI}, and \A{VII}, reaching below the HB. The spatial coverage of these observations varies, covering between 0.06 (\A{II}) and 1.2 (\A{V}) times the half-light area. The resulting SFHs for these objects are all extended and have comparable duration, halting star formation between 5 and 7 Gyr ago \citep{Weisz:2015a}.

\citet{Weisz:2014c} analyzed a deeper CMD of \A{II}, reaching below the oMSTO, obtained from \hubble/ACS data (F475W, F814W). They report that \A{II} had two distinct periods of elevated star formation, the first at $t>12.5$~Gyr ($\sim 50\%$ of the total stellar mass) and the second between 10 and 5 Gyr ago ($\sim 40\%$ of the total stellar mass), separated by a period of lower intensity star formation and with a final quenching around 5 Gyr ago. The inferred metallicity evolution goes from [Fe/H] $\sim$ $-$1.2 $\sim$ 13 Gyr ago to $\sim -0.6$ $\sim 5$ Gyr ago.

By using \hubble/ACS (F475W, F814W) imaging that extends below the oMSTO from the ISLAndS program (Appendix~\ref{sec:important_obs}), \citet{Skillman:2017} derived the lifetime SFH of 6 satellites of M31, 5 of which are classical dSphs (\A{I}, \A{II}, \A{III}, \A{XV}, and \A{XXVIII}). They found that the SFHs of their sample of satellites are more homogeneous compared to MW satellites of similar luminosity. They reported extended SFHs, with quenching times between 9 and 6 Gyr ago, and no strong trends with luminosity or present-day distance from M31.

Similarly deep \hubble/ACS imaging (F606W, F814W) was used by \citet{Collins:2022} to study the SFH of the ultra-diffuse galaxy \A{XIX}. They found that \A{XIX} formed $>50\%$ of its stellar mass prior to reionization, $\sim 90\%$ of its stellar mass by 9.7 Gyr ago, and had no sign of meaningful star formation in the last 8 Gyr. In light of this rapid SFH truncation, the authors argue against bursty star formation episodes as an explanation for \A{XIX}'s large radius and low surface brightness.

The study of \citet{Makarova:2021} derived the SFH of three M31 satellites: \A{VII}, \A{XXVIII}, and \A{XXIX}. The \hubble/ACS observations range in depth from the oMSTO (\A{XXVIII}) to the lower RGB (\A{VII} and \A{XXIX}). The authors reported that both \A{VII} and \A{XXIX} formed most of their stars in an initial star formation episode, with 80\% of the total stellar mass being formed by 12~Gyr ago in \A{VII} and by 10~Gyr ago in \A{XXIX}. For \A{XXVIII}, the authors reported that half of the stellar mass was formed at ancient ($t\gtrsim12$~Gyr) ages, while the remaining stellar mass was formed at intermediate ages of 6--8 Gyr ago.

Other classical dSphs of M31 have been studied through modeling of their LPV population. By analyzing Isaac Newton Telescope (INT) Wide Field Camera imaging of an $11.26 \times 22.55$~arcmin$^2$ field in \A{I}, \citet{Saremi:2021} identified 59 LPV candidates over 2 half-light radii of the galaxy and used their photometric properties to reconstruct the SFH of \A{I}. They reported a major event of star formation peaking approximately 6.6~Gyr ago and a quenching time of 4~Gyr ago. This SFH determination is substantially more extended than solutions obtained from photometry reaching the HB \citep{Weisz:2014a} and the oMSTO \citep{Skillman:2017,Savino:2025}. Similar methodologies were used to derive the SFH of \A{VII} \citep[55 LPV candidates]{Navabi:2021} and \A{IX} \citep[50 LPV candidates]{Abdollahi:2023}. The SFH of \A{VII} is reported to consist of a main prolonged star formation episode peaking approximately 6 Gyr ago. Most of the star formation in \A{VII} is reported to have ceased between 5--6 Gyr ago, although some low level star formation is reported at younger ages in the central regions of the galaxy. For \A{IX}, the LPV-based SFH spans epochs as recent as 500 Myr ago, peaking at approximately 6~Gyr ago, with 90\% of the stellar mass in place by $\sim3$--$4$~Gyr ago. A radial star formation trend was detected, such that younger stars preferentially occupy the inner regions of the galaxy.

To provide a homogeneous characterization of the M31 satellites, \citet{Weisz:2019} derived the SFHs of 20 dwarf galaxies around M31, including 14 classical dwarfs. They used \hubble/ACS imaging reaching below the HB. From the analysis of the ensemble SFHs, they reported several broad trends: (i) faint satellites ($M_V>-8.5$) tend to have formed most of their stars before 10--12 Gyr ago; (ii) most of the analyzed satellites quenched at an intermediate time, between 3 and 9 Gyr ago; (iii) no satellite quenched within the last 3 Gyr; (iv) the M31 satellites comprise two populations, one best described by a rapidly declining SFH and the other more compatible with a relatively constant SFR.

Much deeper \hubble/ACS imaging (F475W or F606W and F814W) from the \hubble\ M31 Satellite Treasury program (Appendix~\ref{sec:important_obs}) was used by \citet{Savino:2025} to derive the SFHs of a sample of 36 dwarf galaxies, including 23 dSph satellites of M31, i.e., the 20 classical dSphs plus \A{X}, \A{XVI}, and \A{XXIV}, which are just fainter than the classical threshold ($-7.6\le M_V\le-7.3$). To date, this study represents the largest compilation of homogeneous SFHs in the M31 satellites with CMDs that reach the oMSTO. Some of the main conclusions derived from this study are: (i) all the classical satellites of M31 in the sample show extended SFHs, although the measured quenching epochs are, on average, older than reported in previous studies, with most satellites forming 90\% of their stellar mass by 7--8 Gyr ago; (ii) the SFH duration of the satellites significantly correlates with both the galaxy luminosity and their present-day distance from M31, such that knowledge of these two quantities is sufficient to predict the quenching epoch to within $\sim 2$ Gyr (see discussion in \S \ref{sec:environment}); (iii) no systematic difference is noted between the SFH of satellites belonging to the ``Great Plane of Andromeda'' (GPoA) \citep[e.g.,][]{Ibata:2013} and those outside the plane; (iv) as previously suggested by \citet{Weisz:2019}, the classical satellites of M31 separate into two populations, the first one characterized by rapid mass assembly and quenching 8--10 Gyr ago, and the second more appropriately described by a constant SFR. \citet{Savino:2025} showed that this first population of galaxies is not present in cosmological simulations of LG-like environments such as TNG50 and FIRE-2.  These SFHs have also begun to be used in tandem with dynamical modeling. For example, \citet{Pickett:2026} combined the \citet{Savino:2025} SFHs with Jeans modeling of seven M31 dSphs and found that the anomalously low central dark matter densities of several of them cannot be attributed to SFH-driven dark matter heating alone (see also \S \ref{sec:bursts}).

The four dE satellites of M31 (M32, NGC~205, NGC~147, and NGC~185) are among the most massive satellites in the LG \citep[e.g.,][]{McConnachie:2012}, with $M_{\star}\sim3$--$8\times10^{8}$~\msun\ (based on the luminosities of \citealt{Savino:2022} and $M/L_V\approx2$). Galaxy-by-galaxy summaries of their SFHs are provided in Appendix~\ref{sec:app_m31_de}.  All four contain stars that span a wide range of ages, and their SFHs can be viewed as the luminous extension of the M31 dSph population.  \citet{Savino:2025} find $\tau_{80}$ values ranging from $\sim10$~Gyr ago for NGC~185, comparable to the more luminous M31 dSphs, to $\sim3$~Gyr ago for NGC~205, consistent with the trend toward more extended SFHs at higher luminosity across the M31 satellite system, albeit with considerable scatter at fixed luminosity.  The details of their SFHs differ.  NGC~185 built up its stellar mass rapidly, forming $\sim70$\% by 12.5~Gyr ago, whereas NGC~147 formed only $\sim40$\% by that time and the majority between $\sim5$ and 7~Gyr ago \citep{Geha:2015}.  M32 shows a prominent intermediate-age (2--5~Gyr) episode at approximately solar metallicity that accounts for a large fraction of its stellar mass \citep{Monachesi:2012, Savino:2025}.  NGC~205 and NGC~185 also retain small amounts of \hi{}, and low-level star formation persisted in both until a few hundred Myr ago \citep[e.g.,][]{Monaco:2009, Weisz:2014a, HamedaniGolshan:2017}.  We caution that the available oMSTO-depth CMDs cover only small fractions of these galaxies, and spatial gradients in their SFHs remain largely unexplored.  Appendix~\ref{sec:app_m31_de} also summarizes the SFHs of two gas-rich M31 satellites, IC~10 and Pisces~I (LGS-3), which have historically been discussed alongside the isolated dwarfs (\S \ref{sec:isolated}).  Both are still forming stars: IC~10, a starburst, formed roughly half of its stellar mass within the last $\sim4$~Gyr \citep{Weisz:2014a}, and Pisces~I has continued forming stars at a low level to the present \citep{Hidalgo:2011, Savino:2025}.

\subsubsection{Ultra-Faint Satellites} \label{sec:m31_ufds}
There are currently 18 known galaxies associated with M31 that have $M_V>-7.7$ \citep[e.g.,][]{Savino:2022, McQuinn:2023, Smith:2025, Jones:2026}, which classifies them as UFDs according to the definition of \citet{Simon:2019}. Due to the distances involved ($500$--$900$ kpc) and to current depth of observations, these systems are relatively massive UFDs \citep[$M_{\star} \sim 0.3-1.5 \times 10^5$~\msun;][]{Collins:2022b,MartinezDelgado:2022,Savino:2022,Savino:2023,McQuinn:2023,Collins:2024,Ogami:2024,Smith:2025}. These ``more luminous'' UFDs are rare in the MW halo, which makes the M31 population the best available sample to characterize the transition regime between classical dwarfs and the larger population of fainter UFDs. All M31 UFDs for which \hi{} observations exist are gas-poor \citep[i.e., \hi\ non-detections,][]{Putman:2021}.

CMD based SFHs of these galaxies have revealed that they are predominantly ancient, in line with the ancient ages measured in lower-mass MW UFDs. However, as discussed below, multiple studies have highlighted that the SFHs of many M31 UFD satellites show signs of appreciable star formation activity after the end of reionization. Due to the compact apparent size of these objects, all the observations described in this section cover a large ($> 50 \%$) fraction of the stellar mass in these galaxies.

The first CMD-based SFHs of UFDs in the Andromeda system were made by \citet{Weisz:2014a}, who used \hubble/WFPC2 (F606W, F814W) observations of And~{\sc XI}, And~{\sc XII}, and  And~{\sc XIII}, resulting in photometry that reached down to the base of the RGB. From CMD-fitting, they reported that And~{XI} and And~{XIII} are mostly ancient, having formed $\sim$ 80\% and $\sim$ 90\% of their stellar mass, respectively, by $\sim 12.5$~Gyr ago. Conversely, the SFH for And~{\sc XII} showed a much slower mass build-up, with $\sim$ 60\% of the stellar mass formed in the last $\sim$ 8 Gyr.

Deeper \hubble/ACS observations (F475W, F814W), reaching down below the oMSTO, were used by \citet{Weisz:2014c} to study a field centered on And~{\sc XVI}. With a magnitude of $M_V=-7.5\pm 0.3$, And~{\sc XVI} sits right at the boundary between UFDs and classical dwarfs. The SFH of \citet{Weisz:2014c} showed that And~{\sc XVI} built up only $\sim$ 40\%  of its stellar mass by $\sim$ 10 Gyr ago, making it incompatible with reionization-driven quenching, and kept forming stars to a much later quenching time around $5$--$6$ Gyr ago. The AMR from the best-fit CMD shows a mild evolution of [Fe/H] from $-$1.5 $\sim 13$ Gyr ago to $-$1.3 $\sim 6$ Gyr ago. Later re-analyses of the same data by \citet{Monelli:2016}, \citet{Skillman:2017}, and \citet{Savino:2025} provided a similar picture, although the quenching time provided by \citet{Savino:2025} is $\sim$ 8 Gyr ago.

An analogous conclusion is reached by the study of \citet{McQuinn:2023}, who used \hubble/ACS imaging (F606W, F814W), reaching $\sim2$~mag below the HB, to study the stellar population of Peg W, another massive UFD ($M_V = -7.20\pm0.17$) located $\sim350$~kpc from M31, just beyond our nominal 300~kpc satellite boundary (\S \ref{sec:concepts_classification}). Similarly to And~{\sc XVI}, Peg~W appears to have formed only $\sim50$\% of its stars by the end of reionization, quenching at a much later lookback time of 7.4~Gyr ago. \citet{McQuinn:2023} also tentatively suggested that some star formation may have happened in Peg~W as recently as 500 Myr ago, although the currently available data are not sufficient to conclusively establish the presence of such a young stellar population.

As part of their \hubble/ACS (F606W, F814W) survey of the M31 satellites, \citet{Weisz:2019} measured the SFH of 6 UFDs (And~{\sc XI}, And~{\sc XII}, And~{\sc XIII}, And~{\sc XX}, And~{\sc XXII}, and And~{\sc XXVI}). The photometry only reaches down to the HB, which complicates the recovery of old SFHs. The authors reported intermediate quenching epochs between 7--9~Gyr ago for most of the UFDs, while the quenching epoch of \A{XII} is reported at 3.4 Gyr ago.

Additional \hubble/ACS imaging (F606W, F814W) of the same fields was acquired as part of the \hubble\ Treasury survey of M31 satellites. This allowed \citet{Savino:2023} to perform the first homogeneous study of a sample of M31 UFDs using photometry that reached below the oMSTO. The six UFDs studied were And~{\sc XI}, And~{\sc XII}, And~{\sc XIII}, And~{\sc XX}, And~{\sc XXII}, and And~{\sc XXVI}, which with $-7.0<M_V\le-6.0$,  are slightly fainter than \A{XVI} and Peg~W. According to \citet{Savino:2023}, five of the six targets show prominent ancient star formation, building up most of their stellar mass before the end of cosmic reionization. However, only some of these galaxies (And~{\sc XI}, And~{\sc XII}, and And~{\sc XX}) appear to be compatible with being pure relics of reionization, while And~{\sc XXII} and And~{\sc XXVI} sustained meaningful star formation activity at later times, forming 20-40\% of their stellar mass over the course of 1--3 Gyr after reionization.

The remaining galaxy in the sample, And~{\sc XIII}, shows a remarkable SFH. This galaxy appears to have formed only $\sim$ 10\% of its stars at ancient times and built up $\sim$ 75\% in a star formation event between redshifts of 3 and 2. The SFH of And~{\sc XIII} represents the first of its kind for a galaxy of such low stellar mass ($M_* \sim 10^5 M_{\odot}$).

An updated version of the SFHs from \citet{Savino:2023} was presented in \citet{Savino:2025}, which also added a SFH for And~{\sc X} ($M_V = -7.3$; \citealt{Savino:2022}). Comparable to other UFDs in M31, And~{\sc X} also formed most of its stars prior to the end of cosmic reionization but approximately 30\% of the stellar population formed in the subsequent few Gyr.

The work of \citet{Jones:2026} presents \hubble/ACS (F606W/F814W) imaging of two additional UFDs, Peg V ($M_V=-5.3\pm 0.4$) and Psc VII ($M_V=-6.1\pm 0.2$), along with a reanalysis of archival imaging of Peg W. Consisting of 1 orbit of \hubble\ imaging, the photometry reaches down to the base of the RGB. The SFH analysis of \citet{Jones:2026} confirms that Peg W shows signs of extended star formation past the epoch of reionization. The stellar populations of Peg V and Psc VII are compatible with being older than 10 Gyr, but the depth and sample size of the CMD prevent a more precise characterization of their quenching epoch.

In aggregate, these results paint a picture of diversity, with luminous UFDs in M31 showing a variety of SFHs in a relatively narrow mass interval (within a factor of 4). Expectations from galaxy formation models are that the ability of low-mass galaxies to resist reionization-driven feedback and sustain star formation is primarily driven by the dark-matter halo mass and accretion history, rather than stellar mass \citep[e.g.,][]{Gnedin:2000, Hoeft:2006, Bovill:2009, Wheeler:2015, Dayal:2018, Benitez-Llambay:2020}. The diversity of SFHs observed in M31's massive UFDs can therefore be interpreted as the result of large scatter in the stellar-mass to halo-mass relation, and/or of a diversity of halo accretion histories, for galaxies in the $-8<M_V<-6$ luminosity range, which is also in accordance with galaxy formation models. This scenario results in a ``fuzzy" boundary between reionization relics and classical dwarfs, with galaxies of comparable stellar mass having substantially different sensitivity to cosmic UV feedback. While this interpretation is in accordance with $\Lambda$CDM formation models \citep[e.g.,][]{Fitts:2017}, these data are still not sufficient to rule out that the SFHs of the bright UFDs in M31 are instead the result of intrinsic differences in the early environment and/or accretion history between M31 and the MW. A critical dataset to assess this scenario would be to obtain oMSTO-based SFHs for the low-luminosity ($M_V>-6$) UFDs of M31. We discuss trends with UFDs and important new observations in \S \ref{sec:sfh_trends} and \ref{sec:future_formation}, respectively.

\subsection{Isolated Local Group Galaxies} \label{sec:isolated}

Because of their proximity, isolated galaxies within and near the LG are among the best-studied in terms of resolved stellar populations.  We define isolated and truly isolated galaxies, and discuss the ambiguities in these classifications, in \S \ref{sec:concepts_classification}.  Our definition updates some earlier classifications.  For example, \citet{Gallart:2015} described Pisces~I (LGS-3) as one of six ``relatively isolated'' LG dwarfs observed by LCID, although it lies within 300~kpc of M31 \citep{Savino:2022}, a distance within which they note that dwarfs may have orbited their host more than once (their Figure~3).  Orbit reconstructions based on its proper motion further suggest that Pisces~I may previously have been a satellite of M33 \citep{Patel:2026}.  We therefore treat both Pisces~I and IC~10 ($\sim250$~kpc from M31; \citealt{Sanna:2008}) as M31 satellites.  Similarly, \citet{Nagarajan:2022} count Leo~T ($\sim420$~kpc from the MW) among the MW satellites but Phoenix ($\sim400$~kpc) as a field dwarf, whereas we treat both as isolated.

This section synthesizes the SFH studies of these galaxies.  Galaxy-by-galaxy summaries are provided in Appendix~\ref{sec:app_isolated}, organized roughly in order of increasing distance from the LG barycenter following \citet{McConnachie:2012}.

Isolated galaxies are typically gas-rich, and thus, actively star-forming.
As such, they present SFHs very different from those of typical satellite galaxies.  While all isolated galaxies are, by definition, more distant than MW satellite galaxies, space telescopes enable photometry reaching below the oMSTO to provide lifetime SFHs.  In the discussion of individual galaxies (Appendix~\ref{sec:app_isolated}), we emphasize the results of studies of lifetime SFHs.  However, for completeness, we also include references to other resolved star studies (typically shallower photometry not reaching the oMSTO) that may have had different science aims.   These datasets often produce a measurement of the recent SFH, which we cover in greater depth in \S \ref{sec:bursts}.

Truly isolated, low-mass galaxies are especially important as probes of the effects of reionization.  While low-mass satellite galaxies are almost universally quenched at early times (e.g., Figure~\ref{fig:UFDs}), isolated galaxies of similar stellar masses are able to retain their gas and
continue forming stars for longer periods \citep[e.g.,][]{Grebel:2003, Weisz:2014d}.  Characterizing the impact of reionization on low-mass galaxies remains an important goal in understanding galaxy evolution, and those in isolation are particularly useful as they are subject to the fewest confounding effects of environment.

Among the most important developments of the past decade for this population is that isolation is no longer a statistical inference.  Proper motions from \hubble\ and \textit{Gaia} \citep{Bennet:2024, Bennet:2025} now place the majority of these systems (e.g., IC~1613, Phoenix, Pegasus, Leo~T, Leo~A, WLM, Tucana, SagDIG, VV~124) on first infall into the LG. Thus, the SFHs of these galaxies can be interpreted with reasonable confidence as records of low-mass galaxy evolution in the absence of a massive host.

With that context, the SFHs of the isolated galaxies exhibit three notable characteristics.  First, and in contrast to satellite galaxies of similar stellar mass, virtually all of the gas-rich systems have formed stars continuously, or nearly so, over the lifetime of the Universe.  Where the data reach the oMSTO, the SFHs of many are broadly consistent with a roughly constant SFR (e.g., IC~1613, Leo~T), often with modest enhancements at intermediate ages, while at least one, WLM, shows evidence for a slowdown or pause \citep[e.g.,][]{McQuinn:2024a}.  Second, several galaxies show a pronounced \textit{delayed} onset of their dominant star formation.  Leo~A formed $>90$\% of its stars within the last $\sim8$~Gyr \citep{Cole:2007}, Aquarius formed the bulk of its stars in an episode $\sim7$~Gyr ago \citep{Cole:2014}, and Pegasus shows a dominant episode $\sim8$~Gyr ago \citep{Savino:2025}.  Because these systems appear to be on first infall, such intermediate-age peaks are unlikely to be due to interactions with the MW or M31. Dwarf-dwarf interactions or internal gas cycling remain plausible explanations \citep[e.g.,][]{Bennet:2025}, but are challenging to verify.  Third, the two gas-poor isolated galaxies, Cetus and Tucana, are entirely ancient, having quenched $\sim8$--$9$~Gyr ago \citep{Monelli:2010_b, Monelli:2010_a}.  With stellar masses well above the regime in which reionization is expected to be effective, both require another quenching mechanism.  For Cetus, the orbit of \citet{Bennet:2025} suggests an interaction with M31 more than $\sim6$~Gyr ago that may have truncated its star formation.  In contrast, Tucana has no evidence of a past interaction with a massive host. It remains the clearest example in the LG of a low-mass galaxy quenched by something other than environment or reionization, with stellar feedback from a final burst of star formation one plausible candidate \citep[e.g.,][]{Savino:2019}.

Two other trends are also clear.  First, nearly every galaxy with spatially extended data shows the ``outside-in'' pattern of an exclusively old outer population and centrally concentrated younger stars, accompanied by a negative radial metallicity gradient \citep[e.g.,][]{Hidalgo:2009, Hidalgo:2013, Albers:2019, Cohen:2025}.  Second, where CMD-based AMRs have been compared to spectroscopic metallicities, the agreement is generally good \citep[e.g.,][]{Cole:2007, Cole:2014, Taibi:2018, Taibi:2020, Fu:2024}.  One final note: several of the most interesting systems (e.g., NGC~6822, SagDIG, VV~124) still lack a SFH from a CMD that reaches the oMSTO, in some cases because their large angular size or high foreground extinction make such observations expensive with \hubble.  \jwst\ should remedy this at least for SagDIG as it has recently acquired oMSTO depth imaging (GO-5255, PI Skillman).

\subsection{Galaxies Beyond the Local Group}
\label{sec:BLG}

Most of the observations of galaxies outside of the LG environment are characterized by relatively shallow CMDs (with the notable exception of Leo~P, see Appendix~\ref{sec:app_blg}).
In what follows, we synthesize results from large, systematic SFH surveys beyond the LG.  Individual surveys and galaxies are summarized in Appendix~\ref{sec:app_blg}.  A representative collection of published lifetime SFHs from resolved stars beyond the LG is compiled in Table~\ref{tab:sfh_compilation}.

The literature beyond the LG differs from that within it in two important respects.  First, as mentioned above, except for Leo~P, which \jwst\ has resolved to the oMSTO \citep{McQuinn:2024b}, no galaxy beyond the LG has a published CMD that reaches the oMSTO.  Most SFHs are based on CMDs that reach only $\sim1$--$2$~mag below the TRGB, and a smaller number reach the RC. Their lifetime SFHs therefore have coarse time resolution at ages $\gtrsim1$--$2$~Gyr (\S \ref{sec:shallow_cmds}).  As a result, the most secure measurement these data provide is the recent ($\lesssim1$~Gyr) SFH.  Second, the sample is dominated by a few large, uniformly analyzed surveys (e.g., ANGST, \citealt{Weisz:2011}; the starburst duration project, \citealt{McQuinn:2010a}; LEGUS, \citealt{Cignoni:2019}; LUVIT, \citealt{Choi:2025}; GLOW, \citealt{Cohen:2026b}) rather than by deep studies of individual objects.  The internal uniformity of each of these samples is a strength for population-level inference, even if the depth of any single SFH is modest.  However, the surveys differ from one another in their data (e.g., filters, depth) and in how the SFHs are measured (e.g., CMD-fitting code, stellar models, treatment of uncertainties), such that comparisons between them require care.

\begin{figure}[pos=tbp]
    \centering
    \includegraphics[width=0.8\linewidth]{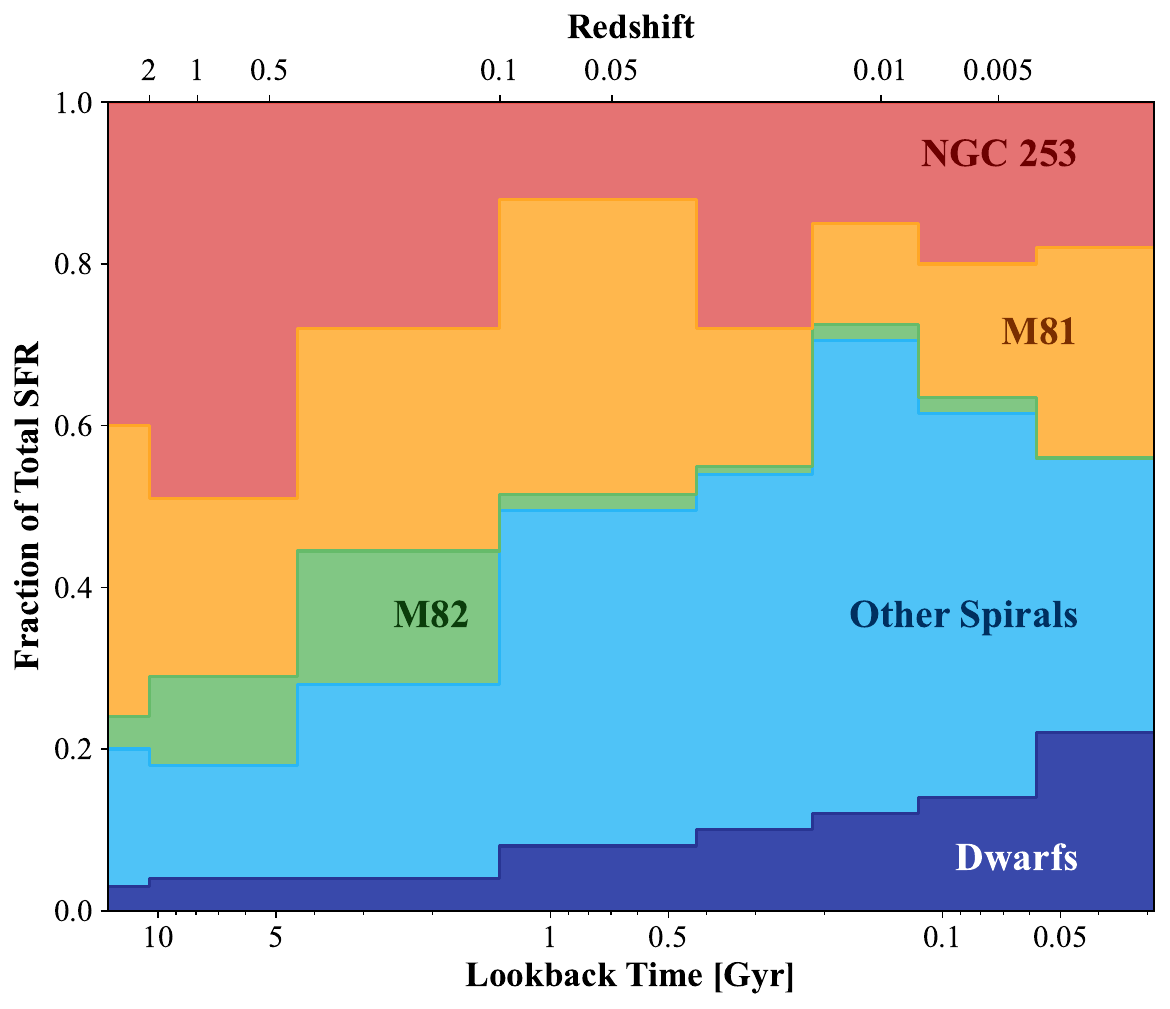}
    \caption{The fractional contribution of several components to the total SFH of the ANGST survey volume ($D\lesssim4$~Mpc) as computed by \citet{Williams:2011a}. Massive disk galaxies make the majority of the stellar mass budget at ages more than $\sim1$~Gyr ago. At more recent times, lower-mass spirals (e.g., NGC~300) and dwarf galaxies make up an increasing fraction of the star formation budget. An expansion to more galaxies (e.g., Cen~A) and the separation of disk and halo components is a clear and important future direction. }
    \label{fig:angst_disk_sfhs}
\end{figure}

With these caveats in mind, we summarize the main results of these programs.  For low-mass galaxies, the ANGST sample of $\sim60$ dwarfs within $\sim4$~Mpc established that the majority formed the bulk of their stellar mass prior to $z\sim1$ regardless of present-day morphology, that the mean SFHs of dIrrs, dTrans, and dSphs diverge only within the last few Gyr, and that the SFHs are too complex to be described by simple parametric forms \citep{Weisz:2011}.  The same data revealed a density-morphology relation within the LV, such that the transition from gas-rich to gas-poor cannot be driven by internal feedback alone.  A second result is the diversity of individual histories at fixed mass, including several galaxies with strongly delayed or recently reignited star formation \citep[e.g., NGC~3109, UGCA~92, KDG~215;][]{Weisz:2011, Makarova:2012, Cannon:2018} that resemble the delayed-onset systems in the LG (\S \ref{sec:isolated}).  We caution, however, that many of these are based on single-orbit CMDs, such that deeper imaging is needed to confirm their SFHs.   Third, most dwarf galaxies show outside-in radial age gradients, i.e., older outer regions, though their strength ranges from pronounced to flat and correlates with the global SFH (\S \ref{sec:spatial}). \citet{Cohen:2026b} measured such gradients from CMD-based SFHs for 42 systems, extending to the LV the outside-in pattern seen in the isolated LG dwarfs (\S \ref{sec:isolated}).  The uniformly reduced archival \hubble\ photometry, \hi{} imaging, and TRGB distances for 37 of these galaxies have been assembled into a public atlas by the GLOW project \citep{McQuinn:2026}, which should facilitate population-level studies that combine SFHs with the gas and chemical properties of these systems.

For more massive galaxies, resolved star SFHs remain limited to a handful of systems within $\sim4$~Mpc (e.g., NGC~300, NGC~2403, NGC~2976, M81, M82, NGC~5128), in part due to the paucity of nearby $L_{\star}$ galaxies.  These studies have primarily addressed spatial questions, such as inside-out disk growth \citep[e.g.,][]{Gogarten:2010, Williams:2013}, the absence of an age gradient across spiral arms in M81 \citep{Choi:2015}, star formation within the galactic wind of M82 \citep[e.g.,][]{Rao:2025}, and the age structure of the halo of NGC~5128 \citep{Rejkuba:2011, Rejkuba:2022}.  An important integrated result from ANGST is that massive disk galaxies dominate the stellar mass budget of the $\lesssim4$~Mpc volume at all epochs, with dwarf galaxies contributing appreciably only within the last $\sim1$~Gyr \citep[][Figure~\ref{fig:angst_disk_sfhs}]{Williams:2011a}.  As we discuss in \S \ref{sec:future_formation}, \jwst\ and \textit{Roman} are poised to substantially expand this sample.

\section{General Trends in Star Formation Histories}
\label{sec:sfh_trends}

The extensive efforts invested in measuring SFHs of nearby galaxies over the past $\sim20$~years (\S \ref{sec:formation}) have produced an increasingly detailed picture of stellar mass assembly in the local Universe over cosmic time.  In the following sections, we discuss some of the broad trends that can be gleaned from an aggregation of these SFHs.  These topics include the impact of reionization, the diversity of low-mass galaxy SFHs, environmental trends in quenching, and the evidence for bursty star formation.  Our analysis is necessarily weighted toward low-mass galaxies, which have been the focus of most SFH efforts of the past $\sim20$~years.

\subsection{The Impact of Reionization on Low-Mass Galaxy Formation}

As noted in \S \ref{sec:mw_ufds}, theoretical models of galaxy formation have long predicted that the heating of gas in the early Universe by cosmic reionization should affect the growth of galaxies in the lowest-mass halos \citep[e.g.,][]{Efstathiou:1992, Gnedin:2000}.  Reionization became an increasingly important mechanism for explaining the `missing satellites problem' \citep{Klypin:1999, Moore:1999} in the 1990s and early 2000s, which was considered a strong challenge to the $\Lambda$CDM cosmological paradigm \citep[e.g.,][]{Bullock:2000, Benson:2002, Somerville:2002}.  The discovery of UFDs orbiting the MW in the early 2000s \citep[e.g.,][]{Willman:2005a, Belokurov:2006a,Zucker:2006a} confirmed the existence of extremely faint galaxies that appeared to have ancient stellar populations \citep[e.g.,][]{Brown:2012, Okamoto:2012, Brown:2014, Weisz:2014a} consistent with predictions from some galaxy formation models that included the effects of reionization \citep[e.g.,][]{Bovill:2009, Sawala:2016, Wetzel:2016, Bullock:2017}.

\begin{figure}[pos=tbp]
    \centering
    \includegraphics[width=1.0\linewidth]{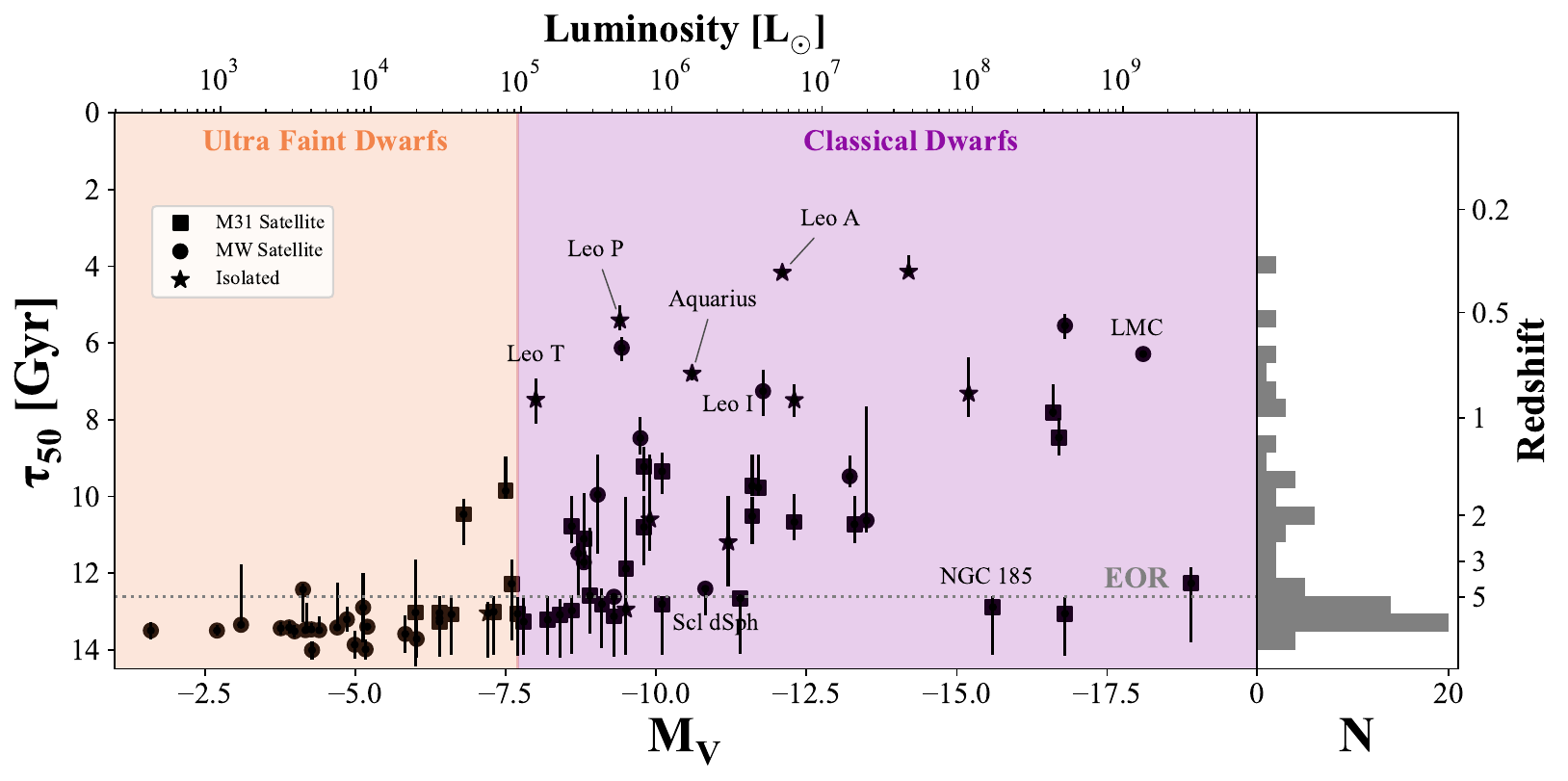}
    \caption{The relationship between galaxy luminosity (or stellar mass) and median mass-weighted stellar population age ($\tau_{50}$) determined from each galaxy's cumulative SFH.  Galaxies included on this diagram all have CMDs that reach, or nearly reach, below the oMSTO.  The epoch of reionization is indicated by the dashed grey line.   We indicate the generally accepted division between classical (purple shading) and ultra-faint dwarfs (orange shading) at $M_V=-7.7$ as specified in  \citet{Simon:2019}. We adopt the \citet{Planck:2020} cosmology to convert between lookback time and redshift. Symbols distinguish MW satellites, M31 satellites, and isolated galaxies, as defined in \S \ref{sec:concepts_classification}. By historical convention, \A{XXVIII} and \A{XXXIII} (340--370~kpc from M31) are shown as M31 satellites. Pisces~I, historically classified as isolated, lies within 300~kpc of M31 and is also shown as an M31 satellite.  IC~10 and NGC~6822 are omitted, as they lack oMSTO-depth SFHs. Data for the plot is from a variety of sources: \citet{Cole:2007}, \citet{Hidalgo:2009}, \citet{Monelli:2010_a}, \citet{Monelli:2010_b}, \citet{Hidalgo:2011}, \citet{Clementini:2012}, \citet{deBoer:2012}, \citet{McConnachie:2012}, \citet{Brown:2014}, \citet{Cole:2014}, \citet{deBoer:2014}, \citet{Skillman:2014}, \citet{Weisz:2014a}, \citet{deBoer:2015}, \citet{Geha:2015}, \citet{Bettinelli:2018}, \citet{Mazzi:2021}, \citet{Ruiz-Lara:2021}, \citet{Rusakov:2021}, \citet{Sacchi:2021}, \citet{Simon:2021}, \citet{Massana:2022}, \citet{McQuinn:2024a}, \citet{McQuinn:2024b}, \citet[Gold sample]{Durbin:2025}, \citet{Savino:2025}, \citet{Perez-Garcia:2026}.}
    \label{fig:T50_vs_Lum}
\end{figure}

\begin{figure}[pos=tbp]
    \centering
    \includegraphics[width=1.0\linewidth]{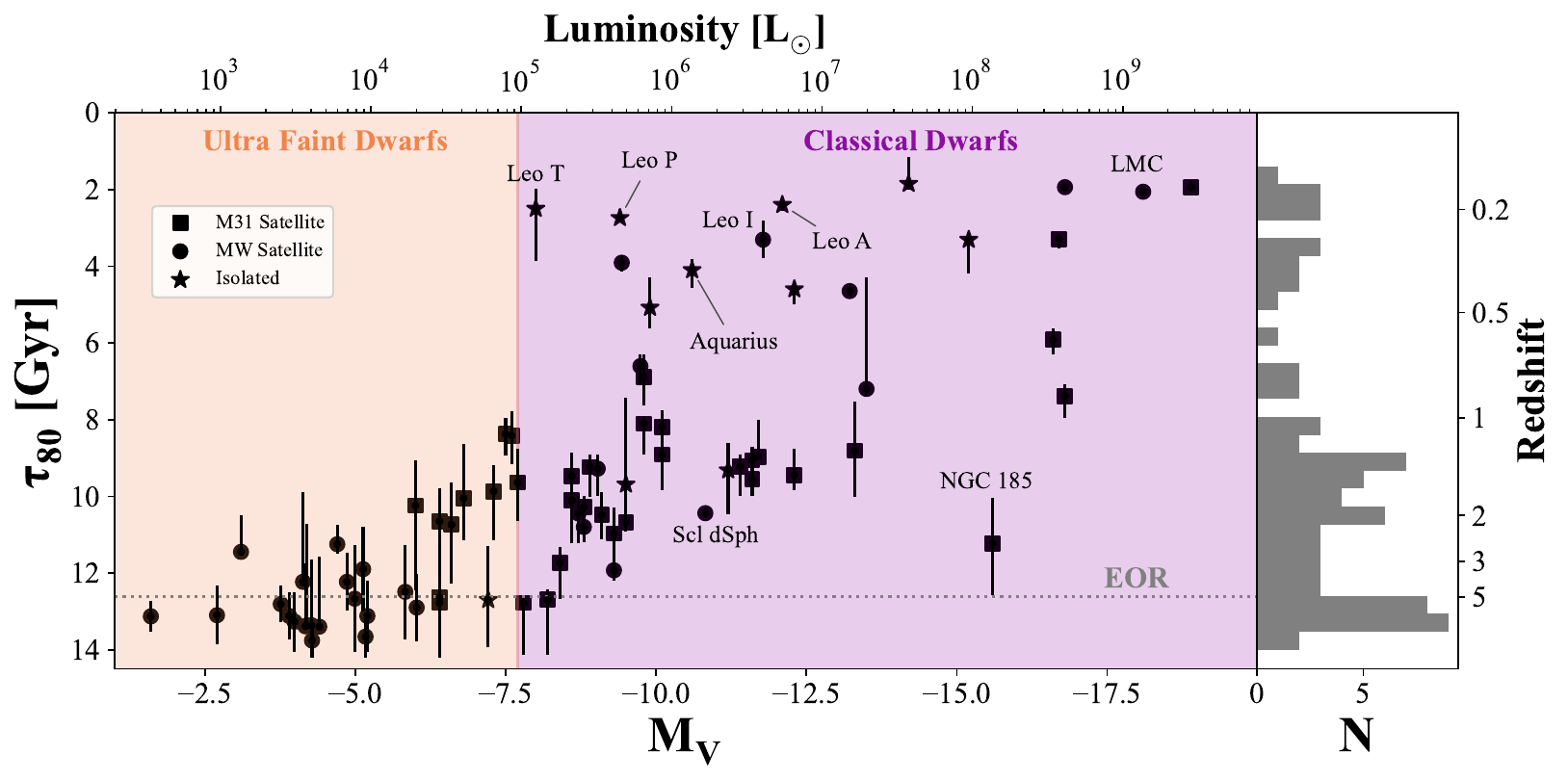}
    \caption{Same as Figure~\ref{fig:T50_vs_Lum}, but showing $\tau_{80}$, the lookback time when a galaxy had formed 80\% of its total stellar mass, which is a proxy for the quenching epoch.}
    \label{fig:T80_vs_Lum}
\end{figure}

Since their discovery, substantial efforts have been made to determine whether the SFHs of UFDs were consistent with being quenched by reionization and to determine the mass at which galaxies no longer appeared to be strongly affected by cosmic reionization.   Figures~\ref{fig:T50_vs_Lum} and \ref{fig:T80_vs_Lum} illustrate these trends.  Specifically, they show $\tau_{50}$ and $\tau_{80}$ as a function of galaxy luminosity for systems with SFHs that have been measured from CMDs that reach the oMSTO.

Figure~\ref{fig:T50_vs_Lum} shows that the majority of galaxies with $M_V \gtrsim -10$ had formed $\sim50$\% of their stellar mass prior to the conclusion of cosmic reionization.  For galaxies with $M_V \gtrsim -7.7$, the adopted definition of UFDs from \citet{Simon:2019}, only two, both M31 satellites (\A{XIII} and \A{XVI}), are not consistent with having formed 50\% of their stars before reionization. Leo~T, which lies just brighter than this threshold ($M_V\approx-8.0$), also formed most of its stars after reionization.  The MW satellite CVn~I ($M_V\approx-8.8$) is a similar case: its SFH, measured from deep \hubble/ACS imaging, places both its peak star formation and its $\tau_{50}$ at $\sim11.8$~Gyr ago, i.e., after the end of reionization \citep{Perez-Garcia:2026}.  This figure clearly shows that virtually all of the faintest galaxies host sizable fractions of stars from the pre-reionization era.

Figure~\ref{fig:T80_vs_Lum} highlights a far more nuanced picture of when these galaxies finally quenched star formation. The majority of galaxies with $M_V \gtrsim -6$ appear to have halted star formation by the end of reionization, but there are a handful of exceptions.  For slightly brighter galaxies ($-6 \gtrsim M_V \gtrsim -8$), only a minority of galaxies had quenched by the end of cosmic reionization, with several exhibiting quenching times 2--4 Gyr after reionization concluded.  For $M_V \lesssim -8$, the regime of classical satellite galaxies, only \A{XXIX} (and, within its uncertainties, \A{XV}) has a quenching time consistent with reionization. \A{XVII} ($M_V=-7.8$), which is just brighter than the UFD threshold, also quenched by the end of reionization.

Two caveats apply to interpreting the trends in Figures~\ref{fig:T50_vs_Lum} and \ref{fig:T80_vs_Lum}.  First, the requirement that CMDs reach (or nearly reach) the oMSTO restricts the sample almost entirely to the LG and its immediate periphery, raising the question of whether the inferred reionization-quenching trends reflect general early-Universe galaxy formation or LG-specific physics.  The progenitor region of the LG at $z\sim 7$, however, spans roughly $7$~comoving~Mpc and is large enough to be representative in terms of mean matter density and dark-matter halo counts \citep{Boylan-Kolchin:2016}, so the sample-variance concern is weaker than the present-day LG size might suggest.  Second, the orbital interpretation is important. As noted in \S \ref{sec:isolated}, precise proper motions and reconstructed orbital histories now place virtually all of the isolated LG dwarfs in this sample (IC~1613, Phoenix, Pegasus, Leo~T, Leo~A, WLM, Aquarius, and Tucana) on first infall into the LG \citep[e.g.,][]{Bennet:2024, Bennet:2025}, so their SFHs are largely unperturbed by past environmental interactions.  Together, these results suggest that the Figure~\ref{fig:T50_vs_Lum} and Figure~\ref{fig:T80_vs_Lum} trends are meaningful constraints on intrinsic low-mass galaxy formation rather than artifacts of an LG-specific environment.  Nevertheless, direct confirmation of this picture will require expanding the sample of oMSTO-depth SFHs to isolated low-mass galaxies beyond the LG, which would provide an independent test of whether the reionization-quenching and isolated-dwarf trends seen here are general or specifically tied to LG-scale environments.

These figures suggest several conclusions.  First, there is no present-day luminosity at which all galaxies appear to be abruptly quenched by reionization.  Instead, there is a decreasing fraction of galaxies that are quenched at the time of reionization as a function of galaxy luminosity, with sizable scatter in the trend.  This may reflect the theoretical expectation that reionization-induced quenching is driven by dark matter halo mass at the time of reionization, which is only loosely correlated with present-day luminosity.

Second, the fact that many galaxies do not quench until several Gyr after reionization is supportive of scenarios in which reionization along with other mechanisms (e.g., supernova feedback, gas stripping) appears to drive quenching -- there are very few systems that rapidly quench immediately after reionization ends.  There are some suggestions that the diversity of quenching times may reflect inhomogeneous reionization, but such inhomogeneities are expected to be several hundred Myr, not a few Gyr \citep[e.g.,][]{Lunnan:2012, Ocvirk:2016, Katz:2020}.  For example, the recent hydrodynamic zoom-in simulations of isolated UFDs embedded in an inhomogeneous reionization field reach a similar conclusion \citep{Deng:2026}.  They find that star formation continues for only several hundred Myr after the ionization front arrives, fueled by gas that remains self-shielded from the UV background.  The resulting diversity in stellar mass, age spread, and chemical enrichment is driven primarily by the halo mass at the time of reionization, rather than by when reionization reached each galaxy.

Third, the often adopted working definition of UFD ($M_V > -7.7$) does not correspond to a notable change in galaxy SFH properties.  While it may be useful to have an operational divide (as we employ in this review) between UFD and classical dwarf galaxies, there is no single luminosity that can capture a divide in the complex interplay between stellar mass assembly and reionization-driven quenching.  Fourth, and finally, it is not possible to rule out environmental trends in quenching.  That is, while most galaxies with $M_V \gtrsim -9$ have $\tau_{50} \gtrsim12.5$~Gyr, there is a much larger spread in values of $\tau_{80}$ for M31 satellites than for those around the MW.  Of course, selection effects may play a role. There are many more luminous M31 systems in this range than MW satellites, and only a few M31 satellites with $M_V \gtrsim -6$ are known (e.g., Peg~V; \S\ref{sec:m31_ufds}), none of which yet has an oMSTO-depth SFH.  Several theoretical models indicate that present-day satellites were located in low-density regions at the time of reionization.  However, by the time they quenched they may have already fallen into a host's halo. Deep CMDs and SFHs of M31 satellites with $M_V \gtrsim -6$ and of isolated galaxies with $M_V \gtrsim -9$ are essential to understanding the possible role of environment in ultimately quenching the faintest galaxies.

It is an interesting exercise to ask whether Figures~\ref{fig:T50_vs_Lum} and \ref{fig:T80_vs_Lum} \textit{require} reionization.  Taken on their own, the data show only that the faintest galaxies formed most of their stars within the first $\sim1$~Gyr of cosmic time and that many ceased forming stars within the first few Gyr.  Reionization, which concluded by $z\sim6$, is one natural explanation for this timing.  However, other mechanisms can, in principle, produce the same signature.  For example, stellar feedback could exhaust or expel the gas of sufficiently low-mass halos \citep[e.g.,][]{Dekel:1986}.  Similarly, a dwarf galaxy accreted by a larger system at early times could have been quenched by its environment rather than by the UV background.  Within the LG, these scenarios are challenging to disentangle \citep[e.g.,][]{Weisz:2014d}, as nearly all UFDs with SFHs are satellites whose individual orbits at $z\gtrsim6$ are essentially unconstrained (\S\ref{sec:orbital_histories}).

Nonetheless, three lines of evidence constrain these alternatives.  First, for galaxies with $M_V\gtrsim-10$, $\tau_{50}$ is remarkably uniform across $\sim3$--$4$ decades in luminosity, whereas feedback- or environment-driven quenching would be expected to depend more strongly on halo mass and orbit.  Second, the few isolated, low-mass galaxies with oMSTO- or SGB-depth SFHs (e.g., Leo~T, Leo~A, Aquarius) formed most of their stars \textit{after} reionization, consistent with reionization suppressing, but not halting, star formation in low-mass halos.  Leo~P is a particularly clean example. Despite a stellar mass at the epoch of reionization comparable to those of UFDs, it formed stars early, underwent an extended pause, and then resumed forming stars \citep{McQuinn:2024b}.  Reionization did not appear to quench it, though it is hard to rule out re-ignition from post-reionization gas accretion \citep[e.g.,][]{Wright:2019}.

Third, quenching by the host environment may be statistically disfavored, even without knowledge of individual orbits.  As one example, \citet{RodriguezWimberly:2019} extended galaxy-by-galaxy comparisons of quenching and infall times \citep[e.g.,][]{Rocha:2012, Weisz:2015a} to the UFD population as a whole.  They used the ELVIS simulations and found a $\lesssim0.1$\% probability that the six MW UFDs in the \hubble\ sample of \citet{Brown:2014} were all accreted early enough to have been quenched by the MW environment.  Including pre-processing by SMC-mass hosts raises this to only $\lesssim1$\%.  However, this result addresses only the host environment, not internal feedback. It is also least constraining for systems that quench several Gyr after reionization, such as many M31 UFDs (\S\ref{sec:m31_ufds}; \citealt{Weisz:2014c, Monelli:2016, Weisz:2019, McQuinn:2023, Savino:2023, Savino:2025}).  

Taken at face value, these lines of evidence suggest that reionization sets the early conditions for star formation in low-mass halos, while feedback and environment may determine when star formation finally ceases.  However, none of these lines of evidence is decisive on its own, and a definitive answer likely requires oMSTO-depth SFHs for a statistical sample of truly isolated UFDs.  The first such measurement, for Tucana~B, is underway with \jwst\ (GO-4748, PI: Sand, cf.\ \S \ref{sec:future_formation}), but assembling a larger sample remains challenging.

Beyond quenching, the SFHs of LG galaxies have been used to predict their optical and UV luminosities during the reionization era, providing a direct bridge to the high-redshift populations now probed by \hubble\ and \jwst. This near-field perspective was first demonstrated by \citet{Boylan-Kolchin:2014}, who combined the observed counts of MW satellites with the ELVIS simulations to show that reionization scenarios requiring substantial star formation in the smallest halos ($M_{\rm vir}\approx10^8$~\msun\ at $z\sim8$) would overproduce the number of surviving MW satellites seen today. This implies that star formation became sharply inefficient below $\sim10^9$~\msun, so that the high-$z$ rest-frame UV luminosity function (UVLF) must turn over near $M_{\rm UV}\approx-14$.

\citet{Weisz:2014b} introduced an empirical ``near-far'' approach, by reconstructing the rest-frame UVLF of LG progenitors over cosmic time down to $M_{\rm UV}\sim-5$ at $z\sim5$, which is orders of magnitude fainter than what is possible with direct imaging. They found a faint-end slope that steepens from $\alpha\approx-1.2$ at $z<1$ to $\alpha\approx-1.6$ at $4<z<5$. \citet{Boylan-Kolchin:2015} applied a similar technique to select LG galaxies and demonstrated that during the reionization-era most present-day LG dwarf galaxies would have been fainter than \hubble\ blank-field limits.  They suggested that \jwst\ should detect the $z\sim7$ progenitors of LMC-like galaxies, and that the progenitors of systems such as Fornax and Sagittarius were plausibly contributors to the UV ionizing background in the early Universe. Pushing to the extreme faint end, \citet{Weisz:2017} found that the ancestors of the faintest known MW UFDs ($M_V\sim-2$ at the present day) had $M_{\rm UV}\sim-3$ to $-6$ during the reionization era. This finding implies that if the high-$z$ UVLF remains steep ($\alpha\lesssim-2$) to these faint luminosities, UFDs would have produced the majority of the ionizing UV flux such that only $\sim10$--$50$\% of the total UV light
at the time of reionization is
directly detectable by \hubble\ and \jwst.

These near-far connections rest on the assumption that the present-day fossil record of LG galaxies faithfully encodes the high-$z$ progenitor population. \citet{Gandhi:2024} used cosmological simulations to confirm that the fossil record of surviving low-mass galaxies yields an essentially unbiased reconstruction of the proto-LG stellar mass function and UVLF at $z>6$. Together with the recognition that the LG progenitor region is a representative patch of the early Universe \citep{Boylan-Kolchin:2016}, these results establish that CMD-based SFHs can be used as a quantitative probe of the reionization-era galaxy population, including the galaxies that may have driven reionization, at masses far below the reach of direct high-$z$ surveys.

\subsection{The Diversity of Low-Mass Galaxy Star Formation Histories} \label{sec:diversity}

In the seminal review of LG dwarf galaxies, \citet{Mateo:1998} concluded that ``No two dwarf galaxies have the same star formation history!'' This view is now largely supported by hydrodynamical simulations of low-mass galaxy formation, which also show a great diversity in SFH that is linked to halo mass and accretion history, gas inflows, feedback, stochasticity, and the effects of environment \citep[e.g.,][]{Stinson:2007, Governato:2010, Munshi:2013, El-Badry:2016, Fitts:2017, Wright:2019, Gutcke:2021}.

With the rich set of SFH measurements made over the past two decades, we examine our empirical knowledge of diversity in low-mass galaxy SFHs.  To mitigate the effects of reionization, we focus our discussion of SFH diversity on classical galaxies with $M_V \lesssim -8$.

The diversity of classical dwarf galaxy SFHs is apparent in Figures~\ref{fig:T50_vs_Lum} and \ref{fig:T80_vs_Lum}.  At any fixed luminosity, there is a clear spread in both $\tau_{50}$ and $\tau_{80}$ of several Gyr.  Values of $\tau_{50}$ change weakly with galaxy luminosity, whereas values of $\tau_{80}$ are generally younger for more luminous systems.

\begin{figure}[pos=tbp]
    \centering
    \includegraphics[width=1.0\linewidth]{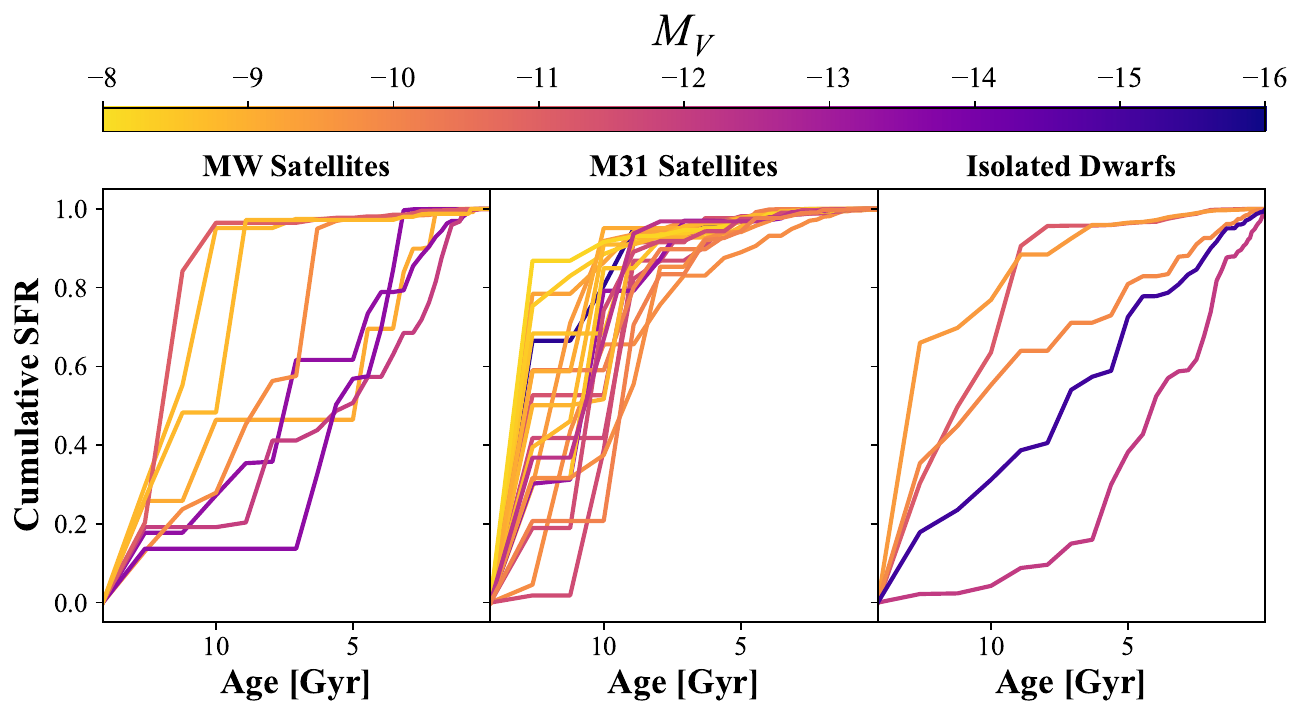}
    \caption{The SFHs of classical LG galaxies with $-8\ge M_V\ge -16$ in different environments color-coded by galaxy luminosity as selected from various literature compilations. All SFHs were measured from oMSTO-depth CMDs. \textbf{Left:} MW satellites with data from \citet{Weisz:2014a}; \textbf{Middle:} M31 satellites with data from \citet{Savino:2025}; \textbf{Right:} Isolated Local Group galaxies from the LCID program  \citep[e.g.,][]{Gallart:2015}. Pisces~I, part of the original LCID sample, is shown among the M31 satellites (\S \ref{sec:concepts_classification}).   Following historical convention, the M31 panel also includes \A{XXVIII} and \A{XXXIII}, which are located 340--370~kpc from M31.  M31 ellipticals NGC~147, NGC~205, and M32 ($-16.8\le M_V\le-16.6$; \citealt{Savino:2022}) fall just outside the plotted luminosity range. IC~10 is not shown, as it lacks an oMSTO-depth SFH.}
    \label{fig:Diversity}
\end{figure}

Beyond luminosity, an important factor for diversity is environment.  Figure~\ref{fig:Diversity} illustrates our current knowledge of SFH diversity for classical dwarf galaxies as a function of environment (MW satellites, M31 satellites, isolated galaxies), measured from CMDs that reach a galaxy's oMSTO.

This figure shows several interesting trends.  First, both the MW satellites and isolated dwarf galaxies show a diverse range of SFHs from mostly ancient systems to galaxies that formed a substantial fraction of their stars at intermediate or younger ages.   A second trend, within the MW and isolated galaxy groups, is a correlation such that more luminous systems generally have more extended SFHs, albeit with some scatter.  This affirms previous findings for MW satellites.

Based on well-constrained SFHs of LG galaxies, \citet{Gallart:2015} suggested that the diversity in dwarf galaxy SFHs may be a manifestation of galaxy halo assembly bias.  That is, a galaxy with an extended or `slow' SFH was likely in a low-density region in the early Universe, while those that formed a majority of their stars at early times, i.e., `fast' SFHs, were likely in higher density regions of the early Universe.  This finding is, in some sense, a confirmation of the morphology-density relation, though with more emphasis on holistic stellar mass assembly as opposed to quenching and gas content.

Interestingly, \citet{Bennet:2025} have shown that Tucana is on first infall to the LG.  With a SFH that is purely ancient and a likely location in a low-density region of the early Universe, Tucana provides an intriguing counterexample to the slow-fast scenario.  \citet{Shread:2026} predict the existence of a much larger population of quenched, field galaxies within the LV, which, if correct, would provide interesting insights into halo assembly bias and self-quenching mechanisms in low-mass galaxies.

A third, and perhaps most striking, trend is the difference between the SFHs of the M31 satellites and either the MW or isolated systems.  Comparatively, the M31 satellite SFHs are in some ways less diverse.  Although they show a broad range of initial stellar mass growth, virtually all M31 satellites in Figure~\ref{fig:Diversity} quench at intermediate ages (i.e., no later than $\sim5$~Gyr ago).  With the exception of Pisces~I, a gas-rich dTrans that formed $\sim10$\% of its stellar mass within the last $\sim5$~Gyr \citep{Savino:2025}, there are no systems with significant late-time star formation.  Moreover, the dependence of SFH on luminosity is less apparent in Figure~\ref{fig:Diversity} than among the MW satellites.  Interestingly, \citet{Savino:2025} found that luminosity and present-day distance from M31 together predict the quenching epoch to within $\sim2$~Gyr.  They also identified two subpopulations of M31 satellites based on their SFHs.  Nearly half of them have SFH trends that are similar to MW satellites, while the other half show prominent star formation $\gtrsim12$~Gyr ago followed by quenching $\gtrsim$8--10~Gyr ago, which is not common among MW satellites.  However, they noted that membership in these two subpopulations shows no apparent trend with luminosity or distance from M31.

The M31 panel of Figure~\ref{fig:Diversity} omits IC~10, a gas-rich, star-forming M31 satellite in this luminosity range, because high foreground extinction has so far precluded an oMSTO-depth CMD.  Its SFH, measured from a shallower CMD, indicates that roughly half of its stellar mass formed within the last $\sim4$~Gyr \citep{Weisz:2014a}.  However, among the dSphs, the contrast with the MW persists: Fornax and Leo~I formed stars until $\lesssim1$~Gyr ago (Appendix~\ref{sec:app_mw_classical}), whereas none of the M31 dSphs in Figure~\ref{fig:Diversity} formed a significant fraction of their stars within the last $\sim5$~Gyr.

The cause of the stark differences between the M31 SFHs and the other two populations is not clear.  Qualitatively, it may be related to the putative recent major merger that is often invoked to explain the complex and prominent tidal debris around M31.  As one example, \citet{DSouza:2018} suggest that the progenitor of M32 entered the halo of M31 $\sim$ 5 Gyr ago, which coincides with the typical quenching time of M31 satellites.
In this scenario, multiple groups of satellites may be involved, which could plausibly explain the two subpopulations identified by \citet{Savino:2025}.

The relationship between the SFHs of M31 and its satellites is not clearly established. M31 itself appeared to experience an increase in star formation in the region of its GSS $\sim4$--$5$~Gyr ago, while the disk shows a burst of activity $\sim2$--$4$~Gyr ago.  Whether these timescales are connected to features in the SFHs of the satellite population remains to be determined.

\subsection{Environmental Trends in Low-Mass Galaxy Quenching} \label{sec:environment}
The effect of environment on local galaxy evolution is anchored by the morphology-density relationship.  The presence of red, gas-poor galaxies near massive hosts and blue, gas-rich galaxies preferentially found in the field are strong indicators that environment is an important driver of galaxy evolution.

At the same time, the morphology-density relationship alone is a coarse tool for determining when and how galaxies transitioned from gas-rich to gas-poor. Additional timing information (e.g., $\tau_{80}$) provided from CMD analysis may help to constrain the role of various quenching mechanisms (e.g., astration, feedback, ram pressure, tidal stripping, mergers).

\begin{figure}[pos=tbp]
    \centering
    \includegraphics[width=1.0\linewidth]{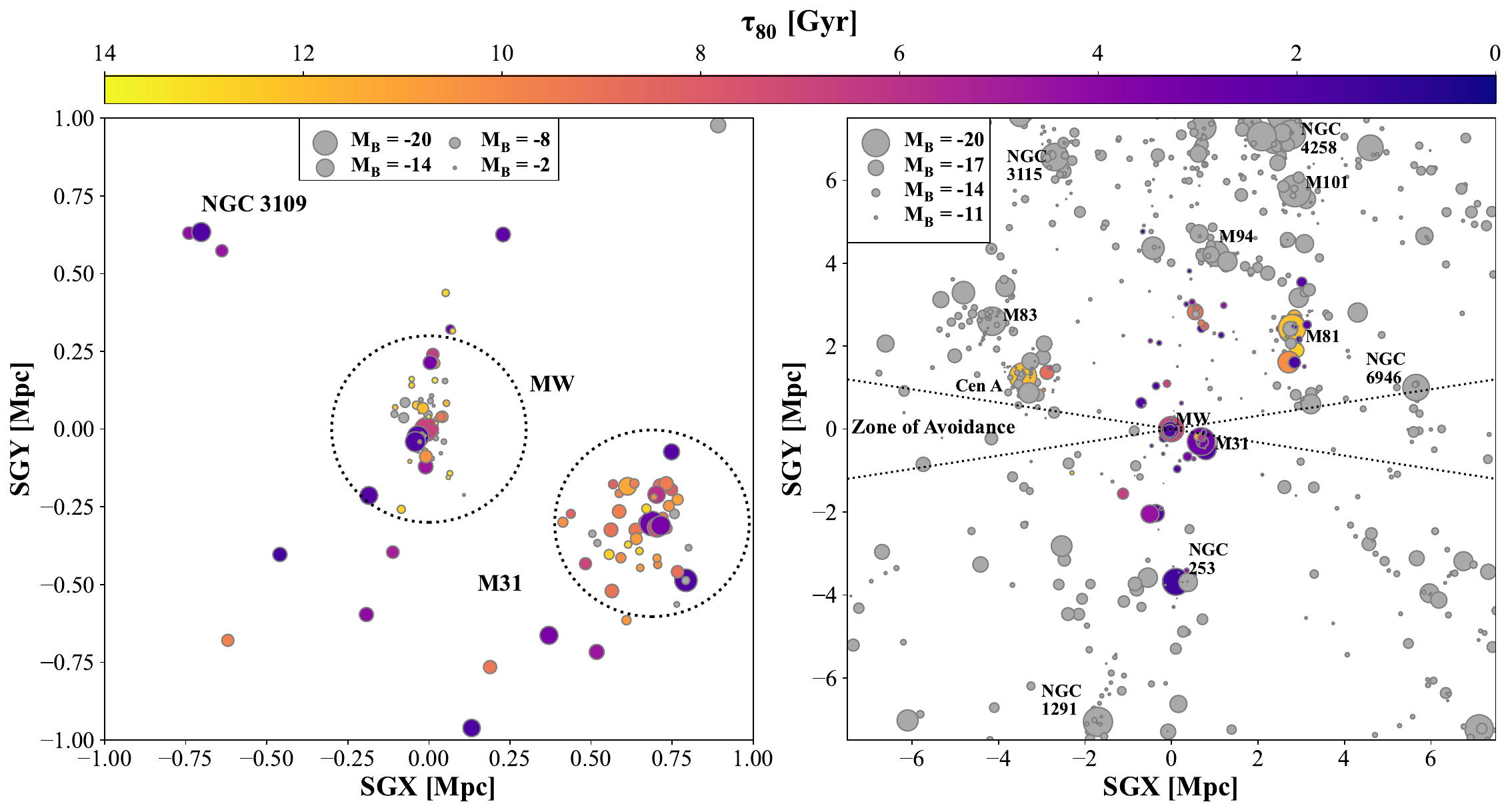}
    \caption{Galaxy quenching in the local Universe. The distribution of known nearby galaxies in the SGX-SGY supergalactic plane, from the catalogs of \citet[updated to 2025]{Karachentsev:2013} and \citet{Pace:2024}. The symbol sizes are proportional to each galaxy's luminosity. Galaxies for which any determination of the lifetime SFH exists in the literature are color-coded according to their $\tau_{80}$ values, a proxy for quenching epoch. For completeness, we also include the MW, whose illustrative $\tau_{80}$ value is derived from the SGB sample of \citet{Xiang:2022}. \textbf{Left:} A map zoomed-in on the Local Group showing galaxies within $\sim1$~Mpc, centered on the MW. The dotted circles mark a radius of 300~kpc around the MW and M31, our nominal boundary between satellites and isolated galaxies (\S \ref{sec:concepts_classification}). \textbf{Right:} A similar map of known galaxies within a cube 15~Mpc on a side, centered on the MW. We include the name of select notable nearby bright galaxies for reference. The dotted lines indicate the approximate region corresponding to the MW zone of avoidance (ZOA, $|b|<10^\circ$). SFH data from \citet{Cole:2007}, \citet{Weisz:2008}, \citet{Hidalgo:2009}, \citet{McQuinn:2010a}, \citet{Monelli:2010_a}, \citet{Monelli:2010_b}, \citet{Hidalgo:2011}, \citet{Rejkuba:2011}, \citet{Weisz:2011}, \citet{Williams:2011a}, \citet{Cannon:2012}, \citet{Clementini:2012}, \citet{deBoer:2012}, \citet{Makarova:2012}, \citet{Brown:2014}, \citet{Cole:2014}, \citet{deBoer:2014}, \citet{Karachentsev:2014}, \citet{Skillman:2014}, \citet{Weisz:2014a}, \citet{deBoer:2015}, \citet{Geha:2015}, \citet{McQuinn:2015c}, \citet{Makarova:2017}, \citet{Williams:2017}, \citet{Bettinelli:2018}, \citet{Cannon:2018}, \citet{Hargis:2020}, \citet{Mazzi:2021}, \citet{Ruiz-Lara:2021}, \citet{Rusakov:2021}, \citet{Sacchi:2021}, \citet{Simon:2021}, \citet{Massana:2022}, \citet{Makarova:2023}, \citet{McQuinn:2023}, \citet{McQuinn:2024a}, \citet{McQuinn:2024b}, \citet{McQuinn:2024c},  \citet{Durbin:2025}, \citet{Savino:2025}, \citet{Perez-Garcia:2026}. }
    \label{fig:LVMap}
\end{figure}

Figure~\ref{fig:LVMap} shows the spatial distribution of LG (left) and LV (right) galaxies color-coded by values of $\tau_{80}$ taken from literature SFHs indicated in the figure caption.  The LG remains the best set of systems in the local Universe for studying environmental effects.  The selection function is well-understood and most galaxies have CMD-based SFHs.  In general, galaxies that are not currently satellites of the MW or M31 have younger values of $\tau_{80}$ compared to known satellites. Cetus and Tucana are well-known exceptions. Tucana appears to be on first infall and is thus a truly isolated, quenched galaxy. Cetus may have interacted with M31 more than $\sim6$~Gyr ago, near the limit of reliable orbit reconstruction \citep{Bennet:2025}.

\begin{figure}[pos=tbp]
    \centering
    \includegraphics[width=1.0\linewidth]{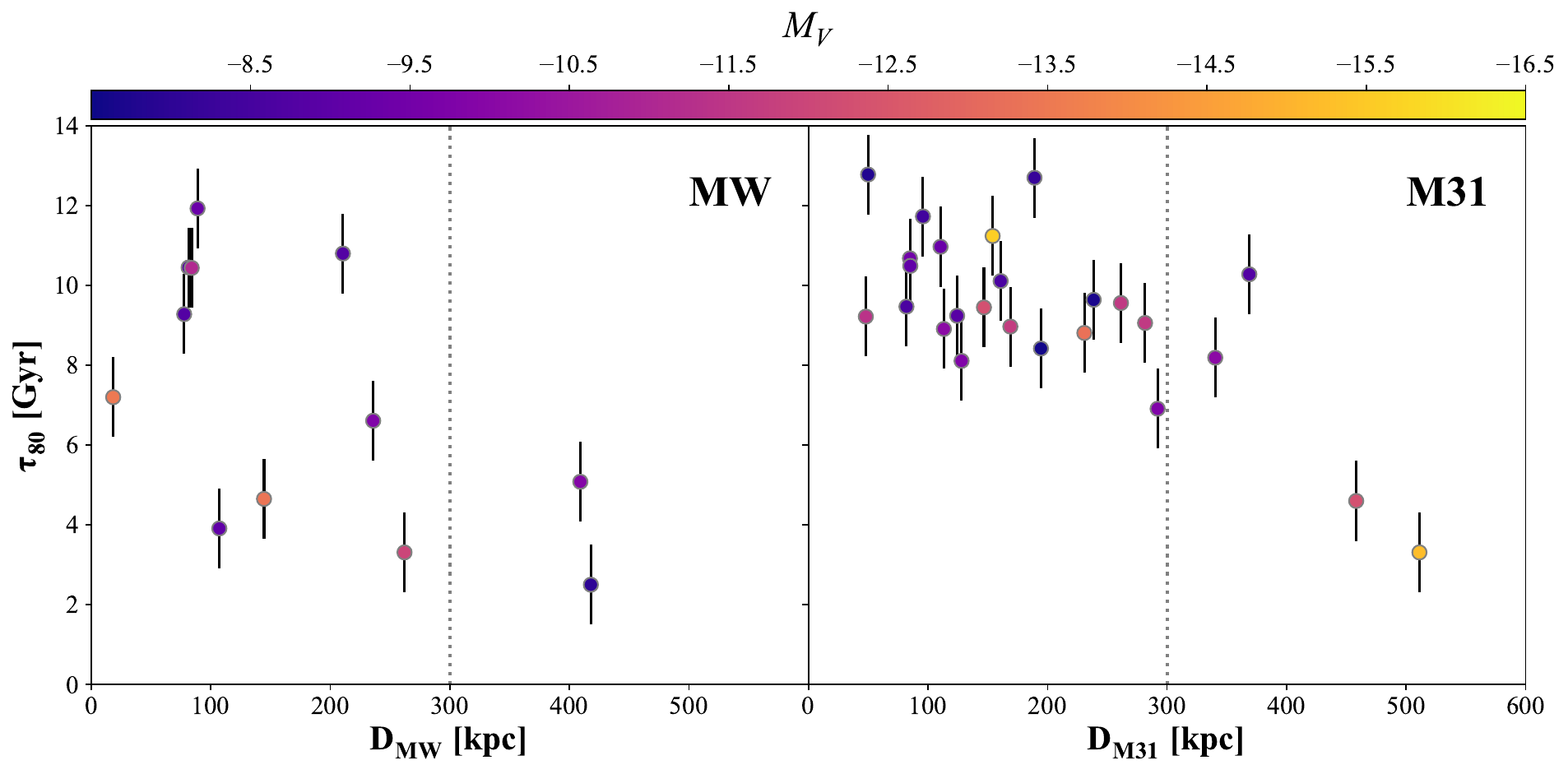}
    \caption{Measured value of $\tau_{80}$, as a function of galactocentric distance for LG galaxies with $-16.5<M_V<-7.5$ in the vicinity of the MW (left panel) and M31 (right panel). Galaxies have been color-coded by absolute luminosity. The vertical dotted line marks 300~kpc, our nominal boundary between satellites and isolated galaxies (\S \ref{sec:concepts_classification}). Isolated galaxies (Phoenix and Leo~T in the left panel; Pegasus and IC~1613 in the right panel) are plotted by their distance from the nearest host. \A{XXVIII} and \A{XXXIII} also lie beyond 300~kpc but are treated as M31 satellites by historical convention (\S\ref{sec:concepts_classification}). IC~10 and NGC~6822 are excluded, as they lack oMSTO-depth CMDs.  For CVn~I, $\tau_{80}$ is from \citet{Perez-Garcia:2026} (see Appendix~\ref{sec:CVnI_sfhs}).}
    \label{fig:sfh_vs_D}
\end{figure}

Within the MW and M31 satellite populations, there have long been suggestions that the quenching time is related to distance from the host galaxy \citep[e.g.,][]{VandenBergh:1994b}. SFHs measured by \citet{Savino:2025} for M31 satellites confirm this observation.  Specifically, as noted in \S \ref{sec:m31_classical}, for M31 satellites \citet{Savino:2025} found that the median stellar age and quenching epoch are correlated with galaxy luminosity and galactocentric distance (Figure~\ref{fig:sfh_vs_D}).  They also showed that the combination of galaxy luminosity and present-day distance from M31 predicts the satellite galaxy quenching epoch to within 1.8 Gyr.  As part of this study, \citet{Savino:2025} also noted the presence of a similar trend in MW satellites, though formal quantification is limited due to the low number of MW satellites with deep CMD-based SFHs.

The right panel of Figure~\ref{fig:LVMap} shows trends in quenching beyond the LG. A clear result is that galaxies that are not obviously associated with a group tend to have younger $\tau_{80}$ values compared to those in groups.  In part, this affirms that blue galaxies are generally more prevalent in the field, though selection effects may play a role as quenched field galaxies have also proven more challenging to find, and a sizable population is predicted to exist \citep{Shread:2026}. Among the known field galaxies, qualitatively, lower-luminosity star-forming galaxies generally appear to have younger $\tau_{80}$ values, though sample sizes are quite small.

There are only a handful of galaxy groups outside the LG for which CMD-based SFHs for even some galaxies have been measured.  The M81 Group is among the best-studied in terms of CMD-based SFHs, with $\sim10$--$15$ member galaxies having published measurements \citep[e.g.,][]{Weisz:2008}.  From its stellar mass, M81 is believed to host a rich satellite population \citep[e.g.,][]{Smercina:2020, Smercina:2022}.  However, its distance (3.6 Mpc) has made it challenging to systematically measure CMD-based SFHs.  Any trends in current data are heavily affected by selection effects.

Low-mass galaxies in other nearby groups (e.g., Cen~A, Sculptor) have even more sparse coverage with CMDs.  The large number of grey galaxies in the right panel of Figure~\ref{fig:LVMap} shows that there is tremendous opportunity for future observing programs to expand our knowledge of low-mass galaxy formation throughout the LV.

\subsection{Bursty Star Formation}
\label{sec:bursts}

A sudden increase in a galaxy's rate or intensity of star formation relative to a smooth, continuous mode has historically been referred to as a burst, flash, or single episode of star formation \citep[e.g.,][]{Burbidge:1958, Tinsley:1968, Searle:1972, Searle:1973}.  The importance of ``bursty'' star formation has grown over the subsequent decades, as galaxy SFHs, galactic winds, dark-matter ``cores'' in cosmological simulations, chemical enrichment models, the circumgalactic medium, the stellar IMF, and integrated-light SFR indicators across cosmic time \citep[e.g.,][]{Stinson:2007, Lee:2009b, McQuinn:2009, Governato:2010,Kennicutt:2012, Weisz:2012a, Conroy:2013, Bullock:2017, Sparre:2017, Tumlinson:2017, Weinberg:2017, Faucher-Giguere:2018, Zheng:2024} all appear to be linked to or influenced by short-lived, intense episodes of star formation.  Bursty star formation also appears to be central to interpreting the properties of the highest-redshift galaxies observed by \jwst\ \citep[e.g.,][]{Sun:2023b, Sun:2023a, Looser:2024, Mintz:2026, Stark:2026}.

For resolved galaxy studies, SFHs that are not smooth and are instead characterized by rapid increases or decreases in SFR have been known virtually since the first CMDs were formally modeled in the early 1990s \citep[e.g.,][]{Tosi:1991, Smecker-Hane:1994, Tolstoy:1996, Gallart:1996, Dohm-Palmer:1997, Dohm-Palmer:1998, Dolphin:1997, Gallart:1999, Harris:2001}.  Bursty star formation in local galaxies as determined from CMD analysis has received increasing attention in the literature.  In this section, we summarize the current evidence for bursty star formation based on formal CMD modeling of nearby galaxies.

\subsubsection{Definitions and Introduction to Concepts}

\begin{figure}[pos=tbp]
    \centering
    \includegraphics[width=0.96\linewidth]{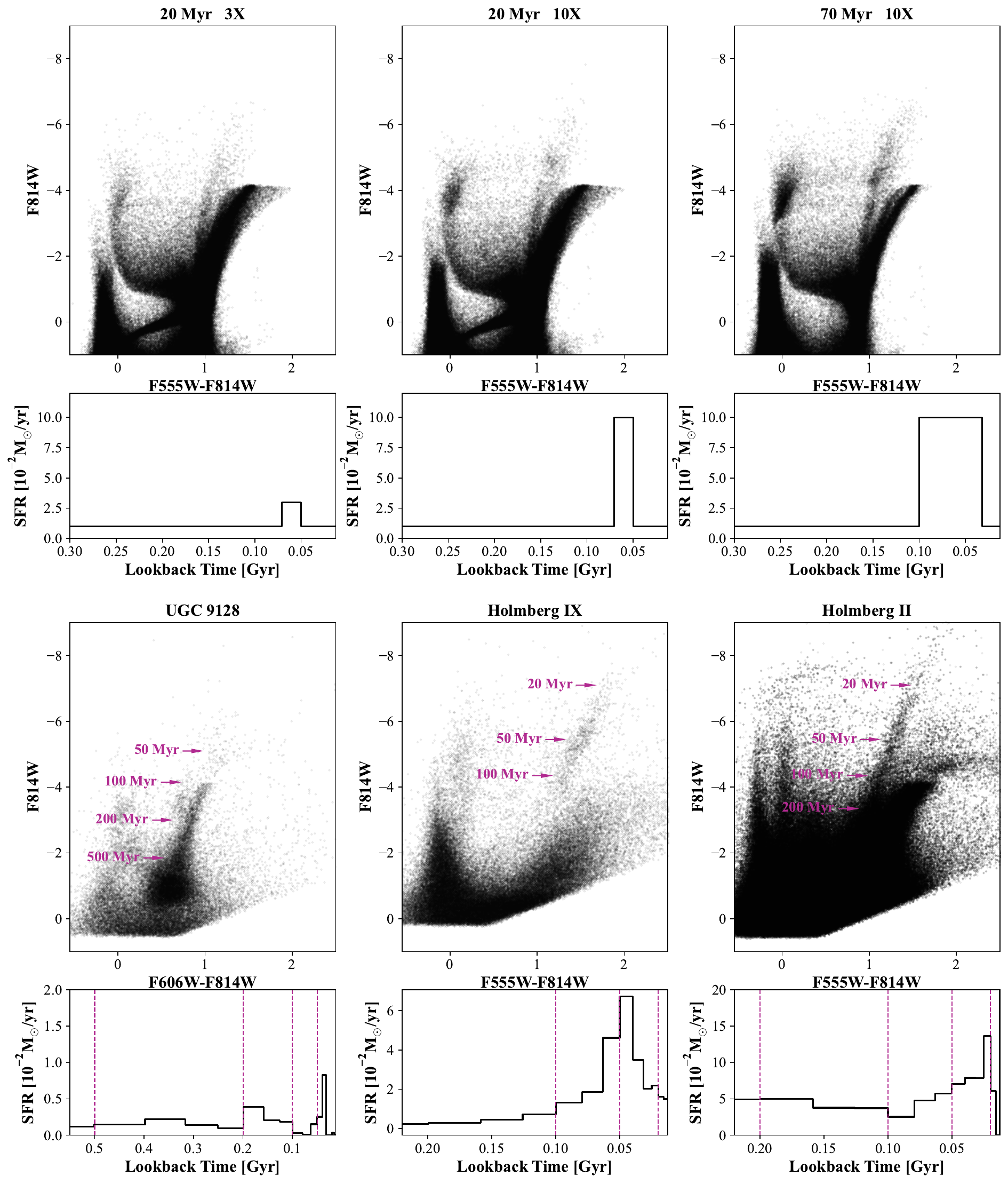}
    \caption{\textbf{Upper}: Synthetic CMDs of three stellar populations that experienced a recent burst of star formation. Compared to the long-term average, the star formation rate was enhanced by three times for 20 Myr (left), by ten times for 20 Myr (center), and by ten times for 70 Myr (right). The corresponding SFHs are reported under the CMDs. \textbf{Lower}: Deep optical CMDs of three Local Volume dwarf galaxies: Sextans A (left), Holmberg IX (center), and Holmberg II (right), all of which show signs of a recent star formation burst. We have highlighted the approximate magnitude of the helium-burning ignition point for stellar population of different ages. The corresponding SFHs are reported under the CMDs, where we have highlighted the same timescales reported on the CMDs. Data from \citet{Dalcanton:2009,Weisz:2011}.}
    \label{fig:Recent_Burst}
\end{figure}

\begin{figure}[pos=tbp]
    \centering
    \includegraphics[width=1.0\linewidth]{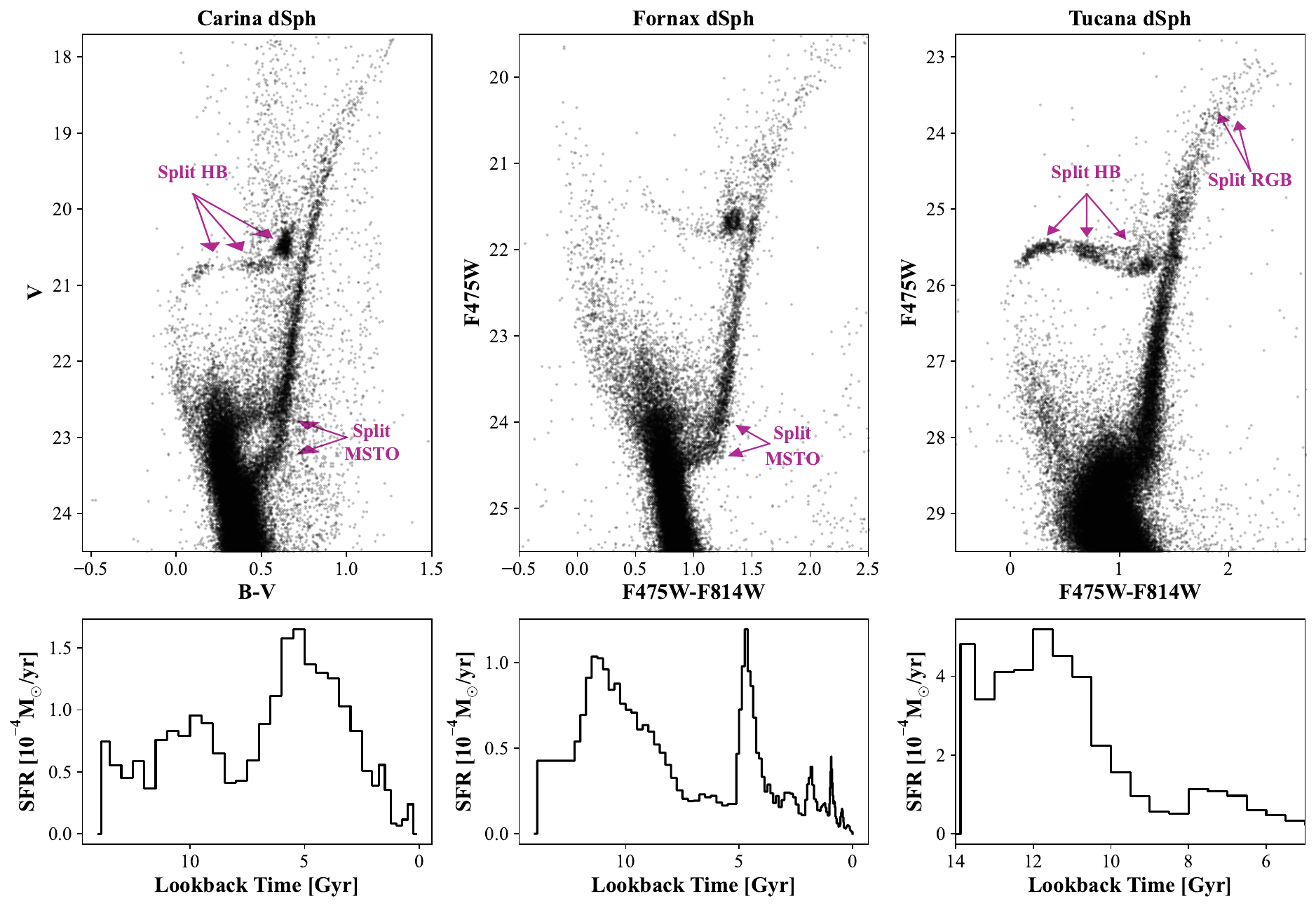}
    \caption{\textbf{Upper}: Deep optical CMDs of three Local Group dSph galaxies: Carina (left), Fornax (center), and Tucana (right). The presence of discrete features in the MSTO, HB, and RGB of these galaxies have been highlighted. \textbf{Lower}: The corresponding lifetime SFHs of these galaxies, in which the presence of episodic star formation activity is readily appreciable. Data from \citet{deBoer:2014}, \citet{Rusakov:2021}, \citet{Monelli:2010_b}, and \citet{Savino:2019}.}
    \label{fig:Fossil_Burst}
\end{figure}

The vocabulary used to describe enhanced star formation is only loosely defined in the literature.  Establishing a more concrete definition is one goal of this review.  Here, we adopt three working definitions.  A \emph{burst of star formation} (also a ``starburst'' or ``flash'') is a temporally localized, intense episode of star formation that significantly exceeds the long-term average of the host galaxy.  Its duration and amplitude can vary widely. We make no attempt to define a single quantitative threshold.  A \emph{starburst galaxy} is one whose current SFR, often concentrated in the central region at a level $\gtrsim 10\times$ the long-term mean for its stellar mass, would exhaust the available neutral-gas reservoir through astration on a timescale much shorter than the age of the Universe \citep[i.e., the ``Roberts time'';][]{Roberts:1963,Kennicutt:1994}.  \emph{Bursty star formation} describes a galaxy whose SFH consists of repeated episodes of enhanced star formation separated by periods of relative quiescence, as opposed to a smoothly varying or simply parameterized (e.g., $\tau$-model) history.  As emphasized by \citet{Knapen:2009}, none of these labels has a uniquely well-defined boundary, and observationally the degree to which a SFH appears bursty depends on the time resolution of the measurement.

Starburst galaxies and bursty star formation are important for several reasons.  First, roughly $\sim20$\% of all high-mass star formation at low redshift takes place in starburst galaxies \citep{Brinchmann:2004,Heckman:2005}, where it drives the galactic winds responsible for chemical enrichment of the circumgalactic and intergalactic media and for setting the escape fraction of ionizing radiation \citep[e.g.,][]{Chevalier:1985,Heckman:1990,Veilleux:2005,Heckman:2015}.  Second, in lower-mass galaxies, repeated burst-driven outflows are thought to redistribute mass in the inner dark-matter halo, transforming central cusps into cores \citep[e.g.,][]{Navarro:1996,Governato:2010,Governato:2012,Pontzen:2012,DiCintio:2014,Read:2016,Tollet:2016}.  Simulations suggest that the SFH may regulate the efficiency of core formation.  For example, in the NIHAO and FIRE-2 suites, galaxies with burstier SFHs, and those that formed more of their stellar mass after reionization, develop shallower inner dark matter profiles at fixed stellar-to-halo mass ratio \citep[e.g.,][]{Sarrato-Alos:2026}.  However, the observational picture is less clear.  In one study, \citet{Pickett:2026} found that five of the seven M31 dSphs they modeled have anomalously low central dark matter densities, and that dark matter heating by star formation, as inferred from their CMD-based SFHs, is disfavored as the sole explanation.  They instead suggest that tides and scatter in halo concentration may contribute, which predicts that these galaxies should be on more plunging orbits, a prediction that proper motions of M31 satellites can test (\S \ref{sec:orbital_histories}).  Third, bursts are also frequently triggered by galaxy interactions, in which tidal forces drive gas toward galaxy centers and enhance cloud-cloud collisions \citep[e.g.,][]{Larson:1978,Keel:1985,Kennicutt:1987,Mihos:1996}.  Finally, at high redshift, strongly bursty star formation has been proposed as one explanation for the abundance of UV-bright galaxies seen by \jwst\ \citep[e.g.,][]{Faucher-Giguere:2018, Sun:2023b,Sun:2023a}.

A starburst galaxy is the most extreme manifestation of bursty star formation.  However, a galaxy can be bursty without being a starburst galaxy.  Cosmological zoom-in simulations of low-mass galaxies routinely produce strongly bursty SFHs on $\sim10$--$100$~Myr timescales \citep[e.g.,][]{Stinson:2007,Hopkins:2014,Hopkins:2018b,El-Badry:2016,Sparre:2017,Emami:2021}. The classical dimensionless measure is the Scalo $b$ parameter, $b \equiv {\rm SFR}_{\rm present} / \langle {\rm SFR} \rangle_{\rm past}$, which is, by construction, large for galaxies undergoing a present-day enhancement relative to their lifetime average \citep{Scalo:1986}; values of $b \gtrsim 1$--$3$ are commonly used as a working definition of a starburst \citep[e.g.,][]{Kennicutt:1998}.  Various other quantitative metrics, such as the variance of the SFR averaged over a chosen timescale, have been proposed to characterize the degree of burstiness \citep[e.g.,][]{Sparre:2017,Caplar:2019,Iyer:2020}.  Another widely-used metric is the ratio of SFR indicators, which are sensitive to different timescales (e.g., H$\alpha$ for $\sim 5$~Myr; FUV for $\sim 100$~Myr).  In principle, they can flag bursty behavior in galaxies \citep[e.g.,][]{Lee:2009b,Weisz:2012a,Emami:2019}, but they also are limited by degeneracies with dust and putative IMF variations.  Moreover, they cannot reconstruct the temporal pattern itself.  SFHs from CMD modeling are the only observational route that can deliver both temporal resolution and an absolute timeline.

One subtlety in interpreting bursty star formation in low-mass galaxies is that some apparent burstiness may arise from sampling rather than from a fundamentally different mode of star formation.  If giant molecular cloud (GMC) sizes and star formation efficiencies are approximately mass-invariant, then the global SFR is proportional to the number of active GMCs at any given time \citep[e.g.,][]{Gerola:1978, Krumholz:2005}. Large galaxies have many active GMCs and average out toward a smooth SFR, while small galaxies have few and can fluctuate stochastically down to zero \citep[e.g.,][]{Searle:1973, Gerola:1978, Lee:2009b,Fumagalli:2011,daSilva:2014,Krumholz:2015}.  The interesting astrophysical question is how much a particular low-mass galaxy's observed burstiness departs from this baseline stochastic expectation \citep[e.g.,][]{Gerola:1980, Orban:2008}.  Low-mass galaxies have the additional complication that intense feedback from a burst can globally heat the interstellar medium (ISM) and temporarily quench further star formation, producing the duty cycle of burst-and-pause that is the defining characteristic of bursty star formation in these systems.

\subsubsection{Studying Star Formation Bursts with Resolved Stellar Populations}

CMDs offer multiple complementary diagnostics of past bursts of star formation, each sensitive to a different age regime.  For recent bursts ($\lesssim 1$~Gyr), the simplest signature is a step function in the MS. For a fixed IMF, the location of the step encodes the age of the burst, its amplitude, and the mass of stars formed.  An associated population of MSTO and SGB stars crossing the CMD provides a complementary constraint for somewhat older bursts where the upper-MS lifetime has already elapsed.  Perhaps the most direct constraints on the timing, duration, amplitude, and SFR of recent bursts come from BHeB and RHeB stars.  Although the color/temperature extension and morphology of the blue loop remain sensitive to poorly modeled physics such as mass loss, rotation, convection, and overshooting (see \S\ref{sec:stellar_evolution}), the underlying luminosity--age relation is comparatively robust, which is what makes BHeB and RHeB stars useful as burst chronometers \citep{Dohm-Palmer:1997, Weisz:2008, McQuinn:2012}.  Figure~\ref{fig:Recent_Burst} shows three examples of CMDs displaying clear BHeB and RHeB burst signatures alongside their reconstructed SFHs.  We note that the observed width of any burst in the recovered SFH is effectively a convolution of its true intrinsic duration with the temporal smearing introduced by the SFH-recovery method itself (see \S\ref{sec:time_resolution}); even a $\delta$-function input SFH passed through a CMD-fitting code is recovered with a finite width, which sets a floor on the burst durations that can be claimed from any single measurement \citep[e.g.,][]{Aparicio:2016}.  Compounding this, the achievable time resolution degrades strongly with lookback time (see \S\ref{sec:time_resolution}): recent SFHs ($\lesssim 1$~Gyr) can be resolved at $\sim 10$--$20\%$ of the lookback time, while at ancient lookback times even the best CMD-based SFHs rarely achieve an age resolution better than $\sim 0.5$--$1$~Gyr.  A direct corollary is that actively star-forming dwarfs may appear more bursty than quenched dSphs in part because their recent SFHs are measured at much finer temporal resolution. Equivalent short-duration bursts at ancient epochs would be smoothed out or missed entirely.  Cross-galaxy comparisons of burstiness must therefore be interpreted alongside the time resolution of the underlying SFHs.

Interpreting CHeB sequences is complicated by long-standing model mismatches between predicted and observed CHeB distributions \citep[e.g.,][]{McQuinn:2011}, and by two sources of contamination that can mimic young BHeB populations.  The first is reddened MS stars, which can be identified by adding UV imaging to extend the color baseline \citep{Choi:2025}.  The second is OBe stars (rapidly rotating MS stars surrounded by a disk, exhibiting emission lines and redder optical colors than ordinary MS stars; e.g., \citealt{Gull:2022, Gull:2026}).  H$\alpha$ imaging can also help distinguish genuine BHeB stars from both sources of contamination.  Be star contamination of stellar population studies is also discussed in \citet{Peters:2020}.

Older ($\gtrsim 1$~Gyr) ``fossil'' bursts can be identified through discrete features in the older MSTOs and SGBs, structure in the HB morphology, and/or potentially splits in the RGB \citep[e.g.,][]{Savino:2018}.  Figure~\ref{fig:Fossil_Burst} shows examples of these fossil-burst signatures in the CMDs of LG galaxies Carina, Fornax, and Tucana, illustrating how the episodic nature of star formation is directly visible (see \S \ref{sec:M31bursts}, \S
\ref{sec:fossils}).

In addition to temporal patterns, spatial information from resolved star observations enables the construction of spatially resolved SFHs \citep[e.g.,][]{Dohm-Palmer:1997, Dohm-Palmer:1998, Dohm-Palmer:2002, Weisz:2009a, McQuinn:2012, Lescinskaite:2022}.  Spatial mapping makes it possible to distinguish global bursts from isolated or propagating star formation; the one-to-one mapping between BHeB CMD position and stellar age makes BHeBs particularly powerful for spatial reconstruction of recent burst activity.

\subsubsection{Evidence for Bursty Star Formation in Nearby Galaxies}

\paragraph[\hubble\ observations of M31 and M33]{\hubble\ observations of M31 and M33:}
\label{sec:M31bursts}

The most prominent fossil burst signature in M31's CMD-based SFH is an enhancement of star formation between $\sim 4$ and 2~Gyr ago.  \citet{Bernard:2012} first detected this feature in a deep \hubble/ACS field of M31's outer disk warp ($\sim 25$~kpc along the major axis), where it followed a multi-Gyr lull and produced most of the remaining stellar mass.  Independent confirmation came from PHAT-based analyses of low-extinction disk regions \citep{Williams:2015}, archival halo and substructure fields \citep{Bernard:2015a}, and the lifetime SFH over the full PHAT footprint \citep{Williams:2017}, with the same enhancement appearing in every disk-like region examined.  Independent halo evidence on a similar timescale comes from AGB-based SFHs of six inner-halo regions, which show elevated star formation between 3 and 5~Gyr ago \citep{Lee:2024}.  Two important caveats apply. First, \citet{Bernard:2015b} re-analyzed the original warp field along with two new southern-disk fields at 20 and 23~kpc and found that the new fields did not show a comparable enhancement, so the event is not strictly universal across the disk.  Second, the historical interpretation as a tidal signature of an M31--M33 encounter is in tension with current proper-motion constraints (\S\ref{sec:orbital_histories}), which disfavor a close passage within the last $\sim 6$~Gyr.

On recent timescales ($\lesssim 500$~Myr), M31 shows no clear globally synchronized burst.  The spatially global recent SFR is roughly constant (see \S \ref{sec:m31}), with a notable decline at very recent times as we discuss below.  However, spatially resolved SFHs reveal substantial sub-galactic structure, as discussed in \S \ref{sec:m31} and shown in Figure~\ref{fig:m31_sfh_map}.  Specifically, analysis of the PHAT regions shows that the recent ($\lesssim 500$~Myr) star formation of M31 is concentrated in three ring structures at $\sim 5$, 10, and 15~kpc, each with independent temporal evolution \citep{Lewis:2015}. The 10~kpc ring has been active for at least the last 400~Myr, the inner 5~kpc ring was most prominent $\sim 200$~Myr ago, and the outer 15~kpc ring is dominated by stars younger than 80~Myr.

\begin{figure}[pos=tbp]
    \centering
    \includegraphics[width=0.9\linewidth]{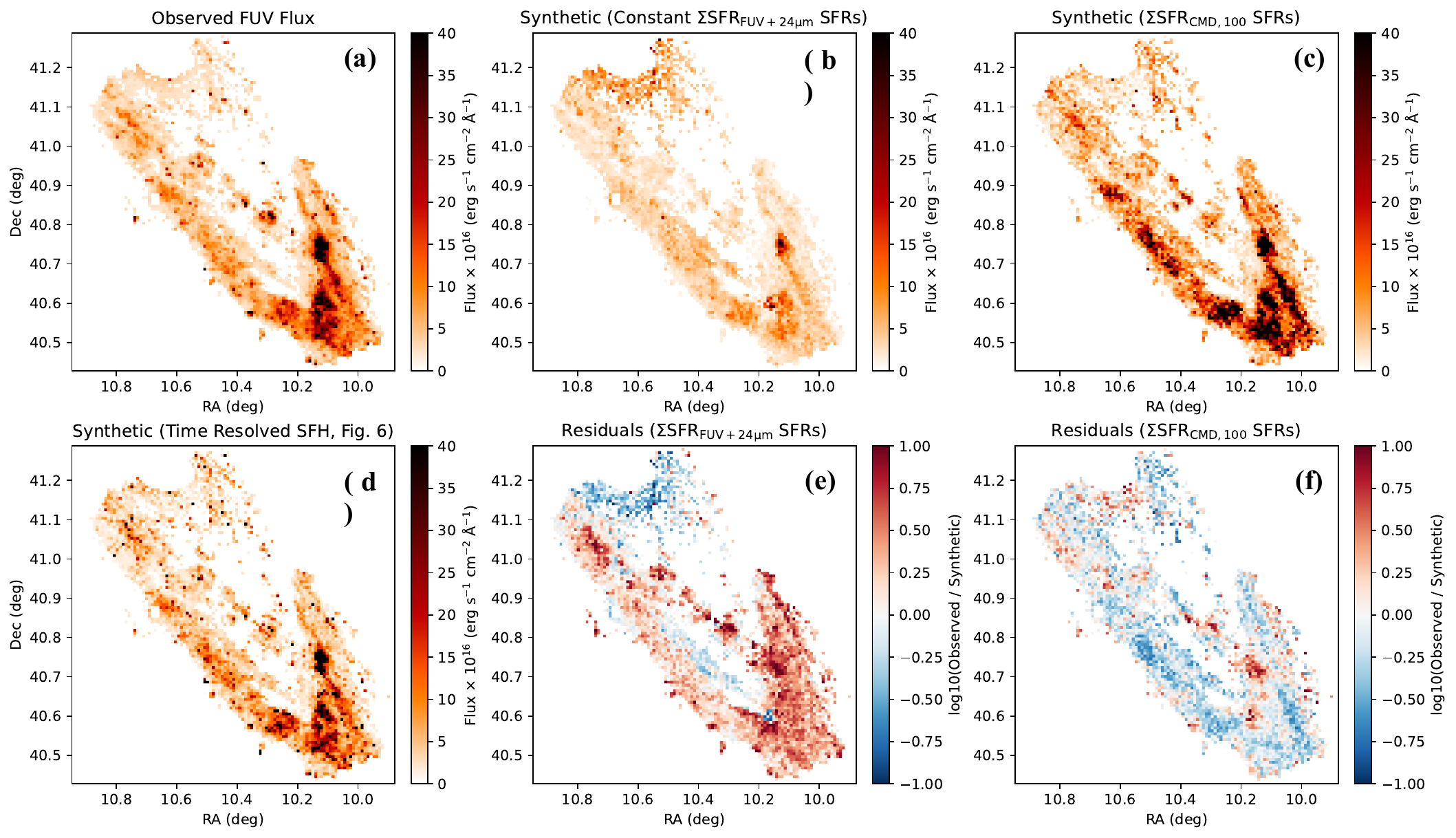}
    \caption{Figure adapted from \citet{Wainer:2026}. A comparison of spatially resolved FUV flux (top left) and predictions for the FUV fluxes based on various assumptions about the underlying SFH.  The three synthetic plots (top middle, top right, bottom left) show the predicted FUV fluxes for M31 assuming a constant SFH over the past 100 Myr with a SFR derived from the spatially global FUV$+24~\mu$m flux, a constant SFR over the past 100 Myr determined from CMD fitting, and CMD-based SFH over the past 100 Myr of each region.  Notably, only the CMD-based SFH approach reproduces the observed FUV flux, indicating that temporal variations in the SFH are important for interpreting integrated FUV fluxes of galaxies. The residuals in the bottom row (middle and right) quantify the mismatches between the two assumed constant SFR scenarios. }
    \label{fig:m31_fuv_map}
\end{figure}

\citet{Wainer:2026} combined the PHAT and PHAST datasets and found a steady decline in the global SFR over the last $\sim 500$~Myr, from $\sim 0.67~M_\odot$~yr$^{-1}$ averaged over the last 100~Myr to $\sim 0.43~M_\odot$~yr$^{-1}$ over the last 20~Myr.  This is consistent with the multi-Gyr declining trend identified from the full PHAT lifetime SFH \citep{Williams:2017}.
Interestingly, the widely used integrated-light tracer (FUV$+24~\mu$m) appears to under-predict the CMD-based SFR by a factor of $\sim 2$ \citep{Lewis:2015, Wainer:2026}, which is attributed to short term variations in the recent SFH that the CMD-based reconstruction captures but integrated tracers smooth over. An illustration of this finding is shown in Figure~\ref{fig:m31_fuv_map}.
 \citet{Wainer:2026} also identified a localized quenching event near M32 in the last $\sim 15$~Myr, providing direct evidence that environment-driven modulation of star formation can be tracked at $\sim 10$~Myr resolution.  These results are a clean illustration of the broader point made in \S\ref{sec:bursts}: an essentially constant global SFR can decompose into structured, time-varying components, some of which may appear as localized bursts or quenching events.

\hubble\ observations of M33 in the PHATTER program \citep{Lazzarini:2022} similarly identified structures in the SFH maps associated with small-scale and galaxy-scale fluctuations in the SFR (Figure~\ref{fig:m33_sfh_map}).  Specifically, the star formation pattern was dominantly in a two-armed spiral more than 80~Myr ago and has more recently become flocculent.  These observations also revealed that M33 has a central bar \citep{Smercina:2023}.  Taken together, these studies of M31 and M33 show that the global SFRs of spiral galaxies can evolve smoothly (roughly constant in M33 and steadily declining in M31), with star formation organized into separate structures with independent spatial identities and temporal evolution.  Comparable evidence for spatially varying star formation in the lower-mass Magellanic Clouds is discussed in \S \ref{sec:burst_spatial}.

\paragraph[Resolved star observations of Local Volume massive galaxies]{Resolved star observations of Local Volume massive galaxies:}

Recent ($\lesssim 1$~Gyr) SFHs of nearby massive galaxies beyond the LG have been reconstructed from CMDs that are generally no deeper than the RC.  The emerging picture is that a roughly constant disk SFR is typical.  For example, \citet{Gogarten:2010} measured six radial bins in NGC~300 and found rates constant to within a factor of two. \citet{Williams:2013} reached the same conclusion for ANGST observations of NGC~2403. \citet{Sacchi:2019} likewise reported flat recent SFHs across four radial regions of the Sculptor Group spiral NGC~7793. The M81 outer-disk ANGST field is consistent with a gradually declining but otherwise smooth recent SFH \citep{Williams:2009}.

Departures from this baseline are common in systems that appear disturbed, presumably due to recent interactions. The M81 group provides the best-studied example, as it appears to be undergoing tidal disruption as revealed by \hi{} filamentary structures linking M81, M82, and NGC~3077 \citep{Yun:1994}.  For M81 itself, \citet{Choi:2015} found no detectable age gradient across the spiral arm in 20 archival \hubble\ fields, disfavoring a stationary density wave in favor of a kinematic spiral pattern driven by interaction with the group.  \citet{Williams:2010} reported that three radial fields of the M81 Group spiral NGC~2976 share a $\gtrsim 10\times$ enhancement in SFR between 0.4 and 0.6~Gyr ago, plausibly tied to a $>1$~Gyr-old M81 Group encounter that simultaneously stripped the outer gas and concentrated star formation in the inner disk, leaving the galaxy in a post-starburst state. In contrast, M82 has an ongoing starburst that is widely attributed to the same encounter. \citet{Williams:2011a} found that the present prototypical starburst accounts for only a small fraction of M82's lifetime stellar mass. Outflow-driven star formation has also been characterized in M82's halo. \citet{Rao:2025} identified two distinct recent episodes in the `Southern Arcs,' $\sim5$~kpc from the disk along the minor axis: a burst $\sim150$--$70$~Myr ago that formed $\sim85$\% of their stellar mass, coincident with the starburst in the disk, and a second beginning $\sim30$~Myr ago. Young, metal-enhanced stars in these arcs plausibly formed from outflow--cloud interaction \citep{Suwannajak:2024}. A population in the M82 Tail likely formed from stripped gas \citep{Rao:2026}.  A combination of resolved star and \textit{Galaxy Evolution Explorer} (\textit{GALEX}) UV imaging also revealed recent star formation in the M81--M82 \hi{} tidal bridge \citep{deMello:2008}.

Similar enhancements are seen elsewhere in the LV and often attributed to interactions.  \citet{Bortolini:2024} identified enhanced episodes at $\sim$ 10 and $\sim$ 100 Myr in the disturbed M101 Group spiral NGC~5474. The nearest giant elliptical, Cen~A, shows comparable evidence for recent enhanced star formation. \citet{Rejkuba:2022} reported elevated inner-halo ($R \lesssim 30$~kpc) star formation $\sim 2$--$3$~Gyr ago from a combined analysis of nine archival \hubble\ halo fields. \citet{Aghdam:2024} detected enhancements at $\sim 800$~Myr and $\sim 3.8$~Gyr ago in LPVs across two halo fields, with the combined evidence pointing to a $\lesssim 4$~Gyr-old minor merger.

Taken together, the resolved star evidence favors a baseline of roughly constant disk SFR punctuated by enhanced episodes following interactions.  However, we note that the links to interactions depend heavily on morphological and circumstantial evidence rather than on independent interaction-timescale constraints.

\paragraph[Fossil bursts in Low-Mass Galaxies]{Fossil bursts in Low-Mass Galaxies:}
\label{sec:fossils}

The clearest evidence for bursty star formation $\gtrsim1$~Gyr ago comes from LG dwarf galaxies, where oMSTO-depth CMDs have revealed discrete episodes of star formation.  Lifetime SFHs for the systems discussed here are summarized in \S\ref{sec:mw_classical}, \S\ref{sec:m31_ecosystem}, and \S\ref{sec:isolated}, with galaxy-by-galaxy details in Appendix~\ref{sec:individual_sfhs}. Here, we synthesize what they collectively imply about the prevalence and nature of bursty star formation in dwarf galaxies.  

A central methodological lesson from the past decade is that the apparent smoothness of a fossil SFH is largely a function of data quality. SFHs that look smooth at modest depth or photometric SNR can resolve into discrete episodes once sufficiently deep, high-SNR, and wide-area data become available.  Figure~\ref{fig:Fossil_Burst} shows three representative LG dSphs: Fornax, whose discrete intermediate-age episodes were sharpened by deep \hubble/ACS imaging \citep{Rusakov:2021}; Carina, with its widely-cited multi-episode SFH \citep[e.g.,][]{Smecker-Hane:1994, deBoer:2014}; and Tucana, whose discrete MSTOs, RGBs, and HB morphology motivated the identification of fossil bursts at different age resolutions \citep{Monelli:2010_b, Savino:2018, Savino:2019}.

The most recent demonstration of this principle is Ursa~Minor. It was long thought to be among the simplest LG dSphs, with an almost entirely ancient stellar population \citep[e.g.,][]{Mighell:1999, Carrera:2002}.  However,  \citet{Sato:2025} used deep Subaru/Hyper Suprime-Cam (HSC) imaging to identify two distinct age and metallicity populations whose mean ages differ by $\sim 1$~Gyr, consistent with the chemodynamically distinct components seen spectroscopically by \citet{Pace:2020}.  Sculptor and Draco currently appear comparatively smooth at the SFH level \citep[e.g.,][]{Bettinelli:2018, Bettinelli:2019, Ruiz-Lara:2021}, but the \citet{Sato:2025} and \citet{Rusakov:2021} results suggest that this may reflect the limits of current data quality rather than a genuine absence of discrete fossil-burst structure.

The most massive LG satellites, the LMC and SMC, present the most data-rich examples of fossil-burst SFHs and provide the cleanest case for interaction-driven episodic star formation.  Detailed SFH measurements from ground-based wide-field surveys (e.g., MCPS, VMC, SMASH, STEP, YMCA) and spatially-resolved \hubble\ programs (Scylla) have identified multi-Gyr-scale enhancements in both galaxies (\S\ref{sec:lmc} for LMC, \S\ref{sec:smc} for SMC).  Both clouds show enhancements at $\sim 2$~Gyr ago and $\sim 500$~Myr ago, an apparent synchronization first noted by \citet{Harris:2009} that has since been reinforced in multiple subsequent analyses \citep[e.g.,][]{Weisz:2013b, Rubele:2018, Mazzi:2021, Massana:2020, Massana:2022, Ruiz-Lara:2020b, Cohen:2024b, Cohen:2024a}.  These have been interpreted as the fossil signatures of recurring LMC--SMC close encounters.  Detailed SFHs of the Magellanic Bridge \citep{Ficara:2026} provide an additional, spatially-distinct probe of the most recent interaction.  The Magellanic burst synchronization is also one of the few fossil-SFH-derived calibration points for proper-motion-based orbital reconstructions of the Cloud system (see \S\ref{sec:orbital_histories}), making it both a fossil-burst record and a dynamical anchor.  The Magellanic Clouds  represent the empirical link between the dwarf-galaxy fossil-burst record and the interaction-driven enhancements identified above in the LV sample.

Isolated LG dwarfs (\S\ref{sec:isolated}) provide perspectives in environments removed from a massive host.  The LCID program showed that fossil bursts in isolated dwarfs are typically more prolonged and less discrete than in MW satellites, with star formation often continuing at low levels to the present epoch \citep{Monelli:2010_b, Monelli:2010_a, Hidalgo:2009, Hidalgo:2011, Skillman:2014, Cole:2014}.  WLM shows a similarly prolonged pattern in \hubble\ data \citep{Albers:2019}, although deeper \jwst\ imaging reveals an extended post-reionization pause before star formation re-ignited \citep{McQuinn:2024a}.  The discovery of a quenched, classical-mass dwarf in isolation \citep{Hai:2026} demonstrates that environmental quenching is not the only path to truncating fossil-burst sequences.  Of course, these bursts could be even shorter and more pronounced than reported, given that their CMDs have lower SNRs than typical for MW satellites.

An intriguing pattern of intermediate age fossil bursts is emerging in the mass range just above that of the ultrafaint galaxies. At this extreme low-mass end of star forming galaxies, \citet{McQuinn:2024b} measured the SFH of Leo~P to oMSTO depth with \jwst\ and revealed an extended post-reionization pause in star formation.  By comparing to three other galaxies with similar quality CMDs at relatively low masses, they identified this pattern in all four low-mass dwarfs, which is qualitatively distinct from the more continuous SFHs of more massive star-forming dwarfs.  Together, the isolated dwarf galaxy fossil burst record analyzed by \citet{McQuinn:2024b} hints at a mass-dependent trend, with prolonged-burst behavior in $\sim 10^8$--$10^9$~\msun\ systems transitioning toward post-reionization pauses at $\lesssim 10^8$~\msun.

\begin{figure}[pos=tbp]
    \centering
    \includegraphics[width=0.7\linewidth]{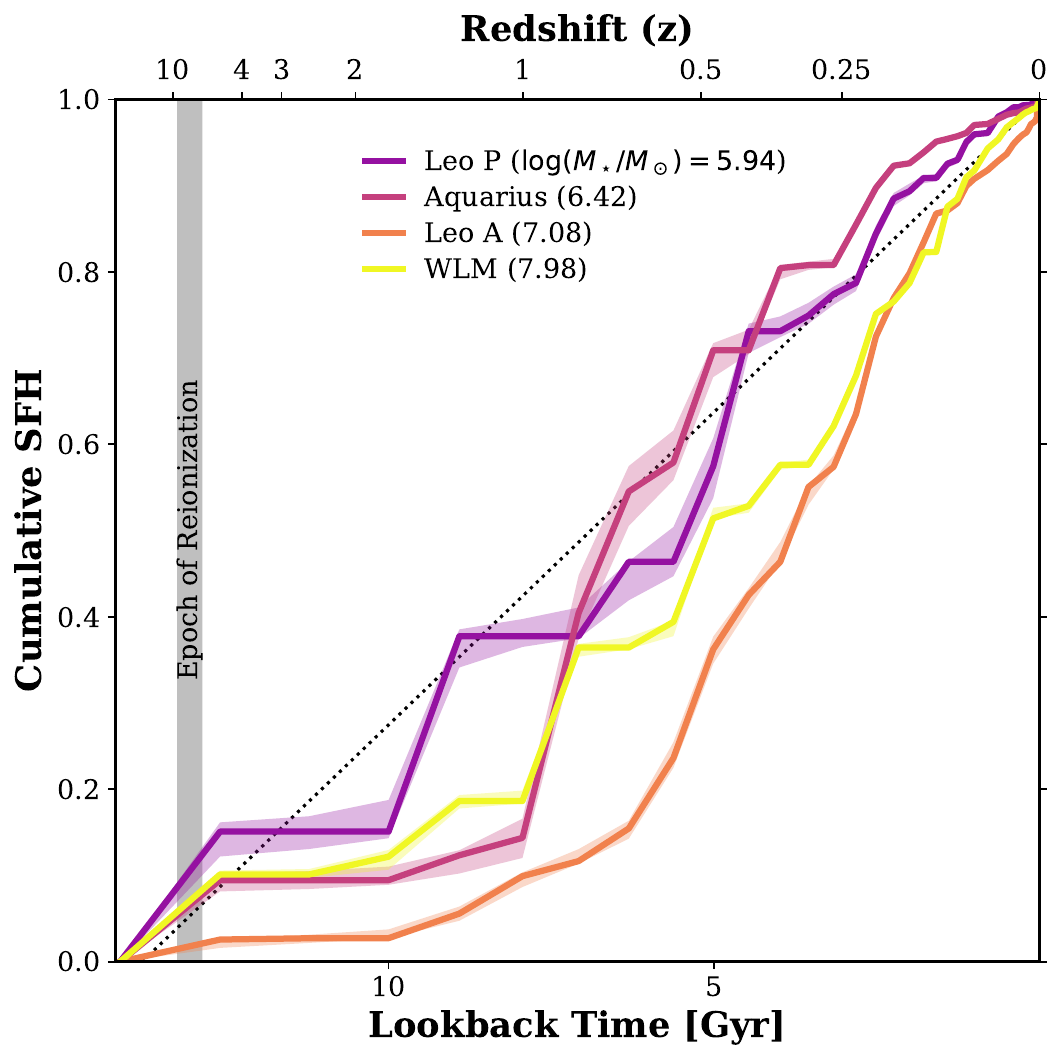}
    \caption{Figure adapted from \citet{McQuinn:2024b} showing
    cumulative SFHs of four gas-rich, isolated dwarf galaxies spanning $\sim 2$~dex in stellar mass, measured from oMSTO-depth, \hubble- or \jwst-based CMDs \citep{Cole:2007, Cole:2014, McQuinn:2024a, McQuinn:2024b}. The observations for all four galaxies were reduced and analyzed identically in \citet{McQuinn:2024b} so that the comparison is as direct as possible. The stellar masses have been slightly altered to agree with their values in the Local Volume Database \citep{Pace:2024}. The vertical shaded grey band marks the epoch of reionization. The diagonal dotted line reflects a constant SFH.  While all four galaxies show a pause in star formation immediately following the epoch of reionization, there is a lack of mass ordering in the fraction of total mass formed early or the delay before reigniting, in contrast with the expectation that lower-mass systems might exhibit more prolonged pauses following an initial burst of star formation. This suggests that additional factors modulate this picture, e.g., present day stellar mass is not a perfect guide to total galaxy mass.}
    \label{fig:isolated_sfhs}
\end{figure}

This phenomenon is illustrated in Figure~\ref{fig:isolated_sfhs}, which shows the cumulative SFHs of four gas-rich, isolated dwarfs spanning $\sim 2$~dex in stellar mass.
Although all four galaxies show the post-reionization pause, there is no clear pattern for a mass dependence for the fraction of stars formed before reionization or the duration of the pause following reionization. This scatter in the SFHs
highlights how additional variables (e.g., dark matter halo mass, gas accretion history) may play a role in this low-mass transition range of galaxy stellar mass growth. Simulations are capable of producing these re-ignition types of events \citep[e.g.,][]{Fitts:2017, Wright:2019, Rey:2020, Applebaum:2021}, but the mass range
and frequency of occurrence is not well established.

The M31 satellite system (\S\ref{sec:m31_ecosystem}) reveals systematic differences from the MW. At fixed luminosity, M31 dSphs tend to have shorter, earlier star formation than their MW counterparts and lack the late-quenching tail ($\tau_{80}\lesssim5$~Gyr) seen in several MW classicals \citep{Skillman:2017, Savino:2025}.  In contrast, M31 UFDs show more extended SFHs than MW UFDs \citep{Savino:2023}, with And~XVI providing a notable example of a low-mass system that appears to have escaped early reionization-driven quenching \citep{Weisz:2014c, Monelli:2016}.

\paragraph[Recent burstiness in star-forming dwarf galaxies]{Recent burstiness in star-forming dwarf galaxies:}

Systematic study of recent ($\lesssim 1$~Gyr) bursty star formation in dwarf galaxies began with \citet{Weisz:2008}. They derived SFHs for nine M81 Group dIrrs from shallow \hubble/ACS imaging and used the Scalo $b$ parameter (see \S\ref{sec:bursts}) to characterize their temporal patterns.  They showed that none of these galaxies were well-described by either a single-burst or constant-SFR models.

The STARBurst IRregular Dwarf Survey \citep[STARBIRDS;][]{McQuinn:2010a, McQuinn:2010b, McQuinn:2012} undertook a more systematic approach to studying bursty star formation in nearby low-mass galaxies. STARBIRDS used archival \hubble/WFPC2 and \hubble/ACS imaging for 18 nearby (D$\lesssim6$~Mpc) starburst dwarf galaxies.  Starburst episodes were operationally defined by a $b$-parameter threshold applied to the recovered SFH, allowing the program to objectively identify current and fossil bursts across the sample.  STARBIRDS established that burst durations are typically $\gtrsim 100$~Myr.  This ruled out rapid self-quenching on massive-star lifetimes. It also found that bursts produced only $\sim 3$--$26\%$ of a galaxy's stellar mass, a direct empirical counterpoint to early simulations in which star formation was modeled as a series of short, complete burst-and-quench cycles \citep[e.g.,][]{Stinson:2007}.  The program also produced the first systematic spatial map of burst patterns in nearby dwarfs (see below).  \citet{McQuinn:2015c} extended the same methodology to the Survey of \hi{} in Extremely Low-mass Dwarfs (SHIELD) sample of very low-mass dwarfs.

A different conclusion was reached by \citet{Cignoni:2019}, who analyzed 23 LEGUS dwarfs over the last 100~Myr and reported no evidence for periods of zero SFR. \citet{Choi:2025} analyzed 10 lower-mass ($M_\star<10^{7.5}$~\msun) LUVIT dwarfs over the last $\sim1$~Gyr and found episodic star formation, with SFRs that exceed the 1~Gyr average by factors of a few to several. The LEGUS results suggest that not all star formation in low-mass galaxies proceeds through complete burst-and-pause cycles, and both sets of results are consistent with the stochastic GMC-sampling expectation discussed in \S\ref{sec:bursts}. The apparent tension with STARBIRDS may partly reflect methodology, as the surveys measure burstiness relative to different baselines.

\jwst\ is beginning to extend these recent-burst analyses to fainter limits and more distant systems.  \citet{McQuinn:2024a} produced the first \jwst\ resolved star SFH (for WLM) and demonstrated that \jwst/NIRCam SFHs are consistent with \hubble\ results, providing a methodological foundation for future CMD-based burstiness work in regimes where \hubble\ is reaching its limits.

\paragraph[Spatial patterns of bursts in low-mass galaxies]{Spatial patterns of bursts in low-mass galaxies:} \label{sec:burst_spatial}

The spatial structure of star formation in low-mass galaxies provides a complementary test of how bursts originate and propagate.  The foundational technique, which was to use well-populated BHeB sequences to reconstruct spatially-resolved recent SFHs, was developed for Sextans~A and GR~8 \citep{Dohm-Palmer:1997, Dohm-Palmer:1998, Dohm-Palmer:2002}. It was later extended to M81 Group dwarf galaxies IC~2574 and Holmberg~II by \citet{Weisz:2009b, Weisz:2009a}, who demonstrated explicit spatiotemporal sequences of star formation associated with expanding \hi{} shells.  Their analyses showed that stellar-feedback energy is sufficient to drive the observed \hi{} morphology even when no obvious young cluster occupies the hole interior, indicating that feedback-driven ISM structure can outlive its progenitors and that recent star formation propagates spatially in response to cleared \hi{} cavities.  \citet{Warren:2011} confirmed this general conclusion with studies of kpc-size \hi{} holes in five different galaxies where the calculated feedback energy was sufficient to produce the holes.

At the survey level, \citet{McQuinn:2012} mapped the spatial distributions of bursts in the STARBIRDS sample and found substantial galaxy-to-galaxy diversity, from strongly centrally-concentrated bursts to spatially distributed activity covering the entire galaxy area.  \citet{Sabbi:2008} reached comparable conclusions for the M81 Group tidal-dwarf candidate Holmberg~IX.  \citet{Sacchi:2018} extended the analysis to the LEGUS starburst dwarf NGC~4449. \citet{Cignoni:2019} reported on the LEGUS sample as a whole: about half of their galaxies showed evidence of spatial progression over the last 60~Myr, most were not strongly centralized, and offsets between current star-formation sites and RGB centroids were common --- broadly consistent with the \citet{McQuinn:2012} picture of spatial diversity rather than universal centralization.  In the lowest-mass regime, \citet{Lescinskaite:2022} for Leo~A and \citet{McQuinn:2024b} for Leo~P show that even strongly star-forming dwarfs at $M_\star \sim 10^6$~\msun\ exhibit spatially-coherent recent star-formation patterns at the scales BHeBs can resolve.

At larger stellar mass, the LMC and SMC provide the most data-rich spatial SFHs of any dwarf systems.  Wide-area maps based on programs such as MCPS, VMC, SMASH, and STEP resolve recent and intermediate-age star formation across both clouds at sub-kpc scales, while \hubble/Scylla imaging \citep{Cohen:2024b, Cohen:2024a} extends these maps to the dense central regions at much higher angular resolution.  A central spatial result is the identification of enhanced star formation along LMC spiral arms coincident in time with the $\sim 2$~Gyr and $\sim 500$~Myr peaks in the global SFH \citep{Ruiz-Lara:2020b}. This appears to be the spatial counterpart of the synchronized burst record described above.  It provides direct evidence that LMC--SMC close encounters left coherent fossil signatures in both the timing and the geometry of recent star formation.  Together, the Magellanic Clouds illustrate that bursts can persist on multi-Gyr timescales as identifiable structures in the spatial SFH map, providing an empirical link between global burst timing and the spatial geometry of the trigger.

The totality of all these studies bears directly on a key prediction of feedback-dominated low-mass galaxy simulations \citep[e.g.,][]{El-Badry:2016, El-Badry:2017a}.  Some of these simulations produce strongly centralized bursts whose stellar populations subsequently migrate outward with age.  The data above (along with the recent FEAST \jwst\ starburst-dwarf studies of NGC~4449 and the interacting NGC~4485/4490 pair; \citealp{Correnti:2025, Bortolini:2025}) do not show universal central concentration.  This suggests that there may be several modes responsible for producing the observed diversity in bursts.  It is important to note that interpreting such spatial information also requires accounting for the generic dispersal of hierarchically-clustered young stellar populations.  Notably, \citet{Bastian:2011} used BHeB-age-dated stars across a sample of nearby non-starburst dwarfs to show that young stellar substructure dissipates on $\sim 100$~Myr timescales, so older populations will appear systematically more distributed than younger ones even in the absence of any feedback-driven migration.  Distinguishing feedback-driven outward migration from generic structural dissolution therefore requires structural analyses (e.g., two-point correlations, hierarchical-cluster identification) rather than radial profiles alone.  The corresponding fossil burst test comes from the radial age and metallicity gradients seen in many LG dwarfs as described in detail in \S\ref{sec:mw_classical} and \S\ref{sec:isolated}.

\section{Stellar Initial Mass Function Measurements}
\label{sec:imf}

The stellar IMF specifies the relative number of stars as a function of their birth mass.  From the time of its initial conception \citep{Salpeter:1955}, the IMF has become central to our interpretation of resolved and unresolved stellar populations, from the Solar neighborhood to the most distant galaxies known \citep[e.g.,][]{Bastian:2010, Conroy:2013, Weinberg:2017, Kroupa:2013, Hopkins:2018, Kruckow:2018, Smith:2020, Zonoozi:2025, Kroupa:2026}.  Though virtually all papers agree on the foundational importance of the IMF, measurements of the IMF have been accompanied by significant controversy over the past several decades.  Disagreements over its functional form, how to measure it, and how to interpret measurements (e.g., is the IMF statistically the same throughout the Universe or does it systematically vary as a function of environment?; \citealt{Scalo:1986, Salpeter:2005, Kroupa:2013, Hopkins:2018}) are a few examples of these tensions.  Much of the same debate continues today, as described in the detailed IMF review of \citet{Hopkins:2018}.

There are dozens of excellent reviews and books on the IMF that cover its history, practical challenges in measuring it, connecting it to theoretical expectations, and philosophical discussions about the nature of the IMF (e.g., what does it mean for the IMF to be ``universal''?, does the IMF differ on scales of individual clusters vs.\ entire galaxies; e.g., \citealt{Scalo:1986, Kennicutt:1998, Larson:1998, Massey:1998, Scalo:1998, Scalo:2005, Bastian:2010, Kroupa:2013, Offner:2014, Hopkins:2018, Hennebelle:2024, Kroupa:2026}).  We will not recap all the excellent work from decades of IMF study in this review. Instead, we refer the interested reader to these works directly.

The SFHs discussed in \S\S \ref{sec:formation}--\ref{sec:sfh_trends} show that many nearby galaxies formed their stars under conditions that are not represented by the stellar populations that anchor the Galactic IMF, i.e., the Solar-neighborhood disk and young star clusters.  For example, most UFDs formed the bulk of their stars within the first $\sim1$~Gyr of cosmic time at [Fe/H]~$\lesssim-2$.  The MW's own old, metal-poor populations are either dynamically evolved (e.g., GCs) or sparsely sampled (e.g., the stellar halo).  Because the stars in UFDs are dynamically unevolved and can be counted directly, these galaxies provide a unique test of whether the IMF depends on metallicity or cosmic time \citep[e.g.,][]{Wyse:2002}.

It is only within the last $\sim15$~years that IMF studies have \textit{systematically} expanded to LG galaxies. In this section, we (i) discuss progress made in measuring the IMF directly from resolved star counts in nearby galaxies since the seminal review of \citet{Bastian:2010} and (ii) assess whether these measurements support significant departures from the commonly adopted IMFs based on measurements in our Galaxy.   In this review, we do not attempt to summarize studies that aim to measure the maximum mass or minimum mass of a star.  While of great importance to the IMF, these quantities are generally inferred through means other than stellar populations analysis (e.g., spectroscopy for massive stars; e.g., \citealt{Massey:2003}) or are generally inaccessible outside the MW (e.g., the hydrogen burning limit for low-mass stars). Instead, as in most of the relevant literature, we focus largely on the slope(s) or shape characteristics of the IMF.

\subsection{Definitions and Introduction to Concepts}

The principle of measuring the IMF is simple: count the relative number of stars that formed from the same parent gas cloud at the time of their formation as a function of their birth mass.  In practice, it is challenging, and perhaps impossible, to observe an entire stellar population (i.e., from the lowest mass to the highest mass star) exactly at its birth.  Instead, studies measure the present day stellar mass function (PDMF) of a stellar population and then translate it into an IMF measurement by adopting a number of assumptions.

The challenges of interpreting a PDMF as an IMF are well-documented \citep[e.g.,][]{Scalo:1986, Bastian:2010, Kroupa:2013, Hopkins:2018, Hennebelle:2024, Kroupa:2026}.  SFH, stellar multiplicity, uncertainties in stellar interior models and atmospheres (e.g., mapping luminosity to birth mass), and dynamics (e.g., mass segregation, evaporation, migration) are just some of the factors that complicate the mapping of the PDMF to an IMF.  Moreover, observational completeness and the inability to observe (e.g., due to crowding, finite lifetimes) the entire range of viable stellar masses in a single environment (e.g., $\sim0.08$ to $\sim300$ \msun) provide additional challenges.  Many of these issues are discussed at length in the review articles referenced above.

For the most part, we separate our discussion of IMF studies into low- and high-mass sections, around a dividing line of $\sim$ 1 M$_{\odot}$; there are a few cases that transcend this division outside the MW. Though we limit our literature review to papers that were not already included in the IMF review of \citet{Bastian:2010} we do include the work of \citet{Wyse:2002} as it was the only IMF measurement in a MW satellite available to \citet{Bastian:2010}.  With more studies in MW satellites since that time, it is now possible to appreciate the pioneering work of \citet{Wyse:2002} in a broader context.

\subsection{Low-Mass IMF}

\begin{figure}[pos=tbp]
    \centering
    \includegraphics[width=1.0\linewidth]{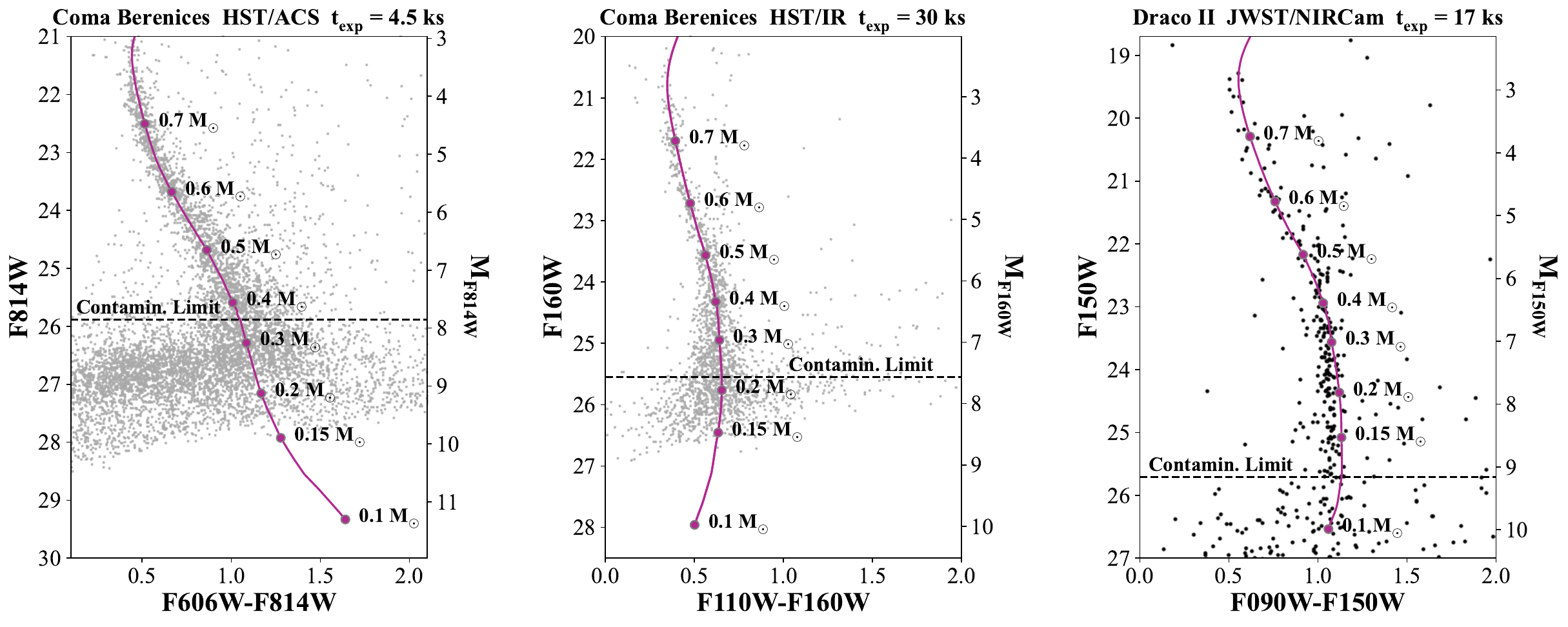}
    \caption{Examples of the type of data that is used to measure the low-mass IMF in UFDs around the MW.  Main sequence masses from the BaSTI models \citep{Hidalgo:2018} are overplotted to indicate the range of masses covered. \textbf{Left:} An optical CMD of Coma Berenices based on deep \hubble/ACS imaging \citep{Brown:2012, Gennaro:2018}.  Though the photometry formally extends to $\sim0.2$\msun, significant contamination from background galaxies limits the reliable range of masses to only $\sim0.37$\msun. \textbf{Middle:} A near-IR CMD of Coma Berenices based on deep \hubble/WFC3 IR imaging \citep{Gennaro:2018b}.  Compared to the optical, the IR CMD reaches lower masses (red stars are brighter in the IR) and exhibits far less contamination due to background galaxies.  This combination resulted in the best-constrained \hubble-based sub-Solar IMF measurement made for a UFD outside the MW.   \textbf{Right:} A near-IR CMD of Draco~II from deep \jwst/NIRCam imaging \citep{Weisz:2023a, Weisz:2024}. This CMD is among the deepest available for any galaxy outside the MW, but there are too few stars from which to make a reliable IMF measurement. }
    \label{fig:IMFCameras}
\end{figure}

The low-mass IMF ($M_{\star} \lesssim 1$\msun) is typically measured by counting stars fainter than the oMSTO. In principle, for an IMF that does not depend on age or metallicity, the mass function of these unevolved MS stars has the benefit of not being substantially affected by a galaxy's SFH, although details of the MDF may affect the magnitude-mass conversion.  However, lower MS stars are faint, and acquiring suitably deep data can be challenging, as illustrated in Figure~\ref{fig:IMFCameras}.  Even when such deep data can be acquired, a major issue has been in identifying and accounting for contaminants (e.g., MW foreground stars, compact background galaxies) that can complicate and potentially bias low-mass IMF measurements, which we discuss below.

Most low-mass IMF measurements in nearby galaxies focus on field populations, not clusters.   While the approximately coeval, mono-metallic stars in clusters make for more straightforward PDMF to IMF conversions, they pose additional complications for low-mass stars such as crowding and mass-dependent ejection or evaporation of low-mass stars. ASTs and the inclusion of dynamical modeling alongside IMF determinations may help to mitigate these issues.

\begin{figure}[pos=tbp]
    \centering
    \includegraphics[width=0.8\linewidth]{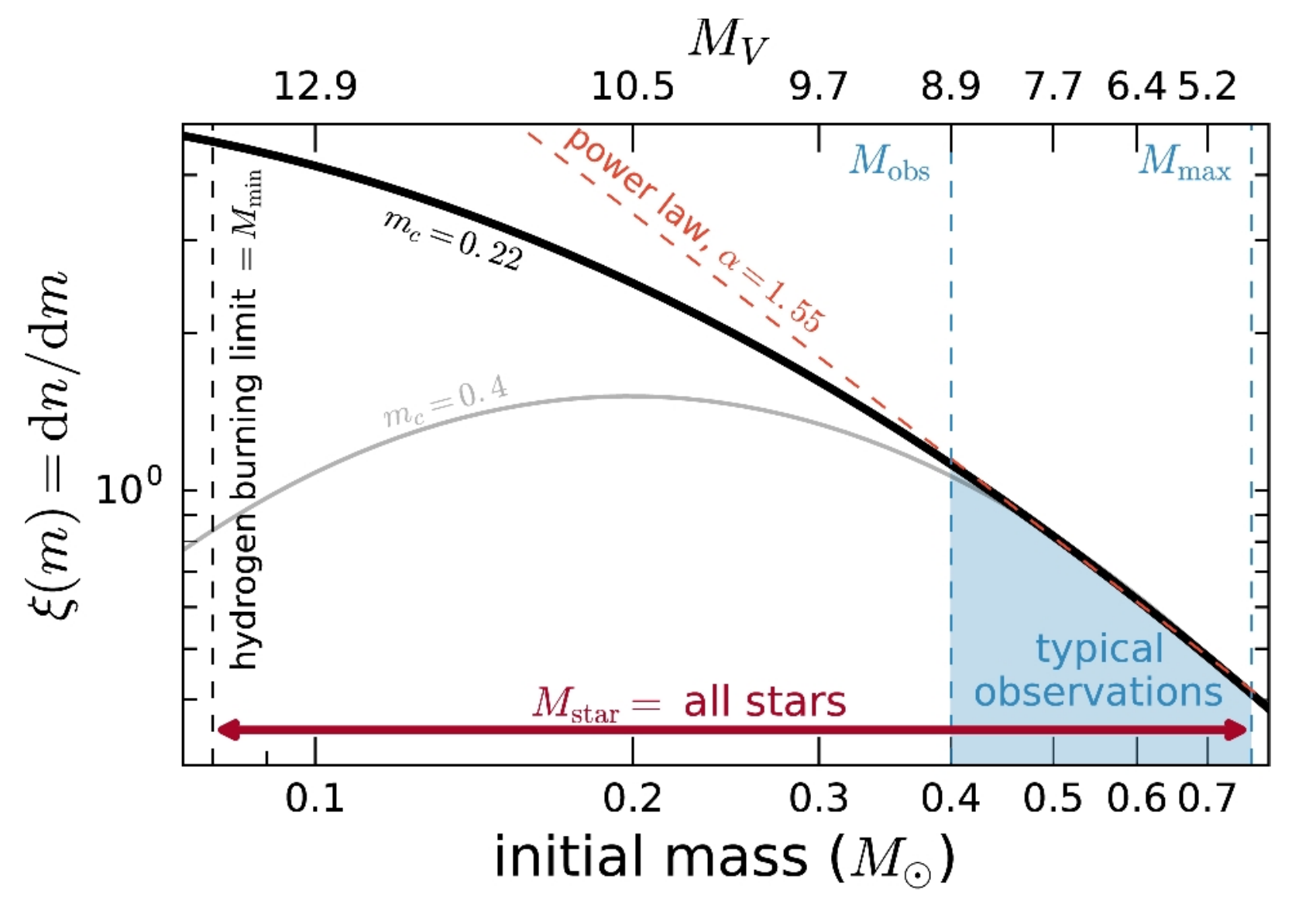}
    \caption{A schematic figure from \citet{El-Badry:2017} illustrating the need for deep data for measuring the low-mass IMF from resolved star counts.  The black line represents a log-normal IMF with Galactic parameters from \citet{Chabrier:2003}.  The grey line is for a log-normal with a larger value of $m_c$.  The blue shaded region illustrates the typical depth of \hubble\ data in most MW satellites, i.e., the data sample $M\sim0.4-0.8$\msun.  Constraining a MW-like log-normal IMF requires sampling to at least 0.2 \msun, which has only been achieved with a few datasets to date.  Fitting a single-sloped power-law model to a log-normal IMF results in a slope that is systematically flatter than a Salpeter IMF ($\alpha=-2.35$).}
    \label{fig:imf_schematic}
\end{figure}

The low-mass IMF is commonly characterized by three functional forms.  The simplest is a single-sloped power-law that has a slope and minimum and maximum mass boundaries.  A Salpeter IMF has a power-law exponent of $\alpha=-2.35$ in terms of linear mass \citep[i.e., $\frac{dn}{dm} \propto\ m^{\alpha}$;][]{Salpeter:1955}.  The lowest luminosity star used in modeling is usually set by the faint limit of the usable data, while the most luminous star included is generally set by the MSTO magnitude.  It has long been established that the low-mass Galactic IMF is not well-described by a single-sloped power-law \citep[e.g.,][]{Miller:1979, Kroupa:1993, Chabrier:2003}.  Nevertheless, this model is often used in extragalactic contexts due to its simplicity in comparing relative IMF measurements.  However, as pointed out by \citet{El-Badry:2017} and illustrated in Figure~\ref{fig:imf_schematic}, applying the single-slope power law IMF to an intrinsically log-normal IMF can lead to recovered values that depend on the depth of the observation, even for the same underlying population.

There are other adopted functional forms for the low-mass IMF, e.g.,  a broken power law or a log-normal, that are motivated by Galactic datasets.  The most commonly adopted Galactic broken power-law IMF is that of \citet{Kroupa:2001}.  One parameterization of the Kroupa broken power law IMF has two slopes in the sub-Solar mass regimes, a break point between them, and a minimum and maximum mass \citep{Kroupa:2001}.  In extragalactic studies, the breakpoint is usually set to the Galactic value, though it is possible the breakpoint could vary as a function of environment, which is a largely untested theoretical prediction \citep[e.g.,][]{Hennebelle:2024}.

The sub-Solar log-normal IMF was originally introduced by \citet{Miller:1979}, and the most commonly cited Galactic parameters are from \citet{Chabrier:2003, Chabrier:2005}. Because most of the studies discussed below fit for the binary fraction separately, we compare to the \citet{Chabrier:2005} single-star values ($m_c=0.20$~\msun, $\sigma=0.55$) unless otherwise noted. The log-normal IMF is parameterized by a characteristic mass ($m_c$) and a width ($\sigma$) as well as a minimum and a maximum mass.  It is (somewhat arbitrarily) joined to a single power typically with $\alpha=-2.3$ at $\sim$ 1 \msun\ to connect the low- and high-mass regimes.  Constraining all parameters for a log-normal IMF typically requires data that sample slightly below $m_c$ \citep[e.g.,][]{El-Badry:2017, Gennaro:2018b}.  The work of \citet{Gennaro:2020} on the low-mass IMF in the Orion cluster in the MW demonstrates the power of extremely deep data for IMF constraints.  Unfortunately, acquiring this quality of data is virtually impossible outside select regions in the MW.

Modern low-mass IMF studies in nearby galaxies use CMDs and not single-band luminosity functions, which were the historical standard for IMF measurements \citep{Scalo:2005, Bastian:2010}.  Critically,  color information from CMDs provides constraints on the contribution of unresolved binary stars and contaminants (e.g., compact background galaxies, MW foreground stars), which are challenging to capture from single-band luminosity functions.  Figure~\ref{fig:IMFCameras} shows examples of deep \hubble\ and \jwst\ CMDs that can be used to measure the low-mass IMF in nearby galaxies.  The challenge in both routinely obtaining data that goes so faint and of distinguishing between the faintest stars and contaminants should be apparent.

\subsubsection{Ursa Minor}

The first measurement of an extragalactic low-mass IMF using resolved star counts was in Ursa~Minor \citep{Wyse:2002}.  The scientific aim of this study was to measure the sub-solar IMF in a dark matter dominated galaxy with an ancient, metal-poor stellar population ($\gtrsim$ 10 Gyr, [Fe/H] $\sim$ $-2$) to search for differences with comparably ancient, metal-poor GCs in the MW.
This pioneering study obtained deep \hubble/WFPC2, Space Telescope Imaging Spectrograph, and NICMOS imaging of a single field in Ursa~Minor that extended to $\sim0.3$ \msun.  They used ASTs to correct for completeness, qualitatively assessed the impact of binary stars, and adopted the stellar models of \citet{Baraffe:1997}. They fit a single-sloped power-law model and found a linear exponent value of $\alpha=-1.8$, which they found to be identical to MW halo GCs M92 and M15, but flatter (i.e., fewer of the lowest-mass stars) compared to a Salpeter IMF ($\alpha=-2.35$).  This study noted that the similarity of the IMFs between Ursa~Minor and the two MW GCs suggested that star formation may not have been very different in these environments (i.e., high-density GC vs. low-density, dark matter dominated dwarf galaxy). This measurement also provided some of the first insight into the IMF in the early Universe, since Ursa~Minor formed nearly all of its stars $\gtrsim10$~Gyr ago.

\subsubsection{Magellanic Clouds} \label{sec:mc_imf}
\citet{Kalirai:2013} used deep \hubble/ACS F606W and F814W imaging to measure the low-mass IMF in the outskirts of the SMC.  Because this field contained the lower MS of both the SMC and 47 Tuc, members of 47 Tuc were removed using proper motion criteria.
\citet{Kalirai:2013} measured the IMF using a single power-law mass distribution over the stellar mass range of 0.37 to 0.93 \msun, adopted a binary fraction of 30\%, corrected for incompleteness using ASTs, and employed the Dartmouth stellar models \citep{Dotter:2008}.  The resulting fit produced an IMF slope of $\alpha=-1.90^{+0.15}_{-0.10}$ ($3\sigma$ uncertainties), which is much flatter than a Salpeter IMF ($\alpha=-2.35$), and is similar to the results of \citet{Wyse:2002}.  Modest changes to the binary fraction and adopted completeness limit did not change this result beyond the quoted uncertainty.  The high degree of precision for this measurement is due to the large dynamic range of masses and the large number of stars \citep[e.g.,][]{Weisz:2013}.  The mass range in this study was limited such that they did not have the constraining power to test for the suitability of a log-normal vs. a single power-law IMF.

\citet{Legnardi:2025} measured the low-mass IMF in the SMC using deep \jwst/NIRCam F115W and F322W2 imaging in a field that is adjacent to that of \citet{Kalirai:2013}.  They fit single-sloped and broken power-law IMF models over a mass range of 0.22 to 0.95 \msun.  Though their photometry formally extends to $\sim0.1$~\msun, they set the faint limit equal to the 50\% completeness limit of their data (0.22 \msun), which also mitigated the worst effects of contamination. They included a model for the binary fraction, but only consider mass ratios with $q>0.6$ as they reported that binaries with lower mass ratios cannot be readily detected in their data. They also adopt the externally measured MDF from \citet{Mucciarelli:2023}.  They used the BaSTI stellar models \citep{Hidalgo:2018} to convert their stellar luminosity function to masses and then performed a least squares fit to measure the IMF parameters.  For a single-sloped power law, they reported $\alpha = -1.99 \pm 0.08$, which is nearly identical to \citet{Kalirai:2013}.  For a broken power-law, they reported $\alpha =-1.69 \pm 0.15$ for $M<0.5$~\msun\ and $\alpha =-2.02 \pm 0.29$ for $M>0.5$~\msun.  The higher-mass slope agreed with a Galactic \citet{Kroupa:2001} IMF, but the lower-mass slope is steeper.  They also reported a total binary fraction of $0.4$, based on an extrapolation of their model binary fraction parameter to all mass ratios.  They did not attempt to fit a log-normal IMF.

\citet{Cohen:2026} measured the low-mass IMF in the outskirts of the SMC using deep \jwst/NIRCam F150W and F322W2 imaging.  Their field was designed to overlap with existing \hubble\ imaging from \citet{Noel:2009} in order to use their measured age distribution, AMR, and ground-based spectroscopy of \citet{Carrera:2008} as informative priors for their IMF modeling.
They fit three functional forms of the IMF (single power-law, broken-power law, log-normal) from $\sim0.8$ to $0.16 $\msun.  The lower mass limit is set by the 50\% completeness limit of their data.  This is one of the deepest extragalactic low-mass IMF measurements to date.  In addition to including a binary fraction in their model, they also adopted literature line-of-sight distance and extinction distributions and allowed their values to vary.  They included detailed models of foreground and background contaminants.  They used the Bayesian fitting method \citep{Gennaro:2026} that is an expansion of the techniques introduced by \citet{Gennaro:2018} and the BaSTI stellar evolution models with empirical color corrections to improve data-model agreement on the lower MS.  For a single power-law, they reported a value of $\alpha = -1.61\pm0.03$, which is flatter than other SMC studies.

For the first time, the depth of this dataset enabled a formal model selection comparison between different functional forms of the IMF. Using a Kolmogorov--Smirnov test, \citet{Cohen:2026} showed that their SMC results clearly disfavor a single-slope power-law model, just as in the MW. For a broken power-law, they found $\alpha_1 = -1.44\pm0.04$ and $\alpha_2 = -2.17\pm0.11$, which are consistent with Galactic values.   They performed a test by fitting the data and allowing the breakpoint of the broken power-law IMF to vary and found no significant difference in the results. Finally, for a log-normal IMF they find $m_c = 0.12\pm0.03$ \msun\ and $\sigma=0.62\pm0.06$, which are lower than the Galactic values for $m_c$ and higher values for $\sigma$.  They suggest these values are consistent with some other recent results in metal-poor star clusters and simulations, but indicate that sufficiently deep data in other extragalactic systems are needed to establish empirical trends.  For all function forms of the IMF, they find a binary fraction of $\sim0.2\pm0.05$.

\citet{Legnardi:2026} measured the low-mass IMF in a field in the outskirts of the LMC, near star cluster NGC~1846, using deep \jwst/NIRCam F115W and F322W2 imaging.  Following the methodology of \citet{Legnardi:2025}, they measured the binary fraction directly from the data for mass ratios of $q>0.6$ ($f_{\rm bin}=0.15\pm0.01$), which corresponds to a total binary fraction of $0.34\pm0.02$ for a flat mass-ratio distribution.  This is consistent with values measured in the SMC and MW.  They fit a single power-law IMF over 0.17 to 0.82~\msun\ and reported $\alpha=-1.49\pm0.16$.  This is shallower than a Salpeter IMF and the SMC values of \citet{Kalirai:2013} and \citet{Legnardi:2025}, but consistent with that of \citet{Cohen:2026}, whose fit spans a nearly identical mass range.  To date, this is the only field-star measurement of the low-mass IMF in the LMC.  

Taken at face value, the agreement at matched mass ranges in the SMC and LMC suggests that neither the low-mass IMF nor the binary fraction differs strongly between the two Clouds.  However, because single power-law fits to a log-normal IMF depend on the mass range considered, fitting a more realistic functional form is an important future direction for the LMC.

\subsubsection{Ultra-Faint Dwarf Galaxies}

The low-mass IMF has been measured in several ancient, metal-poor UFDs. \citet{Geha:2013} first measured the low-mass IMF of 2 MW UFDs (Hercules and Leo~IV) using deep \hubble/ACS F606W and F814W imaging from \citet{Brown:2012}.   They fit both power-law and log-normal IMF models over a mass range of $\sim0.5$ to $\sim0.8$ \msun, used ASTs to correct for completeness, fit for the binary fraction assuming a uniform mass ratio distribution, and adopted stellar models of \citet{VandenBerg:2012}.

For a single-sloped power-law, they find IMF slopes that are much flatter than Salpeter or Kroupa.  Specifically, they find $\alpha=-1.2^{+0.4}_{-0.5}$ in Hercules and $\alpha=-1.3^{+0.8}_{-0.8}$ in Leo~IV.  Because the data were not deep enough to robustly constrain a log-normal IMF, they fixed the width of the log-normal to $\sigma=0.69$, the width of the \citet{Chabrier:2003} single-star IMF, and fit for the $m_c$ and binary fraction simultaneously.  For Hercules, they found $m_c=0.4^{+0.9}_{-0.3}$ and a binary fraction of $0.48^{+0.2}_{-0.12}$.  For Leo~IV, they found $m_c=0.4^{+2.1}_{-0.3}$ and a binary fraction of $0.47^{+0.37}_{-0.17}$.  The resulting log-normal fits are consistent with the Chabrier IMF found in the MW, though the uncertainties are large due to the limited depth of the data.

\citet{Gennaro:2018} expanded upon the IMF study of \citet{Geha:2013} by using a Bayesian analysis technique and analyzed all 6 MW UFDs with deep \hubble/ACS data observed by \citet{Brown:2012}.  Specifically, \citet{Gennaro:2018} applied an Approximate Bayesian Computational MCMC modeling technique to deep \hubble/ACS F606W and F814W observations of Leo~IV, Hercules, Coma~Berenices, Ursa~Major, Canes~Venatici~II, and Bootes~I.  The data cover a range of stellar masses from $\sim0.45$ to $0.8$ \msun.  \citet{Gennaro:2018} adopted the Victoria stellar models \citep{VandenBerg:2012, VandenBerg:2014}, an unresolved binary model with a uniform mass ratio distribution, completeness corrections from ASTs, and the age and MDFs as reported in \citet{Brown:2014}. They fit both single-sloped power-law and log-normal IMF models.  For the single power-law model, they found slopes that range from $\alpha=-1$ to $-1.9$, which are flatter than Salpeter ($\alpha=-2.35$) in all 6 galaxies.  They found binary fractions that range from $\sim0.05$ to $\sim0.3$.  For the log-normal model, \citet{Gennaro:2018} found values of $m_c=0.3-0.6$ \msun, $\sigma\sim0.6-0.7$, and binary fractions ranging from $\sim0.06$ to $0.6$.  For all 6 galaxies, the reported values of $m_c$ and $\sigma$ are larger than those of the \citet{Chabrier:2003} system IMF ($m_c=0.22$~\msun, $\sigma=0.57$).  The order-of-magnitude range in the binary fractions include the \citet{Chabrier:2003} value of 0.2.  In general, \citet{Gennaro:2018} find that power-law IMFs in all 6 UFDs are strongly and systematically `bottom light' compared to the MW.  The log-normal IMFs in the UFDs show a more modest deficit of low-mass stars compared to a MW IMF, mostly because the characteristic mass in all of the UFDs is larger than in the MW.

\citet{Gennaro:2018b} undertook a more comprehensive measurement of the low-mass IMF in Coma~Berenices using deep \hubble/WFC3/IR F110W and F160W imaging.  As shown in Figure~\ref{fig:IMFCameras}, these data formally reach stars with masses of $M$ $\sim$ 0.17 \msun, but contamination due to background galaxies limits the usable portion of the data to $M$ $\gtrsim 0.23$ \msun.  This is still $\sim0.1$ \msun\ below the adopted limit for the optical-based IMF measurement in Coma~Berenices.
The analysis of this deep IR CMD largely follows that of \citet{Gennaro:2018}, i.e., completeness from ASTs, Bayesian modeling, binary star treatment, and stellar models from \citet{VandenBerg:2014} with empirical corrections. They fit several IMF parameterizations including single-sloped and broken power-laws and a log-normal model.  For the single-power law model, \citet{Gennaro:2018b} found a slope of $\alpha=-1.45^{+0.29}_{-0.3}$, which is similar to results from analysis of the optical data in \citet{Gennaro:2018}.  For the broken-power law and log-normal IMF models, \citet{Gennaro:2018b} found values for Coma~Berenices that are statistically consistent with Galactic values.    \citet{Gennaro:2018b} noted large, and statistically significant, discrepancies in the binary fractions between the analyses of the optical and IR datasets.  They discussed several plausible causes (e.g., shallower optical data, under-estimated uncertainties from the ASTs), including that the final dataset only included $\sim1000$ usable stars, but did not identify a sole cause of the tension.

\citet{Filion:2022} measured the low-mass IMF of Bootes from deep \hubble/ACS F606W and F814W imaging that extends to $m\sim0.3$\msun.  Like \citet{Gennaro:2018}, they used an approximate Bayesian computational technique with MCMC sampling to fit a single-slope, broken-power law, and a log-normal IMF model to the data.  They used ASTs to quantify completeness and uncertainties, undertook a detailed selection of membership \citep{Filion:2020}, and included unresolved binary stars in their model.  They noted that their approach to including binaries is slightly different from the one used by \citet{Geha:2013} and \citet{Gennaro:2018}.  They provided a simple analytic conversion between the two formulations of binary fraction. They found a single power IMF model has a slope of $\alpha=-1.95_{-0.28}^{+0.32}$ and a binary fraction of $0.61_{-0.14}^{+0.11}$ (where we have converted the binary fraction to the scale used by \citealt{Geha:2013} and \citealt{Gennaro:2018}).    The power-law slope is similar to that of \citet{Gennaro:2018}, both of which are substantially flatter than Salpeter.  However, the binary fractions of \citet{Filion:2022} and \citet{Gennaro:2018} differ by a factor of $\sim2$, even after being corrected to the same scale. For the log-normal IMF model, \citet{Filion:2022} found $m_c=0.17^{+0.05}_{-0.11}$, a width of $\sigma=0.49^{+0.13}_{-0.17}$, and a binary fraction of $0.52^{+0.19}_{-0.18}$.  The values of $m_c$ and $\sigma$ are consistent with the MW \citet{Chabrier:2005} IMF.  Similarly, their broken powerlaw fit is consistent with the MW \citet{Kroupa:2001} IMF.

As part of their study of wide binaries in the MW UFD Ret~{\sc II}, \citet{Safarzadeh:2022} measured the low-mass IMF using \hubble/ACS F606W and F814W imaging from the same program that measured its SFH \citep{Simon:2023} over the mass range of 0.34 to 0.78\msun.  They modeled the binary star distribution alongside a single-sloped power-law IMF and adopted the PARSEC stellar models \citep{Bressan:2012}. They did not explicitly use ASTs to account for completeness, and instead only used a region on the CMD that is designed to minimize the effects of incompleteness and contamination from background galaxies. They found an IMF slope of $\alpha=-1.15_{-0.14}^{+0.27}$. This is much flatter than a Salpeter IMF and is broadly consistent with results from \citet{Geha:2013} and \citet{Gennaro:2018}.  They estimated a binary fraction of $\sim0.5$.

\citet{Filion:2024} measured the IMF of two more UFDs, Ursa~Maj~{\sc II} and Ret~{\sc II}, with deep archival \hubble\ F606W and F814W imaging using the same methodology as \citet{Filion:2022}.  They fit single-sloped and broken power-laws, as well as a log-normal model over the mass range of $\sim$ 0.2 to $\sim$ 1 \msun\ in each galaxy.  For both galaxies, they found single-sloped power-law values that are moderately shallower than Salpeter with values of $\alpha=-1.79^{+0.29}_{-0.27}$ ($1\sigma$) for Ret~II and  $\alpha=-2.19^{+0.36}_{-0.29}$ for Ursa~Maj~II.  For the broken power-law and log-normal models, they found parameters for Ret~{\sc II} that are consistent with Galactic values and that are formally more bottom heavy for Ursa~Maj~{\sc II}, although possible contamination from background galaxies makes the latter less secure.  For a log-normal IMF they found $m_c=0.2_{-0.15}^{+0.06}$, $\sigma=0.65_{-0.20}^{+0.25}$, and a binary fraction of 0.14$_{-0.14}^{+0.07}$ in Ret~II and $m_c=0.15_{-0.10}^{+0.05}$, $\sigma=0.53_{-0.26}^{+0.13}$, and a binary fraction of 0.41$_{-0.17}^{+0.33}$ in Ursa~Maj~II.  The binary fractions have been converted to the scale of \citet{Geha:2013} and \citet{Gennaro:2018}.  \citet{Filion:2024} noted minor tensions with their results and the lower mass slope in the broken power-law for Ret~II, but found that the log-normal values are consistent with \citet{Chabrier:2005} values.  They also found that their binary fractions are consistent with other measurements in UFDs.

\begin{figure}[pos=tbp]
    \centering
    \includegraphics[width=1.0\linewidth]{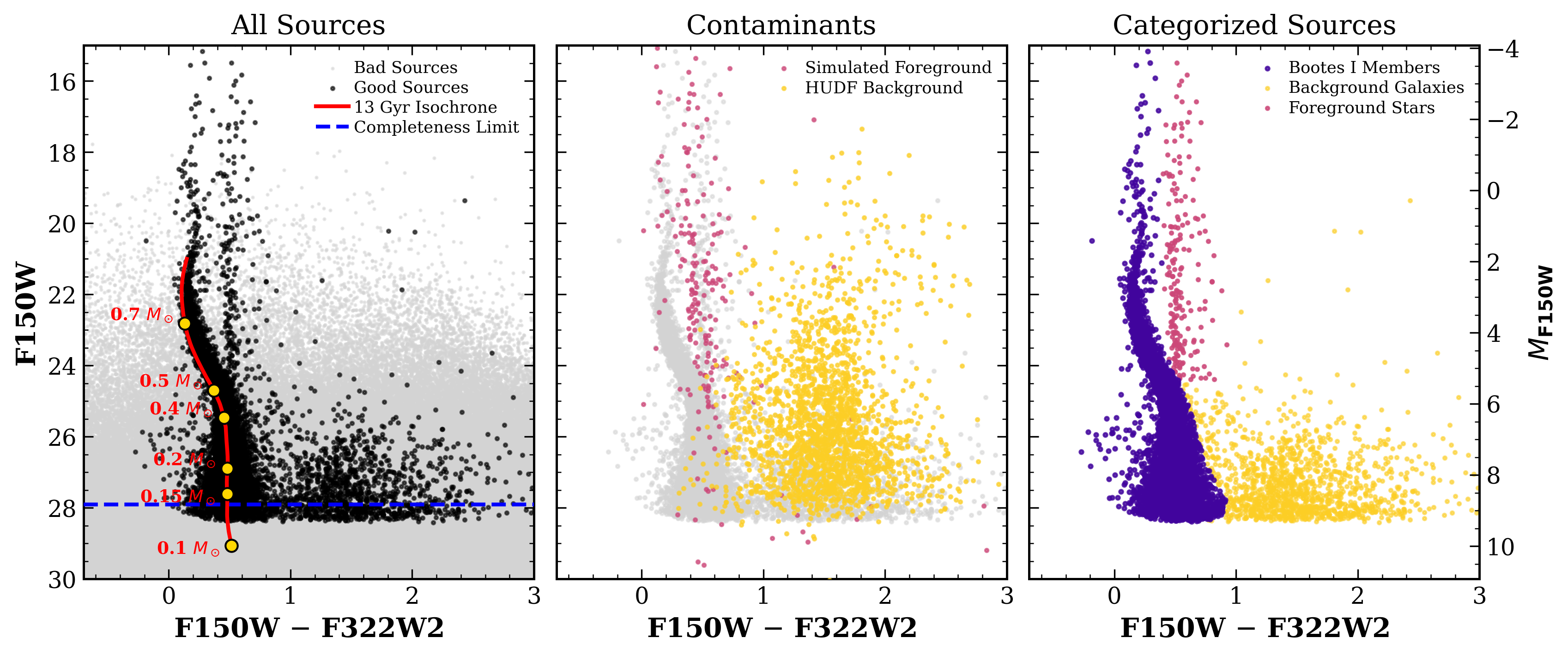}
    \caption{The deep \jwst\ CMD of Bootes~I used to measure the low-mass IMF.  Figure adapted from \citet{Ding:2026}.  \textbf{Left:} The selection of point sources (black) relative to all sources (grey).  The lower-MS and select MS stellar masses from the BaSTI models are indicated in red and gold.  We have not made color corrections to the models following \citet{Ding:2026}.  The blue dashed line indicates the 50\% completeness limit, corresponding to the minimum mass used for IMF fitting. \textbf{Middle:} Indications of foreground stars (magenta) from Trilegal simulations and background galaxies (yellow) from the \hubble\ ultra deep field, as described in \citet{Ding:2026}. \textbf{Right:} The statistical classification of likely members of Bootes~I (purple), likely foreground stars (pink), and likely background galaxies (yellow) from \citet{Ding:2026}.  Overall, this figure highlights the ability of \jwst\ to acquire data needed to robustly measure low-mass IMF and statistically mitigate sources of contamination, which proved quite challenging, even with \hubble \citep[e.g.,][]{Gennaro:2018b}.}
    \label{fig:bootes_imf}
\end{figure}

\citet{Ding:2026} measured the low-mass IMF of MW UFD Bootes~I with deep data acquired with \jwst. To determine the IMF, they followed the general methodology described in \citet{Gennaro:2018}.    Figure~\ref{fig:bootes_imf} shows the deep \jwst\ CMD used for this measurement.
They fit three functional forms of the IMF (single power-law, broken-power law, log-normal) to MS stars from $\sim$ 0.8 to $0.15$ \msun, which is the $\sim$ 50\% completeness limit of their data.  They included a model to account for unresolved binaries, adopted the spectroscopic MDF from \citet{Geha:2026}, assumed a single age population of $\sim13.3$~Gyr, which is consistent with its SFH and also adopted oxygen-enhanced stellar models that were used to measure the SFH \citep{Brown:2014}.  Similar to \citet{Cohen:2026}, they made empirical color corrections to the model to improve agreement with the data.  For their single-slope power-law model, they find $\alpha=-1.60^{+0.04}_{-0.04}$ and a binary fraction of $0.54^{+0.04}_{-0.04}$.

Like \citet{Cohen:2026}, they found that a single-sloped power law provides a poor description of the data.  They note that because their data extend well below the characteristic mass of the Galactic IMF, it is unsurprising that the best-fit single slope is shallower than Salpeter \citep[see also][]{El-Badry:2017}.  For a broken power-law, they found $\alpha_1 = -1.31\pm0.07$ (below the break), $\alpha_2 = -1.99^{+0.16}_{-0.22}$ (above the break), a break point of $0.47^{+0.09}_{-0.07}$ \msun, and a binary fraction of $0.60^{+0.08}_{-0.08}$.  Finally, for the log-normal model, they found $m_c = 0.18\pm0.02$\msun, $\sigma=0.56\pm0.06$, and a binary fraction of $0.51^{+0.03}_{-0.03}$.  They reported that both the broken power-law and log-normal parameters are consistent with Galactic IMF, though constraints on the broken power-law model are weaker due to degeneracies in the slopes and the break mass.  They also noted consistency with past IMF studies of Bootes~I, though with much improved constraints due to the superior depth of this dataset.

\subsubsection{Systematic Variations in the Low-Mass IMF}
\label{sec:low_mass_imf_trends}

\citet{Geha:2013} first suggested that the low-mass IMF measured in UFDs may be part of a broader trend in systematic IMF variations.  By comparing their single power-law fits for two UFDs with IMF measurements in the MW and in massive elliptical galaxies \citep[see][and references therein]{Smith:2020}, they found that the low-mass IMF is more bottom-light (i.e., fewer low-mass stars) with decreasing galactic velocity dispersion and metallicity.

Using a larger sample of UFDs,
\citet{Gennaro:2018} explored correlations between various galaxy properties (e.g., stellar mass, metallicity, velocity dispersion) and their IMF parameters.  They only found a statistically significant correlation between the slope of a single sloped power-law IMF and metallicity such that higher metallicities have steeper IMF slopes, which confirmed previous results from \citet{Geha:2013}.  \citet{Gennaro:2018} also noted that IMF slope and binary fraction, as well as $m_c$ and binary fraction, exhibit a modest correlation, but with sizable scatter.  The origin of this correlation is unclear, and it is challenging to interpret for two reasons.  First, there are only broad constraints on the binary fraction in any individual galaxy.  Second, the inferred binary fractions could be affected by the choice of binary model parameterization, by contamination (e.g., from compact background galaxies), and by metallicity effects, which can broaden the lower MS.  As noted above, the IMF measurement in Coma~Berenices by \citet{Gennaro:2018b} yielded a single-sloped power-law IMF that is flatter than Salpeter, but a log-normal IMF that is consistent with MW values.  This can be interpreted as an indication that a single power-law is not a good representation of the sub-Solar IMF.  Similarly, comparing any measurements to the Galactic value for a single power-law IMF will not yield a meaningful result, as suggested by \citet{El-Badry:2017}.

\begin{figure}[pos=tbp]
    \centering
    \includegraphics[width=1.0\linewidth]{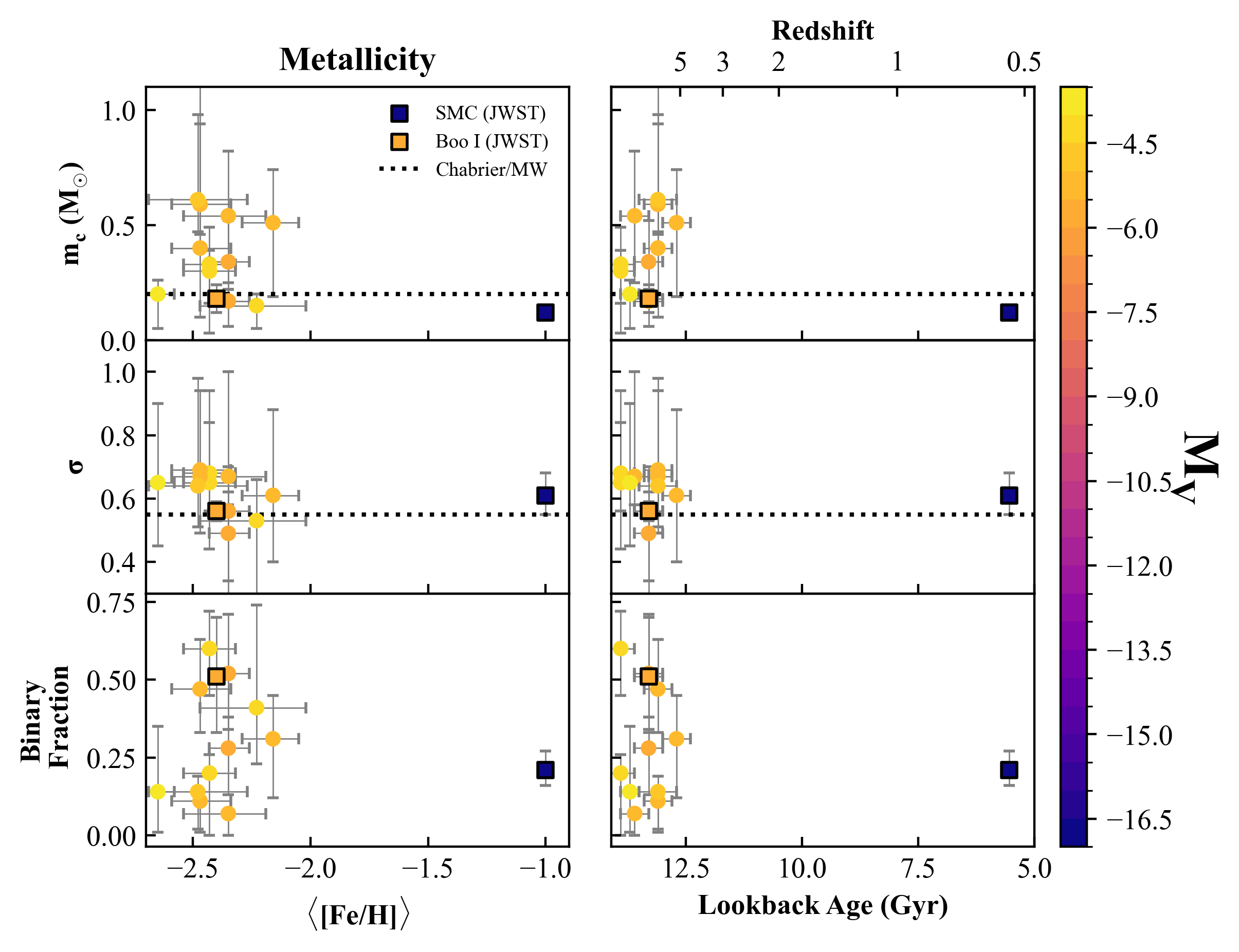}
    \caption{Log-Normal IMF parameters measured from deep imaging of MW satellite galaxies as a function of galaxy mean metallicity and mass-weighted mean age. \hubble-based studies are shown as circles, while the two \jwst-based measurements (SMC; \citealt{Cohen:2026} and Bootes~I, \citealt{Ding:2026}) are shown as squares.  Although not all studies find parameters that are consistent with MW values, at this time, there is no evidence for \textit{systematic} deviations from the Galactic values of $m_c$ and $\sigma$ (dotted lines) for a log-normal IMF as a function of metallicity or age/redshift.  The \jwst-based measurement in Bootes~I is very close to MW values, while the SMC measurement favors a somewhat lower $m_c$ and broader $\sigma$.  The large scatter in the binary fraction, including discrepancies between different studies of the same system, warrants further investigation.}
    \label{fig:low-mass-imf-log-normal}
\end{figure}

\citet{Baumgardt:2023} measured the stellar mass functions of 120 star clusters in the MW and Magellanic Clouds using archival \hubble\ imaging, with the low-mass IMF constrained by dynamically unevolved MW GCs, and used $N$-body simulations to account for dynamical effects in their interpretation of the PDMF as the IMF.  Including the IMF results from \citet{Gennaro:2018} in their population-level analysis, \citet{Baumgardt:2023} did not identify clear trends in the low-mass IMF as a function of environmental parameters (e.g., metallicity, stellar density), some of which span several orders-of-magnitude in dynamic range.

\citet{Cohen:2026} considered their single and broken power-law low-mass IMF measurement alongside findings from other MW satellites.  For the single power-law, they reported a value that is consistent with virtually all other single slope power-law measurements in the literature, including UFDs.  However, they cautioned that this IMF parameterization is clearly a poor representation of the data.  They also considered trends in broken power-law IMF measurements and found no systematic correlations as a function of mass range of the study.  \citet{Cohen:2026} did highlight that an improved understanding of binarity was important to future progress in low-mass IMF studies, as well as the need for a large sample of systems in order to establish or reject the presence of systematic trends.  The subsequent LMC measurement of \citet{Legnardi:2026} is consistent with this picture (see \S \ref{sec:mc_imf}).

Finally, \citet{Ding:2026} emphasized the consistency between both their log-normal and broken power-law IMF fits to Bootes~I and the Galactic values, despite Bootes~I being far older (13.3~Gyr) and more metal-poor ([Fe/H]$=-2.4$) than the MW stars used to make the Galactic IMF measurements.

In Figure~\ref{fig:low-mass-imf-log-normal}, we consider trends between log-normal IMF parameters, mean galaxy age, and mean metallicity for the dwarf galaxy measurements discussed above. We adopt mean spectroscopic metallicities from the literature \citep[e.g.,][]{Kirby:2013, Brown:2014} and mass-weighted ages from CMD-based SFHs \citep[e.g.,][]{Brown:2014, Simon:2023}.  Our motivations for considering only a log-normal IMF are (a) that it is more physically motivated than a single-sloped power-law IMF\footnote{In addition to its incompatibility with the shape of the low-mass stellar mass function in the Galaxy, the commonly adopted value of the Salpeter IMF, $\alpha=-2.35$, is known to be incorrect as \citet{Salpeter:1955} assumed the MW to be only 6 Gyr old. Assuming a correct age of the MW yields a power-law index of $\alpha=-2.05$ \citep{Salpeter:2005, Zinnecker:2005}.} and (b) that it has fewer parameters that could depend on environment than a broken power-law, making for a cleaner comparison.  See \citet{Hennebelle:2024} for discussion of both points.

From Figure~\ref{fig:low-mass-imf-log-normal}, we see that the values of $m_c$ and $\sigma$ show no obvious, systematic departures from Galactic values as a function of metallicity or age.  Though the SMC value for $m_c$ is lower than the Galactic value, it is hard to interpret this as evidence for a \textit{systematic} trend. The trend with age (redshift) is similar.

Limiting this to only UFDs, taken at face value, this figure implies that for galaxies known to have formed prior to and/or during the reionization-era, there do not appear to be any obvious departures from a Galactic-type IMF. Indeed, the most secure measurement, from Bootes~I, is consistent with the MW within $1\sigma$. It is clear from current measurements in the LG that the low-mass IMF in the early Universe appears to be MW-like, though, aside from Bootes~I, current uncertainties on these measurements remain large.

The IMF measurement in the SMC by \citet{Cohen:2026} is particularly instructive in searching for IMF variations over time and metallicity.  The location of their field in the SMC was designed to allow them to use the SFH and AMR measured from oMSTO depth \hubble\ imaging from \citet{Noel:2009}.  This field shows a nearly constant SFH over all cosmic time and an AMR that ranges from [M/H] $=-1.8$ to $-0.75$.  Yet, despite these large variations in age, and modest variations in metallicity, they find a log-normal IMF that differs only modestly from a MW-like log-normal (a lower $m_c$ and a slightly broader $\sigma$).  That is, if there were substantial deviations from a MW-like log-normal as a function of age/redshift or metallicity, it would be reasonable to expect that the composite, or average IMF they measured may not be MW-like. Although IMF variations could in principle average out in a field population, producing MW-like values in this way would require a certain degree of fine-tuning.

It is important to note that the uncertainties on the log-normal IMF parameters in Figure~\ref{fig:low-mass-imf-log-normal} are generally large, owing to relatively shallow data available for most galaxies.  However, given  recent measurements in Bootes and the SMC, there is emerging evidence that the low-mass IMF departs at most modestly from a MW-like log-normal over a large dynamic range of parameter space, e.g., to metallicities as low as [Fe/H] $\sim$ $-2.4$.  Extensions to even more metal-poor UFDs may help bring these results closer to the metal-free regime, for which the characteristic mass may increase \citep[e.g.,][]{Hennebelle:2024}.

As we highlight in \S \ref{sec:future}, deeper data (i.e., comparable in quality to that of \citealt{Cohen:2026} or \citealt{Ding:2026}) for a larger sample of galaxies are needed to confirm the presence or absence of even modest deviations from the low-mass Galactic IMF in nearby galaxies.

\subsection{High-Mass IMF}

In nearby galaxies, high-mass IMF ($M_{\star} \gtrsim 1$~\msun) measurements are generally based on counting resolved stars either in young star clusters or in the field.
The coeval nature of clusters can mitigate the effects of SFH.  Young clusters ($\lesssim 20$~Myr) are preferred as there may not be enough time for significant mass-dependent ejection of stars with $M$ $\gtrsim$ 1 \msun\ to have occurred, though reports of OB stars outside of clusters may provide evidence to the contrary \citep[e.g.,][]{Blaauw:1961,Gies:1986,Gies:1987}.  There is debate as to whether massive stars can form in the field, absent other lower mass stars, or whether they are all products of dynamical ejection \citep[e.g.,][]{Kobulnicky:2022,Carretero:2023}. Within clusters, issues such as mass segregation, a high degree of stellar multiplicity, the effects of observational crowding, and spatially varying observational completeness can complicate PDMF measurements and their interpretation as an IMF \citep[e.g., see discussions in][]{Scalo:1986, Bastian:2010, Kroupa:2013, Hopkins:2018, Kroupa:2026}.

Though the field high-mass IMF has been the subject of numerous historical studies, interpreting the PDMF of a field population as an IMF requires external knowledge of the population's SFH, which is often not available.  Without knowledge of the SFH, or at least marginalizing over an unknown SFH when measuring the IMF from a field population, the resulting IMF can appear artificially steep \citep{Miller:1979, Elmegreen:2007}.  Despite this challenge, field populations do have several advantages over clusters for IMF studies.  For example, they are often less crowded than young clusters and can provide for IMF constraints in environments in which young clusters are sparse (e.g., dwarf galaxies).

\subsubsection{M31}
\label{sec:M31_IMF}

One of the main aims of the \hubble\ multicycle PHAT survey \citep{Dalcanton:2012b} was to make a precise measurement of the high-mass IMF in a spiral galaxy outside the MW.  PHAT obtained homogeneous imaging and employed uniform data analysis and IMF measurement techniques in order to mitigate many of the systematic issues that have hampered previous high-mass IMF measurements.  Within PHAT, the strategy was to resolve stars in a large sample of young star clusters that enabled a `global' IMF measurement as well as the search for trends between inferred IMF and other galactic properties (e.g., mass, metallicity, star formation density).

\begin{figure}[pos=tbp]
    \centering
    \includegraphics[width=1.0\linewidth]{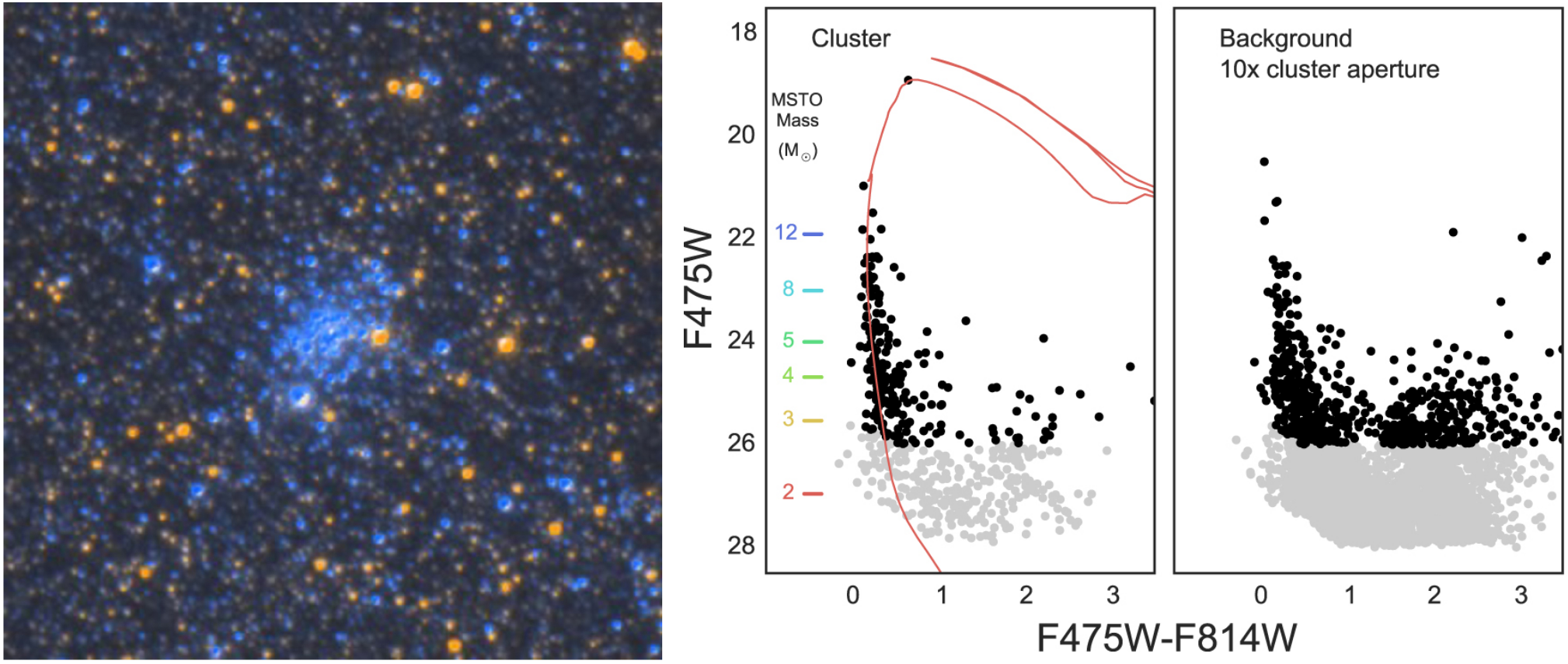}
    \caption{Example of a typical star cluster observed with \hubble\ that was used to measure the high-mass IMF in M31 as part of the PHAT survey \citep{Dalcanton:2012b, Weisz:2015b}.  \textbf{Left:} A color optical image of star cluster AP 94, and the field surrounding it, as observed with \hubble/ACS in the F475W and F814W filters.  The image corresponds to $\sim60$~pc on each side. \textbf{Right:} The optical CMD of the cluster (left) and a surrounding background field that covers 10 times the area of the cluster (right), which is used to statistically account for non-cluster members in the IMF measurement process.  MSTO masses from the Padova models are indicated \citep{Girardi:2010}.  The IMF measurement is limited to masses $\gtrsim3$\msun, due to completeness at the faint-end.  Assuming a power-law model, the measured IMF slope of this single cluster is $\alpha=-2.39^{+0.32}_{-0.30}$.  From similar IMF measurements of 85 clusters in M31 as part of PHAT, \citet{Weisz:2015b} measured an ensemble IMF slope of $\alpha=-2.45^{+0.03}_{-0.06}$ which is slightly steeper than Salpeter or Kroupa IMFs.  \citet{Wainer:2024} undertook a similar study in M33 as part of PHATTER and found a similar result of $\alpha=-2.49\pm0.18$.  }
    \label{fig:phat_imf}
\end{figure}

PHAT resolved $\sim2800$ star clusters across $\sim1/3$ of M31's star-forming disk.  It is one of the most complete (resolved or unresolved) star cluster surveys in the literature and extends to lower cluster masses than any other $L_{\star}$ galaxy outside the MW \citep{Johnson:2015, Johnson:2015b}.
From this extensive cluster catalog, \citet{Weisz:2015b} measured the IMF in 85 young ($<25$~Myr), intermediate mass ($10^3-10^4$ \msun) star clusters by modeling their optical CMDs for stars with MS masses $\gtrsim2$\msun.  Figure~\ref{fig:phat_imf} shows an optical image and CMD of a typical star cluster used in the PHAT IMF study.  \citet{Weisz:2015b} used  spatial modeling of completeness via ASTs, a model for binary star fraction, the Padova stellar evolution models \citep{Girardi:2010}, and a statistical model of background contamination.   They fit for the IMF slope only, leaving the maximum stellar mass set to 100 \msun\ and the minimum mass set to the 50\% completeness limit of the data in each cluster.

Owing to the small number of stars, IMF constraints on a per cluster basis were only $\pm0.3$~dex.  However, \citet{Weisz:2015b} employed a hierarchical Bayesian model that used each cluster's posterior distribution (IMF slope, age, extinction, mass) to constrain the global IMF.  Specifically, they fit for a mean IMF slope and variance in the IMF slope between the clusters, while simultaneously searching for trends with age, mass, and galactic radius.  Finally, they used an extensive suite of artificial clusters, with known input parameters, to verify the ability of this entire process to recover the true IMF slope and other trends (e.g., with age).

The study found a mean IMF slope of $\alpha=-2.45^{+0.03}_{-0.06}$ with a cluster-to-cluster scatter that is consistent with zero.  There were no significant trends with cluster age, mass, or location in M31.  As discussed in \citet{Weisz:2015b}, this is a slightly steeper IMF than the canonical MW IMFs and, for example, implies the production of $\sim25$\% fewer high-mass stars.  To date, this is the most comprehensive high-mass IMF study from resolved stars in the literature.

\subsubsection{M33}
\label{sec:M33_IMF}

The PHATTER survey \citep{Williams:2021} measured the high-mass IMF in M33 with a methodology similar to that used by PHAT in M31. M33 has a higher SFR density than M31 \citep[e.g.,][]{Lazzarini:2022}, which provides an opportunity to search for IMF variations in a more actively star-forming galaxy.

Within the inner $\sim14$ square kpc covered by PHATTER, \citet{Johnson:2022} cataloged $\sim1200$ star clusters all of which are resolved into stars by \hubble, with a quality similar to that shown in Figure~\ref{fig:phat_imf}.  From this cluster catalog, \citet{Wainer:2024} measured the IMF slope in 9 of the higher-mass ($M\gtrsim10^{3.6}$ \msun), young star clusters ($\lesssim 25$~Myr) in the sample using their optical CMDs. Their CMD data quality, fitting procedure, and related details are virtually identical to that used in \citet{Weisz:2015b}.  Their typical precision on the IMF slope in each cluster is $\sim0.3$~dex.

Using Bayesian hierarchical modeling, \citet{Wainer:2024} found a mean high-mass IMF slope of $\alpha=-2.49\pm0.18$ with no indication of significant cluster-to-cluster scatter or trends with age or radius.  The M33 high-mass IMF slope is similar to what is found in M31, though the larger uncertainty on the mean (owing to a smaller number of suitable clusters) is formally consistent with widely used high-mass IMF slope values from Galactic studies (e.g., Salpeter, Kroupa).

\subsubsection{Magellanic Clouds}

Though the LMC and SMC have historically been targets for high-mass IMF studies \citep[e.g.,][]{Massey:1998, Scalo:1998, Scalo:2005}, many legacy IMF studies in these galaxies focused on field populations using luminosity functions and minimal accounting for SFH effects \citep[e.g.,][]{Scalo:1998, Scalo:2005, Bastian:2010}. More recently, there have been targeted IMF studies of star clusters using \hubble-based CMDs and more sophisticated analysis techniques, most of which were covered in the IMF review of \citet{Bastian:2010}.  Here, we summarize IMF studies
published after the \citet{Bastian:2010} review that
employ detailed quantitative modeling of CMDs to infer the IMF.

\citet{Cignoni:2015} used deep optical \hubble\ imaging from the HTTP survey \citep{Sabbi:2013, Sabbi:2016} to assess the IMF in NGC~2070, the young star cluster at the heart of 30 Dor.  They used F555W-F775W CMDs that were spatially divided due to varying completeness.  The CMDs were fit down to 0.5 $M_{\odot}$.  While the main goal of this paper was to model the SFH, they also tested for the impact of IMF variations on the robustness of their solution.  They concluded that there were no deviations from a Kroupa IMF.

\citet{Baumgardt:2023} measured the IMF for 120 clusters in the MW, LMC, and SMC using homogeneously reduced \hubble\ imaging data and a uniform measurement technique.  While the low-mass IMF measurements are limited to MW GCs, which are beyond the scope of this review, they also measured the high-mass IMF ($M_{\star} > 1$ \msun) for 14 intermediate-age and young clusters in the LMC and SMC.  They used ASTs to account for completeness and applied the stellar models of both Dartmouth \citep{Dotter:2008}, with varying degrees of $\alpha$-enhancement, and PARSEC \citep{Bressan:2012}. The inferred masses of each star were used as input into a maximum likelihood fitting routine as described in \citet{Clauset:2009}. They did not adopt a statistical model of unresolved binary stars.  They also ran a series of $N$-body simulations to model how the dynamical evolution of the clusters affects the interpretation of the PDMF as the IMF.

For the high-mass IMF, they find an average value of $\alpha=-2.3\pm0.2$.  Taken at face value, this IMF measurement is compatible with both the canonical Galactic IMFs, as well as those measured in M31 and M33.  \citet{Baumgardt:2023} note that there are no clear trends between high-mass IMF and cluster property.  One caveat to this study is that some of the intermediate age clusters have a fairly limited dynamical range in mass, but precise IMF slope measurements.  For example, LMC cluster Hodge 6 has an age of $\sim2.5$~Gyr and the high-mass IMF was fit over the available mass range of 1--1.37 \msun\ to find an IMF slope of $\alpha=-2.48 \pm 0.21$.  Given the limited mass range, achieving this degree of precision, in theory, would require $>10^4$ stars \citep{Weisz:2013}.  The paper does not specify the number of stars used.  Not all clusters in the study have such limited dynamical ranges in mass.

Although entirely based on spectroscopy, given its high degree of importance as a local starburst, we also include the high-mass IMF of 30 Dor by \citet{Schneider:2018}.  Using a sample of 247 single stars, as determined from repeated ground-based spectroscopic measurements as part of the VLT-FLAMES
Tarantula Survey, they inferred masses of individual stars and simultaneously measured the SFH and IMF for $M\gtrsim15$~\msun.  This lower limit was set by the VFTS magnitude limit, above which sample completeness is independent of magnitude, and they corrected for the remaining selection effects (e.g., as a function of spectral type and proximity to R136).  First, they reported the presence of a single star with $M\sim200$~\msun, providing a constraint on the upper mass limit of the IMF, $M_{\rm max}$.  Second, over the range of 15--200~\msun, they find a high-mass IMF slope of $\alpha=-1.90^{+0.37}_{-0.26}$, which is marginally flatter than the Salpeter value of $\alpha=-2.35$.  They find that increasing the lower limit of this range to 30 and 60~\msun\ produces consistent results.  Interestingly, Figure 1 of \citet{Schneider:2018} shows that their plotted PDMF is consistent with a Salpeter slope between 15 and 30~\msun, while there is a `bump' in the mass function starting at $\sim30$~\msun\ that appears to be driving the flatter slopes (i.e., there is an excess of stars above $\sim30$~\msun\ relative to predictions from Salpeter).  They discussed various reasons for this excess (e.g., runaway lower mass stars, stellar collisions), but did not come to a firm conclusion, other than a flatter high-mass IMF slope.  In a related paper, \citet{Farr:2018} re-fit the stellar mass distribution used by \citet{Schneider:2018} using a hierarchical Bayesian model and found a slope of $\alpha=-2.11^{+0.19}_{-0.17}$, which is also marginally flatter than Salpeter.  The resultant flatter IMF slope in a region of high star formation intensity has also been reported in some other environments, such as the Arches cluster in the MW \citep[e.g.,][]{Hosek:2019}.  Subsequent measurements of the IMF in regions in and around 30~Dor based on optical spectroscopy \citep[e.g.,][]{Bestenlehner:2020, Massey:2025b} find results that are similar to that of \citet{Schneider:2018}, though we note that different techniques are used among these groups.

One largely unexplored issue is the degree to which IMF measurements only from spectroscopy compare to those derived only from imaging.  Whether such approaches produce consistent results is not clearly quantified and will be important to understand given the vastly different techniques, selection effects, systematics, etc., inherent to each method.

\subsubsection{Outskirts of Nearby Galaxies}

Several studies have used resolved stars to investigate the high-mass IMF in the outskirts of nearby galaxies.  These regions are particularly interesting as they enable the search for variations in the IMF in areas with low SFRs, low gas surface densities, and typically, low metallicities.

The general approach to these studies is to model the young MS population on an optical CMD by adopting a fixed SFH and allowing the IMF to vary.  The parameters for the IMF are the slope of a single power-law and the maximum mass of a star.  As discussed in \citet{Weisz:2013}, the maximum stellar mass constrained by stars alive today is only a lower limit on the most massive stars ever formed in a given stellar population and may not be directly comparable to the true upper mass limit of the IMF.

\citet{Bruzzese:2015} used this technique to measure the high-mass IMF in the outskirts of NGC~2915, a nearby blue compact dwarf galaxy ($D=3.8$~Mpc).  They constructed CMDs from deep \hubble/ACS F475W, F606W, and F814W imaging, accounted for completeness with ASTs, adopted the Geneva stellar models \citep{Schaerer:1993}, and considered stars with $M\gtrsim4$\msun.  They did not include unresolved binaries in their modeling, but argue that the impact of these binaries is small compared to other sources of uncertainty.   By holding the SFH constant and fitting for the IMF, they found a best-fit value of $\alpha=-2.85$ and very broad constraints on the maximum stellar mass.  Their simulations indicate uncertainties on the IMF slope are $\pm0.36$. These uncertainties do not include any variations in the SFH. Their best fit IMF is slightly steeper than the canonical Salpeter or Kroupa values (i.e., $\alpha=-2.35$), but at modest statistical significance ($\sim1.4\sigma$).  This IMF slope is statistically consistent with the IMF slopes ($\alpha=-2.45$) in M31 and M33 \citep{Weisz:2015b, Wainer:2024}.

\citet{Watts:2018} apply the same technique to the outskirts of isolated, gas-rich dwarf galaxy DDO~154 ($D=4.0$~Mpc).  Here, they model the upper MS on CMDs constructed from deep \hubble/ACS F475W, F606W, and F814W imaging in a field located $\sim5$~kpc from the galaxy's optical center.  The only change in the modeling approach is the adoption of the PARSEC stellar models \citep{Bressan:2012}.  For a fixed, constant SFH, they find an IMF slope of $\alpha=-2.45$ and a maximum stellar mass that is not well-constrained.  They indicate an approximate uncertainty on the IMF slope of $\pm$0.37.  Though the authors reported it as deficient in high-mass stars, this measurement is statistically consistent with Salpeter/Kroupa IMF values, as well as those measured in M31 and M33.

\citet{Bruzzese:2020} measured the IMF in the outer disk of M83 using a similar technique.  Specifically, they used deep \hubble/ACS F435W, F606W, F814W imaging of 4 fields in M83, all at the edge (or beyond) of the main star-forming disk.  By holding the SFH constant, they find a high-mass IMF slope of $\alpha=-2.35\pm0.3$ and an upper mass limit of $M=25_{-3}^{+17}$ \msun.  The IMF slope is consistent with the Salpeter/Kroupa IMF values, as well as those found in M31 and M33.   The upper mass limit is clearly far lower than what is commonly adopted in standard IMF assumptions, i.e., values of usually $M_{\star}\gtrsim100$\msun.

\subsubsection{Systematic Variations in the High-Mass IMF}
\label{sec:high_mass_imf_trends}

Studies of the high-mass IMF slope in M31, M33, and the LMC/SMC each report no significant trends or variations within each galaxy, aside from the possibility of a marginally flatter IMF slope in 30 Dor reported by \citet{Schneider:2018}.   Taken at face value, the ensemble unweighted high-mass IMF slopes in M31 ($\alpha=-2.45^{+0.03}_{-0.06}$), M33 ($\alpha=-2.49\pm0.18$), and the LMC/SMC ($\alpha=-2.30\pm0.20$) are all consistent with a single-valued slope that is close to the M31 value. These analyses are supportive of a ``universal'' high-mass IMF in the sense that all clusters could have IMFs drawn from the same underlying distribution.  However, there is no clear consensus on what defines a ``universal'' IMF, as has been extensively discussed in the literature over the past few decades \citep[e.g.,][]{Scalo:1986, Scalo:1998, Bastian:2010, Kroupa:2013, Hopkins:2018}.

The most robust extragalactic determination of the high-mass IMF slope is in M31.  It is marginally steeper than canonical MW IMFs (e.g., Salpeter, Kroupa).  However, it is unclear if this difference is physical in origin or due to methodological differences.  Whereas results in M31 use modern stellar models, artificial clusters, detailed completeness corrections, etc. the canonical high-mass IMF values (e.g., in the MW) were established using older data and IMF measurement techniques. A clear need to move this field forward is a uniform measurement (from data reduction through IMF modeling) of the high-mass IMF throughout the local Universe, including the MW.

IMF studies in the outskirts of nearby galaxies represent a promising step forward, as they extend IMF measurements to substantially different environments.  The results of the current studies in the outskirts of galaxies are generally compatible with an M31-like high-mass IMF slope, largely because of the sizable statistical uncertainties.  Expanding these efforts would be valuable, although future studies would benefit from marginalizing over possible variations in the recent SFH, which are well-established as degenerate with the high-mass IMF slope.  The result would likely be larger uncertainties on the high-mass IMF slope for these field populations.

\section{Proper Motions and Orbital Histories}
\label{sec:orbital_histories}

Galaxies in our most immediate neighborhood ($D \lesssim 1.5$ Mpc) are the only extragalactic stellar systems for which we can currently measure apparent proper motions (i.e., the tangential motion of astrophysical objects, projected onto the plane of the sky). Assuming a known distance to the galaxy, the proper motion can then be converted into a physical transverse velocity. This can, in turn, be combined with spectroscopic radial velocities to obtain 3D kinematics and, when coupled with a gravitational potential model, the galaxy's 3D position and velocity can be integrated back in time to reconstruct its orbital history. The determination of precise proper motions for galaxies in the local Universe, and the consequent determination of orbital histories, is valuable for a broad range of science cases, from studying the mass distribution within the MW to understanding the origin and evolution of dwarf satellite galaxies to probing the process of hierarchical mass assembly. Orbital histories also provide the environmental context for interpreting SFHs.  For example, whether a satellite quenched at infall or at a pericentric passage through its host's halo can only be assessed if its orbit is known (see \S \ref{sec:sfh_trends}).  

In the following sections, we provide an overview of selected major science themes connected to proper motions of nearby galaxies. Due to the available proper motion measurements, the discussion is mainly focused on the MW and its satellite system. We include results from the M31 system and other dwarfs in the LG, which represent the current frontier of extragalactic proper motion science.

\subsection{Definitions and Introduction to Concepts}
\begin{figure}[pos=tbp]
    \centering
    \includegraphics[width=1.\linewidth]{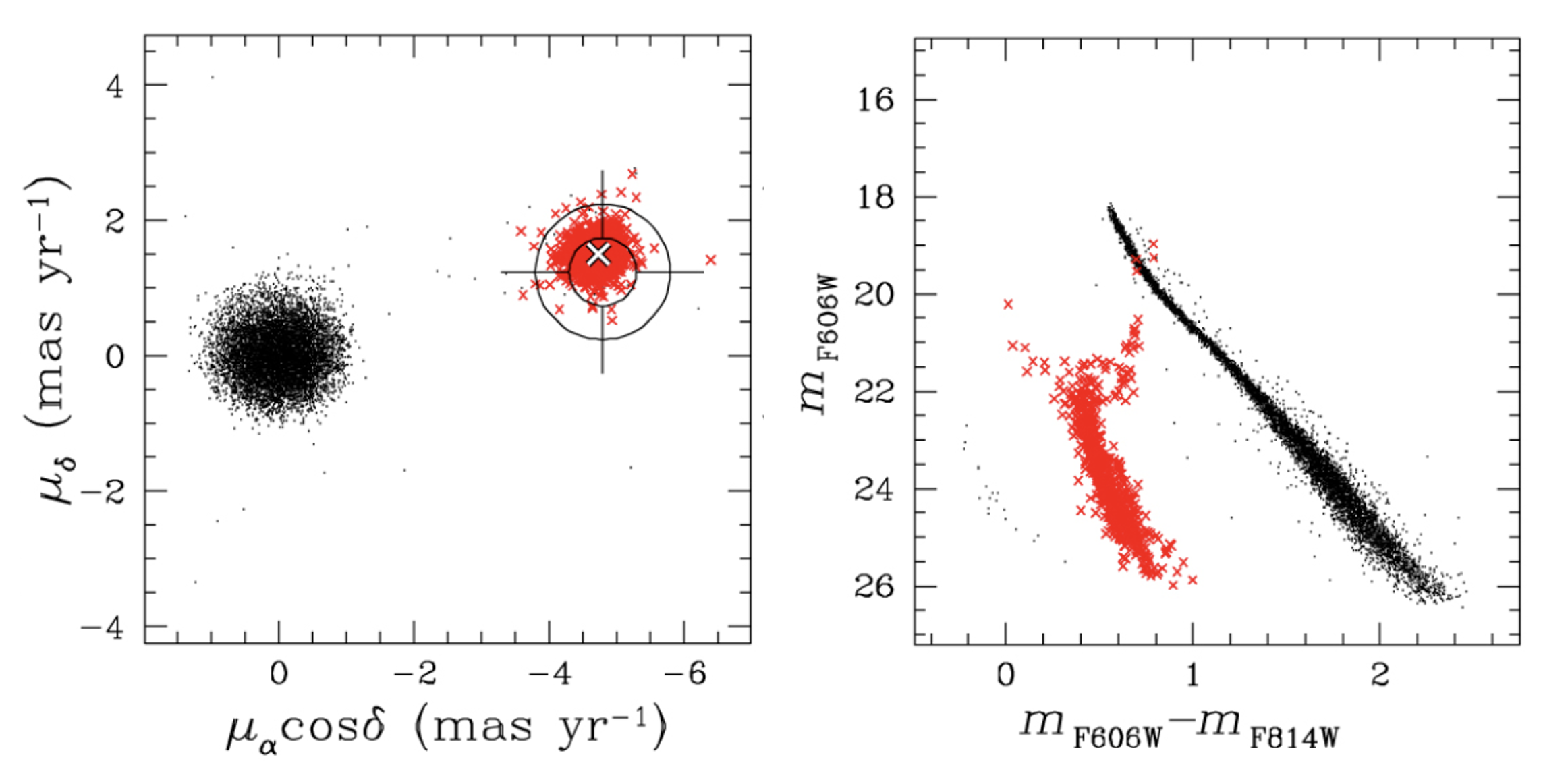}
    \caption{An example of how proper motions can be used to differentiate between spatially co-located stellar populations, in this case MW globular cluster 47 Tuc, which is located in front of the SMC. \textbf{Left}: The distribution, in relative proper motion space, of stars belonging to the globular cluster 47 Tuc (black points) and the nearby galaxy SMC (red crosses). \textbf{Right:} \hubble\ CMD of the stars belonging to the two kinematical components. Adapted from \citet{Bellini:2017}.}
    \label{fig:SMC_Selection}
\end{figure}

The measurement of apparent proper motions typically involves comparing images taken at two or more epochs to measure the small shifts in the positions of the target stars in a galaxy with respect to a known astrometric reference frame. The astrometric reference frame is usually set by motionless background objects in the field, either quasars \citep[e.g.,][]{Cudworth:1986,Tinney:1997} or distant galaxies \citep[e.g.,][]{Mahmud:2008,Sohn:2012}, although it is possible to use nearby stars with known tangential velocities \citep[e.g.,][]{Pryor:2010,Warfield:2023, Warfield:2026}. When measured with sufficient precision, the resulting proper motions can be used to identify member stars versus contamination (e.g., Figure~\ref{fig:SMC_Selection}), to derive the systemic proper motion of the target stellar system and, in conjunction with radial velocities, its 3D kinematical vector, and even to study the internal kinematics of the target stellar system. We highlight that, while the precision in relative proper motions is sufficient for tasks such as determining membership probability or studying the internal kinematics of nearby systems, determining bulk proper motions with respect to the Sun also requires absolute astrometric accuracy, which in turn is affected by the reliability of the astrometric reference frame.

Due to the small angular velocities of astrophysical systems at extragalactic distances, measuring proper motions is a challenging task that involves measuring positional shifts of a small fraction of a detector pixel. This is facilitated by: i) high-resolution imaging, which allows the detection of small astrometric shifts; ii) good knowledge of the PSF, which allows the detection of sub-pixel source position shifts; iii) a long time-baseline, which results in larger astrometric shifts; iv) deep imaging, which increases the number of observed target stars (and potentially reference objects), tightening the statistical constraints on the systemic velocity of the galaxy; and v) using the same camera and filter combination for all astrometric epochs or, if not possible, cameras operating at similar wavelengths/resolutions to minimize systematics in the geometric transformations and astrometric reference frame.

Early proper motion measurements of nearby galaxies were derived using ground-based observations of MW satellites, leveraging long temporal baselines of decades. Even so, the limited depth and resolution of these observations, coupled with poorly constrained systematics linked to the analysis of photographic plates, severely limited the accuracy of the inferred velocities \citep[e.g.,][]{Cudworth:1986,Jones:1994,Schweitzer:1995,Ibata:1997}.

A significant advance in our ability to measure proper motions occurred with the use of large space observatories, with their higher angular resolution and exquisite photometric stability. For decades, \hubble\ has been the primary workhorse to obtain precise proper motions of individual satellite galaxies in the LG \citep[e.g.,][]{Anderson:2000,Piatek:2002,Piatek:2003,Piatek:2005,Piatek:2006,Piatek:2007, Piatek:2008, Piatek:2016, Kallivayalil:2006b, Kallivayalil:2006a, Kallivayalil:2013, Pryor:2010, Pryor:2015, Lepine:2011, Massari:2013, Sohn:2013, Sohn:2017, Sohn:2020,Casetti-Dinescu:2022}.

A second paradigm shift in the field came with the launch of the \textit{Gaia} satellite, which surveyed the entire sky from 2014 to 2025 and builds on the pioneering astrometry of its predecessor, \textit{Hipparcos} \citep{Perryman:1997}.  \textit{Gaia} has substantially improved the precision and homogeneity of proper motion measurements within the MW halo \citep[e.g.,][]{Gaia-Helmi:2018, Fritz:2018, Simon:2018, Kallivayalil:2018, Massari:2018a, Pace:2019, Carlin:2018, Patel:2020, McConnachie:2020a, McConnachie:2020b, Battaglia:2022, Li:2021, Pace:2022}. At the time of this review, precise proper motions have been measured for $>60$ satellites within 300~kpc of the MW \citep[e.g.,][]{Pace:2024}. We note that, for relatively nearby stellar systems, ground-based proper motions remain a viable option, especially if using long temporal baselines and observing large sky areas \citep[e.g.,][]{Mendez:2010,Casetti-Dinescu:2016,Casetti-Dinescu:2018, Fritz:2018b}.

Outside the halo of the MW, \textit{Gaia} is able to measure tangential velocities in some classical dwarf galaxies \citep[e.g.,][]{McConnachie:2021,Battaglia:2022, Bennet:2024}, but \hubble\ remains a highly competitive observatory, due to its higher resolution and sensitivity and the fact that it continues to operate (whereas the \textit{Gaia} mission has ended, though several data releases are still expected). As of this writing, \hubble\ proper motion measurements of galaxies outside the MW halo have been made for a few members of the M31 system \citep[e.g.,][]{Sohn:2012, VanderMarel:2019, Sohn:2020, Warfield:2023, Casetti-Dinescu:2024, Casetti-Dinescu:2025, Patel:2025}, and several galaxies at the periphery of the LG \citep{Bennet:2024, Bennet:2025}. In principle, \jwst\ offers even higher angular resolution and depth than \hubble, making it an appealing facility for measuring proper motions in distant galaxies. However, as of this writing, proper motion measurements using \jwst\ are still in early stages \citep{Libralato:2023,Nardiello:2023}. In Table~\ref{tab:PM}, we provide a summary of recent compilations of orbital histories for large samples of LG dwarf galaxies, along with the dark matter halo masses used within the gravitational potential model.

\begin{threeparttable}[]
    \centering
    \caption{A summary of recent compilations for orbital history reconstructions of LG dwarf galaxies.  We include the data source, number of dwarf galaxies in the paper, the assumed mass(es) of the MW, and we indicate whether additional gravitational potentials were included.}
    \begin{tabular}{lccccl}
    \toprule
    Study &Proper Motion& $N_{\rm Dwarfs}$& $M_{h}^{\rm MW}$ & $M_{h}^{\rm LMC}$  & Other \\
          &Source&         & $10^{12}$ \msun &  $10^{11}$ \msun & Potentials \\
    \toprule
    \citet{Gaia-Helmi:2018} &\textit{Gaia} DR2& 10 & 1.37, 1.9 & 0 & -\\
    \citet{Fritz:2018}&\textit{Gaia} DR2&39 & 0.8, 1.6&0& -\\
    \citet{Simon:2018}&\textit{Gaia} DR2& 17 & 0.9 & 0 & -\\
    \citet{Armstrong:2021} \tnote{a}&\textit{Gaia} DR2& 17 & 0.6 -- 2.4 & 0  & -\\
    \citet{Patel:2020}\tnote{b}&\textit{Gaia} DR2& 20 & 1, 1.5& 0.8, 1.8, 2.5 & SMC \\
    \citet{Li:2021}&\textit{Gaia} DR3& 46 & 0.19 -- 1.4& 0& -\\
    \citet{McConnachie:2021}\tnote{c}&\textit{Gaia} DR2 & 7 & 1.3, 1.6& 0& M31\\
    \citet{Battaglia:2022}\tnote{d} &\textit{Gaia} DR3&51& 0.88, 1.6 & 0, 1.5 &-\\
    \citet{Correa-Magnus:2022}\tnote{e}& \textit{Gaia} DR3&62&1.1,1.3&1 -- 2&-\\
    \citet{Pace:2022}&\textit{Gaia} DR3& 52 & 1.3 & 1.38 &-\\
    \citet{Bennet:2024}\tnote{f} &\textit{Gaia} DR3/\hubble& 12& 1 & 1.8 & M31, M33\\
    \citet{Bennet:2025}\tnote{f}&\hubble &6& 1&1.8 &M31, M33\\
    \citet{Patel:2025}\tnote{g,h} & \hubble/VLBA & 10&0 &0 & M31, M33\\

    \toprule
    \end{tabular}
        \begin{tablenotes}[para]\footnotesize
        \item[a] \raggedright The MW potential includes cosmological dark matter accretion\\
        \item[b] \raggedright $M_{h}^{SMC} = 5, 30 \times 10^9 M_{\odot}$ \\
        \item[c] \raggedright $M_{h}^{M31} = 1.3, 1.6, 2.6\times 10^{12} M_{\odot}$ \\
        \item[d] \raggedright The MW potential includes a time-dependent response to the LMC potential\\
        \item[e] \raggedright The mass values for the MW and the LMC are not fixed parameters but rather fitted constraints from a dynamical model of MW dwarf satellites and globular clusters.\\
        \item[f] \raggedright $M_{h}^{M31} = 2\times 10^{12} M_{\odot}$. $M_{h}^{M33} = 2.5\times 10^{11} M_{\odot}$ \\
        \item[g] \raggedright The dwarf galaxies in this sample are satellites of M31, therefore no MW/LMC potential is included. \\
        \item[h]\raggedright  $M_{h}^{M31} = 1.5, 3 \times 10^{12} M_{\odot}$. $M_{h}^{M33} = 0.5, 1, 2.5 \times 10^{11} M_{\odot}$

    \end{tablenotes}
    \label{tab:PM}
\end{threeparttable}

\subsection{Orbital Properties of MW Satellites}
As \textit{Gaia} allowed large homogeneous studies of the MW satellite kinematics, several authors have compared the inferred distribution of orbital properties of the dwarf galaxies around the MW to expectations from cosmological models. For instance, $N$-body simulations of dark matter halos predict that galaxy-hosting sub-halos around a MW-sized halo should preferentially be on high-eccentricity orbits, with median eccentricities, $\epsilon$, typically reported around 0.6-0.9 \citep[e.g.,][]{VandenBosch:1999,Wetzel:2011,Barber:2014}. By contrast, \citet{Gaia-Helmi:2018} analyzed the kinematics of 10 MW satellites, 9 of which were classical dSph galaxies, and concluded that their sample had relatively circular orbits, with $\epsilon \lesssim 0.6$ (median $\epsilon \sim 0.45$). In contrast, \citet{Simon:2018} reports a median $\epsilon$ of 0.79 for their sample of 17 UFDs. While this could be a hint of differences in the orbits of dSphs and UFDs, the two studies adopted substantially different assumptions for their orbital reconstructions (e.g., potential model).

Among the compilations that contain both dSphs and UFDs \citep{Fritz:2018,Li:2021,Battaglia:2022,Pace:2022}, the median eccentricities of the two groups typically differ by less than 0.2, although larger differences are reported for particularly light MW masses \citep[$\lesssim 0.8 \times 10^{11}$ \msun,][]{Li:2021}. Overall, there is no strong evidence for an intrinsic difference between the orbital elongations of UFDs and classical dSphs, especially considering the many observational effects that could be playing into the comparison (e.g., small number statistics in the dSph sample, limited precision in the UFD proper motions, asymmetries in the potential).

Regarding the typical eccentricity of the satellite population as a whole, published compilations imply median $\epsilon$ values that range from 0.4 to 0.9 \citep{Gaia-Helmi:2018,Fritz:2018,Simon:2018,Li:2021,Pace:2022,Battaglia:2022}. In particular, there is a tight correlation between the inferred median $\epsilon$ and the assumed dark matter mass of the MW, with heavier MW halos resulting in more circular orbits. The influence of the LMC, both as an external gravity source and through its effect on the MW halo, also has a substantial impact on the elongation of the orbits \citep{Gomez:2015,Patel:2020,Battaglia:2022}. This means that establishing whether the orbits of the MW satellites are as radial as expected from cosmological models will require further improvements in our understanding of the MW potential and a careful inclusion of time-dependent perturbations in the potential model.

Another ongoing debate relates to the position of known MW satellites along their orbits (referred to as the orbital phase). Orbital reconstructions from \citet{Fritz:2018,Simon:2018,Li:2021} concluded that a larger-than-expected fraction of dwarfs are currently close to their orbital pericenters, whereas simple orbital mechanics arguments suggest that most dwarfs should be observed near apocenter. \citet{Fritz:2018} argues that the observed excess of satellites near pericenter points to the existence of an undiscovered population of UFDs that are close to apocenter, and have not been detected yet due to their large distances and low luminosities. \citet{Li:2021} reach a similar conclusion, although they find that such a population can explain only part of the excess.

Other authors have examined the distribution of orbital positions in the known MW satellites and argued against a pericentric excess. \citet{Correa-Magnus:2022} recovered a flat distribution in orbital phases in their dynamical model of the MW dwarfs, although they stress that their modeling framework is implicitly biased towards those configurations. \citet{Pace:2022} showed that, under the metric used by \citet{Fritz:2018} to calculate orbital positions, the MW satellites observed near pericenter are not in excess of what is expected from either $N$-body dark matter-only simulations or Monte-Carlo simulations of MW orbits, especially when limiting the comparison to long-term MW satellites. However, while these results seem to rule out a strong tension with the satellite number counts at pericenter, \citet{Pace:2022} note that there is a lack of MW satellites observed close to apocenter, which again favors the view that our current census of MW UFDs is incomplete.

\subsection{Satellite Disruption}

\subsubsection{dSphs and UFDs}
Having access to the orbital histories of the MW satellites provides an opportunity to investigate whether --- and if so, how many of --- these galaxies might be tidally disrupting, and to probe how the process of satellite disruption may manifest through some of the observed satellite properties (e.g., ellipticity).
There are several approaches to tackling the question of satellite disruption. Early orbital studies of bright dSphs looked for alignment between the satellite ellipticity and the orbital motion \citep[e.g.,][]{Piatek:2003,Piatek:2005,Pryor:2015}. The results were mostly inconclusive, likely due to the limited precision of the available proper motions.

More recently, \citet{Pace:2022} used \textit{Gaia} EDR3-based proper motions to investigate the same question for a large sample of MW satellites. Their analysis found that there is an excess of satellites whose isophotal ellipticity aligns with the orbital motion, and this excess is particularly significant for the highly elongated satellites. This could be a sign of tidal disruption, but other non-disruptive processes, such as tidal torques \citep[e.g.,][]{Pereira:2008}, could be responsible.  We also note that elongation by itself is not evidence of tidal influence. Dark matter halos are expected to be triaxial, and the isolated dwarf galaxies of the LG, which are presumably the least tidally affected, are themselves typically quite elliptical \citep[e.g.,][]{Higgs:2021}. It is noteworthy that, while a satellite's elongation vector appears connected to its orbital motion, no relation has been found between ellipticity (or its alignment to orbital motion) and pericentric distance \citep{Battaglia:2022,Pace:2022}.

Some of the known MW satellites show additional evidence for disruption. For example, several satellite orbits have small orbital pericenters ($r_{peri}\lesssim 30$~kpc, as reported by at least one of the literature compilations of \citealt{Fritz:2018,Battaglia:2022,Pace:2022}), which would favor the hypothesis that these galaxies are currently being disrupted \citep{Fritz:2018,Simon:2018,Carlin:2018,Battaglia:2022,Pace:2022}. Notably, some of these dwarfs are known to present morphological disturbances in their stellar distribution, or to be associated with stellar streams \citep[e.g.,][]{Majewski:2003,Roderick:2015,Drlica-Wagner:2015,Longeard:2018,Carlin:2018,Massari:2018a}. The most obvious example is represented by the Sagittarius dSph \citep{Ibata:1994}, whose prominent stream has been known and studied for decades \citep[e.g.,][]{Mateo:1996,Ibata:2001,Newberg:2002,Majewski:2003,Belokurov:2006a}.

Among the UFDs, Boo~III and Tuc~III are generally regarded as disrupting systems, due to their particularly small pericenters and obvious tidal features \citep[e.g.,][]{Drlica-Wagner:2015, Carlin:2018, Erkal:2018}. The case of Tuc~III is noteworthy as it appears that a close encounter with the LMC might have played a role in the disruption of this satellite \citep[e.g.,][]{Erkal:2018}. The low-surface-brightness dwarfs Antlia~II and Crater~II are also likely to be disrupting, given their structural properties \citep[e.g.,][]{Vivas:2022,Coppi:2024,Limberg:2025,Atzberger:2026,Vivas:2026}, orbital motions \citep[e.g.,][]{Fritz:2018, Battaglia:2022}, and internal kinematic gradients \citep[e.g.,][]{Ji:2021}, which support proposed scenarios to explain their low stellar densities and small velocity dispersions \citep[e.g.,][]{Torrealba:2016,Caldwell:2017,Torrealba:2019}. Other disrupting galaxy candidates, based on their small pericentric distances ($r_{peri}\lesssim30$--$40$~kpc) and/or the ratio of their tidal to half-light radii, include Wil~1, Seg~1, Seg~2, Dra~II, Tri~II, Her, CVn~I, Tuc~IV, Tuc~V, Boo~I, and Gru~II \citep{Fritz:2018,Battaglia:2022,Pace:2022}. Interestingly, the Carina dSph shows clear morphological disturbances \citep[e.g.,][]{Battaglia:2012}, yet is on an orbit with a relatively large pericenter \citep[$r_{peri}\gtrsim50$~kpc, e.g.,][]{Fritz:2018,Battaglia:2022,Pace:2022}, which might indicate a different origin for its morphology.  Carina also has a  peculiar, episodic SFH \citep[e.g.,][see \S \ref{sec:Carina_sfhs}]{Smecker-Hane:1996,deBoer:2014,Savino:2015}.

\subsubsection{The Magellanic Clouds and M33}
Among the most notable examples of environmental interaction in the LG, prominent morphological disturbances have long been known to exist in the most massive satellites of the MW and M31, i.e., the Magellanic pair and M33 \citep[e.g.,][]{Hindman:1963,deVaucouleurs:1972,Gardiner:1992,McConnachie:2009,Putman:2009,Nidever:2010,Mackey:2016,Belokurov:2017}. These features have been extensively used to study the disruption mechanisms of these massive satellites and to put constraints on their orbital histories.

The LMC/SMC pair is surrounded by a number of complex stellar and gaseous structures spanning over a hundred degrees on the sky. These include the Magellanic stream, the Magellanic Bridge and the Leading Arm \citep[see the review of][for a comprehensive description of the Magellanic system]{Donghia:2016}. \citet{Besla:2026} provide a recent review of the orbital dynamics of the Clouds, including their mutual interaction and the response of the MW halo to the LMC. 

Until the early 2000s, consensus models for the formation of these structures posited that they emerged over several orbital passages of the Magellanic pair around the MW, and their formation was predominantly driven by the tidal/ram pressure disturbance of the MW. However, improved proper motion measurements of the Magellanic Clouds \citep{Kallivayalil:2006b,Kallivayalil:2006a,Piatek:2008,Kallivayalil:2013} revealed that these galaxies are likely on their first encounter with the MW or, at most, are on long-period orbits and have experienced fewer than 3 pericentric passages \citep[e.g.,][]{Vasiliev:2024}. This revision of the LMC/SMC orbit, coupled with an improved consensus mass for the LMC \citep[e.g.,][]{Boylan-Kolchin:2011,Penarrubia:2016,Erkal:2019,Watkins:2024}, and the apparent misalignment between the orbit of the Clouds and the Magellanic stream, required new formation scenarios for the Magellanic system. Currently, most formation scenarios rely on a strong interaction, and possibly a collision, between the LMC and the SMC to explain the initial removal of the gas and stars, while the tidal field and ram pressure of the MW contribute to shaping this material in the configuration we observe today \citep[e.g.,][]{Besla:2007,Besla:2010,Besla:2012,Diaz:2011,Pardy:2018,Tepper-Garcia:2019,Lucchini:2020}. New models, in which the MW interaction is the main mechanism for the Magellanic stream formation, are also being explored \citep[e.g.,][]{Hammer:2015,Wang:2019,Wang:2022}.

Similarly, M33 displays morphological perturbations, featuring significantly warped gaseous and stellar disks \citep[e.g.,][]{Rogstad:1976,Deul:1987,Corbelli:1997,McConnachie:2009,Putman:2009,Lewis:2013,Corbelli:2014,Kam:2017}. Early models for the origin of these features invoked a tidal mechanism, stemming from a close (50--100 kpc) pericentric passage of M33 around M31, happening within the last 3--4 Gyr \citep[e.g.,][]{Bekki:2008,McConnachie:2009,Putman:2009,Semczuk:2018}. However, these models were severely limited by the lack of direct measurements for the transverse velocity of M31, which was often left free to vary or inferred through indirect methods.

More recently, measurements of the proper motion of M31 have become available, first through \hubble\ \citep{Sohn:2012}, then through \textit{Gaia} \citep{VanderMarel:2019,Salomon:2021}. When these measurements are combined with Very Long Baseline Array (VLBA) proper motions of two H$_2$O masers in M33 \citep{Brunthaler:2005}, a major conclusion is that M33's orbit around M31 appears to be significantly less bound than previously thought, with orbital reconstructions suggesting that either M33 is currently at its first infall or that any previous pericenter happened more than 6 Gyr ago and at relatively large (100--200 kpc) distances \citep[e.g.,][]{Patel:2017a,VanderMarel:2019,Salomon:2021,Corbelli:2024,Patel:2025}. This poses a significant challenge for the hypothesis that irregularity in M33's morphology is tidally induced. Among the possible solutions, \citet{Tepper-Garcia:2020} proposed that environmental mass loss in the M33 halo can still result in a close (50 kpc) pericenter, even adopting the recent proper motion measurements of M31. However, we note that the analysis of \citet{Corbelli:2024}, which also includes tidal stripping of M33, reached the opposite conclusion, favoring first infall and excluding any close pericentric passage. At present, the origin of M33's disturbed gaseous and stellar morphology remains unclear, and a possible link to the interaction with M31 needs to be corroborated by further dynamical modeling. We also note that alternative scenarios have been suggested, such as accretion of satellites belonging to the M33 system, interactions with other satellites of M31, or gas inflow from a nearby cosmic filament \citep[e.g.,][]{Shaya:2013,Patel:2017a,VanderMarel:2019,Corbelli:2024}. At present, no strong evidence exists to corroborate any of these hypotheses.

\subsection{Connections with the SFH}
A powerful application of measuring orbital histories in nearby satellite galaxies is to explore possible connections between the orbit of the galaxy and its SFH, as measured from resolved stars. This is usually done by comparing specific orbital events, such as past pericentric passages or infall within the host's virial radius, with features in the SFH, such as cessation of star formation or discrete star formation bursts. In practice, this is a challenging exercise as it requires excellent reconstructions of both the SFH and the orbit. Given that the uncertainties in both dramatically increase as we consider larger lookback times, meaningful comparisons are only possible within the last 3--5 Gyr \citep[e.g.,][]{Hidalgo:2011, Armstrong:2021}.

For the MW dSphs, several studies have examined the correlation between their SFHs and potential interactions with the MW or the LMC. Early analyses \citep{Piatek:2003,Dinescu:2004,Piatek:2007,Sohn:2013} yielded mixed results, in most cases limited by the quality of available SFHs and proper motions. For instance, \citet{Dinescu:2004} analyzed the orbit of Fornax and suggested that gas removal and quenching in this galaxy, approximately 200 Myr ago, could have been triggered by a passage through the LMC gaseous tail. However, this hypothesis was later refuted by \citet{Piatek:2007} using improved proper motion measurements of the LMC.

\begin{figure}[pos=tbp]
\centering
    \includegraphics[width=1.\linewidth]{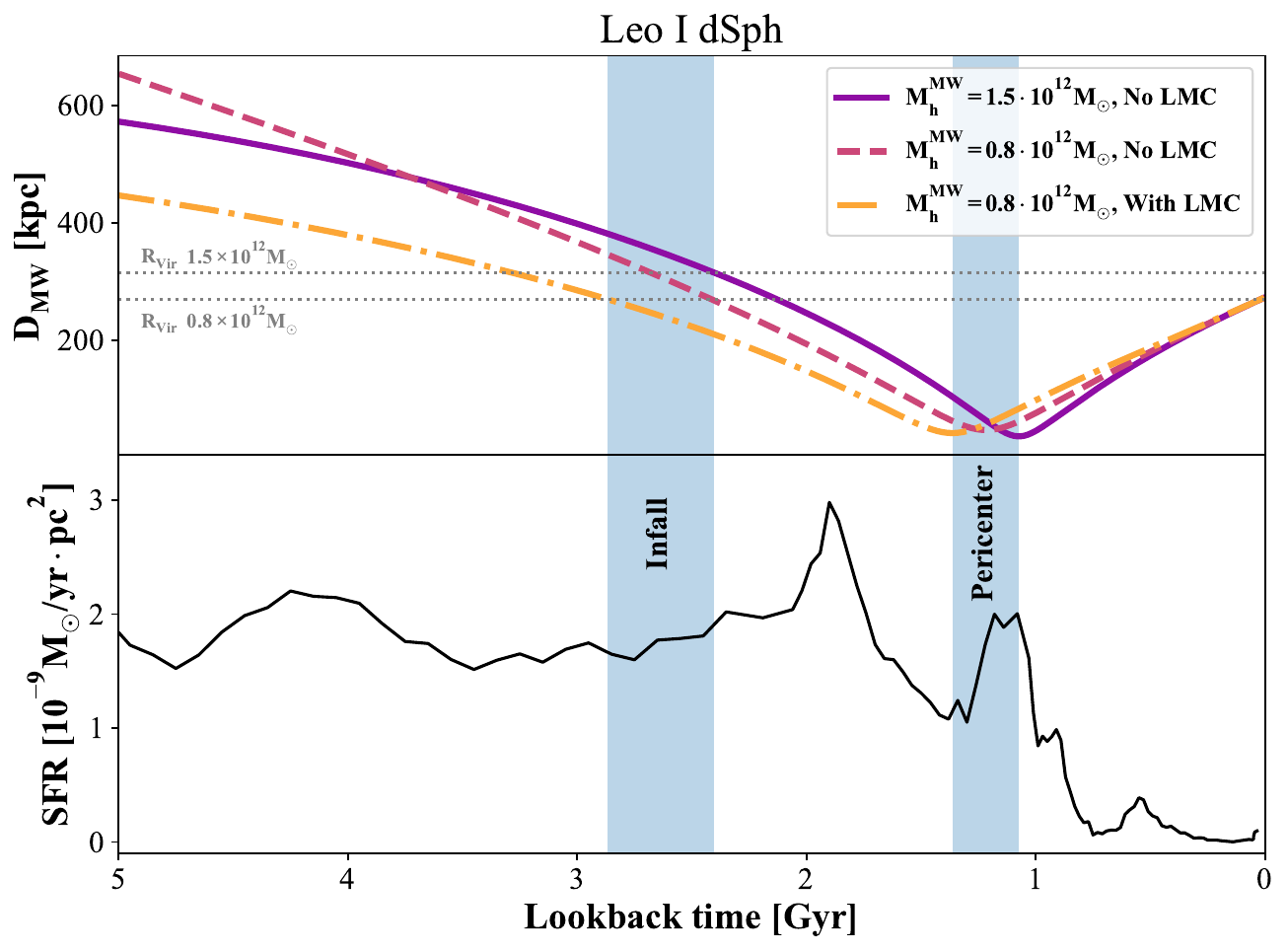}
    \caption{An illustration of the connection between orbital history and SFH for the Leo I dSph. \textbf{Upper:} Distance of Leo I from the MW as a function of lookback time, for three potential models (heavy MW in purple, light MW in magenta, and light MW + LMC in orange), from \citet{Battaglia:2022}. The gray dotted lines mark the virial radii associated with the assumed MW masses. \textbf{Lower:} SFR density of Leo I as a function of lookback time, from \citet{Ruiz-Lara:2021}. For the different orbital solutions, the epoch of infall (i.e., Leo I crossing the MW virial radius) and the epoch of pericenter (i.e., closest MW approach) are marked in cyan. Roughly 1 Gyr after crossing the MW halo, Leo I experienced a notable increase in star formation rate. A second episode of star formation happened coincident with the pericentric passage, followed by rapid decline in star formation activity. See \citet{Bennet:2024} for an alternative version of this figure.}
    \label{fig:LeoI}
\end{figure}

As of today, the clearest connection between star formation and orbital history in a MW dSph is for Leo~I (Figure~\ref{fig:LeoI}). By deriving a precise \hubble-based proper motion for this galaxy, \citet{Sohn:2013} inferred that Leo~I is currently completing its first passage around the MW, with the closest approach having occurred  $\sim1$ Gyr ago. They noted that this matches Leo I's SFH, which shows that the galaxy underwent sustained star formation within the past several Gyr, before a rapid cessation of activity, approximately 1 Gyr ago (see \S \ref{sec:LeoI_sfhs}). This result was later confirmed through more detailed determinations of the SFH \citep{Ruiz-Lara:2021} and of the orbital motion \citep{Fritz:2018,Gaia-Helmi:2018,Battaglia:2022, Bennet:2024}. It seems plausible that the quenching of Leo I can be attributed to its recent infall within the MW halo.

Other tentative connections between orbital motion and SFH events have also been made for other MW satellites. By using \textit{Gaia} DR2 proper motions and an \hubble-based SFH, \citet{Rusakov:2021} suggest a possible association between past pericentric passages of the Fornax dSph around the MW and two major star formation events detected in the stellar population of this galaxy. \citet{Battaglia:2022} analyzed \textit{Gaia} DR3-based orbital histories to revisit the orbit-SFH connection for Fornax, Carina, Leo I, and NGC 6822. Besides confirming the connection highlighted by \citet{Sohn:2013} in Leo I, \citet{Battaglia:2022} identified potential connections between the last pericenter passage of Fornax and Carina and specific features in their SFH, but also noted that most fluctuations in the star formation activity of these galaxies cannot be easily explained by orbital motion.

The LMC and the SMC also represent a very interesting case regarding the connection between orbital history and SFH. Several studies of the stellar populations in the Magellanic Clouds have reported that both galaxies have experienced a substantial increase in star formation activity in the last 3--5 Gyr \citep[e.g.,][see \S \ref{sec:lmc}]{Harris:2004,Harris:2009,Noel:2009,Cignoni:2012,Weisz:2013b,Meschin:2014,Ruiz-Lara:2020b,Mazzi:2021,Massana:2022}. With the realization that the LMC and SMC are likely to be on their first infall, \citet{Besla:2007} suggested that the recent star formation burst of both galaxies might be connected to accretion of the Clouds onto the MW. However, subsequent orbital models showed that the evolutionary history of the Magellanic Clouds is likely more complex, with interaction between the LMC and the SMC also having a strong influence \citep[e.g.,][]{Besla:2010,Diaz:2011,Besla:2012,Zivick:2018}. It is plausible that some of the features in the SFH of the LMC and SMC may be the result of the tidal influence of the two galaxies on one another. This is, for instance, the scenario presented by \citet{Massana:2022}, who argue for a high degree of synchronicity between the recent star formation bursts of the LMC and the SMC.  Extending this approach, \citet{Burhenne:2026} used high-spatial-resolution \hubble/Scylla SFHs to identify global and local bursts in both Clouds and to compare their timing and amplitude against several published orbital models.  They found that for models predicting a shared dynamical trigger across both galaxies, the corresponding burst systematically begins earlier in the SMC and produces a larger SFR enhancement than in the LMC --- providing a direct fossil-SFH constraint on which families of LMC--SMC encounter histories are consistent with the resolved star record.

A similar synchronicity argument has been put forward to motivate a tidally induced burst of star formation in M31 and M33. Both galaxies have been found to have experienced a significant event of star formation between 2 and 4 Gyr ago \citep[e.g.,][]{Williams:2009a,Bernard:2012,Williams:2015,Williams:2017}. This, connected with the orbital models of the time \citep{McConnachie:2009} along with the disturbed morphology of M33 \citep[e.g.,][]{McConnachie:2009,Putman:2009,Corbelli:2014,Kam:2017,Smercina:2023} was suggested as evidence of strong interactions between the two galaxies during that interval \citep{Bernard:2012,Williams:2015}. However, it is worth noting that more recent orbital models strongly disfavor close encounters between M31 and M33 within the last 6 Gyr
 \citep[e.g.,][]{Patel:2017a,Corbelli:2024, Patel:2025}. At this time, it remains unclear whether the recent star formation activity of M31 and M33 is truly connected and what the true nature of their past interactions, or lack thereof, may be.

 Within the fainter M31 satellite system, recent work by \citet{Patel:2026} presents evidence that the dwarf galaxy Pisces~I (LGS-3) may have experienced an environmentally induced episode of star formation. The analysis of precise \hubble-based proper motions reveals a high likelihood that Pisces~I experienced a close encounter (60--200 kpc) with M33, approximately 2 Gyr ago. The timing of this pericentric passage coincides remarkably well with an observed increase in the SFR of Pisces~I, which happened between 2.5 and 1.5 Gyr ago \citep[][see \S \ref{sec:PiscesI_sfhs}]{Hidalgo:2011}. The case of Pisces~I appears similar to that of the MW satellite Leo~I \citep[e.g.,][]{Sohn:2013, Ruiz-Lara:2021} and suggests that even less massive spiral galaxies such as M33 might act as a meaningful source of environmental perturbations on the star formation of nearby dwarf galaxies.

Another interesting case regarding the SFH-orbit connection is that of the dSphs Cetus (see \S \ref{sec:Cetus_sfhs}) and Tucana (see \S \ref{sec:Tucana_sfhs}). These two galaxies have long been known to show no signs of recent star formation \citep{Castellani:1996,Saviane:1996,Whiting:1999} with the last significant star formation happening more than 8 Gyr ago \citep[e.g.,][]{Monelli:2010_b,Monelli:2010_a}. At the same time, the relative isolation of these galaxies from both the MW and M31 has sparked discussion on the cause of their quenching, on whether these galaxies have been isolated throughout their life or whether they had past interactions with a massive member of the LG \citep[e.g.,][]{Fraternali:2009,Teyssier:2012,Buck:2019,Santos-Santos:2023}. The recent study of \citet{Bennet:2025} has, for the first time, measured the proper motions of both Cetus and Tucana, providing valuable insights into the orbital history of these objects. While the analysis is complicated by the old lookback times at which Cetus and Tucana could have had environmental interactions, the results from \citet{Bennet:2025} paint a relatively different picture for the two galaxies. Within the orbital uncertainties, Cetus has a non-negligible probability to have experienced a close passage ($\lesssim 250$ kpc) to M31 between 5 and 10 Gyr ago, which is compatible with its quenching from CMD-based SFHs \citep{Monelli:2010_a}. Conversely, Tucana does not appear to be in proximity of either the MW or M31 at any point in its lifetime. It is therefore unlikely that the SFH of this galaxy can be explained by it being a backsplash galaxy.

\subsection{Kinematical Associations and Satellite Substructures}
\subsubsection{Streams of Satellites and the Vast Polar Structure}

A powerful application of extragalactic proper motion measurements is to probe the existence and stability of kinematical associations among members of satellite systems. For instance, by examining the positions and radial velocities of MW satellite galaxies and GCs, several authors hypothesized that some of these objects might belong to kinematically coherent streams, possibly of tidal origin \citep[e.g.,][]{Lynden-Bell:1995}. Early \hubble\ proper motions of the MW dSphs were subsequently used to argue against the existence of such tight orbital associations \citep{Piatek:2002,Piatek:2003,Piatek:2005,Piatek:2006,Piatek:2007}.

Nevertheless, it has long been recognized that there appears to be a degree of clustering in the orbital properties of MW satellites. A significant fraction of these objects have orbital planes that are roughly perpendicular to the MW disk. This has led to the claim that these satellites belong to a planar association, which is commonly known as the Vast Polar Structure \citep[VPOS, e.g.,][]{Lynden-Bell:1976,Kroupa:2005,Metz:2007,Pawlowski:2013}. Over the years, the existence and kinematical coherence of the VPOS have been looked at with ever more accurate proper motion determinations. Most studies do confirm a degree of clustering in the orbital planes of many MW satellites, with a strong tendency toward high inclinations with respect to the MW disk \citep[e.g.,][]{Pryor:2015,Fritz:2018,Gaia-Helmi:2018,Pace:2019,Li:2021,Taibi:2024}. However, the exact degree of kinematical coherence is still under debate, with some studies indicating a large scatter in the orbits of proposed VPOS members or even instances of counter-rotation \citep[e.g.,][]{Lipnicky:2017,Maji:2017,Sohn:2017,Gaia-Helmi:2018,Sawala:2023}.  The most recent census, based on orbital poles for 73 MW dwarfs with proper motions, identifies 22 on-plane members, including the LMC and 9 of its satellites, nearly all of which are co-rotating \citep{Martinez-Garcia:2026b}.

Among the possible origins proposed for the VPOS, there is satellite accretion from cosmic filaments \citep[e.g.,][]{Zentner:2005,Libeskind:2005,Libeskind:2011,Lovell:2011,Pawlowski:2012b,Libeskind:2015}, as well as tidal disruption of a larger system \citep[e.g.,][]{Pawlowski:2011,Fouquet:2012,Akib:2025}, and accretion of a small galaxy group \citep[e.g., ][]{Li:2008b,Donghia:2008,Jerjen:2025}. But none of them has yet been conclusively established \citep[e.g.,][]{Pawlowski:2018}. A hypothesis that has received recent attention is whether the emergence of the VPOS could be connected to the accretion of the LMC, which is indeed well-aligned with the proposed planar structure \citep[e.g.,][]{Pawlowski:2013b,Pawlowski:2015}. The idea that the LMC could be associated with satellite substructures around the MW is not new (e.g., see the ``Great Magellanic Galaxy'' scenario from \citealt{Lynden-Bell:1976}), but the topic has gained renewed traction due to: i) the recognized influence of the LMC on the MW's potential and its satellites \citep[e.g.,][]{Garavito-Camargo:2021,Vasiliev:2023}; and ii) the discovery that some candidate members of the VPOS are also likely satellites of the LMC \citep[e.g.,][see \S~\ref{sec:LMC_Satellites}]{Kallivayalil:2018,Patel:2020,Pace:2022}. However, the connection between the VPOS and the LMC is not as straightforward as it might first appear. For instance, it has been shown that, even when excluding the LMC satellites from the analysis and accounting for the gravitational effect of the LMC, the VPOS signature remains appreciable within the MW satellite population \citep[e.g.,][]{Correa-Magnus:2022,Pawlowski:2022}. A recent orbital modeling by \citet{Martinez-Garcia:2025} supports the idea that the VPOS has been kinematically coherent for the last 5 Gyr, well before the expected accretion of the LMC and its satellites in ``first-infall'' scenarios. Interestingly, a scenario in which the LMC is in its second MW passage, such as the one proposed by \citet{Vasiliev:2024}, would move the infall time of the LMC system to much older epochs and could expand the list of potential LMC satellites, thus providing a plausible connection between the VPOS and the Magellanic Cloud group.

\subsubsection{The Great Plane of Andromeda}

As in the MW satellite system, there has been open debate about the presence and cosmological significance of large-scale substructures within the satellite system of M31. The feature that has received the most attention is the so-called GPoA, a proposed thin, possibly co-rotating, planar structure that appears to contain roughly half of the known M31 satellites \citep[e.g.,][]{Koch:2006,McConnachie:2006,Conn:2013,Ibata:2013,Santos-Santos:2020,Savino:2022,Doliva-Dolinsky:2026}. To date, the incidence, stability, and lifetime of such a feature, along with other proposed satellite planes \citep[e.g.,][]{Pawlowski:2012,Muller:2018,Muller:2021} remain a subject of ongoing discussion \citep[e.g.,][]{Kroupa:2005,Zentner:2005,Libeskind:2005, Sawala:2016,Fernando:2017, Fernando:2018, Pawlowski:2018,Boylan-Kolchin:2021b,Pawlowski:2021a,Pawlowski:2021b,Samuel:2021,Sawala:2023}. Another similarly intriguing feature, albeit one that has received less attention, is the strong lopsidedness observed in the M31 satellite population, with roughly 80\% of the known satellites currently residing on one side of M31 \citep{McConnachie:2006,Conn:2013,Savino:2022}. While several hypotheses have been put forward to explain this strong asymmetry, the cause of this lopsidedness remains poorly constrained \citep[e.g.,][]{McConnachie:2006,Patel:2025}, although it does not appear to be an observational selection effect \citep{Doliva-Dolinsky:2022, Doliva-Dolinsky:2026}.

Unlike with the MW satellites, the status of proper motion measurements for the M31 satellites, which would allow us to gain more insight into the nature of these substructures, remains piecemeal. While a handful of satellites have published systemic proper motions \citep{Brunthaler:2005,Brunthaler:2007,VanderMarel:2019,Sohn:2020,Patel:2025}, the only ones that have been studied in the context of the GPoA are NGC~147, NGC~185, IC~10, and \A{III}. The study of \citet{Sohn:2020} derived the first \hubble-based proper motions for the dE satellites NGC 147 and NGC 185, and demonstrated that their orbital vectors are consistent with them co-orbiting within the GPoA. In particular, the orbital pole of NGC 147 is almost perfectly coincident with the normal vector to the GPoA, while the orbital pole of NGC 185 is offset by approximately $16^{\circ}$, but it is still within uncertainties. The study also examined the literature proper motions of IC~10 \citep{Brunthaler:2007}, finding that, while the galaxy currently resides within the GPoA, its angular momentum vector is oriented at a very large angle from the GPoA, implying a chance alignment.

The alignment of the orbits of NGC~147 and NGC~185 with the GPoA has been interpreted as evidence that the GPoA is a long-lived and stable feature, at odds with expectations from $\Lambda$CDM \citep{Pawlowski:2021b}. However, as noted by the authors themselves, definitive conclusions cannot be drawn from the orbits of only two members, and only by measuring the proper motions of more M31 satellites will we be able to conclusively determine the nature of this proposed substructure.

More recently, the proper motion of \A{III} was derived by \citet{Casetti-Dinescu:2024} using \hubble/ACS and \hubble/WFPC2 images. The resulting orbit shows that \A{III} is broadly consistent with being dynamically associated with the GPoA. However, the case of \A{III} is less clear compared to NGC~147 and NGC~185 as, while the present-day motion of \A{III} is roughly in line with the GPoA, most orbital solutions correspond to misalignments greater than $30^{\circ}$.

\subsubsection{Proposed Galaxy Pairs}
Having access to the 6D kinematics of dwarf galaxies also allows us to probe whether select galaxy pairs share a similar orbital history or even whether they form a gravitationally bound system. Besides the immediate example of the LMC/SMC pair, which has been discussed extensively in the previous sections, other candidate galaxy pairs have been studied through analysis of their orbital motion.  Even without proper motions, \citet{Fattahi:2013} used positions, line-of-sight velocities, and luminosities to show that $\sim30$\% of the MW and M31 satellites brighter than $M_V=-8$ are likely in physical pairs, including Draco and Ursa~Minor and \A{I} and \A{III}, a far higher fraction than predicted by models.

Among the classical satellites of the MW, the Draco and Ursa Minor dSphs have been proposed to be a physically associated pair \citep{Gaia-Helmi:2018, Pawlowski:2021}. Integration of their orbits using \textit{Gaia} proper motion shows that these two MW satellites share a strikingly similar orbital history, with a large fraction of orbital realizations maintaining relative distances below $\sim$50 kpc for several Gyr of lookback time. A second association between MW satellites has been proposed by \citet{Richstein:2022} for the Pisces II and Pegasus III UFDs. Using \textit{Gaia} EDR3 proper motions, this study raises the possibility that this UFD pair might be gravitationally bound and, through orbit integration, shows that the relative distance between these galaxies has likely remained below $\sim$ 40 kpc for approximately 2 Gyr. Another proposed association pertains to a larger number of objects, namely the MW satellites Crater II, Leo II, Leo IV, Leo V, and the GC Crater I \citep[e.g.,][]{Torrealba:2016,Fritz:2018,Julio:2024}. \citet{Fritz:2018} compared the orbital poles of this proposed subgroup, concluding that a common dynamical origin for all the proposed members of the group appears unlikely, but associations between a smaller number of objects cannot be ruled out. A follow-up study by \citet{Julio:2024} used a combination of \textit{Gaia} and \hubble-based proper motions to analyze the orbital alignment and angular momentum of this group of stellar systems, concluding that, with the exception of Crater II, a common accretion scenario appears plausible. However, we note that the orbital motion of Leo IV and Leo V changes significantly depending on whether \textit{Gaia} or \hubble\ proper motions are considered, indicating that more robust proper motion measurements are needed to conclusively confirm this association.

Within the M31 satellite system, orbital histories have been used to study the possible association between NGC~147 and NGC~185. Due to their similar morphological properties, radial velocities, and small separation on the sky, this pair of luminous galaxies has long been suggested to form a bound pair \citep[e.g.,][]{VandenBergh:1998,Geha:2010,Arias:2016}. However, using \hubble\ proper motions, \citet{Sohn:2020} showed that the two galaxies are not gravitationally bound. Orbital integration shows that, while NGC 147 and NGC 185 are currently separated by less than 100 kpc, they only approached each other recently and they have likely entered the M31 system at different times and on different trajectories. Recently, a possible association has been proposed between the dwarf galaxies Pisces~I and \A{XVI}, with orbital reconstructions showing that these galaxies have remarkably similar orbits and there is a high probability that they might have completed a passage around each other in the last 1 Gyr \citep{Patel:2026}.

\subsubsection{Satellites of Satellites: The Case of the LMC}
\label{sec:LMC_Satellites}
Under the hierarchical structure assembly paradigm, the sub-halo distribution around a host dark matter halo is expected to be self-similar across a large range of host halo masses. It is therefore expected that large satellite galaxies, such as the LMC, were accreted along with a small satellite population of their own \citep[e.g.,][]{Donghia:2008,Sales:2013,Dooley:2017,Jahn:2019}. From an observational perspective, it was with the detection of several faint satellites close in projection to the LMC-SMC system \citep[e.g.,][]{Drlica-Wagner:2015,Koposov:2015,Martin:2015,Bechtol:2015,Laevens:2015b,Torrealba:2018,Kim:2015c} that support for this scenario started emerging. However, clearly identifying which of the Galactic UFDs were accreted with the LMC-SMC system and which were not is a challenging task.

Initial attempts could only make use of spatial positions and radial velocities, and typically relied on statistical association methods, paired with tailored simulations \citep{Deason:2015,Yozin:2015,Jethwa:2016,Dooley:2017,Sales:2017}. As \textit{Gaia} provided reliable proper motions, more direct methods based on 6D kinematics and orbital integration became possible. A large number of studies have been conducted to try to identify the most likely LMC satellites \citep[e.g.,][]{Kallivayalil:2018,Simon:2018,Pace:2019,Fritz:2019,Erkal:2020a,Patel:2020,Battaglia:2022,Correa-Magnus:2022,Pace:2022,Pace:2025}. The precise list of LMC satellite candidates varies somewhat among studies. However, there is a general consensus that several UFDs are likely LMC satellites. These are Hor I, Car II, Car III, Hyi I, Ret II, and Phx II. A recent spectroscopic analysis of Pic II also identifies it as a LMC satellite \citep{Pace:2025}. Other, more debated, associations are those of Tuc II, Hor II, and Eri III. The potential association between the LMC and more luminous classical dwarfs (e.g., Fornax and Carina) has also been long discussed \citep[e.g.,][]{Lynden-Bell:1976, Nichols:2011,Jahn:2019,Pardy:2020}. But current orbital reconstructions, in which the LMC is most likely in first infall, do not seem to support this connection \citep{Erkal:2018, Patel:2020, Battaglia:2022, Pace:2022}. Interestingly, under scenarios in which the LMC is on its second MW passage, several of the classical MW satellites, such as Carina, Draco, Fornax, and Ursa Minor have a non-negligible chance of being associated with the LMC/SMC pair \citep[e.g.,][]{Vasiliev:2024}.
\begin{figure}[pos=tbp]
\centering
    \includegraphics[width=1.\linewidth]{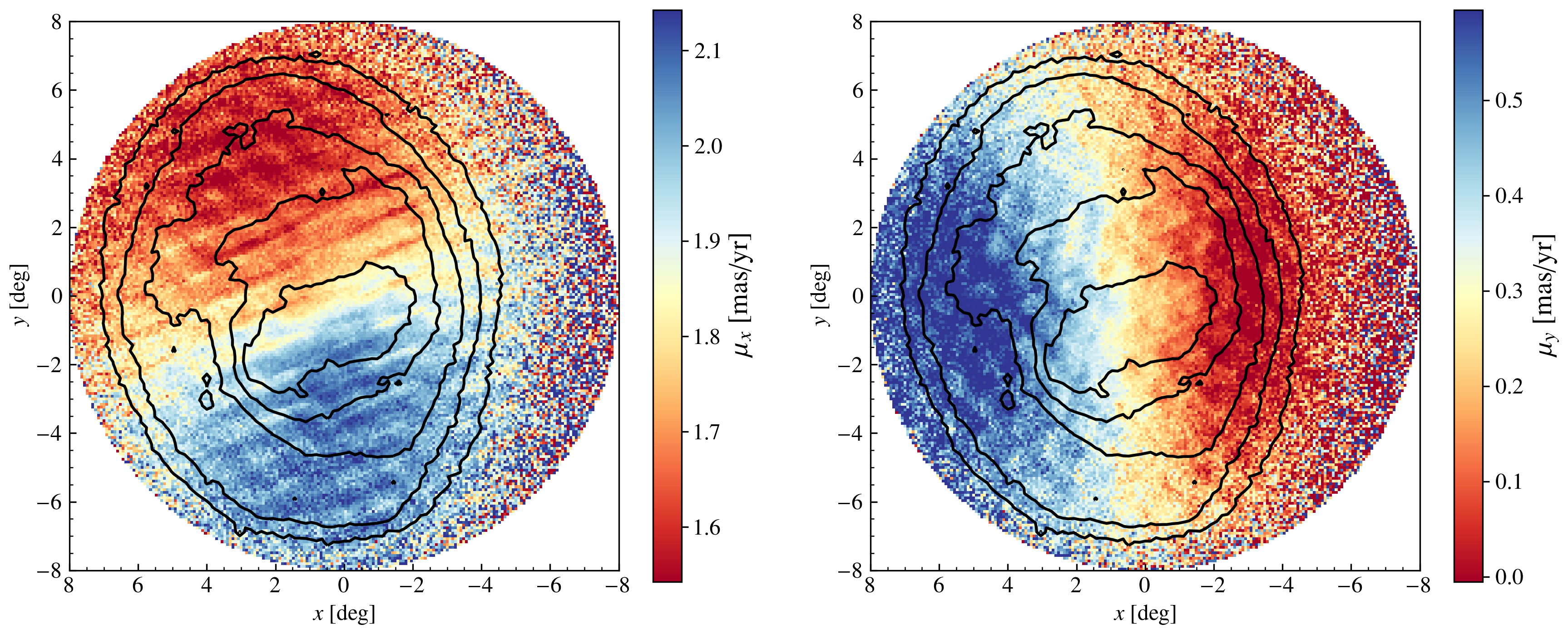}
    \caption{An illustration of the ability to measure transverse motions of individual stars in nearby galaxies. The maps show the median \textit{Gaia} DR2 proper motion of LMC stars in each pixel along the $x$ (left; $\mu_x$) and $y$ (right; $\mu_y$) axes of a projected coordinate system centered on the LMC. Black contours mark logarithmically spaced levels of stellar density. The gradient in $\mu_x$ along $y$ and in $\mu_y$ along $x$ is the signature of the rotating LMC disk. Figure adapted from \citet{Gaia-Helmi:2018}.}
    \label{fig:LMC_Rotation}
\end{figure}

\subsection{Internal Kinematics of Galaxies}
An exciting application enabled by the accumulation of large high-quality astrometric datasets is to use proper motions to study the internal kinematics of nearby galaxies, as illustrated in Figure~\ref{fig:LMC_Rotation}. Thanks to the combination of multiple \hubble\ astrometric fields, or to the large-area measurements made by \textit{Gaia}, clean proper motion-based rotation curves have been routinely detected for the LMC \citep[e.g.,][]{Piatek:2008,Kallivayalil:2013,VanderMarel:2014,VanderMarel:2016,Gaia-Helmi:2018}, as well as for M31 and M33 \citep{VanderMarel:2019}. A consistent result of these studies is the general agreement between proper motion-based rotation velocities and line-of-sight velocities. However, having access to the tangential rotation curve in addition to the radial velocity measurements allows a more complete characterization of the 3D kinematics of the galaxies, resulting in better constraints on the geometric model of the disk \citep[e.g., inclination;][]{VanderMarel:2014}. Interestingly, the impact of the central bar on the kinematics of the LMC disk can also be discerned from proper motion data \citep{Gaia-Helmi:2018}, providing information about the orbital dynamics in this type of structure.

Similarly, extensive datasets have also been analyzed for the SMC. However, the velocity field of this galaxy is more kinematically complex. Little to no coherent stellar rotation is detected in this galaxy \citep{Piatek:2008,Kallivayalil:2013,VanderMarel:2016,Gaia-Helmi:2018,Niederhofer:2018,Zivick:2018}, in accordance with radial velocity surveys \citep[e.g.,][]{Harris:2006}. On the other hand, large systemic motions are detected in different regions of the galaxy and have been plausibly linked to tidally induced disturbances from the neighboring LMC and MW \citep{Gaia-Helmi:2018,Niederhofer:2018,Zivick:2018}.  It is plausible that this complexity is linked to the hypothesis that the SMC may be two distinct systems \citep{Murray:2024a}.

There is also promising work on deriving the internal 3D kinematics in even smaller galaxies. This was pioneered, using a combination of \hubble\ and \textit{Gaia} data, in the Sculptor dSph \citep{Massari:2018b} and later in the Draco dSph \citep{Massari:2020}. A notable aspect of these studies was the use of the 3D kinematics of member stars to derive the first direct measurement of orbital anisotropy in an external galaxy. The value of this quantity has important implications for the derivation of dynamical mass profiles. In spite of the large uncertainties, due to the limited number of stars with suitable data, both Sculptor and Draco appear to have a meaningful degree of radial anisotropy in their kinematics \citep[e.g.,][]{Massari:2020, Vitral:2024, Vitral:2026}.

A later study by \citet{Martinez-Garcia:2021} expanded this approach to six MW dSphs. The authors discuss evidence for rotation in three galaxies (Carina, Fornax, and Sculptor) and conclude that, with the exception of Carina, all galaxies in the sample are predominantly pressure supported. The study also explored potential trends between the degree of rotational support and other properties (ellipticity, galactocentric distance), finding only low-significance correlations. Further work on the orbital anisotropy of MW dSphs was carried out by \citet{delPino:2022} using a combination of \textit{Gaia} and \hubble\ data, and a slightly longer baseline compared to \citet{Massari:2018b,Massari:2020}. While the study analyzed four galaxies (Draco, Sculptor, Sextans, and Fornax), the proper motion data only allow a robust anisotropy measurement for Sculptor and Draco. Both systems show radially anisotropic kinematics. Finally, the work of \citet{Vitral:2024} and \citet{Vitral:2026} leveraged long time-baseline \hubble\ imaging (18 and 20~yr, respectively) and radial velocity data to measure the spatially resolved 3D velocity dispersion profiles of Draco and Sculptor. They applied Jeans modeling to the kinematic data to constrain the dynamical mass profile of these galaxies. While in the case of Sculptor, uncertainties in the inclination allow for a large family of dynamical models, the data for Draco robustly support radial anisotropy, a cuspy dark-matter profile, and a modest degree of rotation.

\subsection{The M31-MW Merger}
Following the recognition of M31 as an external galaxy of similar nature to the MW, its negative line-of-sight velocity \citep[$V_{los} \approx -300$ km/s,][]{Slipher:1913,VandenBergh:1999b} revealed that the two galaxies are currently moving towards each other. However, it is only through detailed knowledge of M31's 3D velocity vector that an accurate prediction of the MW-M31 interaction can be made. While models of the MW-M31 orbits and interaction have long been a subject of interest \citep[e.g.,][]{Innanen:1977,Dubinski:1996,Cox:2008}, a very limited knowledge of M31's transverse motion allowed only for relatively uncertain predictions or required strong assumptions about the orbital motion of M31.

The measurement of M31's proper motion by \citet{Sohn:2012} provided the first direct estimate of M31's transverse velocity \citep{VanderMarel:2012a} and enabled a more robust forecast for the future of the MW-M31 system. Using a combination of $N$-body models and semi-analytic orbit calculations, \citet{VanderMarel:2012b} modeled a future interaction of the MW-M31-M33 system, demonstrating that the MW-M31 pair is on a highly radial orbit. In their models, M31 reaches a first close pericenter with the MW in $3.87^{+0.42}_{-0.32}$ Gyr and ultimately merges with the MW in $5.86^{+1.61}_{-0.72}$ Gyr. They also estimate a 41\% probability that the first MW-M31 pericenter will result in a direct collision between the two galaxies. The results from their $N$-body simulations show that the expected merger remnant will have typical structural characteristics of an elliptical galaxy.

Using \textit{Gaia} DR2 measurements, combined with earlier \hubble\ data, \citet{VanderMarel:2019} updated the proper motion of M31, resulting in a modest increase in M31's tangential motion. Incorporating these new measurements in their orbital modeling shifts the epoch of the first pericenter relative to the MW back to $t=4.5$ Gyr and substantially increases the pericentric distance to $\sim 130$ kpc, compared to the $\sim 31$ kpc of the \citet{VanderMarel:2012b} model. The larger pericenter drastically decreases the probability that the MW and M31 are on a direct collisional trajectory. The authors also note that including the gravitational influence of M33 and the LMC further delays the MW-M31 encounter by $\sim 1$ Gyr but reduces the pericentric distance to $\sim 75$ kpc.

More recently, \citet{Sawala:2025} presented an alternative orbital solution, which was calculated by modeling the MW-M31-M33-LMC system and included updated \textit{Gaia} EDR3 proper motions for M31 \citep{Salomon:2021}. Notably, the combined virial mass of the MW and M31 was set at $2.3 \times 10^{12} M_{\odot}$, compared to the value of $3 \times 10^{12} M_{\odot}$ used in \citet{VanderMarel:2012b} and \citet{VanderMarel:2019}. From their simulation suite, they estimated only a $\sim 50$\% probability that the MW and M31 will merge within the next 10~Gyr, with the median merging epoch of those outcomes being $t=7.6$~Gyr. In the remaining $\sim50$\% of their simulations, the MW and M31 attain pericentric distances $\gtrsim200$~kpc and continue to evolve in relative isolation for $>10$~Gyr.

\subsection{The Importance of the Gravitational Potential}
As a final remark, we highlight an important issue that should be kept in mind when comparing different orbital determination studies. Calculating the orbit of a dwarf galaxy around its host (the MW or M31 for all current studies) requires knowledge of two ingredients: the 6D vector (position and velocity) of the dwarf (and of other potentially relevant galaxies) with respect to the host's center of mass, and the gravitational potential within which the dwarf is moving. Early determinations of the 6D coordinates were generally coarse due to poor proper motion precision and accuracy and were ultimately the limiting factor in determining accurate orbital histories \citep[e.g.,][]{Ibata:1997,Piatek:2002,Dinescu:2004}. Even today, proper motion errors for many MW satellites remain a meaningful contributor to the uncertainty in their orbital histories. However, as our proper motion measurements have improved (and will keep improving due to, e.g., \textit{Gaia} future data releases), our knowledge of other factors such as the distance and gravitational potential has become the major source of uncertainties for the orbital histories of at least most MW satellites \citep[e.g.,][]{Dsouza:2022,Pace:2022}.

For the MW and M31, the simplest gravitational potential that is usually assumed consists of a static model composed of a bulge, one or more stellar disks, and a dark matter halo, while all other galaxies are assumed to be massless. Even in such a simple model, several parameters need to be assumed with regard to the mass, the symmetry, and the density profile of the various mass components. In particular, the total mass of the halo plays a critical role. Under certain assumptions (e.g., bound satellites, dynamical equilibrium) the kinematics of the satellites themselves can be used to constrain the total mass of the host \citep[e.g.,][]{Piatek:2002,Dinescu:2004,Watkins:2010,Sohn:2013,Li:2017,Patel:2017b,Patel:2017a,Gaia-Helmi:2018,Patel:2018,Callingham:2019}. However, if the focus is an agnostic reconstruction of orbital histories (without assumptions on, e.g., the bound status), independent knowledge of the potential is required.

\begin{figure}[pos=tbp]
    \centering
    \includegraphics[width=1.\linewidth]{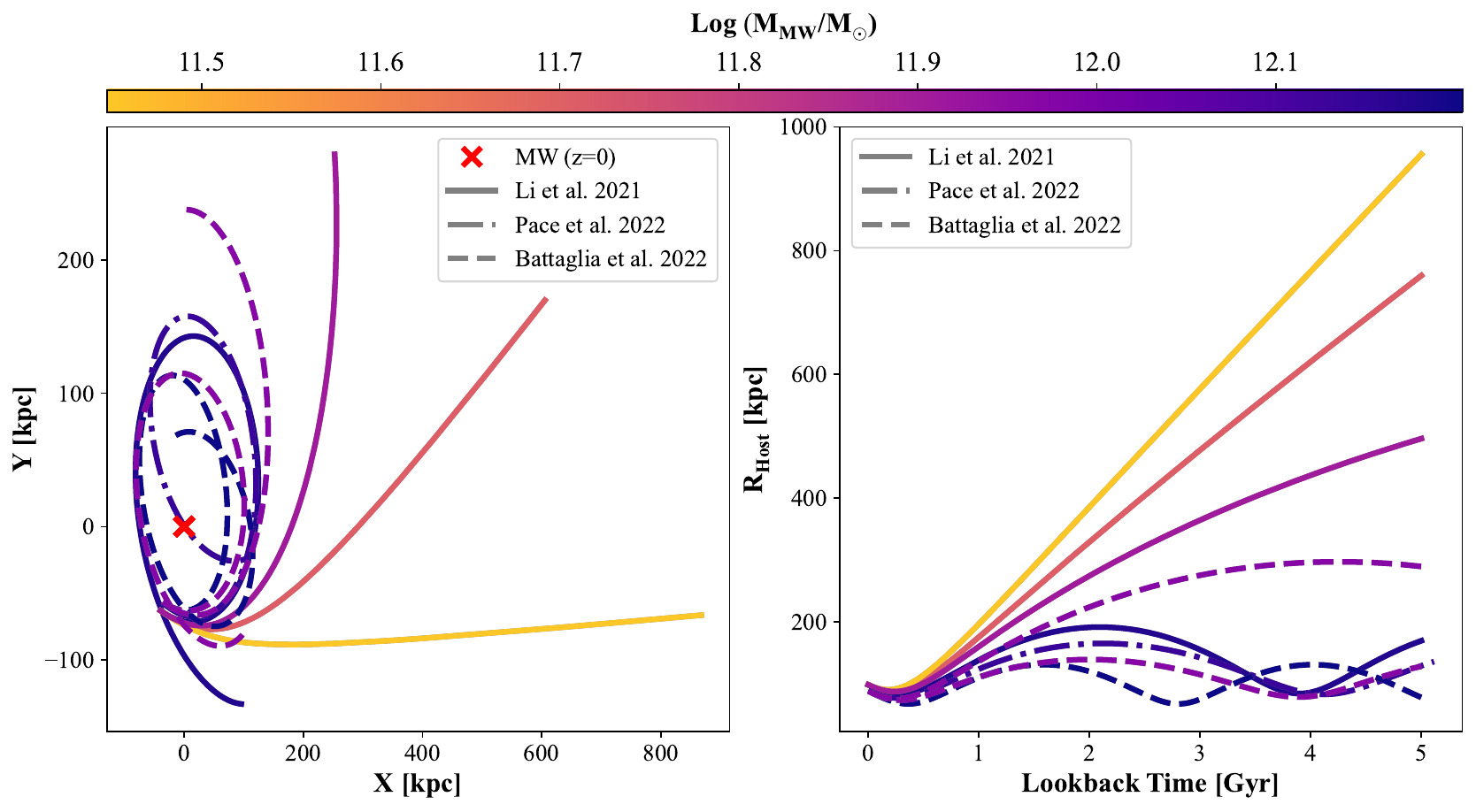}
    \caption{A compilation of published orbits for the Sextans dSph galaxy, extending to 5 Gyr ago. All orbital models are based on \textit{Gaia} EDR3 proper motions. Shown are orbits projected on the plane of the MW disk (left panel), and the MW galactocentric distance as a function of time (right).  Solid lines mark orbits from \citet{Li:2021}, dashed lines mark orbits from \citet{Battaglia:2022}, and dotted-dashed lines show orbits from \citet{Pace:2022}. The color-coding represents the total MW mass used to reconstruct the orbit. The red X indicates the present-day position of the MW.}
    \label{fig:Orbit_Sextans}
\end{figure}

For the MW, there are several lines of evidence that point to a dark matter halo with virial mass between $0.8 \times 10^{12}$ \msun\ and $2 \times 10^{12}$ \msun\ \citep[e.g.,][]{Smith:2007,Li:2008,Busha:2011,Deason:2012,Piffl:2014,Bowden:2015,Penarrubia:2016,McMillan:2017,Posti:2019,Watkins:2019,Cautun:2020,Fritz:2020, Hayati:2024}. However, current uncertainties in the MW mass can still have a meaningful impact on estimated satellite orbits (e.g., see Figure~\ref{fig:Orbit_Sextans} for a comparison of different published orbits for the Sextans dSph). In particular, changing the mass of the MW has a strong effect on the orbital apocenters, the orbital periods, and the estimated infall times \citep[e.g.,][]{Besla:2007,Sohn:2013,Gaia-Helmi:2018,Fritz:2018, Patel:2020}.

\begin{figure}[pos=tbp]
    \centering
    \includegraphics[width=1.\linewidth]{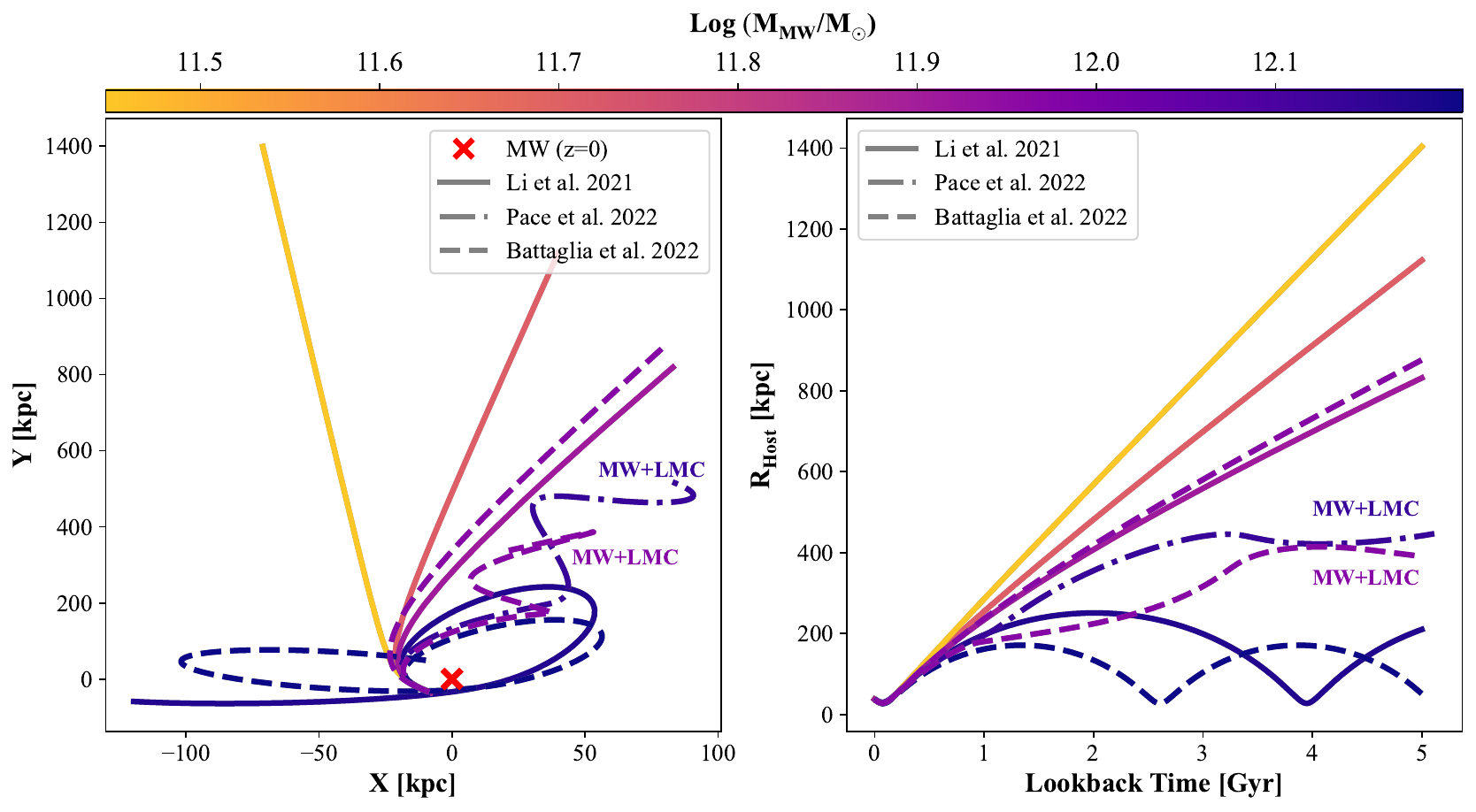}
    \caption{Same as Figure~\ref{fig:Orbit_Sextans}, but for the Carina II UFD. We highlight the two orbital solutions that include the gravitational effect of the LMC.}
    \label{fig:Orbit_Carina}
\end{figure}

A further complication is that some galaxies around the host are massive enough to have a meaningful impact on the total gravitational potential. In particular, it has been shown that the inclusion of the LMC's gravitational potential has a substantial impact on the inferred orbits of other MW satellites (e.g., see Figure~\ref{fig:Orbit_Carina} for the case of Carina~II), regardless of their position in the MW halo \citep[e.g.,][]{Erkal:2020a,Patel:2020,Battaglia:2022,Pace:2022}. In the case of M31, the effect of M33 is likely to be of similar amplitude \citep[e.g.,][]{Bennet:2024,Patel:2025,Bennet:2025}. A further complication is that the gravitational influence of these large satellites can affect dynamical mass estimates of the MW and M31 \citep[e.g.,][]{Erkal:2020b,Kravtsov:2024}. While the effect of galaxies such as the LMC and M33 can be included in modern orbital reconstruction, the precise mass assumed is going to impact the recovered orbital properties of the satellites \citep[e.g.,][]{Patel:2020,Patel:2025}.  \citet{Martinez-Garcia:2026a} revisited the orbits of the full MW satellite population using updated \hubble\ and \textit{Gaia} proper motions in a potential that includes the LMC.  For each galaxy, they quantified how the inferred pericenters, infall times, and LMC associations change relative to MW-only models. While significant progress has been made to improve the estimates\footnote{Notably, in the last 20 years, the consensus estimate for the total mass of the LMC has been revised upwards by roughly an order of magnitude.}, there remain uncertainties of order unity for the halo masses of these large satellites \citep[e.g.,][]{Corbelli:2014,VanderMarel:2014,Penarrubia:2016,Fune:2017,Kam:2017,Laporte:2018,Shao:2018,Garavito-Camargo:2019,Erkal:2019,Erkal:2020a,Koposov:2023}. The effect of lower-mass galaxies like the SMC has also been studied and, while they appear to be of less importance, they might still be relevant, for example, for satellites that passed sufficiently close to the Magellanic Clouds \citep{Patel:2020}.

Lastly, we note that it is unlikely that the potential of the host galaxy is static in time. For instance, several studies have shown that the gravitational influence of the LMC has an effect on the MW dark matter halo, creating density perturbations, inducing a reflex motion in the MW center of mass, and leaving an imprint in the observed kinematics of the satellites \citep[e.g.,][]{Gomez:2015,Garavito-Camargo:2019,Garavito-Camargo:2021,Cunningham:2020,Vasiliev:2021}. This effect is sometimes, but not always, included in orbital calculations \citep[e.g.,][]{Battaglia:2022}.

In addition to the influence of the nearby galaxies, hierarchical structure assembly implies that the host halo mass itself has been evolving over time. The growth of dark matter halos is seldom included in orbital calculations, as that would require detailed knowledge or strong assumptions regarding the mass accretion history of the galaxy. However, the impact of the host's mass growth on orbit integration is going to depend on the considered lookback time. According to cosmological models, it is likely that most of the mass, in galaxies like the MW or M31, has been in place for the last few Gyr \citep[e.g.,][]{Wechsler:2002, Zhao:2009, Fakhouri:2010, Wechsler:2018, Santistevan:2020}. This means that orbits in the last 3--5 Gyr should be relatively unaffected \citep[e.g.,][]{Armstrong:2021}. As orbits are integrated further back in time, the halo growth becomes increasingly relevant. For even earlier lookback times, the effect of past major accretions \citep[e.g., the Gaia-Sausage-Enceladus, GSE;][]{Belokurov:2018, Helmi:2018} needs to be taken into account. Overall, the non-static nature of the host's halo mass introduces a further degree of uncertainty in orbital calculations at lookback times $\gtrsim5$~Gyr ago \citep[e.g.,][]{Santistevan:2023}.

The important lesson from this section is that numerous choices can be made regarding the potential in which orbits are integrated. Orbital calculations in the literature span a wide range of potential models (e.g., see Table~\ref{tab:PM}), which should be kept in mind as a caveat when comparing different studies, especially older analyses with more recent ones. More broadly, this topic can at least partially explain the range of conclusions found in the literature on specific questions, even when using comparable proper motion measurements.

\section{Future Outlooks}
\label{sec:future}

The last $\sim20$ years have been a golden era for stellar populations studies in resolved galaxies.  We expect the next decades to be equally, if not more, fruitful.  The wealth of deep, wide-field, high-angular resolution imaging surveys (e.g., with Rubin, \textit{Roman}, \textit{Euclid} ) has the potential to resolve stars in thousands of known local galaxies and discover a comparable number of new systems.
  Deep targeted imaging (e.g., with \jwst, ELT/MICADO, \textit{HWO}) can extend resolved stellar populations studies to greater distances, fainter stars, and new environments with unprecedented astrometric precision enabling fundamentally new science.

In the following sections, we briefly discuss promising areas for growth in resolved star science.  We begin with the two outstanding observational challenges that deserve careful consideration when planning future resolved star science, crowding (\S \ref{sec:crowding_future}) and contamination (\S \ref{sec:sg_future}).  We then summarize the capabilities of major facilities expected to be online in the 2030s in \S \ref{sec:facilities}.  With the main observational challenges and capabilities defined, we then discuss areas for scientific growth, with a primary focus on topics covered in the review.

 \begin{table*}[]
    \centering
    \begin{threeparttable}
    \caption{Major current and planned facilities that are well-suited for imaging studies of resolved stars in nearby galaxies divided into space- and ground-based categories.}
    \begin{tabular}{lcccccc}
    \toprule
    Facility & Instrument & Wavelength Range & Field of View & PSF FWHM & Plate Scale  \\
             &          & ($\mu m$)  & (deg$^2$)&($^{\prime\prime}$) & ($''/\text{pixel}$)  \\
    \toprule
    \multicolumn{6}{c}{\textbf{Space-Based Facilities}} \\
    \toprule
    \hubble\tnote{a} & WFC3 / UVIS & $0.2 - 1.0$ & $0.0020$ & $0.067 - 0.080$ & $0.04$ \\
 & WFC3 / IR & $0.8 - 1.7$ & $0.0013$ & $0.12 - 0.16$ & $0.13$ \\
 & ACS / WFC & $0.35 - 1.1$ & $0.0031$ & $0.07 - 0.14$ & $0.05$ \\
    \toprule
    \jwst\tnote{b} & NIRCam (SW) & $0.6 - 2.3$ & $0.0027$ & $0.031 - 0.066$ & $0.031$ \\
 & NIRCam (LW) & $2.4 - 5.0$ & $0.0027$ & $0.085 - 0.160$ & $0.063$ \\
 & NIRISS & $0.8 - 5.0$ & $0.0013$ & $0.04 - 0.16$ & $0.066$ \\
 \toprule
 Euclid\tnote{c} & VIS & $0.55 - 0.90$ & $0.57$ & $\sim0.13 - 0.16$ & $0.10$ \\
 & NISP & $0.92 - 2.02$ & $0.57$ & $0.30 - 0.50$ & $0.30$ \\
 \toprule
 Roman\tnote{d} & WFI  & $0.48 - 2.3$ & $0.281$ & $\sim0.06 - 0.17$ & $0.11$ \\
 \toprule
 CSST\tnote{e} & MSC & $0.25 - 1.0$ & $1.1$ & $\sim 0.15$ & $0.073$ \\
  & MCI & $0.25 - 1.0$ & $0.015625$ & $\le 0.18$ & $0.05$ \\
 \toprule
 UVEX\tnote{f} & FUV Imager & $0.139 - 0.190$ & $12.25$ & $2.25$ & $1.03$ \\
 & NUV Imager & $0.203 - 0.270$ & $12.25$ & $2.25$ & $1.03$ \\
\toprule
 CASTOR\tnote{g} & Imager (UV) & $0.15 - 0.30$ & $0.25$ & $\sim 0.15$ & $0.10$ \\
 & Imager ($u'$) & $0.30 - 0.40$ & $0.25$ & $\sim 0.15$ & $0.10$ \\
 & Imager ($g$) & $0.40 - 0.55$ & $0.25$ & $\sim 0.15$ & $0.10$ \\
\toprule
Lazuli\tnote{h} & WCC & $0.4 - 1.7$ & $0.117$ & $0.033 - 0.14$ & $0.017 - 0.021$ \\
\toprule
HWO\tnote{i} & Wide-Field Imager & $0.2 - 2.5$ & $0.0017$ & $0.007 - 0.07$ & $\sim 0.016$ \\
    \toprule
    \multicolumn{6}{c}{\textbf{Ground-Based Facilities}} \\
    \toprule
 Subaru\tnote{j} & HSC  & $0.38 - 1.1$ & $1.77$ & $0.60 - 0.80$ & $0.168$ \\
 \toprule
 LBT\tnote{k} & SHARK-VIS & $0.4 - 1.0$ & $\sim 0.000008$ & $0.015 - 0.025$ & $0.0065$ \\
 & SHARK-NIR & $0.96 - 1.7$ & $0.000025$ & $0.025 - 0.045$ & $0.0145$ \\
 \toprule
 Gemini South\tnote{l} & GSAOI & $0.9 - 2.4$ & $0.0005$ & $0.04 - 0.08$ & $0.02$ \\
 \toprule
 Rubin\tnote{m} & LSST Camera & $0.32 - 1.05$ & $9.62$ & $0.70$ (median) & $0.20$ \\
 \toprule
 VLT\tnote{n} & MAVIS & $0.37 - 0.95$ & $0.00007$ & $0.018 - 0.026$ & $0.0074$ \\
 \toprule
 ELT\tnote{o} & MICADO (LRI) & $0.8 - 2.45$ & $0.0002$ & $0.006 - 0.010$ & $0.004$ \\
& MICADO (HRI) & $0.8 - 2.45$ & $0.00003$ & $0.006- 0.010$ & $0.0015$ \\
\toprule
GMT\tnote{p} & GMTIFS (Imager) & $0.9 - 2.5$ & $0.000032$ & $0.010 - 0.015$ & $0.005$ \\
    \toprule
TMT\tnote{q} & IRIS (Imager) & $0.84 - 2.4$ & $0.000089$ & $0.007 - 0.015$ & $0.004$ \\
\toprule
    \end{tabular}
\begin{tablenotes}[para]\footnotesize
        \item[a]  \citet {Dressel:2012, Avila:2018}
        \item[b] \citet{Rieke:2005, Doyon:2012}
        \item[c] \citet{Laureijs:2011, Cropper:2018, Maciaszek:2022, Mellier:2025}
        \item[d] \citet{Spergel:2015, Akeson:2019}
        \item[e] \citet{Zhan:2011, Zhan:2021}
        \item[f] \citet{Kulkarni:2021} and Harrison et al.\ (submitted); PSF is the mission image-quality requirement (half-power diameter $\leq2.25''$), common to both imagers.
        \item[g] \citet{Cote:2025}
        \item[h] \citet{Roy:2026}
        \item[i] \citet{NationalAcademies:2021}
        \item[j] \citet{Miyazaki:2018}
        \item[k] \citet{Farinato:2014, Pedichini:2017}
        \item[l] \citet{McGregor:2004, Carrasco:2012}
        \item[m] \citet{Ivezic:2019}
        \item[n] \citet{Rigaut:2020}
        \item[o] \citet{Davies:2016, Davies:2018}
        \item[p] \citet{McGregor:2012}
        \item[q] \citet{Larkin:2014}
    \end{tablenotes}
    \label{tab:facilities}
    \end{threeparttable}
\end{table*}

\subsection{The Importance of Crowding} \label{sec:crowding_future}

As discussed in \S \ref{sec:challenges}, a main challenge to resolved galaxy studies is the effect of crowding, which can limit the depth, precision, and accuracy of resolved star observations.  In this section,  we consider how crowding may affect future studies of resolved galaxies drawing from current knowledge in the literature.

\begin{figure}[pos=tbp]
    \centering
    \includegraphics[scale=1.5]{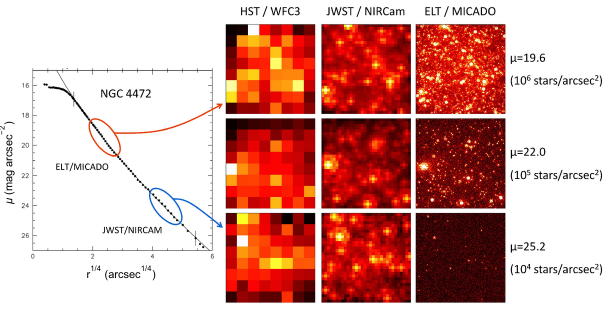}
    \caption{An illustration of the impact of angular resolution on crowding, adapted from the MICADO instrument paper of \citet{Davies:2018}.  \textbf{Left:} The surface brightness profile of Virgo cluster galaxy NGC~4472.  \textbf{Right:} The results of stellar population simulations showing the impact of crowding at various surface brightnesses.  The stellar densities indicated are for stars as faint as 0.1 \msun. MICADO is expected to resolve stars down the RGB at all surface brightnesses indicated, whereas \jwst\ will only do so in the lower surface brightness regime and will be limited to the very brightest stars at the highest surface brightnesses (e.g., AGB stars).  \hubble/WFC3 IR, and facilities with similar angular resolution (e.g., \textit{Roman}) are not expected to resolve individual stars in this type of system.\label{fig:ELT_Mosaic_Images}}

\end{figure}

Figure~\ref{fig:ELT_Mosaic_Images} qualitatively illustrates the effects of crowding on a simulated image of a Virgo cluster elliptical galaxy,  NGC~4472 (M49).  Select simulated fields at different surface brightnesses are shown for \hubble/WFC3/IR, \jwst/NIRCam, and ELT/MICADO.  The effects of crowding are visually apparent.  Compared to MICADO, \jwst will resolve only the brightest stars in the highest surface brightness fields, and WFC3/IR can resolve almost none.  The resolving power for many space-based wide-field facilities will be comparable to WFC3/IR.

Unprecedented insights into the effects of crowding can be gleaned from the PHAT survey.  It provides a large number of stars, a wide range of stellar densities and wavelengths, and uses multiple cameras with different PSF characteristics and plate scales.  As discussed at length in several PHAT and PHAST papers \citep[e.g.,][]{Dalcanton:2012b, Rosenfield:2012, Williams:2014, Chen:2025}, across the disk of M31, only the UV observations (F275W, F336W) were essentially unaffected by crowding (i.e., the limiting magnitude was set by photon counts) owing the lower density of UV bright sources and \hubble's improved resolving power at bluer wavelengths. The UV was found to be crowding-limited only in the bulge of M31 \citep{Rosenfield:2012}. In the optical (F475W, F814W), the resulting CMD depths changed by $\sim5$ magnitudes across the survey area and become photon-limited at $\sim6$ stars arcsec$^{-2}$.  The IR (F110W, F160W) was even more affected by crowding with CMD depths changing by $\sim8$~mag across the survey area.  WFC3/IR never reached the theoretical photon-counting limit even at stellar densities as low as $\sim2$ stars arcsec$^{-2}$.  The larger impact of crowding at redder wavelengths is due to an increasingly broad PSF and the higher densities of RGB stars.  One mitigation strategy that emerged from the PHAT survey was to use the higher angular resolution optical imaging to ``de-blend'' the lower resolution IR imaging. This resulted in a $\sim1$--$1.5$~magnitude gain in depth in the IR CMDs \citep[e.g.,][]{Williams:2014}.  We stress that, to achieve this type of gain, the actual method of performing the PSF photometry matters.  Codes that perform source finding and photometry simultaneously for images/PSFs of different resolutions can realize this type of depth increase due to an effective `deblending'.  However, codes that operate on images independently and attempt source matching afterward will not see improvement.

\begin{figure}[pos=tbp]
    \centering
\includegraphics[width=1.0\linewidth]{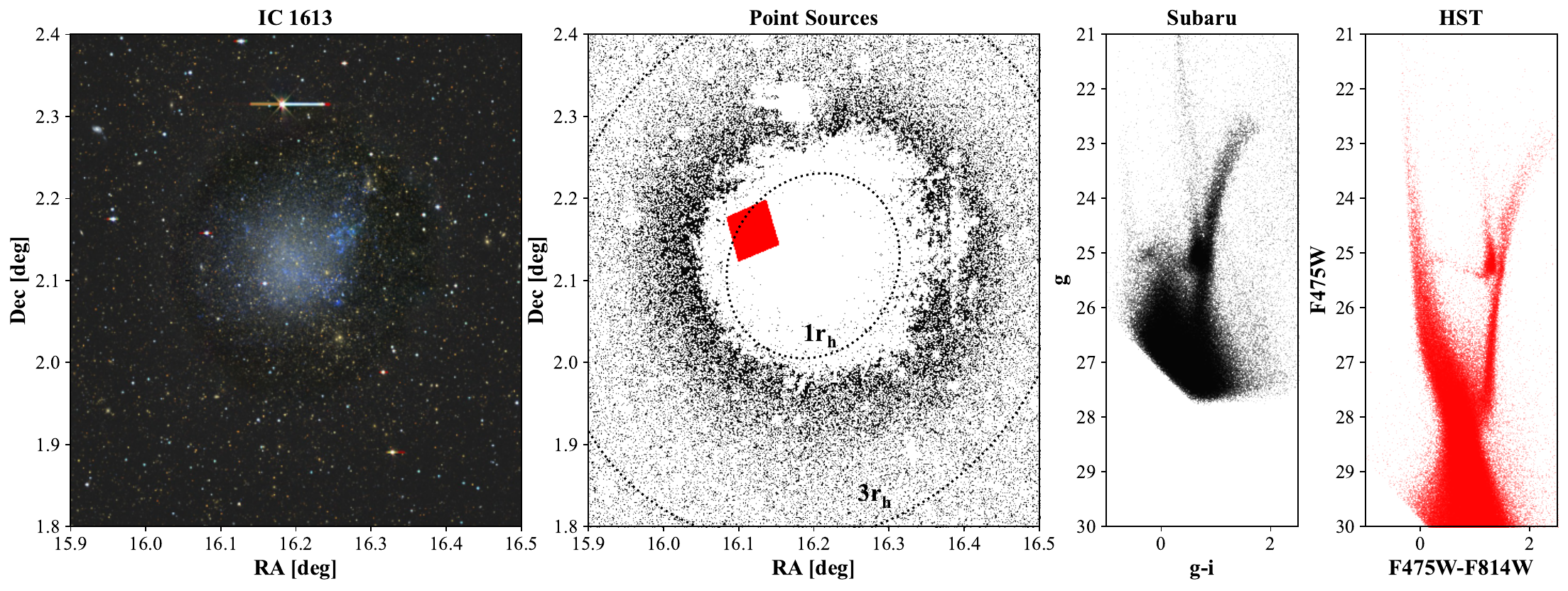}
    \caption{An illustration of crowding in the LSST-era. \textbf{Left:} Color image of the LG dIrr galaxy IC 1613, from the Dark Energy Spectroscopic Instrument (DESI) Legacy Imaging Surveys. Image credit: Legacy Surveys / D.~Lang (Perimeter Institute). \textbf{Center:} Detected point sources, over the same sky area from the Subaru/HSC catalog of \citet{Pucha:2019} (black) and the \hubble/ACS catalog of \citet{Skillman:2014} (red). \textbf{Right:} Optical CMD derived from the Subaru (black) and \hubble\ imaging (red), respectively.  Although the Subaru/HSC extends to $\sim6~r_h$ and reaches below the HB, it is entirely limited by crowding inside $\sim1.5~r_h$ ($r_h=6.8^\prime$; \citealt{McConnachie:2012}).  LSST is expected to provide data similar in quality to HSC.
    \label{fig:hsc_ic1613}}

\end{figure}

One illustrative example of expected crowding characteristics for 8m-class ground-based facilities comes from Subaru/HSC imaging of LG isolated dwarf galaxy IC~1613.  Figure~\ref{fig:hsc_ic1613} shows the CMD and spatial distribution of HSC sources in IC~1613 from \citet{Pucha:2019}, which was reduced using PSF photometry in a version of the LSST pipeline.  Due to crowding, virtually no sources passed the photometric quality cuts in the inner $\sim1.5 \, r_h$.  In comparison, a single \hubble\ observation within the central $\sim1.5 \, r_h$ can reach well-below the oMSTO.  This comparison highlights the strengths and limitations of our most powerful ground-based, wide-area imaging facilities, including LSST, which is expected to have similar performance to HSC.  Even where the HSC data formally reach below the HB, the HB is contaminated by blends of fainter MS stars that crowding scatters to brighter magnitudes (Figure~\ref{fig:hsc_ic1613}).  This effect is rarely captured by exposure time calculators (ETCs) and generally requires ASTs to quantify, though \citet{Savino:2024} provide guidelines for approximating the effects of crowding in the \jwst/NIRCam ETC.

Because surface brightness, and hence crowding, can vary widely across a typical nearby galaxy, it will be important to coordinate resolved star observations using multiple facilities.  Resolving stars in higher surface brightness central regions is best served with higher-angular resolution facilities (e.g., \hubble, \jwst, MICADO), while resolving the lower-density outskirts (e.g., stellar halos) can be the focus of wide-area space-based facilities (e.g., \textit{Roman}, \textit{Euclid}, \textit{CSST}) throughout the LV, and even by ground-based facilities (e.g., Rubin, HSC) for galaxies within a few Mpc.

\subsection{The Importance of Contamination Removal} \label{sec:sg_future}

As highlighted in \S \ref{sec:challenges}, the ability to separate member stars from contaminants using imaging is certain to improve in the next decade.  The four considerations for improved contamination removal are imaging depth, angular resolution, number of bands available, and, in some cases, proper motion determinations.  As illustrated in Figure~\ref{fig:sg_separation}, improved SNR and angular resolution can each provide for better constraints on object shapes, which can lead to better morphological discrimination between stars and galaxies \citep[e.g.,][]{Slater:2020}.  Multi-color data can be used, for example, to define stellar sequences in color-color space \citep[e.g.,][]{Smercina:2018}.  For the nearest systems, proper motions may be able to remove MW foreground stars as sources of contamination \citep[e.g.,][]{Momany:2014,Sanderson:2026}.

The most immediate improvements from next-generation surveys will be for the study of objects closest to the MW.  The high SNRs provided by the full stack of LSST data should effectively remove many contaminants relative to current surveys.   The anticipated depth and angular resolution of wide-field space-based facilities should extend this to even fainter magnitudes.  The separation of MW foreground stars from member stars of MW satellites at fainter magnitudes (e.g., $m>22$) will remain moderately challenging, although star-by-star proper motions, e.g., with wide-area space-based imaging, may be one path forward \citep[e.g.,][]{Han:2023, Sanderson:2026}.  Studies that require extremely faint depths, e.g., the low-mass IMF, will either require \jwst\ or similarly high-angular resolution imaging (e.g., MICADO) or possibly star-by-star proper motion decontamination (e.g., with \textit{Roman}).

Improvements in star-galaxy separation in more distant systems are mixed.  Pathfinder work with GHOSTS \citep[e.g.,][]{Radburn-Smith:2011} and more recent studies that combine HSC and \hubble\ imaging in the M81 Group and around M83 \citep[e.g.,][]{Smercina:2020, Bell_e:2022, Bell:2026} have previewed what is possible.  One challenge in this regime is depth.  For example, the TRGB in an M81 Group galaxy is located at $m_I \sim 24$, and typically depths needed to robustly identify the RGB are $\sim2$~magnitudes fainter.  Ground-based facilities (e.g., Rubin, HSC) will formally reach these depths, but at modest to low SNRs and with marginal effective angular resolutions, which raise concerns about crowding and photometric upscatter (e.g., Figure~\ref{fig:hsc_ic1613}).  HSC-based studies of the M81 Group and M83 referenced above rely on using higher-angular resolution \hubble\ imaging to inform star-galaxy classification in the ground-based data.  Wide-field space-based facilities (e.g., \textit{Roman}, \textit{Euclid}) will represent a substantial step forward over ground-based imaging at these distances, since their factor-of-several gain in angular resolution alleviates the morphological confusion between compact background galaxies and stars that dominates the ground-based systematics.  Their performance will still fall short of \hubble, as illustrated by GHOSTS \citep[e.g.,][]{Radburn-Smith:2011} and recent work in the M81 Group \citep[e.g.,][]{Bell_e:2022}.  At issue is that at bluer wavelengths ($\lesssim 1\,\mu$m), which provide stronger star-galaxy discriminatory power, \textit{Roman}'s PSF, particularly at blue wavelengths, is severely under-sampled (Table~\ref{tab:facilities}). A large number of dithers are needed to overcome this limitation if the preferred bluer filters are used, driving up the exposure times.  At redder wavelengths ($\gtrsim 1\,\mu$m), \textit{Roman}'s angular resolution and PSF characteristics (i.e., light concentrated in a small number of central pixels) are similar to those of \hubble/WFC3/IR. They represent a vast improvement over ground-based imaging, but remain far less powerful than \hubble/ACS.  The work on the low-mass IMF with Coma~Berenices by \citet{Gennaro:2018b} illustrates the strengths and limitations of star-galaxy separation with \hubble/WFC3/IR.  

\textit{Euclid}'s VIS filter will likely provide improved star-galaxy separation owing to its higher angular resolution than ground-based imaging, though its nominal survey depths restrict its main utility to galaxies within a few Mpc.  With its multiple bands and good angular resolution, the \textit{CSST} should provide useful star-galaxy separation as well, although this may ultimately depend on its PSF performance and adopted observing strategy (e.g., the lack of dithers may be a hindrance).  It is possible that the highest-fidelity star-galaxy separation at these distances will continue to rely on coordinated (or archival) use of \hubble\ and/or \jwst\ imaging, particularly for faint substructures and the outer extents of stellar halos.  Even so, \textit{Euclid} and, in particular, \textit{Roman} should be transformative for wide-area resolved population work.  More broadly, there is additional room for algorithmic progress on this problem.

\subsection{Summary of Important Major Facilities} \label{sec:facilities}

Here, we summarize the broad capabilities of facilities that we expect to make substantial contributions to resolved star science in the coming years and decades.   We list the broad characteristics of major facilities in Table~\ref{tab:facilities} and, for guidance, we also refer to the CMD sensitivity limits shown in Figure~\ref{fig:Telescopes}.

\subsubsection{Space-Based}

\paragraph[\textit{HST}]{\textit{HST}:}  For over 30 years, \hubble\ has been the workhorse of resolved stellar populations studies.  Following the introductions of \jwst\ and \textit{Euclid}, \hubble\ has already transitioned from the sole space-based workhorse for resolved stellar populations to one of a suite of facilities. This trend will continue with \textit{Roman}. One of \hubble's unique strengths is at blue-optical and UV wavelengths, which are absent from \jwst, \textit{Euclid}, and \textit{Roman}. The study of massive stars, recent star formation,  dust, and recent efforts to measure photometric metallicities at blue wavelengths are uniquely suited to \hubble.  Its astrometric capabilities are well-established and it remains an excellent choice for proper motion studies, particularly for galaxies with first epoch \hubble\ imaging.  The only potential replacements on the horizon for \hubble's sensitivity and angular resolution are MAVIS, \textit{HWO}, and potentially \textit{Lazuli}. It is imperative to keep \hubble\ operating as long as possible so as to not lose access to these observational capabilities.

\paragraph[\textit{JWST}]{\textit{JWST}:} \jwst\ is now the premier facility for resolved star studies.  Its angular resolution, sensitivity, wide array of filters, and moderate field-of-view are unmatched.  There are several science areas for which \jwst\ will be particularly transformative, including measuring the low-mass IMF, resolving the crowded regions of disks in nearby  $L_{\star}$ and giant elliptical galaxies ($D\lesssim10$~Mpc) to the depth of the RC, constructing CMDs that reach the oMSTO in galaxies at $D\sim2-3$~Mpc, studying luminous evolved red stars (e.g., RHeBs, AGBs) and measuring SFHs in galaxies out to 10s of Mpc and precise proper motions of galaxies out to $\sim5$~Mpc.   Obtaining first-epoch \jwst\ imaging for long-term proper motion studies is critical to maximizing its scientific return. A less appreciated feature of \jwst\ is that it is an excellent facility for star-galaxy separation owing to its exquisite angular resolution and broad near-IR wavelength coverage.  This feature, along with its precise astrometry, can be useful when removing contamination from the study of faint stars (e.g., for the low-mass IMF in MW satellites).

\jwst\ does have several limitations. Like \hubble, \jwst\ is not particularly efficient at mapping large areas.  Optimal use of \jwst\ may involve targeted areal coverage with \jwst\ (e.g., crowded, star-forming disks) coupled with larger area surveys in lower density regions (e.g., stellar halos) with \textit{Euclid}  or \textit{Roman}.  \jwst's PSF is highly under-sampled in its bluer filters (e.g., F070W), which can adversely affect photometry and astrometry without careful planning.  \jwst\ also saturates easily and care needs to be taken when bright stars are near or within the field of view.

Several recently selected \jwst\ Cycle 5 programs illustrate how the community is using these capabilities for the science discussed in this review.  They include deep NIRCam imaging to measure the ancient SFHs of isolated low-mass galaxies beyond $\sim1$~Mpc from any massive host (GO-9601, PI McQuinn), including the isolated UFD Sculptor~A (GO-10160, PI Hai); a multi-band NIRCam and Mid-Infrared Instrument (MIRI) map of $\sim40$ million stars in the center of M33 for IMF, SFH, and dust-production studies (GO-9622, PI Rosolowsky); the first resolved star chemodynamical study of a giant elliptical, in the halo of Cen~A (GO-10246, PI Sandford); and an archival program to measure proper motions of faint stars by aligning \jwst\ imaging to the \textit{Gaia} reference frame (AR-12274, PI McKinnon; \S \ref{sec:orbital_histories_future}).

   \paragraph[\textit{Wide-Field Optical/IR Facilities}]{\textit{Wide-Field Optical/IR Facilities}:}  \textit{Euclid}, launched in 2023, is the first of the new generation of wide-field space telescopes.  It is broadly similar to \textit{Roman} in its wide field-of-view and optical to near-IR wavelength coverage, albeit with lower angular resolution in the near-IR \citep{Laureijs:2011, Mellier:2025}.
A clear strength of \textit{Euclid}  is that, as part of its prime mission, it is acquiring a tremendous amount of data that can be used for resolved star studies of nearby galaxies. Its main survey covers $\sim14,000$ deg$^2$ to modest depths ($\sim26.2$ in its VIS filter and $\sim24.5$ in the Near Infrared Spectrometer and Photometer (NISP) Y, J, H filters), along with $\sim50$~deg$^2$ of deep fields. This survey is $\sim3\times$ larger than \textit{Roman}'s High-Latitude Wide-Area Survey (HLWAS), with comparable depths in the VIS filter, making them highly complementary. \textit{Euclid} 's near-IR depths are notably shallower than \textit{Roman} and \textit{Euclid}  will not excel in certain areas such as science cases that require faint and/or crowded regions or require excellent star-galaxy separation at faint magnitudes. Like \textit{Roman}, \textit{Euclid}  is likely to excel in regions of low-to-modest crowding over large areas (e.g., stellar halos of galaxies out to $\sim5$~Mpc). The potential of \textit{Euclid}  for resolved stellar populations data has already been demonstrated in the literature (See Appendix~\ref{sec:important_obs}), e.g., by tracing the RGB population of IC~10 to $\sim8$~kpc \citep{Annibali:2026}, and by the discovery in its first data release (DR1) of Fornax-7, a $\sim170$~\msun\ stellar system located near the Fornax dSph that may be a satellite of a satellite \citep{Saifollahi:2026}.

   \textit{Roman} \citep{Spergel:2015, Akeson:2019}, launched in August 2026 but not yet taking science data at the time of writing, has $\sim200\times$ the areal coverage of \hubble\ and it is designed to efficiently survey large areas.  It has an angular resolution similar to WFC3/IR but with better sensitivity and broader wavelength coverage.  It has the potential to efficiently resolve the stellar populations of nearby galaxies over large regions of low-to-modest crowding, such as the stellar halos of some nearby galaxies, angularly large systems (e.g., some MW satellites, some isolated dwarf galaxies), and the disks of more massive galaxies in the LG (e.g., M31, M33).  It will struggle to resolve stars in regions of high crowding. Figure~\ref{fig:ELT_Mosaic_Images} shows an example of simulated images of resolved stars in Virgo with different facilities.  The wide-field facilities discussed here should have resolving power similar to the \hubble/WFC3/IR inset.

    Resolved star observations with \textit{Roman} will require careful planning.  Its PSF at all wavelengths is under-sampled and some filters are sub-optimal for some resolved star science (e.g., similar to WFC3 IR's F160W band, prominent molecular bands in \textit{Roman}'s F129 and F158 filters may make separation of AGB C and M stars more challenging).  \textit{Roman} also has the potential for measuring proper motions for large numbers of faint stars for galaxies within the LG, though careful planning (e.g., dither patterns, time baseline) is important. \textit{Roman} will not be as good at star-galaxy separation as \jwst, but it should provide an improvement over wide-area ground-based imaging.

    \textit{Roman}'s HLWAS will include several nearby galaxies and offers the opportunity to discover faint new systems.  The HLWAS consists of $\sim2400$~deg$^2$ in Y, J, and H bands to a depth of 26.5 and an additional $\sim2400$~deg$^2$ only in H-band to a depth of 26.2.  It will also acquire $\sim20$~deg$^2$ of deep fields to 27.7 in Z and Y bands.  Additional nearby galaxy science is selected through competitive peer review, and the first such programs have now been chosen.

The first \textit{Roman} General Investigator programs provide a preview of the near-field science the mission will enable.  Notably, there is a program to map the resolved stellar populations of the M31 and M33 disks and inner halos (General Astrophysics Survey program GAS-2002; PI Gilbert; see \S \ref{sec:future_formation}).  Other relevant programs generally focus on resolved star studies of dwarf galaxies in the HLWAS and the LV, including their SFHs, structures, and the characterization of faint galaxy candidates (19000, PIs McQuinn and Mao; 19014, PI Mutlu-Pakdil; 19022, PIs Garling and Kallivayalil).

\textit{CSST} has the potential to make a significant impact on resolved galaxy studies. It includes two imaging instruments, the Main Survey Camera (MSC) and the Multichannel Imager (MCI), both of which feature high-angular resolution imaging ($\sim0.15^{\prime\prime}$)  from the near-UV through near-IR as summarized in Table~\ref{tab:facilities}.  Over the 7-year mission lifetime, the \textit{CSST} will leverage MSC's wide-field (1.1 deg$^2$) to image $\sim17,500$~deg$^2$ to $\gtrsim26$~mag in $g$-band and $\gtrsim25.5$~mag in the near-UV, $u$, $r$, $i$, and $z$-bands.  The $\sim400$~deg$^2$ \textit{CSST} deep field will reach at least $\sim1$~mag deeper than the wide survey, while the MCI extreme-deep fields (six fields totaling $\sim300$~arcmin$^2$) will reach $\gtrsim29$--$30$~mag. Its footprint will include $\sim50$--$100$ galaxies within $\sim4$~Mpc that can be resolved into stars \citep{CSST:2026}.  It will also survey the entire M31 disk and halo aimed at a variety of science such as dust mapping, substructure discovery and characterization, and proper motions \citep[e.g.,][]{Zhan:2021, Qu:2023, Fu_Z:2023, Chen:2024}.   The \textit{CSST} is planned to launch in 2027.  Following a nominal proprietary period, \textit{CSST} data are slated to be made publicly available.

Finally, the \textit{Lazuli} space telescope has many characteristics (see Table~\ref{tab:facilities}) that would enable resolved stellar population studies in nearby galaxies.  As described in \citet{Roy:2026}, the main science cases for Lazuli do not include the type of resolved galaxy science we discuss in this review, though any of its observations that include nearby galaxies may be useful, assuming the appropriate tools (e.g., crowded field photometry) are available.

\paragraph[UVEX]{UVEX:}  The Ultraviolet Explorer (UVEX; \citealt{Kulkarni:2021}; Harrison et al., submitted) is a NASA mid-sized explorer mission that is slated to launch in 2030.  It provides coarse angular resolution ($\lesssim2^{\prime\prime}$) in two far- and near-UV bands over an extremely wide-field ($\sim12$~deg$^2$).  Its relevance for resolved stars is that it will undertake a highly cadenced survey of the entire main bodies of the LMC and SMC over a two year period, providing for both deep UV CMDs from the stack of all images ($m_{UV} \sim 27$) and time series data on hot stellar populations.    Its primary science aims in the LMC and SMC are to catalog the population of stripped stars, characterize hot massive star properties, and improve our knowledge of stellar physics using eclipsing binary systems with at least one hot companion (\citealp{Kulkarni:2021}; Harrison et al., submitted).

\paragraph[CASTOR]{CASTOR:} The Cosmological Advanced Survey Telescope for Optical and UV Research (\textit{CASTOR}; \citealt{Cote:2025}) is a proposed Canadian-led 1-m class space telescope that would provide \hubble-like angular resolution ($\sim0.15^{\prime\prime}$) imaging in the UV and blue-optical over a $\sim0.25$~deg$^2$ field of view.  It would offer the combination of wide field and UV sensitivity that is well-suited to mapping young and hot stellar populations (e.g., upper MS, BHeB, and HB stars) across nearby galaxies and their halos, complementing the redder wavelength coverage of \textit{Roman} and \textit{Euclid}.

\paragraph[HWO]{HWO:} \textit{HWO} is a next-generation telescope concept endorsed by the Astro 2020 decadal survey \citep{NationalAcademies:2021}.  Though the exact specifications of \textit{HWO} are still under exploration, we broadly expect that it will feature a $\gtrsim6$~m aperture with imaging and spectroscopic capabilities that span the UV through near-IR.  \textit{HWO} is anticipated to have an exceptionally high-angular resolution over a sizable field-of-view, such as those listed in Table~\ref{tab:facilities}.  Its capabilities make it well-suited for an incredibly broad range of resolved stellar science \citep[e.g., SFHs, proper motions, IMF, distances;][]{Anand:2025b, Smercina:2025}.

\subsubsection{Ground-Based}

\paragraph[Rubin \& HSC]{Rubin \& HSC:}  LSST on the Rubin observatory \citep{Ivezic:2019} will be transformative for many aspects of resolved star studies.  By surveying $\sim18,000$~deg$^2$ of the southern sky over $\sim10$~years, Rubin will reach faint magnitude limits across 5 optical filters: $u\sim26.1$, $g\sim27.4$, $r\sim27.5$, $i\sim26.8$, $z\sim26.1$, $y\sim24.9$ AB mag.

Rubin promises to touch virtually all aspects of resolved star science.  Its depth and cadence will provide a definitive census of stellar populations in galaxies from the nearest MW satellites out to several Mpc over large areas \citep[e.g.,][]{Mutlu:2021, Shread:2026}.  The cadenced data are expected to provide good sampling of RR Lyrae light curves out to $\sim600$~kpc and for higher mass eclipsing systems to even larger distances.  It promises to affect science from detailed stellar physics to the large scale distribution of galaxies in the LV to the stellar populations in nearby galaxy halos.  A first glimpse is provided by the Rubin Data Preview 2 imaging of the UFD Virgo~III \citep{Pai:2026}, in which $\sim10$~hr of LSSTCam exposure, i.e., close to the 10-year survey depth, increased the number of candidate member stars by a factor of $\sim4$ over the Subaru HSC discovery data \citep{Homma:2024}.

Rubin's effective angular resolving power ($\gtrsim0.7^{\prime\prime}$) will limit its impact to relatively uncrowded regions \citep[e.g.,][]{Wainer:2025} and it will only have marginal star-galaxy separation power compared to space-based imaging (e.g., Figure~\ref{fig:sg_separation}).  It may be that joint analysis of LSST and space-based data (e.g., \textit{Euclid}, \textit{Roman}) will be necessary for maximal science impact.   Rubin data volumes will be so large that new approaches to reduction and analysis of resolved stars in even moderately crowded fields may be warranted.

Subaru's HSC is very similar to the Rubin Observatory, but located in the northern hemisphere.  It has comparable sensitivity, angular resolution, and wavelength coverage, albeit with a smaller field of view.  HSC is not designed to be a survey facility like Rubin and instead is used for targeted observations.  HSC's impressive capabilities for resolved star studies have already been demonstrated in the literature, and in some sense, provide a preview for LSST data in the southern sky.  HSC shares many of the same limitations as Rubin (e.g., due to moderate angular resolution).  However,  one particularly noteworthy feature of HSC is that it boasts dozens of filters, including certain narrow bands that are particularly useful for resolved star science.  For example, HSC has Ca H\&K and MgB filters that  enable photometric metallicities and efficient removal of MW foreground stars, respectively.  Such narrow band filters have proven incredibly effective with HSC, and may be a good future addition to Rubin \citep[e.g.,][]{Kahn:2019}.

\paragraph[VLT/MAVIS]{VLT/MAVIS:} MAVIS (Multi-Conjugate Adaptive Optics [MCAO] Assisted Visible Imager and Spectrograph) is a new instrument being built for Europe's VLT \citep{Rigaut:2020}.  It will operate over optical wavelengths and features an angular resolution slightly better than \jwst, albeit with a much smaller field of view.  MAVIS appears best suited for targeted follow up observations in the crowded regions of nearby galaxies (e.g., high-mass IMF studies in young star clusters).

\paragraph[Extremely Large Telescopes]{Extremely Large Telescopes:} The 30m class telescopes (ESO's ELT, the Giant Magellan Telescope [GMT], and the Thirty Meter Telescope [TMT]) will all feature revolutionary high-angular resolution imagers.  As the GMT and TMT are still being developed and built, we highlight the general capabilities of these facilities by describing relevant imaging for the ELT and list the characteristics of similar prospective capabilities for the GMT and TMT in Table~\ref{tab:facilities}.

The Multi-AO Imaging Camera for Deep Observations (ELT/MICADO; \citealt{Davies:2018}) promises to be among the most transformative ground-based imagers for resolved stars of this century.  It features sensitivity and angular resolution in the near-IR that will be superior to \jwst, and over a moderate $1\times1^{\prime}$ field of view.  For example, as illustrated in Figure~\ref{fig:ELT_Mosaic_Images}, MICADO applied to the Virgo cluster is expected to reach $\sim1$~pc scales, meaning that it can resolve stars near the cores of giant ellipticals such as M87 \citep[e.g.,][]{Davies:2018}.  In comparison, \jwst\ can only resolve stars beyond a few $r_{\rm eff}$ \citep[e.g.,][]{Anand:2025}.  Like \jwst\ and \hubble, MICADO does not excel at surveying large areas, and instead should be thought of as a deep field facility capable of targeted observations of resolved stars out to $\sim100$~Mpc.

\subsection{Star Formation Histories} \label{sec:future_formation}

While there is understandably a large degree of excitement on the discovery of new stellar systems in the LG and beyond, there is a less clear understanding of how to proceed with characterization of these new systems, as well as how and why we should be improving our knowledge of galaxies that have already been reasonably well-studied.  In the following section, we outline several directions for new and improved SFHs of resolved galaxies throughout the LV, and include broad suggestions for efficient observing strategies.

\subsubsection{Local Group}

\paragraph[Milky Way Dwarf Galaxies]{Milky Way Dwarf Galaxies:} There are several future prospects for improving the SFHs of MW dwarf galaxies.  First, wide-area surveys (e.g., LSST, \textit{Euclid}, \textit{Roman}, \textit{CSST}) are expected to uncover many new MW satellites, particularly at very low luminosities \citep[e.g.,][]{Jethwa:2018, Newton:2018, Drlica-Wagner:2020, Nadler:2020, Manwadkar:2022, Tan:2026, Rusterucci:2026}.  Injection-recovery tests on simulated LSST data, combined with an empirical model of the MW satellite population, suggest that LSST could discover $\sim50$ new MW dwarfs, mostly fainter than $M_V\approx-2$. The metal-sensitive $u$ band appears to improve the recovery of the most extended systems, as it better isolates metal-poor member stars from the MW foreground. Pairing LSST with space-based star-galaxy separation (e.g., from \textit{Euclid} or \textit{Roman}) could reveal even fainter satellites \citep{Rusterucci:2026}.
For most of these new systems, the discovery CMDs are unlikely to be deep enough on their own to measure robust SFHs.  While the data may have sufficiently high SNRs, they are likely to fall below the $\sim100$~MSTO star threshold required for a robust SFH.  Targeted deeper follow-up imaging (e.g., with \hubble\ or \jwst), and the inclusion of the MS kink in CMD modeling where feasible, may make it possible to recover SFHs.  However, the degree to which the MS kink improves SFH fidelity in the few-stars regime remains to be tested.

SFHs for larger numbers of very faint systems from different groups (e.g., LMC vs.\ MW satellites) may improve our knowledge of how reionization impacts growth of the lowest-mass galaxies.  The wide areal coverage of these datasets should also enable (a) mapping the full angular extent of faint galaxies to determine the existence and characteristics of stellar halos \citep[e.g.,][]{Chiti:2021} and (b) the measurement of population gradients in their stellar populations.

One challenge that is becoming increasingly important is the ability to distinguish between faint star clusters and \textit{bona fide} UFDs.  The line between the two has become increasingly blurred over the past decade (e.g., objects overlap in the size-luminosity plane).  This issue is likely to be amplified by ever-fainter new discoveries.  In addition, the new class of ``ultra-faint compact systems'' provides a new source of ambiguity between faint galaxies and star clusters \citep[e.g.,][]{Cerny:2023c}, a topic that has long been of great interest in this community \citep[e.g.,][]{Willman:2012}.  There may be ways in which resolved star imaging can help with classifying such systems.  For example, deep CMDs that reach the MS kink may be useful for SFH determination or at least in searching for the presence of multiple populations associated with Galactic GCs.  Searching for variations in the low-mass PDMF due to mass segregation may be another route.  Similarly, the addition of deep Ca H\&K imaging \citep[e.g.,][]{Fu:2023} may provide a handle on metallicity spreads.  Whether these are genuine fossil galaxies, stripped relic galaxies, or low-mass star clusters has fairly significant implications in the area of the galaxy-halo connection, galaxy formation in the early Universe, and/or the long timescales of survival for low-mass clusters in a strong tidal field.

\begin{figure}[pos=tbp]
    \centering
    \includegraphics[width=0.7\linewidth]{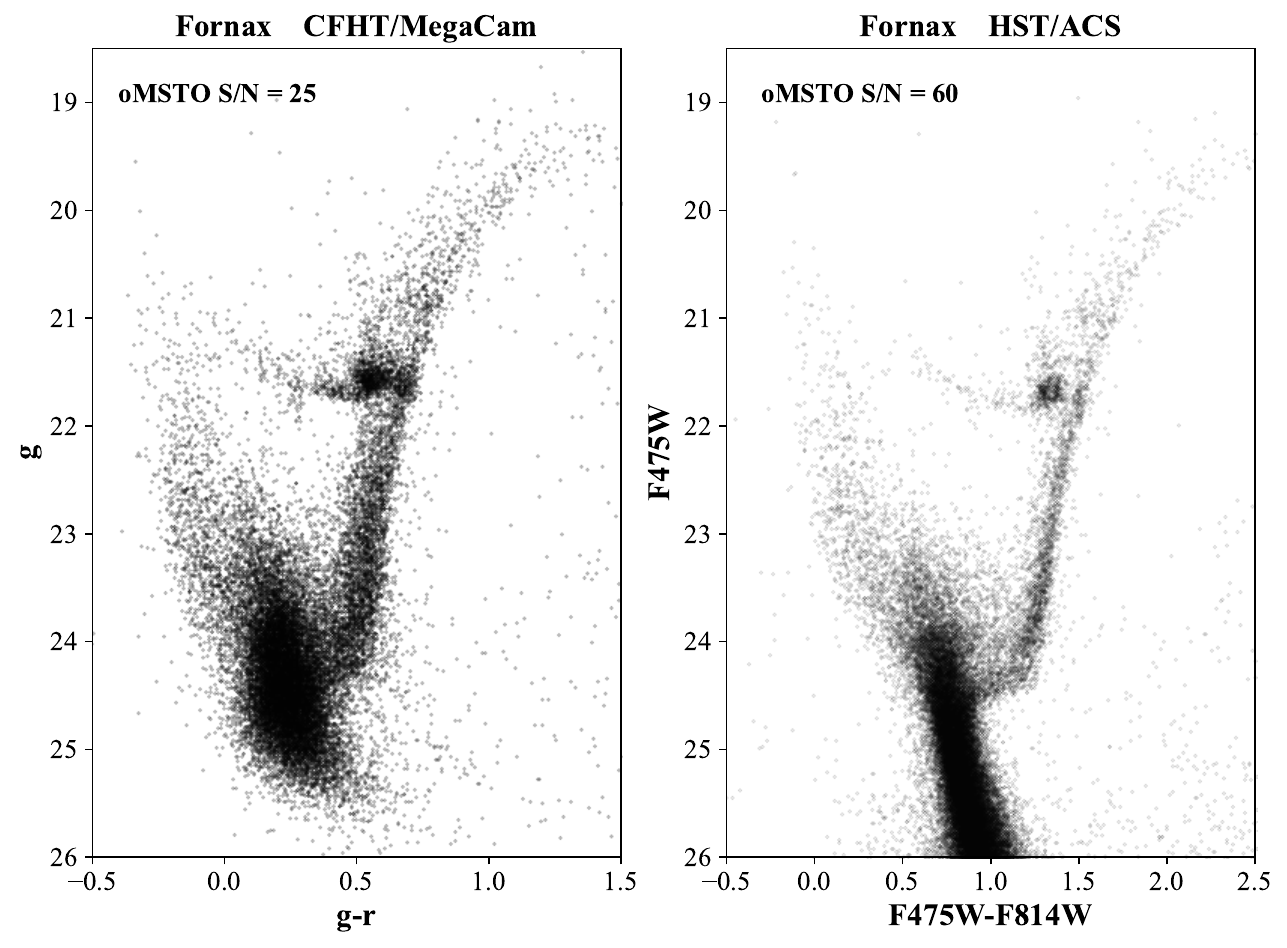}
    \caption{An illustration showing the anticipated improvement in the CMD quality for MW satellites from Rubin and HSC imaging.  \textbf{Left:} A Canada-France-Hawaii Telescope (CFHT)-based CMD of Fornax from \citet{deBoer:2012b} with SNR$=25$ at the oMSTO; \textbf{Right:} A \hubble-based CMD of Fornax from \citet{Rusakov:2021} with SNR$=60$ at the oMSTO.  Compared to CFHT, the \hubble-CMD shows several tight features such as the RC and the split ancient SGBs and MSTOs.  Rubin/LSST is expected to provide similarly high SNR CMDs for MW satellites in the southern hemisphere, while targeted HSC imaging can provide comparable data in the northern sky.  These figures do not illustrate issues with differentiating member stars from MW foreground stars and background galaxies.}
    \label{fig:Fnx_SNR}
\end{figure}

Second, for classical MW satellites (including objects such as Crater~II and Antlia~II), 8m-class ground-based facilities (e.g., Rubin, HSC) will provide CMDs that extend well-below the oMSTO over the entire spatial extent of all known classical satellites.  Figure~\ref{fig:Fnx_SNR} and recent HSC observations of Ursa~Minor by \citet{Sato:2025} provide good examples of expected data quality.  These observations will provide for truly global SFHs of these systems, including the measurement of gradients in the SFH and AMR out to large radii.  The anticipated SNR of the data should be sufficiently high to enable the resolution of detailed bursts of star formation in these systems at all cosmic epochs.

A third area is the characterization of faint, wide-area features, such as streams or tidal tails associated with known galaxies.  Measuring SFHs and AMRs over the full spatial extent of these systems will help to provide a more comprehensive picture of how and when the MW stellar halo was assembled through hierarchical accretion \citep[e.g.,][]{deBoer:2015, Gallart:2019, Ruiz-Lara:2020a}.  A key challenge is that, in the low surface brightness regime of tidal tails and streams, the CMD is often dominated by MW foreground stars and unresolved background galaxies.  Measuring a SFH in this regime requires either aggressive membership selection (e.g., using proper motions, narrowband metallicity indicators, spectroscopy) or explicit forward modeling of the contaminant populations as part of the CMD fit, e.g., by including an empirical CMD of an adjacent control field in the model \citep[e.g.,][]{deBoer:2015}.  Neither approach has been widely tested at the depths and areas that upcoming surveys will provide. We regard this as a critical next step.

However, contamination removal (e.g., MW foreground stars, background galaxies) is likely to remain challenging \citep[see \S \ref{sec:sg_future}; e.g.,][]{Soumagnac:2015, Drlica-Wagner:2020, Tsiane:2025}.  \textit{Roman} is the most likely facility to mitigate this problem at MW-satellite distances.  Its angular resolution, multi-band imaging, and expected SNR for stars at magnitudes relevant for these systems should substantially reduce confusion between compact background galaxies and faint member stars relative to ground-based surveys.  \textit{Euclid}'s VIS imaging offers a complementary trade: coarser angular resolution in a single broad optical band, but over a $\sim3\times$ larger area, such that it will provide morphological star-galaxy separation for many more MW satellites than \textit{Roman}, albeit with less discriminating power for any individual system.  Mitigating contamination issues is particularly important for the low-density regions (e.g., streams, outer regions of classical satellites, extended UFDs) as the sparse populations can be overwhelmed by MW foreground stars and compact background galaxies, which can not only lead to poor detection of new systems, but also to inaccurate characterization (e.g., compact background galaxies being mistaken as an ancient metal-rich population).

Though spectroscopic membership and/or metallicities may be challenging due to the faintness of target stars of interest (e.g., $m_r \gtrsim 23.5$), it is possible to obtain narrowband imaging \citep[e.g.,Ca H\&K;][]{Majewski:2000, Starkenburg:2017, Almeida-Fernandes:2022, Fu:2023, Chiti:2026} with facilities such as \hubble\ or HSC that may help to distinguish member stars from contaminants.  Space-based facilities should provide deep imaging with less contamination, due to their higher angular resolution and potential for determining membership based on proper motion measurements \citep[e.g.,][]{Sanderson:2026}.

Alongside deep, wide area CMDs, the addition of metallicity information, e.g., via spectroscopy or narrowband imaging (e.g., Ca H\&K), should improve our knowledge of SFHs (particularly AMRs), gradients, and help to identify coherent subpopulations, as highlighted in  \citet{Sato:2025}.  Photometric data from Ca H\&K imaging programs \citep[e.g.,][]{Starkenburg:2017, Longeard:2018, Longeard:2020, Longeard:2022, Fu:2022, Longeard:2023, Fu:2023, Barbosa:2025, Atzberger:2026} that have targeted MW dwarf galaxies provide a highly complementary avenue for stellar population characterization alongside broadband CMDs.  Formally modeling broad and narrowband CMDs in tandem is an interesting future direction.

By the mid-2030s, a combination of wide-area imaging efforts could deliver, for every known MW dwarf galaxy with $\gtrsim100$ MSTO stars, broadband CMDs that reach high SNRs at the oMSTO over the full spatial extent, including faint tidal features (e.g., streams) that could be associated with these galaxies.  In the best case, each system could have SFHs measured to their theoretical precision limit (e.g., set by the number of stars available and the age-metallicity degeneracy), including the resolution of relic bursts and SFH/AMR gradients, with stellar-evolution model systematics as the dominant residual uncertainty.   The addition of deep Ca H\&K narrowband imaging to every known system would help to improve metallicity and AMR characterization for larger numbers of stars than are readily accessible with spectroscopy.  With such exquisite and complete imaging data in-hand, a concerted effort at uniform, deep spectroscopy (e.g., for kinematics, chemical abundances, gradients, dark matter modeling) across the community should be a high priority.

\paragraph[M31 Ecosystem]{M31 Ecosystem:}

Galaxies associated with M31 represent our best opportunity to study an entire galactic ecosystem to a level of detail comparable to the MW (e.g., access to faint galaxies and streams, deep CMDs, orbital histories, chemical abundances) for the foreseeable future.

To construct deep, spatially comprehensive CMDs across the M31 ecosystem, investments from a variety of facilities are needed.    The M31 ecosystem features a large dynamic range in surface brightness (crowding) and a wide variety of target characteristics (e.g., luminosity, angular size).   Although most known galaxies in the M31 ecosystem have a published SFH \citep[see \S \ref{sec:formation};][]{Savino:2025}, there is ample room for improvement on two complementary axes: the most crowded regions (M31's disk, the central regions of the dEs) lack high-fidelity deep CMDs needed for robust ancient SFHs and require very high-angular resolution imaging (e.g., \jwst, \textit{HWO}).  Many wide-area targets, including M31 itself, lack oMSTO-depth coverage.  Existing wide-area CMDs are generally too shallow for SFH work \citep[e.g.,][]{McConnachie:2018} and affected by contamination, including from M32.

Unfortunately, at this time, observations of M31, and its associated galaxies, are not planned as part of LSST (which can only access the northern sky to a declination of $\sim+33^{\circ}$, whereas M31 is at $\sim+41^{\circ}$), with \textit{Euclid}, or the \textit{Roman} HLWAS, though any of these facilities could provide transformational science if they did observe it in the future.  Fortunately, a \textit{Roman} General Astrophysics Survey of M31 and M33 (GAS-2002, PI Gilbert; \S \ref{sec:facilities}) will now provide wide-field, resolved star imaging of both disks and their inner halos.  However, most systems associated with M31 are not accessible from the southern hemisphere (e.g., VLT, ELT).  As a result, other approaches are needed to obtain CMDs and improve SFH knowledge, some of which we outline below.  We suggest that observations of the M31 system would benefit from a coordinated approach across facilities.  There are notable examples of such coordination, e.g., the \hubble\ programs that placed oMSTO-depth pointings on substructures identified in the PAndAS map \citep[e.g.,][]{Richardson:2008, Bernard:2015a}, but they have been the exception rather than the rule over the last $\sim20$~years.  Beyond community coordination, this type of cross-facility investment may require the development of observatory policies or programs that are underused or may not currently exist.

In more concrete terms, we discuss areas of progress from most to least massive galaxies.  M31 and M33 have large areas of their star-forming disks mapped out with \hubble\ via PHAT, PHAST, and PHATTER, as described in \S \ref{sec:formation}. Optical CMDs from these programs typically reach below the RC and provide the tightest constraints on recent and intermediate age SFHs. The southern part of M31, covered by PHAST, lacks near-IR imaging, the addition of which is particularly well-suited for \textit{Roman} observations.    The ancient SFHs of the disks of these galaxies are of particular interest.  Deep \jwst\ CMDs in M31's disk have been obtained as a Cycle 3 program (e.g., \jwst-GO-4735, PI Sandford), but it is unclear if they can reach the oMSTO due to the effects of crowding.  Nevertheless, they should provide improved knowledge of the disk SFH back to older times than are currently available.  Resolving the crowded bulge and nucleus of M31 \citep[e.g.,][]{Lauer:2012, Rosenfield:2012, Dong:2018} to depths near the oMSTO may be possible with \textit{HWO}, or perhaps the TMT, but it is likely beyond the capabilities of \jwst\  \citep[e.g.,][]{Smercina:2025}.  A complementary path to the SFHs of crowded disks is to model luminous AGB stars, which remain accessible where the oMSTO is not.  \citet{Lee:2024} applied this approach to ground-based near-IR CMDs of M31's outer disk and found good agreement with the SFH measured from the much deeper optical PHAT CMDs over the past $\sim8$~Gyr.  \citet{Lee:2026} extended this approach to WLM and found that the SFH and AMR measured from AGB stars alone agree well with those from the oMSTO-depth \jwst\ CMDs of \citet{McQuinn:2024a} and \citet{Cohen:2025}.  Given the need for hundreds of AGB stars for a robust SFH, it remains to be seen whether AGB and RHeB stars alone can be used for spatially resolved recent SFHs of disk populations on fine spatial scales (e.g., tens of pc).  Moreover, comparisons of AGB models show large variations in predicted AGB star counts and lifetimes \citep[e.g.,][]{Ventura:2026}.  However, not all AGB models are equally well matched to observations, so such model-to-model spreads may overstate the true systematic uncertainties.  While near-IR imaging of AGB stars provides a path to expanding the accessible volume for studies of stellar populations in resolved galaxies, a coordinated community AGB calibration and stress-testing effort, anchored against oMSTO-based SFHs in well-characterized systems, will serve to further improve the fidelity of AGB star SFHs.

Measuring wide-area SFHs of the outer disks and halos of M31 and M33 would be scientifically transformative. M31 appears to have a qualitatively different assembly history from the MW that may involve a recent major merger \citep[e.g.,][]{DSouza:2018, Hammer:2018}. Moreover, these are the only systems in their mass range for which the oMSTO is accessible with virtually any current or planned facility.  When combined with full phase space information from proper motions and radial velocities, a comprehensive mapping of M31 to its oMSTO may be our only opportunity to robustly constrain the entire assembly history of another $L_{\star}$ galaxy for the foreseeable future. In M31, it is critical to sample both the smooth stellar halo component as well as a large number of substructures, as both are required to piece together its comprehensive assembly history.  M33 appears to have less pronounced features in its halo compared to M31, and it may be possible to reconstruct its history using sparser spatial coverage.

A key question is what type of data are needed to measure wide-area lifetime SFHs in M31 and M33.  Of course, there is a desire for CMDs that reach the oMSTO ($m_{V, 0} \sim 28.5$ or $m_{J, 0}\sim27.8$).  But obtaining such deep CMDs over the entire halo, i.e., hundreds of square degrees, is costly for any facility.  \textit{Roman}, through its Cycle 1 imaging of M31 and M33, will provide a first step in this direction.  Another promising facility is the \textit{CSST}, which has a large field of view and comparable sensitivity and resolution.  However, even with the \textit{CSST}, it would likely take hundreds of hours, if not more, to construct such deep CMDs across all of M31's stellar halo.

Possibly more efficient alternative strategies include targeted sampling of deep imaging combined with wider areal coverage to a modest depth.  For example, select pointings could be used to obtain oMSTO depth CMDs in a few positions in the smooth halo and along prominent substructures (e.g., the GSS) for scientific measurements and also to calibrate shallower CMDs (e.g., HB depth) that are more efficient to acquire over much larger areas.  Similarly, improved understanding and usage of near-IR AGB star CMDs for SFHs and AMRs may obviate the need for HB-depth CMDs for SFH work and even shallower near-IR CMDs could be efficiently acquired over most of the stellar halo.  While \textit{Roman} and the \textit{CSST} are clear candidates for mapping the stellar halos of M31 and M33, ground-based facilities such as HSC should not be discounted as they can provide CMDs that reach below the HB in modest amounts of integration time.  Contamination removal is a clear issue, and the pairing of space- and ground-based observations may provide one efficient solution (see \S \ref{sec:sg_future}).

The diversity of M31 satellite properties also warrants a strategic approach to observations for measuring SFHs. The four brightest M31 satellites (M32, NGC~205, NGC~185, NGC~147) have limited SFH information from deep CMDs \citep[e.g.,][]{Monachesi:2012}.  Single \hubble\ pointings reach the oMSTO in NGC~185 and NGC~147, while NGC~205 and M32 have marginally shallower CMDs \citep[e.g.,][]{Geha:2015, Savino:2025}. Given crowding concerns and M31's location in the north, imaging with \jwst, \textit{HWO}, or perhaps the TMT, may be needed to reach the oMSTOs in the center of any of these systems. \textit{Roman} and the \textit{CSST} are most likely to contribute to construction of moderately deep CMDs in the outskirts of these systems.  Such wide-field imaging may also help to determine the full extent of these systems, including the presence of low-surface brightness streams or tidal features.

UFDs in M31 are targets of high scientific value.  Already, a sizable fraction of M31 UFDs show different patterns of star formation compared to MW UFDs, as discussed in \S \ref{sec:formation}.  This has potential implications for our understanding of the interplay between reionization and galaxy formation.  As more UFDs associated with M31 or M33 are discovered \citep[e.g.,][]{MartinezDelgado:2022, Collins:2022b, McQuinn:2023, Collins:2024, Jones:2026}, oMSTO depth follow-up imaging will be necessary to robustly measure their SFHs. These will likely require \hubble\ or \jwst\ (e.g., GO-7119, PI Sand), \textit{HWO}, or the TMT to achieve the necessary depth as well as mitigate contamination, which can be a considerable problem for some M31 systems, particularly those at lower Galactic latitudes.

Many wide-field surveys (e.g., Rubin, \textit{Euclid}, and the \textit{Roman} HLWAS) are not planning to target M31 as part of their community programs, meaning that alternative paths are needed to further its characterization.  At this time, advances in measuring the SFHs of M31 and its satellites over the next decade will require substantial and coordinated community efforts and investments of telescope time across multiple facilities.

\paragraph[Isolated LG Galaxies]{Isolated LG Galaxies:}

LG galaxies that are not satellites are critical for our understanding of secular evolution of low-mass galaxies, deciphering the role of environment in affecting backsplash and satellite systems, and constraining cosmological models of low-mass galaxy formation and the interplay between stellar feedback and dark matter.  While many of these systems have published SFHs (see \S \ref{sec:formation}), they often cover small areas of each galaxy.  Spatially comprehensive SFHs constitute an important next step.

Reaching the oMSTO in galaxies just beyond the edge of the MW halo (i.e., $D\sim400-600$~kpc; e.g., Phoenix, NGC~6822, Leo~T) presents an observational dilemma for space- vs. ground-based approaches.  At a distance of $\sim500$~kpc, the oMSTO is $m_{\rm V, 0} \sim 27.5$, which is technically achievable for 8m ground-based facilities (e.g., HSC, Rubin) outside perhaps the crowded central regions \citep[e.g.,][]{Tantalo:2022, Tantalo:2025, Pucha:2019}. As discussed in \S \ref{sec:crowding_future} and \S \ref{sec:sg_future}, the central regions require higher angular resolution facilities while the outer regions are accessible to wide-field facilities. \textit{Euclid} is not deep enough to reach the oMSTO \citep[e.g.,][]{Hunt:2025}. \textit{Roman} or the \textit{CSST} may be the most efficient middle option, with angular resolution sufficient to overcome crowding at these distances and improved contamination removal relative to ground-based facilities.  A SFH of NGC~6822 from near-IR observations of AGB stars may be an option as the data are already available from ground-based facilities \citep[e.g.,][]{Sibbons:2012}. Phoenix and Leo~T have too few AGB stars for robust SFHs from this approach \citep[e.g.,][]{Menzies:2008}.  Galaxies at larger distances ($D\sim0.6-1$~Mpc; e.g., IC~1613, Leo A, Peg~DIG) face a more acute version of the same observational challenge and will require a cross-facility approach as discussed in \S \ref{sec:crowding_future}.

Finally, we note that a handful of \hubble\ programs (e.g., LUVIT, GO-17833) have pursued panchromatic UV/optical/IR observations of LG star-forming dwarf galaxies in order to connect properties of the stellar populations with high-resolution gas and ISM observations \citep[e.g.,][]{Koch:2025}, for example, to understand the impact of star formation on gas turbulence timescales \citep[e.g.,][]{Stilp:2013, Hunter:2023, Hunter:2025a}.  The depth and wavelength coverage for some of these observations are comparable to PHAT and PHATTER, though other galaxies still lack near-IR coverage.  Overall, these datasets will bring the rich science generated in M31 and M33 to bear on very different ISM conditions in these lower-mass and lower-metallicity systems, previewing the power of these types of datasets for star-forming galaxies throughout the LV.

By the mid-2030s, it will be technically possible to measure the lifetime SFHs of all isolated LG dwarfs from CMDs that reach, or nearly reach, the oMSTO across the full extent of each galaxy.  However, realizing this picture will require deliberate community coordination across multiple facilities (e.g., \jwst, \textit{Roman}, \textit{CSST}, MICADO), as many of these targets are not part of the planned footprints of the major wide-field surveys.

\begin{figure}[pos=tbp]
    \centering
\includegraphics[width=1.0\linewidth]{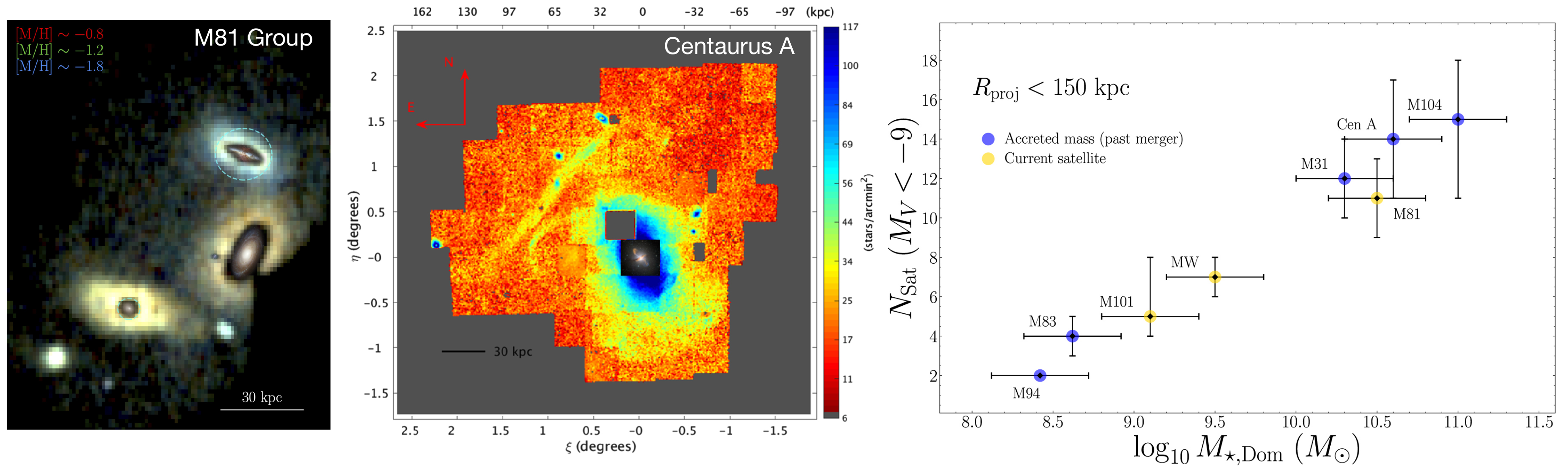}
    \caption{Examples of current resolved stellar maps and emerging trends for major galaxy groups in the LV. \textbf{Left:}  A map of estimated stellar metallicities in the M81 Group derived using ground-based Subaru/HSC imaging by \citet{Smercina:2020}; \textbf{Middle:} A stellar density map of Cen~A, one of the closest giant elliptical galaxies, illustrating the rich substructures it hosts.  This map is based on  Magellan/Megacam by \citet{Crnojevic:2016b}; \textbf{Right:}  The number of classical satellites versus stellar mass of either the host galaxy's halo (blue) or the most massive existing satellite (yellow) estimated from various resolved star observations. This trend and figure are from  \citet{Smercina:2022}, who remarked on the tightness of this trend and noted that it is not reproduced by current state-of-the-art cosmological simulations.
    \label{fig:LV_halos}}

\end{figure}

\subsubsection{Galaxies in the Local Volume}

Among the most exciting and consequential advances in CMD-based galaxy SFHs is the expansion to new environments beyond the LG.   Surveys such as ANGST and LEGUS have  pioneered important SFH measurements for galaxies out to $\sim5$~Mpc (see Figure~\ref{fig:LVMap}), while there has been a less systematic approach to galaxies from $\sim5$--$10$~Mpc.  Most SFHs of galaxies in the LV are of low-mass systems, as they dominate the local galaxy number count \citep[e.g.,][]{Karachentsev:2013, Ohlson:2024, Shread:2026}.   We first highlight scientific prospects in several areas and conclude with some observational considerations.

\paragraph[Galaxy Disks]{Galaxy Disks:} \label{sec:future:disks} A first area for growth is in measuring the SFHs of disks in massive galaxies. There is a paucity of well-constrained disk SFHs, as described in \S \ref{sec:formation}.  Currently, PHAT and PHAST have set a benchmark for detailed recent ($\lesssim 1$~Gyr) SFHs in the disk of M31, but it remains unclear how representative some of these results are of disk galaxies, e.g., the long-lived ring structures, the mismatches between CMD-based SFRs and integrated light-based SFRs, discrepancies between dust extinction and emission measures, etc. \citep[e.g.,][]{Lewis:2015, Dalcanton:2015}.

One concrete target would be to replicate PHAT/PHAST-style analysis of spatially resolved recent, and perhaps ancient, SFHs across the disks of massive LV galaxies. CMDs that reach the RC across the disks of nearby $L_{\star}$ galaxies would provide some constraint on the ancient SFH of the disk, while brighter stars on these CMDs would provide for recent ($\lesssim 1$~Gyr) star formation. Very high-angular resolution imaging (e.g., \jwst, MICADO, HWO) is required for these observations. There are open questions about what fraction of a galaxy's disk needs to be resolved and how many galaxies should be targeted, but these will depend on the science goals at hand.  It is quite clear that moderately deep CMDs across galaxy disks will lead to rich science returns across multiple fields (e.g., clusters, IMF, ISM, stellar migration, morphologies, galaxy assembly timescales, calibrating SFR indicators, bursty star formation).  There are $\sim10$ nearby massive galaxies (i.e., $L\gtrsim L_{\star}$ systems within $\sim10$~Mpc; \citealt{Karachentsev:2013}), for which these types of datasets, and detailed SFHs of disks are possible. The acquisition of these datasets would have both high immediate and long-term scientific impact and enable legacy science for a wide variety of topics.

\paragraph[Stellar Halos]{Stellar Halos:} \label{sec:future_halos} A second important direction is comprehensive mapping of stellar halos.  Resolved stellar populations provide some of the most direct evidence that galaxies assemble hierarchically.  Over the past two decades, wide-area maps of resolved stars around the MW, M31, and the nearest $L_{\star}$ groups have revealed stellar halos rich in accreted substructure, from coherent streams to bound and disrupting satellites \citep[e.g.,][]{Radburn-Smith:2011, Crnojevic:2016b, Okamoto:2015, Smercina:2018, Smercina:2020, McConnachie:2018}.  A central lesson from this work is the diversity of halo properties. Host galaxies with similar stellar masses, such as the MW and M31, show markedly different halo morphologies and substructure, indicating that the specific accretion history of an individual galaxy is important.

Figure~\ref{fig:LV_halos} shows examples of contemporary maps of stellar halos based on wide-area ground-based datasets of resolved stars.  These datasets provide a handle on stellar metallicity gradients, estimated halo mass, and the presence of substructures.  As illustrated in the right-hand panel, these datasets have revealed a surprisingly tight relationship between the number of classical satellites and the estimated stellar mass of the host galaxy's halo or most massive existing satellite, a trend that is not found in current generations of cosmological simulations \citep{Smercina:2022}.  More recently, \citet{Bell_e:2022} note a surprising absence of faint satellites in the M81 Group, given that the group's estimated total mass is larger than that of the LG.  A better understanding of how stellar halos and their satellite populations have assembled \citep[e.g.,][]{Smercina:2018, Mutlu:2024}, derived from CMD-based SFHs of a larger sample of nearby galaxies, may yield new insight into these puzzles.

Obtaining data for CMD-based SFHs in
stellar halos can be observationally challenging as surveys of the nearest halos may have to cover tens to hundreds of square degrees.  To cover such large areas in reasonable amounts of observing time, CMDs would likely reach no more than $\sim2$~mag below the TRGB.  The properties of AGB and RGB stars would then be used to infer SFHs, AMRs, and quenching timescales \citep[e.g.,][]{Harmsen:2023, Lee:2024, Lee:2026}.  Within a few Mpc, it is possible to make some progress with 6--8m class ground-based facilities as illustrated by studies around the Cen~A, M81, and NGC~253 Groups \citep[e.g.,][]{Okamoto:2015, Crnojevic:2016b, Smercina:2017, Crnojevic:2019, Okamoto:2019, Smercina:2020, Mutlu:2022, Bell_e:2022, Bell:2026}.  Going further down the luminosity function will be important for identifying fainter substructures that can then be linked to SFHs (\S \ref{sec:sg_future}).

Tiling with \textit{Roman} is desirable, but it is observationally expensive over such large areas even when reaching only a few magnitudes below the TRGB.  Many LV massive galaxies will fall into \textit{Euclid}'s footprint, while
 \textit{CSST}'s areal coverage and angular resolution seem well-suited to mapping out stellar halos.  Newly discovered galaxies in these groups are good targets for follow-up imaging with very high-angular resolution facilities \citep[e.g.,][]{Gozman:2024}.

Three of the nearest major galaxy groups (M81, NGC~253, Cen~A) are the most obvious targets for significant observational efforts. These three systems differ from each other and from the LG, and as the next closest groups they are the natural targets for the highest-quality resolved star datasets.  With appropriate planning it may be possible to have robust SFHs (from AGB stars), radial velocities, and bulk proper motions for many of the brightest galaxies in these groups.

It is worth emphasizing that Cen~A, the nearest giant elliptical galaxy, is a particularly high-value target.  Building on the work of \citet{Crnojevic:2016b}, a comprehensive, wide-area resolved star-based characterization of Cen~A and its rich substructures using a space-based facility would provide our most detailed constraints on how such massive galaxies form and evolve, placing it on similar footing as M31.   Of course, Cen~A has a large angular size, high surface brightness in the inner regions, and a distance ($D\sim3.6$~Mpc) that all make it challenging to map.  However, it is truly a unique environment in the LV with exceptional scientific importance that has a small sampling of moderate-depth CMDs from \hubble\ \citep{Rejkuba:2005, Rejkuba:2014}.  An upcoming \jwst\ program will provide resolved stellar spectroscopy and deep CMDs at several radial locations providing new constraints on its formation history (GO-10246, PI Sandford).

\paragraph[Low-Mass Galaxies]{Low-Mass Galaxies:} \label{sec:future_low_mass} A third direction is measuring the SFHs of very low-mass galaxies in the field.  The faintest systems (e.g., Tucana B, Leo~P, Pavo) may provide new insight into reionization with minimal ambiguity about the role of a massive host in quenching.   \jwst\ provides the clearest path toward acquiring deep CMDs of low-mass galaxies in the LV. Systems at these distances are sufficiently crowded and faint such that other major facilities (e.g., \textit{Roman}, LSST, HSC, \textit{Euclid}) cannot reach their oMSTOs. For galaxies within $\sim2$~Mpc, \jwst\ can reach the oMSTO for galaxies with modest exposure times, as demonstrated by \citet{McQuinn:2024b}. New \jwst\ observations of Tucana~B (GO-4748, PI Sand) will make it the first UFD outside the LG to have a SFH based on the oMSTO, while other comparable faint systems are being targeted by \hubble\ (e.g., GO-18062, PI Jones) and \jwst\ as well (e.g., GO-10160, PI Hai).  Measurements of oMSTO-depth SFHs for several classical systems in this distance range are also underway, using archival \jwst\ imaging of Sex~A (AR-6118, PI Newman) and new \jwst\ imaging of Sex~B (GO-9601, PI McQuinn).  These measurements should provide new constraints on the secular evolution of low-mass galaxies in a low-density environment, expanding work by the LCID program and others outside the LG.
Detailed SFHs of the rare population of quenched low-mass galaxies located in isolation (e.g., KKR~25, KKR-03, Hedgehog) will lead to new insights into mechanisms that drive galaxy self-quenching.  Upcoming \jwst\ programs are starting to investigate the SFHs of these types of systems (e.g., GO-11277, PI Danieli). Though only a handful of these systems are currently known, simulations indicate a sizable population should be uncovered in future dwarf galaxy searches \citep{Shread:2026}.

SFHs of low-mass galaxies in groups \citep[e.g., UFDs in M81;][]{Bell_e:2022, Gozman:2024} are also of interest for studying quenching processes.  However, as we discuss below, there are several challenges facing these measurements given the distances to nearest groups and the CMD characteristics of low-mass galaxies.

\paragraph[Observational Considerations]{Observational Considerations:} As discussed in \S\ref{sec:crowding_future}, it will be challenging to obtain a sizable population of galaxies in the LV with CMDs that reach the oMSTO.  For example, the oMSTO at the distance of the M81 group ($D\sim3.5$~Mpc) is at $m_{\rm V} \sim 31.7$ or $m_J \sim 31$, which is an expensive observation for \jwst, MICADO, \textit{HWO}, etc. to make routinely, even in absence of crowding considerations.

Even reaching the RC with \jwst\ in the LV will require modest-to-large integration times.  For example,  \citet{Lee:2026} computed that for \jwst\ to reach 1 mag below the RC in galaxies at $7$ and $13$~Mpc with \jwst\ in F115W and F200W requires total integration times of $\sim8$ and $\sim149$ hours per galaxy, respectively, without any considerations for crowding.  Space-based wide-field facilities are less sensitive than \jwst, meaning that they are unlikely to reach the RC throughout most of the LV.  Instead, CMDs that include the upper RGB and AGB, along with any luminous young stars (e.g., RHeBs, MS), will be the norm in this distance range.

Near-IR observations of brighter, less-crowded AGB stars may provide an efficient way to measure SFHs in all but the lowest-mass LV galaxies.  Although AGB star evolution is governed by complex physics \citep[e.g.,][]{Herwig:2005, Hofner:2018}, they are known to be highly age-sensitive and moderately metallicity sensitive \citep[e.g.,][]{Marigo:2017} and they have been reasonably well-calibrated to oMSTO ages in the LMC and SMC \citep{Pastorelli:2019, Pastorelli:2020}, which at least provides for ages that are self-consistent with MSTO stars.  Ratios of AGB-to-RGB stars can provide empirical estimates of $\tau_{90}$ \citep[e.g.,][]{Harmsen:2023, Ventura:2026}. More recently, efforts to measure detailed SFHs solely from AGB stars have shown promising results \citep{Lee:2024, Lee:2026}.   In terms of observational efficiency \citet{Lee:2026} estimate that only 4 and 12 minutes of \jwst\ integration time in F115W/F277W is needed to measure well-constrained SFHs in galaxies at $7$
 and $13$~Mpc, respectively, which is orders of magnitude more efficient than acquiring RC-depth CMDs.

SFHs from AGB stars face their own challenges. For example, at least several hundred are needed for statistically robust SFHs, which sets an approximate lower stellar mass limit of $M_{\star}\sim10^7$\msun\ for robust SFHs.  On the theoretical side, improved calibrations of AGB models, particularly at Solar and super-Solar metallicities are needed \citep[e.g.,][]{Boyer:2019, Karakas:2020}.  On the observational side, continued vetting of AGB star-based SFHs versus oMSTO-based SFHs will be important to understanding the strengths and limitations of this new approach.

If the community can mobilize the necessary multi-facility investment, by the mid-2030s the disks (e.g., with \jwst) and halos (e.g., with \textit{Roman} and the \textit{CSST}) of most massive galaxies within $\sim10$~Mpc could have robustly measured SFHs and AMRs over at least a large fraction of their total extent.  \textit{Euclid}'s contribution will be most valuable for the nearest few-Mpc systems (e.g., M81), where the top few magnitudes of the RGB and the brighter AGB are accessible. Driving the science out to the full $\sim10$~Mpc volume of massive galaxy halos in resolved stars will require \textit{Roman}, the \textit{CSST}, and very-high-angular-resolution facilities (e.g., \jwst, MICADO) as the primary workhorses, with \textit{Euclid} playing a supporting role at closer distances.

Measuring robust, lifetime SFHs for low-mass galaxies throughout the LV, particularly those with $M_{\star}\lesssim10^7$\msun, will be challenging. Although we expect there to be  thousands of new low-mass galaxies in the LV by virtue of a steep stellar mass function \citep[e.g.,][]{Shread:2026}, the majority of low-mass systems in the LV will have too few AGB stars, if any, for robust measurements using this technique.  Brighter CMD features (e.g., RGB, RC, HB) provide limited age information for such faint systems, due to a paucity of stars and/or absence of features (e.g., very metal-poor galaxies will not have a RC).  Moreover, SFHs of such faint systems from moderately deep imaging (i.e., 1--2 mag below the HB) have sizable uncertainties at ancient times, and in many cases, appear to be biased to younger ages when compared to oMSTO-based SFHs of the same galaxies \citep[e.g.,][]{Weisz:2019, Savino:2025}. Beyond a few Mpc (roughly the limit to which reaching the oMSTO is feasible), it is not entirely clear what the best strategy is for measuring robust lifetime SFHs of the faintest systems.

\subsubsection{Galaxies Beyond the Local Volume}

Perhaps the newest frontier in this area is the prospect of measuring SFHs of galaxies beyond the LV. As highlighted in Table~\ref{tab:sfh_compilation}, only a handful of coarse SFHs based on resolved stars in this distance regime have been published.

However, what previously seemed inconceivable, i.e., CMD-based SFHs at such large distances, may now become routine thanks to \jwst.  As part of establishing the local cosmic distance ladder, numerous studies have shown that \jwst\ can resolve the upper RGB and both carbon and oxygen rich AGB stars in disk galaxies out to $\sim40$~Mpc with integration times of several hours or less \citep[e.g.,][]{LiS:2024, Freedman:2025, Lee:2025, Li:2025}. Modeling AGB star distributions on near-IR CMDs enables robust SFH determinations \citep[e.g.,][]{Lee:2024, Lee:2026}, subject to the caveats previously discussed.

At distances of $\gtrsim10$~Mpc, qualitatively new environments become accessible for resolved stellar populations studies.  For example, galaxies in clusters such as Virgo and Fornax can be resolved into individual AGB and RGB stars with \jwst, providing our first window into massive galaxy clusters using resolved stellar populations (e.g., \citealt{Anand:2025}, \jwst-GO-7763 PI Weisz). In comparison, \hubble\ resolved the upper RGB of a single intracluster field in $\sim40$~orbits \citep{Williams:2007}. Similarly, larger populations of star-forming disk galaxies, field ellipticals, ultra-diffuse systems can have their brightest stellar populations resolved, some of which are already being pursued with \jwst\ (e.g., GO-8277, PI Smercina).

Given the current \jwst\ data quality at $\sim40$~Mpc, it is certainly possible to measure AGB-star SFHs of galaxy halos, and perhaps outer disks, to at least this distance. Within $\sim50$~Mpc, there are $\gtrsim10^4$ known galaxies \citep{Ohlson:2024}.  Many more are likely to be discovered in the coming decade \citep[e.g.,][]{Shread:2026}.  However, galaxies with $M_\star \lesssim 10^8$~\msun\ are expected to have few AGB stars and may be systematically excluded from SFH studies that rely on AGB stars \citep{Lee:2024, Lee:2026}.  The volume ($D\sim50$~Mpc, V$\sim5\times10^5$~Mpc$^3$) enabled by \jwst\ and AGB stars enables SFH measurements that go far beyond what is routinely possible now with \hubble\ ($D\sim5$~Mpc, V$\sim5\times10^2$~Mpc$^3$).  This represents a rich opportunity to probe a variety of new environments and establish large statistical samples with well-measured CMD-based SFHs.

Measuring SFHs out to 100 Mpc may even be possible. \jwst\ observations (GO-6876, PI Freedman) targeting resolved AGB stars in the Coma cluster ($D\sim100$~Mpc) should provide a sense of empirical sensitivity and crowding characteristics. As illustrated in Figure~\ref{fig:Telescopes}, MICADO and \textit{HWO} have the potential to outperform \jwst\ in this regard, and could, in principle, routinely provide access to age-sensitive resolved AGB stars in the near-IR in thousands of galaxies out to $D\sim100$~Mpc, though empirical validation of their performance will be critical.

\subsubsection{Bursty Star Formation} \label{sec:burst_future}

A first, and arguably most important, direction for improving our understanding of bursty star formation is to develop a concrete definition of a burst.  As discussed in \S \ref{sec:bursts}, there are numerous qualitative criteria in the literature used to identify or describe bursts.  However, there are far fewer quantitative metrics, which are necessary for both empirical identification of bursts and also to compare with predictions and expectations from simulations.

A second direction concerns the data, for which there are several clear areas for improvement.  The most immediate is simply to increase the sample of galaxies for which high fidelity SFHs are available, as discussed in \S \ref{sec:future_formation}.  For bursts at recent times ($\lesssim 1$~Gyr ago), this involves CMDs that reach approximately the RC in depth, which includes the BHeB and RHeB sequences and/or the use of AGB and RHeB stars in the near-IR.  Much of these data are already available for low-mass galaxies through programs such as ANGST, ANGRRR, LEGUS, STARBIRDS, and LUVIT, and what is missing is a uniform analysis across a large sample of galaxies.  As discussed previously, higher-mass galaxies generally lack suitable datasets at this time.

For fossil bursts, CMDs that reach the oMSTO with high SNR are required (e.g., see Figure~\ref{fig:Fnx_SNR}), which will enable the resolution of ancient bursts to a precision of $\sim0.5$--$1$~Gyr.  With sufficient planning, by the mid-2030s, every LG galaxy with a sufficient number of stars should have its fossil-burst record resolved.

A third direction is the development of a framework for interpreting bursts from a collection of galaxies.  Currently, even if the recent SFHs for dozens of galaxies were measured, there is not a clear path for making statistical statements about, for example, the properties of bursts (e.g., duration, amplitude, frequency) across a population of galaxies --- at fixed stellar mass, over a range of stellar masses, or as a function of ISM properties.  Similarly, population-level models need to be developed for spatially resolved bursts --- both recent bursts traced through BHeBs and fossil bursts, where chemodynamical information from spectroscopy may also be available.  Establishing population-level trends, as opposed to studies of single galaxies or small samples, is necessary for comparison with cosmological simulations of galaxy formation.  Some possible avenues may draw from recent work on modeling the distribution of SFHs and bursts from the integrated light spectral energy distributions of more distant galaxies \citep[e.g.,][]{Leja:2019, Lower:2020, Tacchella:2020}.

Finally, we note that improved models of BHeBs and RHeBs would be helpful for interpreting bursts.  Well-established mismatches in BHeB and RHeB models, as discussed in \S \ref{sec:bursts}, are typically attributed to systematic color offsets, while the age-luminosity relationships are assumed to be robust.  Further development and exploration of stellar physics for BHeBs and RHeBs is necessary to confirm the robustness of this assumption.

\subsection{The IMF}

There are several clear directions for improving our knowledge of the low- and high-mass IMF in nearby galaxies.  The main areas of improvement in both cases are increases in sample size, expansion of the types of environments sampled, continued improvement in measurement techniques, and a better understanding of binary stars.   For the low-mass IMF, handling contamination will also be important. As recently emphasized by \citet{Hennebelle:2024}, there are theoretical reasons why aspects of the functional form of the IMF, specifically stellar masses at which the IMF slope may change in a broken power-law model, may depend on environmental parameters (e.g., metallicity).  The possibility of more flexible, physically motivated, IMF parameterizations is another area in which further empirical investigation is needed.

\subsubsection{Low-Mass IMF}

As shown in Figure~\ref{fig:low-mass-imf-log-normal}, current data do not support systematic deviations from a Chabrier-like log-normal IMF in MW satellites.  Most constraints are too imprecise.  To date, three galaxies (Coma~Berenices, \citealt{Gennaro:2018b};  the SMC, \citealt{Cohen:2026}; Bootes~I, \citealt{Ding:2026}) have published log-normal IMF measurements from sufficiently deep data. Expanding low-mass IMF studies will require comparably deep data to these studies in order to robustly determine log-normal IMF parameters.  Studies should aim for samples of tens of galaxies that span a large dynamic range in metallicity, SFH, and dark matter content in order to establish trends that have clear statistical significance.

Another important test will be to measure the IMF in multiple fields in the same galaxy (e.g., with several pointings in the SMC or LMC at different radii).  Measurements at different locations can help to test if there are variations as a function of environment (e.g., by sampling different metallicity and age populations due to established population gradients) and/or  to diagnose potential systematics in the measurements (e.g., if binary fractions vary substantially between two different fields that are otherwise identical in SFH and AMR).

\jwst\ is particularly well-suited for expanding low-mass IMF measurements.  It can efficiently reach the lowest mass stars ($M\lesssim 0.2$ \msun) needed to constrain a log-normal IMF.  Critically, compared to \hubble, \jwst\ has a superior ability to distinguish between stars and background galaxies, a main source of contamination in low-mass IMF studies.  Deep \jwst\ data in the SMC and Bootes~I \citep{Shariat:2025, Ding:2026, Cohen:2026} provide a clear example of the data quality that is required for robust low-mass IMF measurements and stellar binarity, as shown in Figure~\ref{fig:bootes_imf}.  Similar datasets for a larger sample are needed to establish trends to a clear degree of statistical significance for a log-normal IMF and to better understand binarity of low-mass stars, both in its relation to the IMF and to improve our knowledge of star formation physics.

Beyond \jwst, MICADO has the potential to make low-mass IMF measurements, but its smaller field of view may require more tiling. \textit{Roman} may also be useful for expanding this science area, particularly for large angular diameter systems, though it will not reach the lowest-mass stars as easily as \jwst.  However, \textit{Roman's} IR optimization and wide field-of-view are good for efficiently collecting large numbers of lower MS stars.  Its angular resolution will provide modest star-galaxy separation capabilities (i.e., not as good as \jwst, not as limited as Rubin), although it has the potential to reduce contamination through per-star proper motion measurements, if multiple epochs are carefully planned \citep[e.g.,][]{Sanderson:2026}.  \textit{Euclid} observations are likely to be too shallow for IMF science. The \textit{CSST} MCI data may be suitable for low-mass IMF studies, but its limited near-IR capabilities mean more integration time is required to reach the lowest-mass stars.

Improving our knowledge of the low-mass IMF will require a better understanding of the impact of binary stars on the measurements.  Constraints on the binary fraction in current studies exhibit incredible scatter, with no clear correlation with any galaxy property (e.g., Figure~\ref{fig:low-mass-imf-log-normal}).  \citet{Gennaro:2018b} noted that the binary fraction from deeper \hubble\ IR data does not agree with determinations from shallower data. \citet{Cohen:2026} also highlighted trends in binarity in low-mass IMF studies as a puzzle. \citet{Shariat:2025} and \citet{Ding:2026} demonstrated the ability of deep \jwst\ CMDs for placing tight constraints on binary star properties in a single UFD, but there are currently no other measurements for comparison on equally well-measured footing. It is difficult to discern whether these results suggest that current implementations of statistical binary star models are insufficient for IMF work or if there are genuinely system-to-system changes in the properties of low-mass binary stars.  Spectroscopic constraints on binarity may be useful, though further investigation is needed for how spectroscopically measured binary fractions (which usually focus on RGB stars) compare to those measured from the lower MS.

Finally, in the limit that the IMF is environmentally dependent (e.g., the log-normal characteristic mass and/or width varies with metallicity; see related discussion of this in \citealt{Hennebelle:2024}), it may be necessary to implement more sophisticated IMF models.  For example, the SMC IMF study of \citet{Cohen:2026} includes stars that span a large range in age and a modest range in metallicity.  If the IMF varies as a function of either of these parameters, for example, it is not accurate to then measure a single IMF for this composite population.  Instead, a separate $m_c$ and $\sigma$, and perhaps binary star parameterization, would be needed for each metallicity and/or age population.  Again, there is no strong evidence that such IMF variations exist from current datasets, but even considering the hypothesis of an environmentally dependent IMF \citep[e.g.,][]{Kroupa:2013, Kroupa:2026} requires a more sophisticated treatment than is currently used.

\subsubsection{High-Mass IMF}

There are several exciting directions for future high-mass IMF studies.  One is the expansion of studies to the outskirts and low-density regions of nearby galaxies \citep[e.g.,][]{Bruzzese:2015, Watts:2018, Bruzzese:2020}.  These low-density and low-metallicity environments represent departures from the environments where the high-mass IMF has been measured so far.  Measuring the IMF of the field populations in these systems will require increased sample size, to detect statistically robust trends, and more sophisticated modeling approaches (e.g., to mitigate the IMF-SFH degeneracy; \citealt{Elmegreen:2007}).

Another direction is in expanding the PHAT and PHATTER-style IMF measurements in resolved star clusters to other galaxies \citep[e.g.,][]{Weisz:2015b, Wainer:2024}.   Deep, wide-area surveys that include, or will include, the LMC and SMC (e.g., with SMASH, DELVE, Rubin, \textit{Euclid}) should enable systematic IMF studies across these systems.   In the longer term, the sensitivity and angular resolution of facilities such as MAVIS, MICADO, and \textit{HWO} should enable M31 and M33-like IMF studies in star clusters in dozens of nearby galaxies, some of which have much higher star formation intensities than galaxies found in the LG (e.g., M83, NGC~253), providing new opportunities to search for potential high-mass IMF variations.

Like the low-mass IMF, issues such as contamination, membership, stellar multiplicity, and the fidelity of stellar evolution models are all critical to making accurate IMF measurements in tandem with the acquisition of more data \citep[e.g.,][]{Offner:2023, Marchant:2024}.  An improved understanding of binary stars in the intermediate and massive star regimes through spectroscopy and/or cadenced imaging (e.g., for eclipsing systems) should help to improve our knowledge of the IMF, particularly given that a large fraction of massive stars are thought to exist in binary systems \citep[e.g.,][]{Kobulnicky:2007, Kobulnicky:2012, Sana:2012, Offner:2023, Marchant:2024, Sana:2025}.  Similarly, understanding if/how massive stars can form in isolation will improve our knowledge of star formation and provide critical data on the IMF \citep[e.g.,][]{Lamb:2016}.

Though we have focused mainly on the IMF slope in this review, another important parameter for the high-mass IMF is the maximum mass of a star that can form, which may vary as a function of environment (e.g., SFR, metallicity).  Improved constraints on this quantity can come from optical spectroscopy \citep[e.g.,][]{Schneider:2018, Marchant:2024, Massey:2025a}, and possibly from UV imaging  (e.g., with \hubble, \textit{CSST}, \textit{UVEX}, \textit{CASTOR}), the latter of which can provide improved estimates of stellar mass that are generally consistent with those from optical spectroscopy \citep[e.g.,][]{Gull:2022}.

\subsection{Orbital Histories} \label{sec:orbital_histories_future}
The next few decades promise to deliver exciting new avenues for proper-motion-based science.
Here, we provide a summary of selected areas in which we anticipate major improvements.

\subsubsection{Improved orbital histories through long-baseline astrometry}

Within the broad area of resolved stellar populations, proper motion measurements have a unique advantage in that they can improve by the mere passage of time. As the first high-resolution \hubble/ACS imaging in nearby galaxies dates as far back as the early 2000s, space-based astrometric datasets with 20-30 year baselines will soon be a concrete reality for many galaxies within the LG. This will be complemented by upcoming data releases from the \textit{Gaia} mission, which, by DR5, will more than double the precision in its proper motion determinations.  This will result in increasingly precise orbital solutions, particularly for Galactic UFDs and for larger galaxies outside the MW halo. However, as proper motions increase in precision, an increasing fraction of the error budget in the orbital determinations will come from other sources, such as the gravitational potential assumptions and a galaxy's distance \citep[e.g.,][]{Pace:2022}. It will therefore be of paramount importance that effort is focused on improving our knowledge of these parameters.

From a technical standpoint, and for targets that have existing \hubble\ astrometric epochs, follow-up with \hubble will be the safest choice to obtain reliable multi-decade proper motion measurements, as this will minimize systematics due to cross-facility mismatches in resolution, filter, geometric distortions, etc.\footnote{We note that, although most \hubble\ astrometric imaging in the last two decades has been acquired with the ACS/WFC camera, WFC3/UVIS imaging has comparable resolution and can use similar broadband filters, and will therefore still be a suitable choice for the acquisition of additional epochs.} However, as \hubble\ has accrued over 35 years of service, it is unclear how long its imaging capabilities will remain available.

To truly leverage the power of multi-decade astrometric baselines, it is therefore a priority to improve our ability to measure proper motions across observatories. The development of tools such as GaiaHUB and BP3M \citep{delPino:2022,McKinnon:2024} has already demonstrated how \hubble\ and \textit{Gaia} observations can be successfully combined for this purpose, and the same approach is now being extended to \jwst\ through the \texttt{GaiaWebb} tool (\jwst\ AR-12274, PI McKinnon), which aligns \jwst\ images to the \textit{Gaia} reference frame to deliver proper motions for stars far fainter than \textit{Gaia} alone can measure. This achievement is facilitated by both \hubble\ and \textit{Gaia} astrometry being routinely measured in similar optical bands. In this regard, combining existing optical imaging with future observations of \textit{Euclid}/VIS and the \textit{CSST} is a clear path to the derivation of long-baseline proper motions. As we progress in the 2030s, the optical high-resolution imaging of facilities such as \textit{HWO} and MAVIS will provide the opportunity for even longer baselines.

The use of near-IR imaging from facilities such as \jwst\ and \textit{Roman}, in combination with existing \textit{Gaia} and \hubble\ imaging, will present untested challenges related to the different wavelength regimes involved. Among the possible challenges, the determination of a robust astrometric reference frame is an important factor. Observing programs are already in place to prototype \hubble-\jwst\ proper motion measurements (e.g., \jwst\ programs GO-4570, PI Warfield; GO-9225, PI Bennet). Positive results from these early experiments have the potential to significantly broaden the opportunities for proper motion measurements over the next two decades.

In the even longer term, MICADO and \textit{HWO} will provide near-IR capabilities that meet, and likely exceed, those of near-term IR missions such as \jwst\ and \textit{Roman}.  To enable multi-decade baselines, it would be advantageous to begin establishing proper motion baselines now in the near-IR.

\subsubsection{Reconstructing the assembly history of the M31 system}
Orbital reconstruction studies are no longer confined to MW satellites, as they have progressively extended to include M31 satellites. Within the next $\sim 15$ years, we anticipate this trend will continue. While the census of M31 satellites with orbital solutions is steadily increasing \citep[e.g.,][]{Sohn:2020,Casetti-Dinescu:2024,Casetti-Dinescu:2025,Patel:2025}, and additional programs are already in place to acquire more data (e.g., HST-GO-18029, PI Bennet), a large fraction of the M31 companions still lack 6D kinematics. Even in targets with already measured proper motions, the large distances involved make current proper motion uncertainties a dominant contributor to the total error budget. Near-future follow-ups with \hubble\ and \jwst\ have the potential to address both of these issues. As the temporal baseline from the first epoch gradually increases, proper motion measurements will become feasible for increasingly fainter galaxies and improve in precision for the ones with existing measurements. Newly discovered satellites, which lack suitable first-epoch astrometry, will require space-based imaging to be acquired as soon as possible, and could potentially have orbital histories determined by the end of the 2030s.

With appropriate proper motion observations, it is possible to obtain a complete census of satellite orbits in the M31 system. These data will shed new light on the environmental history of our nearest $L_{\star}$ system. Orbits of M31 satellites will clarify the dynamics and stability of the peculiar structures seen within their spatial distribution, such as the GPoA and the strong lopsidedness of the satellite system. Recent studies are already pursuing this avenue \citep[e.g.,][]{Sohn:2020, Casetti-Dinescu:2024, Patel:2025}, but larger satellite samples and more precise orbits are needed to establish a statistical sample. Proper motion measurements will also enable us to link the SFH of the M31 satellites to their orbital histories, allowing the exploration of connections between star formation, quenching, infall timescales, and other environmental perturbations \citep[e.g.,][]{Sohn:2013,Battaglia:2022,Patel:2026}.

A second opportunity involves obtaining relative proper motions for a large sample of bright stars in the M31 and M33 halos. While all the upcoming wide-area surveys of the 2020s have the depth and resolution to deliver this, the M31 system is not in the LSST or \textit{Euclid} footprints. And, although it will not be targeted by the \textit{Roman} community surveys, the Cycle~1 \textit{Roman} imaging of M31 and M33 will obtain two epochs separated by $\gtrsim4$~yr over the disks and inner halos. The M31 region is expected to be in the survey footprint of \textit{CSST} \citep{CSST:2026}, although the number and timing of its imaging epochs over M31 have not yet been specified. \textit{CSST} astrometry can, in principle, be combined with existing \textit{Gaia} data and/or pencil-beam \hubble\ fields to enable the derivation of proper motions, but the resulting measurement will be either confined to the limited area of the \hubble\ data or restricted to the brightest giants ($M_G \lesssim -4.5$) that are observable by \textit{Gaia}. A dedicated, wider-area \textit{Roman} campaign, a second-epoch follow-up with \textit{CSST}, or \textit{Euclid} imaging beyond its primary mission are therefore the only foreseeable channels to obtain large-area proper motions of fainter RGB stars. If executed during the nominal mission timelines, such observations have the potential to deliver spatially resolved proper motion maps of the stellar halos of M31 and M33\footnote{While the tangential velocity precision that \textit{Roman} or \textit{CSST} can deliver on single M31 stars is expected to be in the hundreds of km/s, statistically averaged measurements over $\sim 10^3$ stars should be sufficient to obtain a relatively fine tessellation of the M31/M33 halo region \citep[e.g.,][]{Sanderson:2026}}. Combined with radial velocities and abundances from spectroscopic surveys such as DESI and the Subaru Prime Focus Spectrograph, such a dataset would enable a deep chemodynamical characterization of the many stellar substructures that inhabit the M31/M33 halo region.

If community efforts are focused on the opportunities described above, it is possible that within the next $\sim 15$ years, we will have an extremely detailed dynamical census of the halo, stellar substructures, and satellites that currently inhabit the M31/M33 system. We emphasize that the M31 ecosystem is the only such environment outside the MW that can be studied at this level of detail, and establishing such rich datasets will have legacy value for decades.

\subsubsection{Pushing Orbital History Reconstruction into the Local Volume}

Even within the LG, measurements of extragalactic proper motions are technically demanding tasks, requiring astrometric precisions of fractions of a \hubble\ pixel and epoch differences of many years. For galaxies more distant than $\sim 1$ Mpc, the apparent motion of their stars in the plane of the sky becomes so minuscule that proper motion measurements are practically impossible with current instrumentation and datasets. However, the field is rapidly approaching a point at which the first proper motions outside the LG will finally be in reach.

On the one hand, the time elapsed since some nearby galaxies were first imaged with \hubble\ is becoming long enough that purely \hubble-based proper motions are gradually increasing in feasibility. In fact, observing programs already exist to capture a second astrometric epoch in the nearby galaxy groups of M81 and Cen A ($D\sim 3-4$ Mpc; $\sim 20$ yr baseline; GO-17513, GO-17925, PI Bennet), and in the nearby isolated dwarf Leo P ($D\sim 1.6$ Mpc; $\sim 10$ yr baseline; GO-17501, PI Bennet). In this regard, the ultimate lifetime of \hubble\ will dictate the range and precision of the proper motions we will be able to obtain.

On the other hand, new and upcoming observatories will boost our angular resolution and dramatically improve our ability to perform precision astrometry. At its shorter wavelengths, \jwst\ already delivers twice the spatial resolution of \hubble. In the 2030s and beyond, facilities such as MICADO, MAVIS, and \textit{HWO} will enable resolving powers an order of magnitude better than \hubble. A natural synergy will be to combine new high-resolution images from these telescopes with early \hubble\ images, capitalizing on both resolving power and long time-baseline. This will require overcoming the aforementioned challenges of cross-observatory proper motions and, even then, the final accuracy of the measurements is likely to be limited by the lower resolving power of \hubble. Pushing beyond this limit will require additional epochs observed by the new-generation telescopes. For this to happen, it is imperative that a first set of imaging is acquired as soon as possible, so that sufficient time can elapse before additional epochs are observed. With \jwst\ already in its fifth year of operation, securing NIRCam imaging of nearby galaxies and galaxy groups would be a promising first step in this direction.

From a science standpoint, we identify two major sets of targets in the LV that could benefit from proper motion measurements. The first is low-mass galaxies in relatively low-density environments. These galaxies have long been considered excellent targets to study galaxy evolution in the absence of strong environmental perturbations. However, while these galaxies are isolated today, a deeper understanding of their orbital histories is necessary to pin down the importance of past environmental encounters. For instance, there has been a long-standing debate on whether quenched dSphs at the edge of the LG, such as Cetus, Tucana, and \A{XVIII}, are truly isolated or are backsplash galaxies of the MW and M31 \citep[e.g.,][]{Fraternali:2009,Teyssier:2012,Buck:2019,Santos-Santos:2023,Bennet:2025}, which would have important implications for the mechanisms behind their star formation cessation. Further away from the LG, there is a growing body of known low-mass, quenched objects that appear to have remained even more isolated, such as KKR25, Hedgehog, Tucana B, Sculptor A, and CVn~C \citep[e.g.,][]{Makarov:2012,Sand:2022,LiJ:2024,Sand:2024, Hai:2026}. Having access to the 6D kinematics of these systems, and whether they might have interacted with a nearby neighbor, is a critical missing piece to disentangle the role of environmental factors in the quenching of these objects.  Importantly, reconstructing the orbital histories of isolated galaxies requires modeling all major relevant potentials (e.g., LG, M81 Group, Sculptor Group) that may have influenced the target galaxy.

The other promising avenue involves obtaining proper motion measurements for the members of nearby galaxy groups, such as Sculptor, M81, Cen A, and M83. With the long baseline from early \hubble\ imaging and the high resolution of \jwst, it should be possible, within the next 5--10 years, to determine the systemic velocity of the entire galaxy group with respect to the LG\footnote{While resolution and temporal baseline are dictated by the available instrumentation and by the archival datasets, the observing design can also play a role in the precision of proper motion measurements. High-resolution imaging over large halo areas in the brightest group members can increase the sample of member stars, as well as reference sources (quasars, background galaxies, \textit{Gaia} stars), in turn improving uncertainties.}. Not only will these measurements be essential for interpreting the environmental histories of field galaxies within 1 to 5 Mpc, but also the 6D kinematics of nearby cosmic structures will provide for improved understanding of the local dynamical flow and cosmic structure assembly history of the local Universe \citep[e.g.,][]{Shaya:1995, Courtois:2012, Shaya:2017, Valade:2024}. Over the next decade(s), the exceptional resolution of facilities like MICADO, MAVIS, and \textit{HWO} should gradually bring the internal kinematics of these galaxy groups within reach. As proper motion uncertainties approach $\sim 50-100$ km s$^{-1}$, orbital histories of individual group members will become feasible, based on what has been possible with \hubble\ within the LG at these measurement precisions. Supplementing similar studies in the MW and M31, this will open the door to comparative studies of how hierarchical satellite assembly proceeds across different $L_{\star}$ hosts within the LV. We stress that, along with the new proper motion determinations, accurate distances will be required to properly explore the dynamical history of nearby groups. Even modest  distance uncertainties can translate to positional errors of a few hundred kpc at the distances of these targets, which could compromise the fidelity of the orbital histories.

\subsubsection{Internal dynamics of nearby galaxies}
An area that is likely to progress substantially in the near future is the study of the internal kinematics of nearby galaxies using 3D velocities from proper motions and spectroscopy. For the MW satellites, early work on this subject has shown promising results \citep[e.g.,][]{Massari:2018b,delPino:2022,Vitral:2024}. However, the constraining power of current datasets remains limited due to large proper motion uncertainties and the small number of stars considered. Progress in this area will require higher proper motion precision, deeper limits for precision astrometry, and extensive spatial coverage of the target galaxies. Facilities such as Rubin, \textit{Roman}, and \textit{Euclid}, as well as future \textit{Gaia} data releases, promise to deliver all of the above. The large sky areas covered by these observatories capture the full stellar population of MW satellites out to the tidal radius. The high astrometric precision of \textit{Gaia}, \textit{Roman}, and \textit{Euclid}, in combination with the steadily increasing time-baseline from the first \hubble\ epochs, is expected to meaningfully lower proper motion uncertainties for the bright stars in these systems. At the same time, both \textit{Roman} and Rubin have sufficient depth to access much fainter stars. At the end of the nominal LSST survey, the Rubin observatory is expected to measure proper motions of order $\sim 0.1$ mas yr$^{-1}$ down to magnitudes of $r\sim 22$ \citep{Eyer:2012}. For the classical MW satellites, this will increase the sample of stars with individual proper motions by approximately one order of magnitude. Applications of these datasets include a precise determination of the dark matter density profile, with implications for the core-cusp problem;  a better characterization of distinct chemodynamical populations that are known to inhabit these galaxies; and, potentially, in larger galaxies such as the LMC, detailed and spatially resolved phase-space maps that to date have only been possible within the MW itself.

A second promising avenue related to the 3D kinematics of nearby galaxies will be the use of very high-resolution imaging to obtain proper motions in high-density stellar structures within the LG. While the resolution of \hubble\ and \jwst\ has unlocked resolved stellar imaging throughout most of the LG, there are regions where crowding is too high for these observatories to obtain deep photometry or astrometry. Examples are the bulge of M31, the central regions of M33, and the cores of dEs such as M32 and NGC~205. To obtain precision astrometry of stars within these regions, the higher spatial resolution of instruments such as MICADO, MAVIS, or \textit{HWO} is required. Near-IR imaging on the ELTs will be a particularly good option, because the near-IR wavelengths can mitigate dust extinction in these regions and because the long lifetime of ground-based facilities enables long-term observing campaigns. Resolved kinematics in the inner regions of galaxies such as M31, M32, or M33 can enable the measurement of dynamical mass profiles and the detection and characterization of central black holes, as recently demonstrated by two decades of \hubble\ proper motions in the dense core of the MW globular cluster $\omega$~Cen, which revealed fast-moving stars that imply an intermediate-mass black hole \citep{Haberle:2024}. Such measurements can also provide a valuable view into the chemodynamical properties of what are thought to be the earliest regions that participate in the galaxy formation process.

\section{Summary}

Over the past two decades, the study of resolved stellar populations beyond the MW has matured from a collection of heterogeneous, often qualitative analyses into a quantitative and increasingly homogeneous enterprise. Substantial investments with \hubble, the addition of \jwst, wide-area ground- and space-based surveys, and \textit{Gaia} have together delivered panchromatic photometry and precise astrometry for hundreds of millions of individually resolved stars across hundreds of galaxies, which span $\sim9$ decades in luminosity, a wide range of metallicities, and a diverse set of environments. Over the same period, the methods used to measure the IMF, SFHs, and proper motions have become mature and robust. The result has been remarkable progress across the four science areas reviewed in this paper, which we distill into the following summary points and future directions.

\subsection*{Summary Points}

\begin{enumerate}

\item Two foundational challenges frame virtually all resolved star studies.  The first is crowding, which approximately tracks the product of surface brightness and angular resolution.  The second is contamination: MW foreground stars can dominate in nearby low-surface-brightness systems, while unresolved (compact) background galaxies can dominate at the faint magnitudes required for the low-mass IMF and for stellar halos throughout the LV. Because no single facility can resolve stars across the full extent of a typical galaxy, progress will rely on coordinated multi-facility strategies, likely led by coordinated efforts by interested members of the community.

\item The impact of reionization on low-mass galaxies appears to be a continuum rather than a threshold. That is, there is no present day luminosity at which all galaxies abruptly quench. Instead, the fraction of galaxies quenched by the end of reionization generally declines with luminosity, albeit with substantial scatter. The commonly adopted UFD--classical dwarf boundary at $M_V=-7.7$ does not appear to correspond to a physical transition in SFH properties.

\item The SFHs of low-mass galaxies are remarkably diverse. Systems of nearly identical present day stellar mass can have significantly different SFHs. Environment plays a clear role. The SFHs of M31 dSphs differ systematically from their MW counterparts in that they form earlier, lack the late-time star formation seen in several MW dSphs (e.g., Fornax, Leo~I), and appear to separate into two subpopulations.  M31 satellite quenching epochs correlate with both luminosity and present-day distance from the host.

\item Bursty star formation is well-documented in low-mass galaxies. For ancient, fossil bursts, the apparent smoothness of a SFH is often an artifact of limited data quality. Higher SNR data at the oMSTO have revealed discrete episodes in many systems (e.g., Fornax, Carina, Tucana), although some (e.g., Sculptor, Draco) still appear comparatively smooth. At recent times ($\lesssim 1$~Gyr), burst durations are typically $\gtrsim100$~Myr and produce only a few to $\sim25$\% of a galaxy's stellar mass. There is no single, universal short-duration burst-and-quench cycle, though a larger sample of non-bursting galaxies still needs to be studied, as substantial selection effects remain. The quantitative connection to integrated-light SFR tracers and to predictions from cosmological simulations is still being established.  Similarly, there are only a handful of studies of spatially localized bursts in galaxies of moderate (LMC, M33, NGC~300) and higher stellar mass (M31).

\item Beyond the LG, wide-area resolved star maps of the nearest massive groups (M81, Cen~A, NGC~253) reveal a surprisingly tight relationship between the stellar mass of a host's stellar halo (or of its most massive satellite) and the number of classical satellites it hosts. Current cosmological simulations do not reproduce this trend.  The addition of CMD analysis, in order to formally provide measurements of ages, metallicities, and stellar masses, is a promising area for growth and new insight.

\item Resolved star counts provide the tightest extragalactic constraints on the high-mass IMF to date.  Studies by PHAT and PHATTER of young clusters in M31 and M33 reported a high-mass IMF slope of $\alpha \approx -2.45 \pm 0.1$. These values are broadly consistent with measurements in the LMC (outside of 30 Dor), the SMC, and the outskirts of nearby galaxies, and, at most, are marginally steeper than canonical MW values. There is no compelling evidence for significant environmental variation of the high-mass IMF in the LG.

\item  To date, the low-mass IMF across LG galaxies is well-described by a MW-like log-normal, even at metallicities as low as [Fe/H] $\sim-2.4$ and for ancient, reionization-era galaxies. No systematic variation with environment has been established. Single-power-law comparisons can be misleading as they are heavily reliant on data quality. Unresolved binarity remains an important and poorly understood source of uncertainty.  Only three systems (Coma~Berenices, the SMC, and Bootes~I) have published log-normal low-mass IMF measurements from sufficiently deep data, and larger samples are needed to establish the presence or lack of variation.

\item The LMC is massive enough to perturb the orbits of multiple MW satellites and hosts its own satellite system. As proper-motion precision has improved, the dominant uncertainty in orbital reconstructions has shifted from the measurements themselves to assumptions about the gravitational potential, particularly the masses of the MW and LMC.  Proper motions are now able to connect the orbital histories of nearby galaxies to their SFHs (e.g., the quenching of Leo~I near infall and pericenter).  They also show that most nominally isolated LG dwarfs are on first infall.

\end{enumerate}

\subsection*{Future Directions}

\begin{enumerate}

\item Construct spatially comprehensive lifetime SFHs for all MW dwarf galaxies:  With deep, wide-area imaging in hand, systematics from distance/extinction and stellar-evolution models are likely to be the dominant residual source of error.

\item Map wide-area lifetime SFHs of the M31 and M33 disks and halos to the oMSTO.  Because the M31 system lies largely outside the LSST, \textit{Euclid}, and \textit{Roman} community survey footprints, this will require deliberate multi-facility coordination on a scale that has so far been achieved only for select substructures.

\item Extend systematic SFH measurements across the stellar halos of LV massive galaxies and, with \jwst, into galaxy cluster environments (Virgo, Fornax, and, perhaps, eventually, Coma).  It is important to further develop and apply resolved AGB stars for efficiency at distances beyond the LV.  This requires a parallel, community-wide effort to calibrate and stress-test AGB models, particularly at high metallicity, against oMSTO-based SFHs.

\item Develop a quantitative framework for bursty star formation. Establish a single, concrete, measurable definition of a ``burst,'' and population-level metrics to quantify the durations, amplitudes, and frequencies of bursts that can be compared across galaxies and against cosmological simulations.  Improved models of BHeBs and RHeBs are needed to improve recent burst measurements. High SNR oMSTO-depth CMDs are needed to resolve the fossil-burst record in LG galaxies.

\item Expand IMF measurements across mass and environment.  For the low-mass IMF, it is important to acquire data deep enough to constrain log-normal parameters for galaxies spanning a wide range of age and metallicity.  For the high-mass IMF, imaging of resolved young clusters across a broader range of star formation intensities and metallicities will help to identify potential variations.  A more realistic treatment of stellar multiplicity is essential in both regimes.

\item Where not already available, acquire first-epoch, high-resolution imaging {\bf now} in nearby galaxies/groups (e.g., M81, Cen~A, Sculptor, M83) so that precise systemic proper motions and orbital histories outside the LG become feasible within the next 15--20 years (or beyond). At the same time, it is important to exploit the growing time baselines and large stellar samples from \textit{Gaia}, \textit{Euclid}, \textit{Roman}, and Rubin to measure internal 3D kinematics and dark matter profiles, and to search for black holes, in LG dwarfs.  An equally important goal is to establish how cross-facility (e.g., \hubble, \jwst, \textit{Gaia}, \textit{Roman}, \textit{Euclid}, \textit{CSST}) proper motions can routinely be made in the face of systematics, as they will be essential for proper motions that require multi-decade baselines.

\item Continue to advance the methods that underpin every measurement above, such as crowded field photometry and contamination removal.  The promise of the future datasets (e.g., with Rubin, \textit{Roman}, \textit{CSST}) will rely on continued algorithmic development (e.g., machine learning, joint space- and ground-based analysis) to push reliable star-galaxy classification and photometry in crowded fields to fainter limits.

\end{enumerate}

\section*{Declaration of competing interest}

The authors declare that they have no known competing financial interests or personal relationships that could have appeared to influence the work reported in this paper.

\section*{Declaration of generative AI and AI-assisted technologies in the manuscript preparation process}

The authors generated all original concepts, text, and new visualizations for this article.  AI assistance was used in the following ways.  First, the authors made use of Anthropic's Claude Opus 4.8 and 5.0 in July and August of 2026 to help improve the construction of Figure~\ref{fig:fire_m11q} and the associated animation.  Claude was used to write code to query the relevant FIRE simulation data, extract star particle properties (e.g., age, location, mass, abundance information), connect the simulation data to isochrones, and stitch together individual still frames for the movie. Second, during our final internal review of the manuscript in September and October 2026, the authors used Anthropic's Claude Opus 5.5 to review the manuscript for spelling, grammar, clarity, proofreading, identifying redundancy, checking for consistent acronym usage, and resolving LaTeX errors.  After using this tool, the authors manually reviewed and edited any suggested content changes as needed and take full responsibility for the content of the published article.

\section*{CRediT authorship contribution statement}

\textbf{Daniel R.\ Weisz:} Conceptualization, Writing -- original draft, Writing -- review \& editing, Visualization. \textbf{Alessandro Savino:} Conceptualization, Writing -- original draft, Writing -- review \& editing, Visualization. \textbf{Evan D.\ Skillman:} Conceptualization, Writing -- original draft, Writing -- review \& editing.

\section*{Acknowledgments}

The authors thank the referee, Alan McConnachie, for his thorough, constructive, and critical comments, which resulted in an improved final paper. The authors are greatly indebted to the generations of astronomers whose passion and dedication have resulted in tremendous growth, scientific insights, and future prospects for resolved stellar populations in nearby galaxies.  The authors are grateful to several people who provided data used to make figures and tables in this review including Gagandeep Anand, John Cannon, Michele Cignoni, Gisella Clementini, Roger Cohen, Andrew Cole, Michelle Collins, Thomas de Boer, Meredith Durbin, Carme Gallart, Mario Gennaro, Sebastian Hidalgo, Hefan Li, Kristen McQuinn, Lidia Makarova, Pol Massana, Matteo Monelli, Andrew Pace, Ekta Patel, Raga Pucha, Tomas Ruiz-Lara, Vadim Rusakov, Elena Sacchi, Josh Simon, and Ben Williams.   The authors are thankful for the feedback and critical comments on various drafts or specific topics provided by Nate Bastian, Giuseppina Battaglia, Eric Bell, Michele Cignoni, Roger Cohen, Andrew Cole, Mario Gennaro, Sebastian Hidalgo, Rob Kennicutt, Kristen McQuinn, Burcin Mutlu-Pakdil, Andrew Pace, Ekta Patel, and Ben Williams.
DRW appreciates the kind hospitality of the Minnesota Institute for Astrophysics for hosting him for a portion of his sabbatical while this review was being written in Fall 2024.

Figure~\ref{fig:cmd_pedagogy} is based on data obtained from the ESO Science Archive Facility with DOI: \url{https://doi.org/10.18727/archive/64}, from the VMC survey (ESO programme 179.B-2003).

Figure~\ref{fig:hsc_ic1613} makes use of imaging from the DESI Legacy Imaging Surveys. The Legacy Surveys consist of three individual and complementary projects: the Dark Energy Camera Legacy Survey (DECaLS; Proposal ID \#2014B-0404; PIs: David Schlegel and Arjun Dey), the Beijing-Arizona Sky Survey (BASS; NOAO Prop. ID \#2015A-0801; PIs: Zhou Xu and Xiaohui Fan), and the Mayall z-band Legacy Survey (MzLS; Prop. ID \#2016A-0453; PI: Arjun Dey). DECaLS, BASS and MzLS together include data obtained, respectively, at the Blanco telescope, Cerro Tololo Inter-American Observatory, NSF's NOIRLab; the Bok telescope, Steward Observatory, University of Arizona; and the Mayall telescope, Kitt Peak National Observatory, NOIRLab. Pipeline processing and analyses of the data were supported by NOIRLab and the Lawrence Berkeley National Laboratory (LBNL). The Legacy Surveys project is honored to be permitted to conduct astronomical research on Iolkam Du'ag (Kitt Peak), a mountain with particular significance to the Tohono O'odham Nation.
NOIRLab is operated by the Association of Universities for Research in Astronomy (AURA) under a cooperative agreement with the National Science Foundation. LBNL is managed by the Regents of the University of California under contract to the U.S. Department of Energy.
This project used data obtained with the Dark Energy Camera (DECam), which was constructed by the Dark Energy Survey (DES) collaboration. Funding for the DES Projects has been provided by the U.S. Department of Energy, the U.S. National Science Foundation, the Ministry of Science and Education of Spain, the Science and Technology Facilities Council of the United Kingdom, the Higher Education Funding Council for England, the National Center for Supercomputing Applications at the University of Illinois at Urbana-Champaign, the Kavli Institute of Cosmological Physics at the University of Chicago, Center for Cosmology and Astro-Particle Physics at the Ohio State University, the Mitchell Institute for Fundamental Physics and Astronomy at Texas A\&M University, Financiadora de Estudos e Projetos, Fundacao Carlos Chagas Filho de Amparo, Financiadora de Estudos e Projetos, Fundacao Carlos Chagas Filho de Amparo a Pesquisa do Estado do Rio de Janeiro, Conselho Nacional de Desenvolvimento Cientifico e Tecnologico and the Ministerio da Ciencia, Tecnologia e Inovacao, the Deutsche Forschungsgemeinschaft and the Collaborating Institutions in the Dark Energy Survey. The Collaborating Institutions are Argonne National Laboratory, the University of California at Santa Cruz, the University of Cambridge, Centro de Investigaciones Energeticas, Medioambientales y Tecnologicas-Madrid, the University of Chicago, University College London, the DES-Brazil Consortium, the University of Edinburgh, the Eidgenossische Technische Hochschule (ETH) Zurich, Fermi National Accelerator Laboratory, the University of Illinois at Urbana-Champaign, the Institut de Ciencies de l'Espai (IEEC/CSIC), the Institut de Fisica d'Altes Energies, Lawrence Berkeley National Laboratory, the Ludwig Maximilians Universitat Munchen and the associated Excellence Cluster Universe, the University of Michigan, NSF's NOIRLab, the University of Nottingham, the Ohio State University, the University of Pennsylvania, the University of Portsmouth, SLAC National Accelerator Laboratory, Stanford University, the University of Sussex, and Texas A\&M University.
BASS is a key project of the Telescope Access Program (TAP), which has been funded by the National Astronomical Observatories of China, the Chinese Academy of Sciences (the Strategic Priority Research Program ``The Emergence of Cosmological Structures'' Grant \# XDB09000000), and the Special Fund for Astronomy from the Ministry of Finance. The BASS is also supported by the External Cooperation Program of Chinese Academy of Sciences (Grant \# 114A11KYSB20160057), and Chinese National Natural Science Foundation (Grant \# 12120101003, \# 11433005).
The Legacy Survey team makes use of data products from the Near-Earth Object Wide-field Infrared Survey Explorer (NEOWISE), which is a project of the Jet Propulsion Laboratory/California Institute of Technology. NEOWISE is funded by the National Aeronautics and Space Administration.
The Legacy Surveys imaging of the DESI footprint is supported by the Director, Office of Science, Office of High Energy Physics of the U.S. Department of Energy under Contract No. DE-AC02-05CH1123, by the National Energy Research Scientific Computing Center, a DOE Office of Science User Facility under the same contract; and by the U.S. National Science Foundation, Division of Astronomical Sciences under Contract No. AST-0950945 to NOAO.

\section*{Funding}

This research did not receive any specific grant from funding agencies in the public, commercial, or not-for-profit sectors.

\section*{Data availability}

No new data were generated in support of this review. The data underlying the figures and tables are drawn from the published literature cited herein and, where indicated, were kindly provided by the colleagues named in the Acknowledgments.

\appendix

\section{Summary of Important Observations}
\label{sec:important_obs}

In this appendix, we summarize many of the major observational investments that have targeted the resolved stellar populations of nearby galaxies or happen to include them in the survey's footprint (e.g., \textit{Gaia}, \textit{Euclid}).  For clarity of organization, we divide the programs based on the facility (e.g., \hubble) or type of facility (e.g., ground-based telescopes) used to acquire the data.  We further distinguish between surveys (i.e., programs that involve many observations, usually of different targets or multiple fields within a single galaxy) and observations of individual targets of particularly high science value (e.g., the most metal-poor galaxy).  There is some overlap between these two areas (e.g., M31 is a single object, but most major observations of it came in the form of surveys).

Programs are approximately chronologically ordered, with the parenthetical date indicating the year data was first acquired.  While we generally focus on programs within the last $\sim20$~years, we do include select legacy datasets that have greatly contributed to scientific output during the period of this review, even if the data were acquired earlier.  Very new programs or facilities that have yet to take data and/or publish papers related to resolved stellar populations analysis are generally not included in this appendix.

For each entry in the following subsections, we provide a brief summary of the observations and highlight some of the main science goals and/or outcomes associated with the data. We link to the publicly available data products and/or the project website, when possible.

\subsection{Hubble Space Telescope}

\hubble\ has been the workhorse facility for extragalactic resolved stellar populations studies over the past $\sim20$~years.  Its high-angular resolution, excellent sensitivity, and broad wavelength coverage have made nearby galaxy studies one of the largest areas of data collection over \hubble's lifetime.  Below, we summarize many of the major projects in resolved stellar populations undertaken by \hubble.

\subsubsection{HSTPROMO (2000)} The High-resolution Space Telescope Proper Motion Collaboration (HSTPROMO) is a science collaboration that uses a variety of sensitive, high-angular resolution imaging observations to study the motions and dynamics of stars, stellar systems, galaxies, and black holes using high-precision astrometry.  It is the oldest continuous science program that makes use of resolved stars to study nearby galaxies.  The majority of published results from HSTPROMO have made use of \hubble\ observations, though over the last decade it has expanded to include data from \textit{Gaia} and \jwst.  A brief overview of HSTPROMO is given in \citet{Vandermarel:2014b}.  To date, HSTPROMO has acquired data from $>70$ \hubble\ programs, $>10$ \jwst\ programs, and a variety of \textit{Gaia} data.  An overview of the collaboration and its publications is maintained on the \href{https://www.stsci.edu/~marel/hstpromo.html}{HSTPROMO website}.

Some of the main published science results from HSTPROMO on nearby galaxies include the bulk and internal motions of the Magellanic Clouds \citep[e.g.,][]{Kallivayalil:2006b, Kallivayalil:2006a, Kallivayalil:2013, VanderMarel:2014, Libralato:2023}, the 3D orbit of M31 \citep[e.g.,][]{Sohn:2012, VanderMarel:2012b, VanderMarel:2012a, VanderMarel:2019}, measurements of the orbital histories of dozens of dwarf galaxies in the LG \citep[e.g.,][]{Sohn:2013, Sohn:2017, Fritz:2018, Sohn:2020, Bennet:2024, Patel:2026, Warfield:2026}, and the measurement of internal motions of stars in MW satellite galaxies Draco and Sculptor \citep{Vitral:2024,Vitral:2026}.   HSTPROMO has also begun to investigate the promise of \textit{Roman} for proper motion studies \citep{McKinnon:2026}. Projects done by or in tandem with HSTPROMO have contributed to significant technical advancements that enable high-precision astrometry \citep[e.g.,][]{Anderson:2000, Anderson:2003, Anderson:2010, Anderson:2022, Warfield:2026}.

\subsubsection{M31's Stellar Halo (2002)} A series of large \hubble\ programs obtained deep optical ACS imaging of several fields in the stellar halo and outer disk of M31 in order to measure its SFH.  These were the first CMDs to reach below the oMSTO in an $L_{\star}$ galaxy outside the MW.  The first program (GO-9453, PI Brown) used 126 orbits to acquire ACS F606W and F814W imaging of a single field along the minor axis of M31, 11~kpc from the nucleus.  A second program (GO-10265, PI Brown) used 107 orbits to acquire ACS F606W and F814W imaging that reached below the oMSTO in M31's outer disk ($\sim25$~kpc from the nucleus along the major axis) and tidal stream.  Finally, a third program (GO-10816, PI Brown) was allocated 128 orbits to acquire ACS F606W and F814W imaging reaching below the oMSTO in four fields in M31's stellar halo; three were observed (one at 21~kpc and two at 35~kpc from M31's nucleus) before ACS/WFC failed in early 2007.  The goal of each program was to measure the SFH of each field in order to reconstruct the assembly history of M31's stellar halo.  These programs produced several papers on the SFH of M31's halo and streams, GCs, and RR Lyrae populations \citep[e.g.,][]{Brown:2003, Brown:2004a, Brown:2004b, Brown:2006, Brown:2008, Jeffery:2011}.  In parallel, a series of \hubble/ACS programs led by A.~Ferguson (GO-9458, GO-10218) placed oMSTO-depth pointings on specific substructures identified in the wide-field INT Wide Field Camera and PAndAS maps of M31's halo \citep[e.g.,][]{Ferguson:2002, McConnachie:2009}, in order to measure the ages and metallicities of the streams, shells, and outer disk features in the context of the wide-field map \citep[e.g.,][]{Richardson:2008, Bernard:2012, Bernard:2015a}.  The small \hubble\ field of view relative to the size of M31 means that pointed programs of this type are most powerful when explicitly coordinated with wide-field imaging, a strategy we highlight in \S \ref{sec:future}.  \href{https://archive.stsci.edu/prepds/andromeda}{Data from these programs} is available on the Barbara A. Mikulski Archive for Space Telescopes (MAST).

\subsubsection{GHOSTS (2003)} The Galaxy Halos, Outer disks, Substructure, Thick disks, and Star clusters survey (GHOSTS) \citep{Radburn-Smith:2011} consisted of three \hubble\ programs (HST-GO-9765, 10523, 10889;
    PI: de Jong, R.) which acquired 18 orbits as a Snapshot (SNAP) program, 36 orbits as a SNAP program; and 128 orbits as a GO program, respectively.  All data were taken with \hubble/ACS in the F606W and F814W filters.  The main science goal of GHOSTS was to study the resolved stellar populations in the halos of 14 disc galaxies within 17 Mpc.

    The project began with single orbit \hubble/ACS observations in the halos of 18 nearby galaxies, showing the power of \hubble to deliver resolved star observations in these galaxies \citep{Seth:2005_a}, particularly to allow the measurement of the scale heights of halo populations in edge-on disk galaxies \citep{Seth:2005_b}.  The second program provided multiple fields in galaxies within 5 Mpc, allowing detailed measurements of different stellar populations at very low surface brightness \citep[e.g.,][]{deJong:2007}. The third program preferentially targeted edge-on galaxies \citep{Radburn-Smith:2011}. A detailed study of NGC~7793 showed that the older stellar population was much more extended than the younger population, supporting the view that extended stellar disks are grown through migration \citep{Radburn-Smith:2012}.  A detailed study of the stellar halos of six highly inclined galaxies revealed large galaxy-to-galaxy variations in the halo populations, reflecting the stochasticity of the galaxy assembly process \citep{Monachesi:2016}.  This theme was expanded upon in \citet{Bell:2017} and \citet{Harmsen:2017}.  \href{https://archive.stsci.edu/prepds/ghosts}{GHOSTS data} is available on MAST.

\subsubsection{LCID (2005)} The Local Cosmology from Isolated Dwarfs (LCID) program (HST-GO-10505, 10590; PIs: Gallart, C., Cole, A.) consists of 113 orbits of \hubble/ACS F475W and F814W observations of five dwarf galaxies in the periphery of the LG obtained in cycle 14, along with deep archival \hubble/WFPC2 observations of a sixth galaxy, Phoenix in F555W and F814W. The purpose of these observations was to constrain the onset of star formation in isolated dwarfs of different morphological types (dwarf Spheroidal, dwarf Irregular, and dwarf Transition).

The studies appeared as analyses of individual galaxies' SFHs, starting with the analysis of Leo~A by \citet{Cole:2007}. This study confirmed earlier indications that Leo~A experienced a delayed onset of the majority of its star formation, creating the hypothesis that isolated dwarf galaxies evolve more slowly than those in denser environments. Subsequent studies of the dSphs Cetus \citep{Monelli:2010_a} and Tucana \citep{Monelli:2010_b} showed that the duration of the early star formation of these two galaxies differed by roughly one Gyr, indicating that the star formation in Cetus was not strongly affected by the radiation from the epoch of reionization. The SFH of the transition dwarf Pisces~I was presented in \citet{Hidalgo:2011} and revealed a significant early episode of star formation with continuing low level star formation up to the current epoch, remarkably similar to that observed in the dwarf transition galaxy Phoenix \citep{Hidalgo:2009}.  Finally, the SFH of the dIrr IC~1613, along with a comparison to the other LCID galaxies, was published in \citet{Skillman:2014}.  It was highlighted that the nearly constant star formation over the age of the universe provides a challenge for models where star formation is tied to the infall rate.

Other LCID studies include the identification of the RGB bump \citep{Monelli:2010_c}, the identification of BSS populations in Cetus and Tucana \citep{Monelli:2012}, and the radial gradients in the LCID galaxies \citep{Hidalgo:2013}.  The deep \hubble\ observations allowed the identification and characterization of the short period variable star populations \citep{Bernard:2008, Bernard:2009, Bernard:2010, Bernard:2013}.  \citet{Gallart:2015} provides the main overview of the program along with new insight into halo assembly bias for low-mass galaxies.

 \subsubsection{ANGST (2006)} The ACS Nearby Galaxy Survey Treasury (ANGST, HST-GO-10915 and HST-DD-11307; \citealt{Dalcanton:2009}) is a 295 orbit cycle 15 \hubble\ Treasury program that acquired optical imaging (F475W, F606W, F814W) of 69 galaxies within 4~Mpc using ACS and WFPC2.  The galaxies span a range of $\sim10^4$ in stellar mass.  The primary science goals of ANGST include measuring the SFH of the LV, calibrating stellar evolution models for bright stars, stellar radial gradients, and connecting recent SFHs to properties of the ISM.  Some of the major results from ANGST include the calibration of AGB star models \citep{Girardi:2010}, uniformly measured SFHs of nearby dwarf galaxies \citep{Weisz:2011}, and investigations of radial gradients in nearby galaxies \citep{Gogarten:2010}. Combined with archival imaging, ANGST provides uniformly measured photometric data for $>14$ million stars.  \href{https://archive.stsci.edu/prepds/angst/}{ANGST data products} can be downloaded from MAST.

 ANGST is related to two other \hubble\ programs that target LV galaxies.  The first is a Cycle 15 \hubble\ archival legacy program titled the Archive of Nearby Galaxies: Reduce, Reuse, Recycle (ANGRRR).  This program provides a uniform optical and ultra-violet photometry database of resolved stars in nearby galaxies within 5 Mpc. \href{https://archive.stsci.edu/prepds/angrrr/}{ANGRRR data products} can be downloaded from MAST.

The second is a cycle 17 \hubble\ SNAP program that uses 62 orbits to acquire WFC3/IR imaging of bright resolved stars in 23 nearby galaxies (HST-GO-11719, PI Dalcanton; \citealt{Dalcanton:2012a}).  Some results from this program include a quantification of how young, evolved, IR-bright stars (e.g., RHeBs, AGB) contribute to the SEDs of star-forming galaxies at high-redshifts \citep{Melbourne:2012} and improved calibrations for theoretical models of low-thermally pulsating AGB stars \citep{Rosenfield:2014}.

\subsubsection{PHAT (2010)} The Panchromatic Hubble Andromeda Treasury (PHAT) program (HST-GO-12055, 12116, PI Dalcanton; \citealt{Dalcanton:2012b, Williams:2014, Williams:2023}) is a Cycle 18 828 orbit \hubble\ multi-cycle Treasury survey of $\sim$1/3 of M31's star-forming disk in 6 filters, including the near-UV (F275W, F336W), optical (F475W, F814W), and near-IR (F110W, F160W).  It is the largest single program targeting the resolved stellar contents of nearby galaxies in the history of \hubble. Some of the main science goals of PHAT include measuring the high-mass stellar IMF, providing resolved star imaging for a rich and diverse set of star clusters, characterizing star formation as a function of galactic position, creating spatially resolved spectral energy distributions for millions of individual stars in various phases of evolution, probing the interactions between stars and gas on parsec scales across a large portion of a disk galaxy, and providing an unprecedented legacy dataset for future use by the community.  PHAT underwent 3 data releases over the course of $\sim10$~years, each the product of important new technical developments in crowded-field photometry and image processing techniques.  The final data release includes $\sim138$~million stars.  \href{https://archive.stsci.edu/hlsp/phat}{Data products from PHAT} are available on MAST.

PHAT has produced a number of science results across a wide range of topics including the spatially resolved SFH of an $L_{\star}$ galaxy \citep{Lewis:2015, Williams:2017}, insights into star cluster formation and evolution \citep{Fouesneau:2014, Johnson:2016, Johnson:2017}, a study of rare phases of evolution in M31's bulge \citep{Rosenfield:2012}, a parsec-scale dust map across M31's disk \citep{Dalcanton:2015}, the most precise measurement of the high-mass IMF to date \citep{Weisz:2015b}, a global burst of star formation in M31's disk \citep{Williams:2015}, an analysis of stellar populations associated with core-collapse supernovae \citep{Williams:2018}, and a study of dust around hot massive stars in M31 \citep[e.g.,][]{Lindberg:2024}.

\subsubsection{Milky Way Ultra-Faint Dwarf Galaxies (2011)} This survey (GO-12549, PI: T. Brown; \citealt{Brown:2012, Brown:2014}) is a 113 orbit Cycle 19 program that obtained deep ACS F606W and F814W imaging of 6 UFD galaxies around the MW. The primary goal of this program is the study of the SFH of what, at the time, some suggested might be a new class of galaxies, specifically testing the hypothesis of cosmologically-driven synchronization. The result of this program \citep{Brown:2014} found ubiquitous ancient ages ($>$12.5Gyr) for the stellar populations in the sample, providing strong empirical support to the idea that reionization is a major driver of very-low mass galaxy evolution. Subsequent publications related to this dataset included studies of the stellar IMF of UFD stellar populations \citep[e.g.,][]{Geha:2013, Gennaro:2018}.

\subsubsection{HTTP (2012)} The Hubble Tarantula Treasury Project (HTTP; GO-12939, PI: Sabbi; \citealt{Sabbi:2013, Sabbi:2016}) is a 60 orbit cycle 20 \hubble\ imaging survey of the Tarantula Nebula in the LMC.  Using ACS and UVIS, HTTP acquired new UV, optical, and IR imaging in 6 filters (F275W, F336W, F555W, F658N, F110W, F160W) that overlapped with existing \hubble\ F775W imaging.  The science aims of this program include the study of massive star formation in a region of low metallicity and high star formation intensity (30 Dor), improve our knowledge of stellar evolution (e.g., massive stars, cool stars, pre-MS stars), measure the SFH of the Tarantula Nebula, determine proper motions of massive stars, and much more. HTTP has produced a number of publications touching a variety of science such as the dust law, pre-main sequence stars, star formation, and star cluster dynamics \citep[e.g.,][]{Sabbi:2012, Cignoni:2015, DeMarchi:2016, Sabbi:2016, Ksoll:2018}.  \href{https://archive.stsci.edu/hlsp/http}{HTTP data} is available on MAST.

\subsubsection{LEGUS (2013)} The Legacy ExtraGalactic UV Survey (LEGUS) is a 164 orbit \hubble\ Cycle 21 Treasury program that obtained five-band WFC3/UVIS UV and optical imaging (F275W, F336W, F438W, F555W, F814W) and three-band ACS optical imaging (F435W, F606W, F814W) of 50 star-forming galaxies with distances of $\sim3.5$--$12$~Mpc and that span a factor of $\sim10^4$ in stellar mass \citep{Calzetti:2015}.  The primary science goals of the program include investigating the hierarchy of star formation, the formation and evolution of star clusters, the upper-end of the high-mass stellar IMF, calibrating SFR indicators, constraining the progenitors of core collapse supernovae, and investigating multiple populations in star clusters.    These science areas involve the use of both resolved and unresolved stellar populations.   LEGUS has produced several publications based on resolved stellar populations primarily focused on measurements of recent SFHs from multi-wavelength CMDs \citep[e.g.,][]{Cignoni:2018, Sacchi:2018, Cignoni:2019, Sacchi:2019} as well as a database of resolved stars \citep{Sabbi:2018}.  \href{https://archive.stsci.edu/prepds/legus/}{LEGUS data products} can be downloaded from MAST.

\subsubsection{ISLAndS (2013)} The Initial Star formation and Lives of the ANDromeda Satellites (ISLAndS) program (HST-GO-13028, HST-GO-13739; PI: Skillman; \citealt{Skillman:2017}) consists of 111 orbits of \hubble/ACS F475W and F814W imaging of six Andromeda satellites obtained in cycles 20 and 22. The purpose of these observations was to determine if the first epoch of star formation occurred at the same time in the M31 satellites as in the MW satellites.  The imaging reached below the oMSTO in all six galaxies.  The focus of the program was on the more massive dSphs in order to have strong constraints on the SFHs, i.e., not limited by small number statistics.   The \hubble observations of And XVI showed it to be a UFD galaxy that was not quenched by reionization, which is a departure from the trend established by the MW UFDs \citep{Brown:2014, Weisz:2014c, Monelli:2016}. Comparing M31 and MW satellite dSph galaxies of similar mass, the most significant difference appears to be a lack of late-quenching (${\tau}_q$ $\le$ 5 {Gyr}) in the M31 satellite sample \citep{Skillman:2017}.  The deep \hubble observations allowed the identification and characterization of the short period variable star populations \citep[e.g.,][]{Martinez-Vasquez:2017}.

\subsubsection{Milky Way Faint Satellite Treasury Survey (2016)} This is a 164 orbit Cycle 24 program (HST-GO-14734, PI Kallivayalil), targeting 30 dwarf galaxies within 420~kpc from the MW (primarily UFDs). The program acquired \hubble/ACS F606W and F814W imaging, reaching several magnitudes below the oMSTO of all targets. The primary goal of the program is to establish the first astrometric epoch for galaxies that had no previous suitable \hubble\ observations. In turn, this will enable the determination of accurate proper motions, from which the orbital history and the internal kinematic of the targets can be studied. An additional goal of this program is the study of the targets' SFH, which is enabled by the deep 2-band photometry. The data from this program have been used to study the structural properties of the MW satellites \citep{Richstein:2022}, constrain their chemical evolution \citep{Alexander:2023}, and study their SFH \citep{Sacchi:2021}. The latter study suggested differences in the quenching epochs of UFDs associated with different kinematic groups (e.g., MW vs LMC satellites), which may carry implications for the homogeneity of reionization.

\subsubsection{PHATTER (2017)} The Panchromatic Hubble Andromeda Treasury: Triangulum Extended Region program (PHATTER; HST-GO-14610, PI Dalcanton; \citealt{Williams:2021}) is a 108 orbit cycle 24 \hubble\ Treasury program targeting the inner 14 kpc$^2$ of M33.  PHATTER largely followed the PHAT observing strategy and obtained imaging in 6 filters including the near-UV (F275W, F336W), optical (F475W, F814W), and near-IR (F110W, F160W). The program shares many of the same science goals as PHAT, though aimed at quantifying the effects of environment as M33 is a lower mass and lower metallicity system than M31. The main PHATTER data release includes photometry for $\sim22$~million stars.   PHATTER has produced several publications on M33 including a catalog and analysis of its star cluster populations \citep{Johnson:2022, Wainer:2022}, a spatially resolved recent SFH \citep{Lazzarini:2022}, and a determination of its galactic structure \citep{Smercina:2023}.  \href{https://archive.stsci.edu/hlsp/phatter}{PHATTER} data products are available on MAST.

\subsubsection{LUVIT (2018)} The Local Ultraviolet to Infrared Treasury (LUVIT) program (HST-GO-15275 PI: Gilbert and HST-GO-16162 PI: Boyer; \citealt{Gilbert:2025}) consists of two \hubble\ programs targeting 22 low-mass, nearby galaxies ($D<3.5$~Mpc).  The first (GO-15275) is a cycle 25 WFC3/UVIS 58 orbit program that obtained F275W and F336W in order to study hot, massive stellar populations.  The second (GO-16162) is a  50 orbit cycle 28 optical (WFC3/UVIS) and near-IR (WFC3/IR) survey of an overlapping sample in broad and medium bands aimed primarily at cool, evolved stars.  Along with archival optical \hubble data, this program provides panchromatic near-UV through near-IR coverage for all 22 systems.  This program aims to study star formation, hot stars, evolved stars, dust production and ISM physics at low-metallicities.  The survey strategy and broadband catalogs were published in \citet{Gilbert:2025}. \citet{Gull:2022} used LUVIT data to study massive stars in Leo~A and \citet{Choi:2025} measured recent SFHs of several LUVIT galaxies using UV and optical CMDs.

\subsubsection{M31 Satellites Treasury Program (2019)} The \hubble\ M31  Satellite Galaxy Survey Treasury is a 244 orbit Cycle 27 Treasury program (HST-GO-15902, PI Weisz; \citealt{Savino:2025}) that obtained \hubble/ACS imaging in F606W and F814W of 23 dwarf galaxies within $\sim300$~kpc of M31, with the goal of obtaining photometry deep enough to reach the oMSTO. In addition, the program has re-analyzed archival observations of virtually any galaxy within 500~kpc of M31, assembling a homogeneous dataset that covers virtually the entire satellite system of M31. The scientific goals of the program are centered around studying the SFH and quenching of M31 satellites, laying the foundations for measuring the 6D properties of the sample, by updated distances and by establishing a first proper motion epoch, in order to reconstruct the first star formation and assembly history of a complete satellite system outside the MW. This program used RR Lyrae to make a homogeneous 3D map of the M31 satellite system (the first such map, based on TRGB distances, was presented by \citealt{McConnachie:2006}) and showed that M31's satellites are asymmetrically distributed \citep{Savino:2022}, found that M31 UFDs have more extended SFHs than their MW counterparts \citep{Savino:2023}, and provided the first lifetime SFHs of a satellite system outside the MW measured from oMSTO depth CMDs \citep{Savino:2025}.  \href{https://archive.stsci.edu/hlsp/m31-satellites}{Data products} from this program are available via MAST.

\subsubsection{GULP (2020)} The Galaxy
UV Legacy Project (GULP; GO-16316, PI Sabbi) is an 84 orbit Cycle 28 \hubble\ Treasury survey of 27 nearby star-forming galaxies, ranging from low-metallicity dwarfs to Solar-metallicity spirals.  It supplements archival \hubble\ imaging by acquiring new UVIS imaging, mainly in F218W, to ensure all galaxies have uniform UV and optical wavelength coverage in 7 bands (F218W, F275W, F336W,
F438W/F435W, F555W, F814W, and F657N).  It also adds bluer UV coverage using the ACS Solar Blind Channel (F150LP).  The science aims of the program include calibrating massive star evolutionary models, exploring environmental variations in high-mass IMF, studying the role of stellar feedback in star-forming galaxies, and generating a holistic picture of massive star evolution from their birth to their death. An early publication from GULP is aimed at the hierarchical nature of star formation in NGC~4449 \citep{Meena:2025}.  \citet{Sabbi:2026} provides an overview of the survey.

\subsubsection{Scylla (2020)} Scylla is a pure parallel \hubble\ program (\hubble-GO-15891, GO-16235, GO-16786, PI Murray; \citealt{Murray:2024b}) that ran in tandem with the \hubble\ UV Legacy Library of Young Stars as Essential Standards (ULLYSES) survey from cycles 27-31 \citep[\hubble-DD-16090-16859, DD-16930, DD-17297, DD-17296, PI Roman-Duval;][]{Roman-Duval:2025}.  It used 342 orbits with ACS and WFC3 (UVIS and IR) to target 96 non-contiguous fields in the central regions of LMC and SMC in seven broadband filters (F225W, F275W, F336W, F475W, F814W, F110W, F160W) in the near-UV, optical, and near-IR.  The main goals of Scylla are to determine variations in dust properties throughout the LMC and SMC, map the multidimensional structure of the ISM in the LMC and SMC, and to measure detailed SFHs from CMDs that reach well-below the oMSTO.  \citet{Murray:2024b} provides an overview of the Scylla survey.  Data from Scylla have been used to measure the global and spatially resolved SFHs of the SMC and LMC \citep{Cohen:2024b, Cohen:2024a}, to constrain line-of-sight extinction properties for $>1$ million stars \citep{Lindberg:2025}, to search for bursts of star formation \citep{Burhenne:2026} and connect them to chemical enrichment \citep{Escala:2026}.  \href{https://archive.stsci.edu/hlsp/scylla}{Data from Scylla} is available on MAST.

\subsubsection{PHAST (2021)} The Panchromatic Hubble Andromeda Southern Treasury (PHAST; HST-GO-16778, PI Williams; \citealt{Chen:2025}) is a 195 orbit \hubble\ Treasury survey of the southern portion of M31's star-forming disk in 4 filters in the near-UV (F275W, F336W) and optical (F475W, F814W).  Along with PHAT it provides nearly-complete coverage of M31's star-forming disk with \hubble\ at near-UV and optical wavelengths.  The first release of the PHAST stellar catalog contains $\sim96$~million stars.  All imaging has been reduced and an initial combined PHAT and PHAST image of M31 has been released, along with structural analysis of M31's stellar populations \citep{Chen:2025}. \citet{Wainer:2026} completed an analysis of the spatially resolved recent SFH of M31 using the combined PHAT and PHAST datasets.  \href{https://archive.stsci.edu/hlsp/phast}{PHAST data products} are available on MAST.

\subsubsection{GLOW (2026)} The Galaxies Losing Oxygen via Winds (GLOW) project \citep{McQuinn:2026} is an archival program that has assembled a uniformly processed atlas of 37 gas-rich, star-forming, low-mass galaxies within $\sim6$~Mpc.  The atlas combines resolved star photometry from archival \hubble\ optical imaging with Very Large Array (VLA) \hi{}, \textit{Spitzer} 3.6~$\mu$m, and ground-based optical imaging, along with TRGB distances, direct-method gas-phase abundances, and environmental measures for every galaxy.  Its primary aim is to characterize the star formation--chemical enrichment cycle of low-mass galaxies on a galaxy-by-galaxy basis. It also provides a homogeneous CMD-based stellar mass and SFH sample for the LV that complements ANGST and LUVIT \citep[e.g.,][]{Cohen:2026b}.

\subsection{James Webb Space Telescope}

Launched in December 2021, \jwst\ has rapidly emerged as a premier facility for resolved star studies.  Like \hubble, \jwst\ features excellent angular resolution and sensitivity, while extending coverage into near-IR wavelengths.  Among the first science observations taken by \jwst\ were of resolved stars in the MW GC M92, which were used both to help with early calibrations of \jwst\ and to showcase its transformative potential for resolved star studies (e.g., easy access to lower MS stars).  With an anticipated lifetime of up to $\sim20$~years, \jwst\ should enable resolved stellar populations studies that both build upon \hubble and enable studies of qualitatively new environments.  Here, we summarize many of the major \jwst\ studies of resolved galaxies to date.  As with \hubble, all raw data (i.e., images) are available on MAST. We provide links to catalogs when possible.

\subsubsection{GTO and Commissioning observations of Resolved Stars (2022)} As part of its guaranteed time observations, \jwst\ acquired several observations of resolved stellar populations in nearby galaxies.  These include NGC~6822 (20.8 hr, GTO-1234, PI Meixner), I~Zw~18 (6.5 hr, GTO-1233, PI Meixner; see Appendix~\ref{sec:individual_galaxies} for details), NGC~346 (30.3 hr, GTO-1227, PI Meixner), LMC-N79 (16.9 hr, GTO-1235, PI Meixner), select MW (4.8 hr, GTO-1304, PI van der Marel) and M31 (5.5 hr, GTO-1305, PI van der Marel) satellites, with NIRCam and/or MIRI for imaging in a variety of filters.  Several of these programs have published results including the CMDs and the embedded stellar populations of NGC~6822 \citep{Nally:2024}, the stellar content of NGC~346 \citep{Habel:2024, Jones:2024b}, and the stellar populations of I~Zw~18 \citep[e.g.,][]{Hirschauer:2024, Bortolini:2024b}.

Beyond the GTO programs, \citet{Libralato:2023} make use of NIRISS imaging of the LMC from the \jwst\ commissioning program to provide effective PSFs and distortion solutions, as well as demonstrate the potential of NIRISS for proper motion measurements.

\subsubsection{JWST-STARS (2022)}  The \jwst\ Early Release Science (ERS) Program for Resolved Stellar Populations (JWST-STARS) is a 27.5~hr cycle 1 \jwst\ director's discretionary time program.  It uses \jwst/NIRCam and NIRISS to acquire multi-band IR imaging of three nearby stellar systems: M92 (MW GC), \dracoii\ (UFD), and WLM (star-forming dwarf galaxy).  The main goals of the program are to develop NIRCam and NIRISS specific modules for DOLPHOT, a widely-used crowded-field stellar photometry package, and to serve as a pathfinder for \jwst\ observations of resolved stellar systems.  The program has published several science and technical papers including early calibrations of NIRCam's zero points \citep{Boyer2022}, the first SFH of a nearby galaxy (WLM) based on \jwst\ imaging \citep{McQuinn:2024a}, the efficient identification of AGB stars in \jwst\ filters \citep{Boyer2024}, a detailed stress test of the \jwst\ exposure time calculator \citep{Savino:2024}, RR Lyrae in \jwst\ filters \citep[e.g.,][]{Slaughter:2026}, and a description of the DOLPHOT NIRCam and NIRISS modules \citep{Weisz:2024}.  The survey strategy is described in \citet{Weisz:2023a} and \href{https://archive.stsci.edu/hlsp/jwststars}{data products from this program} are available on MAST and \jwst\ tutorials are available on the \href{https://ers-stars.github.io}{ERS team webpage}.

\subsubsection{FEAST (2023)} The Feedback in Emerging extrAgalactic Star clusTers (FEAST; GO-1783, PI Adamo) is a 52.0 hr cycle 1 \jwst\ program.  It has acquired a broad array of \jwst/NIRCam and MIRI imaging, along with NIRSpec spectroscopy from a related program (GO-3503), of select galaxies (NGC~628, M83, M51, NGC~4449, NGC4485-NGC4490) within 11~Mpc.  Though many of the science goals for FEAST focus on integrated light data (e.g., unresolved star clusters, H~II regions) in tandem with data from \hubble, MUSE, and ALMA, a portion of FEAST's data includes the study of galaxies through their resolved stellar content.  Publications based on FEAST's resolved stellar populations program include the young stellar populations of starburst dwarf galaxy NGC~4449 \citep{Correnti:2025} and interacting galaxies NGC~4485 and NGC~4490 \citep{Bortolini:2025}.   More details are available on \href{https://feast-survey.github.io}{the FEAST team webpage}.

\subsection{Ground-Based}

Here, we summarize major ground surveys that have specifically targeted nearby galaxies in order to study their resolved stellar populations and/or discover new systems.  We provide a brief overview of each survey and highlight major science aims and results.  We link to available data when possible.  We include a handful of particularly high impact ground-based datasets that pre-date the nominal review periods.

\subsubsection{MCPS (1996)}

The Magellanic Cloud Photometric Survey (MCPS) used the Great Circle Camera on the Las Campanas Swope Telescope to obtain U, B, V, I imaging of the LMC and SMC.  It covered $\sim64$~deg$^2$ in the central region of the LMC and $\sim18$~deg$^2$ in the central region of the SMC.  Typical CMDs reached $\sim20$th magnitude in regions of low stellar density, which is brighter than the oMSTO.  The MCPS was the first wide-area digital survey of resolved stars in the Magellanic Clouds.  It produced a number of seminal science papers, including the first wide-area detailed internal reddening maps \citep{Zaritsky:2002, Zaritsky:2004} and the first spatially resolved SFHs of the LMC and SMC \citep{Harris:2004, Harris:2009}.  The MCPS catalogs can be found on the Strasbourg astronomical Data Center (CDS) for the \href{https://doi.org/10.26093/cds/vizier.51230855}{SMC} and the \href{https://cdsarc.cds.unistra.fr/viz-bin/cat/J/AJ/128/1606}{LMC}.

\subsubsection{LGGS (2000)}

The Local Group Galaxy Survey (LGGS) used the 4m telescope at Kitt Peak National
Observatory (KPNO; $35' \times 35'$ field-of-view) and the 4m telescope at the Cerro Tololo Inter-American Observatory (CTIO; $15' \times 15'$ field-of-view) to obtain broad- (UBVRI) and narrow-band ($H\alpha$, [SII], [OIII]) imaging for 9 galaxies in or near the LG \citep[][]{Massey:2006, Massey:2007a}. It is one of the first CCD-based multiband surveys of nearby galaxies. The science goals of the program were largely aimed at the study of massive stars (e.g., 20 \msun) and their surroundings with high photometric precision ($\sim0.02$~mag).  The survey depths range from $\sim22$--$24$~mag depending on the band, galaxy distance, extinction (internal and foreground), and degree of crowding.  Data from this program have been used for a wide variety of science from study of massive star evolution \citep[e.g.,][]{Massey:2007b, Drout:2009, Massey:2009,Drout:2012, Neugent:2020} to star clusters and the clustering of star formation \citep[e.g.,][]{Bastian:2007, Caldwell:2009}.  Photometric catalogs are available from the \href{http://www2.lowell.edu/users/massey/lgsurvey/}{LGGS website}.

\subsubsection{PAndAS (2008)} The Pan-Andromeda Archaeological Survey (PAndAS) \citep{McConnachie:2009,McConnachie:2018} used the 1-square-degree field-of-view of the MegaPrime/MegaCam camera on the 3.6 m Canada-France-Hawaii Telescope (CFHT) to create mosaic images in g and i bands of M31 and M33 and their surroundings. The survey covers more than 400~deg$^2$ and extends to a maximum projected radius from M31’s center of 150 kpc. PAndAS built on earlier Isaac Newton Telescope \citep[e.g.,][]{Irwin:2005} and CFHT \citep{Ibata:2007} wide-field photometric studies.

The PAndAS project continued the ongoing discovery and characterization of M31 satellite companion galaxies \citep{Martin:2009, Richardson:2011, Conn:2012, Martin:2013, Crnojevic:2014, Martin:2016a, Doliva-Dolinsky:2022}.  It also cataloged and characterized M31's GCs \citep{Mackey:2010, Cockcroft:2011, Mackey:2013, Veljanoski:2013, Huxor:2014, Mackey:2019A, Mackey:2019B}.

The PAndAS project provided a fantastic overall view of the structure of the M31/M33 system \citep{McConnachie:2009, McConnachie:2010, Carlberg:2011, Lewis:2013, Cockcroft:2013, Ibata:2014, Bate:2014, Martin:2014, Conn:2016, McMonigal:2016, McConnachie:2018, Doliva-Dolinsky:2022, Doliva-Dolinsky:2023}, with special attention to the thin plane of dwarf galaxies rotating around M31 \citep{Ibata:2013, Conn:2013}.  \href{https://www.cfht.hawaii.edu/Science/PandAS/}{PAndAS data products} are hosted by the Canadian Astronomy Data Center.

\subsubsection{Subaru Survey of Faint MW Satellites (2008)} \citet{Okamoto:2012} used Subaru / Suprime Cam to obtain among the first imaging of MW UFDs to extend below the oMSTO over a large area.  They imaged three UFDs (Boo~I, CVn~II, and Leo~IV) and the faint classical dwarf CVn~I in the V and I bands.  The data spatially extended out to tens of arcminutes from each galaxy's center.  The CMD of Boo~I extended below the oMSTO.  The CMDs of the other three systems marginally reached the oMSTO, once photometric quality criteria were considered.  CMDs were used to qualitatively discuss the stellar populations of each galaxy.

\subsubsection{A MegaCam Survey of Outer Halo Satellites (2009)} This survey used MegaCam on the CFHT, the Magellan-Clay telescope, and archival DECam imaging to measure $g$- and $r$-band photometry for 58 stellar overdensities located $\sim25$--$400$~kpc from the MW. The survey depths are $g\sim25.6$ and $r\sim25.3$.  The main science goal of this survey is to provide homogeneous photometric and structural parameters for a large sample of the more distant MW satellites.  An overview of the survey is given in \citet{Munoz:2018} and photometric and structural parameters are presented in \citet{munoz:2018b}.  Data from this survey have also been used in other contexts such as discovering an ultra-faint star cluster in the direction of the Ursa~Minor dSph \citep{Munoz:2012}, studying BSSs in GCs and MW satellite galaxies \citep{Santana:2013}, and measuring the SFH of Carina \citep{Santana:2016}.  \href{https://cdsarc.cds.unistra.fr/viz-bin/cat/J/ApJ/860/65}{Data products} for this survey are available via CDS.

\subsubsection{VMC (2009)} The VISTA Magellanic Cloud (VMC) survey \citep{Cioni:2011} is a public ESO survey of the Magellanic Cloud system, carried out on the VISTA telescope from 2009 to 2018. The survey obtained $YJK_s$ time-series photometry, down to $K_s \sim 21$ of 170 deg$^2$, over a region comprising the LMC, the SMC, and the major tidal features of the Magellanic system. The primary goals of the survey are the reconstruction of a spatially resolved SFH of the Magellanic system, the derivation of precise 3D morphologies for the different stellar components in/around the LMC/SMC, and enabling a vast range of ancillary science.  A deep extension, VMCDeep, has obtained additional $J$ and $K_s$ imaging of the crowded LMC bar and SMC core ($\sim1.8$~deg$^2$) that reaches below the RC and extends the proper motion and variability time baseline to $\sim11$~yr \citep{Cioni:2026}.

The main results from this program include detailed SFH maps across the VMC footprint \citep{Rubele:2012,Rubele:2015,Rubele:2018, Mazzi:2021}, an extensive study of the variable star population in the Magellanic Clouds, which resulted in distance calibrations \citep{Ripepi:2012, Ripepi:2016, Marconi:2017, Ragosta:2019,Ripepi:2015, Ripepi:2014, Sicignano:2024} and a reconstruction of the 2D and 3D morphology of the LMC and SMC \citep{Moretti:2014, Muraveva:2018, Cusano:2021,Ripepi:2022}. Additional results include studies of star clusters \citep{Piatti:2014,Piatti:2015,Piatti:2016}, the derivation of proper motions \citep{Cioni:2014,Cioni:2016,Niederhofer:2018,Schmidt:2020,Niederhofer:2021,Niederhofer:2022,Schmidt:2022}, investigations on the mass-loss, chemistry and variability of AGB stars \citep{Gullieuszik:2012,Groenewegen:2020}, the study of metallicity maps and gradients in both clouds \citep{Choudhury:2020,Choudhury:2021}, spatial characterization of the RC population \citep{Subramanian:2017,Tatton:2021}, and investigations on the hierarchical nature of star formation in different regions of the clouds \citep{Sun:2017b,Sun:2017a,Sun:2018, Miller:2022}. VMC has recently completed its 7th data release \citep{Cioni:2025}.  Data products are linked on the main \href{https://star.herts.ac.uk/~mcioni/vmc/}{VMC project page}.

\subsubsection{Solo dwarfs (2012)} The Solitary Local dwarf galaxy survey (Solo dwarfs) is a wide-field survey that provides g and i band imaging and homogeneous photometry of all known isolated dwarf galaxies that are beyond 300~kpc from the MW or M31 and within 3~Mpc. Imaging in the north was acquired with CFHT and imaging in the south was obtained with Magellan/Megacam.  This survey includes 42 galaxies.  \citet{Higgs:2016} provides an overview of the survey.  Other publications include measurements of structural parameters \citet{Higgs:2021}, determination of the origin of isolated dwarf galaxies \citet{McConnachie:2021}, and comparisons with MW satellite systems \citet{Higgs:2021b}.

\subsubsection{STEP (2012)}

The SMC in Time: Evolution of a Prototype interacting late-type dwarf galaxy (STEP) was a guaranteed time program on the VLT Survey Telescope (VST).  The STEP survey covered $\sim50$~deg$^2$ in the main
body of the SMC and the Magellanic Bridge using in g, r,
i and $H\alpha$ bands to a limiting magnitude of $g\sim24$.   Even in the most crowded regions, STEP CMDs reach the oMSTO with a SNR $\sim10$ in the broadband filters.  The main goals of STEP were to measure SFHs, use variable stars to map 3D structures, study the cluster populations, and identify pre-MS stars.

The survey overview and first results from STEP were presented in \citet{Ripepi:2014b}.  \citet{Gatto:2021} provide a detailed structural analysis for 170 star clusters in the SMC.  \citet{Ficara:2026} present a SFH of the Magellanic Bridge over $\sim14$~deg$^2$.  \href{https://cdsarc.cds.unistra.fr/viz-bin/cat/J/A+A/709/A10}{Photometric catalogs} associated with STEP can be found on CDS.

\subsubsection{SMASH (2013)} The Survey of the Magellanic Stellar History \citep[SMASH;][]{Nidever:2017} is a DECam photometric survey that mapped 480 deg$^2$ around the LMC and SMC over a total area of 2400 deg$^2$ with a filling factor of $\sim20$\%. The survey obtained $ugriz$ photometry for 360 million sources, down to a limiting magnitude of $\sim 24$. The survey's main goals revolve around the detection and characterization of the stellar structures associated with the Magellanic system, including: i) mapping the extended stellar disks and halos of the Clouds; ii) searching for stellar counterparts to notable \hi{} features (Magellanic stream, leading arm); iii) detecting stellar streams and substructures not associated with \hi{} detection; iv) deriving resolved SFH maps across the survey footprint.

Some of the results from the SMASH survey include the discovery of low-luminosity stellar systems, such as Hydra~II \citep{Martin:2015} and SMASH-1 \citep{Martin:2016b}, the detection of an extended stellar component around the Magellanic Clouds \citep{Nidever:2019, Massana:2020}, a reconstruction of the 3D morphology of the LMC \citep{Choi:2018a}, the discovery and characterization of stellar over-densities in the LMC and SMC \citep{Choi:2018b, Martinez-Delgado:2019}, SFH measurements for several regions throughout the Magellanic systems \citep{Ruiz-Lara:2020b,Massana:2022}. Other community science enabled by the SMASH dataset included studies of the star clusters in the Magellanic Clouds \citep{Asad:2017,Piatti:2021,Piatti:2022} and updated reddening maps of the LMC and SMC \citep{Bell:2019,Bell:2020,Bell:2022,Chen:2022}.  \href{https://datalab.noirlab.edu/data/smash}{SMASH data products} are hosted by NOIRLab.

\subsubsection{MADCASH (2016)} The Magellanic Analog Dwarf Companions and Stellar Halos (MADCASH) is a survey that uses observations of resolved stars to search for satellite galaxies of LMC-like central galaxies within $D\sim4$~Mpc.  Subaru/HSC is used to acquire g and i band imaging for northern targets, while DECam is used to obtain comparable imaging for southern targets.  \citet{Carlin:2016} report the discovery of a new satellite galaxy (MADCASH-1) of NGC~2403, while \citet{Carlin:2021} present \hubble follow up imaging of MADCASH-1 and MADCASH-2, a new dwarf galaxy in the vicinity of NGC~4214. The program recently published a census of dwarf galaxies around NGC~2403 \citep{Carlin:2024} and, through a collaboration with DELVE (see below), have also published a census of satellites around NGC~3109 \cite{Doliva-Dolinsky:2025}.

\subsubsection{YMCA} The Yes, Magellanic Clouds Again (YMCA) survey was an imaging campaign with the VST that covered $\sim110$~deg$^2$ of the outer, low-density Magellanic system in the $g$ and $i$ bands down to $\sim24$th magnitude.  CMDs in all regions reached below the oMSTO.  The main science aim was to measure the SFHs of the outer regions of the LMC and SMC in order to constrain their interaction history.  Using YMCA data, \citet{Gatto:2022} uncovered a new faint stellar system located near the LMC.  \citet{Gatto:2024} provides an overview of the YMCA survey.  \href{https://cdsarc.cds.unistra.fr/viz-bin/cat/J/A+A/690/A164}{YMCA catalogs} are available via CDS.

\subsubsection{LBT-SONG (2021)} The Large Binocular Telescope Search for Nearby Galaxies (LBT-SONG\footnote{Because this program repurposed older data for a new science goal, we provide an estimated start date when the data were used for faint galaxy searches.}) is a survey that uses LBT imaging originally obtained to identify failed supernovae \citep[e.g.,][]{Kochanek:2008} to search for satellites around 19 host galaxies of intermediate mass ($\sim 5 \times 10^8 - 5\times 10^{10}~\text{M}_{\odot}$) with distance of $\sim4$--$11$~Mpc.  Deep 4-band LBT imaging provided depths of $\sim25.0$~mag in U-band and $\sim27.0$~mag in the B, V, and R bands.  For hosts within $\sim5$~Mpc, the search for satellites used resolved stars.  No new satellite systems were identified \citep{Garling:2021}.  For more distant host galaxies, unresolved light was used for searches and a handful of new candidate systems have been reported \citep{Davis:2021, Davis:2024}.

\subsection{Wide-Area Surveys}

There are increasing numbers of imaging surveys that cover large areas of the sky. Although the primary goals of these surveys are not centered on nearby galaxy studies, they have proven to be useful datasets for the discovery and characterization of galaxies and star clusters in the local Universe. Here, we summarize several of these important surveys that have led to resolved star studies in galaxies outside the MW.  The revolutionary success of the Sloan Digital Sky Survey (SDSS) in discovering LG galaxies has already been extensively covered in the literature \citep[e.g.,][]{Tolstoy:2009, Simon:2019} and we do not include a description of it here.

\subsubsection{Pan-STARRS 3$\pi$ Survey (2010)} The Panoramic Survey Telescope and Rapid Response
System (Pan-STARRS) is a pair of 1.8-m telescopes (PS1 and PS2) on Haleakal\={a}.  The Pan-STARRS1 3$\pi$ Steradian Survey \citep{Chambers:2016}, conducted with PS1, is a wide-area optical and near-IR imaging survey that covers $\sim30,000$ deg$^2$.  The typical angular resolution of the imaging is $\sim1-1.3^{\prime\prime}$ and the depth of the survey is $m_g \sim 23$, $m_r \sim 23$, $m_i \sim 23$, $m_z \sim 22$, $m_y \sim 21$~mag.  The 3$\pi$ survey featured a wide variety of science goals including mapping stars and galaxy discovery in the MW and LG, time domain exploration of the sky, surveying the Solar system, and cosmology.  Pan-STARRS data were used to discover a number of new stellar systems in the LG \citep[e.g.,][]{Martin:2013c, Martin:2013b, Laevens:2014, Laevens:2015b, Laevens:2015a}.  \href{https://archive.stsci.edu/panstarrs/}{Pan-STARRS data products} can be found on MAST.

\subsubsection{DES (2013)} The Dark Energy Survey (DES) is a wide-area optical imaging survey that used DECam to cover $\sim5000$~deg$^2$ of the southern sky.  The survey reaches a $10\sigma$ depth of $m_g \sim 24.7$, $m_r \sim 24.4$, $m_i \sim 23.8$, $m_z \sim 23.1$, $m_Y = 21.7$~mag with a typical angular resolution of $\sim1^{\prime\prime}$.  The main science goals of the survey include weak lensing, galaxy clustering, large scale structure, and supernovae distances.  The deep and wide-area nature of the survey also enabled the discovery of a number of extremely faint satellite galaxies and stars clusters associated with the MW or LMC \citep[e.g.,][]{Bechtol:2015, Drlica-Wagner:2015, Kim:2015a, Kim:2015b, Kim:2015c, Koposov:2015}.  DES data products can be found on the \href{https://www.darkenergysurvey.org}{DES program website}.

\subsubsection{Gaia (2013)} The \textit{Gaia} mission provides deep imaging, proper motions, and spectroscopy across the entire sky, including a holistic view of the LMC, SMC, the Magellanic Bridge, and a number of MW satellites. \citet{Gaia-Luri:2021} provides a comprehensive overview of the CMDs, structural properties, and kinematics of the entire Magellanic system based on the \textit{Gaia} eDR3 data release. Using a combination of photometric and astrometric quality cuts, they identify $\sim11.1$~million stars in the LMC and $\sim1.7$~million in the SMC.  The limiting magnitude of their data is $m_G\sim20$.  This rich dataset has been foundational to a number of Gaia-specific explorations of the LMC and SMC, including studies of their bulk and star-by-star kinematics \citep[e.g.,][]{Gaia-Helmi:2018, Katz:2023}, detection and characterization of stellar variability \citep[e.g.,][]{Eyer:2023, Ripepi:2023}, and a reference for other surveys \citep[e.g.,][]{Drlica-Wagner:2021}. Beyond the LMC and SMC, \textit{Gaia} has been instrumental in measuring the proper motions and kinematics for a number of MW satellites and LG galaxies \citep[e.g.,][]{VanderMarel:2016, Gaia-Helmi:2018, Fritz:2018, Simon:2018, VanderMarel:2019, McConnachie:2020b, Li:2021, Battaglia:2022, Pace:2022}.  \textit{Gaia} data is available from the \href{https://gea.esac.esa.int/archive/}{ESA archive}, while many of the database queries needed to extract catalogs are provided directly in relevant papers.

    \subsubsection{Skymapper (2014)} Skymapper is a wide-area imaging survey that has mapped $\sim26,000$ deg$^2$ of the southern sky in 6 filters, $griz$ and two medium near-UV filters, $u$ and $v$ that help measure stellar temperature, metallicity, and surface gravity \citep[e.g.,][]{Keller:2007}.  The survey is optimized for stellar astrophysics, primarily of MW stars, but it also enables a broader range of extragalactic science.  As of data release 4, the $10\sigma$ survey depths range from $u\sim18.5$~mag to $g\sim20.5$~mag.  Though most published Skymapper science focuses on stars in the MW, Skymapper photometry has been used to map stellar populations in the LMC \citep[e.g.,][]{Wan:2020, Oh:2023} and characterize the extended stellar halos of MW satellites \citep[e.g.,][]{Chiti:2020, Chiti:2021}. Skymapper provided its 4th data release in 2024 \citep{Onken:2024}.  Data products can be accessed through the \href{https://skymapper.anu.edu.au}{Skymapper webpage}.

\subsubsection{DESI Legacy Imaging Surveys (2014)} In support of the Dark Energy Spectroscopic Instrument (DESI), there have been coordinated, multi-telescope surveys with DECam, the Mayall Telescope, and the Bok Telescope to provide deep optical imaging for spectroscopic targeting.  The aggregate of these imaging surveys is DESI Legacy Imaging Survey (LS; \citealt{Dey:2019}).  In total, LS provides $g$, $r$, and $z$ band optical imaging over $\sim14000$ deg$^2$ with $5\sigma$ point source depths of $m_g\sim24$, $m_r\sim23.5$, and $m_z\sim23$ with a typical angular resolution of $\sim1^{\prime\prime}$.  While the main focus of LS is the investigation of dark energy and cosmology, the LS provides excellent data on resolved stars in nearby galaxies, notably the discovery of faint dwarf galaxies and stellar streams from resolved stars or integrated light \citep[e.g.,][]{MartinezDelgado:2021, Collins:2022b, Karachentsev:2022, Sand:2022, Jones:2023, McQuinn:2023, LiJ:2024, MartinezDelgado:2024, McQuinn:2024c, Sand:2024, Hunter:2025b, Hunter:2026}.  LS data products can be accessed through their \href{https://www.legacysurvey.org}{program website}.

\subsubsection{PRISTINE (2015)} The Pristine survey uses the narrowband Ca H\&K imaging on the CFHT to infer photometric metallicities over $\sim6000$ deg$^2$ of the northern sky.  The main driver of this program is to efficiently identify extremely metal-poor stars that belong to the MW.  However, the depth, large areal coverage, and high information content of the data have enabled a number of other science areas, including the study of MW satellite galaxies \citep{Longeard:2018, Longeard:2020, Longeard:2021, Longeard:2022, Longeard:2023, Taibi:2026}. Pristine data are now calibrated to \textit{Gaia} spectrophotometric metallicities \citep{Martin:2024}.  \href{https://cdsarc.cds.unistra.fr/viz-bin/cat/J/A+A/692/A115}{Pristine data} are available on CDS.

\subsubsection{UNIONS (2015)} The Ultraviolet Near-Infrared Optical Northern Survey (UNIONS\footnote{Although UNIONS uses imaging from Pan-STARRS taken as early as 2009, the foundational work for UNIONS stems from the Canada-France Imaging Survey which first collected new data in 2015 \citep{Ibata:2017}.}) is a multi-telescope (CFHT, Pan-STARRS, Subaru) program that is acquiring optical imaging $\gtrsim6000$ deg$^{2}$ of the northern sky in ugriz filters.  It aims to reach 10$\sigma$ depths of $u\sim23.7$, $g\sim24.5$, $r\sim24.2$, $i\sim23.8$, and $z\sim23.3$.
UNIONS has numerous science goals including large-scale structure, weak lensing, and finding the faintest galaxies. It also provides critical northern sky optical wavelength coverage for \textit{Euclid}.  UNIONS is described in detail in \citet{Gwyn:2025}.  UNIONS data have been used to uncover several faint stellar systems in the LG \citep[e.g.,][]{Smith:2023, Smith:2024, Smith:2025}.  UNIONS data products can be found on their \href{https://www.skysurvey.cc}{collaboration website}.

 \subsubsection{DELVE (2019)} The DECam Local Volume Exploration Survey (DELVE) is a large optical imaging survey that has used DECam to cover $\sim20,000$ deg$^2$ of the southern sky in griz filters to depths of $\sim23.5$~mag \citep{Drlica-Wagner:2021, Drlica-Wagner:2022}.  It uses uniform data reduction to combine new DELVE imaging with archival data from DES, SMASH, and DeCALS. DELVE consists of four components: (i) a wide-area survey of $\sim20,000$ deg$^2$ in griz filters to depths of $\sim23.5$~mag; (ii) a deep, contiguous area survey of $\sim$2,200 deg$^2$ of the periphery of the Magellanic Clouds to a depth of gri $\sim24.2$; (iii) deep imaging over 135 deg$^2$ targeting four Magellanic Cloud analogs in the LV to depths of $\sim25.5$~mag in g and i; (iv) the Mapping the Ancient Galaxy in CaHK (MAGIC) survey that covers 5400 deg$^2$ in a metallicity sensitive narrowband Ca H \& K filter \citep{Chiti:2026}.  DELVE data has been used in the discovery of many faint stellar systems around the MW and other nearby galaxies \citep[e.g.,][]{Mau:2020, Cerny:2021a, Cerny:2021b, Cerny:2023b, Cerny:2023a, McNanna:2024, Cerny:2025, Doliva-Dolinsky:2025, Fielder:2025, Tan:2025, Medoff:2026, Tan:2026b}.  \citet{Tan:2026} published a selection function for MW satellites based on the DELVE survey.  The narrowband survey has recently produced its first results \citep{Barbosa:2025, Placco:2025, Atzberger:2026}.  DELVE has also been combined with additional data to enhance its scientific scope \citep[e.g., targeted u-band data for stellar metallicities;][]{Pan:2025}. More details on DELVE can be found on their \href{https://delve-survey.github.io}{project website} and data products are available through \href{https://datalab.noirlab.edu/data/delve}{NOIRLab}.

\subsubsection{Euclid (2023)} \textit{Euclid}  is a wide-area, space-based imaging survey, launched in 2023, that is covering $\sim14000$ deg$^2$ of the extragalactic sky in 1 broad optical band (VIS) and 3 near-IR bands (Y, J, H) with good angular resolution of $\sim0.1^{\prime\prime}$ in the optical and $\sim0.3^{\prime\prime}$ in the near-IR \citep[e.g.,][]{Mellier:2025}.  The nominal wide-area survey depths are $\sim26.2$ in the optical and $\sim24.5$ in the near-IR, with smaller area deep field ($\sim50$ deg$^2$) that will reach $\sim2$~mag deeper.  Although \textit{Euclid}'s primary science goals are related to dark energy and cosmology, the nature of the wide-area survey means it is providing extensive imaging of nearby galaxies.  The scientific potential of \textit{Euclid}  for resolved star science in nearby galaxies has been showcased by its Early Release Observations of several nearby galaxies \citep{Hunt:2025, Annibali:2026}, and the first Data Release (DR1) has already yielded the discovery of an ultra-faint stellar system near the Fornax dSph \citep{Saifollahi:2026}. \textit{Euclid} has begun releasing data publicly and will continue to do so over the course of the next several years.  \href{https://www.cosmos.esa.int/web/euclid/q1-data}{\textit{Euclid} data products} are available via the ESA archive.%

\subsection{Examples of Individual High-value Galaxies}
\label{sec:individual_galaxies}

Beyond the programs described above, a handful of nearby galaxies have been the targets of substantial investments of observational resources owing to their particularly high scientific value for a particular scientific endeavor (e.g., the study of reionization in an isolated galaxy, the massive stellar populations of an extremely metal-poor galaxy).  We select a few representative examples of targeted systems in order to illustrate their science value and the type of observational investments that have been made for resolved star studies. Some observations of important systems that could be included in this subsection (e.g., M31) are covered elsewhere in this appendix.

\subsubsection{Within the Local Group}

\paragraph[Coma~Berenices]{Coma~Berenices:} Coma~Berenices is a MW satellite that has the best constrained low-mass IMF measurement made to date outside the MW based on \hubble\ data.  It was first discovered in 2007 \citep{Belokurov:2007}.  Using deep \hubble/ACS observations from a \hubble\ survey of MW UFDs (GO-12549; \citealt{Brown:2012}), \citet{Gennaro:2018} found that its sub-Solar IMF was significantly different than that of the sub-Solar Galactic IMF.  However, the \hubble/ACS imaging was limited in depth and suffered from challenges with star-galaxy separation. Using 44 orbits of \hubble/WFC3/IR imaging (GO-13449), Coma~Berenices was imaged with a $2\times2$ mosaic to a depth of $m_{F160W}\sim27$~mag ($\sim0.17$~\msun\ on the MS at 50\% completeness) in order to make an improved measurement of the low-mass IMF \citep{Gennaro:2018b}.

\paragraph[Eridanus II]{Eridanus II:}  Discovered in 2015, Eri~II is the most luminous UFD near the MW that lies beyond our nominal 300~kpc boundary \citep[\S \ref{sec:concepts_classification};][]{Koposov:2015, Crnojevic:2016a}.  It has the distinction of hosting a central GC that is $\sim4$\% of the galaxy's total luminosity. Eri~II has been the target of numerous studies of its resolved stellar populations, including multiple \hubble programs (13 orbits, GO-14234; 13 orbits GO-14224; 4 orbits, GO-15901) that have resulted in several measurements of its SFH, characterization of its GC, and a measurement of its MDF based on narrowband imaging \citep{Gallart:2021, Simon:2021, Fu:2022, Weisz:2023b}.

\paragraph[Reticulum II]{Reticulum II:} Discovered in 2015 \citep{Bechtol:2015}, Ret~II is an ultra-faint MW satellite galaxy.  Spectroscopic studies revealed that $\sim70$\% of its stars are enhanced in r-process elements \citep[e.g.,][]{Ji:2016, Roederer:2016}, the formation of which is linked to explosive phenomena such as merging neutron stars \citep[e.g.,][]{Chornock:2017, Drout:2017}.   \citet{Simon:2023} used 12 orbits of deep \hubble/ACS to tile a large area of Ret~II (GO-14766) in order to measure its SFH and place constraints on the enrichment pathways responsible for the observed r-process enhancements.

\paragraph[Bootes~I]{Bootes~I:} Bootes~I is a UFD galaxy located in close proximity to the MW \citep{Belokurov:2006b} that is the target of a Cycle 2 \jwst\ program (\jwst-GO-3849, PI Gennaro), primarily to tightly constrain the low-mass IMF in an extremely ancient ($\sim13$~Gyr) and metal-poor ([Fe/H]$=-2.4$) environment.  Its IMF has previously been studied with \hubble\ imaging (GO-12549; \citealt{Brown:2012} and GO-15317, PI Platais; \citealt{Gennaro:2018, Filion:2022}), but those data were too shallow to provide robust constraints on log-normal IMF parameters.  Using $\sim19,000$~sec of integration time in the F150W and F322W2 filters, this program reaches a lower stellar mass of $\sim0.15$\msun, enabling one of the most robust extragalactic constraints on low-mass binary stars and the low-mass IMF to date \citep{Shariat:2025, Ding:2026}.

\subsubsection{Outside the Local Group}

\paragraph[I Zw 18]{I Zw 18:} Since its discovery \citep{Zwicky:1966}, the stellar populations of I Zw 18 have been of high scientific value. With its low metallicity and high SFR, I Zw 18 has long been considered one of the best local analogs to metal-poor galaxies in the early Universe.
Historically, I Zw 18 has served as the prototype of the blue compact dwarf (BCD) class \citep[e.g.,][]{Searle:1972} and was later identified as a Wolf-Rayet galaxy on the basis of broad stellar He~{\sc ii} and C~{\sc iv} emission \citep{Izotov:1997}; contemporary resolved star work focuses less on these morphological and spectroscopic labels and more on directly measuring its SFH.

Early \hubble\ imaging suggested that I Zw 18 may have only young populations \citep[e.g.,][]{Aloisi:1999, Ostlin:2000, Izotov:2004}, indicating that it may only be forming its first stars now out of pristine gas, though some analyses claimed a TRGB detection at the faint limit of the data \citep[e.g.,][]{Momany:2005}.  Using 49 \hubble/ACS orbits from GO-9400 and GO-10586, \citet{Aloisi:2007} placed I Zw 18 at a larger distance (18.2$\pm1.5$~Mpc) than previously thought and revealed a clear RGB, a sign of an older stellar population.  I Zw 18 continues to be the subject of significant observational investment, including 6.5 hours of NIRCam and MIRI \jwst\ imaging as part of GTO-1233 \citep{Hirschauer:2024}. More broadly, I Zw 18 illustrates a recurring pattern in extragalactic resolved star work, in which shallow data are suggestive of exclusively young populations, whereas deeper data obtained at later times reveal an ancient red-giant population.

\paragraph[Centaurus~A]{Centaurus~A:} Cen~A ($D\sim3.8$~Mpc) is one of the closest examples of a massive elliptical galaxy.  Its resolved stellar populations have been studied via a combination of wide-area ground-based surveys and targeted deep \hubble fields.  Notably, PISCeS (Panoramic Imaging Survey
of Centaurus and Sculptor) used the Megacam on Magellan to obtain deep, wide-field optical imaging that covers $\sim15$ deg$^2$, 100--200~kpc in projected distance along the major and minor axes, of Cen~A in the g and r bands to depth of $\sim1.5$~mag below the TRGB \citep[e.g.,][]{Crnojevic:2016b, Crnojevic:2019}.  This survey has provided the most comprehensive map to date of a giant elliptical galaxy based on resolved stars and has characterized both the smooth halo and identified new substructures.  In parallel, several \hubble imaging programs over the past two decades have placed dozens of fields on Cen~A and its satellites in order to better characterize their stellar populations \citep[e.g.,][]{Rejkuba:2005, Crnojevic:2010, Rejkuba:2011, Crnojevic:2012,Rejkuba:2014, Crnojevic:2019, Rejkuba:2022}.

\paragraph[Leo P]{Leo P:} First discovered in 2013, Leo~P is thought to be the nearest example of an isolated,  metal-poor, gas-rich, low-mass star-forming galaxy \citep{Rhode:2013, Giovanelli:2013, Skillman:2013, McQuinn:2013}.  Leo~P provides unique insight into a broad array of science from the physics of extremely metal-poor massive stars to the impact of reionization on galaxy formation.  Accordingly,  Leo P's stellar populations have been the target of significant observational investments including 17 \hubble/ACS orbits (GO-13376) to establish its distance and broadly characterize its age \citep[][]{McQuinn:2015a} and 35.8~hours (GO-1617) of \jwst/NIRCam imaging to make a precise measurement of its SFH \citep{McQuinn:2024b}.

\paragraph[M82]{M82:}  M82 ($\sim3.6$~Mpc) is one of the nearest examples of an $L_{\star}$ starburst galaxy.  Intense star formation in its central region, possibly triggered by a recent interaction with M81, appears to be responsible for a kpc-scale outflow along its minor axis.  M82 represents one of our best opportunities to study the cause and impact of nuclear outflows, and their effects on local and larger scale environment (e.g., the circumgalactic medium).  The resolved stellar populations of M82 have been extensively studied with ground- and space-based facilities \citep[e.g.,][]{Dalcanton:2009, Okamoto:2019} including several programs with \hubble\ that have measured the global SFH as well as age-dating star formation along the path of the outflow \citep[e.g.,][]{Williams:2011a, Velguth:2024, Rao:2025, Rao:2026}.  A $\sim65$~hr Cycle 3 \jwst\ program (\jwst-GO-5145, PI Smercina) has imaged a large, contiguous area of M82's disk with the aim of measuring the first global and spatially resolved SFH of a prototypical starburst galaxy.

\section{Star Formation Histories of Individual Galaxies}
\label{sec:individual_sfhs}

This appendix collects galaxy-by-galaxy summaries of published CMD-based SFHs.  For each galaxy, we describe the relevant observational setup of each study (e.g., filters, depth, spatial coverage) and its main SFH findings.  The population-level syntheses of these results are presented in \S \ref{sec:formation}. A tabulated compilation of all published lifetime SFHs is given in Appendix~\ref{sec:sfh_table}.

\subsection{Milky Way Classical Satellites}
\label{sec:app_mw_classical}

\subsubsection{Sculptor}
\label{sec:Sculptor_sfhs}
Sculptor was the first dSph discovered nearly $\sim100$ years ago by \citet{Shapley:1938}.
Sculptor is located at a distance of $\sim80$~kpc \citep{Nagarajan:2022} and is gas-poor \citep{Putman:2021}.

\citet{deBoer:2012} measured the first wide-area ($\sim1^{\circ}$ in projected radius) SFH of Sculptor using deep ground-based B, V, I CMDs that reach below the oMSTO.   A novelty of this analysis was to include spectroscopic values of [Fe/H] and [$\alpha$/Fe] for RGB stars from the literature.  Using only MSTO stars, they found that $\sim85$\% of star formation in Sculptor occurred $>10$~Gyr ago and $\sim60$\% $>12$~Gyr ago.  Interestingly, they found a modest tail of star formation that extended to younger ages and higher metallicities, such that $\sim15$\% of Sculptor's stellar mass formed 6--10~Gyr ago.  They report an age gradient such that this population of young stars is centrally concentrated and that the outer regions are entirely ancient.  \citet{deBoer:2012} report an AMR that increases from [Fe/H]$=-2.4$  $\sim12$~Gyr ago to [Fe/H]$=-1.1$ $\sim6$~Gyr ago, which broadly agrees with spectroscopic metallicities \citep[e.g.,][]{Tolstoy:2004, Kirby:2009, Starkenburg:2010, Hill:2019, Shetrone:2025}.

\citet{Weisz:2014a} measured the SFHs of two centrally located \hubble/WFPC2 fields (F555W, F814W) that extend several magnitudes below the oMSTO. They found that $\sim80$\% of the stellar mass formed by $\sim11.2$~Gyr ago and $96$\% by $\sim10$~Gyr ago.  Younger star formation is consistent with the expected contribution from BSSs.  A limitation of this work is the small areal coverage, as the two \hubble\ fields only cover $\sim4$\% of Sculptor's half-light radius.  They did not report an AMR.

\citet{Savino:2018} measured the SFH of Sculptor using the data from \citet{deBoer:2012} but their CMD analysis included more phases of evolution including the MSTO, HB, and RGB.  Notably, this study takes variable RGB mass-loss into account, enabling a more robust inclusion of the HB and RGB. \citet{Savino:2018} found a SFH of Sculptor that is very similar to \citet{deBoer:2012}.  They identified two main SFH episodes from 11--14~Gyr ago and 8--11~Gyr ago, with a small tail of star formation that extends to $\sim6$~Gyr ago.  One advantage of their CMD-fitting code, \texttt{MORGOTH}, that was highlighted by \citet{Savino:2018} is improved age resolution of the SFH.  Because it uses the HB, the detailed morphology of which is strongly age sensitive, it is possible to achieve higher time resolution than from just the MSTO.  For Sculptor, \citet{Savino:2018} reported an age resolution of 0.5~Gyr at all ages versus 1~Gyr found by \citet{deBoer:2012}.  They recovered an AMR that increases from [Fe/H]$=-2.4$  $\sim14$~Gyr ago to [Fe/H]$=-1.3$ $\sim6$~Gyr ago, which is similar to \citet{deBoer:2012}.

\citet{Bettinelli:2019} measured the SFH of Sculptor from deep, wide-field ($\sim0.7^{\circ}$ in projected radius) g and r band photometry taken with DECam.  The resulting CMDs reach $\sim2$~mag below the oMSTO, making this an even deeper CMD (i.e., higher SNR at the oMSTO) than that analyzed by \citet{deBoer:2012}.   \citet{Bettinelli:2019} found that $\sim70$\% of Sculptor's stellar mass formed $>12$~Gyr ago and that it completely stopped forming stars 11.3~Gyr ago.  They did not find any low level star formation at younger ages, a point of contrast with \citet{deBoer:2012} and \citet{Savino:2018}, but in good agreement with \citet{Weisz:2014a}.  They noted that the inner region of Sculptor formed stars for $\sim1.5$~Gyr, while the outer region only formed stars for $\sim0.5$~Gyr.  The measured age gradient increases with radius, though it is not as steep as the gradient reported by \citet{deBoer:2012}.  They concluded that Sculptor cannot be a fossil of reionization.  They suggested that the discrepancy with the extended star formation reported by \citet{deBoer:2012} and \citet{Savino:2018} may be due to the lower SNR of the \citet{deBoer:2011} Sculptor CMD.  They recovered an AMR that increases from [Fe/H]$=-2.1$  $\sim13$~Gyr ago to [Fe/H]$=-1.7$ $\sim11$~Gyr ago, with some evidence for metal-rich stars being more centrally concentrated.   The CMD-based AMR and metallicities are in reasonable agreement with other CMD-based AMRs as well as spectroscopic metallicities \citep[e.g.,][]{Kirby:2009, Kirby:2011, Walker:2023, Shetrone:2025}.

\citet{Yang:2024} identified a small number of luminous putative intermediate-mass CHeB stars in Sculptor using \textit{Gaia} data.  They infer an age of $\sim3$~Gyr, which is significantly younger than what the SFHs report.  However, the deep CMDs from \citet{deBoer:2012} and \citet{Bettinelli:2019} do not show a MS population that would correspond to this age.

\subsubsection{Fornax}
\label{sec:Fornax_sfhs}
Fornax was the second classical MW satellite discovered after Sculptor \citep{Shapley:1938b}.  Located at a distance of $\sim140$~kpc \citep[e.g.,][]{Nagarajan:2022, Oakes:2022}, Fornax is a gas-poor galaxy \citep{Putman:2021} that formed stars until at least $\sim0.5$~Gyr ago \citep{Buonanno:1999}, with tentative evidence for stars younger than $\sim100$~Myr \citep{Coleman:2008}.  One unusual aspect of Fornax is its large number of GCs (6) relative to its stellar mass \citep[e.g.,][]{Mackey:2003, Larsen:2012}.

\citet{deBoer:2012b} acquired B, V, and I-band imaging that covered the central $\sim0.8$~deg$^2$ of Fornax.  The resulting CMDs reached $\sim0.5$~mag below the oMSTO.  This dataset overlapped with spectroscopic measurements of [Fe/H] for $\sim900$ and [$\alpha$/Fe] for $\sim600$ stars, which they used in tandem with the CMD fitting.  They measured a SFH over the entire area that is continuous from 14 to 0.25 Gyr ago, with the majority of SFH taking place between 10 and 1 Gyr ago.  From the oldest ages, the SFR rises and reaches a peak at $\sim9$~Gyr ago.  This is followed by a lull and a second peak $\sim6$~Gyr ago.  The SFR then declines before entirely stopping with the most recent star formation a few hundred Myr ago.  Approximately $50\%$ of the stellar mass of Fornax formed from $\sim2$--$8$~Gyr ago.  The corresponding AMR is such that the mean
metallicity increased from [Fe/H] $\sim -2.5$ to [Fe/H] $\sim -1.5$ from 12 to 8 Gyr ago.  Subsequently, it slowly increased to a value of [Fe/H] $\sim -0.8$ by $\sim3$~Gyr ago.  They reported clear gradients in age and metallicity. Younger ($\lesssim2$~Gyr), more metal-rich stars appear more centrally concentrated compared to older, more metal-poor stars.

\citet{delPino:2013} used deep V- and I-band ground based imaging taken with the VLT of 3 fields to measure a global and spatially extended SFH of Fornax over a non-contiguous area of $\sim140$ arcmin$^2$.  One field was at the center of the galaxy, while the other two were located at the core radius. Each CMD reached at least 1 magnitude below the oMSTO.  They found that Fornax hosts complex populations in terms of age, metallicity, and spatial variations in both.  Its SFH shows activity at all ages $\gtrsim1$~Gyr ago, with larger enhancements $\sim7$~Gyr ago in the central region and $\sim10$~Gyr ago in the outer fields.  There are multiple smaller peaks and troughs of star formation at $\sim3$ and $\sim6$~Gyr ago, with a rapid decline in activity more recently than that.  They suggested that some, or perhaps most, of this population may be BSSs.  They measured an AMR that ranges from [M/H] $\sim-1.6$ in the outer fields at $\sim12$~Gyr ago to [M/H] $\sim-0.8$ for the youngest populations in the inner field.  While the SFH and AMR are in general agreement with \citet{deBoer:2012b}, they found differences such that the main star formation activity measured by \citet{delPino:2013} is systematically older by a few Gyr.   Building on their SFH, \citet{delPino:2015} use their SFH, along with kinematics and metallicity information, to hypothesize that Fornax is the remnant of a merger between two dwarf galaxies.

\citet{Weisz:2014a} measured the SFH of a single, central \hubble/WFPC2 field in Fornax.  The \hubble-based CMD extends well below the oMSTO, but subtends only a small fraction of the galaxy's area.  Like the wider-area studies, \citet{Weisz:2014a} reported a SFH that is continuous and nearly constant, with notable enhancements in star formation $\sim7$--$8$~Gyr ago and $\sim3$--$4$~Gyr ago.  They found that $\sim50$\% of the stellar mass is formed by $\sim7$~Gyr ago and $80\%$ by $\sim3$~Gyr ago.  They did not report an AMR.

\citet{Rusakov:2021} measured the SFHs of two deep \hubble/ACS (F475W, F814W) fields in Fornax, one near the center and one near the core radius.  They both extend several magnitudes below the oMSTO. Overall, they find that the SFH of Fornax has a single
large star formation episode at early times ($\gtrsim10$~Gyr ago), a second strong episode $\sim4.6$~Gyr ago, and several small peaks between $\sim2$ and $0.2$~Gyr ago.  The bulk of the stellar mass formed in the earliest episode.

The outer field is marginally older having formed 50\% of its stellar mass by $\sim10$~Gyr ago compared to $\sim9$~Gyr ago for the inner field.  There is a similar offset in timing for when 80\% of the stellar mass formed. The outer field reached this mark by $\sim5.5$~Gyr ago and the inner field by $\sim4.5$~Gyr ago.  \citet{Rusakov:2021} highlighted the two strong peaks (ancient and intermediate age) in their SFH solutions along with the lull in between.  These are not present in the other contemporary SFHs of Fornax.  \citet{Rusakov:2021} suggested this is because their superior data quality is able to resolve the continuous star formation measured in the other studies into finer detail.

They reported an AMR that changes from [Fe/H] $\sim-1.6$ at $\sim13$~Gyr ago to [Fe/H] $\sim -0.7$ at $\sim1$~Gyr ago to [Fe/H] $\sim -0.4$ at $\sim100$--$200$~Myr ago.  They found an AMR with low dispersion, which they suggest is evidence of gas being well-mixed at all epochs.

In addition, they found an age gradient such that stars associated with the 4.6~Gyr burst and those $\lesssim2$~Gyr in age are located centrally and that older stars ($\gtrsim 10$~Gyr) dominate the outskirts.  For metallicity, they find Fornax's metal-rich populations are centrally concentrated, while the metal-poor population is predominantly in the outer region.

For papers that reported AMRs, their metallicities, and radial metallicity trends, are in reasonably good agreement with those from spectroscopic studies \citep[e.g.,][]{Battaglia:2006, Letarte:2010, Starkenburg:2010, Kirby:2011, Lemasle:2014, Hasselquist:2021, Walker:2023, Shetrone:2025}.

\subsubsection{Leo~I}
\label{sec:LeoI_sfhs}
Leo~{\sc I} was discovered in the 1950s \citep{Harrington:1950, Wilson:1955}. At a distance of $\sim240$~kpc from the MW \citep[e.g.,][]{Nagarajan:2022}, Leo~{\sc I} is among the most distant MW satellites. It has been confirmed to be gas-poor \citep{Putman:2021}.  Analysis of \hubble\ imaging in the late 1990s and early 2000s showed that Leo~{\sc I} hosted a population of stars $<$1~Gyr in age \citep{Gallart:1999, Hernandez:2000, Dolphin:2002b}.

Due to its large distance from the MW and high velocity, Leo~{\sc I} has provided a tight constraint on the dark matter mass of the MW  \citep{Boylan-Kolchin:2013}. Leo~{\sc I} may also be one of the nearest dwarf galaxies to host a central black hole.  \citet{Bustamante-Rosell:2021} reported the detection of a super massive black hole in Leo~{\sc I} based on kinematics of its central region.  However, re-analysis by \citet{Pascale:2024} suggests that at most it is an intermediate-mass black hole, and perhaps there may be no black hole at all due to a high central density of dark matter \citep{Pascale:2025}.

\citet{Weisz:2014a} used a single central \hubble/WFPC2 (F555W, F814W) field to measure the SFH of Leo~I.  The CMD extends below the oMSTO with modest SNR.  They found $\sim20$\% of the stellar mass in this field formed by $\sim12$~Gyr ago, followed by no or low-level star formation from $\sim9$--$12$~Gyr ago.  They found two prominent enhancements in Leo~I's SFH from $\sim7$--$9$~Gyr ago and $\sim1$--$3$~Gyr ago.  Leo~I formed $\sim50$\% of its stellar mass by $\sim5$~Gyr ago and quenched within the most recent $\sim1$~Gyr.   They did not report an AMR.  The results of this work are very similar to older SFH measurements that used the same \hubble\ data \citep{Gallart:1999, Dolphin:2002b}.

\citet{Ruiz-Lara:2021} used deeper and wider-area \hubble/ACS (F435W, F814W) and WFC3/UVIS (F438W, F814W) imaging to measure a more detailed SFH of Leo~I.  In the central ACS field, they found a SFH that is broadly similar to that of \citet{Weisz:2014a}, but with much improved precision and age resolution.  They reported an early star formation episode that accounts for $\sim30$\% of the stellar mass formed by 12~Gyr ago followed by a period of little-to-no star formation from $\sim10$--$6$~Gyr ago.  They identify multiple enhancements of star formation in Leo~I at $\sim$13, 5.5, 2.0, and 1.0 Gyr ago.  They find that it quenched within the last $\sim1$~Gyr. They attribute the more recent peaks to external events in the lifetime of Leo~I such as the possible accretion of a smaller galaxy and infall on the MW's halo. The outer regions of Leo~I are older.  The SFH of the WFC3/UVIS field shows that Leo~I formed $\sim60$\% of its stellar mass by $\sim12$~Gyr ago and $\sim80$\% by $\sim8$~Gyr ago.  Interestingly, the SFH of this outer region shows small peaks at $\sim5.5$, $\sim2.0$, and $\sim1.0$~Gyr ago, the same as the inner field.  \citet{Ruiz-Lara:2021} find that the inner $\sim200$~pc of Leo~I has a nearly uniform SFH, while the remainder shows a positive age gradient with a mean age of $\sim6.8$~Gyr in the central region and $\sim10.4$~Gyr in their outermost region.

\citet{Ruiz-Lara:2021} measured a global and spatially resolved AMR. For the entire galaxy, they found a metallicity evolution of [Fe/H]$=-1.9$ at $\sim13$~Gyr ago to [Fe/H]$=-0.75$ for the youngest stars.  Radially, they find that the mean metallicity of the central region is [Fe/H] $\sim-1.5$ and [Fe/H] $\sim-1.7$ for their outermost region.   Their mean metallicities appear to be in reasonable agreement with spectroscopic studies \citep[e.g.,][]{Koch:2007a, Kirby:2011, Walker:2023}.

\subsubsection{Leo~II}
\label{sec:LeoII_sfhs}
Leo~II was discovered around the same time as Leo~I in the 1950s \citep{Harrington:1950, Wilson:1955}.  It has been confirmed as gas-poor \citep{Putman:2021}. Leo~II is one of the least studied MW satellites in terms of stellar populations.  In fact, a distance with precision on par with other LG galaxies was only published in 2025, when \citet{Monelli:2025} used ground-based observations of RR Lyrae to determine a distance of $\sim209\pm4$~kpc.

There are a handful of legacy SFHs of Leo~{\sc II} that predate the period of this review \citep[e.g.,][]{Mighell:1996, Hernandez:2000, Dolphin:2002b}.  \citet{Komiyama:2007} used wide-field Subaru imaging to undertake a qualitative study of its stellar populations. Two other studies estimated photometric metallicities from RGB stars in the near-IR \citep[][]{Gullieuszik:2008} and using \hubble\ medium and broadband optical imaging \citep[F390M, F555W, F814W;][]{Ross:2015}.  They each report a mean metallicity consistent with [Fe/H]$\sim-1.7$.  These are in reasonable agreement with spectroscopic metallicities \citep[e.g.,][]{Koch:2007b, Kirby:2011, Walker:2023}.

The only contemporary SFH of Leo~II comes from \citet{Weisz:2014a}, who used a single, central \hubble/WFPC2 (F555W, F814W) field to measure its SFH.  The CMD extended below the oMSTO.  Leo~II's SFH shows distinct phases including an initial episode that formed $\sim20\%$ of its stellar mass by $\sim12$~Gyr ago, a second strong enhancement from $\sim8$--$10$~Gyr ago, and the strongest episode of star formation from $\sim6$--$7$~Gyr ago, followed by abrupt quenching $\sim6$~Gyr ago.  The main takeaway from this result is that Leo~II appears to be dominated by intermediate age populations.  They did not report an AMR or any gradient information.

\citet{Monelli:2025} undertook a qualitative analysis of Leo~II's stellar population using its variable stars from ground-based data and a deep CMD based on archival \hubble/WFC3 (F555W, F814W) imaging.  The populations of RR Lyrae and anomalous Cepheids point to an extended SFH with a metallicity spread.  Their qualitative analysis of the \hubble\ CMD suggests that Leo~II had a complex SFH with a strong star formation enhancement from 5--6~Gyr ago.

\subsubsection{Ursa~Minor}
\label{sec:UrsaMinor_sfhs}
Ursa~Minor was discovered in the 1950s \citep{Wilson:1955}.  It is a gas-poor \citep{Putman:2021} system located $\sim70$~kpc from the MW \citep{Garofalo:2025}.  Along with Draco, Ursa~Minor is one of the best-studied MW satellites in terms of its dark matter content and stellar populations.  SFHs from the late 1990s and early 2000s have shown that Ursa~Minor is almost entirely ancient (i.e., 13 Gyr old) making it, along with Draco, a quintessential `red and dead' galaxy \citep[e.g.,][]{Mighell:1999, Carrera:2002, Dolphin:2002b}.

The first wide-field SFH of Ursa~Minor is from \citet{Carrera:2002}.   They used deep B,V,R,I imaging using the Isaac Newton Telescope that covers 0.75 deg$^2$.  All their CMDs reach below the oMSTO.  They find that $\sim80$\% of the stellar mass formed by $\sim13$~Gyr ago, assuming an oldest allowable age of 15 Gyr.  They report a weak, positive age gradient and no metallicity gradient.  Their mean metallicity of [Fe/H] $\approx -1.9$ is consistent with spectroscopic studies \citep[e.g.,][]{Cohen:2010, Kirby:2011, Pace:2020, Sato:2025}.

More recently, \citet{Weisz:2014a} analyzed the CMD from a single \hubble/WFPC2 (F555W, F814W) pointing near the galaxy's center.  The CMD extends well-below the oMSTO.  \citet{Weisz:2014a} found that $\sim50$\% of the stellar mass formed by $\sim11$~Gyr ago and $\sim90$\% by $\sim9$~Gyr ago.  They found no evidence for intermediate or young star formation, though they did identify a clear blue plume population, which they consider to be BSSs.   They did not report an AMR nor any spatial information.

\citet{YangY:2024} applied a deep learning approach to measure the SFH from the same \hubble/WFPC2 data as \citet{Weisz:2014a}.  They found a more extended SFH with 50\% of the stellar mass formed by $\sim9$~Gyr ago and 90\% by $\sim4$~Gyr ago.  They attributed some of the intermediate and younger star formation to (a) the large spread in Ursa~Minor's metallicity which is reflected in the width of its RGB and (b) BSSs.  They note that it is not obvious from the CMD what features are driving the fit to younger ages.

\citet{Sato:2025} measured the SFH of Ursa~Minor from deep, wide-area optical data acquired with Subaru/HSC.  Spatially, the data cover most of the northern extent of the galaxy from the center to $\sim\ 6\ r_h$. The CMDs reach $\sim2$~mag below the oMSTO in the g and i bands. Using their new genetic annealing algorithm for measuring SFHs, they found that $\sim50$\% of the stellar mass formed by $\sim12.5$~Gyr ago, and $\sim80\%$ of the stellar mass formed by $\sim11.5$~Gyr ago, with a small tail to ages as young as $\sim9$~Gyr ago.  The latter value is $\sim2$ and $\sim6.5$~Gyr older than the times by which 80\% of the stellar mass formed in the small \hubble\ field analyzed by \citet{Weisz:2014a} ($\sim9.3$~Gyr ago) and in the analysis of \citet{YangY:2024} ($\sim5$~Gyr ago), respectively.  Overall, the \citet{Sato:2025} SFH is similar to that of \citet{Carrera:2002}.

\citet{Sato:2025} reported no clear age gradient.  However, they did identify two distinct metallicity populations, the first at [Fe/H] $=-2.5$ and the second at [Fe/H] $=-2.2$, which are similar to the two chemodynamically distinct populations reported by \citet{Pace:2020}. They found that the more metal-rich population formed $\sim1$~Gyr later than the metal-poor population.  The metal-rich population also appears to have a gradient such that it is more centrally concentrated, while the metal-poor population is distributed more uniformly over the extent of the galaxy.  Their CMD-based MDFs are in good agreement with spectroscopic abundance measurements when matched to the same area \citep[$\lesssim2\,r_h$;][]{Cohen:2010, Kirby:2011, Pace:2020, Walker:2023, Shetrone:2025}.  Overall, the \citet{Sato:2025} analysis of Ursa~Minor is among the most comprehensive and detailed (i.e., the combination of CMDs and chemo-dynamics) available for a MW satellite at the present and sets an important standard showing what is possible in future studies of nearly all LG dwarf galaxies.

\citet{Yang:2024} reported the detection of candidate young ($\sim1$--$3$~Gyr) CHeB stars. But, the deep CMDs from \citet{Carrera:2002} and \citet{Sato:2025} do not show a corresponding MS population. Instead, both studies suggested that the blue plume population is consistent with being entirely BSSs.  Separate analysis from \citet{Mapelli:2007} also suggests the blue plume is likely BSSs.

\subsubsection{Draco}
\label{sec:Draco_sfhs}
Draco was discovered in the 1950s \citep{Wilson:1955}.  It is located $\sim80$~kpc from the MW and is confirmed to be gas-poor  \citep[e.g.,][]{Putman:2021}. Along with Ursa~Minor, Draco is one of the best-studied MW satellites in terms of its dark matter content and stellar populations.  SFHs from the late 1990s and early 2000s have shown that Draco is almost entirely ancient (i.e., 13 Gyr old) making it, along with Ursa~Minor, a quintessential `red and dead' galaxy \citep[e.g.,][]{Mighell:1999, Aparicio:2001, Dolphin:2002b}.

The only wide-field SFH of Draco is from \citet{Aparicio:2001}.  They used a deep B-R CMD constructed from imaging using the Isaac Newton Telescope that covers 1 deg$^2$.  Their CMD reaches below the oMSTO. They found that $\sim90$\% of the stellar mass formed by $\sim10$~Gyr ago, assuming an oldest allowable age of 15 Gyr.  They also identified a small population of stars ($\sim5$\% of the stellar mass) that formed 2--3~Gyr ago, which they discussed as either BSSs or young populations. The SFH of \citet{Aparicio:2001}, using only SGB stars, also recovers this small, young population supporting the view that they could be \textit{bona fide} younger stars.

\citet{Aparicio:2001} find no evidence for age or metallicity gradients. They report a mean metallicity of [Fe/H]$\sim-1.8$, which is in reasonable agreement with spectroscopic studies \citep[e.g.,][]{Kirby:2011, Walker:2023, Shetrone:2025}.

\citet{Weisz:2014a} analyzed the CMD from a single \hubble/WFPC2 pointing near the galaxy's center.  The CMD extends well-below the oMSTO.  \citet{Weisz:2014a} found that $\sim50$\% of the stellar mass formed by $\sim11$~Gyr ago and $\sim90$\% by $\sim10$~Gyr ago.  They found no evidence for intermediate or young star formation, though they did identify a clear blue plume population, which they considered to be BSSs.   They did not report an AMR or have the spatial coverage for age or metallicity gradients.

Analysis by \citet{Mapelli:2007} contended this blue plume population must be BSSs, but identification of putative CHeB stars in Draco by \citet{Yang:2024} supports a $\sim2$~Gyr stellar population in Draco.  The study of \citet{Aparicio:2001} can only rule out a population younger than $\sim2$~Gyr old, but is inconclusive about whether the blue plume is BSSs or a sparse younger population.

\subsubsection{Carina}
\label{sec:Carina_sfhs}
Carina is located $\sim100$~kpc \citep{Nagarajan:2022} from the MW.  Despite its proximity, it was not discovered until the late 20th century owing to its low surface brightness \citep{Cannon:1977}.  It is confirmed to be a gas-poor system \citep{Putman:2021} and there is some evidence that it is disrupting \citep[e.g.,][]{Munoz:2006, Battaglia:2012}.

Carina's stellar populations have long been of interest to the community as even early deep CMDs showed visually distinct MSTO and SGB populations, one of the clearest known indications of multiple distinct episodes of star formation in a nearby galaxy \citep{Smecker-Hane:1994, Mighell:1997, Hurley-Keller:1998, Rizzi:2003, Monelli:2003}.

\citet{Small:2013} measured the SFH of Carina by applying a new genetic CMD modeling algorithm to deep B and V ground-based observations presented in \citet{Bono:2010}.  The CMD used for the SFH determination covers the central $40^\prime \times 55^\prime$ area and extends $\sim0.5$~mag below the oMSTO.  They report a SFH that has two main episodes, one at $\sim13$ Gyr~ago with a metallicity of [Fe/H] $\sim-2.3$ that accounts for 34\% of the stellar mass formed and a second burst at $\sim6$ Gyr~ago with a metallicity of [Fe/H] $\sim-1.8$ (46\% of the stellar mass).  They reported declining star formation to $\sim3$~Gyr ago, and attribute a low-level of star formation at younger ages (e.g., 1 Gyr ago) to BSSs.  They did not report any gradient information.

\citet{deBoer:2014} measured the SFH of Carina from deep, ground-based B and V band imaging that covers past the tidal radius: $28^{\prime}$ along the minor axis and $1.6^{\circ}$ along the major axis.  The resulting CMD extends $\sim0.5$~mag below the oMSTO.  They included $\sim300$ stars with spectroscopic [Fe/H] measurements in their analysis.  Within the tidal radius, they found the SFH of Carina was characterized by two distinct populations, one star formation event $>8$~Gyr ago and low-metallicity ([Fe/H] $\lesssim -2.0$) that formed $\sim40\%$ of the galaxy's stellar mass.  They identify a second episode from $\sim2$--$7$~Gyr ago that accounts for $\sim60$\% of the stellar mass formed. The mean metallicity during this event reached a value of [Fe/H] $\sim-1.0$. Over 75\% of the stellar mass formed near the center at intermediate ages.  They noted a clear and strong age gradient such that younger and metal-rich stars ([Fe/H] $\sim-1.7$) are located in the center and that the outer radii have older and more metal-poor stars ([Fe/H] $\sim-2.1$).  They noted that Carina's SFH does not seem compatible with simple (i.e., one zone) chemical evolution models.

\citet{Weisz:2014a} measured the SFH of Carina for a single central \hubble/WFPC2 field (F555W,F814W).  The CMD reaches $\sim2$ mag below the oMSTO.  They reported a complex SFH with two distinct episodes of star formation $\gtrsim8$~Gyr ago and $\sim2$--$5$~Gyr ago, with a lull or gap in between.  Each episode contains $\sim50$\% of the total stellar mass formed.  Overall, they report that $50$\% of the stellar mass formed by $\sim8$~Gyr ago and 80\% of the stellar mass formed by $\sim4$~Gyr ago, which is similar to previous results.  They did not report an AMR.

\citet{Santana:2016} measured the SFH of Carina using deep g and r band imaging that spans over $2^{\circ}$, spectroscopic metallicities for 430 stars, and the CMD modeling methodology of \citet{deBoer:2012}.  Their CMD reaches $\sim1.5$~mag below the oMSTO.  They provided both a global SFH as well as spatially resolved SFHs.  For the global SFH, they found two main episodes of star formation.  In the first, $\sim55$\% of Carina's stellar mass formed $\sim10$--$13$~Gyr ago with metallicities of [Fe/H] $\sim -3.0$ to $-2.0$. A second episode occurred $2$--$8$~Gyr ago in which $\sim45$\% of its stellar mass formed.  The metallicity over this period increased from [Fe/H] $\sim -2.0$ to $-1.0$.  They report a gap in star formation between the two episodes from $\sim8$--$10$~Gyr ago.  The results are qualitatively similar to \citet{deBoer:2014}, although the older population in this study is more dominant.  An inferred small amount of recent star formation is attributed to BSSs.  For their spatially resolved SFH, similar to other studies, they reported a positive age gradient and a negative metallicity gradient, both of which are more pronounced at larger radii. \citet{Yang:2024} suggest that at least some of the blue plume stars may be genuinely younger stars (e.g., 1--3~Gyr in age) and not BSSs.

The AMRs reported by all CMD modeling studies agree well (i.e., within $\sim0.3$~dex) with spectroscopic observations in Carina \citep[e.g.,][]{Koch:2006b, Kirby:2011, Venn:2012, Walker:2023, Shetrone:2025}.

Two studies that employ slightly less rigorous modeling reported the presence of multiple distinct star formation events in Carina.  \citet{Savino:2015} re-measured the SFH of Carina using the data from \citet{deBoer:2014} and detailed models of the HB including mass-loss.  They did not undertake a full exploration of parameter space, but rather demonstrated the existence of four separate bursts of star formation $\sim2$, $5$, $8.6$, and $11.5$~Gyr ago.  Critically, this paper also showed that modeling the HB can, in principle, provide more precise time information than the MSTO populations.

\citet{Norris:2017} used the deep B and I-band CMD from \citet{Stetson:2011} along with select synthetic CMDs to identify multiple star formation events.  Based on qualitative comparisons of real and synthetic CMDs, they found four separate star formation episodes with [Fe/H] $=-1.85$, $-1.5$, $-1.2$, and $-1.15$ at ages of $\sim13$, $\sim7$, $\sim3.5$, and $\sim1.5$~Gyr ago.

\subsubsection{Sextans}
\label{sec:Sextans_sfhs}
Due to its low surface brightness, Sextans was not discovered until 1990, making it one of the most recently discovered classical MW satellites \citep{Irwin:1990}. It is a gas-poor galaxy \citep{Putman:2021} located $\sim80$~kpc from the MW \citep[e.g.,][]{Medina:2018, Vivas:2019}.  In terms of SFH measurements, it is among the least studied of the MW satellites.  For example, it is one of the few classical MW satellites without a \hubble-based SFH; existing \hubble\ imaging was obtained for proper-motion measurements.

\citet{Lee:2009} measured the first SFH of Sextans.  They constructed a CMD that reached the oMSTO based on CFHT V and I optical imaging \citep{Lee:2003} that covered the central $42^{\prime}\times28^{\prime}$.  They found a purely ancient population in which $84$\% of its stellar mass formed by $\sim11$~Gyr ago, though we note they consider lookback times to 15~Gyr, which is older than the current age of the Universe.  They found marginal evidence for an age gradient such that the $\sim10$~Gyr population appears to be more centrally concentrated, while the older populations are distributed throughout the galaxy. They found that the mean metallicity changes from [Fe/H] $=-1.6$ to [Fe/H] $=-1.8$ from the central to outer regions of their data. They identified a blue plume population that is most likely BSSs as they do not identify other evidence of intermediate age or recent star formation.

\citet{Bettinelli:2018} measured the SFH of Sextans using a CMD that extends below the oMSTO based on B and I imaging taken with Subaru Suprime-Cam.  The data cover a $34^\prime\times27^\prime$ area centered on the galaxy.  Their SFH is nearly identical to that of \citet{Lee:2009}, although the improved depth and contemporary CMD modeling techniques provide more precision.  \citet{Bettinelli:2018} found that $\sim90$\% of the stellar mass formed by $\sim11.5$~Gyr ago, assuming an oldest possible age of 13.5~Gyr ago.  They suggested that the main burst of star formation occurred within a $\sim0.6$~Gyr period, based on temporal resolution tests.  They found no evidence of a radial age gradient, and both their AMR and radial metallicity trends are consistent with a constant value of [Fe/H] $=-1.6$.  The CMD-based metallicities are broadly consistent with results of spectroscopic analyses \citep[e.g.,][]{Kirby:2011, Walker:2023, Shetrone:2025}.

\citet{Bettinelli:2018} suggest the blue plume is composed entirely of BSSs.  However, as with several other classical MW dSphs, \citet{Yang:2024} identify a small population of candidate CHeB stars, which would indicate much more recent star formation ($\sim1$--$3$~Gyr ago) than suggested from the deep CMD analysis of \citet{Bettinelli:2018}.

Based on their SFH, \citet{Bettinelli:2018} suggest that Sextans is a fossil from the pre-reionization era. However, it is unclear if reionization alone could have halted Sextans star formation. In general, classical MW satellites, like Sextans, are thought to be too massive to have been quenched by reionization.  \citet{Bettinelli:2018} argue that supernova feedback, along with reionization, is likely to be responsible for quenching Sextans.

\subsubsection{Sagittarius}
\label{sec:Sagittarius_sfhs}
The Sagittarius dwarf galaxy is a tidally disrupting dwarf galaxy in the halo of the MW.  Its low surface brightness and location of $\sim19$~kpc behind the Galactic bulge contributed to its relatively late discovery \citep{Ibata:1994, Ibata:1995}.  Its associated stars are stretched across the MW by tens of kpc \citep[e.g.,][]{Ibata:2020, Muraveva:2025}.  It does not appear to have significant amounts of cold, neutral gas associated with it \citep[e.g.,][]{Putman:2021}.

Its close proximity, disrupting nature, and sizable GC population have made Sagittarius a popular target for the study of galaxy accretion.  These studies include both the core of the galaxy as well as the stream.   It is useful to recap some of the legacy SFH studies for context before summarizing more contemporary work.

Photometric studies in the 1990s used CMDs to identify multiple distinct stellar populations in Sagittarius.  For example, as part of the Sagittarius Dwarf Galaxy Survey, a wide-field survey of select fields across Sagittarius, \citet{Bellazzini:1999} semi-empirically inferred a low-level of ancient star formation, a main enhancement from 8--10 Gyr ago, and star formation that ended $\sim2$~Gyr ago, with possible low-level star formation until $\sim0.5$~Gyr ago.  They posit a mean metallicity range of [Fe/H] $\sim-1.3$ to $-0.7$ during the main burst of star formation. \citet{Layden:2000} used optical imaging of RR Lyrae to infer multiple episodes of star formation with ages of $\sim$11, 5, and 3--0.5~Gyr ago with metallicities of [Fe/H] values of $-1.3$, $-0.7$, and $-0.4$, respectively.  \citet{Dolphin:2002b} was the first to measure the SFH of Sagittarius based on \hubble data.  Their analysis of 6 WFPC2 fields (F555W, F814W) shows a significant enhancement 8--11~Gyr ago, along with metallicity evolution from [M/H] $\sim -1.0$ at older ages to [M/H] $=0$ at recent times.

\cite{Siegel:2007} measured the SFH near the core of Sagittarius using deep \hubble/ACS data taken as part of the ACS Galactic Globular Cluster Survey \citep{Sarajedini:2007}.  The primary target of this field is the cluster M54, which is itself the nuclear cluster of Sagittarius, but field stars from Sagittarius are also in the ACS field providing an opportunity to measure the field SFH.  Their F606W and F814W CMD reaches $\sim2$~mag below the oMSTO.  \citet{Siegel:2007} identify multiple episodes of star formation in the Sagittarius field, including activity from 9--13 Gyr at [M/H] $\sim -1.4$, at 4 and 6 Gyr ago with [M/H] $\sim -0.6$ to $-0.4$, $\sim2.3$~Gyr ago with [M/H] $\sim 0.0$, and a small amount of star formation $\sim 0.1 - 0.8$~Gyr ago with [M/H] $\sim 0.6$.

\citet{Weisz:2014a} measured a SFH for Sagittarius from two \hubble/WFPC2 (F555W, F814W) fields that reach well-below the oMSTO and are located near the center of the galaxy.  They found low levels of ancient star formation such that $\sim20$\% of the stellar mass formed prior to $\sim9$~Gyr ago, and that 70\% of the stellar mass of Sagittarius formed at intermediate ages from $\sim3$--$7$~Gyr ago.  They found little star formation younger than $\sim2$--$3$ Gyr ago.  They did not report an AMR.

\citet{deBoer:2015} measured the SFH of Sagittarius stream using deep photometry and spectroscopy from the SDSS Stripe 82 region.  Using a combination of color and spatial criteria, they constructed CMDs of a bright and faint stream around the main body of the galaxy.  Their data span $\sim20^{\circ}$ on the sky and their $g-i$ CMDs reach well-below the oMSTO.  They supplemented their CMDs with SDSS spectroscopic metallicities of RGB stars and employed a SFH derivation technique described in \citet{deBoer:2012}.  The SFHs of both streams are dominated by ancient populations, with the bright stream exhibiting a more extended SFH.  Stellar populations in the faint stream are almost entirely $>10$~Gyr in age. The bright stream shows star formation extending for several Gyr with $\sim50$\% of the stellar mass forming by $\sim10$~Gyr ago followed by a sharp decline such that 80\% of the stellar mass formed by $\sim7.5$~Gyr ago.

\citet{deBoer:2015} measured an AMR for the bright and faint streams.  They found AMRs with little scatter with the metallicity of the faint stream changing from [Fe/H] $\sim-1.9$ at 13~Gyr ago to [Fe/H] $\sim-0.9$ by $\sim8$~Gyr ago.  The bright stream shows a similar evolution, though with a much more prominent metal-rich population. Their reported metallicities from joint CMD and spectral-abundance fitting are in broad agreement with other spectroscopic abundance studies of Sagittarius \citep[e.g.,][]{McWilliam:2013, Hasselquist:2017, Hasselquist:2021}, though given the large angular size of Sagittarius spatial selection effects make a direct comparison challenging.

In an interesting related study, \citet{Ruiz-Lara:2020a} measured the SFH of the MW within $\sim2$~kpc of the Sun using \textit{Gaia} photometry and found correlations between similarly timed peaks in the MW's SFH and the SFH of Sagittarius from \citet{Siegel:2007} and \citet{deBoer:2015}.  They suggest this may be evidence of the influence of the MW and Sagittarius on one another.

\subsubsection{CVn~I}
\label{sec:CVnI_sfhs}
Canes~Venatici~I was discovered by \citet{Zucker:2006b} in their search for dwarf galaxies in SDSS imaging.  It was the brightest MW satellite ($M_V\sim-8.8$; \citealt{munoz:2018b}) discovered within the SDSS footprint.  It is notable for being fairly diffuse and located in the outer halo of the MW ($D\sim220$~kpc; \citealt{Zucker:2006b}).  It has been confirmed to be gas-poor \citep{Putman:2021}.

There are only three published SFH measurements of CVn~I.  The first is from \citet{deJong:2008}, which modeled the SDSS $g-r$ discovery CMD.  The data reach a depth just below the HB.  They found that CVn~I appears to be predominantly ancient, but with hints of a low-level, younger ($\sim3$~Gyr) population.  For the older population they found a metallicity of [Fe/H]$\sim -2$ and that the younger population has [Fe/H]$\sim -1.5$.

\citet{Weisz:2014a} measured the SFH of CVn~I from a single central \hubble/WFPC2 field (F606W, F814W) from a CMD that reached below the oMSTO.  They found that  CVn~I formed $\sim60$\% of its stellar mass by $\sim12.5$~Gyr ago and $\sim$40\% between 8 and 12.5~Gyr.  They reported no evidence of younger star formation, but have limited spatial sampling.  They did not publish an AMR.

\citet{Perez-Garcia:2026} measured the SFH of CVn~I from deep \hubble/ACS imaging (F606W, F814W; GO-15476, PI: Weisz) of two adjacent fields that lie along the major axis and within the half-light radius.  Their CMD contains $\sim11{,}000$ stars and extends $\sim2$--$2.5$~mag below the oMSTO, making it the deepest CMD of CVn~I to date.  They modeled the CMD with the new DirSFH code and the BaSTI-IAC stellar models \citep{Hidalgo:2018}.  They found that the SFH is dominated by a single episode of star formation that peaked $\sim11.8$~Gyr ago, which is also when 50\% of the stellar mass had formed; 90\% had formed by $\sim10$~Gyr ago.  They also found a weaker episode $\sim8$~Gyr ago, but no evidence for younger star formation.  Their AMR shows that the full metallicity range ([Z/H] $\sim-2.4$ to $-1.5$, with a peak at $\sim-1.8$) was established early.  When converted to [Fe/H] assuming that [$\alpha$/Fe] declines with increasing metallicity, their MDF is consistent with spectroscopic metallicities of RGB stars.  They found no radial gradient in age or metallicity across their fields.

\citet{Perez-Garcia:2026} interpret the ``blue plume'' as BSSs rather than young stars: its frequency relative to HB stars is consistent with the $F_{\rm BSS}$--$M_V$ relation of \citet{Momany:2007}, it is spatially uniform across both fields, and removing it (together with likely field contaminants) from the fit eliminates the apparent young populations without otherwise changing the SFH, as expected if these populations lack counterparts elsewhere in the CMD.  Their SFH shows no evidence for the younger populations suggested by earlier, shallower studies \citep[e.g.,][]{deJong:2008}.  Overall, the SFHs of \citet{Weisz:2014a} and \citet{Perez-Garcia:2026} are broadly consistent: both are predominantly ancient and show no evidence for star formation in the past $\sim8$~Gyr.  However, they differ in detail.  The SFH of \citet{Weisz:2014a} has $\tau_{50}\approx12.8$~Gyr and $\tau_{80}\approx8.8$~Gyr, whereas that of \citet{Perez-Garcia:2026} has a younger $\tau_{50}$ ($\approx11.8$~Gyr) and an older $\tau_{80}$ ($\approx10.8$~Gyr), i.e., a shorter main episode of star formation.  Differences in CMD depth, spatial coverage, and modeling choices (e.g., CMD-fitting codes and stellar models) may all contribute.

\citet{Perez-Garcia:2026} note that a peak of star formation after the end of reionization makes CVn~I one of the youngest of the predominantly ancient MW satellites, in contrast to Draco, Ursa~Minor, and Sextans, which formed their stars earlier.  They suggest that the halo of CVn~I evolved just above a redshift-dependent critical mass for star formation until $z\sim2$, when gas loss driven by supernova feedback, acting together with the UV background (and perhaps an early pericentric passage about the MW), pushed it below this threshold and ended star formation.  However, as they caution, their fields only sample the central region of CVn~I, which contains $\sim18$\% of its stellar mass.  If CVn~I has an outside-in age gradient similar to those of other MW satellites (\S\ref{sec:mw_classical}; see also \citealt{Riggs:2024}), its global SFH may be older than that measured from the central fields.  Wide-area imaging that reaches the oMSTO is needed to test this possibility.

\subsection{M31 Dwarf Ellipticals and Gas-Rich Satellites}
\label{sec:app_m31_de}

\subsubsection{M32}
\label{sec:M32_sfhs}
M32 is a compact dE located in close proximity to M31 \citep[21~kpc, ][]{Savino:2022}. It has a stellar mass of $\sim 8\times 10^8 M_{\odot}$ (based on its luminosity from \citealt{Savino:2022} and $M/L_V\approx2$) and no detectable \hi{} mass \citep{Putman:2021}. It was first discovered by Guillaume Le Gentil in 1749 and first resolved into individual stars by Baade \citep[e.g.,][]{Baade:1944, Jones:1991, VandenBergh:2000}. Due to its morphology and concentrated light profile, M32 represents a unique object in the LG, whose formation mechanism is still unclear \citep[e.g.,][]{DSouza:2018}.

Using the HRC channel of \hubble/ACS (F435W, F555W), \citet{Monachesi:2012} studied a $26\times29$~arcsec$^2$ region located $\sim 2^{\prime}$ ($\sim 4\, R_{eff}$) from M32's center. Their SFH showed a prominent episode of star formation 2--5 Gyr ago, responsible for 40\% of M32's stellar mass in the field. Approximately 55\% of the stellar mass formed in the 5--14 Gyr period. They reported only a few percent of the stellar mass having ages younger than 2~Gyr. The CMD-fit suggests little-to-no metallicity evolution, with the majority of the stellar population being close to Solar metallicity.

SFH measurements over a wider field were obtained by \citet{Weisz:2014a} and \citet{Savino:2025}, who used \hubble/WFPC2 (F555W, F814W) and the WFC channel of \hubble/ACS (F606W, F814W), respectively \citep[the latter field contains the][HRC field]{Monachesi:2012}. Both SFH determinations qualitatively agree with the SFH of \citet{Monachesi:2012}, having a major star formation episode approximately 3~Gyr ago and only a small amount of more recent star formation. Both the SFHs from \citet{Weisz:2014a} and \citet{Savino:2025} contain a higher amount of old star formation, with $\sim$75\% of the total stellar mass having formed more than 5~Gyr ago. It is worth noting that in all the datasets described above, the severe stellar crowding strongly limits the ability to robustly recover the oldest star formation episodes.

\subsubsection{NGC~205}
\label{sec:NGC205_sfhs}
NGC~205 is a massive dE in close proximity to M31 \citep[58~kpc, ][]{Savino:2022} that was first discovered by Charles Messier in 1773 \citep[e.g.,][]{Jones:1991, VandenBergh:2000}. It has a stellar mass of $\sim 7\times 10^8 M_{\odot}$ (based on its luminosity from \citealt{Savino:2022} and $M/L_V\approx2$) and a detected \hi{} mass of $4 \times 10^5 M_{\odot}$ \citep{Putman:2021}. NGC~205 displays a number of morphological peculiarities, such as pronounced tidal distortion, conspicuous dust patches, and a centrally concentrated blue stellar population \citep[e.g.,][]{Hodge:1973,Bertola:1995,Choi:2002,Marleau:2006}.

\citet{Monaco:2009} used the HRC channel of \hubble/ACS (F555W, F814W) to study the recent SFH in the inner $\sim 120$ pc of NGC~205. Their photometry reaches down to the MSTO of a $\sim 650$ Myr stellar population. They found that the MS luminosity function is consistent with a roughly constant rate of star formation between $\sim 60$ Myr and $\sim 300$ Myr ago, with evidence of continuous star formation back to at least $\sim 650$ Myr ago and the presence of a secondary star formation event $\sim 25$ Myr ago.

Using three \hubble/WFPC2 (F555W, F814W) fields, \citet{Weisz:2014a} derived a lifetime SFH of NGC~205 over a wider area, comparable to the half-light area. They measured a wide range of ages for this galaxy, with star formation active throughout the lifetime of this galaxy. Virtually identical conclusions are reached, using deeper \hubble/ACS (F606W, F814W) data, by \citet{Savino:2025}.  Quantitatively, \citet{Savino:2025} find $\tau_{50}\sim8.5$~Gyr and $\tau_{80}\sim3.3$~Gyr.  In their SFH, $\sim25$\% of the stellar mass in the observed field formed more than 10~Gyr ago and $\sim25$\% formed within the last $\sim5$~Gyr. Only $\sim3$\% formed within the last 2~Gyr.

\subsubsection{NGC~147/NGC~185}
\label{sec:NGCs147185_sfhs}
NGC~147 and NGC~185 are a pair of dEs located at 107~kpc and 154~kpc, respectively, from M31 \citep{Savino:2022}. NGC~185 was discovered in 1787 by William Herschel and NGC~147 was discovered in 1829 by his son, John Herschel \citep{Herschel:1789, Herschel:1833}.

NGC~147 has a stellar mass of $\sim 7\times 10^8 M_{\odot}$ and no detectable \hi{}, while NGC~185 has a stellar mass of $\sim 3\times 10^8 M_{\odot}$ and a detected \hi{} mass of $\sim 1\times 10^5 M_{\odot}$ (stellar masses based on the luminosities of \citealt{Savino:2022} and $M/L_V\approx2$; \hi{} masses from \citealt{Putman:2021}). Their similar morphology, luminosity, and projected separation have long prompted the idea that these two galaxies form a gravitationally bound pair \citep[e.g.,][]{VandenBergh:1998,Arias:2016}, although this hypothesis seems to be rejected by recent kinematical data \citep[e.g.,][]{Geha:2010,Watkins:2013,Sohn:2020}. We discuss them as a pair here, following historical convention. A thorough comparison of the SFH in the two galaxies is hampered by the fact that high-resolution space observations can only cover a small area compared to the total size of these galaxies. However, from available data it appears the two galaxies have followed somewhat different evolutionary paths.

Using \hubble/WFPC2 (F555W, F814W) data close to the galaxies' photometric centers, \citet{Weisz:2014a} measured extended SFHs for both NGC~147 and NGC~185. Their SFHs for the two systems are broadly comparable with each other. \citet{Weisz:2015a} reported that NGC~147 formed 90\% of its stellar mass by $\sim 2.7$~Gyr ago and NGC~185 formed 90\% of its stellar mass by $\sim 3.6$~Gyr ago. However, the depth of the data only reached below the HB, which complicates the measurement of the ancient SFH.

Using deeper \hubble/ACS (F606W, F814W) imaging that reaches below the oMSTO, \citet{Geha:2015} showed that, while both NGC~147 and NGC~185 contain stellar populations over a large range in ages, the details of their stellar mass build-up differ. They reported that NGC~147 formed approximately 40\% of its stellar mass by 12.5~Gyr ago, with most of the remaining star formation taking place between 5 and 7~Gyr ago. Conversely, NGC~185 had a much more rapid stellar mass build up, forming 70\% of its stellar mass by 12.5~Gyr ago, and forming most of the remaining stellar mass by a lookback time of 8~Gyr. Interestingly, in spite of its earlier star formation, NGC~185 does possess low levels of present-day star formation, while NGC~147 does not. Later reanalysis of the same dataset by \citet{Savino:2025} provided a similar picture. They reported that NGC~147 reached 90\% of its total star formation 5.7~Gyr ago, while NGC~185 reached it 9.3~Gyr ago.

By analyzing the K-band magnitude of LPV stars, \citet{HamedaniGolshan:2017} reconstructed the lifetime SFR evolution over the central $6.5^\prime \times 6.5^\prime$ of both galaxies (i.e., a substantial fraction of the stellar mass in these systems). They confirmed findings from earlier studies, showing that NGC~185 had a more rapid stellar mass build up, which peaked approximately 8.3~Gyr ago, while NGC~147 had a prominent star formation episode between 7~Gyr and 3~Gyr ago. According to their measurements, both NGC~147 and NGC~185 had low level star formation until recent times, ceasing 300~Myr and 200~Myr ago, respectively. In the case of NGC~185, they also reported a significant gradient in recent star formation, with star formation happening in the last 1~Gyr, only within the central 500~pc of the galaxy.

\subsubsection{IC~10}
\label{sec:IC10_sfhs}

   At a distance of $\sim250$~kpc from M31 \citep{Sanna:2008}, IC~10 is located within our nominal 300~kpc boundary and is thus an M31 satellite (\S \ref{sec:concepts_classification}).  With a stellar mass of $8.6\times10^7~M_{\odot}$, it is a gas-rich galaxy, with a ratio of M$_{HI}$/M$_*$ of 0.6.  IC~10 is classified as a starburst galaxy, showing a large number of young, massive Wolf-Rayet stars \citep[e.g.,][]{Massey:1995}.  \citet{Nidever:2013} discovered a companion \hi{} cloud with an \hi{} mass of roughly 1\% that of IC~10, which they interpret as the result of a recent interaction with a low-mass galaxy.  This could explain both the starburst nature of IC~10 and the previous reports of counter-rotating \hi{}. A proper motion determination from maser observations \citep{Brunthaler:2007} allowed \citet{Nidever:2013} to reconstruct an orbital path which further constrains other explanations for the current starburst. In that reconstruction, IC~10 passed within 80 kpc of M31 about 1.8 Gyr ago. A more recent proper motion from combining \textit{Gaia} and \hubble\ observations (but less precise than afforded by the maser observation) enabled \citet{Bennet:2024} to determine a pericenter distance of 130 kpc from M31, which occurred $\sim1.2$~Gyr ago. Thus, although IC~10 is currently $\sim250$~kpc from M31, it was much closer in the recent past.

Because of its low Galactic latitude ($-$3$^\circ$) and resulting high foreground extinction (A$_V\approx4.3$~mag from foreground dust maps, but $\approx2.3$~mag in the low-extinction window targeted by \hubble; \citealt{Weisz:2014a}), optical studies are greatly limited in their possible photometric depth.  \hubble/WFPC2 observations reaching to V and I (F555W, F814W) depths of 26.7 and 25.3 mag.\ (corresponding to an absolute V mag of $\sim$ 0), produced a CMD with over 29,000 stars.  From these observations, \citet{Weisz:2014a} presented a SFH showing $\sim30$\% of its stellar mass forming in roughly the first Gyr, followed by a quiescent period lasting until about 5 Gyr ago, followed by relatively constant star formation.  A closer look at the details of the SFH indicates enhanced star formation between $\sim$ 2.5 and 1.5 Gyr ago.  It is tempting to connect this enhanced star formation to the close passage by M31 as indicated by the reconstructed orbits from \citet{Nidever:2013, Bennet:2024}.

From CFHT optical and UKIRT near-IR observations, \citet{Gerbrandt:2015} produced density maps of the
young, intermediate, and old stellar populations, finding no evidence of a lower-mass interacting
galaxy as hypothesized by \citet{Nidever:2013}.  Using \textit{Euclid} Early Release Observations, \citet{Annibali:2026} traced the RGB population of IC~10 to $\sim8$~kpc from its center and to $\mu\sim29$~mag~arcsec$^{-2}$, revealing an extended outer component that contains a substantial fraction of the old stellar mass, tentative tidal overdensities aligned with its orbit around M31, and a new TRGB distance of $762\pm20$~kpc.

\citet{DellAgli:2018} combined the UKIRT observations from \citet{Gerbrandt:2015} with Spitzer
observations from \citet{Boyer:2015a} to study the AGB and RSG stars in IC~10.  From the AGB stars, they derived a SFH which compares well with that of \citet{Weisz:2014a}, in particular with enhanced star formation 2.5 Gyr ago, although their derived SFH is generally closer to a constant SFR than that of \citet{Weisz:2014a}.

\subsubsection{LGS-3 (Pisces I)}
\label{sec:PiscesI_sfhs}

Originally discovered as LGS-3\footnote{LGS-3 was the third out of five ``Local Group Suspected'' (LGS) galaxies found near M31.  It is the only one of the five to be confirmed as a true galaxy.} \citep{Kowal:1978}, Pisces~I is located at a distance of 290 kpc from M31 \citep{Savino:2022}. It is therefore an M31 satellite by our definition (\S \ref{sec:concepts_classification}), although LCID considered it relatively isolated \citep{Hidalgo:2011, Gallart:2015}.  Because of this history, it is the only M31 satellite of comparable luminosity with its own entry in this appendix. Pisces~I is equally close to M33, and there have been significant discussions in the literature concerning the true ``host'' galaxy for Pisces~I \citep[see discussion in][]{Bennet:2025, Patel:2026}. From very precise proper motion observations, \citet{Bennet:2025} place Pisces~I on its first infall into the virial radius of M31.

With a stellar mass of 10$^6$ M$_{\odot}$ and an M$_{HI}$/M$_*$ of 0.4, it is a gas-rich, low-mass star-forming galaxy.  Because of its detectable \hi{} and a lack of \hii{} regions, Pisces~I has been classified as a transition-type galaxy.  Pisces~I lies in the plane of co-rotating satellite galaxies of M31 as identified by \citet{Ibata:2013}.

\citet{Hidalgo:2011} used deep \hubble/ACS observations (F475W, F814W) of Pisces~I to determine a lifetime SFH. They found an initial strong episode of star formation $\sim12$~Gyr ago with a duration of roughly 1.4 Gyr.  After that initial episode, which formed roughly 90\% of the stellar mass of the galaxy, star formation continued at a greatly reduced level until the present. The SFH also shows a modest increase between $\sim2.5$ and $\sim1.5$~Gyr ago, which \citet{Patel:2026} associate with a pericentric passage around M33.  The later star formation is accompanied by chemical enrichment raising the mean metallicity from [M/H] $\sim$ $-$1.7 at 11 Gyr ago to [M/H] $\sim$ $-$1.2 at 2 Gyr ago.   The chemical enrichment has continued to a present-day value of [M/H] $\sim$ $-$1.0.

\citet{Hidalgo:2013} showed that Pisces~I (and the other transition galaxy studied, Phoenix) has a strong radial population gradient in the sense of an exclusively old stellar halo with recent central star formation.   This results in a negative mean metallicity radial gradient and a decreasing stellar scale length with time.

From the compendium of \hubble/WFPC2 photometry, \citet{Weisz:2014a} used a F555W$-$F814W CMD that reached the HB to derive a SFH for LGS-3 which rose relatively quickly and has shown a decreasing SFR throughout most of its history, in general agreement with the SFH derived from the deeper \hubble/ACS photometry by
\citet{Hidalgo:2011}.  They did not publish an AMR.

In their comprehensive study of M31 satellite galaxies, \citet{Savino:2025} reanalyzed the \hubble/ACS imaging presented in \citet{Hidalgo:2011} and recovered a similar SFH dominated by early star formation ending about 10 Gyr ago and followed by relatively constant star formation at a lower amplitude.  Pisces~I fits well in their derived relationship between absolute magnitude, distance from M31, and quenching time.

\subsection{Isolated Local Group Galaxies}
\label{sec:app_isolated}

\subsubsection{IC~1613 (DDO~8)}
\label{sec:IC1613_sfhs}

At distances of 730 kpc from the MW and 510 kpc from M31 \citep{Bernard:2010, Savino:2022},
IC~1613 is an isolated galaxy.  With a stellar mass of $\sim10^8$ M$_{\odot}$, it is a gas-rich galaxy, with a ratio of M$_{HI}$/M$_*$ of 0.7. From their proper motion determination, \citet{Bennet:2024} reconstructed an orbital trajectory for IC~1613 showing it to be clearly on its first infall into the LG.  Because of its high Galactic latitude ($-$61$^{\circ}$), with an accompanying low foreground extinction, and its low metallicity, with an accompanying low internal extinction, IC~1613 is a frequent target to calibrate distance measurement techniques \citep[e.g.,][]{Freedman:2009, Gorski:2011, Scowcroft:2013, Hatt:2017, Madore:2018}.  There are many studies of its resolved stellar populations.

The first detailed study of the stellar populations of IC~1613 with the \hubble\ was made by \citet{Cole:1999}.  A central field was observed with WFPC2 (F439W, F555W, F814W) which contained 64,000 stars and was crowding-limited. From the brightest MS and BHeB stars it was determined that star formation was roughly constant over the last Gyr.  This was followed by a deeper \hubble/WFPC2 (F555W, F814W) study in an outer field which nearly reached the oMSTO and observed 25,000 stars \citep{Skillman:2003}.  The SFH for this field showed a period of enhanced star formation of a factor of 2 - 3 over the period 3--6 Gyr ago on top of a nearly constant lifetime SFH.  \citet{Skillman:2014} present very deep \hubble/ACS observations (F475W, F814W) of a field at a similar radius to the \citet{Skillman:2003} field which reached below the oMSTO and observed 166,000 stars.  In these deeper observations, the SFR is observed to be relatively constant (within a factor of two) over the entire lifetime of the galaxy.  All three \hubble\ observations were reduced in a uniform manner and the SFHs from the two shallower \hubble/WFPC2 fields were found to be consistent with the nearly constant lifetime SFH for the deep ACS field.  This nearly constant SFH was noted to be in contrast to the early peaked SFHs most commonly found in simulations and pointed to the possibility of star formation prescriptions in the simulations which are inherently too efficient.

\citet{Cole:1999} also produced a first AMR for IC~1613 with [Fe/H] rising from $\sim$ $-$2 at the earliest times to a current value of $-$0.8.  The deep \hubble/ACS photometry from \citet{Skillman:2014} enabled the determination of an AMR in rough agreement with that determined by \citet{Cole:1999}, but starting at [M/H] $=$ $-$1.6 at earliest times and rising to a present value of $-$0.7.

\citet{DellAgli:2016} used ground-based J, H, and K band photometry \citep{Sibbons:2015} and \textit{Spitzer} mid-IR photometry \citep{Jackson:2007b, Boyer:2015a}, compared the AGB star distributions with those predicted from the SFH and AMR of \citet{Skillman:2014}, and found good agreement overall. \citet{Hashemi:2019} used the methodology developed in \citet{Javadi:2017} and the AGB star photometry listed above to derive a SFH for IC~1613. They found a SFH generally consistent with previous studies with no evidence of a dominant period of star formation over the past 5 Gyr.  They further infer that the constant star formation over recent times suggests that IC~1613 may have evolved in isolation for at least that long, in concordance with the recent proper motion measurement.

In their comprehensive study of M31 satellite galaxies, \citet{Savino:2025} reanalyzed the \hubble/ACS imaging of the field obtained by \citet{Skillman:2014} and recovered a similar SFH of nearly constant star formation over the entire history of the galaxy.  IC~1613 fits well in their derived relationship between absolute magnitude, distance from M31, and quenching time.

\subsubsection{Phoenix Dwarf Galaxy}
\label{sec:Phoenix_sfhs}

At a distance of 560 kpc from the LG barycenter, Phoenix is an isolated galaxy.  With a stellar mass of $\sim8\times10^5$ M$_{\odot}$, it is a gas-rich galaxy, with a ratio of M$_{HI}$/M$_*$ of 0.16. Although \citet{Teyssier:2012} found Phoenix to be highly likely to be associated with the MW, from their proper motion determination, \citet{Bennet:2024} reconstructed an orbital trajectory for Phoenix showing it to be clearly on its first infall into the LG.

A detailed SFH from \hubble/WFPC2 observations (F555W, F814W) was first presented by \citet{Young:2007}.  It showed an early episode of enhanced star formation followed by relatively constant star formation.  They combined the optical imaging with VLA \hi{} imaging and showed the relative distributions of the neutral ISM associated with Phoenix and its stars.
From kinematic and spatial arguments, they concluded that the neutral ISM, which is offset from the center of the galaxy but still within the extent of the stellar distribution, was likely ejected from the center of the galaxy and also still likely to be gravitationally bound to the galaxy.
Thus, the isolated nature of Phoenix (now confirmed through proper motion studies) renders it an ideal example of how feedback from star formation can distort the ISM distribution of a low-mass galaxy.

\citet{Hidalgo:2009} added \hubble/WFPC2 observations (F555W, F814W) of an outer field of Phoenix to determine its radial gradients in SFH and AMR.
They found that Phoenix was dominated by early star formation in the first few Gyrs and that the star formation gradually became more centralized with time.  Concordant with the radial gradient in SFH, a significant metallicity gradient is observed where the outer parts, with only old stars, show only low metallicity ([M/H] $\sim$ $-$1.8), while the inner parts, with stars at all ages, show a metallicity spread from [M/H] $\sim$ $-$1.8 to $-$1.1. Note that the idea that stellar feedback could lead to extreme stellar migration in low-mass galaxies \citep[e.g.,][and references therein]{El-Badry:2016}, which can result in similar predictions about gradients, was not considered at that time.
\citet{Hidalgo:2013} found the radial properties of Phoenix to be very similar to those of Pisces~I, the other ``transition'' galaxy in their sample.

From their compendium of \hubble/WFPC2 photometry, \citet{Weisz:2014a} derived SFHs for the two Phoenix fields cited above, and their results are in general agreement with those in \citet{Hidalgo:2013}.

The central field \hubble/WFPC2 observations of Phoenix were re-analyzed by \citet{YangY:2024} using their machine learning approach to determining SFHs.  Their SFH compares well with that determined by \citet{Weisz:2014a} although it appears more constant at early times, i.e., it does not show an early episode of enhanced star formation.

\citet{Menzies:2008} and \citet{Boyer:2009} used ground-based and \textit{Spitzer} near-IR imaging (respectively) to identify AGB stars in Phoenix. From a comparison with several other nearby low-mass galaxies, \citet{Boyer:2009} conclude that mass loss can still be traced by dust at low metallicities.  This conclusion was reinforced and expanded upon in \citet{Boyer:2015b,Boyer:2015a}.

\citet{Ordonez:2014} presented a detailed study of the RR Lyrae stars in Phoenix recovered from the \hubble/WFPC2 observations.  They derive an AMR which is in good agreement with the AMR derived by \citet{Hidalgo:2013}.

\subsubsection{NGC~6822 (DDO 209)}
\label{sec:NGC6822_sfhs}

At a distance of $\sim470$~kpc from the MW \citep{Rich:2014}, and a significantly
larger distance to M31, at first blush,
NGC~6822 appears to be an isolated galaxy \citep[e.g.,][]{Mateo:1998}.  Note that, due primarily to its relatively low Galactic latitude ($-$18$^\circ$) with its resultant significant foreground reddening (A$_V\approx0.6$), NGC~6822 has historically had a significant range in reported distances. With a stellar mass of $\sim10^8$ M$_{\odot}$, it is a gas-rich galaxy, with a ratio of M$_{HI}$/M$_*$ of 1.3. From their proper motion determination, \citet{Bennet:2024} reconstructed an orbital trajectory for NGC~6822 showing it to have just barely entered the virial radius of the MW approximately 2.8 Gyr ago.  Thus, NGC~6822 is likely a backsplash galaxy.

NGC~6822 is known for several distinguishing characteristics.  First, the very extended \hi{} distribution is roughly perpendicular to the position angle of the central stellar distribution \citep{Davies:1972, deBlok:2000}.  This misalignment continues well out into an extended stellar halo as traced by AGB stars \citep{Battinelli:2006}. Radial velocities of halo carbon stars showed the stellar halo to be kinematically disconnected from the \hi{} disk, leading \citet{Demers:2006} to conclude that NGC~6822 is a polar-ring galaxy, and therefore would have experienced an interaction within the last few Gyr.  Second, there is a hole in its \hi{} distribution which is more than 1 kpc across \citep{deBlok:2000}.  Third, NGC~6822 was suspected of having a close companion galaxy \citep{deBlok:2000, deBlok:2006}.

NGC~6822 presents significant challenges to obtaining a
detailed lifetime SFH.  First, NGC~6822 covers a very large angular size.
With an optical diameter of roughly $18^{\prime}$, space-based imaging historically had been limited to small portions of the galaxy \citep[e.g.,][]{Cannon:2012, McQuinn:2010b}.
Recently, this changed dramatically with the 0.7$^\circ$ $\times$ 0.7$^\circ$ early release \textit{Euclid} near-IR imaging presented in \citet[][]{Hunt:2025}, although the photometry only reaches to the RC.  Secondly, at both small Galactic longitude (25$^\circ$) and Galactic latitude ($-$18$^\circ$) there is significant foreground extinction (A$_V\approx0.6$) and foreground stellar contamination.

Until recently, wide-field studies of the SFH of NGC~6822 were limited to ground based observations and thus, were limited in photometric depth to the HB or brighter.   \citet{Tantalo:2022} analyzed a massive compilation of ground-based observations presenting a wide field study of resolved stars.  By using g, r, and i photometry, \citet{Tantalo:2022} were able to separate the Galactic foreground stars from the NGC~6822 stars to produce a clean sample for studying galaxy properties.  This work recovered and clearly defined the misalignment between the old halo stars and the younger disk stars as observed previously. \citet{Tantalo:2025} added the \hubble\ imaging from \citet{Cannon:2012} (discussed below) to their photometry compilation and through comparison with stellar models characterized the build-up of metals with time.

At present, SFHs from space-based observations are limited to
relatively shallow CMDs which are just barely sufficient to produce lifetime SFHs.  The \hubble/ACS (F475W, F814W) study by \citet{Cannon:2012} observed six fields spread along the \hi{} major axis. These observations were interpreted to argue against the presence of a small companion galaxy as proposed by \citet{deBlok:2000, deBlok:2006}.  One of the fields was centered on the giant \hi{} hole and it was determined that feedback from recent star formation provided sufficient energy to create that \hi{} hole.

\citet{Fusco:2014} re-analyzed these \hubble observations and derived SFHs up to a look-back time of 5 Gyr.  Four of the six fields observed showed declining SFRs over the last 1.5 Gyr, while two fields showed enhanced star formation in the last 0.5 Gyr. They determined that there was no evidence of a spatial gradient in metallicity, although there was strong evidence of an AMR and that the mean stellar metallicity and spread in metallicity are in agreement with the spectroscopic determinations by \citet{Kirby:2013}.

In their compendium of \hubble/WFPC2 photometry, \citet{Weisz:2014a} derived SFHs for three NGC~6822 fields which were originally obtained in order to study stellar clusters.  These fields are relatively shallow, and thus their derived SFHs carried significant uncertainties, but all are broadly consistent with constant star formation.  \citet{Weisz:2014a} combined all three into a single, unweighted SFH and found evidence of an initial burst, followed by a quiescent period lasting until 6 Gyr ago, and then followed by relatively constant star formation.
One of these three \hubble/WFPC2 observations of NGC~6822 was re-analyzed by \citet{YangY:2024} using their machine learning approach to determining SFHs.  Their SFH is also broadly consistent with a constant star formation, but differs from that determined by \citet{Weisz:2014a} in that it shows a generally decreasing SFR.

\citet{Nally:2024} present a $\sim5^{\prime}\times6^{\prime}$ mosaic of \jwst/NIRCam observations (F115W, F200W, F356W, F444W) centered on the main body of NGC~6822.  With regard to stellar populations, these observations are comparably deep to the \hubble observations of \citet{Cannon:2012}.  Photometry of over half a million point sources was recorded in both F115W and F200W.  These should be sufficient for a new SFH, significantly less impacted by reddening than those obtained in the optical.  \citet{Nally:2026} subsequently combined these NIRCam data with MIRI imaging to characterize the dust-producing AGB population of NGC~6822, providing a near- and mid-IR census of its evolved stars that complements the optical SFHs.

\citet{Hunt:2025} present \textit{Euclid}  early release observations of six galaxies, including NGC~6822, representing the first truly wide-field optical view of NGC~6822 from space.  They only performed aperture photometry, which is limited in depth compared to PSF photometry in crowded-fields.  As such, the photometry presented therein is too shallow to allow a lifetime SFH, but the \textit{Euclid}  imaging holds promise for deeper photometry.

\subsubsection{Cetus Dwarf Spheroidal}
\label{sec:Cetus_sfhs}

At a distance of $\sim600$~kpc from the LG barycenter, and distances of over 750 kpc from the MW and $\sim650$~kpc from M31, Cetus is currently an isolated galaxy roughly 60\% of the distance to the zero-velocity surface of the LG \citep{McConnachie:2012}.  With a stellar mass of $3\times10^6$ M$_{\odot}$, it is a gas-poor, quiescent galaxy, with non-detections in \hi{} \citep{Spekkens:2014} and H$\alpha$ \citep{Kaisin:2012}. From a statistical argument, \citet{Teyssier:2012} found Cetus to have a high likelihood ($\ge$ 50\%) of having passed through the virial radius of the MW. From proper motions, \citet{Bennet:2025} find that Cetus may have interacted with M31 more than 6 Gyr ago (they emphasize that this is around the limit to which orbital reconstruction is reliable), and thus may be a backsplash galaxy.

\citet{Monelli:2010_a} produced a lifetime SFH from deep \hubble/ACS observations (F475W, F814W) of a central field.
The resultant SFH shows a single, dominant episode of star formation lasting from $\sim13$~Gyr ago until $\sim9$~Gyr ago with a peak around 12 Gyr ago. The derived SFH indicates that star formation in Cetus was completely quenched by $\sim8$~Gyr ago.

Because the peak of star formation occurs significantly after the epoch of reionization \citep[see analysis and comparison to Tucana in][]{Monelli:2010_b} and star formation continues for a few Gyr after reionization, it is highly unlikely that reionization played a significant role in the quenching of star formation in Cetus.
With a stellar mass that is significantly above those of galaxies that are expected to be quenched by reionization, this is perhaps no surprise, but this then raises the question of how star formation came to be totally quenched.  The leading candidates are a close passage by one of the massive LG galaxies or quenching by feedback.  If Cetus is a backsplash galaxy, as the orbits of \citet{Bennet:2025} allow, its star formation may have been truncated by an interaction with M31.

The AMR derived by \citet{Monelli:2010_a} shows that Cetus is generally characterized as very metal-poor with increasing metallicity from [Fe/H] $\sim$ $-$1.8 at earliest times to [Fe/H] $\sim$ $-$1.5 at the time of quenching.  This range in metallicity agrees well with the spectroscopic measurements of \citet{Taibi:2018}.

\subsubsection{Pegasus Dwarf Irregular (DDO 216)}
\label{sec:Pegasus_sfhs}

At a distance of $\sim620$~kpc from the LG barycenter, Pegasus is an isolated galaxy with distances of $\gtrsim900$~kpc to the MW and  $\gtrsim450$~kpc to M31 \citep{McConnachie:2012}.  With a stellar mass of  $7\times 10^6$ M$_{\odot}$, it is a gas-rich galaxy, with M$_{HI}$/M$_*$ of 0.9. From statistical modeling, \citet{Buck:2019} assigned a probability of 79\% that Pegasus is a backsplash galaxy from an earlier encounter with M31.  \citet{Bennet:2024} used a combination of \textit{Gaia} and \hubble\ proper motion observations to determine an orbit which places Pegasus  on its first infall with only a 5\% chance of previously interacting with M31. \citet{Bennet:2025} greatly improved the accuracy of the proper motion of Pegasus and concluded that it could not be a backsplash galaxy of M31, but there was a small, but non-zero, probability of an earlier interaction with the MW. By comparing the distribution of stars and \hi{} gas in Pegasus, \citet{McConnachie:2007} deduced that Pegasus is undergoing ram-pressure stripping of its ISM by the intergroup medium of the LG.   Nonetheless, Pegasus still has sufficient neutral gas for active star formation.

Early \hubble/WFPC2 observations (F555W, F814W) of Pegasus by \citet{Gallagher:1998} allowed the determination of its recent SFH and revealed gradients in the distributions of its different aged populations (MS, BHeB, RGB) in the sense that the younger stars were much more centrally concentrated than the older stars.  \citet{Weisz:2014a} reanalyzed these observations and found a lifetime SFH consistent with constant star formation over the age of the universe with very large associated uncertainties due to the shallow depth of the photometry.

Deep \hubble/ACS observations (F475W, F814W) of Pegasus have recently been analyzed by \citet{Savino:2025}.  The derived SFH shows an early episode of star formation accounting for $\sim20$\% of the total star formation in Pegasus.  This is followed by a relatively quiescent period up until $\sim8$~Gyr ago when there was a dominant episode of star formation accounting for $\sim50$\% of the stellar mass of Pegasus.  This has been followed by relatively constant star formation up to the present.  The earliest star formation has minute statistical uncertainties but substantial systematic uncertainties, meaning the details of this dominant episode may still carry significant uncertainty.

\citet{Ren:2024} derived a SFH for Pegasus based on ground-based near-IR observations.  These were compared to the SFH derived in the \citet{Weisz:2014a} compendium and, although uncertainties were not included in the comparison, it appears that \citet{Ren:2024} found a slower growth with more stars being formed at later times.

Pegasus has a prominent GC and the age of that cluster has been determined to be consistent with the earliest episode of star formation in Pegasus \citep[$\sim$ 12 Gyr,][]{Cole:2017}.

\subsubsection{Leo~T}
\label{sec:LeoT_sfhs}

Leo~T, a relatively recent addition to the LG found in SDSS imaging \citep{Irwin:2007}, is located beyond our nominal 300~kpc boundary (\S \ref{sec:concepts_classification}), at a distance of $\sim420$~kpc from the MW and $\sim990$~kpc from M31, with a resulting distance of $\sim650$~kpc from the LG barycenter \citep{McConnachie:2012}.  With a stellar mass of $2\times 10^5$ M$_{\odot}$, it is a gas-rich galaxy, with M$_{HI}$/M$_* =  2.0$. \citet{Teyssier:2012} found Leo~T to have a high likelihood to have previously passed through the halo of the MW, while \citet{Buck:2019} found a probability of 51\% that Leo~T is a backsplash galaxy.  However, from an orbit reconstructed from proper motion measurements, \citet{Bennet:2024} found Leo~T to be on its first infall into the LG with a probability of 0\% of being a backsplash galaxy.  This conclusion was confirmed with significantly more precise proper motion measurements in \citet{Bennet:2025}. The \hi{} distribution is limited to the size of the stellar distribution \citep{RyanWeber:2008, Adams:2018}, which has been speculatively attributed to an interaction with the MW; this now seems most likely due to ram-pressure stripping by the intergroup medium.  Leo~T is especially interesting as it is the lowest-mass gas-rich galaxy known \citep[e.g.,][]{RyanWeber:2008, Putman:2021}.  Its luminosity ($M_V \sim -8$) lies right on the boundary between UFD and classical dwarf galaxies.

Soon after its discovery, \citet{deJong:2008b} used the Large Binocular Telescope to acquire g and r band resolved star imaging of Leo~T.  Similar to most low-mass, star-forming galaxies, the younger populations were found to be more centrally concentrated than the older populations.  From a HB-depth CMD, they derived a rudimentary SFH showing star formation dating back to $\gtrsim10$~Gyr ago and persisting until at least a few 100 Myr ago.

\citet{Weisz:2012b} measured the first detailed lifetime SFH from \hubble/WFPC2 observations (F606W, F814W) which extended below the oMSTO.
Their SFH is in general agreement with that derived by \citet{deJong:2008b} and is consistent with nearly constant star formation from the first Gyr of history up until $\sim$ 25 Myr ago.

\citet{Clementini:2012} used the same \hubble\ observations to study the short period variable stars and to independently derive a SFH.  Eleven anomalous Cepheid variables and one RR Lyrae star were recovered.  The RR Lyrae yielded a distance in agreement with previous measurements.  Their SFH recovered dominant episodes of star formation at 2 and 9 Gyr ago, but the uncertainties on the amplitudes of these events are roughly consistent with constant star formation.

Leo~T is included in the compilation of SFHs from \hubble/WFPC2 imaging of LG galaxies by \citet{Weisz:2014a}.  The re-derived SFH is highlighted in their Figure 4, and shows that when constrained to only the statistical uncertainties, there is a pause in the star formation that lasts from $\sim$ 12 to $\sim$ 8 Gyr ago.  The systematic uncertainties render this feature insignificant and the SFH is consistent with a nearly constant SFR over the age of the Universe, but since this pause of similar time and duration is seen in many other low-mass star forming galaxies \citep[e.g.,][]{Cole:2007, McQuinn:2024b}, it is worth noting here.  Note that with under 4000 total stars measured, the statistical uncertainties, especially at the older ages, are non-negligible.  Deeper and wider field observations which detected more stars would reduce this uncertainty.

\subsubsection{WLM (DDO 221)}
\label{sec:WLM_sfhs}

At a distance of $\sim790$~kpc from the LG barycenter, WLM is an isolated galaxy with distances of $\sim930$~kpc to the MW and $\sim840$~kpc to M31 and lies very close to the zero-velocity surface of the LG \citep{McConnachie:2012}.  With a stellar mass of $4\times 10^7$ M$_{\odot}$, it is a gas-rich galaxy, with M$_{HI}$/M$_* = 1.4$. \citet{Teyssier:2012} found WLM to have less than a 1\% chance of being associated with the MW and \citet{Buck:2019} found WLM to have zero probability of being a backsplash galaxy. From proper motions, \citet{Bennet:2024} found WLM to be on its first infall into the LG.  Despite its relatively isolated status, \citet{Yang:2022} suggest there is evidence in the \hi{} observations of ram-pressure stripping on WLM from an inter-group medium, though \citet{Rybarczyk:2026} find that this conclusion is likely due to a flaw in the data reduction technique used and that there is no clear evidence in the gas observations for ram pressure stripping.

Observations of WLM have played an important historical role in the development of our understanding of SFHs of nearby galaxies \citep{Tosi:1989, Ferraro:1989} with the development of new analysis code and investigations of spatial variations.
Through \hubble\ and recent
\jwst\ observations, the SFH of WLM has become well-studied.

Early \hubble/WFPC2 observations (F555W, F814W) of two fields were reanalyzed in the compendium study by \citet{Weisz:2014a}, and WLM was found to have star formation at all ages. However, a significant fraction of the star formation appeared to have occurred in the last 3 Gyr.  Deep \hubble/ACS and \hubble/UVIS observations (F475W, F814W) of an ``inner'' (but not central) field and ``outer'' field, respectively, were presented by \citet{Albers:2019}.  They found that the inner field SFH was not dominated by early star formation and, overall, was dominated by star formation in the last 5 Gyr.  The outer field showed essentially the opposite with early star formation dominating and later star formation lacking (most of the stars in the outer field were formed before 5 Gyr ago).  Potentially interesting,  both fields show AMRs increasing from [Fe/H] $\sim$ $-$1.9 to [Fe/H] $\sim$ $-$0.8 over the lifetime of the galaxy (but with the outer field rising faster than the inner field in keeping with expectations of the SFHs). \citet{Albers:2019} noted that their derived AMR was in good agreement with that derived by \citet{Leaman:2013} from spectroscopic measurements of WLM RGB stars.

WLM was a target of the \jwst\ Early Release Science program on resolved stellar populations \citep{Weisz:2023a}.  Deep \jwst/NIRCam observations (F090W, F150W) were obtained along the main body of WLM along with parallel NIRISS observations in the halo.  \citet{McQuinn:2024a} presented an analysis of the NIRCam observations and compared them to a comparable analysis of the \hubble/ACS observations.  The derived SFHs of the fields (which have some degree of overlap) were found to be in good agreement.  Both fields showed a small amount of star formation prior to the epoch of reionization, followed by a relative pause in star formation, and then a re-ignition of star formation roughly $\sim$ 8 Gyr ago.  The amplitude of the pause in star formation is comparable to that seen in Aquarius \citep{Cole:2014}, but less than that observed in Leo~A \citep{Cole:2007}.  \citet{McQuinn:2024a} used their SFH and AMR to generate synthetic photometry to determine what the MDF should be for the RGB stars in WLM.  Their result of <[Fe/H]> $=$ $-0.9$ and $\sigma$ $= 0.3$ was found to be in good agreement with the values from stellar spectroscopy of $-1.1$ and $0.3$ \citep{Leaman:2009} and $-1.3$ and $0.4$ \citep{Leaman:2013}. \citet{Garling:2025} used the WLM SFH from \citet{McQuinn:2024a} derived
using MATCH as a test case for their new open-source SFH code {\it StarFormationHistories}.

\citet{Cohen:2025} have combined \jwst/NIRCam and NIRISS observations with \hubble\ observations to conduct a definitive study of the spatial variations in the SFH of WLM.  As expected from studies of other low-mass star-forming galaxies \citep[e.g.,][]{Hidalgo:2013} and confirming the work of \citet{Albers:2019}, there is a strong radial gradient in the mean age of the stellar population in the sense that the outer parts of the galaxy are dominated by old stars.  Specifically, in the inner fields $\sim$ 50\% of the stars formed within the last 4 Gyr; while, in contrast, in the outer fields $\sim$ 50\% of the stars had already formed before 8 Gyr ago.  Intriguingly, \citet{Cohen:2025} find evidence for an azimuthal variation in the SFHs in the sense that the stars on the leading edge of the galaxy (as determined from its proper motion) are, on average, younger than those on the trailing edge.

\citet{Lee:2026} used the AGB stars from wide-field ($\sim$ 15 $\times$ 10 arcmin) Magellan Baade/FourStar Infrared Camera near-IR observations (J, K$_S$) to produce a global SFH that is in excellent agreement with that derived from \jwst observations.  The SFH solution also produces an AMR in agreement with the \jwst AMR and individual stellar metallicity determinations.  Interestingly, the near-IR AGB observations reveal a stellar overdensity with an age of $\sim$ 8 Gyr which is interpreted to be a early time merger candidate.

\subsubsection{Leo~A (DDO~69)}
\label{sec:LeoA_sfhs}
At a distance of $\sim940$~kpc from the LG barycenter, Leo~A is an isolated galaxy with distances of $\sim800$~kpc to the MW and $\sim1200$~kpc to M31 and lies very close to the zero-velocity surface of the LG \citep{McConnachie:2012}.  With a stellar mass of $6\times 10^6$ M$_{\odot}$, it is a gas-rich galaxy, with M$_{HI}$/M$_* = 1.8$. \citet{Teyssier:2012} found Leo~A to have a 2\% chance of being associated with the MW and \citet{Buck:2019} found Leo~A to have zero probability of being a backsplash galaxy. From proper motions, \citet{Bennet:2024} found Leo~A to be on its first infall into the LG.  Perhaps as a sign of its isolation, Leo~A appears to have a large, extended, low-surface brightness halo \citep{Stonkute:2014, Stonkute:2018}.

\citet{Tolstoy:1998} first drew attention to the potentially unique (at the time!) character of the SFH of Leo~A.  Based on relatively shallow \hubble/WFPC2 (F439W, F555W, F814W) photometry, they found that the central regions of Leo~A were dominated by stars less than 2 Gyr old and limited early star formation to less than 10\% of the total stars. \citet{Dolphin:2002} followed up with ground based observations of the variable stars in Leo~A, finding 8 RR Lyrae stars and providing proof of star formation more than 10 Gyr ago.  This was followed by \citet{Schulte:2002} who obtained deeper \hubble/WFPC2 (F439W, F555W, F814W) observations of an outer field of Leo~A which showed a larger fraction of older stars than the inner field and evidence of star formation at least as old as 9 Gyr.

A definitive SFH of Leo~A came from deep \hubble/ACS observations (F475W, F814W) reaching below the oMSTO presented by \citet{Cole:2007}.  These observations revealed that more than 90\% of the stars in the ACS field were formed less than 8 Gyr ago and that star formation in Leo~A reached a peak between 3 and 1 Gyr ago.  These observations confirmed the initial analysis by \citet{Tolstoy:1998} that Leo~A is a predominantly young galaxy.  \citet{Cole:2007} speculated that the unusual SFH of Leo~A could be attributable to early truncation by either UV heating from the epoch of reionization or supernovae feedback, followed by a long quiescent period while a cold ISM was re-established. An especially interesting feature of Leo~A's SFH is the jump in SFR about 6 Gyr ago \citep[seen also in the re-analysis presented in][]{Skillman:2014}.  \citet{Cole:2007} point out the interesting question of what may have triggered the re-ignition of star formation in Leo~A and offer simply that a rare dwarf-dwarf interaction may have been responsible.

These observations were also used to study the short-period variable stars and only 10 RR Lyrae were found, further supporting the weakness of the early star formation \citep{Bernard:2013}.  Interestingly, the AMR derived by \citet{Cole:2007} shows almost no signs of chemical enrichment during its lifetime.  The SFH was well fit by a constant metallicity of [Fe/H] $\sim -1.4$.  This was supported by spectroscopic observations of individual stars, yielding $<$[Fe/H]$>$ $=$ $-$1.58 \citep{Kirby:2013} and $<$[Fe/H]$>$ $=$ $-$1.67 \citep{Kirby:2017}, which are only slightly lower than the current ISM metallicity measured from \hii{} regions \citep{vanZee:2006, RuizEscobedo:2018} and young massive stars \citep{Urbaneja:2023}.

Leo~A is one of ten galaxies which \citet{Choi:2025} use to demonstrate that adding UV observations (specifically \hubble/WFC3 F275W, F336W) to preexisting \hubble\ optical observations results in improved precision in the derived SFHs over the last Gyr.  Because of its proximity and the depth of existing photometry, Leo~A provides the most diagnostic power in this investigation.  The main source of the improvement is the ability of the UV observations to clearly separate the BHeB stars from the MS stars \citep[in optical only studies, BHeB stars could be incorrectly attributed to reddened MS stars, due, in part, to the inadequacy of the BHeB evolution models, e.g.,][]{McQuinn:2011, Cignoni:2018, Lescinskaite:2022}.  The SFH derived for the last Gyr shows a relatively high SFR over the last 5 Myr accounting for close to 10\% of the stars formed in the last Gyr within that short period of time.

The very recent SFH (ages $\le$ 300 Myr) has been studied by \citet{Lescinskaite:2022}.  They combined Subaru/SuprimeCam observations in B, V, and I \citep{Stonkute:2014} and R and H$\alpha$ \citep{Stonkute:2022} (with a depth in R of 25 mag., roughly 4 mag.\ deeper than the TRGB) with the \hubble/ACS observations to obtain a galaxy-wide study. Specifically, they studied the numbers and locations of BHeB stars to produce a spatially resolved, recent SFH.  This allowed the reconstruction of the recent star formation and its relation to the large central hole in the \hi{} distribution.

\subsubsection{Aquarius (DDO~210)}
\label{sec:Aquarius_sfhs}

At a distance of $\sim1050$~kpc from the LG barycenter, Aquarius is an isolated galaxy with distances of $\sim1070$~kpc to the MW and $\sim1170$~kpc to M31 and lies very close to the zero-velocity surface of the LG \citep{McConnachie:2012}.  With a stellar mass of $\sim2\times 10^6$ M$_{\odot}$, it is a gas-rich galaxy, with M$_{HI}$/M$_* = 2.6$. Aquarius's isolation is supported by \citet{Teyssier:2012} and \citet{Buck:2019} who both find it to be unassociated with either the MW or M31.  The orbital reconstruction from the proper motions study by \citet{Bennet:2025} confirms that Aquarius is and has been isolated from the MW and M31 for the explorable past and predictable future.

Aquarius is a ``transition'' galaxy \citep{Mateo:1998} as it has detectable \hi{} and shows recent star formation but no observable prominent \hii{} regions indicative of current massive star formation.  However, it is quite distinct from other transition galaxies like LGS-3 and Phoenix as they both have M$_{HI}$/M$_* < 0.5$ which is five times lower than in Aquarius and is likely reflective of the impact of the LG group environment. In other words, galaxies like LGS-3 and Phoenix may well be in the transition from dIrr to dSph due to the loss of ISM, but Aquarius is not. Interferometric imaging of the \hi{} shows a central peak in excess of 15 M$_{\odot}$ pc$^{-2}$, which is at a level commonly associated with star formation; however, the \hi{} distribution is asymmetric and misaligned with stellar distribution, possibly indicating the effect of a recent merger or strong internal feedback \citep{Young:2003, Begum:2004, Iorio:2017}.  The lack of \hii{} regions in the lowest-mass, gas-rich star forming regions could simply be temporarily interrupted star formation due to stochastic effects as opposed to suppressed star formation \citep{Skillman:2003a, Weisz:2011}. This makes the SFH of Aquarius particularly interesting, as it requires distinguishing between a very low but consistent SFR and a recent event causing its low SFR.

Prior to 2014, only shallow, RGB-depth \hubble/WFPC2 observations (F606W, F814W) of Aquarius were available for a SFH determination, and these resulted in large uncertainties at early ages \citep{Weisz:2014a}.
\citet{Cole:2014} presented deep, SGB-depth \hubble/ACS photometry (F475W, F814W) of Aquarius.  The SFH shows early star formation at a low level followed by a dominant episode of star formation at  $\sim7$~Gyr ago.  This is reminiscent of the delayed onset of star formation seen in Leo~A.  The dominant episode is followed by nearly constant star formation up until 2 Gyr ago with a further drop up to the present.  Focusing in on the last Gyr where the time resolution is the highest, \citet{Cole:2014} found a peak in the SFR about 250 Myr ago after which the SFR dropped to a very low rate for the last 100 Myr.  Thus, the irregularities seen in the \hi{} distribution could be related to the period of relatively high SFR 250 Myr ago.  This is also supported by the stellar kinematics study by \citet{Hermosa2020} who suggest that the current state of Aquarius could be explained by the recent accretion of a gas cloud or the re-accretion of gas expelled earlier via feedback.

\citet{Bennet:2025} point out that Aquarius is similar to Leo~A and Leo~I with peaks in their SFHs at intermediate ages that cannot be attributable to interactions with the MW or M31, and thus may be attributable to dwarf-dwarf interactions.

\citet{Choi:2025} highlight the relatively high SFR in the last few Myr with a result that close to 10\% of the star formation in Aquarius in the last Gyr has taken place recently.

The AMR derived by \citet{Cole:2014} starts at [Fe/H] $\sim-2.2$ and rises to [Fe/H] $\sim-1.5$ before the dominant episode of star formation at 7 Gyr ago.  It increases further to $\sim-1.3$ at present, consistent with the stellar metallicities derived by \citet{Kirby:2013, Kirby:2017, Hermosa2020}.

\subsubsection{Tucana}
\label{sec:Tucana_sfhs}

At a distance of $1080$~kpc from the LG barycenter, Tucana is isolated today with distances of $\sim880$~kpc to the MW and $\sim1360$~kpc to M31 and lies very close to the zero-velocity surface of the LG \citep{McConnachie:2012}.  Tucana has a stellar mass of $5.6\times10^5$ M$_{\odot}$ and is undetected in \hi{}, with a limit on the ratio of M$_{HI}$/M$_*$ $\le$ 0.4 \citep[5 $\sigma$,][]{Spekkens:2014}.
\citet{Teyssier:2012} concluded that it was very likely that Tucana has passed through the MW and \citet{Buck:2019} found the opposite.  The high accuracy proper motions measured by \citet{Bennet:2025} reveal that Tucana is truly isolated with no evidence of ever having interacted with the MW or M31.

There was some excitement that \hi{} emission was associated with Tucana when
\citet{Oosterloo:1996} detected an \hi{} cloud in the general direction of Tucana, but the stellar radial velocities measured by \citet{Fraternali:2009} and later \citet{Gregory:2019} ruled out any association with Tucana, deeming the \hi{} detection likely part of the  Magellanic Stream \citep[as originally proposed by][]{Oosterloo:1996}.
Tucana was also in the limelight as a ``massive failure'' in the sense of the too-big-to-fail problem \citep{Boylan-Kolchin:2011a} because of the large stellar velocity dispersion measured by both \citet{Fraternali:2009} and \citet{Gregory:2019}.  However, new observations by \citet{Taibi:2020} and a reanalysis of the existing observations resulted in a lower velocity dispersion and a loss of massive failure status.

The environmental history of Tucana has been discussed many times in the literature.  Soon after its discovery \citep{Lavery:1992}, ground-based CCD observations revealed the absence of both high mass MS stars and AGB stars \citep{Castellani:1996, Saviane:1996}, indicating that Tucana consisted exclusively of old stars.  If Tucana has not been affected by the environment of the LG, then it poses an interesting puzzle to understand its ISM-free nature and cessation of star formation.

Deep \hubble/ACS photometry (F475W, F814W) by \citet{Monelli:2010_b} provided a detailed SFH, confirming that Tucana has exclusively old stars with a dominant episode of star formation roughly 12.5 Gyr ago and a complete cessation by roughly 8 Gyr ago.  From the same observations, there is good evidence that the dominant early episode of star formation consisted of two phases.  The RR Lyrae stars divide into a more extended, more metal-poor population and a second more concentrated, more metal-rich population \citep{Bernard:2008}.  This is reinforced by the presence of two RGB bumps \citep{Monelli:2010_c}.  Although \citet{Monelli:2010_b} did not emphasize it, Tucana also shows a split RGB \citep[as noted by][]{Fu:2024} also indicative of at least two main episodes of star formation and/or chemical enrichment.

\citet{Savino:2019} re-analyzed the \hubble/ACS photometry with an emphasis on improving the SFH by adding a detailed analysis of the structure in the HB.  They demonstrated that the combination of the oMSTO and the HB enables a higher time resolution at older ages.  Tucana's HB has three discernible structures, and these are associated with three episodes of star formation: two corresponding to the previously identified early star formation and a third occurring roughly coincident with the cessation of star formation about 7 Gyr ago.  \citet{Savino:2019} further show a sequence of concentration with age with the third episode of star formation more concentrated than the first two.  It is possible that this
third episode of star formation is causally connected to the cessation of star formation in Tucana, possibly from feedback from this last burst.

The SFH of Tucana is beautifully supported by measurements of the AMR.  \citet{Monelli:2010_b} show a gradual increase from [Fe/H] of $-$1.6 to $-$1.1 from earliest times to $\sim$ 8 Gyr ago. The AMR from \citet{Savino:2019} is in rough agreement, with a population of more metal-poor stars at earliest ages.
The spectroscopic analysis of \citet{Taibi:2020} found a mean [Fe/H] $=$ $-$1.6 with an intrinsic dispersion of 0.4 dex,  in excellent agreement with the AMRs derived photometrically.  This has been improved upon by \citet{Fu:2024} who derived metallicities from \hubble\ narrow-band imaging of the calcium H\&K feature of over 300 stars to produce a mean [Fe/H] of $-$1.55 with a dispersion of 0.55 dex and a gradient of $-$2.1 $\pm$ 0.3 dex kpc$^{-1}$.  They further found that the metallicities of the two sides of the split RGB were distinct, with averages of [Fe/H] of $-$1.8 and $-$1.1 corresponding well with the AMR extremes derived from broadband photometry by \citet{Monelli:2010_b} and \citet{Savino:2019}.

\subsubsection{SagDIG (UKS 1927-177)}
\label{sec:SagDIG_sfhs}

At a distance of $\sim1160$~kpc from the LG barycenter, SagDIG\footnote{The authors are aware that the official abbreviation for the constellation Sagittarius is Sgr, but are bound by the naming used in the discovery paper \citep{Cesarsky:1977}} is isolated today with distances of $\sim1060$~kpc to the MW and $\sim1360$~kpc to M31 and lies very close to the zero-velocity surface of the LG \citep{McConnachie:2012}.  SagDIG has a stellar mass of  $2.5\times10^6$ M$_{\odot}$ and
it is a gas-rich galaxy, with M$_{HI}$/M$_* = 2.5$.
\citet{Teyssier:2012} found that SagDIG has a less than 1\% chance of ever having been associated with the MW and \citet{Buck:2019} concurred, classifying SagDIG as a clean field galaxy.  This was supported by the proper motion measurement of \citet{Bennet:2024} who found SagDIG to be on its first infall into the LG.

Nearby, extremely metal-poor star-forming galaxies are valuable places to study star formation and stellar evolution at the low metallicities characteristic of the early universe \citep[see][]{Kunth:2000}.  SagDIG is an extremely metal-poor galaxy \citep{Skillman:1989, Saviane:2002, Andrade:2025} with an ISM metallicity comparable to that of I~Zw~18 (only 17 times closer).  It has often not been included in extremely metal-poor galaxy lists because it lacked an \hii{} region with high enough surface brightness to detect the temperature-sensitive [O~III] $\lambda$4363 emission line necessary for a direct oxygen abundance measurement, but \citet{Andrade:2025} now report a [O~III] $\lambda$4363 detection.   \citet{Kirby:2017} have measured the average stellar metallicity to be $<[$Fe/H]$>$ $\sim -1.88$, affirming its extremely low metallicity.  Thus, the deep imaging of SagDIG required for a detailed SFH would be important for testing/anchoring stellar evolution models at very low metallicities.

The earliest wide-field ($\sim$ 6 $\times$ 6 arcmin) ground-based (B,V,I) resolved star imaging of SagDIG obtained with the ESO/NTT EMMI instrument \citep{Momany:2002} to roughly 3 mag.\ below the TRGB revealed the presence of both old (RGB) and young (MS) stars with the now familiar pattern of a more concentrated young star population (note that due to a low Galactic latitude of $-$16$^\circ$, a significant foreground population was subtracted statistically).  This was soon followed by shallow (3 orbits) \hubble/ACS imaging (F475W, F606W, F814W) by \citet{Momany:2005}. The \hubble\ CMD, which reaches below the HB, provided a recent SFH, and the presence of the HB is evidence of star formation at the
earliest times.

Unfortunately, to date, there is no SFH of SagDIG based on imaging that reaches the oMSTO.  Although SagDIG represents a desirable isolated, low metallicity target, given its distance and foreground reddening (A$_V$ $=$ 0.34), reaching the photometric depth required for accurately reconstructing the {\em ancient} SFH with \hubble has been prohibitively expensive.  In their compilation of SFHs from \hubble/WFPC2
(F606W, F814W) imaging, \citet{Weisz:2014a} present a SFH from the very shallow imaging in which the uncertainties are consistent with a large range of SFHs from an early dominant burst to largely delayed star formation.  The nominal solution shows three dominant episodes of star formation, one at earliest times, one starting at roughly 6 Gyr ago, and one starting roughly 1 Gyr ago. Hopefully, new, deep \jwst\ observations (JWST-GO-5255) will bring a more definitive SFH for SagDIG.

\subsubsection{VV~124 (UGC~4879)}
\label{sec:VV124_sfhs}

At a distance of $\sim1320$~kpc from the LG barycenter, VV~124 is isolated today with distances of $\sim1370$~kpc to the MW and $\sim1400$~kpc to M31 and lies near, and perhaps just beyond, the nominal zero-velocity surface of the LG \citep{McConnachie:2012}.  VV~124 has a stellar mass of $8.3\times 10^6$ M$_{\odot}$ and
has a marginally detectable ISM, with M$_{HI}$/M$_* = 0.1$.
\citet{Teyssier:2012} found that VV~124 has a less than 1\% chance of ever being associated with the MW and \citet{Buck:2019} classified VV~124 as a definite field galaxy.  \citet{Bennet:2024} obtained a proper motion and found VV~124 to be on its first infall into the LG.

Soon after the discovery that VV~124 had a discrepant velocity, \citet{Kopylov:2008} established a new distance that moved it just to the periphery of the LG.  This discovery was followed closely by LBT and WSRT observations establishing the stellar and \hi{} distributions \citep{Bellazzini:2011a}.  Archival, shallow (1 orbit) \hubble/ACS observations (F606W, F814W) were analyzed by \citet{Bellazzini:2011b} who found centrally concentrated young star formation with two massive star clusters. \citet{Jacobs:2011} derive a SFH from the same observations finding a dominant old episode of star formation followed by a period of quiescence until roughly 1 Gyr ago when star formation recommenced.

\citet{Neeley:2021} analyzed 14 orbits of \hubble/ACS observing (F475W, F814W) with the main goal of studying short period variable stars.  Centrally concentrated classical Cepheid stars confirm the extent of the recent star formation and a very diffuse distribution of RR Lyrae stars confirms the age gradient seen in the earlier studies.  Although these observations are quite deep (but not oMSTO-depth) no SFH was calculated.

Keck spectroscopy by \citet{Kirby:2012} confirmed the overall low metallicity and metallicity spread deduced from the imaging observations with <[Fe/H]> $=$ $-$1.1 and a standard deviation of 0.5 dex.

\subsection{Galaxies Beyond the Local Group}
\label{sec:app_blg}

\subsubsection{Low-Mass Galaxies}

Given the large (and constantly increasing) number of low-mass galaxies beyond the LG with SFHs, it is an organizational challenge to present them in an ordered fashion (especially given the large dispersion in depth of photometry).  Simply organizing by distance tends to mingle galaxies in different environments.  Many galaxies appear in comparative studies that span significant ranges in distance. Here, with apologies to the reader, we present comparative studies (ordered by mean distance), followed by galaxies in prominent nearby structures, and then studies of individual galaxies, with each subgroup ordered roughly by distance.

\paragraph[The ANGST Survey]{The ANGST Survey:}
The majority of galaxies (60) in ANGST (see Appendix~\ref{sec:important_obs}) are low mass galaxies spanning a range of morphological types:
12 dSph/dE, 5 dwarf spiral, 28 dIrr, 12 dSph/dIrr
(transition), and 3 tidal dwarf galaxies. \citet{Weisz:2011} derived SFHs for all of these galaxies using a uniform methodology and compared the results for the different morphological types.
\citet{Weisz:2011} found three significant results for
dwarf galaxies in the ANGST volume: (1) the majority of dwarf galaxies formed the bulk of their mass prior to z
$\sim$ 1, regardless of current morphological type; (2) the mean SFHs of dIrrs, dTrans, and
dSphs are similar over most of cosmic time, diverging only in the last few Gyr — with the clearest differences appearing in the most recent 1 Gyr; and (3) the SFHs are complex and the mean values
are inconsistent with simple SFH models, e.g., single bursts, constant SFRs, or smooth,
exponentially declining SFRs.

Because of a significant range in environment, environmental influences could be explored, and a strong density–morphology relationship was found, such that the dSphs in the sample are less isolated than the dIrrs.
They also found that the transition from a gas-rich to a gas-poor galaxy cannot be solely due to internal mechanisms such
as stellar feedback, and instead is likely the result of external mechanisms, e.g., ram pressure and tidal stripping.  Tidal forces commonly played a role. In terms of their environments, SFHs, and gas fractions, the majority of the dTrans appear to be
low-mass dIrrs that simply lack H$\alpha$ emission, similar to the LG dTrans Aquarius. However, a handful of
dTrans have remarkably low gas fractions, suggesting that they have nearly exhausted their gas supply, analogous
to LG dTrans such as Phoenix.

A dedicated ANGST analysis of the Magellanic-type starburst NGC~4214 by \citet{Williams:2011b} derived a spatially-resolved lifetime SFH that revealed that older stellar populations are systematically more spatially extended than younger ones.

\paragraph[LUVIT]{LUVIT:}  LUVIT (see Appendix~\ref{sec:important_obs}) added \hubble\ UV and near-IR observations of nearby galaxies to pre-existing optical \hubble\ imaging.  In \citet{Choi:2025}, new F275W and F336W observations are added to archival F475W and F814W observations to assess the impact of adding NUV observations on the determination of SFHs for 10 low-mass star forming galaxies.  The primary result is a much better separation of the BHeB stars from the MS stars in the CMD.  This enables more precision in the SFHs in the last Gyr, in part, by reducing the tendency for the SFH code to interpret BHeB stars as reddened MS stars.  Discrepancies between the observed BHeB stars and the model tracks for BHeB stars are noted.  The 10 galaxies studied range in distance from 800 kpc $\le$ D $\le$ 3.5 Mpc.  Overall, the recent ($\le$ 1 Gyr) SFHs show a great variety, and all show significant, episodic deviations from a constant SFR.

\paragraph[The Durations of Starbursts in Dwarfs Project]{The Durations of Starbursts in Dwarfs Project:}

\citet{McQuinn:2010a, McQuinn:2010b, McQuinn:2012} derived SFHs from archival \hubble observations of a sample of 18 dwarf galaxies, all but one within 6 Mpc, for the explicit purpose of identifying bursts of star formation and measuring their durations.  \citet{McQuinn:2010a} used the lifetime SFHs to define a threshold in SFR above which a galaxy would be considered to host a starburst.  Current starbursts (13) and ``fossil'' starbursts (5) were identified in the sample.  Durations in excess of 100 Myr were found to be typical, ruling out rapid global quenching of star formation by bursts as a universal scenario. The identified bursts were responsible for producing $\sim $ 3 to 26\% of the host galaxy's stellar mass. The spatial distributions of the starbursts were found to cover a large range from strongly centrally concentrated bursts to global bursts spread over the entire area of the galaxies. These papers are discussed further in \S\ref{sec:bursts}.

\paragraph[The LEGUS Survey]{The LEGUS Survey:}
As part of LEGUS (see Appendix~\ref{sec:important_obs}), \citet{Cignoni:2018} studied three dIrrs (NGC~1705, NGC~4449, and Ho~II).  These galaxies are at distances of 5.1, 3.8, and 3.4 Mpc, respectively.  The emphasis is on the recent ($\le$ 200 Myr) SFHs.  The LEGUS survey emphasizes the importance of adding UV resolved star photometry to optical observations with the potential to improve the time resolution of recent SFHs.  They highlight the lack of a clear separation in the observations between the upper MS stars and the BHeBs, which limits the ability of fitting to the stellar models which show a clear separation.

\citet{Sacchi:2018} followed with a study of the spatially resolved recent SFH of NGC~4449 (D $\sim$ 3.8 Mpc), and \citet{Cignoni:2019} then followed with a comparative study of the SFHs of 23 star forming dwarf galaxies.  These are discussed in more detail in \S\ref{sec:bursts}.

\paragraph[SHIELD]{SHIELD:} The SHIELD survey \citep{Cannon:2011} consists of 82 galaxies selected from the Arecibo Legacy Fast ALFA (ALFALFA) survey \citep{Giovanelli:2005, Haynes:2011} catalog based on low gas masses, narrow \hi{} line widths, optical identification of counterparts in SDSS imaging, and distance estimates within $\sim$ 11 Mpc.  To date, 42 of the SHIELD galaxies have been observed with \hubble/ACS (F606W, F814W), primarily to obtain accurate distances from their TRGBs, but these \hubble\ observations allow for coarse SFH work.  \citet{McQuinn:2021} presented an overview of 30 of these galaxies. Because of the relatively larger distances of the SHIELD galaxies, typically the entire galaxy is covered with the ACS field of view. Thus, the stellar masses derived from their lifetime SFHs can be compared with the stellar masses inferred from \textit{Spitzer} near-IR imaging, and the comparisons are remarkably good.  The recent SFRs were also compared to H$\alpha$ and NUV derived SFRs, with better agreement with the NUV rates, as expected. The SHIELD galaxies span a decade in specific SFR, reflecting the less biased nature of \hi{} surveys (relative to optical-only surveys).

\paragraph[The NGC~3109 (14$+$12) Association]{The NGC~3109 (14$+$12) Association:}
\label{sec:LeoP_sfhs}
At distances between $\sim$ 1.3 and 1.7 Mpc, just beyond the LG, lies an association of low-mass galaxies in the direction of the constellations Sextans and Antlia which has been called the Ant-Sex group \citep{VandenBergh:1999a}, the (14$+$12) Group \citep{Tully:2002}, and relabeled as an association by \citet{Tully:2006}.  Since its most luminous member is NGC~3109, it also has that label.  The NGC~3109 association represents a treasure of low-mass, isolated galaxies which lies just outside the reach of \hubble\ for oMSTO photometry, but within the grasp of \jwst.

The recent discovery of member Leo~P \citep{Giovanelli:2013} with a distance determination by \citet{McQuinn:2013} increased the membership by one and provided an especially interesting target for many studies due to its proximity, low stellar mass (3 $\times$ 10$^5$ M$_{\odot}$), high gas mass (M$_{HI}$/M$_*$ $=$ 3), and low ISM metallicity \citep[{3\%} of the solar value,][]{McQuinn:2013,Bernstein:2014,Skillman:2013}.  \citet{McQuinn:2015a} derived a SFH for Leo~P from 17 orbits of \hubble/ACS (F475W, F814W) imaging.  The CMD reached two magnitudes below the HB but not to the oMSTO, so the uncertainties on the SFH are relatively large; the resulting SFH is consistent with constant.  36 hours of \jwst\ observations (F090W, F150W, F277W) were analyzed by \citet{McQuinn:2024b} resulting in strongly reduced uncertainties and revealing a significant suppression in star formation coincident with the epoch of reionization.  Interestingly, while this is similar to what is seen in the low-mass isolated galaxies in the LG, the amplitude and duration of the pause in star formation are smaller than those seen in Aquarius, Leo~A, and WLM.  Also of interest, Leo~P is the lowest mass of these four galaxies, and, at the epoch of reionization, had a stellar mass in the range associated with ultra-faint galaxies, which all show complete quenching of star formation.

The other members of the NGC~3109 association have relatively shallow CMDs from \hubble\ imaging. \citet{McQuinn:2010a} analyzed one orbit of \hubble/ACS (F606W, F814W) observations of Antlia and found evidence of a fossil burst of star formation in the recent SFH $\sim$ 200 to 300 Myr ago.  Their lifetime SFH for Antlia shows a dominant early period of star formation followed by a quiescent period until about 4 Gyr ago, but with very coarse time bins.
In the ANGST program, \citet{Weisz:2011} derived SFHs for Antlia, Sextans~A, Sextans~B, and NGC~3109, but all four had large uncertainties consistent with constant star formation.  The possible exception is NGC~3109, which shows evidence of a dominant initial period of star formation followed by a long quiescent period before re-igniting at roughly 2 Gyr ago, though with considerable uncertainties.
\citet{Weisz:2014a} revisited the observations of Sextans~A and Sextans~B in their study of \hubble/WFPC2 (F606W,  F814W) observations of LG galaxies and produced SFHs with significantly smaller uncertainties, in part by distinguishing between statistical and systematic uncertainties.  Both show evidence of a relatively long period of low SFR between an initial dominant period of star formation and a later return to nearly constant star formation.  \jwst\ cycle 5 program GO-9601 (PI K. McQuinn) will acquire oMSTO depth imaging of low-mass galaxies Sex~A, Sex~B, UGC~4879, and Antlia~B.

\paragraph[The Sculptor Filament]{The Sculptor Filament:}

Several low-mass Sculptor filament galaxies had their SFHs reported in \citet{Weisz:2011} (NGC~55, DDO~6, UGCA~438, ESO294-010, ESO410-005, ESO540-G030, ESO540-G032, and Sc22).

Additionally, the Sculptor Dwarf Irregular Galaxy (SDIG) is located in the NGC~7793 sub-group of the Sculptor filament at a distance of $\sim$ 3.2 Mpc.  \citet{Lianou:2013} used a single orbit archival \hubble/ACS observation (F606W, F814W) reaching 2.5 mag.\ below the TRGB to determine a coarse SFH.  The SFR in SDIG has stayed relatively constant over the lifetime of the universe with enhancements about 1 Gyr ago and in the last 100 Myr.  Interactions with the nearest large galaxy (NGC~7793) seem to be an unlikely trigger for the enhanced star formation and it is speculated that an unknown dwarf companion may be responsible.

\paragraph[The M81 Group]{The M81 Group:} \citet{Weisz:2008} used primarily new, and some archival, \hubble/ACS imaging (F555W, F814W) to derive the SFHs of 9 star-forming dwarf galaxies in the M81 Group.  The CMDs reached a depth of $M_V \sim -0.5$, providing little leverage on ancient star formation.  This paper instead focused on quantifying temporal patterns of star formation over the most recent $\sim1$~Gyr, which we discuss further in \S\ref{sec:bursts}.   \citet{Weisz:2008} also discuss the possibility of two systems, Holmberg~IX and Garland, as being tidal dwarf galaxies.  \citet{Sabbi:2008} analyzed the same \hubble/ACS CMD of Holmberg~IX as \citet{Weisz:2008} and found that 20\% of its stellar mass formed within the more recent $\sim200$~Myr and also speculated that Holmberg~IX could be a tidal dwarf galaxy. \citet{Weisz:2009a} and \citet{Weisz:2009b} measured the spatially resolved recent SFHs of IC~2574 and Holmberg~II using their well-populated BHeB sequences in order to study the connection between stellar feedback and the morphology of their ISM (e.g., holes and shells). They found that in all cases, energy from stellar feedback was sufficient to have created all holes in the ISM, even if a clear single-aged stellar cluster wasn't present.  They also identified a clear spatiotemporal sequence of star formation around the expanding supergiant \hi{} shell in IC~2574.

The compendium of SFHs reported in \citet{Weisz:2011} consisted of many low-mass M81 group members (BK3N, F6D1, BK5N, KDG~73, IKN, A0952+69, KDG~52, KK~77, HS~117, KDG~63, DDO~44, DDO~78, F8D1, KDG~64, KDG~61, UGC~4483, DDO~53, Ho~IX, Ho~I, DDO~82, Ho~II, IC~2574), as well as several nearby galaxies that lie outside the M81 group (DDO~181, UGC~8508, NGC~3741, DDO~190).

\citet{Sacchi:2021b} presented deep \hubble/WFC3/UVIS observations (F606W, F814W) of UGC~4483, a member of the M81 Group.  UGC~4483 is of special interest because it is the closest (D $\sim$ 3.4 Mpc) very metal-poor \citep[$\sim$ 7 \% of the solar O/H,][]{Skillman:1994} blue compact dwarf galaxy.  Early \hubble/WFPC2 observations \citep[F606W, F814W,][]{Dolphin:2001b} and \citep[F555W, F814W,][]{Izotov:2002} had been analyzed multiple times \citep{Dolphin:2001b, McQuinn:2010a, Weisz:2011} with coarse SFHs showing early dominant star formation followed by somewhat constant, relatively low-level star formation in the last several Gyr.  The new, deep observations presented by \citet{Sacchi:2021b} far surpass the older observations and reach four magnitudes below the TRGB, and, importantly, below the HB. As a result, a relatively detailed SFH is possible, and the authors divide the galaxy into 4 radial zones so that the SFH can be studied as a function of radius.  Comparable to most low-mass star-forming galaxies, UGC~4483 shows decreasing rates of star formation in its outer regions and increasing SFRs in its central regions with a smooth gradient connecting them.  The evidence of very early star formation is supported by the detection of 6 RR Lyrae stars in the halo of UGC~4483.  The authors point out that the properties of UGC~4483 are consistent with the properties of typical low-mass star forming galaxies up until the onset of the current burst of star formation.

Recently, \citet{Smercina:2026} have presented deep \hubble\ (F606W, F814W) imaging of the M81 Group ultra-diffuse galaxy F8D1.  The deep central field CMD produced a SFH showing a dominant early episode of star formation, followed by relatively constant star formation and two bursts at 2 and 0.5 Gyr ago.  These two recent bursts are also detected in a parallel field 6 kpc from the center. These lifetime SFHs reveal the conversion of a normal star-forming low-mass galaxy into an ultra-diffuse galaxy through tidal processing.

\paragraph[The IC~342 Group]{The IC~342 Group:}
\citet{Weisz:2011} included the SFH of the IC~342 low-mass galaxy KKH~37.

\citet{Makarova:2012} used a single orbit archival \hubble/ACS observation (F606W, F814W) reaching $\sim$ 2 mag.\ below the TRGB to determine a coarse SFH of UGCA~92 (D $\sim$ 3 Mpc).
UGCA~92's SFH starts with an early, dominant period of star formation followed by relative quiescence lasting up until 1 Gyr ago when it seems to have reignited with significant star formation over the last 500 Myr.  Although the photometry is not deep, UGCA~92 has the potential to be an extreme Leo~A type case with an exceptionally long delay before re-igniting.

\paragraph[NGC~1569]{NGC~1569:} \citet{Grocholski:2012} used \hubble/ACS observations (F606W, F814W) of the IC~342 group member NGC~1569 (D $\sim$ 3.1 Mpc) that reach 4 mag.\ deeper than the TRGB to study the SFH of its halo.  NGC~1569 is an important galaxy for the study of star formation in low-mass galaxies because of the presence of two super-star clusters \citep[e.g.,][]{Hunter:2000} and the detection of its galactic wind in x-rays \citep{Martin:2002}.  Previously, \citet{McQuinn:2010a} had used the \hubble/ACS observations (F606W, F814W) from \citet{Grocholski:2008} (which were used to determine a significantly improved new distance of $\sim$ 3.4 Mpc, which was later revised to $\sim$ 3.1 Mpc), to determine a SFH with a focus on the duration of the recent burst.  They had found dominant early star formation followed by relative quiescence and then elevated star formation in last Gyr. The observations are divided into core and outer region samples and the outer region sample is studied in detail.  A comparison of the respective CMDs shows that the recent star formation is heavily concentrated in the core sample.  The derived SFH of the outer region shows it to be dominantly old and metal poor (<[Fe/H]> $\sim$ $-$1.5) while there is some evidence of a younger (age $\le$ 1 Gyr), metal enriched  (<[Fe/H]> $\sim$ $-$0.6) population. Interestingly, the SFH for the outer region of NGC~1569 is in qualitatively good agreement with the global SFH from \citet{McQuinn:2010a}.

\paragraph[Pegasus~W, Leo~M, Leo~K, Canes Venatici~C]{Pegasus~W, Leo~M, Leo~K, Canes Venatici~C:} Pegasus~W, Leo~M, and Leo~K were discovered in DESI Legacy Imaging Survey data \citep{Dey:2019} by \citet{McQuinn:2023, McQuinn:2024c}. CVn~C, previously catalogued as BTS~113 and KKH~78, was recovered in the same data and characterized by \citet{Hai:2026}.  Although Pegasus~W, Leo~M, and Leo~K lie within the LG (0.4--0.9~Mpc), we discuss them here alongside CVn~C because all four were identified in the same survey data and characterized in the same way.  All four galaxies were observed with \hubble/ACS for a single orbit in F606W and F814W.  These observations allowed HB or TRGB distance measurements (0.9, 0.5, 0.4, 8.4 Mpc) which then allowed structural parameters and stellar masses.  Coarse SFHs were calculated with all four showing strong episodes of early star formation followed by quenching at later times.  In Pegasus~W, the star formation continues for several Gyr, which is not typical for an ultra-faint galaxy. Leo~M and Leo~K both are quenched early, consistent with showing a strong impact by the epoch of reionization.  Canes Venatici~C is more massive than the ultra-faint galaxy limit and shows no obvious indication of how it was quenched. The variety of quenching times may indicate different quenching processes for each of the galaxies.  These works provide a glimpse of what will be possible with the anticipated discoveries of low-mass galaxies in the future with Rubin and \textit{Roman}.

\paragraph[KKR~25]{KKR~25:} \citet{Weisz:2011} and \citet{Makarov:2012} both produced SFHs of KKR~25 \citep{Karachentsev:2001}, an isolated dSph galaxy at a distance of $\sim$ 1.9 Mpc, using shallow, single-orbit \hubble/WFPC2 observations (F606W, F814W).  KKR~25 is potentially extremely interesting as it appears to be a dSph in isolation, while the vast majority of dSph galaxies are satellites of massive galaxies.  The SFH of KKR~25 from \citet{Weisz:2011} reflects the relatively large distance and short integration time and is overall consistent with star formation over the entire history of the universe, but the uncertainties are so large that an early, dominant episode of star formation cannot be ruled out.   \citet{Makarov:2012} derive a SFH which is dominated by early (t $\le$ 12 Gyr) star formation but with evidence of some star formation between $\sim$ 4 and $\sim$ 1 Gyr ago.  KKR~25 represents a promising target for future \jwst\ observations to understand its true nature.

\paragraph[KKH~98]{KKH~98:} \citet{Melbourne:2010} compared \hubble\ (F475W, F814W) and Keck laser guide star adaptive optics near-IR (K$^\prime$) observations of the upper CMD of the isolated dwarf galaxy KKH98 (D $\sim$ 2.5 Mpc).  Interestingly, SFHs from both the optical and near-IR observations show a drop in SFR between 1 and 3 Gyr ago in addition to relatively constant star formation over most of the lifetime of the galaxy.  It was concluded that adaptive optics observations have the potential to extend SFH work to larger distances.

\paragraph[Pavo, Corvus~A, Kamino]{Pavo, Corvus~A, Kamino:} \citet{Mutlu:2026} report shallow \hubble\ observations (F606W, F814W) of the faint star-forming galaxies Pavo (D $\sim$ 2.2 Mpc), Corvus~A (D $\sim$ 3.3 Mpc), and Kamino (D $\sim$ 6.0 Mpc).  Pavo and Kamino show relatively constant lifetime SFHs while Corvus~A shows a dominant episode of early star formation accounting for over 60\% of the total stars formed followed by relatively constant star formation after 10 Gyr ago.

\paragraph[KDG~215]{KDG~215:} \citet{Cannon:2018} report shallow \hubble\ observations (F606W, F814W) of KDG~215 (D $\sim$ 5 Mpc) reaching $\sim$ 2.5 mag.\ below the TRGB.  The derived lifetime SFH indicates that roughly 60\% of the stars in KDG~215 have been formed in the last Gyr, starting with a dominant burst 1 Gyr ago.  This represents an unusually delayed bulk of star formation, and, as such, deeper imaging would be valuable.

\paragraph[DDO~68]{DDO~68:} \citet{Sacchi:2016} used \hubble/ACS observations (F606W, F814W) of DDO~68 (D $\sim$ 12 Mpc) to produce a CMD reaching roughly 1 mag.\ below the TRGB.  The galaxy was divided into 4 regions, starting with a central higher surface density region and expanding to more external regions.  SFHs for all four regions were derived for the associated CMDs.  All four regions are consistent with dominant early star formation which decreases with time until a major event started $\sim$ 200 Myr ago.  The enhanced recent star formation is suggested to be a result of a merger.

\paragraph[I~Zw~18]{I~Zw~18:} I~Zw~18 is a particularly well known starburst galaxy because of its very low metallicity \citep{Sargent:1970, Searle:1972}, which held the record for the lowest measured nebular chemical abundances for a long time.  Based on its radial velocity, it was thought to be about 10 Mpc distant.  However, the breakthrough study by \citet{Aloisi:2007}, based on 12 epochs of \hubble/ACS (F606W, F814W) observations, identified 3 Cepheid variable stars and evidence of a TRGB and provided a secure distance of $\sim$ 18 Mpc.  The new distance clarified many issues, especially the degree to which I~Zw~18 could be considered a ``young'' galaxy \citep[see additional discussion in][]{Contreras:2011, Annibali:2013}, but it also meant that a detailed lifetime SFH was beyond the reach of \hubble.  New \jwst/NIRCam observations \citep[F115W, F200W, F356W, F444W][]{Hirschauer:2024} presented and analyzed by \citet{Bortolini:2024b} have improved our view of I~Zw~18, but still only reach 1 magnitude below the TRGB, so the improvement in our knowledge of the lifetime SFH is still quite limited. However, the new \jwst\ observations yield deeper insights into the recent SFH and strengthen the hypothesis that the current starburst in I~Zw~18 has been triggered by its interaction with its ``Component C'' companion.  Note that the new \jwst\ observations represent only $\sim$ 40 minutes of integration and provide just a glimpse of \jwst's capabilities.

\subsubsection{Massive Galaxies}

\paragraph[ANGST]{ANGST:}
The ANGST survey (see Appendix~\ref{sec:important_obs}) also observed a handful of spiral galaxies. Although the survey was designed to observe radial strips in all the spiral galaxies outside of the LG and inside of 4 Mpc, the 2007 failure of the \hubble/ACS resulted in limited observations of the nearby spirals.  The original design goal was to have shallow imaging in the central high surface brightness regions (which are limited in depth by crowding) and deep imaging in the outer disks.  This was achieved in M81 \citep{Williams:2009} and NGC~2976 \citep{Williams:2010}.   The imaging in the outer deep field in M81 reached to one magnitude below the RC; thus, the lifetime SFH of this field is very coarsely sampled in time, but overall is consistent with a gradually declining SFR.  An interesting result is that modeling of the RC and AGB bump indicates that the metallicity increased rapidly to values of one-fifth to one-third of the solar value.
The imaging of NGC~2976 consisted of a shallow field covering the main body of the galaxy and a contiguous outer field which reached just below the RC.  Coarse modeling of the SFH in four different radial fields showed that the star formation was truncating with time from the outside in.

NGC~300 was another ANGST target, but the ACS failure occurred before the deep outer field was obtained so there is only shallow imaging covering the main body of the galaxy.
\citet{Gogarten:2010} divided these observations into six radial bins and found that the outer regions have a higher percentage of younger stars, consistent with inside-out growth. Interestingly, the disk scale length is relatively constant with time, in contrast to the factor of two growth seen in M~33, its morphological twin \citep{Williams:2009a}.

NGC~2403 is similarly a twin to M~33 and was also studied as part of the ANGST program.  \cite{Williams:2013} used deep \hubble/WFPC2 observations (F606W, F814W) of an outer field, combined with a collection of shallow \hubble/ACS observations (primarily F606W, F814W) to study variations in SFH as a function of radius.  Perhaps surprisingly, there is very little evidence of a radial dependence in the SFH.  At all distances from the center, the SFH is well characterized by a dominant early episode of star formation forming $\sim$ 50\% to 70\% of the total stellar mass, followed by relative quiescence until $\sim$ 5 Gyr ago, followed by relatively constant star formation up until the present.  NGC~2403 is compared to M~33 and NGC~300, and found to be very similar to NGC~300.  Both are dissimilar from M~33 in that M~33 shows a significant gradient in its SFH.  The more centrally concentrated star formation in M~33 was speculated to be due to M~33's environment with close proximity to the other large galaxies in the LG in contrast to the relative isolation of the other two galaxies.

An overall goal of the ANGST survey was to determine the cumulative SFH of the 4 Mpc volume surrounding the LG.  \citet{Williams:2011a} attempted that with the observations in hand.  As shown in Figure~\ref{fig:angst_disk_sfhs}, they determined that the large spiral galaxies (e.g., M81, NGC~253, etc.) dominated the star formation in this volume at all times, and that the contribution from the dwarfs is only becoming significant ($\ge$ 5\%) in the last Gyr.  An expansion to more galaxies (e.g., Cen~A, dozens of newly discovered dwarfs, large volumes) will help to provide a more complete picture of stellar mass assembly history of the local Universe.

\paragraph[LEGUS]{LEGUS:} This survey (see Appendix~\ref{sec:important_obs}) also targeted some more massive galaxies.
\citet{Sacchi:2019} studied the spiral galaxy NGC~7793 in the Sculptor Group (D $\sim$ 3.7 Mpc).  They combined new \hubble/UVIS observations (F275W, F336W, F438W, F555W, F814W) with archival ACS observations (F555W, F814W) to create a lifetime SFH.  Because the photometry did not reach down to the oMSTO, the SFH does not resolve the oldest star formation and a single bin is used for the period older than 2 Gyr.  This showed that the average SFR in the oldest bin for the inner galaxy was significantly higher than in any of the more recent time bins.  The middle and outer annuli show a decreased importance of early star formation and peak rates of star formation in the most recent time bins.  Interestingly, this trend of inside-out growth does not extend to a sparsely sampled outermost annulus, but the authors are less confident of the results in this region.

In a more recent LEGUS study, \citet{Bortolini:2024} observed the peculiar dwarf spiral galaxy NGC~5474 with \hubble/ACS (F606W, F814W).  From a CMD that reaches $\sim$ 2 magnitudes below the TRGB, they derive SFHs for the off-center bulge, the disk, and a SW extension.  Due to the depth of photometry, the SFHs resolve the last Gyr and the older star formation is covered with a single time bin.  All three components show an enhanced star formation event around 10--35 Myr ago.  The outer two components show enhanced star formation $\sim100$~Myr ago.  Taken together, the evidence suggests that the peculiar nature of NGC~5474 is due to a separate dwarf galaxy orbiting NGC~5474, which gives the appearance of an off-center bulge, while features in the SFH are consistent with an interaction with M~101.

\paragraph[M81]{M81:}
In addition to the ANGST SFH work cited above,
\citet{Choi:2015} used archival \hubble\ (F435W, F606W) imaging of M81 to focus on the question of whether there is a detectable age gradient across a spiral arm in M81.  Assessing results for 20 fields spanning the spiral arm as defined by the \hi{} distribution, no evidence was found of an age gradient.  Thus, the presence of a stationary density wave was discounted in favor of a kinematic spiral pattern driven by galaxy-galaxy interaction.

\paragraph[M82]{M82:}

M82 hosts the prototype starburst in the LV. Although visually spectacular for that reason, it presents a real challenge for resolved star photometry.  In addition to the challenge of distance (D $\sim$ 3.6 Mpc), the combination of the edge-on orientation of its high surface brightness disk and its high dust content make resolved star photometry in the optical prohibitive.  In the ANGST survey, \citet{Dalcanton:2009} used \hubble/ACS (F435W, F555W, F814W) to produce a 6 field mosaic covering the body of M82.  In the best fields, the photometry reaches 2 magnitudes below the TRGB.  Although \citet{Williams:2011a} calculated a SFH for M82 from the ANGST observations for inclusion of their census  of star formation in the LV, the SFH is coarse with significant uncertainties.  The SFH appears to be dominated by ancient star formation with the present starburst responsible for only a small fraction of the galaxy's stellar mass.

\citet{Suwannajak:2024} used archival \hubble/ACS observations (F555W, F814W) and the ANGST mosaic of an arc in the southwest halo of M82 to discover the presence of young ($\sim$ 100 Myr), metal-enhanced (relative to the halo, <[M/H]> $\sim$ $-0.7$) stars.  They concluded that the stars had formed as a result of the interaction of the outflow with \hi{} clouds in the halo.  They further speculate that this could be a tidal dwarf galaxy in the making.
\citet{Rao:2025} have used new \hubble/ACS observations (F475W, F814W) centered on the Southern Arcs and the ANGST mosaic of M82 to confirm the star formation in its galactic wind. Young stars ($\sim$ 100 Myr) identified from their positions in the CMD are found spread throughout the halo of M82, with a strong concentration in the Southern Arcs of M82's galactic outflow. The recent SFH shows a strong peak in star formation which began 150 Myr ago with a duration of $\sim$ 100 Myr and a second peak which began $\sim$ 30 Myr ago. They discuss two scenarios, shocks in the outflow leading to star formation and dense gas in the outflows leading naturally to star formation.   They dismiss the tidal dwarf galaxy formation scenario based on the relative lack of \hi{} emission.
\citet{Rao:2026} present new, deep \hubble/ACS halo field observations (F475W, F814W) concentrating on the M82 Tail.  Based on the derived SFH for the Tail, it is concluded that it was likely born from clouds of stripped M82 gas.

Fortunately, \jwst\ is dramatically improving our understanding of M82.  There have now been numerous \jwst\ observations of M82 \citep[e.g.,][]{Bolatto:2024}, but, with regard to SFHs, the observations from \jwst-GO-5145 (PI: Smercina) promise to finally provide the long-needed quality of photometry (improved angular resolution, less sensitivity to dust extinction).

\paragraph[Centaurus~A]{Centaurus~A:}

As the nearest giant elliptical galaxy and the host for an active galactic nucleus \citep[e.g.,][]{Ebneter:1983, Harris:2010, DSouza:2018b}, NGC~5128 represents a tremendous observational opportunity at a distance just shy of 4 Mpc, i.e., just on the edge of what is possible for deep resolved stellar population work.  It is highlighted in Appendix~\ref{sec:individual_galaxies} of this review.
\citet{Rejkuba:2011}, using \hubble/ACS observations (F606W, F814W) of an outer halo field first presented in \citet{Rejkuba:2005}, explored the questions of whether NGC~5128 contained stars dating back to the earliest age of the universe and how extended was the duration of its star formation.   From matching
stellar evolution models to the CMD (in particular, alpha-enhanced stellar models) they conclude that the bulk of star formation in the observed halo field must be at least 10 Gyr old.  There is also evidence for a minor component of star formation in the range of 2--4 Gyr ago.

\citet{Rejkuba:2022} combined 9 halo fields of archival \hubble observations (all F606W, F814W) of NGC~5128 to present a comprehensive view of the history of its halo. While the emphasis of this work is the radial structure of NGC~5128, they find evidence in the inner halo (R $\le$ 30 kpc) of star formation 2 -- 3 Gyr ago.

In a complementary approach, \citet{Aghdam:2024} use ESO/VLT near-IR observations (J$_s$, H, K$_s$) of the LPVs in two halo fields of NGC~5128 first presented by \citet{Rejkuba:2003}, supplemented by ESO/NTT observations of the second field.  Almost 400 LPVs are identified in field 1 and almost 700 in field 2.   Using the methodology developed by \citet{Javadi:2011}, they find similar SFHs for the two halo fields.  Comparing to models over a decade range in metallicity, they detected enhanced star formation $\sim$ 800 Myr and $\sim$ 3.8 Gyr ago. Combined with the results from  \citet{Rejkuba:2011, Rejkuba:2022}, there appears to be substantial evidence for recent ($\le$ 4 Gyr ago) enhanced star formation in NGC~5128, which is likely the result of a minor merger event.

\section{Compilation of Published CMD-based Star Formation Histories}
\label{sec:sfh_table}

\begin{landscape}
\setlength{\LTcapwidth}{0.8\linewidth}
\begin{longtable}{llc>{\footnotesize}p{10cm}}
\caption{A representative list of published lifetime SFHs of nearby galaxies. To formulate this list, we required the SFH to (i) be the result of quantitative modeling of resolved-star CMDs or luminosity functions; (ii) include age information back to the approximate age of the Universe, even if older age information is coarse (i.e., not just a recent SFH); (iii) have been published within the last $\sim20$~years. In the Environment column, MW and M31 denote satellites of those hosts (i.e., within 300~kpc; \S \ref{sec:concepts_classification}), and entries beginning with Iso denote isolated galaxies.    Iso/MW and Iso/M31 denote isolated galaxies listed with the MW or M31 system because of their proximity to that host. \label{tab:sfh_compilation} } \\
\hline
\textbf{Galaxy} & \textbf{Environment} & \textbf{No. of SFHs} & \textbf{\normalsize Publications} \\
\hline
\endfirsthead

\multicolumn{4}{c}%
{{\bfseries \tablename\ \thetable{} -- continued from previous page}} \\
\hline
\textbf{Galaxy} & \textbf{Environment} & \textbf{No. of SFHs} & \textbf{\normalsize Publications} \\
\hline
\endhead

\hline \multicolumn{4}{r}{{Continued on next page}} \\
\endfoot

\hline \hline
\endlastfoot

\multicolumn{4}{c}{\textbf{MW System}} \\
\hline
BooI & MW & 3 & \citet{deJong:2008}, \citet{Brown:2014}, \citet{Durbin:2025} \\
BooII & MW & 2 & \citet{deJong:2008}, \citet{Durbin:2025} \\
CardSph & MW & 4 & \citet{Small:2013}, \citet{deBoer:2014}, \citet{Weisz:2014a}, \citet{Santana:2016} \\
CetII & MW & 1 & \citet{Durbin:2025} \\
CoBeI & MW & 3 & \citet{deJong:2008}, \citet{Brown:2014}, \citet{Durbin:2025} \\
ColI & MW & 1 & \citet{Durbin:2025} \\
CVnI & MW & 3 & \citet{deJong:2008}, \citet{Weisz:2014a}, \citet{Perez-Garcia:2026} \\
CVnII & MW & 4 & \citet{deJong:2008}, \citet{Brown:2014}, \citet{Weisz:2014a}, \citet{Durbin:2025} \\
DradSph & MW & 1 & \citet{Weisz:2014a} \\
EriII & Iso/MW & 4 & \citet{Gallart:2021}, \citet{Simon:2021}, \citet{Weisz:2023b}, \citet{Durbin:2025} \\
EriIII & MW & 1 & \citet{Durbin:2025} \\
FnxdSph & MW & 4 & \citet{delPino:2013}, \citet{deBoer:2012b}, \citet{Weisz:2014a}, \citet{Rusakov:2021} \\
GruI & MW & 1 & \citet{Durbin:2025} \\
GruII & MW & 1 & \citet{Durbin:2025} \\
Her & MW & 4 & \citet{deJong:2008}, \citet{Brown:2014}, \citet{Weisz:2014a}, \citet{Durbin:2025} \\
HorI & MW & 3 & \citet{Sacchi:2021}, \citet{Garling:2025}, \citet{Durbin:2025} \\
HorII & MW & 1 & \citet{Durbin:2025} \\
HyaII & MW & 2 & \citet{Sacchi:2021}, \citet{Durbin:2025} \\
HydI & MW & 1 & \citet{Durbin:2025} \\
LeoI & MW & 2 & \citet{Weisz:2014a}, \citet{Ruiz-Lara:2021} \\
LeoII & MW & 1 & \citet{Weisz:2014a} \\
LeoIV & MW & 4 & \citet{deJong:2008}, \citet{Brown:2014}, \citet{Weisz:2014a}, \citet{Durbin:2025} \\
LeoV & MW & 1 & \citet{Durbin:2025} \\
LeoK & Iso/MW & 1 & \citet{McQuinn:2024c} \\
LeoT & Iso/MW & 5 & \citet{deJong:2008}, \citet{deJong:2008b}, \citet{Weisz:2012b}, \citet{Clementini:2012}, \citet{Weisz:2014a} \\
LeoM & Iso/MW & 1 & \citet{McQuinn:2024c} \\
LMC & MW & 7 & \citet{Harris:2009}, \citet{Rubele:2012}, \citet{Weisz:2013b}, \citet{Monteagudo:2018}, \citet{Ruiz-Lara:2020b}, \citet{Mazzi:2021}, \citet{Cohen:2024a} \\
NGC6822 & Iso/MW & 5 & \citet{McQuinn:2010a}, \citet{Cannon:2012}, \citet{Weisz:2014a}, \citet{YangY:2024}, \citet{Cohen:2026b} \\
PegIII & MW & 1 & \citet{Durbin:2025} \\
Phx & Iso/MW & 5 & \citet{Young:2007}, \citet{Hidalgo:2009}, \citet{Weisz:2014a}, \citet{YangY:2024}, \citet{Cohen:2026b} \\
PhxII & MW & 2 & \citet{Sacchi:2021}, \citet{Durbin:2025} \\
PicI & MW & 1 & \citet{Durbin:2025} \\
PscII & MW & 1 & \citet{Durbin:2025} \\
RetII & MW & 3 & \citet{Sacchi:2021}, \citet{Simon:2023}, \citet{Durbin:2025} \\
RetIII & MW & 1 & \citet{Durbin:2025} \\
ScldSph & MW & 4 & \citet{deBoer:2012}, \citet{Weisz:2014a}, \citet{Savino:2018}, \citet{Bettinelli:2019} \\
SexdSph & MW & 2 & \citet{Lee:2009}, \citet{Bettinelli:2018} \\
SagdSph & MW & 3 & \citet{Siegel:2007}, \citet{Weisz:2014a}, \citet{deBoer:2015} \\
SgrII & MW & 2 & \citet{Sacchi:2021}, \citet{Durbin:2025} \\
Segue1 & MW & 2 & \citet{deJong:2008}, \citet{Durbin:2025} \\
Segue2 & MW & 1 & \citet{Durbin:2025} \\
SMC & MW & 5 & \citet{Weisz:2013b}, \citet{Rubele:2015}, \citet{Rubele:2018}, \citet{Massana:2022}, \citet{Cohen:2024b} \\
TriII & MW & 2 & \citet{Sacchi:2021}, \citet{Durbin:2025} \\
TucII & MW & 2 & \citet{Sacchi:2021}, \citet{Durbin:2025} \\
TucIII & MW & 1 & \citet{Durbin:2025} \\
TucIV & MW & 1 & \citet{Durbin:2025} \\
TucV & MW & 1 & \citet{Durbin:2025} \\
UMaI & MW & 3 & \citet{deJong:2008}, \citet{Brown:2014}, \citet{Durbin:2025} \\
UMaII & MW & 2 & \citet{deJong:2008}, \citet{Durbin:2025} \\
UMidSph & MW & 3 & \citet{Weisz:2014a}, \citet{YangY:2024}, \citet{Sato:2025} \\
VirI & MW & 1 & \citet{Durbin:2025} \\
Wil1 & MW & 2 & \citet{deJong:2008}, \citet{Durbin:2025} \\
\hline

\multicolumn{4}{c}{\textbf{M31 System}} \\
\hline
M31 & M31 & 8 & \citet{Brown:2006}, \citet{Brown:2007}, \citet{Brown:2008}, \citet{Bernard:2012}, \citet{Bernard:2015a}, \citet{Bernard:2015b}, \citet{Williams:2015}, \citet{Williams:2017} \\
M32 & M31 & 3 & \citet{Monachesi:2012}, \citet{Weisz:2014a}, \citet{Savino:2025} \\
M33 & M31 & 6 & \citet{Barker:2007a}, \citet{Williams:2009a}, \citet{Barker:2011}, \citet{Bernard:2012}, \citet{Javadi:2017}, \citet{Savino:2025} \\
AndI & M31 & 4 & \citet{Weisz:2014a}, \citet{Skillman:2017}, \citet{Saremi:2021}, \citet{Savino:2025} \\
AndII & M31 & 4 & \citet{Weisz:2014a}, \citet{Weisz:2014c}, \citet{Skillman:2017}, \citet{Savino:2025} \\
AndIII & M31 & 3 & \citet{Weisz:2014a}, \citet{Skillman:2017}, \citet{Savino:2025} \\
AndV & M31 & 2 & \citet{Weisz:2014a}, \citet{Savino:2025} \\
AndVI & M31 & 3 & \citet{Weisz:2014a}, \citet{Makarova:2021}, \citet{Savino:2025} \\
AndVII & M31 & 3 & \citet{Weisz:2014a}, \citet{Navabi:2021}, \citet{Savino:2025} \\
AndIX & M31 & 3 & \citet{Abdollahi:2023}, \citet{Weisz:2019}, \citet{Savino:2025} \\
AndX & M31 & 2 & \citet{Weisz:2019}, \citet{Savino:2025} \\
AndXI & M31 & 4 & \citet{Weisz:2014a}, \citet{Weisz:2019}, \citet{Savino:2023}, \citet{Savino:2025} \\
AndXII & M31 & 4 & \citet{Weisz:2014a}, \citet{Weisz:2019}, \citet{Savino:2023}, \citet{Savino:2025} \\
AndXIII & M31 & 4 & \citet{Weisz:2014a}, \citet{Weisz:2019}, \citet{Savino:2023}, \citet{Savino:2025} \\
AndXIV & M31 & 2 & \citet{Weisz:2019}, \citet{Savino:2025} \\
AndXV & M31 & 2 & \citet{Skillman:2017}, \citet{Savino:2025} \\
AndXVI & M31 & 4 & \citet{Weisz:2014c}, \citet{Monelli:2016}, \citet{Skillman:2017}, \citet{Savino:2025} \\
AndXVII & M31 & 2 & \citet{Weisz:2019}, \citet{Savino:2025} \\
AndXIX & M31 & 2 & \citet{Collins:2022}, \citet{Savino:2025} \\
AndXX & M31 & 3 & \citet{Weisz:2019}, \citet{Savino:2023}, \citet{Savino:2025} \\
AndXXI & M31 & 2 & \citet{Weisz:2019}, \citet{Savino:2025} \\
AndXXII & M31 & 3 & \citet{Weisz:2019}, \citet{Savino:2023}, \citet{Savino:2025} \\
AndXXIII & M31 & 2 & \citet{Weisz:2019}, \citet{Savino:2025} \\
AndXXIV & M31 & 2 & \citet{Weisz:2019}, \citet{Savino:2025} \\
AndXXV & M31 & 2 & \citet{Weisz:2019}, \citet{Savino:2025} \\
AndXXVI & M31 & 3 & \citet{Weisz:2019}, \citet{Savino:2023}, \citet{Savino:2025} \\
AndXXVIII & M31 & 3 & \citet{Skillman:2017}, \citet{Makarova:2021}, \citet{Savino:2025} \\
AndXXIX & M31 & 3 & \citet{Weisz:2019}, \citet{Makarova:2021}, \citet{Savino:2025} \\
AndXXX & M31 & 2 & \citet{Weisz:2019}, \citet{Savino:2025} \\
AndXXXI & M31 & 2 & \citet{Weisz:2019}, \citet{Savino:2025} \\
AndXXXII & M31 & 2 & \citet{Weisz:2019}, \citet{Savino:2025} \\
AndXXXIII & M31 & 2 & \citet{Weisz:2019}, \citet{Savino:2025} \\
LGS3 & M31 & 4 & \citet{Hidalgo:2011}, \citet{Weisz:2014a}, \citet{Savino:2025}, \citet{Cohen:2026b} \\
NGC147 & M31 & 5 & \citet{Weisz:2014a}, \citet{Geha:2015}, \citet{HamedaniGolshan:2017}, \citet{Savino:2025}, \citet{Ren:2024} \\
NGC185 & M31 & 6 & \citet{Weisz:2014a}, \citet{Geha:2015}, \citet{HamedaniGolshan:2017}, \citet{Savino:2025}, \citet{Ren:2024}, \citet{YangY:2024} \\
NGC205 & M31 & 3 & \citet{Weisz:2014a}, \citet{Savino:2025}, \citet{YangY:2024} \\
IC1613 & Iso/M31 & 6 & \citet{Skillman:2014}, \citet{Weisz:2014a}, \citet{Hashemi:2019}, \citet{Savino:2025}, \citet{Ren:2024}, \citet{Cohen:2026b} \\
Pegasus & Iso/M31 & 4 & \citet{Weisz:2014a}, \citet{Savino:2025}, \citet{Ren:2024}, \citet{YangY:2024} \\
IC10 & M31 & 1 & \citet{Weisz:2014a} \\
PegW & Iso/M31 & 1 & \citet{McQuinn:2023} \\
\hline

\multicolumn{4}{c}{\textbf{LG ISOLATED}} \\
\hline
LeoA & Iso & 5 & \citet{Cole:2007}, \citet{Weisz:2014a}, \citet{Ren:2024}, \citet{Cohen:2026b}, \citet{Garling:2025b} \\
CetdSph & Iso & 2 & \citet{Monelli:2010_a}, \citet{Weisz:2014a} \\
TucdSph & Iso & 4 & \citet{Monelli:2010_b}, \citet{Savino:2019}, \citet{Weisz:2014a}, \citet{YangY:2024} \\
Aquarius & Iso & 4 & \citet{Cole:2014}, \citet{Weisz:2014a}, \citet{Cohen:2026b}, \citet{Garling:2025b} \\
WLM & Iso & 9 & \citet{Weisz:2014a}, \citet{Albers:2019}, \citet{McQuinn:2024a}, \citet{Ren:2024}, \citet{Cohen:2025}, \citet{Cohen:2026b}, \citet{Garling:2025}, \citet{Garling:2025b}, \citet{Lee:2026} \\
SagdIr & Iso & 1 & \citet{Weisz:2014a} \\
VV124 & Iso & 1 & \citet{Jacobs:2011} \\
\hline

\multicolumn{4}{c}{\textbf{$D < 2$~Mpc}} \\
\hline
NGC3109 & NGC3109 & 3 & \citet{Williams:2011a}, \citet{Weisz:2011}, \citet{Cohen:2026b} \\
Antlia & NGC3109 & 4 & \citet{McQuinn:2010a}, \citet{Williams:2011a}, \citet{Weisz:2011}, \citet{Cohen:2026b} \\
AntliaB & NGC3109 & 1 & \citet{Hargis:2020} \\
NGC55 & NGC55 & 2 & \citet{Williams:2011a}, \citet{Weisz:2011} \\
ESO410-005 & NGC55 & 3 & \citet{Williams:2011a}, \citet{Weisz:2011}, \citet{Cohen:2026b} \\
ESO294-010 & NGC55 & 3 & \citet{Williams:2011a}, \citet{Weisz:2011}, \citet{Cohen:2026b} \\
KK258 & NGC55 & 1 & \citet{Karachentsev:2014} \\
AndXVIII & Iso & 2 & \citet{Makarova:2017}, \citet{Weisz:2019} \\
SexA & Iso & 6 & \citet{Williams:2011a}, \citet{Weisz:2011}, \citet{Weisz:2014a}, \citet{Ren:2024}, \citet{YangY:2024}, \citet{Cohen:2026b} \\
SexB & Iso & 5 & \citet{Williams:2011a}, \citet{Weisz:2011}, \citet{Weisz:2014a}, \citet{Ren:2024}, \citet{Cohen:2026b} \\
LeoP & Iso & 3 & \citet{McQuinn:2015a}, \citet{McQuinn:2024b}, \citet{Garling:2025b} \\
KKR25 & Iso & 3 & \citet{Williams:2011a}, \citet{Weisz:2011}, \citet{Makarov:2012} \\
IC5152 & Iso & 3 & \citet{Williams:2011a}, \citet{Weisz:2011}, \citet{Cohen:2026b} \\
KK230 & Iso & 2 & \citet{Williams:2011a}, \citet{Weisz:2011} \\
\hline

\multicolumn{4}{c}{\textbf{$D < 5$~Mpc}} \\
\hline
NGC300 & Scl & 2 & \citet{Gogarten:2010}, \citet{Williams:2011a} \\
UGCA438 & Scl & 2 & \citet{Williams:2011a}, \citet{Weisz:2011} \\
NGC253 & Scl & 1 & \citet{Williams:2011a} \\
ESO540-032 & Scl & 3 & \citet{Williams:2011a}, \citet{Weisz:2011}, \citet{Cohen:2026b} \\
NGC625 & Scl & 1 & \citet{McQuinn:2010a} \\
NGC247 & Scl & 1 & \citet{Williams:2011a} \\
DDO6 & Scl & 2 & \citet{Williams:2011a}, \citet{Weisz:2011} \\
ESO540-030 & Scl & 2 & \citet{Williams:2011a}, \citet{Weisz:2011} \\
Scl22 & Scl & 1 & \citet{Weisz:2011} \\
NGC7793 & Scl & 1 & \citet{Sacchi:2019} \\
SDIG & Scl & 1 & \citet{Lianou:2013} \\
M81 & M81 & 2 & \citet{Williams:2009}, \citet{Williams:2011a} \\
NGC3077 & M81 & 1 & \citet{Williams:2011a} \\
KDG52 & M81 & 3 & \citet{Weisz:2008}, \citet{Williams:2011a}, \citet{Weisz:2011} \\
DDO53 & M81 & 4 & \citet{Weisz:2008}, \citet{Weisz:2011}, \citet{Williams:2011a}, \citet{Cohen:2026b} \\
HoI & M81 & 4 & \citet{Weisz:2008}, \citet{Williams:2011a}, \citet{Weisz:2011}, \citet{Cohen:2026b} \\
HoII & M81 & 4 & \citet{Weisz:2008}, \citet{McQuinn:2010a}, \citet{Williams:2011a}, \citet{Weisz:2011} \\
HoIX & M81 & 3 & \citet{Weisz:2008}, \citet{Williams:2011a}, \citet{Weisz:2011} \\
DDO165 & M81 & 4 & \citet{Weisz:2008}, \citet{McQuinn:2010a}, \citet{Weisz:2011}, \citet{Cohen:2026b} \\
IC2574 & M81 & 3 & \citet{Weisz:2008}, \citet{Williams:2011a}, \citet{Weisz:2011} \\
NGC2976 & M81 & 2 & \citet{Williams:2010}, \citet{Williams:2011a} \\
UGC4483 & M81 & 5 & \citet{McQuinn:2010a}, \citet{Williams:2011a}, \citet{Weisz:2011}, \citet{Sacchi:2021b}, \citet{Cohen:2026b} \\
NGC2366 & M81 & 5 & \citet{Weisz:2008}, \citet{McQuinn:2010a}, \citet{Williams:2011a}, \citet{Weisz:2011}, \citet{Cohen:2026b} \\
M82 & M81 & 1 & \citet{Williams:2011a} \\
KDG61 & M81 & 3 & \citet{Makarova:2010}, \citet{Williams:2011a}, \citet{Weisz:2011} \\
KDG64 & M81 & 3 & \citet{Makarova:2010}, \citet{Williams:2011a}, \citet{Weisz:2011} \\
IKN & M81 & 2 & \citet{Williams:2011a}, \citet{Weisz:2011} \\
DDO78 & M81 & 2 & \citet{Williams:2011a}, \citet{Weisz:2011} \\
A0952+69 & M81 & 2 & \citet{Williams:2011a}, \citet{Weisz:2011} \\
HS117 & M81 & 2 & \citet{Williams:2011a}, \citet{Weisz:2011} \\
DDO82 & M81 & 2 & \citet{Williams:2011a}, \citet{Weisz:2011} \\
BK3N & M81 & 2 & \citet{Williams:2011a}, \citet{Weisz:2011} \\
FM1 & M81 & 2 & \citet{Williams:2011a}, \citet{Weisz:2011} \\
KK77 & M81 & 2 & \citet{Williams:2011a}, \citet{Weisz:2011} \\
DDO44 & M81 & 2 & \citet{Williams:2011a}, \citet{Weisz:2011} \\
F8D1 & M81 & 3 & \citet{Williams:2011a}, \citet{Weisz:2011}, \citet{Smercina:2026} \\
BK5N & M81 & 2 & \citet{Williams:2011a}, \citet{Weisz:2011} \\
NGC2403 & M81 & 2 & \citet{Williams:2011a}, \citet{Williams:2013} \\
MCG9-020-131 & M81 & 1 & \citet{Williams:2011a} \\
KDG73 & M81 & 1 & \citet{Weisz:2011} \\
KDG63 & M81 & 2 & \citet{Williams:2011a}, \citet{Weisz:2011} \\
NGC4214 & M94 & 4 & \citet{McQuinn:2010a}, \citet{Williams:2011a}, \citet{Weisz:2011}, \citet{Williams:2011b} \\
NGC4163 & M94 & 4 & \citet{McQuinn:2010a}, \citet{Williams:2011a}, \citet{Weisz:2011}, \citet{Cohen:2026b} \\
NGC4449 & M94 & 1 & \citet{McQuinn:2010a} \\
DDO99 & M94 & 3 & \citet{Williams:2011a}, \citet{Weisz:2011}, \citet{Cohen:2026b} \\
DDO125 & M94 & 3 & \citet{Williams:2011a}, \citet{Weisz:2011}, \citet{Cohen:2026b} \\
UGC8833 & M94 & 2 & \citet{Williams:2011a}, \citet{Weisz:2011} \\
DDO113 & NGC4214 & 2 & \citet{Williams:2011a}, \citet{Weisz:2011} \\
UGCA292 & NGC4214 & 3 & \citet{Williams:2011a}, \citet{Weisz:2011}, \citet{Cohen:2026b} \\
CenA & CenA & 3 & \citet{Rejkuba:2011}, \citet{Rejkuba:2022}, \citet{Aghdam:2024} \\
ESO321-014 & CenA & 2 & \citet{Williams:2011a}, \citet{Weisz:2011} \\
ESO325-011 & CenA & 1 & \citet{Weisz:2011} \\
ESO383-087 & CenA & 1 & \citet{Weisz:2011} \\
NGC5253 & M83 & 1 & \citet{McQuinn:2010a} \\
ESO269-037 & NGC4945 & 1 & \citet{Williams:2011a} \\
NGC1569 & IC342 & 2 & \citet{McQuinn:2010a}, \citet{Grocholski:2012} \\
KKH37 & IC342 & 2 & \citet{Williams:2011a}, \citet{Weisz:2011} \\
UGCA92 & IC342 & 1 & \citet{Makarova:2012} \\
Pavo & Iso & 1 & \citet{Mutlu:2026}\\
Corvus A & Iso & 1 & \citet{Mutlu:2026}\\
KKH98 & Iso & 3 & \citet{Melbourne:2010}, \citet{Williams:2011a}, \citet{Weisz:2011} \\
ESO006-001 & Iso & 1 & \citet{Makarova:2023} \\
NGC404 & Iso & 1 & \citet{Williams:2010b} \\
UGC6456 & Iso & 1 & \citet{McQuinn:2010a} \\
NGC6789 & Iso & 1 & \citet{McQuinn:2010a} \\
NGC4068 & Iso & 2 & \citet{McQuinn:2010a}, \citet{Cohen:2026b} \\
IC4662 & Iso & 2 & \citet{McQuinn:2010a}, \citet{Cohen:2026b} \\
GR8 & Iso & 3 & \citet{Williams:2011a}, \citet{Weisz:2011}, \citet{Cohen:2026b} \\
DDO187 & Iso & 4 & \citet{McQuinn:2010a}, \citet{Williams:2011a}, \citet{Weisz:2011}, \citet{Cohen:2026b} \\
DDO181 & Iso & 3 & \citet{Williams:2011a}, \citet{Weisz:2011}, \citet{Cohen:2026b} \\
NGC3741 & Iso & 3 & \citet{Williams:2011a}, \citet{Weisz:2011}, \citet{Cohen:2026b} \\
DDO190 & Iso & 3 & \citet{Williams:2011a}, \citet{Weisz:2011}, \citet{Cohen:2026b} \\
UGC8508 & Iso & 3 & \citet{Williams:2011a}, \citet{Weisz:2011}, \citet{Cohen:2026b} \\
DDO183 & Iso & 2 & \citet{Williams:2011a}, \citet{Weisz:2011} \\
KKH86 & Iso & 2 & \citet{Williams:2011a}, \citet{Weisz:2011} \\
NGC4190 & Iso & 2 & \citet{Williams:2011a}, \citet{Cohen:2026b} \\
UGC685 & Iso & 1 & \citet{Cohen:2026b} \\
DDO154 & M64 & 1 & \citet{Cohen:2026b} \\
UGC8638 & Iso & 1 & \citet{Cohen:2026b} \\
\hline

\multicolumn{4}{c}{\textbf{$D < 10$~Mpc}} \\
\hline
NGC784 & NGC784 & 2 & \citet{McQuinn:2010a}, \citet{Cohen:2026b} \\
KK13 & NGC672 & 1 & \citet{McQuinn:2015c} \\
KK17 & NGC784 & 1 & \citet{McQuinn:2015c} \\
KK15 & NGC672 & 1 & \citet{McQuinn:2015c} \\
KK16 & NGC784 & 1 & \citet{McQuinn:2015c} \\
AGC112521 & NGC672 & 1 & \citet{McQuinn:2015c} \\
KDG215 & M64 & 1 & \citet{Cannon:2018} \\
NGC5474 & M101 & 1 & \citet{Bortolini:2024}\\
HoIV & M101 & 1 & \citet{Cohen:2026b} \\
Kamino & Iso & 1 & \citet{Mutlu:2026}\\
CVnC & Iso & 1 & \citet{Hai:2026} \\
ESO154-023 & Iso & 1 & \citet{McQuinn:2010a} \\
AGC174585 & Iso & 1 & \citet{McQuinn:2015c} \\
AGC182595 & Iso & 1 & \citet{McQuinn:2015c} \\
AGC748778 & Iso & 1 & \citet{McQuinn:2015c} \\
AGC749241 & Iso & 1 & \citet{McQuinn:2015c} \\
UGC1281 & Iso & 1 & \citet{Cohen:2026b} \\
NGC3738 & Iso & 1 & \citet{Cohen:2026b} \\
UGCA281 & Iso & 1 & \citet{Cohen:2026b} \\
UGCA320 & Iso & 1 & \citet{Cohen:2026b} \\
\hline

\multicolumn{4}{c}{\textbf{$D > 10$~Mpc}} \\
\hline
AGC174605 & Iso & 1 & \citet{McQuinn:2015c} \\
AGC731457 & Iso & 1 & \citet{McQuinn:2015c} \\
AGC749237 & Iso & 1 & \citet{McQuinn:2015c} \\
SBS1415+437 & Iso & 1 & \citet{McQuinn:2010a} \\
IZw18 & Iso &1 & \citet{Bortolini:2024b}\\
DDO68 & Iso & 1 & \citet{Sacchi:2016} \\
\end{longtable}
\end{landscape}

\bibliographystyle{cas-model2-names}

\bibliography{citations,telescope_table_citations}

\end{document}